\documentclass[11pt]{article}
\usepackage[T1]{fontenc}
\usepackage[utf8]{inputenc}
\usepackage{lmodern}
\usepackage[a4paper,margin=23mm]{geometry}
\usepackage{amsmath,amssymb,amsthm,mathtools}
\usepackage{booktabs,longtable,array,graphicx}
\usepackage{xcolor,xurl}
\usepackage{xr-hyper}
\usepackage[colorlinks=true,linkcolor=blue!45!black,citecolor=blue!45!black,urlcolor=blue!45!black]{hyperref}
\newtheorem{theorem}{Theorem}
\newtheorem{lemma}[theorem]{Lemma}
\newtheorem{proposition}[theorem]{Proposition}
\newtheorem{corollary}[theorem]{Corollary}

\theoremstyle{remark}
\newtheorem{remark}[theorem]{Remark}
\allowdisplaybreaks
\newcommand{\Psiico}{\Psi_{\mathrm{ico}}}
\newcommand{\Sussex}{Department of Physics \& Astronomy, University of Sussex, Brighton BN1 9QH, UK}
\newcommand{\affiliation}[1]{\\[2pt]{\small\normalfont #1}}

\hypersetup{pdfauthor={Antony Lewis},pdftitle={The most entangled states of four and seven qubits}}
\title{The most entangled states\\of four and seven qubits}
\author{\href{https://orcid.org/0000-0001-5927-6667}{Antony Lewis} \affiliation{\Sussex}}
\date{}
\begin{document}
\maketitle
\begin{abstract}
Pure states of four and seven qubits cannot be maximally entangled across every balanced bipartition; we determine
how entangled they can be on average. For four qubits we prove the conjecture of Higuchi and Sudbery: the mean von
Neumann entropy of the three $2|2$ cuts is at most $1+\tfrac12\log_2 3$ bits, with equality exactly on the
local-unitary orbits of their state $M_4$ and its complex conjugate. For seven qubits the mean purity of the 35
balanced $3|4$ cuts is at least $121/917$ and their mean entropy is at most $2.965356029\ldots$ bits, each with a
unique optimiser up to local unitaries and qubit permutations. The two optimisers are distinct members of one
icosahedral family, so minimum purity and maximum entropy select different states.

The proofs bound moments of the marginal spectra by positive identities on several copies of the state, some of them
spin flipped, and bound the entropy by a polynomial in those moments. The four-qubit maximum has an analytic proof; the
seven-qubit proofs, and some of the four-qubit refinements below, are computer-assisted, with exact arithmetic and
rigorous error bounds. We also prove the
Gour--Wallach conjecture for R\'enyi entropies of all orders, with the maximiser changing from $M_4$ to cluster
states at $q_*\approx1.588$, and obtain sharp Tsallis bounds, mixed-state bounds and quantitative stability.

\end{abstract}

\section{Introduction}\label{sec:fqe-intro}

Multipartite entanglement requires a compromise: a state maximally entangled across one division of its qubits
need not be maximally entangled across another. A state that achieves maximal entanglement across every
bipartition is called \emph{absolutely maximally entangled} (AME). Such states are closely related to
quantum error-correcting codes~\cite{Scott} and ``perfect'' tensors~\cite{PYHP}; Ref.~\cite{AMEreview} reviews them. They exist for two, three, five and six qubits,
but not for four~\cite{HS,HSi26}, seven~\cite{HGS}, or eight or more~\cite{Scott}.
Four and seven qubits are therefore the first two sizes at which no state can maximise
entanglement across every cut simultaneously; at most four of the six two-qubit marginals, and at most 32 of the 35
three-qubit marginals, can be maximally mixed~\cite{HS,HGS,ZNSZ}. Balanced cuts expose this obstruction: if all
balanced marginals are maximally mixed, their smaller reductions are too. Conversely,
states with all one- and two-qubit marginals maximally mixed exist for every $n\ge5$~\cite{Scott},
so small marginals alone do not reveal the seven-qubit obstruction. We determine the greatest
mean entropy across balanced cuts at both sizes; minimisers of the mean balanced purity are the maximally multipartite
entangled states of Facchi et al.~\cite{FFPP}, and highly entangled states have been sought numerically since
Refs.~\cite{BSSB,Borras}. At four qubits this is the question posed by Higuchi and Sudbery in ``How entangled can two couples get?''~\cite{HS}.

Label the qubits $1,2,3,4$. There are three ways to divide them into two
groups of two:
\[
12|34,\qquad 13|24,\qquad 14|23.
\]
A \emph{bipartition}, or \emph{cut}, is balanced when its two groups are as close in size as possible:
$2|2$ here, and $3|4$ for seven qubits. For a pure state $\psi$, the reduced density matrix
$\rho_{12}=\operatorname{tr}_{34}|\psi\rangle\langle\psi|$ describes qubits $1,2$ alone; define
$\rho_{13}$ and $\rho_{14}$ similarly. Its von Neumann entropy
\[
S(\rho)=-\operatorname{tr}\rho\log_2\rho
\]
measures the entanglement across the cut. Operationally, it is the
number of Bell pairs per copy that the two sides can distil from many copies of the pure state, or need to prepare those copies,
using local operations and classical communication~\cite{BBPS}. The complementary pair has the same entropy, so each cut
is counted once. A pair of qubits
has a four-dimensional state space, so $S\le2$ bits, with equality when $\rho$ is maximally mixed,
$I_4/4$, with four equal eigenvalues $1/4$. Entropies are in bits throughout, except in the Tsallis
corollary and in proofs that use $\ln$, which say so.

We give the three cuts equal weight and study the average
\[
E(\psi)=\frac13\bigl[S(\rho_{12})+S(\rho_{13})+S(\rho_{14})\bigr]
       =\frac13\sum_{j=2}^{4}S(\rho_{1j}).
\]
It is the entanglement, in Bell pairs, available across a balanced cut chosen at random, and it is unchanged by
relabelling the qubits. Reaching $E=2$ would require all three reduced states to be
maximally mixed, hence an AME state. Their compatibility---all three come from one wavefunction---prevents
this~\cite{HS}: three maximally mixed cuts would give total purity $P=\sum_j\operatorname{tr}\rho_{1j}^2=3\cdot\tfrac14$,
whereas every pure four-qubit state has $P\ge1$ (Lemma~\ref{lem:fqe-purity}). The best average is
therefore strictly below $2$.

\paragraph{The two main results.} Higuchi and Sudbery (HS) exhibited the state
\[
M_4=\tfrac1{\sqrt6}\bigl(|0011\rangle+|1100\rangle+\omega(|0101\rangle+|1010\rangle)
+\omega^2(|0110\rangle+|1001\rangle)\bigr),\qquad\omega=e^{2\pi i/3},
\]
whose amplitudes sit on the six strings of Hamming weight two, grouped into complementary pairs, one for each
pairing of the qubits, with phases $1,\omega,\omega^2$. An even permutation of the qubits permutes the three pairings
cyclically (or fixes each of them) and only multiplies $M_4$ by a phase (Remark~\ref{rem:fqe-single-orbit}), so the three cuts are
identical: each reduced state has eigenvalues
$(\tfrac12,\tfrac16,\tfrac16,\tfrac16)$. Every cut then has the same entropy, and
$E(M_4)=1+\tfrac12\log_2 3\approx1.792$. They conjectured that this is the maximum~\cite{HS}. Brierley and Higuchi
proved that $M_4$ is a local maximum~\cite{BH}. Derksen et al.\ proved that $M_4$ maximises a different measure, the
geometric measure of entanglement (distance to the nearest product state)~\cite{DFLW}. Gour and Wallach proved the corresponding Tsallis bound among
\emph{critical} states, those whose one-qubit marginals are all maximally mixed, and conjectured the R\'enyi envelope~\cite{GW}; the precise scope of these results is at the end of this section.
Independent single-qubit unitaries are called \emph{local unitaries} (LU). Two states are \emph{LU-equivalent} if
local unitaries map one to the other up to an overall phase, and the states equivalent to a given one form its
\emph{LU orbit}. Since these operations preserve cut entropies, maximisers occur in whole LU orbits.

\medskip\noindent\emph{Theorem~\ref{thm:fqe-entropy-max} (four qubits). For every pure state of four
qubits, $E(\psi)\le1+\tfrac12\log_2 3$. Equality holds if and only if $\psi$ is equivalent to $M_4$ or
to its complex conjugate $\overline{M_4}$ by independent single-qubit unitaries.}
\medskip

At seven qubits a balanced cut puts three qubits on one side, and there are $\binom73=35$ such cuts, one for each
set $S$ of three qubits with reduced state $\rho_S$. Write
$\pi_3=\frac1{35}\sum_{|S|=3}\operatorname{tr}\rho_S^2$ and $E_3=\frac1{35}\sum_{|S|=3}S(\rho_S)$ for the mean
purity and mean entropy of the 35 cuts. An AME state
would have $\pi_3=1/8$ and $E_3=3$ bits.

\medskip\noindent\emph{Theorems~\ref{thm:fqe-seven-purity} and~\ref{thm:fqe-seven-entropy} (seven qubits). Every
pure seven-qubit state has $\pi_3\ge121/917$, with equality exactly on the local-unitary and
permutation orbit of an explicit icosahedral state $\Psiico$. Every pure seven-qubit state has
$E_3\le E_3(\psi_{x_*})=2.965356029\ldots$ bits, with equality exactly on the local-unitary and permutation orbit of
a different member $\psi_{x_*}$ of the same icosahedral family (Section~\ref{sec:fqe-seven-entropy}). With qubit labels fixed,
each equality set consists of $42$ LU orbits.}
\medskip

For four qubits, many states tie for the minimum mean purity, but the
HS cut spectra uniquely maximise the entropy. At seven qubits even the purity minimiser is globally unique (up to
local unitaries and qubit permutations), yet it is not the entropy maximiser. Both optimisers lie in one explicit
one-parameter icosahedral family, and our proofs separate them globally with five-copy certificates. ``Most entangled'' thus depends on the measure, even when only balanced cuts are compared;
at four qubits the R\'enyi entropies show the same effect, with the maximiser changing at $q_*$.

\begin{figure}[t]
\centering\includegraphics[width=\linewidth]{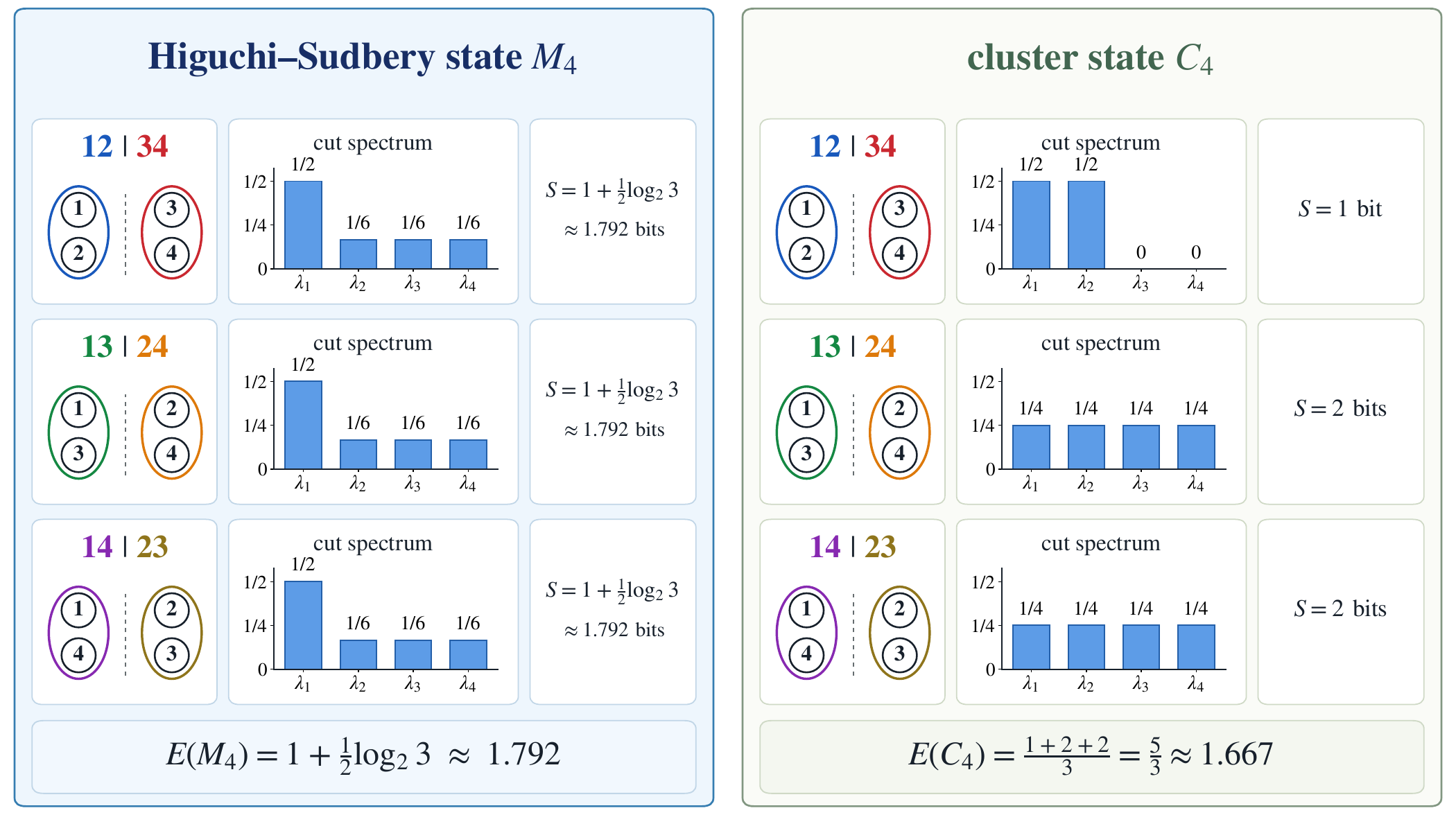}
\caption{Four eigenvalues of each reduced density matrix for the three balanced
$2|2$ cuts (same vertical scale throughout). For $M_4$ each cut has spectrum
$(\tfrac12,\tfrac16,\tfrac16,\tfrac16)$ and entropy
$1+\tfrac12\log_2 3$ bits. For the cluster state $C_4$, the
$12|34$ cut has spectrum $(\tfrac12,\tfrac12,0,0)$ and entropy
$1$ bit, while the other two cuts are maximally mixed and have entropy
$2$ bits, so $E(C_4)=(1+2+2)/3=5/3<E(M_4)$: balancing the three cuts beats perfecting two at the expense of one.}
\label{fig:fqe-cuts}
\end{figure}

\paragraph{Four qubits: compromise between cuts.} To see why a compromise is needed, compare the four-qubit linear cluster state
$C_4=\tfrac12(|0000\rangle+|0011\rangle+|1100\rangle-|1111\rangle)$. Its entropy is $1$ bit for
$12|34$ and $2$ bits for each of the other two cuts, so $E(C_4)=5/3\approx1.667<E(M_4)$. Making two cuts
perfect costs more on the third than is gained. Figure~\ref{fig:fqe-cuts} compares the reduced-state
\emph{spectra}, or lists of eigenvalues, of the two states.

The LU orbits of $M_4$ and $\overline{M_4}$ are distinct~\cite[Prop.~7]{LWB} (Appendix~\ref{sec:fqe-sextic}
gives an elementary separating invariant), but an odd permutation of the qubits exchanges them
(Remark~\ref{rem:fqe-single-orbit}), so the maximiser is unique up to local unitaries and relabelling the qubits.

Purity alone cannot identify the entropy maximiser: the cluster state and $M_4$ both attain
the minimum summed purity $P=1$ but have different entropies, so a proof must also control higher moments of the cut spectra. The entanglement of
a cut depends only on the spectrum of its marginal $\rho_S$, and spectra are what copies of a state measure: $\operatorname{tr}\rho_S^k$ is the expectation value of the operator that
cyclically permutes the subsystems $S$ of $k$ copies of $\psi$. More generally, any homogeneous polynomial of degree $m$
in $|\psi\rangle\langle\psi|$ is an $m$-copy expectation value. A bound is proved by writing the gap between the bound
and the objective as the $m$-copy expectation value of a sum of manifestly positive operators $A^\dagger A$; the
optimisers are the states on which every term vanishes.

The useful positive operators also act on copies replaced by
the spin flip $\tilde\psi=\sigma_y^{\otimes n}\bar\psi$, time reversal ($\sigma_y$ composed with complex conjugation)
applied to every spin-$\tfrac12$. A Pauli string is a tensor product of Pauli matrices and identities; its weight is the number of nonidentity factors.
On density matrices, $\tilde\rho=\sigma_y^{\otimes n}\rho^T\sigma_y^{\otimes n}$ negates
the Pauli-string components of odd weight and keeps those of even weight; on one qubit it reverses the Bloch vector.
For a product state $\tilde\psi$ is orthogonal to $\psi$, while for a Bell pair it equals $\psi$ up to phase (for two qubits
$|\langle\psi|\tilde\psi\rangle|$ is Wootters' concurrence~\cite{Wootters}). Since
$\epsilon\bar u\epsilon^{-1}=u$ for $u\in SU(2)$ and $\epsilon=i\sigma_y$, a flipped copy transforms like an ordinary one, so these operators
remain local-unitary covariant.

The number of copies is set by two things. First, the entropy is not a polynomial. The proof bounds $-t\log_2t$ by a
\emph{majorant}, a polynomial lying above it on $[0,1]$, that touches it at the optimiser's eigenvalues,
and the degree needed grows with the number of touching points (Section~\ref{sec:fqe-entropy_maximum-entropy-bound}). The cuts of $M_4$ have spectrum
$(\tfrac12,\tfrac16,\tfrac16,\tfrac16)$, with two distinct eigenvalues, so a cubic suffices: two copies give $P\ge1$ and
three bound the cubic moments. Second, a few copies may not be enough even for a polynomial objective. Allowing
$m=2,3,\dots$ copies gives a \emph{hierarchy} of ever stronger bounds, but a few copies constrain only low-order
correlations between copies, and they admit moment data that pass every positivity test at that level yet come from no pure state.
At seven qubits both the purity and the entropy need five copies; Section~\ref{sec:fqe-seven-entropy} explains why and
compares the two sizes in Table~\ref{tab:fqe-four-seven}.

\paragraph{The four-qubit proof.} Write $P$ and $T$ for the sums of the second and third
moments of the three two-qubit marginals, and $F=17P-12T-12$. The two-copy spin flip~\cite{GW,WC} gives $P\ge1$
(Lemma~\ref{lem:fqe-purity}); a three-copy sum of squared tensor norms, the sextic identity, gives
$F\ge0$, that is $T\le5/12+(17/12)(P-1)$ (Theorem~\ref{thm:fqe-moment} and Appendix~\ref{sec:fqe-sextic}).
A cubic majorant of $-x\ln x$, its Hermite interpolant (the cubic matching its value and slope at $1/6$ and $1/2$,
which lies above it), turns these
moment bounds into the entropy bound. With $\Delta_H\ge0$ the total gap of the majorant over the twelve cut eigenvalues,
the entropy gap is
\[
3\ln2\,\bigl[E(M_4)-E(\psi)\bigr]=\bigl(\tfrac{15}{2}-6\ln3\bigr)(P-1)+\tfrac32(\ln3-1)F+\Delta_H,
\]
with all three terms nonnegative (Section~\ref{sec:fqe-entropy_maximum-entropy-bound},
equation~\eqref{eq:fqe-entropy-gap}). Equality forces $P=1$ and every cut eigenvalue to be $1/6$ or $1/2$, hence the $M_4$ spectrum
on each cut; the critical normal form identifies the state from just one cut
(Proposition~\ref{prop:fqe-onecut}). The classification of all zeros of the
moment inequality, also given in Appendix~\ref{sec:fqe-sextic}, is not needed for equality.

\paragraph{Relation to earlier work.} Gour and Wallach studied the analogous optimisation for Tsallis and R\'enyi
entropies~\cite{GW}, including a conjecture on R\'enyi entropies that we prove (Corollary~\ref{cor:fqe-renyi-gw}).\footnote{It is distinct from the statement on four-qubit states maximising the absolute value of the
hyperdeterminant that is usually called the Gour--Wallach conjecture, proved by Chen and
\DJ{}okovi\'c~\cite{CD}.} Their Theorem~10(a) proves the Tsallis-entropy bound for exponents $0<q<2$
among the critical states (the normal form of~\cite{VDD}), with the
bound attained by $M_4$. It does not cover every pure state. Their equality clause names only the class of $M_4$;
with labelled qubits $M_4$ and its conjugate are distinct LU classes, exchanged by an odd permutation
(Remark~\ref{rem:fqe-single-orbit}). For $q>2$ their Theorem~9(a) identifies the unrestricted Tsallis maximiser, a
state other than $M_4$ (we prove only that the cluster state beats $M_4$ there), and their Theorem~12 identifies the cluster-state maximisers of the
R\'enyi entropies for $q\ge2$~\cite{GW}; Corollary~\ref{cor:fqe-renyi-two-up} gives a self-contained
elementary proof of that high-order ceiling and its equality cut spectra. For the von Neumann entropy itself, this paper supplies
a short, self-contained analytic proof of the bound for all pure states, the classification of the
equality case, and unrestricted corollaries for other spectral entropies.\footnote{A proof of the unrestricted entropy bound and its attainment at $M_4$, with a Lean verification, appeared
independently while this work was in preparation~\cite{Omega}; it does not address uniqueness of the maximiser or the
equality case, and we make no priority claim for the moment inequality as such.}
At seven qubits, earlier numerical searches found mean three-qubit purities of $0.13387$~\cite{Facchi}
and $0.131952$~\cite{DMFF18}; the latter agrees with our minimum $121/917$ to six digits. \emph{Stabilizer states},
joint eigenstates of $n$ commuting, independent Pauli strings (graph states and the cluster state $C_4$ are examples), do worse: the best,
including that of Zha et al., have purity $19/140\approx0.1357$~\cite{Zha11,HGS}.
Before this work, the best explicit universal lower bound known to us was the
unattainable AME value $1/8$; Rico et al.\ obtained a numerical semidefinite
estimate of $0.127(8)$~\cite{Rico}, without a certificate. It agrees with the exact three-copy bound
$\pi_3\ge125/978\approx0.12781$ proved here (Table~\ref{tab:fqe-seven-levels}), the three-copy level of our hierarchy.

\paragraph{Proofs and verification.} The four-qubit von Neumann bound and its
equality case have a fully analytic proof here. Some extensions, particularly at
seven qubits, use exact-rational or rigorously bounded numerical certificates.
Table~\ref{tab:fqe-guide} is a roadmap of the paper: for each result it gives the optimiser, how it is proved
and where the complete proof is. The supplement uses S-prefixed numbering; the companion
repository~\cite{Companion} contains both documents, the scripts and their
captured outputs, and the Lean formalisation. Appendix~\ref{sec:fqe-verifiers} lists the computer checks
behind each result, and \texttt{verifiers/MAP.md} maps them to scripts. The analytic arguments do not depend on Lean.
Every theorem, proposition, lemma and corollary is proved, analytically or with the exact and computer-assisted checks named in its proof; numerical observations are labelled as such. Table~\ref{tab:fqe-other-n} in Section~\ref{sec:fqe-discussion} collects what is known at other qubit counts.

\begin{table}[htbp]
\centering\small
\begin{tabular}{@{}p{0.19\linewidth}p{0.17\linewidth}p{0.33\linewidth}p{0.23\linewidth}@{}}
\toprule
Question & Optimiser & How proved & Where \\
\midrule
Four qubits: maximum mean entropy & $M_4$, $\overline{M_4}$ &
Analytic (two- and three-copy identities, cubic majorant) &
Sections~\ref{sec:fqe-entropy_maximum-purity-identity}--\ref{sec:fqe-entropy_maximum-equality-and-one-cut-rigidity};
Appendices~\ref{sec:fqe-sextic}, \ref{sec:fqe-normal-form} \\\addlinespace
Four qubits: Tsallis $0<q<2$ & $M_4$ & Analytic (same majorant) &
Section~\ref{sec:fqe-spectral}; Appendix~\ref{sec:fqe-spectral-proofs} \\\addlinespace
Four qubits: R\'enyi $0<q<1$ & $M_4$ & Analytic (from the Tsallis bound) &
Section~\ref{sec:fqe-spectral} \\\addlinespace
Four qubits: R\'enyi $1<q<2$ & $M_4$ for $q<q_*$, cluster for $q>q_*$ &
Analytic with exact finite checks &
Appendix~\ref{sec:fqe-renyi-proofs}; supplement, Section~\ref{sec:sup-renyi} \\\addlinespace
Four qubits: R\'enyi $q\ge2$ & cluster & Analytic (elementary purity bound) &
Appendix~\ref{sec:fqe-renyi-high-order} \\\addlinespace
Four qubits: orbit stability & -- & Analytic with exact finite checks; some constants computer-assisted with rigorous error bounds &
Section~\ref{sec:fqe-stability-summary}; Appendix~\ref{sec:fqe-orbit-stability}; supplement, Sections~\ref{sec:fqe-stability-proofs}, \ref{sec:sup-stability} \\\addlinespace
Seven qubits: minimum mean purity & $\Psiico$ & Computer-assisted with rigorous error bounds (five-copy identity) &
Appendices~\ref{sec:fqe-seven}, \ref{sec:ico}; supplement, Section~\ref{sec:fivecopy} \\\addlinespace
Seven qubits: maximum mean entropy & $\psi_{x_*}\ne\Psiico$ & Computer-assisted with rigorous error bounds (five-copy identity, joint majorant) &
Supplement, Section~\ref{sec:sup-seven-entropy}; uses Theorem~\ref{thm:fqe-seven-purity} \\
\bottomrule
\end{tabular}
\caption{Optimisers, methods and the location of complete proofs. Optimisers are listed up to local unitaries and qubit
permutations; both classes attain the R\'enyi bound at $q=q_*$. Supplementary sections have an S prefix. Exact finite checks use exact arithmetic; computer-assisted
proofs evaluate explicit identities in floating point with rigorous error bounds. Neither uses a numerical search as
evidence. Appendix~\ref{sec:fqe-verifiers} lists the checks behind each result.}
\label{tab:fqe-guide}\label{tab:fqe-proof-guide}
\end{table}

\section{The four-qubit maximum}\label{sec:fqe-main}

Throughout this section, $\psi$ is a unit four-qubit state, $E(\psi)=\frac13\sum_{j=2}^4S(\rho_{1j})$ is its average
$2|2$ entropy in bits, and $M_4$ is the Higuchi--Sudbery state, all as defined in
Section~\ref{sec:fqe-intro}. Write the one-qubit marginals as
$\rho_i=(I+\mathbf r_i\cdot\boldsymbol\sigma)/2$, where $\boldsymbol\sigma$ is the vector of Pauli matrices
and $\mathbf r_i$ is the Bloch vector.

\begin{theorem}[maximum and equality]\label{thm:fqe-entropy-max} For every unit $\psi\in(\mathbb
C^2)^{\otimes4}$,
\[
E(\psi)\le 1+\tfrac12\log_2 3 ,
\]
with equality if and only if $\psi$ is local-unitarily equivalent to $M_4$ or to $\overline{M_4}$.
\end{theorem}

The bound is proved in Sections~\ref{sec:fqe-entropy_maximum-purity-identity}--\ref{sec:fqe-entropy_maximum-entropy-bound}
and the equality case in Section~\ref{sec:fqe-entropy_maximum-equality-and-one-cut-rigidity}; the two
orbits are related by a qubit permutation in Remark~\ref{rem:fqe-single-orbit}.

\begin{table}[htbp]
\centering\small
\begin{tabular}{@{}ll@{}}
\toprule
Symbol (unit states) & Meaning \\
\midrule
$P=\sum_{j=2}^4\operatorname{tr}\rho_{1j}^2$ & Sum of the cut purities \\
$T=\sum_{j=2}^4\operatorname{tr}\rho_{1j}^3$ & Sum of the third moments \\
$F=17P-12T-12$ & Nonnegative defect in the third-moment bound \\
$W_1=\sum_{i=1}^4\|\mathbf r_i\|^2$ & Total squared one-qubit Bloch-vector length \\
$\tau=|\psi^T\sigma_y^{\otimes4}\psi|^2$ & Squared spin-flip overlap (four-tangle) \\
\bottomrule
\end{tabular}
\caption{Notation for the four-qubit proof. A state is critical when all $\mathbf r_i=0$, equivalently
when all one-qubit marginals are maximally mixed. Homogeneous formulas below use $N=\|z\|^2$
and unnormalised marginals; unit states have $N=1$.}
\label{tab:fqe-moment-notation}
\end{table}

\subsection{Purity identity}\label{sec:fqe-entropy_maximum-purity-identity}

The first ingredient bounds the sum of the three cut purities. Its equality condition also
identifies the locus on which an entropy maximiser must lie.

\begin{lemma}[purity identity]\label{lem:fqe-purity} For a unit state, $P=\sum_{j=2}^4\operatorname{tr}\rho_{1j}^2=1+\tfrac12\Delta$ with
$\Delta=\sum_{i=1}^4\|\mathbf r_i\|^2+\tau$, where $\mathbf r_i$ are the one-qubit Bloch
vectors and $\tau=|\psi^T\sigma_y^{\otimes4}\psi|^2$. In particular $P\ge 1$, with equality iff all
$\mathbf r_i=0$ (critical) and $\tau=0$.

\end{lemma}

\noindent Here $\tau$ is the four-tangle of Wong and Christensen~\cite{WC}, and the lemma is
Gour and Wallach's identity $\tau_2=(4\tau_1-\tau_{ABCD})/3$~\cite[Theorem~6]{GW}, with
$\tau_k$ the averaged linear entropies $2(1-\operatorname{tr}\rho^2)$ of the $k$-qubit
marginals. We include the short spin-flip proof to keep the argument self-contained.

\begin{proof} For a \(2\times2\) matrix \(A\), spin flip is \(\sigma_y
A^T\sigma_y=(\operatorname{tr}A)I-A\). Tensoring this identity over the four qubits, a product
operator \(A_1\otimes\cdots\otimes A_4\) is sent to \(\bigotimes_i\bigl((\operatorname{tr}A_i)I-A_i\bigr)\).
By linearity the spin flip \(\tilde\rho=\sigma_y^{\otimes4}\rho^T\sigma_y^{\otimes4}\) of a four-qubit
density matrix is
\(\tilde\rho=\sum_{S\subseteq\{1,\ldots,4\}}(-1)^{|S|}\rho_S\otimes I_{S^c}\), where \(\rho_S\) is the
marginal on the sites in \(S\) and \(\rho_\varnothing=1\). Since
\(\operatorname{tr}[\rho(\rho_S\otimes I)]=\operatorname{tr}\rho_S^2\), \[
\operatorname{tr}(\rho\tilde\rho)=\sum_{S}(-1)^{|S|}\operatorname{tr}\rho_S^2 .
\] For \(\rho=|\psi\rangle\langle\psi|\) we have \(\rho^T=|\bar\psi\rangle\langle\bar\psi|\), hence
\(\tilde\rho=|\tilde\psi\rangle\langle\tilde\psi|\) with \(\tilde\psi=\sigma_y^{\otimes4}\bar\psi\), and
\[
\operatorname{tr}(\rho\tilde\rho)=|\langle\psi|\tilde\psi\rangle|^2
=|\psi^\dagger\sigma_y^{\otimes4}\bar\psi|^2=|\psi^T\sigma_y^{\otimes4}\psi|^2=\tau ,
\] because \(\psi^\dagger\sigma_y^{\otimes4}\bar\psi\) is the complex conjugate of
\(\psi^T\sigma_y^{\otimes4}\psi\) (as \((\sigma_y^{\otimes4})^*=\sigma_y^{\otimes4}\)). For a pure
four-qubit state, complementary purities agree (\(\operatorname{tr}\rho_S^2=\operatorname{tr}\rho_{S^c}^2\)),
so the sum has the two terms \(S=\varnothing,\{1,2,3,4\}\) equal to \(1\), the eight terms with
\(|S|=1,3\) adding to \(-2\sum_i\operatorname{tr}\rho_i^2\), and the six terms with \(|S|=2\)
adding to \(2P\): \(\tau=2-2\sum_i\operatorname{tr}\rho_i^2+2P\). Now
\(\operatorname{tr}\rho_i^2=(1+\|\mathbf r_i\|^2)/2\) gives the result. \end{proof}

\noindent Every term of $P-1=\tfrac12\Delta$ is quadratic in $|\psi\rangle\langle\psi|$ and hence an expectation value in two copies
of the state. The term $\tau$ is the squared overlap of one copy with the spin flip of the other. This two-copy positivity is the first
level of the hierarchy used at seven qubits (defined in Section~\ref{sec:fqe-seven-entropy} and
Appendix~\ref{sec:fqe-seven}).

\subsection{The sharp moment inequality}\label{sec:fqe-entropy_maximum-the-sharp-moment-inequality}

The purity bound alone does not give the sharp entropy bound. We also need to constrain the third
moments of the same marginals. We state this bound homogeneously so that its sum-of-squares identity
applies to unnormalised vectors.

For \(z\in\mathbb C^{16}\), write \(N=\|z\|^2\) and let \(R_{1j}\) be its unnormalised reduced matrix
on qubits \(1,j\). Put \[
P=\sum_{j=2}^4\operatorname{tr}R_{1j}^2,\qquad
T=\sum_{j=2}^4\operatorname{tr}R_{1j}^3,\qquad
F=17NP-12T-12N^3.
\]
For unnormalised \(z\) the letters \(P,T\) denote these homogeneous moments; at \(N=1\) they reduce to the unit-state
\(P,T\) of Table~\ref{tab:fqe-moment-notation}, and \(F=17P-12T-12\).

\begin{theorem}[moment inequality with polarisation remainder]\label{thm:fqe-moment} Let
$W_1=\sum_i\sum_{a=x,y,z}\langle z|\sigma_a^{(i)}|z\rangle^2$. Then
\[
F(z)\ge\frac34NW_1\ge0,
\]
and the constant $3/4$ is attained by every product state.
For unit states this implies $T\le5/12+(17/12)(P-1)$. The state $M_4$ has $(P,T)=(1,5/12)$ and a
product of two Bell pairs has $(3/2,9/8)$; both satisfy $F=0$, so both lie on the line
$T=5/12+(17/12)(P-1)$. Consequently $17/12$ is the least $\kappa$ for which
$T\le5/12+\kappa(P-1)$ holds for all unit states (the Bell point forces $\kappa\ge17/12$).
Every zero $z\ne0$ of $F$ is critical: all its one-qubit marginals are $NI/2$.

\end{theorem}

\begin{proof} The key identity expresses the moment defect as a sum of three kinds of nonnegative terms: one-qubit
polarisation, the squared norms of contractions of two copies of the state with one spin-flipped copy, and the
disagreement between the actions of the one-qubit marginals on the state. Let \(\epsilon=i\sigma_y\), \(\tilde z=\epsilon^{\otimes4}\bar z=\sigma_y^{\otimes4}\bar z\) and
\(w=z\otimes z\otimes\tilde z\), so that each qubit carries three slots \(x,y,t\), one from each copy. At one qubit define
maps from three slots to one by \(L|xyt\rangle=\epsilon_{yt}|x\rangle\) and \(M|xyt\rangle=\epsilon_{xt}|y\rangle\). Let
\(S_i\) be the orthogonal projector onto the symmetric subspace of the three slots of qubit \(i\), and contract the other
three qubits:
\[
U_i=S_i\!\prod_{j\ne i}L_jw,\qquad V_i=S_i\!\prod_{j\ne i}(L_j+M_j)w,
\]
vectors on six qubits. Write \(R_i\) for the unnormalised one-qubit marginal of \(z\) and \(R_i^{(i)}\) for its action on
qubit \(i\) of \(\mathbb C^{16}\). Appendix~\ref{sec:fqe-sextic} proves the identity \[
\boxed{\quad8F=6NW_1+3\sum_i\|V_i-8U_i\|^2+12\sum_{i<j}\bigl\|(R_i^{(i)}-R_j^{(j)})z\bigr\|^2.\quad}
\tag{S}\label{eq:fqe-remainder}
\] Identity~\eqref{eq:fqe-remainder} is a polynomial identity of degree six in the amplitudes of $z$, that is, cubic in
$|z\rangle\langle z|$ (a three-copy identity); we call it the \emph{sextic identity}.
The derivation uses contractions with the two-dimensional antisymmetric tensor, the vanishing of three-vector
antisymmetrisation in dimension two, and elementary swap traces. Nonnegativity and criticality follow at once. At a
product state of norm \(N\) every \(R_i^{(i)}z=Nz\) and \(F=3N^3=\tfrac34NW_1\). \end{proof}

\subsection{Entropy bound}\label{sec:fqe-entropy_maximum-entropy-bound}

We now combine the two moment bounds with a scalar interpolation inequality.
The resulting gap formula makes both the bound and its equality conditions explicit.

\begin{figure}[t]
\centering\includegraphics[width=0.5\linewidth]{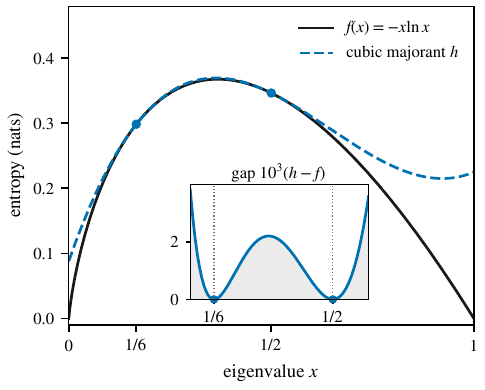}
\caption{The single-eigenvalue entropy contribution $f(x)=-x\ln x$ (in nats) and its cubic majorant $h$, tangent at the eigenvalues $1/6$ and $1/2$
of every cut of $M_4$. The inset magnifies the gap $h-f$ by $10^3$ on $[0.1,0.6]$: it is nonnegative, with double zeros at the nodes.
Summing $h$ over the cut spectra gives an affine upper bound on the entropy in terms of $P$ and $T$;
equality in this scalar bound requires every eigenvalue to be at a node.}
\label{fig:fqe-majorant}
\end{figure}

Let \(f(x)=-x\ln x\), and let \(h\) be its cubic Hermite interpolant at the nodes \(\xi_1=\tfrac16\) and
\(\xi_2=\tfrac12\): it agrees with \(f\) and \(f'\) at both nodes. Its cubic and quadratic coefficients
are \[
a=18(\ln3-1)>0,\qquad b=18-\tfrac{39}2\ln3 .
\] These coefficients are fixed by matching \(f\) and \(f'\) at both nodes.
Since \(f^{(4)}(x)=-2/x^3<0\), the Hermite remainder has the sign
\[
h(x)-f(x)=-\frac{f^{(4)}(\zeta_x)}{4!}
       (x-\tfrac16)^2(x-\tfrac12)^2\ge0\qquad(x>0),
\]
for some \(\zeta_x>0\). Equality occurs only at the two nodes; at the
endpoint \(x=0\), \(h(0)-f(0)=\tfrac12-\tfrac38\ln3>0\).
Thus \(h\) majorises the entropy function on the entire spectral interval
(Figure~\ref{fig:fqe-majorant}). In general a polynomial majorant of $-t\log_2t$ on $[0,1]$ touching it at $k$
interior points has degree at least $2k-1$, attained by the Hermite interpolant
(Proposition~\ref{prop:seven-entropy-majorant-degree} of the supplement). Two tangencies thus need a cubic; at seven
qubits this degree count leads to a two-variable majorant (Section~\ref{sec:fqe-seven-entropy}).

Return now to unit states. Each cut has four eigenvalues \(\lambda\) with \(\sum\lambda=1\),
\(\sum\lambda^2=\mathrm{tr}\rho_{1j}^2\), \(\sum\lambda^3=\mathrm{tr}\rho_{1j}^3\). Summing \(h\) over the 12
eigenvalues of the three cuts, \[
3\ln2\;E(\psi)=\sum_{j,\lambda}f(\lambda)=\kappa_0+bP+aT-\Delta_H,\qquad
\Delta_H=\sum_{j,\lambda}(h-f)(\lambda)\ge0,
\] with a constant \(\kappa_0\) independent of \(\psi\). At \(M_4\) every cut has spectrum
\((\tfrac12,\tfrac16,\tfrac16,\tfrac16)\), so \(\Delta_H=0\), \(P=1\), \(T=\tfrac5{12}\) and
\(E(M_4)=\tfrac12+\tfrac12\log_26=1+\tfrac12\log_23\). Subtracting, and using \(T=(17P-12-F)/12\)
(the definition of \(F\) at \(N=1\)), \[
\begin{aligned}
3\ln2\,\bigl(E(M_4)-E(\psi)\bigr)
&=-b(P-1)-a\bigl(T-\tfrac5{12}\bigr)+\Delta_H\\
&=\bigl(\tfrac{15}{2}-6\ln3\bigr)(P-1)
+\tfrac32(\ln3-1)F+\Delta_H.
\end{aligned}\tag{EG}\label{eq:fqe-entropy-gap}
\] Here \(-b-\tfrac{17}{12}a=\tfrac{15}2-6\ln3\) and \(a/12=\tfrac32(\ln3-1)\). All three terms
are nonnegative: \(\tfrac{15}2-6\ln3>0\) (since \(1<\ln3<5/4\)), \(P\ge1\)
(Lemma~\ref{lem:fqe-purity}), \(F\ge0\) (Theorem~\ref{thm:fqe-moment}), \(\Delta_H\ge0\). This
proves \(E(\psi)\le E(M_4)=1+\tfrac12\log_23\).

\paragraph{Why the bound is tight.} The entropy bound uses only the nonnegativity of $F$, rather than its stronger polarisation remainder. Its unit zeros are exactly the two
$M_4$ LU orbits and the three LU orbits of products of two Bell pairs, one for each pairing of the qubits
(Proposition~\ref{prop:fqe-moment-zeros} of Appendix~\ref{sec:fqe-sextic}). Only the two HS orbits saturate the entropy bound,
since products of two Bell pairs have $P=\tfrac32$. The cubic majorant touches the entropy function exactly at the eigenvalues of $M_4$, and
its coefficient signs make both moment slacks, $P-1$ and $F$, enter~\eqref{eq:fqe-entropy-gap} with nonnegative
weights. The same criterion yields the Tsallis bound for $0<q<2$ and the R\'enyi bound for $0<q<1$
(Proposition~\ref{prop:fqe-majorant}). At Tsallis order $q=2$ all critical states with $\tau=0$ tie.
Appendix~\ref{sec:fqe-correlation} gives a correlation-volume interpretation of the moment inequality. The sextic identity also
controls the spin flip: for unit states $F\ge\frac34W_1+\tau-\tau^{3/2}$ (Theorem~\ref{thm:fqe-general-envelope}), with equality on
product states and, for every $\tau\in[0,1]$, on a family of critical states. With the gap formula~\eqref{eq:fqe-entropy-gap}
this gives the Tsallis and von Neumann stability estimates in $W_1$ and $\tau$ of Appendix~\ref{sec:fqe-spectral-proofs}.

Being cubic in $|\psi\rangle\langle\psi|$, the sextic identity is a certificate at the three-copy level of the same
hierarchy, for the objective $F$ instead of a purity (Appendix~\ref{sec:fqe-seven}).

\subsection{Equality and one-cut rigidity}\label{sec:fqe-entropy_maximum-equality-and-one-cut-rigidity}

\begin{proposition}[one-cut rigidity]\label{prop:fqe-onecut} If $P=1$ and a $2|2$ cut has spectrum
$(1/2,1/6,1/6,1/6)$, then $\psi$ is LU-equivalent to $M_4$ or $\overline{M_4}$.

\end{proposition}

\begin{proof} Relabel the chosen cut as \(12|34\). Lemma~\ref{lem:fqe-purity} gives criticality and
\(\tau=0\). Write two-qubit states in the Bell basis with the phases of the magic basis,
\[
m_1=\tfrac1{\sqrt2}(|00\rangle+|11\rangle),\quad
m_2=\tfrac{i}{\sqrt2}(|01\rangle+|10\rangle),\quad
m_3=\tfrac1{\sqrt2}(|01\rangle-|10\rangle),\quad
m_4=\tfrac{i}{\sqrt2}(|00\rangle-|11\rangle),
\]
and a four-qubit state as \(\psi=\sum_{j,k}Z_{jk}\,m_j\otimes m_k\), with qubits \(1,2\) indexing the rows of \(Z\) and
qubits \(3,4\) the columns. In this basis single-qubit unitaries on each pair act, up to phase, as real four-dimensional rotations, so
local unitaries act by \(Z\mapsto O_1ZO_2^T\) with \(O_1,O_2\in SO(4)\). By the critical normal form
(Theorem~\ref{thm:fqe-critical-normal-form} of Appendix~\ref{sec:fqe-normal-form}), \(\psi\) is LU-equivalent to \(G(v)=\sum_kv_k\,m_k\otimes m_k\), that is
\(Z=\operatorname{diag}(v)\), with \(v\in\mathbb C^4\) a unit vector. The
cut spectrum and \(\tau\) are LU invariant (the latter by
\eqref{eq:fqe-spin-flip-normal-form}). On \(G(v)\) the spectrum of the cut \(12|34\) is
\(\{|v_k|^2\}\) (the \(m_k\) are orthonormal), and \(\tau=|\sum_kv_k^2|^2\) by
\eqref{eq:fqe-spin-flip-normal-form}.
Fix the large coordinate to \(1/\sqrt2\) by a common phase. Then \(\tau=0\) says that the other three
squares have modulus \(1/6\) and sum to \(-1/2\). Their sum has the largest possible modulus,
\(3(1/6)=1/2\), so equality in the triangle inequality forces all three squares to equal \(-1/6\). Thus, up to position of the large coordinate,
\[
 v=(\sqrt3,\pm i,\pm i,\pm i)/\sqrt6.
\]
The rotations \(Z\mapsto O_1ZO_2^T\) reduce these sign choices to two.
Applying the same determinant-one signed permutation on both sides of $G(v)$
moves the large entry to the first position without changing the other entries
except for their order. A diagonal rotation on one side can flip any pair of
small signs. Thus only their parity matters, giving the representatives
$(\sqrt3,i,i,i)/\sqrt6$ and $(\sqrt3,-i,-i,-i)/\sqrt6$.
Expanding the $m_k$ shows that
\[
 M_4=G\bigl((1,1,i\sqrt3,1)/\sqrt6\bigr).
\]
Multiplying by $-i$ and moving its third coordinate to the first gives the second representative.
Complex conjugation gives the first, since each $m_k\otimes m_k$ is a real vector, so $\overline{G(v)}=G(\bar v)$. Both representatives therefore belong to the stated pair
of LU orbits.
Relabelling preserves the pair of orbits, since it permutes the coefficient triple
\((1,\omega,\omega^2)\) up to common phase and conjugation.
\end{proof}

\emph{Proof of equality in Theorem~\ref{thm:fqe-entropy-max}.} Suppose $E(\psi)=E(M_4)$. The
positive coefficients in \eqref{eq:fqe-entropy-gap} force $P=1$ and $\Delta_H=0$.
The strict Hermite remainder above then forces every cut eigenvalue to be
$\tfrac16$ or $\tfrac12$. Four such eigenvalues with sum $1$ are
$(\tfrac12,\tfrac16,\tfrac16,\tfrac16)$, so Proposition~\ref{prop:fqe-onecut} applied to any one cut
shows that $\psi$ is LU-equivalent to $M_4$ or $\overline{M_4}$. This identifies the equality states without classifying all zeros
of $F$ or separately excluding Bell-pair products. Conversely, $M_4$ and its conjugate
have this spectrum on every cut and attain the bound. This proves Theorem~\ref{thm:fqe-entropy-max},
including equality.

\begin{remark}[a single orbit]\label{rem:fqe-single-orbit} The transposition of qubits $2$ and $3$
exchanges $|0011\rangle\leftrightarrow|0101\rangle$ and $|1100\rangle\leftrightarrow|1010\rangle$ and
fixes $|0110\rangle$ and $|1001\rangle$. It therefore sends the coefficient triple
$(1,\omega,\omega^2)$ of $M_4$ to $(\omega,1,\omega^2)=\omega\,(1,\omega^2,\omega)$, that is,
\[
M_4\longmapsto\omega\,\overline{M_4}.
\]
Since $E$ is invariant under qubit
permutations and $\omega$ is a global phase, the two local-unitary orbits of
Theorem~\ref{thm:fqe-entropy-max} are exchanged by a permutation of the qubits: the set of maximisers
is a single orbit under local unitaries together with qubit permutations. This is a special case of
the observation of Lyons, Arnold and Swogger~\cite{LAS}: even permutations of the qubits fix $M_4$ up to
phase, and odd ones map it to $\overline{M_4}$.
\end{remark}

\section{R\'enyi, Tsallis and mixed-state entropies}\label{sec:fqe-spectral}

Different spectral entropies give different weights to the marginal eigenvalues. For $q>0$, $q\ne1$, the R\'enyi~\cite{Renyi} and Tsallis~\cite{Tsallis} entropies
\[
 H_q(\rho)=\frac{\log_2\operatorname{tr}\rho^q}{1-q},\qquad
 S_q(\rho)=\frac{1-\operatorname{tr}\rho^q}{q-1}
\]
tend as $q\to1$ to $S(\rho)$ in bits and $(\ln2)S(\rho)$ in nats,
respectively. As $q\to0$, $H_q$ tends to the base-two logarithm of the rank of $\rho$, which is the Schmidt rank across a pure-state cut; $H_2=-\log_2\operatorname{tr}\rho^2$ is fixed by the
purity, the quantity minimised at seven qubits and the one that randomised-measurement experiments estimate
directly~\cite{Brydges19}; as $q\to\infty$, $H_q$ depends only on the largest eigenvalue. Each is Schur-concave: averaging any pair of eigenvalues cannot lower it. For
pure states it therefore does not increase under deterministic local operations and classical communication~\cite{Nielsen99}:
every member measures how entangled a cut is. Asking which state maximises them all tests whether ``most entangled''
is a robust notion. The answer depends on the order: $M_4$ maximises the mean R\'enyi entropy for $q<q_*\approx1.588$, and the cluster states maximise it for $q>q_*$. For the Tsallis entropies $M_4$ is the maximiser for $0<q<2$; the cluster states tie with it at $q=2$ and beat it for every $q>2$. The
maximiser for $q>2$ is identified by Gour and Wallach~\cite[Theorem~9(a)]{GW}; we do not reprove that result. The cut spectra explain the change. Those of the cluster state are
flat, $(\tfrac12,\tfrac12)$ on one cut and uniform $\tfrac14$ on the other two, so its mean R\'enyi entropy is $5/3$ for
every $q$; the lopsided spectrum $(\tfrac12,\tfrac16,\tfrac16,\tfrac16)$ of $M_4$ gives a mean $H_q$ falling from
$2$ bits as $q\to0$ to $1$ bit as $q\to\infty$. Small $q$ favours full rank and large $q$ a small largest
eigenvalue, hence the crossing at $q_*$. The Tsallis tie at $q=2$ holds because $1-\operatorname{tr}\rho^2$ depends
only on the cut purities, and both states have $P=1$.

The cubic-majorant argument applies to a wider class of spectral functionals. We state that criterion and
apply it to Tsallis entropies and R\'enyi orders below one; orders above one need additional estimates
(Appendix~\ref{sec:fqe-renyi-proofs}). We then transfer the pure-state bounds to mixed-state convex roofs. We retain
the notation \(P,T,F,\tau\) from the maximum proof; proofs not given here are in Appendix~\ref{sec:fqe-spectral-proofs}.

\begin{proposition}[Cubic spectral majorants]\label{prop:fqe-majorant}
For continuous $f:[0,1]\to\mathbb R$, set
\[
\mathcal E_f(\psi)=\frac13\sum_{j=2}^4\operatorname{tr}f(\rho_{1j}).
\]
Suppose $h(x)=ax^3+bx^2+cx+d$ majorises $f$ on $[0,1]$, agrees with $f$
at $1/6$ and $1/2$, and satisfies $a\ge0$, $b+17a/12\le0$.
Then $\mathcal E_f(\psi)\le\mathcal E_f(M_4)$ for every pure four-qubit state.
If both coefficient inequalities are strict, equality holds precisely on the two HS LU orbits.
\end{proposition}
\begin{proof}
Put $\alpha=-b-17a/12$ and
$\Delta_f=\sum_{j,\lambda\in\operatorname{spec}\rho_{1j}}(h-f)(\lambda)\ge0$.
Summing the cubic and subtracting its value at $M_4$ gives
\[
3\bigl(\mathcal E_f(M_4)-\mathcal E_f(\psi)\bigr)
=\alpha(P-1)+\frac a{12}F+\Delta_f.\tag{SF}\label{eq:fqe-spectral}
\]
All three terms are nonnegative, since $P\ge1$ (Lemma~\ref{lem:fqe-purity}) and $F\ge0$
(Theorem~\ref{thm:fqe-moment}). The same two nonnegative moment defects therefore control every functional
satisfying the coefficient conditions. If $a,\alpha>0$, equality requires $P=1$ and $F=0$,
so Proposition~\ref{prop:fqe-moment-zeros} gives the HS orbits; the other zeros of $F$, the products of two Bell pairs,
have $P=3/2$ and are excluded. No strict scalar
majorisation is required. For differentiable $f$, a candidate touching cubic
is its Hermite interpolant at the two nodes.
\end{proof}

\begin{corollary}[Tsallis entropies]\label{cor:fqe-tsallis} For every \(0<q<2\), \(q\ne1\), the mean
Tsallis-\(q\) entropy of the three cuts satisfies \[
\boxed{\quad
\tfrac13\sum_{j=2}^4\frac{1-\operatorname{tr}\rho_{1j}^q}{q-1}
\ \le\ \frac{1-2^{-q}-3\,6^{-q}}{q-1}\quad}
\] with equality iff \(\psi\) is LU-equivalent to \(M_4\) or \(\overline{M_4}\). \end{corollary}
The range is sharp: at \(q=2\) equality is no longer unique, and for every \(q>2\) the cluster state beats \(M_4\)
(proved in Appendix~\ref{sec:fqe-spectral-proofs}, which also gives the corresponding quadratic deficit bounds).

\paragraph{R\'enyi entropies.} With $H_q$ as above, write
\[
\mathcal E_q(\psi)=\frac13\sum_{j=2}^4H_q(\rho_{1j}),\qquad
h_q=\mathcal E_q(M_4)=\frac{\log_2(2^{-q}+3\,6^{-q})}{1-q}.
\]
At $q=1$ these have their continuous von Neumann values in bits.

\begin{corollary}[R\'enyi entropies below one]\label{cor:fqe-renyi} For \(0<q<1\), concavity of
\(\log\) and the Tsallis bound imply \[
\frac13\sum_{j=2}^4
\frac{\log_2\operatorname{tr}\rho_{1j}^q}{1-q}
\le \frac{\log_2(2^{-q}+3\,6^{-q})}{1-q},
\] again with exactly the same LU equality classes. The argument does not assert the analogous R\'enyi
bound for \(q>1\).

\end{corollary}
\noindent Equality needs equality in the Tsallis bound, hence the HS orbits, and in Jensen's inequality for the logarithm
over the three cuts, which holds there because the three cuts of $M_4$ have identical spectra.

Orders above one require additional estimates. The full result is as follows, with $q_*$ the
unique order for which $h_{q_*}=5/3$ (Proposition~\ref{prop:fqe-renyi-crossing}).

\begin{figure}[t]
\centering\includegraphics[width=0.8\linewidth]{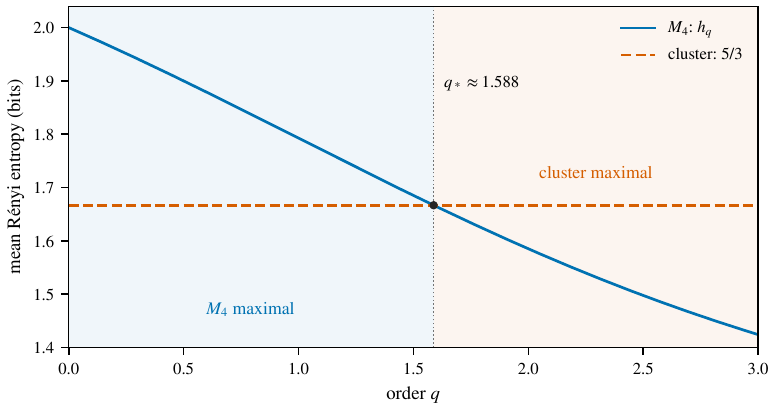}
\caption{Change of R\'enyi-entropy maximiser with the order $q$. The mean entropy of $M_4$
($h_q$, solid) and that of the cluster states ($5/3$, dashed) cross once, at $q_*\approx1.588$
(Proposition~\ref{prop:fqe-renyi-crossing}). The larger of the two, $\max\{h_q,5/3\}$,
is the maximum over all pure four-qubit states (Corollary~\ref{cor:fqe-renyi-gw}):
$M_4$ attains it below $q_*$ and the cluster states above $q_*$ (shaded regions). Values at $q=0$ and $q=1$
are continuous limits.}
\label{fig:fqe-renyi}
\end{figure}

\begin{corollary}[R\'enyi envelope of Gour and Wallach]\label{cor:fqe-renyi-gw}
For every pure four-qubit state and every $q>0$, $\mathcal E_q(\psi)\le\max\{h_q,5/3\}$.
For $q<q_*$ the maximisers are exactly the two HS LU orbits;
for $q>q_*$ they are the cluster states up to local unitaries and qubit permutations (every permutation of the three cuts is induced by a permutation of the qubits).
At $q=q_*$ both classes, and only those classes, attain the bound.
\end{corollary}
Below one the bound follows from the preceding corollary, and at $q=1$ from
Theorem~\ref{thm:fqe-entropy-max}. For $q>1$ it is proved in Appendix~\ref{sec:fqe-renyi-proofs}, which opens with a
plan of the argument; for $1<q<2$ the proof is analytic apart from exact finite checks, namely an exact six-copy
certificate bounding the variance of the three cut purities and exact polynomial sign checks.
Figure~\ref{fig:fqe-renyi} shows the change of maximiser.

\subsection{Mixed-state entanglement}\label{sec:fqe-spectral_corollaries-mixed-state-entanglement}

For a mixed state, the entropy of a marginal also counts classical mixing, which is not entanglement: the
maximally mixed four-qubit state has maximally mixed marginals but no entanglement. We therefore use the
\emph{convex roof}: the least average of a pure-state spectral functional over all ensembles representing the state,
as in the entanglement of formation~\cite{BDSW}.
For a general $f$ obeying the hypotheses below, this construction is an entanglement
monotone when the induced function of the marginal is concave and vanishes on pure marginals, as for the von
Neumann and Tsallis entropies; the ceiling below needs neither property.
In the cut notation below, $\widehat{1j}$ denotes the complementary pair.

\begin{corollary}[Convex roofs]\label{cor:fqe-convex-roof} Let \(f\) satisfy the hypotheses of
Proposition~\ref{prop:fqe-majorant}. For each cut define the convex roof \[
E_f^{1j|\widehat{1j}}(\rho)=
\inf_{\rho=\sum_\ell p_\ell|\psi_\ell\rangle\langle\psi_\ell|}
\sum_\ell p_\ell\operatorname{tr}f(\rho_{1j}^{(\ell)}).
\] Then every mixed four-qubit state satisfies \[
\frac13\sum_{j=2}^4 E_f^{1j|\widehat{1j}}(\rho)
\le\mathcal E_f(M_4).
\] If both coefficient inequalities of Proposition~\ref{prop:fqe-majorant} are strict, equality holds exactly
for pure states in the two \(M_4\) LU orbits. For \(f=-x\log_2x\) these are the three bipartite
entanglements of formation. The same HS ceiling and pure-only equality classification hold for
the Tsallis entropies with $0<q<2$ and the R\'enyi entropies with $0<q<q_*$, including their
von Neumann limits at $q=1$.

\end{corollary}
To prove the ceiling, fix any pure-state ensemble $\rho=\sum_\ell p_\ell|\psi_\ell\rangle\langle\psi_\ell|$.
Each cut's roof is at most that cut's ensemble average, so the mean of the three roofs is at most
$\sum_\ell p_\ell\,\mathcal E_f(\psi_\ell)\le\mathcal E_f(M_4)$ by Proposition~\ref{prop:fqe-majorant} (in the R\'enyi case, with
$\mathcal E_q$ in place of $\mathcal E_f$, by Corollary~\ref{cor:fqe-renyi-gw}). Appendix~\ref{sec:fqe-spectral-proofs} excludes mixed maximisers:
every mixed-state support contains a vector of deficient Schmidt rank (fewer than four nonzero Schmidt coefficients
across a given $2|2$ cut), whereas every pure maximiser has full rank there.

\section{Stability of the four-qubit maximum}\label{sec:fqe-stability-summary}
A nearly maximal four-qubit entropy forces the state close to a maximising orbit. Let $d(\psi)$ be the minimum Hilbert-space
distance from a unit state $\psi$ to the LU orbits of $M_4$ and $\overline{M_4}$, allowing a global phase, and
$\delta(\psi)=1+\frac12\log_23-E(\psi)$ the entropy deficit. Then $\delta\ge d^2/20$ for every pure state
(Theorem~\ref{thm:fqe-orbit-stability}): an entropy within $\varepsilon$ of the maximum guarantees distance at most
$\sqrt{20\varepsilon}$ from a maximiser. Near the orbits the bound improves to $\delta\ge\frac3{10}d^2$ when $d\le1/20$,
and the power two is optimal. On the critical locus, where all one-qubit marginals are maximally mixed, the optimal
bound is $\delta\ge C_*d^2$ with $C_*\approx0.19058$, attained by explicit states (Theorem~\ref{thm:fqe-p1-sharp}),
so no universal coefficient can exceed $C_*$. The coefficient
$\tfrac3{10}$ holds only for $d\le1/20$, whereas the states attaining $C_*$ lie at distance $d\approx0.55$. The proof compares the infidelity to the orbits with a sextic moment
defect, with optimal coefficient one (Theorem~\ref{thm:fqe-global-overlap}). The same estimates transfer
to the Tsallis entropies of Section~\ref{sec:fqe-spectral} (Corollary~\ref{cor:fqe-orbit-spectral}).
Appendix~\ref{sec:fqe-orbit-stability} states these results with their constants and proof strategy.

\section{Seven qubits: compatible moments and entropy}\label{sec:fqe-seven-entropy}

The four-qubit proof bounds compatible marginal moments and then converts those bounds into an entropy bound.
We follow the same strategy for seven qubits, where AME states again do not exist~\cite{HGS}.
Five-copy positivity determines the purity minimum and its equality orbit. For entropy, a majorant involving
each marginal and its spin flip gives a second five-copy certificate. The two optima occur at different
members of one icosahedral family.

\paragraph{The two optimisation problems.} A balanced cut divides the seven qubits into groups of three and four.
It suffices to consider the three-qubit side $S$, whose reduced state $\rho_S$ has eight eigenvalues. There are
$\binom73=35$ such subsets, and we average over all of them:
\[
 E_3(\psi)=\frac1{35}\sum_{|S|=3}S(\rho_S),\qquad
 \pi_3(\psi)=\frac1{35}\sum_{|S|=3}\operatorname{tr}\rho_S^2 .
\]
Every cut has entropy at most $3$ bits and purity at least $1/8$, and reaching either bound on every cut
would require an AME state.

The mean purity has a direct interpretation in terms of correlations. Expanding each marginal in Pauli strings writes $\pi_3$ as $1/8$ plus a
positively weighted sum of the squared expectation values of all Pauli strings acting on one, two or three qubits
(Appendix~\ref{sec:fqe-seven}). An AME state would make all these low-weight correlations vanish. Minimising $\pi_3$
asks how small they can be when compatibility prevents them all from vanishing. Every seven-qubit stabilizer state has
$\pi_3\ge19/140$ (Proposition~\ref{prop:stab7}), strictly above the optimum $121/917$ of
Theorem~\ref{thm:fqe-seven-purity} below; hence the purity minimiser is not a stabilizer state.

\paragraph{The two optima.} Both optimisers are members of an explicit one-parameter family of seven-qubit states
$\psi_x$, defined below. Each has one distinguished qubit $s$; the other six qubits are labelled by the six axes of an
icosahedron. A single mixing weight $x\in[0,1/2]$ parametrises the family. The purity minimiser is the
\emph{icosahedral state} $\Psiico=\psi_{37/524}$, and the entropy maximiser is $\psi_{x_*}$, where $x_*=0.0698\ldots$
is the root of equation~\eqref{eq:fqe-xstar} below.

\begin{theorem}[the seven-qubit purity minimum]\label{thm:fqe-seven-purity}
Every pure seven-qubit state satisfies $\pi_3(\psi)\ge121/917$. Equality holds if and only if $\psi$ is
equivalent to $\Psiico$ under local unitaries and permutations of the qubits.
\end{theorem}
\noindent The proof is computer-assisted: a five-copy identity verified in floating point with rigorous error bounds,
followed by an exact equality analysis. It is outlined below and given in full in Section~\ref{sec:sup-seven-purity} of
the supplement.

\begin{theorem}[seven-qubit mean-entropy maximum and equality]
\label{thm:fqe-seven-entropy}
Every pure seven-qubit state satisfies
\[
 E_3(\psi)\le E_3(\psi_{x_*})=2.965356029477675899\ldots\ \text{bits}.
\]
Equality holds if and only if $\psi$ is local-unitarily and qubit-permutation equivalent to $\psi_{x_*}$.
With fixed qubit labels, the equality set consists of exactly $42$ LU orbits.
\end{theorem}
\noindent The proof is computer-assisted: a five-copy identity verified in floating point with rigorous error bounds,
an exact certificate for the scalar majorant, Theorem~\ref{thm:fqe-seven-purity}, and an exact equality analysis. It is
outlined below and given in full in Section~\ref{sec:sup-seven-entropy} of the supplement.

At four qubits $M_4$ both minimises the summed purity $P=3\pi_2$ and maximises entropy. At seven qubits these
objectives select different states (Table~\ref{tab:fqe-seven-summary}).

\begin{table}[htbp]
\centering\small
\begin{tabular}{@{}lccp{0.3\linewidth}@{}}
\toprule
 & ideal AME value (unattainable) & optimum & optimiser \\
\midrule
$\min\pi_3$ & $0.125$ & $\dfrac{121}{917}\approx0.131952$ & $\Psiico$, unique up to local unitaries and permutations \\[2ex]
$\max E_3$ (bits) & $3$ & $2.96535602947\ldots$ & $\psi_{x_*}$, unique up to local unitaries and permutations \\
\bottomrule
\end{tabular}
\caption{Mean three-qubit purity and entropy of seven-qubit pure states. The AME values are not attained. The states
$\Psiico$ and $\psi_{x_*}$ both lie in the icosahedral family $\psi_x$ defined below.}
\label{tab:fqe-seven-summary}
\end{table}

\begin{figure}[tbp]
\centering\includegraphics[width=\linewidth]{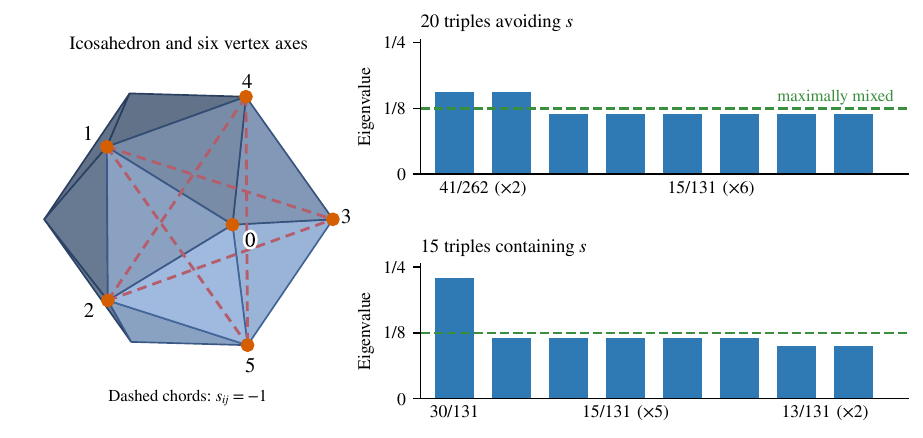}
\caption{Icosahedral geometry and marginal spectra of $\Psiico=\psi_{37/524}$.
Left: six oriented representatives of the axes through opposite icosahedral vertices label qubits
$0,\ldots,5$; the distinguished seventh qubit $s$ is separate. Dashed chords mark pairs whose amplitude sign
$s_{ij}$ is $-1$. Right: spectra of the $20$ three-qubit marginals avoiding $s$ and the $15$ containing it;
multiplicities in parentheses (the values $(4u)^6,(3u+x)^2$ and $8u,(4u)^5,(u+x)^2$ above at $u=15/524$); every marginal in a class has the spectrum shown. The dashed horizontal line marks the maximally mixed eigenvalue
$1/8$. This state minimises the mean purity over all pure states. The entropy maximiser $\psi_{x_*}$,
$x_*=0.0698\ldots$, is a different member of the same family; its spectra have the same multiplicities, and each
eigenvalue differs from the one shown by less than $10^{-3}$, too little to see at this scale.}
\label{fig:fqe-icosahedral}
\end{figure}

\paragraph{An explicit state.} Label the qubits $s,0,1,\dots,5$, with $0,\dots,5$ the axes through opposite vertices of
the icosahedron. Let $e_i$ denote the six-bit string with a single
$1$ at site $i$. Set $s_{ij}=-1$ for $\{i,j\}$ an edge of the pentagon $1\!-\!3\!-\!2\!-\!4\!-\!5\!-\!1$, and
$s_{ij}=+1$ otherwise. For $0\le x\le1/2$ put $\alpha=\sqrt{(1-2x)/30}$, $\beta=\sqrt x$ and (equation~\eqref{eq:ico} of
Appendix~\ref{sec:ico})
\[
\begin{aligned}
  \psi_x ={}& |0\rangle_s\Bigl(\alpha\sum_{i<j}s_{ij}|e_i+e_j\rangle + \beta|111111\rangle\Bigr)\\
         &+ |1\rangle_s\Bigl(\beta|000000\rangle - \alpha\sum_{i<j}s_{ij}|111111-e_i-e_j\rangle\Bigr).
\end{aligned}
\]
Every member has maximally mixed one-qubit marginals, and its three-qubit marginals fall into two spectral classes
(Theorem~\ref{thm:ico}). Writing $u=(1-2x)/30$, the $20$ marginals avoiding $s$ have eigenvalues $4u$ six times and
$3u+x$ twice; the $15$ containing $s$ have eigenvalues $8u$ once, $4u$ five times and $u+x$ twice. Consequently the mean purity $\pi_3(\psi_x)$ is an explicit quadratic in $x$, with minimum $121/917$ at
$x=37/524$: the icosahedral state $\Psiico$ (Figure~\ref{fig:fqe-icosahedral}).

The icosahedron enters through the signs. The matrix $C$ with zero diagonal and entries $s_{ij}$ off it is a symmetric
\emph{conference matrix}: $C^2=5I$. Its eigenvalues are $\pm\sqrt5$, each three times, so $I+C/\sqrt5$ is positive
semidefinite of rank three with unit diagonal. It is therefore the Gram matrix of six unit vectors in $\mathbb R^3$ with
$v_i\cdot v_j=s_{ij}/\sqrt5$: six equiangular lines~\cite{LS} at angle $\arccos(1/\sqrt5)$, which are the axes through opposite
vertices of an icosahedron. The same identity block-diagonalises every two- and three-qubit marginal of $\psi_x$ in
the computational basis, which is how Theorem~\ref{thm:ico} obtains the spectra.

The family is the fixed set of a symmetry group generated by local unitaries, qubit permutations and complex
conjugation (up to signs and phase). Hence a member stationary within the family is stationary on the whole sphere for
the purity, the entropy or any other differentiable functional invariant under these symmetries
(Lemma~\ref{lem:seven-family-symmetric-criticality} of the supplement, which identifies the group).

\paragraph{Why an icosahedron.} An exact purity identity shows why minimising purity selects this geometry. Widen the
family: replace the pair amplitudes $\alpha s_{ij}$ of $\psi_x$ by complex numbers $t_{ij}$ in the $|0\rangle_s$ branch
and by their conjugates in the $|1\rangle_s$ branch. They form a complex symmetric $6\times6$ matrix $Q=(t_{ij})$ with
zero diagonal; put $p_i=\sum_j|t_{ij}|^2$ and $\sigma=\sum_{i<j}|t_{ij}|^2$, so normalisation is $2\sigma+2x=1$. For every
such state the weight-one and weight-three Pauli correlations on the six ordinary qubits cancel between the two branches
(Theorem~\ref{thm:spinflip-double} of the supplement), and (Theorem~\ref{thm:complexsos} of the supplement)
\begin{equation}\label{eq:fqe-complex-sos}
\begin{aligned}
 \pi_3-\frac{121}{917}={}&\frac{(524x-37)^2}{137550}+\frac4{35}\sum_i\Bigl(p_i-\frac\sigma3\Bigr)^2\\
 &+\frac{48}{35}\sum_{i<j}\Bigl(|t_{ij}|^2-\frac\sigma{15}\Bigr)^2+\frac{24}{35}\sum_{i<j}\bigl|(QQ^\dagger)_{ij}\bigr|^2.
\end{aligned}
\end{equation}
At equality $x=37/524$, all $|t_{ij}|$ are equal and the rows of $Q$ are orthogonal: $Q$ is a multiple of a complex
conference matrix of order six, which is unique up to diagonal phases and permutations
(Proposition~\ref{prop:confphase} of the supplement) and is the icosahedral sign pattern above. Minimum purity thus
forces the six rows of pair amplitudes to be orthogonal with equal moduli. The identity is exact only within this
family and does not bound arbitrary states; the global bound is the five-copy certificate of
Theorem~\ref{thm:fqe-seven-purity}, whose equality analysis reduces an arbitrary minimiser to this family and this identity.

\paragraph{Entropy selects a different state.} Since the marginal eigenvalues of $\psi_x$ are affine in $x$, its mean entropy
is explicit: with $f(t)=-t\log_2t$ and $u=(1-2x)/30$,
\[
 E_3(\psi_x)=\tfrac1{35}\bigl[195f(4u)+40f(3u+x)+15f(8u)+30f(u+x)\bigr]
\]
(Appendix~\ref{sec:ico}). It is strictly concave in $x$, and its unique maximum occurs at the root $x_*\in(0,1/2)$ of
\begin{equation}\label{eq:fqe-xstar}
 2^{32}(1-2x)^{15}=3^8(1+8x)^8(1+28x)^7
\end{equation}
(Proposition~\ref{prop:entropy-seven-stationarity} of the supplement),
giving $E_3(\psi_{x_*})=2.965356029\ldots$ bits. This root lies strictly below $37/524$
(Corollary~\ref{cor:ico-entropy-separation}), so $\Psiico$ is not stationary for the entropy, even within the family.
By Theorem~\ref{thm:fqe-seven-purity} no purity minimiser maximises the mean entropy: unlike at four qubits, the two
problems have different solutions. Their objective values are nevertheless very close: $E_3(\Psiico)$ lies only $7.1597\ldots\cdot10^{-6}$ bits
below the maximum, and $\pi_3(\psi_{x_*})$ exceeds $121/917$ by $1.2204\ldots\cdot10^{-6}$ (rigorous interval enclosures).

The marginal moments give a heuristic explanation for this difference; the exact derivative in
Corollary~\ref{cor:ico-entropy-separation} supplies the proof. In the Taylor expansion of $f$ about the mean eigenvalue $1/8$,
the second-moment contribution favours small $\pi_3$, while the positive cubic coefficient favours a large
$b_3=\tfrac1{35}\sum_{|S|=3}\operatorname{tr}\rho_S^3$. At four qubits $M_4$ does both: it minimises $P$ and maximises
the cubic moment $T$ on the slice $P=1$ (Theorem~\ref{thm:fqe-moment}). Along $\psi_x$, instead,
$b_3(\psi_x)=(35008x^3+17328x^2-3084x+709)/31500$ has $b_3'(37/524)=-432/120127<0$ (Appendix~\ref{sec:ico}): moving
$x$ below $37/524$ raises $b_3$ at first order and costs $\pi_3$ only at second order.

The marginal spectra and pair purities of the minimiser agree, to the six digits published, with
those of a seven-qubit state found numerically by Di~Martino, Facchi and Florio~\cite{DMFF18}. This suggests that the two
states are LU-equivalent (Appendix~\ref{sec:ico}); Theorem~\ref{thm:ico} gives an explicit closed-form state.

\paragraph{Copies of the state.} A purity is an expectation value in two copies, and a homogeneous polynomial of
degree $m$ in $|\psi\rangle\langle\psi|$ is an expectation value in $m$ copies. Averaging the projector onto
$\psi^{\otimes m}$ over local unitaries, qubit permutations and complex conjugation gives the \emph{twirled $m$-copy
moment}. Replacing some copies by the spin-flipped state $\tilde\psi=\epsilon^{\otimes n}\bar\psi$ (which differs from $\sigma_y^{\otimes n}\bar\psi$ only by the phase $i^n$),
$\epsilon=i\sigma_y$, gives further positive operators, all linear images of it. Minimising $\pi_3$ over all moment operators subject to these positivity conditions gives lower bounds
that increase with $m$ to the true minimum, a bosonic-extension hierarchy in the spirit of~\cite{DPS}, and close to the copy hierarchy of Yu et al.\ for the pure-state
marginal problem~\cite{YSWNG} (Appendix~\ref{sec:fqe-seven} and Theorem~\ref{thm:spinflip-ppt-hierarchy} of
the supplement). A certificate at level $m$ is an identity
\begin{center}
(objective) $-$ (bound) $=$ a sum of squared norms of linear images of $m$ copies of $\psi$, some spin flipped.
\end{center}
At four qubits $P\ge1$ follows at the two-copy level from Rains' shadow inequalities~\cite{Rains}, and the sextic
identity~\eqref{eq:fqe-remainder} is a three-copy certificate for the different objective $F$ (Appendix~\ref{sec:fqe-seven}). At seven qubits the levels separate: two copies give only the AME value $1/8$, three copies
provably cannot reach $121/917$, four copies provably fall short (numerically by about $7.5\cdot10^{-5}$), and five copies are the first level that attains the sharp purity bound
(Table~\ref{tab:fqe-seven-levels}).

\paragraph{Proof outline for the purity minimum.} The proof has three steps; the entropy proof below repeats them at a
different point of the family.
\begin{enumerate}
\item \emph{Bound.} A five-copy identity writes $\pi_3-121/917$ as a sum of twelve terms, each the expectation value of a
positive operator in five copies of $\psi$, some of them spin flipped, averaged over local unitaries and qubit
permutations. Every term is therefore nonnegative, which gives the bound. Unlike at four qubits, the identity needs
computer assistance: floating-point linear algebra with rigorous error bounds establishes that suitable positive definite
coefficient matrices exist.
\item \emph{Equality moments.} At equality all twelve terms vanish. Their vanishing fixes the averaged five-copy moment of
$\psi$ to that of $\Psiico$ (Theorem~\ref{thm:fivecopy-family-equality} of the supplement): every conjugation-even,
permutation-invariant five-copy invariant of $\psi$ takes its icosahedral value.
\item \emph{Equality state.} A reconstruction theorem (Theorem~\ref{thm:moment-orbit-rigidity} of the supplement) then
shows that $\psi$ differs from $\Psiico$ only by local unitaries and a qubit permutation. Some of these invariants put
$\psi$ in the form of the widened family above, and the identity~\eqref{eq:fqe-complex-sos}
makes the amplitude matrix $Q$ a conference matrix.
With qubit labels fixed, the orbit of $\Psiico$ consists of $42$ LU orbits: $7$ choices of the distinguished qubit times
$6$ inequivalent sign patterns of the conference matrix.
\end{enumerate}
Appendix~\ref{sec:fqe-seven} defines the terms and records the tangent-line step, the invariants used and the sizes of the
computations; Section~\ref{sec:fivecopy} of the supplement gives the full proof.

\paragraph{Proof outline for the entropy maximum.} The proof follows the four-qubit pattern: a polynomial majorant
touching the entropy at the spectra of the maximiser, then a moment certificate. Across the 35 cuts of $\psi_{x_*}$, four
distinct eigenvalues occur. A single one-variable majorant of $-t\log_2t$ sharp on every cut would therefore touch at
four points and need degree at least seven, and hence seven copies (Proposition~\ref{prop:seven-entropy-majorant-degree}
of the supplement). We instead use each marginal $\rho_S$ together
with its spin flip $\widetilde\rho_S=\sigma_y^{\otimes3}\rho_S^T\sigma_y^{\otimes3}$, which is also the marginal of
$\tilde\psi$ on $S$; it has the spectrum of $\rho_S$ but in general different eigenvectors. At $\psi_{x_*}$ the two commute,
and their joint eigenvalue pairs $(\lambda,\mu)$ take only five values. With $u=(1-2x_*)/30$, the cut
eigenvalues of $\psi_{x_*}$ are $\nu_1=3u+x_*$, $\nu_2=8u$, $\nu_3=4u$ and $\nu_4=u+x_*$ (Theorem~\ref{thm:ico}), and the
pairs are $(\nu_3,\nu_3)$, $(\nu_1,\nu_1)$, $(\nu_4,\nu_4)$, $(\nu_2,\nu_3)$ and $(\nu_3,\nu_2)$.
A scalar inequality bounds the average entropy of two eigenvalues by a degree-five polynomial: there is a symmetric
polynomial $H(\lambda,\mu)$ of degree five majorising $\tfrac12[-\lambda\log_2\lambda-\mu\log_2\mu]$, with first-order
contact at those five pairs (Section~\ref{sec:seven-joint-analytic-proof} of the supplement proves it, with exact
one-variable sign checks). For arbitrary marginals the two eigenbases need not coincide. Their squared overlaps
nevertheless form a doubly stochastic matrix, so averaging the scalar inequality with these weights gives, with
$h_{kl}$ the coefficients of $H$,
\[
S(\rho_S)\le
\sum_{k,l}h_{kl}\operatorname{tr}(\rho_S^k\widetilde\rho_S^{\,l}),
\]
even when $\rho_S$ and $\widetilde\rho_S$ do not commute. The right side is a five-copy moment.

It suffices to bound this polynomial for states attaining or exceeding $E_3(\psi_{x_*})$, so the certificate is only
needed on a compact domain containing them, cut out by two technical restrictions (Proposition~\ref{prop:seven-competitor-band}
and Corollary~\ref{cor:seven-entropy-eigen-cap-497} of the supplement). The purity minimum and the entropy--purity slope
for eight eigenvalues give the \emph{purity band} $121/917\le\pi_3<8693/64000$, and
spin-flip compatibility between overlapping cuts caps every marginal eigenvalue at $497/1000$. On this
domain a positive five-copy identity bounds the polynomial by the candidate value $E_3(\psi_{x_*})$
(Theorem~\ref{thm:seven-entropy-localized-identity} of the supplement; Appendix~\ref{sec:fqe-seven} lists its terms).
At equality all its terms vanish, and the equality-moment argument and reconstruction
theorem used for purity identify the orbit (Theorems~\ref{thm:fivecopy-family-equality} and~\ref{thm:moment-orbit-rigidity} of the
supplement).

\paragraph{Stability.} Both optima are quadratically stable: each optimiser is a strict local optimum modulo local
unitaries and phase with an explicit quadratic basin, and compactness extends this to a global lower bound on the objective deficit by a constant
times the squared distance to the orbit under local unitaries and qubit permutations. We do not make the constant explicit (Theorems~\ref{thm:icolocal},
\ref{thm:icoquantbasin}, \ref{thm:entropy-seven-local} and~\ref{thm:entropy-seven-quantbasin} and
Corollary~\ref{cor:seven-global-quadratic} of the supplement).

\paragraph{Comparison with four qubits.} The two sizes differ in the equality set. At four qubits the purity minimum $P=1$ is attained on a positive-dimensional
set, containing $M_4$ and the cluster state, and the entropy needs the cut spectrum to single out $M_4$
(Proposition~\ref{prop:fqe-onecut}). At seven qubits the purity alone selects a single orbit under LU and qubit permutations,
and the two-copy level of the hierarchy is no longer sharp (Section~\ref{sec:fqe-discussion} compares other sizes).

Table~\ref{tab:fqe-four-seven} summarises the common method and the additional constraints needed at seven qubits. The degree rule of
Section~\ref{sec:fqe-entropy_maximum-entropy-bound} explains both majorants: $k=2$ gives the cubic of Section~\ref{sec:fqe-entropy_maximum-entropy-bound}, and the joint
majorant evades $k=4$ by adding the spin-flip variable. It also shows that degree five is the least possible for a joint
majorant. A joint majorant sharp at $\psi_{x_*}$ must touch at the three diagonal pairs $(\lambda,\lambda)$ among its
contacts. On the diagonal it is therefore a majorant of $-t\log_2t$ touching at three distinct points, which needs degree at least five
(Corollary~\ref{cor:seven-joint-degree-five} of the supplement). Within this majorant method five copies are therefore necessary for the entropy; that says nothing about other certificates.

\begin{table}[ht]
\centering\small
\renewcommand{\arraystretch}{1.15}
\begin{tabular}{@{}lll@{}}
\toprule
 & four qubits, $2|2$ cuts & seven qubits, $3|4$ cuts\\
\midrule
mean purity minimum & $\pi_2=1/3$ (two copies) & $\pi_3=121/917$ (five copies)\\
purity equality set & positive-dimensional, contains $M_4$ & one orbit, $\Psiico$\\
entropy maximiser & $M_4$ & $\psi_{x_*}\ne\Psiico$\\
majorant & cubic in $\lambda$, contacts $1/6,1/2$ & degree five in $(\lambda,\mu)$, five contact pairs\\
moment certificate & sextic identity, three copies & certified identity, five copies\\
extra constraints & none & purity band, cap $\lambda\le497/1000$\\
entropy equality & $M_4$ orbit (Theorem~\ref{thm:fqe-entropy-max}) & $\psi_{x_*}$ orbit\\
\bottomrule
\end{tabular}
\caption{The copy-hierarchy method at four and seven qubits. Here $\pi_2=P/3$ and $\pi_3$ are the mean balanced cut
purities. Orbits include local unitaries and qubit permutations; $\mu$ is an eigenvalue of the spin-flipped marginal.}
\label{tab:fqe-four-seven}
\end{table}

\section{Discussion and open problems}\label{sec:fqe-discussion}

\paragraph{Intermediate R\'enyi orders.} Gour and Wallach conjectured that $M_4$ maximises the mean
R\'enyi-$q$ entropy for $q<\alpha_0\approx1.59$ and the cluster states for $q>\alpha_0$, and proved the
cluster-state bound for $q\ge2$~\cite[Fig.~2 and Theorem~12]{GW}; their $\alpha_0$ is the crossing $q_*$ of
Proposition~\ref{prop:fqe-renyi-crossing}, and Corollary~\ref{cor:fqe-renyi-gw} proves the conjecture. For $1<q<2$ a
single-crossing property of individual spectra (Lemma~\ref{lem:fqe-renyi-single-crossing}) transfers two anchor bounds to
all intermediate orders, and Appendix~\ref{sec:fqe-renyi-proofs} gives the argument and the finite checks it uses.

\paragraph{Quantitative stability.} The universal coefficient $1/20$ in
Theorem~\ref{thm:fqe-orbit-stability} is not claimed to be sharp, and whether the critical optimum $C_*$ of
Section~\ref{sec:fqe-stability-summary} holds for all states is open. Numerical tests of locally filtered
critical states have found no smaller $\delta/d^2$
(Remark~\ref{rem:fqe-critical-noncritical} of the supplement). Local filtering does not preserve the entropy deficit,
even though the overlap bound of Theorem~\ref{thm:fqe-global-overlap} holds without a
criticality assumption.

\paragraph{Seven-qubit questions.} The five-copy proofs give sharp purity and
entropy bounds, but an analytic certificate comparable to the four-qubit
sextic identity would better explain the global role of the icosahedral
geometry. The entropy ceiling also holds for mixed-state convex roofs, with equality only for pure states
(Proposition~\ref{prop:seven-mixed-roofs} of the supplement); where the four-qubit argument finds a support vector of
deficient Schmidt rank, the seven-qubit one uses that the optimiser's local-unitary tangent space contains no complex line.

\paragraph{The method.} At both sizes the proof starts from the spectra of the candidate optimiser, builds a
polynomial majorant touching the entropy at those spectra, bounds the required moments by positivity of
multicopy operators with some copies spin flipped, and reads off the equality states from the vanishing
terms. The spin-flip constraints rely on the pseudoreality $\epsilon\bar u\epsilon^{-1}=u$ of $SU(2)$. For $SU(d)$,
$d\ge3$, the defining representation is inequivalent to its conjugate, so this form of the multicopy argument does not extend directly
to qudits.

\paragraph{Purity and entropy at other sizes.} Table~\ref{tab:fqe-other-n} collects what is known. Four qubits is the
smallest size at which the question has content. At five qubits AME states exist, so the mean entropy reaches its
elementary upper bound, exactly at the AME states; their LU classes form a continuum, each LU-equivalent to a point of the
unique $((5,2,3))$ code (five qubits, a two-dimensional code space and distance three), with points identified under a
group of order~24~\cite{Tan}. The minimum purity is
the value of the Rains linear programme~\cite{Rains,HESG}, the two-copy level of the hierarchy, attained by a graph state, at
every $n\le18$ except $7$ and $13$ (Theorems~\ref{thm:lptable}, \ref{thm:fourteenfifteen}, \ref{thm:sixteenseventeen}
and~\ref{thm:eighteen} of the supplement~\cite{Companion}). A graph state is the stabilizer state obtained from
$|+\rangle^{\otimes n}$ by a controlled-$Z$ gate along each edge of a graph on the qubits; every stabilizer state is
locally equivalent to one~\cite{VdN,HEB}, and $C_4$ is locally equivalent to the graph state of the four-vertex path. Seven qubits is the first size at which the two-copy level is
not sharp: the linear-programming value would require the weight distribution of a pure $((7,1,4))_2$ code, which does
not exist~\cite{HGS}. At thirteen qubits it is not sharp either: the three-copy level gives $\min\pi_6\ge0.018171249$, above the
linear-programming value $25/1408$, and an explicit stabilizer state gives $\min\pi_6\le85/4576$, but the exact minimum
is open (Proposition~\ref{prop:thirteeninterval} of the supplement); here the linear-programming value would require a
pure $((13,1,6))_2$ code, which the three-copy bound excludes.

The maximum entropy is known only at the AME sizes
and at four and seven qubits. Elsewhere each purity minimum, or at thirteen qubits its lower bound, gives an explicit
entropy ceiling through a sharp linear inequality: every $d$-dimensional density matrix, $d\ge3$, satisfies
$\log_2d-S(\rho)\ge c_d(d\operatorname{tr}\rho^2-1)$ with $c_d=(d-1)\log_2(d-1)/(d(d-2))$
(Lemma~\ref{lem:seven-entropy-linear-purity} of the supplement). Being linear in the purity, this inequality averages over the
balanced cuts, $E_k\le k-c_{2^k}(2^k\pi_k-1)$, whereas the largest entropy at fixed purity is not concave in the purity and
does not average this way. At thirteen qubits an explicit state also gives $\max E_6>5.8336$
(Corollary~\ref{cor:thirteen-entropy-witness} of the supplement). The ceilings are not attained by the graph-state
purity minimisers. A stabilizer state's marginals are normalised projectors, so unless $n\in\{2,3,5,6\}$ some balanced
marginal is rank-deficient; such a marginal can always be perturbed to raise the entropy at order $t^2|\ln t|$.
Hence no stabilizer state is even a local maximum of the mean balanced entropy
(Corollary~\ref{cor:stabilizer-entropy-noname} of the supplement); in particular neither seven-qubit optimiser is a
stabilizer state.

\begin{table}[htbp]
\centering\small
\begin{tabular}{@{}cccl@{}}
\toprule
$n$ & ideal purity $2^{-k}$ & $\min\pi_k$ & $\max E_k$ (bits)\\
\midrule
2, 3 & $1/2$ & $1/2$ & $1$ (AME)\\
4 & $1/4$ & $1/3$ & $1+\tfrac12\log_23$ (Theorem~\ref{thm:fqe-entropy-max})\\
5 & $1/4$ & $1/4$ & $2$ (AME)\\
6 & $1/8$ & $1/8$ & $3$ (AME)\\
7 & $1/8$ & $121/917$ & $2.96535602947\ldots$ (Theorem~\ref{thm:fqe-seven-entropy})\\
8 & $1/16$ & $3/35$ & $<3.9029$\\
9 & $1/16$ & $1/14$ & $<3.9627$\\
10 & $1/32$ & $1/21$ & $<4.9163$\\
11 & $1/32$ & $1/28$ & $<4.9772$\\
12 & $1/64$ & $5/224$ & $<5.9594$\\
13 & $1/64$ & $[0.018171249,\ 85/4576]$ & $(5.8336,\ 5.9846)$\\
14 & $1/128$ & $331/27456$ & $<6.9702$\\
15 & $1/128$ & $1/104$ & $<6.9874$\\
16 & $1/256$ & $1/156$ & $<7.9800$\\
17 & $1/256$ & $1/208$ & $<7.9928$\\
18 & $1/512$ & $7/2288$ & $<8.9901$\\
\bottomrule
\end{tabular}
\caption{Mean balanced purity and entropy of $n$-qubit pure states. With $k=\lfloor n/2\rfloor$, $\pi_k$ and $E_k$ are
the mean purity and mean von Neumann entropy of the $k$-qubit marginals over all $\binom nk$ subsets of $k$ qubits. The
ideal purity $2^{-k}$ and entropy $k$ are attained exactly at the AME sizes $n=2,3,5,6$. The known purity minima other than
at $n=7$ are Rains linear-programming values attained by graph states (Table~\ref{tab:minima} and
Theorem~\ref{thm:eighteen} of the supplement); at $n=13$ the minimum is open, and the entry is the interval of
Proposition~\ref{prop:thirteeninterval} of the supplement. Single entropy entries without $<$ are exact maxima; entries with $<$
are upper bounds on unknown maxima, the ceilings $k-c_{2^k}(2^k\min\pi_k-1)$ rounded up, with the lower end of the
purity interval in place of $\min\pi_k$ at $n=13$. There the maximum is open; the upper end of its interval is that
ceiling, and the lower end is exceeded by an explicit state (Corollary~\ref{cor:thirteen-entropy-witness} of the
supplement).}
\label{tab:fqe-other-n}
\end{table}

\paragraph{Purity and time reversal.}\label{rem:fqe-spinflip-tradeoff}
A large overlap with the spin-flipped state costs purity: at $n=4$, $\pi_2=\tfrac13+\tfrac16(W_1+\tau)$
(Lemma~\ref{lem:fqe-purity}), and Rains' shadow inequalities give sharp analogues at larger sizes
(Section~\ref{sec:spinflip-tradeoffs} of the supplement~\cite{Companion}).

\paragraph{Formal verification.} The four-qubit maximum, its attainment by $M_4$ and its equality case are formalised in
Lean~4 with Mathlib~\cite{Companion}, together with the moment inequality, the general and critical envelopes, most of the
four-qubit spectral corollaries, the overlap bound with the global orbit-stability estimates, the sharp quadratic
variance bound with the R\'enyi envelope above order one, and the exact
computations for the icosahedral family. Finite checks of two certificates, for the overlap bound and the quadratic
variance bound, use compiled code; the results depending on them inherit that trust assumption, and everything else uses
the Lean kernel alone. The global seven-qubit theorems are not formalised. Section~\ref{sec:fqe-lean} of the supplement
lists the scope, declarations and trust assumptions.

\section*{Use of AI}
All results, documents and scripts were generated by AI agents (mainly Claude Opus and Sonnet 5.5, and OpenAI's GPT 6.1 sol) with human supervision and many cross-checks,
including AI-generated Lean verification of most of the four-qubit results.

\section*{Acknowledgements}
We acknowledge support by the UK STFC grant ST/T000473/1.

\appendix
\section{The sextic from elementary swap contractions}\label{sec:fqe-sextic}

The proof of Theorem~\ref{thm:fqe-moment} has two parts. We first express three tensor contractions as marginal
moments in \eqref{eq:fqe-contractions}. We then eliminate their mixed moment $\Pi_2$ using norms of one-qubit
marginals acting on the state. The derivation uses two-dimensional antisymmetry, partial traces and matrix algebra;
the eleven-type table evaluates the longest contraction. An independent coefficient-by-coefficient check of
\eqref{eq:fqe-remainder} in the sixteen complex amplitudes is also available (SI-1, SI-2,
Appendix~\ref{sec:fqe-verifiers}); it does not use that table.

Write \(R_S=\operatorname{tr}_{S^c}|z\rangle\langle z|\), \(N=\|z\|^2\),
\(P=\sum_{j=2}^4\operatorname{tr}R_{1j}^2\), \(T=\sum_{j=2}^4\operatorname{tr}R_{1j}^3\),
\(\Pi_1=\sum_i\operatorname{tr}R_i^2\) and \(\Pi_2=\sum_{i<j}\operatorname{tr}[R_{ij}(R_i\otimes
R_j)]\). The unnormalised Bloch vectors \(\mathbf r_i=(r_i^a)\) have components \(r_i^a=\langle
z|\sigma_a^{(i)}|z\rangle\); put \(W_1=\sum_i\|\mathbf r_i\|^2=2\Pi_1-4N^2\) and \(F=17NP-12T-12N^3\).

\subsection{One residual tensor and the local moment
map}\label{sec:fqe-sextic_identity-one-residual-tensor-and-the-local-moment-map}

Let \(\epsilon=\begin{pmatrix}0&1\\-1&0\end{pmatrix}=i\sigma_y\) and \(w=z_x\otimes z_y\otimes\tilde z_t\), \(\tilde z=\epsilon^{\otimes4}\bar z\). At one physical site define maps from three qubits to one by \[
L|xyt\rangle=\epsilon_{yt}|x\rangle,\qquad
M|xyt\rangle=\epsilon_{xt}|y\rangle.
\] For each site \(i\), let \(S_i\) symmetrise its three slots (the orthogonal projector onto the symmetric subspace). On the other three sites contract
with \(L\), or with \(L+M\), respectively: \[
U_i=S_i\!\prod_{j\ne i}L_jw,\qquad
V_i=S_i\!\prod_{j\ne i}(L_j+M_j)w.
\] These are ordinary vectors on six qubits (three slots at site \(i\), one at each other site), with the inherited
Hilbert-space norm. Write \(R_i^{(i)}\) for the one-qubit marginal
\(R_i\) acting on qubit \(i\) of \(\mathbb C^{16}\). The exact identity is \[
\boxed{\quad8F=6NW_1+3\sum_i\|V_i-8U_i\|^2+12\sum_{i<j}\bigl\|(R_i^{(i)}-R_j^{(j)})z\bigr\|^2.\quad} \tag{S}
\] It immediately proves the sharp bound \(F\ge\tfrac34NW_1\ge0\). For a product state
\(z=\sqrt N\,a_1\otimes\cdots\otimes a_4\) every \(R_i^{(i)}z=Nz\), so the pair norms vanish, and \(F=3N^3=\tfrac34NW_1\);
the constant \(3/4\) cannot be improved. Proposition~\ref{prop:focus} of the supplement writes \(32F\) as a sum of
site terms alone; (S) keeps the separate \(W_1\) term, which gives this bound and the envelope below.

\emph{How the terms are chosen.} All norms and overlaps of \(U_i,V_i\) are expectation values in \(z^{\otimes3}\) of
slot permutations (Section~\ref{sec:fqe-sextic_identity-the-three-contractions}), so they are polynomials in the
marginal moments. Besides \(N^3,NP,T\), they involve \(\Pi_1\) (equivalently \(W_1\)) and the mixed moment \(\Pi_2\), which
\(F\) does not contain. The pair norms contain exactly this mixed moment. Since \((R_i^{(i)})^2\) acts on qubit \(i\) only,
and a \(2\times2\) matrix of trace \(N\) has \(\operatorname{tr}R^3=\tfrac12(3N\operatorname{tr}R^2-N^3)\),
\[
\bigl\|(R_i^{(i)}-R_j^{(j)})z\bigr\|^2=\operatorname{tr}R_i^3+\operatorname{tr}R_j^3-2\operatorname{tr}[R_{ij}(R_i\otimes R_j)],
\qquad
\sum_{i<j}\bigl\|(R_i^{(i)}-R_j^{(j)})z\bigr\|^2=\tfrac92N\Pi_1-6N^3-2\Pi_2 .
\tag{Pr}\label{eq:fqe-pair-norms}
\]
In \(\sum_i\|V_i-tU_i\|^2\) the coefficient of \(N^3\) is \(-\tfrac43(t-8)(t+4)\). At the root \(t=8\) the \(N^3\)
terms cancel, and the \(\Pi_2\) that remains is removed by the pair norms, leaving the nonnegative multiple
\(6NW_1=18\|U\|^2\); at the other root \(t=-4\) the same elimination leaves a negative multiple of \(NW_1\).

The first tensor is just local polarisation. Contracting the three epsilon pairs gives
\(U_i=-S_i(z_x\otimes(R_i\epsilon^T)_{yt})\). Diagonalise \(R_i\) with eigenvalues \(\lambda,\mu\),
and write \(z=\sqrt\lambda\,|0\rangle|\varphi_0\rangle+\sqrt\mu\,|1\rangle|\varphi_1\rangle\) with orthonormal
\(\varphi_a\) on the remaining sites. Then \(R_i\epsilon^T\) has entries \(-\lambda\) at \((y,t)=(0,1)\) and \(\mu\) at \((1,0)\), the two
Schmidt terms of \(U_i\) are orthogonal, and symmetrising \(|0\rangle(\mu|10\rangle-\lambda|01\rangle)\) and
\(|1\rangle(\mu|10\rangle-\lambda|01\rangle)\) gives \((\mu-\lambda)\) times
\(\tfrac13(|001\rangle+|010\rangle+|100\rangle)\) and \(\tfrac13(|011\rangle+|101\rangle+|110\rangle)\), each of squared norm \(1/3\); the two Schmidt terms carry the weights \(\lambda\) and \(\mu\). Hence \[
\|U_i\|^2=\frac{(\lambda+\mu)(\lambda-\mu)^2}{3}
=\frac N3\|\mathbf r_i\|^2. \tag{U}\label{eq:fqe-local-polarization}
\] Thus the first term in \eqref{eq:fqe-remainder} is \(18\|U\|^2\), an ordinary one-qubit marginal defect. For
nonzero \(z\), equality \(F=0\) forces every \(R_i=NI/2\), hence \(U_i=0\); it then requires \(V_i=0\) for all \(i\).
Section~\ref{sec:fqe-sextic_identity-all-zeros-and-why-the-moment-bound-is-tight} classifies the
remaining zeros.

\subsection{The three contractions}\label{sec:fqe-sextic_identity-the-three-contractions}

Write \(\|U\|^2=\sum_i\|U_i\|^2\), \(\|V\|^2=\sum_i\|V_i\|^2\) and
\(\langle U,V\rangle=\sum_i\langle U_i,V_i\rangle\). We will show \[
\begin{aligned}
\|U\|^2&=(2N\Pi_1-4N^3)/3,\\
\operatorname{Re}\langle U,V\rangle&=(-8N^3-2N\Pi_1+8\Pi_2)/3,\\
\|V\|^2&=(128N^3-226N\Pi_1+136NP+152\Pi_2)/3-32T. \end{aligned}\tag{C}\label{eq:fqe-contractions}
\] These contraction identities imply \eqref{eq:fqe-remainder} by direct substitution:
\(\sum_i\|V_i-8U_i\|^2=\|V\|^2-16\operatorname{Re}\langle U,V\rangle+64\|U\|^2\), and
\eqref{eq:fqe-contractions} turns this into \[
\sum_i\|V_i-8U_i\|^2=\tfrac{136}{3}NP-32T-22N\Pi_1+8\Pi_2 ,
\] in which \(N^3\) has cancelled. Adding twelve times \eqref{eq:fqe-pair-norms} and \(6NW_1=12N\Pi_1-24N^3\) to three
times this gives \(136NP-96T-96N^3=8F\), as claimed. The rest of this subsection
proves \eqref{eq:fqe-contractions}. Every step uses only the invariance of $z^{\otimes3}$ under simultaneous permutations of the
copies, the vanishing of the one-site antisymmetriser and averaging over sites, so \eqref{eq:fqe-contractions} holds as a
linear identity in the twirled three-copy moment with one flipped copy.

\paragraph{Slot permutations.} In this subsection \(p=(12)\), \(q=(13)\), \(r=(23)\) are the
transpositions of the three slots \((x,y,t)\) of one site, acting by permuting the three tensor factors, and \(c\)
is the cycle \(c|a_1a_2a_3\rangle=|a_2a_3a_1\rangle\), with inverse
\(c^{-1}|a_1a_2a_3\rangle=|a_3a_1a_2\rangle\). (Here the letters \(p,q,r\) denote only these
transpositions; in particular \(q\) is not the Tsallis exponent.) Products are composed as operators, right factor first; then \[
cr=q,\qquad pr=c^{-1},\qquad qr=c,\qquad pq=c,\qquad qp=c^{-1}.
\] Antisymmetrising three vectors in dimension two gives zero, so \[
I-p-q-r+c+c^{-1}=0,\qquad S=\tfrac16(I+p+q+r+c+c^{-1})=(p+q+r)/3, \tag{A}\label{eq:fqe-antisym}
\] where \(S\) is the symmetriser of the three slots of one site (a projector).

\paragraph{The kernels.} Let \(\Gamma_3\) denote the partial transpose on slot \(3\), and for an
operator \(A\) on the three slots of a site put
\(\mathfrak t(A)=\bigl[(I\otimes I\otimes\epsilon)^TA(I\otimes I\otimes\epsilon)\bigr]^{\Gamma_3}\). Since
\(\tilde z=\epsilon^{\otimes4}\bar z\), moving \(\epsilon\) from the vector \(\tilde z\) to the operator and
transposing slot \(3\) (which turns \(\bar z\) into \(z\)) gives, for an operator
\(A=\bigotimes_{s=1}^4A_s\) with \(A_s\) acting on the three slots of site \(s\),
\[
\langle w,Aw\rangle=\langle z^{\otimes3},\textstyle\bigotimes_s\mathfrak t(A_s)\,z^{\otimes3}\rangle .
\]
The matrix elements (output \(x'y't'\), input \(xyt\)) of the four needed operators are listed below, with
\(\epsilon_{01}=1=-\epsilon_{10}\); for instance \(L^\dagger L=I\otimes|\epsilon\rangle\langle\epsilon|=I-r\),
where \(|\epsilon\rangle=|01\rangle-|10\rangle\). Conjugation by \(\epsilon\) on slot \(3\) acts on the two
\(\epsilon\)'s that carry the index \(t,t'\) as \(\sum_s\epsilon_{as}\epsilon_{sb}=(\epsilon^2)_{ab}=-\delta_{ab}\);
the two signs cancel, and each such \(\epsilon\)-pair becomes a Kronecker \(\delta\). Then \(\Gamma_3\) exchanges
\(t\) and \(t'\). Hence \[
\begin{aligned}
L^\dagger L&=\delta_{xx'}\epsilon_{y't'}\epsilon_{yt}\ \to\ \delta_{xx'}\delta_{y't}\delta_{yt'}:
&|xyt\rangle&\mapsto|xty\rangle=r,\\
M^\dagger M&=\delta_{yy'}\epsilon_{x't'}\epsilon_{xt}\ \to\ \delta_{yy'}\delta_{x't}\delta_{xt'}:
&|xyt\rangle&\mapsto|tyx\rangle=q,\\
L^\dagger M&=\delta_{x'y}\epsilon_{y't'}\epsilon_{xt}\ \to\ \delta_{x'y}\delta_{y't}\delta_{xt'}:
&|xyt\rangle&\mapsto|ytx\rangle=c,\\
M^\dagger L&=\delta_{xy'}\epsilon_{x't'}\epsilon_{yt}\ \to\ \delta_{xy'}\delta_{x't}\delta_{yt'}:
&|xyt\rangle&\mapsto|txy\rangle=c^{-1}.
\end{aligned}
\] On permutation operators, \(\mathfrak t(I)=I\) and \(\mathfrak t(p)=p\) (slot \(3\) is not involved), while for
\(q=\tfrac12(I+\sum_a\sigma_a\otimes I\otimes\sigma_a)\) and likewise \(r\), the identity
\(\epsilon^T\sigma_a\epsilon=\sigma_y\sigma_a\sigma_y=-\sigma_a^T\) gives \(\mathfrak t(q)=I-q\),
\(\mathfrak t(r)=I-r\). Hence \[
\begin{aligned}
\mathfrak t(S)&=(2I+p-q-r)/3=:S',\\
\mathfrak t((L+M)^\dagger(L+M))&=r+q+c+c^{-1}=-I+p+2q+2r=:K,\\
\mathfrak t(L^\dagger(L+M))&=r+c,\qquad
\mathfrak t(L^\dagger L)=r, \end{aligned}\tag{K}\label{eq:fqe-kernels}
\] using \eqref{eq:fqe-antisym} for \(c+c^{-1}=p+q+r-I\). Since \(S_i^\dagger S_i=S_i\), \[
\|V\|^2=\sum_i\bigl\langle z^{\otimes3},S'_i\textstyle\prod_{j\ne i}K_j\,z^{\otimes3}\bigr\rangle .
\]
The vector \(z^{\otimes3}\) is fixed by every simultaneous slot permutation (the same permutation at all
four sites); these can be multiplied on either side of any word inside its expectation, or used to
conjugate it.

\paragraph{Words and the table.} Expanding \(S'_i\prod_{j\ne i}K_j\) gives a combination of
\emph{words} \((\pi_1,\dots,\pi_4)\), \(\pi_s\in\{I,p,q,r\}\), with \(\pi_s\) acting on the slots of site \(s\).
The coefficient of \(S'\) on \(I,p,q,r\) is \((2,1,-1,-1)/3\) and that of \(K\) is \((-1,1,2,2)\); their
ratios are \((-\tfrac23,\tfrac13,-\tfrac16,-\tfrac16)\). So a word with \(a\) letters \(I\) and
\(\beta_p,\beta_q,\beta_r\) letters \(p,q,r\) has total coefficient (summed over the focused site \(i\)) \[
(-1)^a2^{\beta_q+\beta_r}\bigl[-2a+\beta_p-(\beta_q+\beta_r)/2\bigr]/3 .
\] Conjugation by a global slot permutation (the same permutation on all four sites) permutes the
transposition labels \(p,q,r\) in all six ways and does not change expectations, so the expectation of a word depends only on its \emph{type}
\((a;b,c,d)\): \(a\) is the number of identity letters and \(b\ge c\ge d\) are the multiplicities of the three
transpositions, sorted (together with the placement on the four sites). Let \(\#\) be the number of words of a type (these numbers sum to
\(4^4\)), \(C\) the sum of their coefficients, and \(J\) the mean expectation of the type over its placements
and labellings; then \(\|V\|^2=\sum CJ\). Multiplying every letter on the left by one global transposition \(g\) (written
\(\times g\)) does not change the expectation, and changes \(pq\to c\), \(pr\to c^{-1}\).
The table gives, for each type, a representative word, its reduction, \(J\) and \(C\).
\[
J_*:=\tfrac12(N^3+NP+\Pi_2)-\tfrac34N\Pi_1-\tfrac13T .
\]

{\small
\begin{longtable}{@{}l r l l l r@{}}
\caption{The eleven contraction types in $\|V\|^2$. $\#$ counts the words of each type,
$J$ is their mean expectation over site placements and transposition labels, and $C$ is the sum
of their coefficients, so $\|V\|^2=\sum CJ$. A word acts on the three copies at each of the four
sites; $I$ is the identity and $p,q,r$ are slot transpositions. Types are grouped by the number
of identity letters.}\label{tab:fqe-contractions}\\
\toprule
type & \(\#\) & word & expectation of the word & \(J\) & \(C\) \\
\midrule
\endfirsthead
\multicolumn{6}{@{}l}{\small Table~\ref{tab:fqe-contractions}, continued}\\
\toprule
type & \(\#\) & word & expectation of the word & \(J\) & \(C\) \\
\midrule
\endhead
\bottomrule
\endlastfoot
\((4;0,0,0)\) & 1 & \(IIII\) & \(N^3\) & \(N^3\) & \(-8/3\) \\
\addlinespace
\((3;1,0,0)\) & 12 & \(pIII\) & \(N\operatorname{tr}R_1^2\) & \(N\Pi_1/4\) & \(124/3\) \\
\addlinespace
\((2;2,0,0)\) & 18 & \(ppII\) & \(N\operatorname{tr}R_{12}^2\) & \(NP/3\) & \(-84\) \\
\((2;1,1,0)\) & 36 & \(pqII\) & \(\operatorname{tr}[R_{12}(R_1\otimes R_2)]\) & \(\Pi_2/6\) & \(-136\) \\
\addlinespace
\((1;3,0,0)\) & 12 & \(pppI\) & \(N\operatorname{tr}R_{123}^2=N\operatorname{tr}R_4^2\) & \(N\Pi_1/4\) & \(220/3\) \\
\((1;2,1,0)\) & 72 & \(ppqI\ \overset{\times p}{=}\ IIcp\) & \(\operatorname{tr}[R_{34}^2(R_3\otimes I)]\) & \(\frac{NP+\Pi_2}6-\frac{N\Pi_1}8\) & \(296\) \\
\((1;1,1,1)\) & 24 & \(pqrI\) & see (c) & \(J_*\) & \(64\) \\
\addlinespace
\((0;4,0,0)\) & 3 & \(pppp\ \overset{\times p}{=}\ IIII\) & \(N^3\) & \(N^3\) & \(-20\) \\
\((0;3,1,0)\) & 24 & \(pppq\ \overset{\times p}{=}\ IIIc\) & \(\operatorname{tr}R_4^3\) & \(\frac{3N\Pi_1}8-\frac{N^3}2\) & \(-248/3\) \\
\((0;2,2,0)\) & 18 & \(ppqq\ \overset{\times p}{=}\ IIcc\) & \(\operatorname{tr}R_{34}^3\) & \(T/3\) & \(-48\) \\
\((0;2,1,1)\) & 36 & \(ppqr\ \overset{\times p}{=}\ IIcc^{-1}\) & \(\operatorname{tr}[(R_{34}^{\Gamma})^3]\) & \(J_*\) & \(-16\) \\
\end{longtable}
}

Here \(\operatorname{tr}R_S^k\) is the expectation of the \(k\)-cycle on the sites of \(S\) (and \(\operatorname{tr}R_S^2\)
that of a transposition), as for the usual replica trick, and \(R^\Gamma\) is a partial transpose of a two-qubit matrix on one qubit (see (e)); the entries come from following the index lines
of \(z^{\otimes3}\). The means are obtained as follows.
To make the nontrivial partial-transpose entry explicit, consider the representative
$ppqr$. Multiplication on the left by the global $p$ gives $IIcc^{-1}$.
Trace out sites $1,2$ and write $R=R_{34}$ with entries $R_{ab,a'b'}$.
For partial transpose on the second qubit,
$(R^\Gamma)_{ab,a'b'}=R_{ab',a'b}$, and the opposite cycles give
\[
\begin{aligned}
\langle z^{\otimes3},(IIcc^{-1})z^{\otimes3}\rangle
&=\sum_{a,a',a'';\,b,b',b''}
 R_{ab',a'b}R_{a'b'',a''b'}R_{a''b,ab''}\\
&=\sum_{a,a',a'';\,b,b',b''}
 (R^\Gamma)_{ab,a'b'}(R^\Gamma)_{a'b',a''b''}
 (R^\Gamma)_{a''b'',ab}\\
&=\operatorname{tr}[(R^\Gamma)^3].
\end{aligned}
\]
All six indices run over $\{0,1\}$. This explains why opposite local cycles produce
a partial-transpose moment, whereas equal cycles give $\operatorname{tr}R^3$;
part (e) below evaluates its mean in the marginal invariants.

\begin{enumerate}
\item[(a)] Types \((4;0,0,0)\), \((0;4,0,0)\): \(N^3\) on every word. Types \((3;1,0,0)\), \((1;3,0,0)\), \((0;3,1,0)\):
average over the four sites of \(N\operatorname{tr}R_l^2\) or \(\operatorname{tr}R_l^3\), using
\(\operatorname{tr}R_{123}^2=\operatorname{tr}R_4^2\) for a pure state and
\(\operatorname{tr}R^3=(3N/2)\operatorname{tr}R^2-N^3/2\) for a \(2\times2\) matrix of trace \(N\).
\item[(b)] Types \((2;2,0,0)\), \((0;2,2,0)\): complementary pairs have equal spectra, so over the six pairs
\(\sum_{i<j}\operatorname{tr}R_{ij}^2=2P\) and \(\sum_{i<j}\operatorname{tr}R_{ij}^3=2T\), giving the means
\(NP/3\) and \(T/3\). Type \((2;1,1,0)\): the mean of \(\operatorname{tr}[R_{ij}(R_i\otimes R_j)]\) over the six pairs is \(\Pi_2/6\).
\item[(c)] Type \((1;1,1,1)\): replace the identity letter by \(p+q+r-c-c^{-1}\) (equation~\eqref{eq:fqe-antisym}).
The three transpositions give words of type \((0;2,1,1)\). In the words with \(c\) or \(c^{-1}\), multiplying all letters
by \(c^{-1}\) or \(c\) turns the cycle into \(I\) and permutes the three transpositions, so the type is \((1;1,1,1)\) again.
Thus \(J_{(1;1,1,1)}=3J_{(0;2,1,1)}-2J_{(1;1,1,1)}\): the two types have the same mean.
\end{enumerate}
To evaluate the remaining means, we use the Pauli expansion of a two-qubit matrix, derived for $N=1$ in
Appendix~\ref{sec:fqe-correlation}; the homogeneous forms below follow by scaling.
For the unnormalised \(R_{ij}=\frac14\bigl(NI+\mathbf r_i\!\cdot\!\boldsymbol\sigma_i+\mathbf r_j\!\cdot\!\boldsymbol\sigma_j
+\sum_{a,b}C^{ab}_{ij}\sigma_a\otimes\sigma_b\bigr)\), with \(\mathbf r_i,C_{ij}\) the (unnormalised) expectations
and \(u_{ij}=\|\mathbf r_i\|^2+\|\mathbf r_j\|^2+\|C_{ij}\|_F^2\), \[
\begin{aligned}
&\operatorname{tr}R_{ij}^2=\tfrac14(N^2+u_{ij}),\qquad
\operatorname{tr}[R_{ij}(R_i\otimes R_j)]=\tfrac14\bigl(N^3+N\|\mathbf r_i\|^2+N\|\mathbf r_j\|^2+\mathbf r_i^TC_{ij}\mathbf r_j\bigr),\\
&\operatorname{tr}R_{ij}^3=\tfrac1{16}\bigl(N^3+3Nu_{ij}+6\,\mathbf r_i^TC_{ij}\mathbf r_j-6\det C_{ij}\bigr),\qquad
\operatorname{tr}\bigl[R_{ij}^2(\mathbf r_i\!\cdot\!\boldsymbol\sigma_i\otimes I)\bigr]=\tfrac12\bigl(N\|\mathbf r_i\|^2+\mathbf r_i^TC_{ij}\mathbf r_j\bigr).
\end{aligned}
\]
\begin{enumerate}
\item[(d)] Type \((1;2,1,0)\): writing \(R_i\otimes I=\frac N2I+\frac12\mathbf r_i\!\cdot\!\boldsymbol\sigma_i\otimes I\), the last formula gives the
two-qubit identity \[
\operatorname{tr}[R_{ij}^2(R_i\otimes I)]
=\frac N2\operatorname{tr}R_{ij}^2
+\operatorname{tr}[R_{ij}(R_i\otimes R_j)]
-\frac N2\operatorname{tr}R_j^2 ,
\] since both sides exceed \(\frac N2\operatorname{tr}R_{ij}^2\) by \(\frac14(N\|\mathbf r_i\|^2+\mathbf r_i^TC_{ij}\mathbf r_j)\). Averaging over the twelve ordered pairs
\((i,j)\), whose sums are \(4P\), \(2\Pi_2\), \(3\Pi_1\) for the three terms, gives \(\frac1{12}\bigl(2NP+2\Pi_2-\frac32N\Pi_1\bigr)\), the stated \(J\).
\item[(e)] Type \((0;2,1,1)\): partial transposition of the first qubit (or the second; they differ by a full transpose, which does not change the spectrum) reverses the sign of \(\sigma_y\) there, so
with \(D=\operatorname{diag}(1,-1,1)\) the data become \(\mathbf r_i\to D\mathbf r_i\), \(C_{ij}\to DC_{ij}\) and \(\mathbf r_j\) is unchanged. Hence
\(\mathbf r_i^TC_{ij}\mathbf r_j\) and \(u_{ij}\) are unchanged and \(\det C_{ij}\to-\det C_{ij}\), and the cubic formula above gives
\(\operatorname{tr}[(R_{ij}^{\Gamma})^3]=\operatorname{tr}R_{ij}^3+\tfrac34\det C_{ij}\). Solving the formulas above for
\(\det C_{ij}\) and summing over the six pairs, with \(\sum_{i<j}u_{ij}=8P-6N^2\), \(\sum_{i<j}(\|\mathbf r_i\|^2+\|\mathbf r_j\|^2)=3W_1\) and \(\sum_{i<j}\operatorname{tr}R_{ij}^3=2T\), gives
\(\sum_{i<j}\det C_{ij}=4N^3+4NP+4\Pi_2-6N\Pi_1-\tfrac{16}3T\); hence \(\sum_{i<j}\operatorname{tr}[(R_{ij}^\Gamma)^3]=3(N^3+NP+\Pi_2)-\tfrac92N\Pi_1-2T\), whose mean over six pairs is \(J_*\).
\end{enumerate}
Summing \(CJ\) over the table proves the last line of \eqref{eq:fqe-contractions}.

\paragraph{The other two contractions.} By \eqref{eq:fqe-kernels},
\(\langle U_i,V_i\rangle\) is the expectation of \(S'_i\prod_{j\ne i}(r+c)_j\). Multiply on the right by the global \(r\):
the factors at \(j\ne i\) become \((r+c)r=I+q\), and at \(i\) we get
\(S'_ir_i=\frac13(2r_i-I)+\frac13(c_i^{-1}-c_i)\). Conjugation by the global \(q\) fixes \(q\) and exchanges
\(c\) and \(c^{-1}\), so the expectation of the last term equals its own negative and vanishes. Hence \[
\operatorname{Re}\langle U,V\rangle=\tfrac13\sum_i\Bigl\langle\bigl(2r_i-I\bigr)\prod_{j\ne i}(I+q_j)\Bigr\rangle
=\tfrac13\sum_i\sum_{m=0}^3\ \sum_{A\subseteq\{j\ne i\},\,|A|=m}\bigl\langle(2r_i-I)\,q_A\bigr\rangle ,
\] where \(q_A\) puts \(q\) on the sites of \(A\). There are \(4\binom3m\) pairs \((i,A)\) for each \(m\), and the terms
\(r_iq_A\) have the types \((3;1,0,0)\), \((2;1,1,0)\), \((1;2,1,0)\), \((0;3,1,0)\) for \(m=0,1,2,3\),
and the terms \(I_iq_A\) the types \((4;0,0,0)\), \((3;1,0,0)\), \((2;2,0,0)\), \((1;3,0,0)\), each placement occurring equally often. With
\(J_{(a;b,c,d)}\) the means of the table, \[
\begin{aligned}
\operatorname{Re}\langle U,V\rangle=\tfrac43\Bigl[&2\bigl(J_{(3;1,0,0)}+3J_{(2;1,1,0)}+3J_{(1;2,1,0)}+J_{(0;3,1,0)}\bigr)\\
&-\bigl(J_{(4;0,0,0)}+3J_{(3;1,0,0)}+3J_{(2;2,0,0)}+J_{(1;3,0,0)}\bigr)\Bigr].
\end{aligned}
\] The first bracket is \(\frac12NP+\Pi_2+\frac14N\Pi_1-\frac12N^3\) and the second is \(NP+N\Pi_1+N^3\), which gives
\((-8N^3-2N\Pi_1+8\Pi_2)/3\). Only the first six rows and the row \((0;3,1,0)\) of the table are used.
The same argument for \(\|U\|^2\), with \(L^\dagger L\mapsto r\) on the other sites, leaves
\(\frac13\sum_i\langle 2r_i-I\rangle=\frac13\sum_i(2N\operatorname{tr}R_i^2-N^3)=(2N\Pi_1-4N^3)/3\), agreeing with \eqref{eq:fqe-local-polarization}.
This completes the derivation of \eqref{eq:fqe-contractions}.

\subsection{The critical envelope}\label{sec:fqe-critical-envelope}

On the critical locus $W_1=0$, and the moment bound can be sharpened by the spin-flip invariant. For a critical
normal vector $v$ let $\phi=\sum_kv_km_k\in\mathbb C^4$ be the auxiliary two-qubit vector (Appendix~\ref{sec:fqe-normal-form})
and $\Theta_{\alpha\beta}=\langle\phi|\sigma_\alpha\otimes\sigma_\beta|\phi\rangle$ its Bloch correlation matrix.

\begin{proposition}[critical envelope]\label{prop:fqe-critical-envelope}
Every critical unit four-qubit state satisfies
\[
 \boxed{\quad F=\frac12+\frac74\tau+\frac94\operatorname{perm}\Theta\ \ge\ \tau-\tau^{3/2}.\quad}
 \tag{CE}\label{eq:fqe-critical-envelope}
\]
For $0<\tau\le1$, with $\kappa=\sqrt\tau$, equality holds exactly when
$\Theta=O(\kappa I+(1-\kappa)ss^T)$ with $s\in\{\pm1\}^3/\sqrt3$ and $O$ one of the twelve signed cyclic permutation
matrices whose three signs have product $-1$; such states exist for every $\tau$. Since $P=1+\tau/2$ on this locus, the
cubic moment obeys $T\le\frac5{12}+\frac{17}{12}(P-1)-\frac1{12}(x-x^{3/2})$, $x=2(P-1)$, with equality attained for every
$P\in[1,\tfrac32]$: this is the exact upper boundary of the critical moment region, and it meets the supporting line of
Theorem~\ref{thm:fqe-moment} only at its two ends.
\end{proposition}

\begin{proof}
$F$ and $\tau$ are LU invariant, so take $\psi=G(v)$ with $\|v\|=1$ (Theorem~\ref{thm:fqe-critical-normal-form}).
\emph{The formula.} The twelve magic Bell rays have signed-permutation correlation matrices
(Lemmas~\ref{lem:fqe-critical-correlation}(1) and~\ref{lem:fqe-p1-cuts} of the supplement), so the cut $k\in\mathbb Z/3$ has the four eigenvalues
$(1+\sigma\cdot g^{(k)})/4$, where $g^{(k)}_\beta=\Theta_{\beta+k,\beta}$ is a cyclic diagonal of $\Theta$ and $\sigma$ runs over
the four sign triples with $\sigma_1\sigma_2\sigma_3=-1$. Over these triples $\sum(\sigma\cdot g)=0$,
$\sum(\sigma\cdot g)^2=4|g|^2$ and $\sum(\sigma\cdot g)^3=-24g_1g_2g_3$. The three cyclic diagonals partition the
entries of $\Theta$, and the sum of their products is the even half $e=(\operatorname{perm}\Theta+\det\Theta)/2$ of the
permanent. Hence $P=(3+\|\Theta\|^2)/4$, $T=(3+3\|\Theta\|^2-6e)/16$ and
$F=-\tfrac32+2\|\Theta\|^2+\tfrac92e$. In Schmidt form $\phi=(U\otimes V)(\cos\varphi|00\rangle+\sin\varphi|11\rangle)$,
so $\Theta=R_a\operatorname{diag}(\kappa,-\kappa,1)R_b^T$ with $R_a,R_b\in SO(3)$ the rotations induced by $U,V$ and
$\kappa=\sin2\varphi$ the concurrence of $\phi$, which in the magic basis is $|\sum_kv_k^2|=\sqrt\tau$
by~\eqref{eq:fqe-spin-flip-normal-form}. Thus $\Theta$ has singular values $(1,\kappa,\kappa)$ and $\det\Theta=-\kappa^2$. So $\|\Theta\|^2=1+2\tau$, which gives the formula in~\eqref{eq:fqe-critical-envelope}
and $P=1+\tau/2$ (checks CE-1, CE-2).

\emph{The permanent bound.} Write the polar decomposition $\Theta=OD$, with $O\in O(3)$, $\det O=-1$,
$D=(\Theta^T\Theta)^{1/2}=\kappa I+(1-\kappa)ss^T$ and $ODO^T=\kappa I+(1-\kappa)rr^T$; here $r,s$ are unit top singular
vectors of $\Theta$, the local Bloch directions of $\phi$. For every $3\times3$ matrix $X$,
$\operatorname{perm}X=\langle\omega,X^{\otimes3}\omega\rangle$ with the unit vector
$\omega=6^{-1/2}\sum_\pi e_{\pi1}\otimes e_{\pi2}\otimes e_{\pi3}$. Cauchy--Schwarz gives
\[
 |\operatorname{perm}\Theta|
 =\bigl|\bigl\langle(D^{1/2}O^T)^{\otimes3}\omega,(D^{1/2})^{\otimes3}\omega\bigr\rangle\bigr|
 \le\bigl(\operatorname{perm}(ODO^T)\operatorname{perm}D\bigr)^{1/2}.
\]
For a unit vector $u$ and $y_a=u_a^2$, expanding the permanent by rows,
\[
 \operatorname{perm}\bigl(\kappa I+(1-\kappa)uu^T\bigr)
 =\kappa^3+\kappa^2(1-\kappa)+2\kappa(1-\kappa)^2e_2(y)+6(1-\kappa)^3e_3(y)
 \le\frac{2+3\kappa^2+4\kappa^3}9,
\]
by Maclaurin's inequalities $e_2(y)\le\frac13$, $e_3(y)\le\frac1{27}$ on the simplex (check CE-3); for $\kappa<1$ equality needs
$y=(\frac13,\frac13,\frac13)$. Thus $-\operatorname{perm}\Theta\le(2+3\tau+4\tau^{3/2})/9$, and
$F\ge\frac12+\frac74\tau-\frac14(2+3\tau+4\tau^{3/2})=\tau-\tau^{3/2}$.

\emph{Equality.} Let $\kappa>0$, so $D$ is invertible. Equality forces $r,s\in\{\pm1\}^3/\sqrt3$ when $\kappa<1$, and
equality with a negative sign in Cauchy--Schwarz, $(O^T)^{\otimes3}\omega=-\omega$. As $\omega$ is the symmetric tensor of
$x_1x_2x_3$, this says $x_1x_2x_3$ changes sign under $O^T$; then $O^T$ preserves the union of the coordinate planes, so $O$
is a signed permutation, with sign product $-1$; then $\det O=-\operatorname{sgn}\pi=-1$ makes the permutation $\pi$
even, that is, cyclic (check CE-4). Conversely, for such
$O$ and $s$, $\operatorname{perm}\Theta=\langle(O^T)^{\otimes3}\omega,D^{\otimes3}\omega\rangle=-\operatorname{perm}D$ attains the
bound. Every matrix with singular values $(1,\kappa,\kappa)$ and determinant $-\kappa^2$ has the form
$R_a\operatorname{diag}(\kappa,-\kappa,1)R_b^T$ with $R_a,R_b\in SO(3)$, so it is the correlation matrix of a suitable
$\phi$ in Schmidt form; and every unit $\phi$ is $\sum_kv_km_k$ for a critical $G(v)$.
The moment statement is~\eqref{eq:fqe-critical-envelope} rewritten with $\tau=2(P-1)$ and $F=17P-12T-12$.
\end{proof}

On critical states $W_1=0$, so for unit states \eqref{eq:fqe-critical-envelope} is the critical case of
$F\ge\frac34W_1+\tau-\tau^{3/2}$, which Theorem~\ref{thm:fqe-general-envelope} proves for all states.

\subsection{All zeros and why the moment bound is tight}\label{sec:fqe-sextic_identity-all-zeros-and-why-the-moment-bound-is-tight}

The critical envelope classifies the zeros of $F$. This classification supplements the moment inequality; it is
not used in the proof of Theorem~\ref{thm:fqe-entropy-max}.

\begin{proposition}[complete zero set of the moment inequality]\label{prop:fqe-moment-zeros}
For unit four-qubit states, $F=0$ holds exactly on the two $M_4$ LU orbits and the three LU orbits
of products of Bell pairs, one for each pairing of the labelled qubits.
\end{proposition}

\begin{proof}
A unit zero has $W_1=0$ by Theorem~\ref{thm:fqe-moment}, so it is critical, and since $\tau-\tau^{3/2}>0$ for
$0<\tau<1$, Proposition~\ref{prop:fqe-critical-envelope} leaves $\tau\in\{0,1\}$. At $\tau=1$, $F=0$ is equality in
\eqref{eq:fqe-critical-envelope}, so $\Theta$ is one of the twelve signed cyclic permutation matrices $O$; one cyclic
diagonal then has entries $\pm1$ with product $-1$, and that cut has spectrum $(1,0,0,0)$. The pure cut splits the
state into two two-qubit pure states with maximally mixed one-qubit marginals, that is, two Bell pairs. At $\tau=0$,
$\Theta=rs^T$ with unit $r,s$, and the formula in \eqref{eq:fqe-critical-envelope} reads $F=\frac12+\frac{27}2\prod_ar_as_a$.
Its zero $\prod_ar_as_a=-\frac1{27}$ forces $|r_a|=|s_a|=3^{-1/2}$ by the AM--GM inequality, so every cyclic diagonal
$g^{(k)}$ has entries $\pm\frac13$ with product $-\frac1{27}$ and every cut has spectrum $(\frac12,\frac16,\frac16,\frac16)$;
Proposition~\ref{prop:fqe-onecut} then gives the $M_4$ orbits. Conversely $M_4$, $\overline{M_4}$ and the Bell-pair
products have $F=0$. Section~\ref{sec:sup-sextic-zeros} of the supplement gives an independent proof by a
weighted-variance identity on the critical normal form.
\end{proof}

There is also an elementary way to distinguish the two classes. Let \(D_{1j}\) be the determinant of
the \(1j|\widehat{1j}\) flattening, with qubit indices in increasing order. Every invertible local
transformation multiplies all three determinants by the same factor \(\prod_i(\det A_i)^2\); their
ratios are therefore invariant, even under invertible local filters. Directly, \[
\frac{D_{13}(M_4)}{D_{12}(M_4)}=\frac{1-i\sqrt3}{2},
\] whereas conjugation conjugates this nonreal ratio. Thus the two classes are distinct even under
such filters. LU-inequivalence alone was first proved in~\cite[Prop.~7]{LWB}.

By Proposition~\ref{prop:fqe-moment-zeros}, the supporting line \(17P-12T-12=0\) meets the pure-state
moment region at exactly \((1,5/12)\) and \((3/2,9/8)\); its open line segment is unattainable.

\subsection{The general envelope}\label{sec:fqe-general-envelope}

\begin{theorem}[general envelope and complete equality; exact certificate, FI, FG]\label{thm:fqe-general-envelope}
Every nonzero four-qubit vector satisfies, with $s=z^T\sigma_y^{\otimes4}z$,
\[
 F-\tfrac34NW_1-N|s|^2+|s|^3\ \ge\ \frac{7|s|^2}{8(N+|s|)}\,W_1 .
 \tag{GE}\label{eq:fqe-general-envelope}
\]
For unit states, with $\tau=|s|^2$,
\[
 \boxed{\quad F\ \ge\ \Bigl(\frac34+\frac{7\tau}{8(1+\sqrt\tau)}\Bigr)W_1+\tau-\tau^{3/2}.\quad}
\]
Equality in $F\ge\frac34W_1+\tau-\tau^{3/2}$ holds exactly on fully product states and on the critical
equality family of Proposition~\ref{prop:fqe-critical-envelope}, including its HS endpoint at $\tau=0$ and
Bell-pair-product endpoint at $\tau=1$.
\end{theorem}

The coefficient $\frac78$ of the added $W_1$ term is optimal. On $z=\alpha|0000\rangle+\beta|1111\rangle$ with real
$\alpha,\beta\ge0$, $\alpha^2+\beta^2=1$, one has $s=2\alpha\beta$, $\tau=4\alpha^2\beta^2$, $W_1=4(1-\tau)$ and $F=3+\frac32\tau$, so the boxed bound with $\frac78$ replaced by $C$ leaves the slack
$(\frac72-4C)\tau+(1+4C)\tau^{3/2}$, negative for small $\tau>0$ when $C>\frac78$ (exact check ESH). The proof (Section~\ref{sec:sup-general-envelope} of the supplement) follows $H=F-\frac34NW_1$ along a path of unit states
from a state with $s=0$ to $z$, along which the spin-flip overlap grows from $0$ to $s$; an exact identity for the
derivative of $H$ along this path (exact certificate; checks FI, FG), integrated, gives~\eqref{eq:fqe-general-envelope}, and the zeros of the
remainder give the equality cases.

\section{Critical four-qubit states: an elementary normal form}\label{sec:fqe-normal-form}

The equality proof needs an LU normal form for \emph{critical} states, whose one-qubit marginals are all
maximally mixed. We derive it by elementary linear algebra, including all ranks and singular-value multiplicities. No orbit-closure or Kempf--Ness argument is needed. Exact checks are listed in Appendix~\ref{sec:fqe-verifiers}.

\subsection{Setting}\label{sec:fqe-critical_four_qubit_normal_form-setting}

Let \(m_1,\dots,m_4\) be the magic basis of two qubits, \[
m_1=\tfrac1{\sqrt2}(\lvert00\rangle+\lvert11\rangle),\quad
m_2=\tfrac{i}{\sqrt2}(\lvert01\rangle+\lvert10\rangle),\quad
m_3=\tfrac1{\sqrt2}(\lvert01\rangle-\lvert10\rangle),\quad
m_4=\tfrac{i}{\sqrt2}(\lvert00\rangle-\lvert11\rangle).
\] A four-qubit state is \(\psi=\sum_{j,k}Z_{jk}\,m_j\otimes m_k\) with \(Z\in\mathbb C^{4\times4}\)
(qubits 1,2 index the rows, qubits 3,4 the columns), \(\|Z\|_F=1\). In this basis \(u\otimes u'\in
SU(2)\otimes SU(2)\) acts as a real rotation, \(SU(2)\otimes SU(2)\to SO(4)\) onto, so the local
unitaries act as \(Z\mapsto O_1ZO_2^{T}\), \(O_1,O_2\in SO(4)\). For \(v\in\mathbb C^4\) put \[
G(v)=\sum_{k=1}^4v_k\,m_k\otimes m_k .
\] Expanding the \(m_k\) shows \(G(v)=G_{abcd}\), the family
\(\tfrac{a+d}2(\lvert0000\rangle+\lvert1111\rangle)+\tfrac{a-d}2(\lvert0011\rangle+\lvert1100\rangle)+\tfrac{b+c}2(\lvert0101\rangle+\lvert1010\rangle)+\tfrac{b-c}2(\lvert0110\rangle+\lvert1001\rangle)\)
with \((a,b,c,d)=(v_1,-v_2,v_3,-v_4)\) (CN-2).

\paragraph{Spin flip on the normal form.} The invariant \(\tau=|\psi^T\sigma_y^{\otimes4}\psi|^2\) of
Lemma~\ref{lem:fqe-purity} is LU invariant, because \(u^T\sigma_yu=(\det u)\sigma_y\) for
\(u\in U(2)\): a product unitary multiplies \(\psi^T\sigma_y^{\otimes4}\psi\) by
\(\prod_i\det u_i\). Directly from the definitions, \(\sigma_y^{\otimes2}m_k=s_km_k\) with
\(s=(-,+,-,+)\), and \(m_k^Tm_l=\eta_k\delta_{kl}\) with \(\eta=(+,-,+,-)\) (the \(m_k\) are
orthonormal and real up to the phases \(1,i,1,i\)). Hence
\(m_k^T\sigma_y^{\otimes2}m_l=s_l\eta_k\delta_{kl}=-\delta_{kl}\), and for \(\psi=G(v)\)
\[
\psi^T\sigma_y^{\otimes4}\psi=\sum_{k,l}v_kv_l\,\bigl(m_k^T\sigma_y^{\otimes2}m_l\bigr)^2=\sum_kv_k^2,
\qquad\text{so}\qquad \tau=\Bigl|\sum_kv_k^2\Bigr|^2. \tag{$\tau_G$}\label{eq:fqe-spin-flip-normal-form}
\]

\begin{theorem}[Critical normal form]\label{thm:fqe-critical-normal-form}

For a unit four-qubit state $\psi$ with matrix $Z$ the following are equivalent.

\begin{enumerate}
\item All four one-qubit marginals of $\psi$ equal $I/2$ ($\psi$ is \emph{critical}).
\item $ZZ^\dagger$ and $Z^\dagger Z$ are real matrices. Equivalently, for $Z=X+iY$ with $X,Y$ real,
$XY^T$ and $X^TY$ are symmetric.
\item $Z=O_1\operatorname{diag}(v)\,O_2^T$ with $O_1,O_2\in SO(4)$ and $v\in\mathbb C^4$.
\item $\psi$ is local-unitarily equivalent to $G(v)$ for some $v$ with $\|v\|=1$.

\end{enumerate}
\end{theorem}

\begin{lemma}[Simultaneous real singular values]\label{lem:fqe-real-gram}

For any complex $m\times n$ matrix $Z$, the two Gram matrices $ZZ^\dagger$
and $Z^\dagger Z$ are real if and only if
\[
Z=O_1 D O_2^T,\qquad O_1\in O(m),\quad O_2\in O(n),
\]
where $D$ is a complex rectangular diagonal matrix. This includes all ranks
and multiplicities.
\end{lemma}

\begin{proof} Write \(Z=X+iY\). The Gram conditions are precisely \(XY^T=YX^T\) and \(X^TY=Y^TX\).
Take a real SVD of \(X\) and apply its two orthogonal factors to \(Y\) too. On the nonzero diagonal
of \(X\), write \(d_i>0\). The two symmetry equations give \[
d_iY_{ji}=d_jY_{ij},\qquad d_iY_{ij}=d_jY_{ji},
\qquad (d_i^2-d_j^2)Y_{ij}=0.
\] Thus different positive singular-value blocks do not mix, each equal-value block of \(Y\) is
symmetric, and its entries joining a positive block to either nullspace vanish. Diagonalise each
positive block of \(Y\) by the same real orthogonal transformation on its left and right; this
leaves \(X=dI\) fixed. On the remaining, possibly rectangular nullspace block, take a real SVD of
\(Y\). Both \(X\) and \(Y\) are now diagonal. The converse is immediate from the two Gram products.
\end{proof}

\subsection{Proof of the critical normal form}\label{sec:fqe-critical_four_qubit_normal_form-proof-of-theorem}

\begin{proof}[Proof of Theorem~\ref{thm:fqe-critical-normal-form}]
\textit{Marginals are moment maps.}\label{lem:fqe-moment-map}

In the magic basis the six operators \(i\sigma_a\otimes1\), \(i\,1\otimes\sigma_a\) (\(a=x,y,z\))
have real antisymmetric matrices, and they are linearly independent, hence span \(\mathfrak{so}(4)\)
(CN-1: check of the explicit \(4\times4\) matrices). The reduced state of qubits 1,2 is
\(\rho_{12}=ZZ^\dagger\) in the magic basis. For a real antisymmetric \(L\) and Hermitian
\(H=ZZ^\dagger\), \(\operatorname{tr}(L\,\mathrm{Re}H)=0\), so
\(\langle\sigma\otimes1\rangle=\operatorname{tr}(H(-iL))=\operatorname{tr}(L\,\mathrm{Im}H)\). The
six expectations \(\langle\sigma_a\otimes1\rangle,\langle1\otimes\sigma_a\rangle\) are therefore the
six independent components of the antisymmetric matrix \(\mathrm{Im}(ZZ^\dagger)\) under an invertible
linear change of coordinates, and they vanish iff \(\mathrm{Im}(ZZ^\dagger)=0\). Thus the marginals of qubits 1 and 2
are both \(I/2\) iff \(ZZ^\dagger\) is real. In the same way \(\rho_{34}=(Z^\dagger Z)^T\), so
qubits 3 and 4 are maximally mixed iff \(Z^\dagger Z\) is real. This proves
(1)\(\Leftrightarrow\)(2). (Writing \(ZZ^\dagger=XX^T+YY^T+i(YX^T-XY^T)\) and \(Z^\dagger
Z=X^TX+Y^TY+i(X^TY-Y^TX)\) gives the symmetry form of (2).)

\textbf{(3)\(\Rightarrow\)(2).} \(ZZ^\dagger=O_1\operatorname{diag}(|v_k|^2)O_1^T\) and \(Z^\dagger
Z=O_2\operatorname{diag}(|v_k|^2)O_2^T\) are real.

\textbf{(2)\(\Rightarrow\)(3).} Apply the lemma with \(m=n=4\). If an orthogonal factor has
determinant \(-1\), change its first column's sign and the first diagonal entry's sign; thus both
factors can be taken in \(SO(4)\).

\textbf{(3)\(\Leftrightarrow\)(4).} \((U_{12}\otimes U_{34})\,G(v)\) has matrix
\(O_1\operatorname{diag}(v)O_2^T\) when \(U_{12},U_{34}\) act as \(O_1,O_2\) in the magic basis.
\end{proof}

\subsection{Consequences}\label{sec:fqe-critical_four_qubit_normal_form-consequences}

\begin{enumerate}
\def\labelenumi{\arabic{enumi}.}
\item
  On the unitary slice \(Z=W/2\), the lemma is the real-orthogonal KAK factorisation. Here
\(|v_k|=1/2\) and the \(v_k^2\) are the eigenvalues of \(ZZ^T\).
\item
  The \hyperref[sec:fqe-sextic]{sextic zero classification} and
Proposition~\ref{prop:fqe-onecut} use this theorem.
\end{enumerate}

\section{Correlation geometry of the sextic}\label{sec:fqe-correlation}

We use the notation \(F,P,T,\tau\) of Sections~\ref{sec:fqe-entropy_maximum-purity-identity}
and~\ref{sec:fqe-entropy_maximum-the-sharp-moment-inequality} for a unit pure four-qubit state.

\subsection{Correlation-volume identity}\label{sec:fqe-correlation_geometry-correlation-volume-identity}

For a unit state, write \(r_i^a=\langle\sigma_a^{(i)}\rangle\) and
\(C_{ij}^{ab}=\langle\sigma_a^{(i)}\sigma_b^{(j)}\rangle\), with \(C_{ij}\) a real \(3\times3\)
matrix. Expanding a two-qubit marginal in Pauli matrices, and putting
\(u_j=\|r_1\|^2+\|r_j\|^2+\|C_{1j}\|_F^2\), gives \[
\operatorname{tr}\rho_{1j}^2=\frac{1+u_j}{4},\qquad
\operatorname{tr}\rho_{1j}^3=
\frac{1+3u_j+6r_1^TC_{1j}r_j-6\det C_{1j}}{16}.
\] To see this, write \(\rho_{1j}=\tfrac14(I+X)\) with \(X\) the traceless part, so that
\(\operatorname{tr}\rho^3=\tfrac1{64}(4+3\operatorname{tr}X^2+\operatorname{tr}X^3)\) and
\(\operatorname{tr}X^2=4u_j\). In \(\operatorname{tr}X^3\), the terms with one Pauli matrix on each of the two qubits
in every factor give \(\operatorname{tr}(\sigma_a\sigma_c\sigma_e)\operatorname{tr}(\sigma_b\sigma_d\sigma_f)C^{ab}C^{cd}C^{ef}
=(2i)^2\epsilon_{ace}\epsilon_{bdf}C^{ab}C^{cd}C^{ef}=-24\det C_{1j}\), using
\(\operatorname{tr}(\sigma_a\sigma_c\sigma_e)=2i\epsilon_{ace}\); the terms with one Bloch vector on each qubit and one
factor \(C\) give \(6\cdot4\,r_1^TC_{1j}r_j\) (six orderings); the terms with two Bloch vectors on
one qubit, or three, and those with one Bloch vector and two factors \(C\), are traceless (the last because \(CC^T\) is
symmetric while \(\epsilon_{ace}\) is antisymmetric). Hence \(\operatorname{tr}X^3=24(r_1^TC_{1j}r_j-\det C_{1j})\), which gives the stated formulas.
Lemma~\ref{lem:fqe-purity} implies \(\sum_j u_j=4P-3=1+2(W_1+\tau)\), where
\(W_1=\sum_i\|r_i\|^2\). Substitution yields the compact identity \[
\boxed{\quad 2F=
1+8(W_1+\tau)-9\sum_{j=2}^4
\bigl(r_1^TC_{1j}r_j-\det C_{1j}\bigr).\quad} \tag{CV}\label{eq:fqe-correlation-volume}
\] The sum singles out qubit~$1$ only to label the three balanced cuts $\{1,j\}|\{k,l\}$: by the formulas above each
summand equals $(16\operatorname{tr}\rho_{1j}^3-12\operatorname{tr}\rho_{1j}^2+2)/6$, a function of the cut spectrum
alone, so the right-hand side is permutation invariant. Therefore Theorem~\ref{thm:fqe-moment} is equivalent to an
\emph{oriented correlation-volume monogamy inequality} for three compatible two-qubit marginals.
At \(M_4\) the Bloch vectors and \(\tau\) vanish and each \(\det C_{1j}=-1/27\); at
\(\mathrm{Bell}\otimes\mathrm{Bell}\), \(\tau=1\) and the three determinants are \((-1,0,0)\) up to
ordering. This explains the two tight configurations: the spin-flip invariant compensates for
concentration of the correlation volume in a single Bell pair.

Equation \eqref{eq:fqe-correlation-volume} gives a reduced-state reformulation.
Its nonnegativity still requires the moment inequality. Pairwise density-matrix positivity and known absolute
steering-ellipsoid volume-monogamy bounds do not visibly supply the required \emph{oriented} and
\(\tau\)-sensitive inequality (compare the volume monogamy relation for pure four-qubit states of
Cheng, Milne, Hall and Wiseman~\cite{CMHW}, Eq.~(28): in terms of normalised steering-ellipsoid
volumes \(v_{j|1}=|\det(C_{1j}-r_1r_j^T)|/(1-\|r_1\|^2)^2\) of the ellipsoids that qubit \(1\) steers on
qubit \(j\), it reads \(\sum_j v_{j|1}^{2/3}\le1\), which for \(r_1=0\) is
\(\sum_j|\det C_{1j}|^{2/3}\le1\) and is already strict at \(M_4\), giving \(1/3<1\)). RG-1 and RG-2 (Appendix~\ref{sec:fqe-verifiers}) check the local Pauli
formulas and \eqref{eq:fqe-correlation-volume} exactly.

\paragraph{Replica witness.} Let \(p=(12)\) and \(c=(123)\) permute
the \emph{three copies} of a qubit, and use a subscript \(i\) for the physical qubit on which they
act. The Hermitian replica witness \[
K_{\rm rep}=17\sum_{j=2}^4p_1p_j-6\sum_{j=2}^4(c_1c_j+c_1^{-1}c_j^{-1})-12I
\] satisfies \(F(z)=\langle z^{\otimes3},K_{\rm rep}z^{\otimes3}\rangle\) for unit \(z\) (by the usual
swap/cycle formulas for reduced moments). It is not positive on the
copy-symmetric space. On the sector where all four local triples carry spin \(1/2\),
let \(M\simeq\mathbb C^2\) be the standard \(S_3\) multiplicity space and take the three
diagonal-\(S_3\)-fixed pairings \(\delta_{12}\delta_{34}\), \(\delta_{13}\delta_{24}\),
\(\delta_{14}\delta_{23}\) in \(M^{\otimes4}\). Their Gram matrix has diagonal entries 4 and
off-diagonal entries 2; the matrix of \(K_{\rm rep}\) on these pairings has diagonal entries \(-52\) and
off-diagonal entries 42. The generalised eigenvalues are \(4,-47,-47\) (RG-3, exact). Thus
\(K_{\rm rep}\) is indefinite even on \(\operatorname{Sym}^3((\mathbb
C^2)^{\otimes4})\). By uniqueness of polarisation of a bidegree-\((3,3)\) polynomial, \(F\) cannot
be a sum of squared moduli of \emph{holomorphic} cubic forms in \(z\) alone. This does not contradict \(F(z)\ge0\) on pure-state tensor powers.

\section{Quantitative stability of the maximising LU orbits}
\label{sec:fqe-orbit-stability}

The entropy deficit $\delta(\psi)=E(M_4)-E(\psi)$ controls the
squared Hilbert-space distance to the two HS orbits. The gap identity
\eqref{eq:fqe-entropy-gap} controls $W_1$ and the spin-flip overlap
$\tau$, but both also vanish at cluster states; the sextic moment
defect is needed to distinguish the maximisers.

The defect $Q$ below bounds HS infidelity with sharp coefficient one
(Theorem~\ref{thm:fqe-global-overlap}). This gives a uniform, nonoptimal
entropy-distance bound, while the optimal coefficient on critical
states is known (Theorem~\ref{thm:fqe-orbit-stability};
Appendix~\ref{sec:fqe-critical-orbit-stability}). We also transfer the
bounds to other spectral functionals
(Appendix~\ref{sec:fqe-orbit-consequences}). Complete proofs and
further stability estimates are in
Sections~\ref{sec:fqe-stability-proofs} and~\ref{sec:sup-stability}
of the supplement~\cite{Companion}.
Write
\[
 \mathcal M=\{e^{i\theta}(U_1\otimes\cdots\otimes U_4)\xi:
       U_i\in SU(2),\ \xi\in\{M_4,\overline{M_4}\}\},\qquad
 d(\psi)=\min_{\chi\in\mathcal M}\|\psi-\chi\|.
\]
Thus global phase is included, qubits are labelled, and
$d^2\le2(1-1/\sqrt6)$ for unit states
(Lemma~\ref{lem:fqe-hs-fidelity-floor} of the supplement). Equivalently, if $\Phi(\psi)=\max_{\chi\in\mathcal M}
|\langle\psi,\chi\rangle|^2$, then $d^2=2(1-\sqrt{\Phi})$.
Use $N,P,T,F$ from Theorem~\ref{thm:fqe-moment}, and put
\[
 Q(z)=F(z)+N(P(z)-N^2)=18NP-12T-13N^3,\qquad
 k=\frac{\ln3-1}{2\ln2}>\frac{16}{225}.
\]
This nonnegative homogeneous sextic vanishes on the unit sphere precisely at $\mathcal M$. On unit states also set
$\mathcal D=F+3(W_1+\tau)=F+6(P-1)=Q+5(W_1+\tau)/2$.
The entropy-gap identity \eqref{eq:fqe-entropy-gap} gives
\begin{equation}\label{eq:fqe-orbit-surrogate}
 \delta(\psi):=1+\tfrac12\log_2 3-E(\psi)\ \ge\ k\mathcal D(\psi)\ge k Q(\psi).
\end{equation}
Indeed the left side of \eqref{eq:fqe-entropy-gap} is $3\ln2\,\delta(\psi)$, and its coefficients
$\alpha=15/2-6\ln3$ of $P-1$ and $\beta=3(\ln3-1)/2$ of $F$ satisfy $\alpha>6\beta$, since $\ln3=1.0986\ldots<11/10$.
As $k=\beta/(3\ln2)$, dropping $\Delta_H\ge0$ gives
$\delta\ge(\alpha(P-1)+\beta F)/(3\ln2)\ge k(6(P-1)+F)=k\mathcal D$.
Rational series bounds for $\ln2,\ln3$ give $k>16/225$ (GO-4).

\begin{theorem}[Sharp HS overlap stability]\label{thm:fqe-global-overlap}
For every unit pure four-qubit state,
\begin{equation}\label{eq:fqe-global-overlap-sharp}
 Q(\psi)\ge1-\Phi(\psi)
 \ge\frac{6+\sqrt{6(1+\sqrt{\tau(\psi)})}}{12}\,d(\psi)^2
 \ge\frac{6+\sqrt6}{12}\,d(\psi)^2,\qquad \frac{6+\sqrt6}{12}>\frac7{10}.
\end{equation}
Equality in $Q\ge1-\Phi$ holds on $\mathcal M$ ($Q=0$, $\Phi=1$) and on the three Bell-pair-product LU orbits
($Q=\Phi=1/2$), so the coefficient one is optimal.
\end{theorem}
\noindent[Exact rational certificate, GO-1--GO-3; proof in Section~\ref{sec:fqe-global-overlap} of the supplement.]

\begin{theorem}[Explicit global LU-orbit stability]\label{thm:fqe-orbit-stability}
For every unit pure four-qubit state,
\[
 d(\psi)^2\le\frac{12}{6+\sqrt6}\,Q(\psi),\qquad
 \boxed{\ \delta(\psi)\ge k\,\frac{6+\sqrt6}{12}\,d(\psi)^2\ },\qquad k\,\frac{6+\sqrt6}{12}>\frac1{20}.
\]
For every critical state the optimal bound is $\delta\ge C_*d^2$, with
$C_*\approx0.19058$ (Theorem~\ref{thm:fqe-p1-sharp}).
On an explicit neighbourhood of the two orbits the coefficient improves:
\[
 d(\psi)\le1/20\quad\Longrightarrow\quad
 \delta(\psi)\ge\frac3{10}d(\psi)^2.
\]
Its sharp limiting coefficient is
$c_*=(2-\ln3)/(2\ln2)$, in the precise sense of
\eqref{eq:fqe-entropy-local-sharp}. The power two is optimal.
No optimal global coefficient is claimed; the equality states of
Theorem~\ref{thm:fqe-p1-sharp} show that it is at most $C_*$.
\end{theorem}
\noindent[Uses Theorem~\ref{thm:fqe-global-overlap}; proof in Section~\ref{sec:fqe-stability-proofs} of the supplement.]

The overlap bound is an exact positivity certificate for $Q-(1-\Phi)$ on a $19$-variable closest-orbit normal slice
(Section~\ref{sec:fqe-global-overlap} of the supplement): a degree-six polynomial identity writing it as $277$ rational
Gram forms weighted by the slice's positive constraints, each Gram matrix certified positive by an integer diagonal-dominance test,
plus two explicit nonnegative remainders; the entropy bound then follows from
\eqref{eq:fqe-orbit-surrogate}. The local radius comes from the
transverse quadratic forms of $Q$ and of the entropy at the orbit.
Brierley and Higuchi~\cite{BH} previously proved local maximality of entropy.
The proofs are in Sections~\ref{sec:fqe-stability-proofs} and~\ref{sec:sup-stability} of the supplement.

\begin{table}[htbp]
\centering\small
\begin{tabular}{@{}p{.22\linewidth}p{.22\linewidth}p{.44\linewidth}@{}}
\toprule
Domain & Coefficient in $\delta\ge Cd^2$ & Status and proof \\
\midrule
All unit pure states & $C=1/20$ & Certified, nonoptimal; Theorem~\ref{thm:fqe-global-overlap} and \eqref{eq:fqe-orbit-surrogate}. \\[3pt]
$d\le1/20$ & $C=3/10$ & Explicit local bound; \eqref{eq:fqe-entropy-local-radius} of the supplement. \\[3pt]
$d\to0$ & $c_*=(2-\ln3)/(2\ln2)$ & Sharp limiting coefficient; \eqref{eq:fqe-entropy-local-sharp} of the supplement. \\[3pt]
Critical states & $C=C_*$ & Optimal on the whole critical locus; Theorem~\ref{thm:fqe-p1-sharp}, analytic bounds and interval certificates. \\[3pt]
All unit pure states & $C=C_*$ & Open; the critical equality states give an upper bound on any universal coefficient. \\
\bottomrule
\end{tabular}
\caption{Distance stability: domains, constants and proof methods. The elementary polarisation and
spin-flip defect bounds follow directly from \eqref{eq:fqe-entropy-gap}; they do not yet control $d$.}
\label{tab:fqe-stability-scope}
\end{table}

\subsection{Sharp constants on the critical locus}
\label{sec:fqe-critical-orbit-stability}
A state is critical when all four one-qubit marginals are $I/2$. This locus contains $M_4$, the cluster state and the
Bell-pair products, and on it the constants can be made sharp. Both theorems below are proved in the
supplement~\cite{Companion} (Sections~\ref{sec:fqe-critical-orbit-proof}, \ref{sec:fqe-p1-sharp} and~\ref{sec:fqe-critical-sharp} there).

\begin{theorem}[Sharp critical-state orbit stability]
\label{thm:fqe-critical-orbit-sharp}
For every critical unit pure four-qubit state, with $\Phi,d,Q,k$
as in Appendix~\ref{sec:fqe-orbit-stability},
\begin{align}
 Q(\psi)&\ge\max\{1-\Phi(\psi),\;3/2-2\Phi(\psi)\},
                 \label{eq:fqe-critical-fidelity}\\
 Q(\psi)&\ge\frac{2+\sqrt2}{4}d(\psi)^2,
                 \label{eq:fqe-critical-distance}\\
 \delta(\psi)&\ge k\bigl(\max\{1-\Phi,3/2-2\Phi\}+5\tau/2\bigr)
       \ge k\,\frac{2+\sqrt2}{4}\,d(\psi)^2.
                 \label{eq:fqe-critical-entropy}
\end{align}
The coefficient one in $Q\ge1-\Phi$, and the coefficient
$(2+\sqrt2)/4$ in \eqref{eq:fqe-critical-distance}, are both optimal:
Bell-pair products attain them. The comparison $Q\ge1-\Phi$ holds for all states
(Theorem~\ref{thm:fqe-global-overlap}).
If the critical state also has $\tau=0$, equivalently if $P=1$, then
\begin{equation}\label{eq:fqe-null-critical-sharp}
 \Phi\ge\frac{2+\sqrt3}{6},\qquad
 Q\ge\frac3{4-\sqrt3}(1-\Phi)\ge\frac{3+\sqrt3}{4}d^2.
\end{equation}
These coefficients and the fidelity floor are optimal, attained by cluster states.
\end{theorem}
\noindent[Checks CO-1--CO-3. Uses Theorem~\ref{thm:fqe-global-overlap}; proof in Section~\ref{sec:fqe-critical-orbit-proof} of the supplement.]

\noindent The moment route thus gives the entropy coefficient $k(2+\sqrt2)/4$ on the critical locus. On the slice $P=1$ a critical normal form is described by two auxiliary Bloch vectors
$r,s\in S^2$ (equation~\eqref{eq:fqe-null-critical-bloch} of the supplement), in which Section~\ref{sec:sup-nullsharp} of the
supplement proves a one-parameter sharpening of \eqref{eq:fqe-null-critical-sharp}. The optimal entropy coefficient
on the critical locus is larger. Put $p=\cos^2(\pi/8)=(2+\sqrt2)/4$, $H_{\rm b}(x)=-x\log_2x-(1-x)\log_2(1-x)$ and
\begin{equation}\label{eq:fqe-p1-cstar}
 C_*=\frac{\tfrac12\log_23-\tfrac13\bigl(1+2H_{\rm b}(p)\bigr)}
          {2-\sqrt{(1+1/\sqrt3)(1+\sqrt{2/3})}}
    =0.190578947\ldots
\end{equation}

\begin{theorem}[Sharp entropy stability on the critical locus]
\label{thm:fqe-p1-sharp}
Every critical unit pure four-qubit state satisfies
\[
 \delta(\psi)\ \ge\ C_*\,d(\psi)^2 ,
\]
and equality with $d>0$ holds exactly for states whose critical normal form has
one of the following $144$ Bloch-coordinate pairs, all with $P=1$. In the auxiliary Bloch coordinates $(r,s)\in S^2\times S^2$ of the null-critical normal form they are: one of $r,s$ is a signed
coordinate axis and the other is $(\pm e_b\pm e_c)/\sqrt2$, $b\ne c$ ($6\cdot12$ pairs, twice for the swap).
These pairs attain the bound, so $C_*$ is the optimal coefficient on the critical locus. Consequently every universal coefficient in
$\delta\ge Cd^2$ is at most $C_*$, and the question left open is whether $C_*$ is optimal over all states.
\end{theorem}
\noindent[Interval-certified; proof in Sections~\ref{sec:fqe-p1-sharp} and~\ref{sec:fqe-critical-sharp} of the supplement.]

\noindent The equality states explain the formula~\eqref{eq:fqe-p1-cstar}. For $r=e_1$, $s=(e_2+e_3)/\sqrt2$ one cut is maximally
mixed and the other two have spectrum $(p,p,1-p,1-p)/2$ (Lemma~\ref{lem:fqe-p1-cuts} of the supplement), so the
numerator is the entropy deficit $\delta=1+\frac12\log_23-\frac13\bigl(2+2(1+H_{\rm b}(p))\bigr)$; the closest cube pair gives
$2\sqrt\Phi=\sqrt{(1+1/\sqrt3)(1+\sqrt{2/3})}$, so the denominator is $d^2=2-2\sqrt\Phi$.
The proof represents a critical state by a $3\times3$
correlation matrix in a six-parameter chart, settles $\tau\ge0.36$ analytically, and covers the rest by an
interval-certified bisection together with a local cone lemma at the equality points.

\subsection{Spectral functionals and limitations}
\label{sec:fqe-orbit-consequences}
\begin{corollary}[Spectral-majorant stability]\label{cor:fqe-orbit-spectral}
In Proposition~\ref{prop:fqe-majorant}, suppose $a>0$ and
$\alpha=-b-17a/12>0$. Then
\[
 \mathcal E_f(M_4)-\mathcal E_f(\psi)
 \ge\frac{c_O}{3}\min(\alpha,a/12)d(\psi)^2,\qquad c_O=\frac{6+\sqrt6}{12}.
\]
For critical states, $c_O$ can be replaced by $(2+\sqrt2)/4$.
For $d\le2^{-10}$ the local bound $Q\ge3d^2$ replaces $c_O/3$ by $1$.
In particular this applies to every fixed Tsallis parameter $0<q<2$,
including its von Neumann limit in nats. At $q=2$, the enlarged
maximiser set prevents such stability relative to $\mathcal M$.
The same holds for every defect estimate of this shape: the R\'enyi
defect estimate \eqref{eq:fqe-renyi-defect} in the proof of
Theorem~\ref{thm:fqe-renyi-above-one} of the supplement,
$3m_q\ln2\,(\mathcal E_q(M_4)-\mathcal E_q(\psi))\ge\alpha_q(P-1)+\frac{a_q}{12}F$
with $m_q=2^{-q}+3\cdot6^{-q}$ and $a_q,\alpha_q>0$ the cubic and purity coefficients defined there, gives for $1<q\le5/4$
\[
 \mathcal E_q(M_4)-\mathcal E_q(\psi)
 \ge\frac{c_O\min(\alpha_q,a_q/12)}{3m_q\ln2}\,d(\psi)^2 .
\]
\end{corollary}
\noindent[Uses Theorem~\ref{thm:fqe-global-overlap}.]
\begin{proof} Drop the scalar remainder in \eqref{eq:fqe-spectral}. Since $P\ge1$, $F\ge0$ and $Q=F+P-1$ on unit states,
$\alpha(P-1)+\frac a{12}F\ge\min(\alpha,a/12)\,Q$; this is also the R\'enyi case. Then use $Q\ge c_Od^2$
(Theorem~\ref{thm:fqe-global-overlap}), \eqref{eq:fqe-critical-distance} for critical states, or, for the local
statement, $Q\ge3d^2$ when $d\le2^{-10}$ (Section~\ref{sec:fqe-orbit-local} of the supplement). \end{proof}

\paragraph{Limitations.}
The global coefficient remains a worst-case bound. Bell-pair products have
$F=0$ but $Q=1/2$, and the cluster state has $W_1=\tau=0$ but
$Q=F=1/2$, so neither term of $\mathcal D=F+3(W_1+\tau)$ alone controls the distance. The sharp local entropy
coefficient is $c_*$; the quadratic coefficients of $Q$ at the orbit are $6$ and $20$.
The comparison $Q(\psi)\ge1-\Phi(\psi)$ is sharp (Theorem~\ref{thm:fqe-global-overlap}), but $1-\Phi$ and
$d^2$ are comparable only up to the factor $(1+\sqrt\Phi)/2\in[\frac{6+\sqrt6}{12},1]$, and $\mathcal D\ge Q$ discards the
spin-flip and polarisation terms.

\section{Proofs for R\'enyi, Tsallis and mixed-state entropies}\label{sec:fqe-spectral-proofs}

\begin{proof}[Proof of Corollary~\ref{cor:fqe-tsallis}]The limit \(q\to1\) is the von Neumann entropy bound \emph{in nats} (divide by
\(\ln2\) for bits). For the proof take \(f_q(x)=(x^q-x)/(1-q)\), with \(f_q(0)=0\). On \((0,1)\), \[
f_q'''(x)=q(2-q)x^{q-3}>0,\qquad
f_q^{(4)}(x)=-q(2-q)(3-q)x^{q-4}<0.
\] Thus its Hermite cubic has \(a_q>0\) and majorises \(f_q\), including \(0\) by continuity. Set
\(g(q)=(15-7q)3^{q-1}/(5+3q)\). Direct divided differences give \[
b_q+\tfrac{17}{12}a_q
=\frac{6^{3-q}}{48(1-q)}
\bigl[5+3q-(15-7q)3^{q-1}\bigr]
=\frac{6^{3-q}(5+3q)}{48}\cdot\frac{1-g(q)}{1-q}<0. \tag{T}\label{eq:fqe-tsallis-coefficient}
\] The prefactor is positive, so the sign claim is that \(g>1\) on \((0,1)\) and \(g<1\) on \((1,2)\).
Indeed \(g(0)=g(1)=1\) and \[
(\log g)'=\ln3-\frac{80}{(15-7q)(5+3q)}.
\] Since \((15-7q)(5+3q)=75+10q-21q^2\), and \(16/15<\ln3<5/4\), this derivative is positive at \(q=0\),
increases up to the vertex \(q=5/21\) of the quadratic and then decreases, so it crosses zero exactly once in
\((0,1)\) (it is \(\ln3-\tfrac54<0\) at \(q=1\)) and remains negative on \([1,2)\). Thus \(g>1\) on \((0,1)\)
and \(g<1\) on \((1,2)\), proving the strict sign in \eqref{eq:fqe-tsallis-coefficient} on either
side of \(q=1\). The logarithm bounds follow, for example, from \(\ln3=\sum_{k\ge0}1/((2k+1)4^k)\):
its first two terms give the lower bound \(13/12\), and using \(2k+1\ge3\) in the terms with \(k\ge2\) gives the upper
bound \(13/12+\tfrac13\sum_{k\ge2}4^{-k}=10/9\). At \(q=1\), since \(g(1)=1\), the quotient \((1-g(q))/(1-q)\) tends to \(g'(1)=(\log g)'(1)=\ln3-\tfrac54\),
so the continuous limit is \(b_1+17a_1/12=\tfrac{36\cdot8}{48}(\ln3-\tfrac54)=6\ln3-\tfrac{15}2<0\).
As for the von Neumann entropy, \(f_q^{(4)}<0\) makes the Hermite remainder vanish on \([0,1]\) only at the two nodes, so
equality can also be identified without the zeros of \(F\): \(P=1\) and \(\Delta_f=0\) give every cut the spectrum
\((\tfrac12,\tfrac16,\tfrac16,\tfrac16)\), and Proposition~\ref{prop:fqe-onecut} applies.

\end{proof}

\paragraph{Sharpness of the Tsallis range.} At \(q=2\) the bound is still valid, but equality is not
unique: \(S_2(\rho)=1-\operatorname{tr}\rho^2\) and Lemma~\ref{lem:fqe-purity} give \(\mathcal
E_{S_2}(\psi)=1-P/3\le2/3\), with equality iff all one-qubit marginals are maximally mixed and
\(\tau=0\). The cluster state $C_4$ of the introduction has cut spectra \((1/2,1/2,0,0)\), \((1/4)^4\), \((1/4)^4\), and so also attains \(2/3\) at
\(q=2\). More strongly, \(M_4\) is not a mean Tsallis-\(q\) maximiser for any \(q>2\). The
averaged cut moments are \[
m_q=2^{-q}+3\,6^{-q},\qquad
c_q=(2^{1-q}+2\,4^{1-q})/3,\qquad
m_q-c_q=\frac{2^{-q}}3(1+9\,3^{-q}-8\,2^{-q}).
\] The last parenthesis vanishes at \(q=2\) and has strictly positive derivative
\(8(\ln2)2^{-q}-9(\ln3)3^{-q}\) for \(q\ge2\): indeed \((3/2)^q\ge9/4>9\ln3/(8\ln2)\) follows from
\(4>3\). Therefore \(S_q(C_4)>S_q(M_4)\) for all \(q>2\). This explains why the parameter range
\(0<q<2\) for unique \(M_4\) maximisers ends sharply at 2.

\paragraph{Expanded spectral gap.} In the setting of Proposition~\ref{prop:fqe-majorant}, with $\alpha$ and $\Delta_f$ as in its proof, the purity identity
(Lemma~\ref{lem:fqe-purity}) and the remainder identity~\eqref{eq:fqe-remainder} at \(N=1\), \(z=\psi\), turn
\eqref{eq:fqe-spectral} into the equivalent formula
\[
\begin{aligned}
3\bigl(\mathcal E_f(M_4)-\mathcal E_f(\psi)\bigr)
={}&\left(\frac\alpha2+\frac{a}{16}\right)W_1+\frac\alpha2\tau+\frac a{32}\sum_i\|V_i-8U_i\|^2\\
&+\frac a8\sum_{i<j}\bigl\|(R_i^{(i)}-R_j^{(j)})z\bigr\|^2+\Delta_f,
\end{aligned}\tag{SF+}\label{eq:fqe-spectral-plus}
\]
in which every term is nonnegative. The residual and pair terms sum to $\frac a{12}\bigl(F-\frac34W_1\bigr)$, which
Theorem~\ref{thm:fqe-general-envelope} bounds below by $\frac a{12}(\tau-\tau^{3/2})$.

\paragraph{Tsallis stability.} For \(0<q<2\), with \(\delta_q=\mathcal E_{S_q}(M_4)-\mathcal
E_{S_q}(\psi)\), the identity \eqref{eq:fqe-spectral-plus}, with \(\alpha_q=-b_q-17a_q/12>0\),
and Theorem~\ref{thm:fqe-general-envelope} give
\[
3\delta_q\ge
\left(\frac{\alpha_q}{2}+\frac{a_q}{16}\right)
W_1+\frac{\alpha_q}{2}\tau+\frac{a_q}{12}\bigl(\tau-\tau^{3/2}\bigr).
\] Here \(\alpha_q,a_q>0\) for every fixed \(q\) in this interval (with the von Neumann limit at \(q=1\)), and
\(\tau-\tau^{3/2}\ge0\). At \(q=2\) the bound becomes the identity \(\delta_2=(W_1+\tau)/6\).

\paragraph{Quantitative consequence for von Neumann entropy.} With \(\delta=E(M_4)-E(\psi)\) in bits,
Theorem~\ref{thm:fqe-general-envelope} gives \[
3\ln2\,\delta\ge
\frac{21-15\ln3}{8}W_1
+\Bigl(\frac{15}4-3\ln3\Bigr)\tau+\frac32(\ln3-1)\bigl(\tau-\tau^{3/2}\bigr).
\] Both \(\tau\) terms are nonnegative, and for small \(\tau\) the
coefficient of \(\tau\) rises from \(\frac{15}4-3\ln3\approx0.454\) to \(\frac94-\frac32\ln3\approx0.602\). Dropping $F$ in \eqref{eq:fqe-entropy-gap} gives the simpler joint estimate \[
W_1+\tau
\le\frac{6\ln2}{\tfrac{15}{2}-6\ln3}\,\delta
\approx4.57862\,\delta.
\] Thus near-maximal entropy forces small one-qubit polarisation and spin-flip overlap.
These conditions alone do not identify the state: cluster states also have $W_1=\tau=0$.
Appendix~\ref{sec:fqe-orbit-stability} adds control of the distance to the two HS orbits and transfers it to
other strict spectral majorants.

\begin{proof}[Proof of Corollary~\ref{cor:fqe-convex-roof}]
The ceiling is proved in the main text after the corollary; it remains to show that no mixed state attains it. Fix one \(2|2\) cut. The support of a
mixed \(\rho\) contains independent vectors with \(4\times4\) flattenings
\(A,B\). Their pencil contains a nonzero singular matrix: either a
generator is singular, or both are invertible and the positive-degree
polynomial \(\det(A+tB)\) has a complex root. Let \(\chi\) be the
normalised support vector represented by that matrix. Because \(\rho\)
is positive definite on its support, for some \(p>0\) the operator
\(\rho-p|\chi\rangle\langle\chi|\) is positive. Thus an ensemble for
\(\rho\) can contain \(\chi\) with positive weight. But \(\chi\) has
deficient Schmidt rank, whereas every pure maximiser has full rank on
this cut. Applying the strict pure-state bound to that ensemble makes
the mean of the convex roofs strictly smaller than the ceiling.

For R\'enyi entropies in the stated range, use the pure-state bound of
Corollary~\ref{cor:fqe-renyi-gw} in place of
\eqref{eq:fqe-spectral}; the same support argument proves strictness.
\end{proof}

The same exclusion of mixed maximisers works whenever all pure maximisers have full Schmidt rank on
one fixed square bipartition.

\section{Proofs for R\'enyi orders above one}\label{sec:fqe-renyi-proofs}

The proof separates physical compatibility from a classical entropy problem.
For a triple of four-point spectra, let $p_j$ be the purity of spectrum $j$ and write
\[
 \varepsilon=P-1,\qquad F=17P-12T-12,\qquad
 V_P=\sum_j(p_j-P/3)^2.
\]
Every pure four-qubit state supplies a triple in the relaxation
\[
 \mathcal R=\left\{\varepsilon\ge0,\quad F\ge0,\quad
                  6V_P\le F^2+\tfrac92\varepsilon\right\}.
\]
The first two constraints are the purity and sextic bounds; the third is (QK), Corollary~\ref{thm:fqe-quadratic-variance} of the sharp bound~\eqref{eq:fqe-quadratic-variance}. No spectral cap is imposed.

\subsection{High orders: an elementary bound}\label{sec:fqe-renyi-high-order}
Every order $q\ge2$ is settled by purity alone, with no anchor, crossing or certificate; the rest of this appendix
is needed only for $1<q<2$.

\begin{corollary}[Unrestricted high-order R\'enyi purity deficit]
\label{cor:fqe-renyi-two-up}
For every pure four-qubit state and every $q\ge2$,
\[
\boxed{\quad
\mathcal E_q(\psi)\le\frac53-\frac13\log_2(2P-1)\le\frac53.
\quad}
\tag{R2}\label{eq:fqe-renyi-two-up}
\]
At $q=2$, the first bound is sharp at every $P\in[1,3/2]$ and equality
holds exactly when two of the three cut marginals are maximally mixed.
For $q>2$, equality in the first bound requires and is attained by the
cluster or Bell-pair-product cut spectra. In particular, for all $q\ge2$
the global ceiling $5/3$ has equality exactly at the cluster-type cut
spectra: $(1/2,1/2,0,0)$, $(1/4)^4$, $(1/4)^4$, up to ordering of cuts.
No eigenvalue-cap hypothesis is assumed.
\end{corollary}
\begin{proof}
Put $p_j=\operatorname{tr}\rho_{1j}^2$ and $x_j=4p_j-1\ge0$.
Lemma~\ref{lem:fqe-purity} gives $P=\sum_jp_j\ge1$. Expanding a product
of nonnegative terms yields
\[
64\prod_{j=2}^4p_j=\prod_j(1+x_j)
\ge1+\sum_jx_j=4P-2,
\]
so $\prod_jp_j\ge(2P-1)/32$. Consequently
\[
\mathcal E_2(\psi)=-\frac13\log_2\prod_jp_j
\le\frac53-\frac13\log_2(2P-1).
\]
R\'enyi entropy is nonincreasing in its order, proving the bound for
$q\ge2$. At $q=2$ equality in the product inequality holds exactly
when at most one $x_j$ is positive, equivalently when at least two
cut spectra are uniform-four. The cluster-to-Bell family
\eqref{eq:fqe-quadratic-family} of Section~\ref{sec:fqe-quadratic-certificate}, with cut spectra
$((1+r)/2,(1-r)/2,0,0),(1/4)^4,(1/4)^4$ for $0\le r\le1$, has precisely
the equality form and $P=1+r^2/2$, proving sharpness for every
$P\in[1,3/2]$.

For $q>2$, equality also requires equality in R\'enyi-order
monotonicity on every cut, hence each cut spectrum is uniform on its
support. The two already-uniform-four spectra contribute total purity
$1/2$. Since $P\ge1$, the last uniform support has rank at most two:
rank two gives the cluster pattern, and rank one the Bell-pair-product
pattern. Both attain the displayed bound. Finally the weaker ceiling
$5/3$ requires $P=1$, which excludes the Bell pattern and leaves only
the cluster-type spectra.
\end{proof}

\subsection{Plan of the anchor proof}\label{sec:fqe-renyi-anchor-plan}
\paragraph{Two anchors suffice.}
Set $q_s=1587/1000<q_*<q_u=8/5$. The anchors must bracket $q_*$, because an HS bound transfers only to lower orders
and a cluster bound only to higher ones, and they are rational so that the supporting inequalities are statements
with exact rational data. Since $\mathcal E_q$ decreases in $q$, the cluster bound is easier to prove as $q$ increases towards $q_*$: $q_s$ is the
lower end of the certified enclosure $1587/1000<q_*<1589/1000$ of Proposition~\ref{prop:fqe-renyi-crossing}, and it
exceeds the threshold $q_b$ that the cubic envelope requires
(Corollary~\ref{cor:fqe-renyi-high-f}). The upper anchor $q_u=8/5$, the end of the interval in
that proposition, is a simple order at which the HS supporting inequality has explicit constants
(Lemma~\ref{lem:fqe-hs-polynomial} of the supplement). Theorem~\ref{thm:fqe-renyi-classical}
partitions $\mathcal R$ into just two regions for order transfer:
\[
\begin{array}{c|c|c}
\text{region}&\text{anchor bound}&\text{anchor equality}\\\hline
F<3/20,\quad20\varepsilon+2F<1
 &\mathcal E_{q_u}\le h_{q_u}&\text{three HS spectra}\\
\text{complement in }\mathcal R
 &\mathcal E_{q_s}\le5/3&\text{cluster triple}.
\end{array}
\]
Here $h_q=H_q(1/2,1/6,1/6,1/6)$ and the cluster triple is
$((1/2,1/2,0,0),(1/4)^4,(1/4)^4)$, up to cut order.
Single crossing (Corollary~\ref{cor:fqe-renyi-order-transfer}) transfers an HS
bound to every lower order above one; ordinary order monotonicity transfers
a constant bound to every higher order. Since $h_{q_s}>5/3>h_{q_u}$,
the two anchors give $\mathcal E_q\le\max\{h_q,5/3\}$ for all $q>1$.
With the anchor equality cases, and single crossing strict off HS spectra,
equality is HS below $q_*$, cluster above it, and both at $q_*$. The proof of Corollary~\ref{cor:fqe-renyi-gw} below gives
the order transfer and physical rigidity explicitly. Orders $0<q\le1$
were settled in Section~\ref{sec:fqe-spectral}.

\paragraph{How the anchor proof is organised.}
The subsections below give single crossing, (QV) and (QK), determinant reduction,
the cubic envelope and the Lagrangian dual. Determinant replacement keeps
each cut's second and third moments and does not decrease its entropy, so the
supporting bounds need only two double-root charts, which include spectra
with an eigenvalue above $1/2$. The supplement proves the anchors through the
disjoint partition of Table~\ref{tab:fqe-renyi-coverage} of the supplement, then the supporting
inequalities in order of use.

The physical input (QK) follows from (QV), which has an exact-rational
certificate and an independent analytic proof on critical states (Section~\ref{sec:fqe-critical-quadratic-variance} of the
supplement). The entropy estimates use calculus and explicit polynomial arithmetic, with exact polynomial sign checks.
An independent near-one defect estimate, critical-locus variance refinements and a shorter proof for states with a
rank-two cut, with further purity-controlled bounds, are in Sections~\ref{sec:fqe-renyi-above-one}--\ref{sec:fqe-renyi-filter-limit} of the supplement; the proof here does not use them.

\subsection{The HS--cluster transition}
Put $m_q=2^{-q}+3\,6^{-q}$, so the HS entropy is $h_q=\log_2m_q/(1-q)$.

\begin{proposition}[HS--cluster crossing]
\label{prop:fqe-renyi-crossing}
There is a unique $q_*\in(3/2,8/5)$ satisfying
\[
\frac{\log_2 m_{q_*}}{1-q_*}=\frac53.
\]
For every $q>q_*$, the cluster state has strictly larger mean R\'enyi entropy
than $M_4$; at $q_*$ the two values coincide. Since $1587/1000<q_*<1589/1000$ (RA-4), $q_*$ is the
threshold $\alpha_0\approx1.59$ of the R\'enyi conjecture of Gour and Wallach~\cite[Fig.~2]{GW}. That the two states are the global maximisers below and above $q_*$ is
Corollary~\ref{cor:fqe-renyi-gw}.
\end{proposition}
\begin{proof}
The cluster spectra recalled above give $(1+2+2)/3=5/3$ for every $q>0$.
The R\'enyi entropy of the nonuniform full-support HS spectrum is strictly
decreasing: $g(q)=\ln m_q$ is strictly convex, $g(1)=0$, and its secant slope
$g(q)/(q-1)$ is strictly increasing. Since $1-q<0$ for $q>1$, $h_q>5/3$ if and only if
$m_q<2^{-5(q-1)/3}$. As $m_q=2^{-q}(1+3^{1-q})$, we have
$m_{3/2}<2^{-5/6}$, as $(1+1/\sqrt3)^3=2+10/(3\sqrt3)<4$ reduces to $100<108$; and $m_{8/5}>1/2$, as
$2^{3/5}<47/31<1+3^{-3/5}$ reduces to $8\cdot31^5<47^5$ and $27\cdot16^5<31^5$. So its R\'enyi value crosses $5/3$
strictly between those parameters. Continuity and strict monotonicity finish.
The same monotonicity and the exact enclosures RA-4 of $\log_2m_q$ at $q=1587/1000$ and
$q=1589/1000$ give $h_{1587/1000}>5/3>h_{1589/1000}$, hence $1587/1000<q_*<1589/1000$.
\end{proof}

\subsection{Single crossing in the order}\label{sec:fqe-renyi-single-crossing}

The orders above one are reduced to two anchor orders by a property of single spectra:
as the order increases, the R\'enyi entropy of any four-point spectrum can overtake that of the HS spectrum
at most once, and never falls back below it. For a probability vector $\lambda$ and $q>1$ put
\[
 u_q(\lambda)=\ln\sum_i\lambda_i^q,\qquad v_q(\lambda)=\partial_qu_q(\lambda)=\frac{\sum_i\lambda_i^q\ln\lambda_i}{\sum_i\lambda_i^q},
 \qquad \kappa_q=\frac1q+\frac{\ln3}{\sinh((q-1)\ln3)},
\]
with $0^q\ln0=0$, so that $H_q(\lambda)\ln2=u_q(\lambda)/(1-q)$.
Write $\sigma=(\tfrac12,\tfrac16,\tfrac16,\tfrac16)$ for the HS spectrum.

\begin{lemma}[Single crossing]\label{lem:fqe-renyi-single-crossing}
For every $q>1$ and every four-point probability vector $\lambda$,
\[
 v_q(\lambda)-v_q(\sigma)\le\kappa_q\bigl(u_q(\lambda)-u_q(\sigma)\bigr),
 \tag{SX}\label{eq:fqe-renyi-single-crossing}
\]
with equality only when $\lambda$ is a permutation of $\sigma$.
\end{lemma}

\begin{corollary}[Order transfer]\label{cor:fqe-renyi-order-transfer}
For every four-point spectrum $\lambda$ the function
\[
 G_\lambda(q)=\frac{(q-1)\bigl(H_q(\lambda)-H_q(\sigma)\bigr)}{q\tanh\bigl((q-1)\ln3/2\bigr)}
\]
is nondecreasing on $(1,\infty)$, strictly unless $\lambda$ is a permutation of $\sigma$.
Consequently, for every triple of four-point spectra and $1<q<q_1$:
if $\mathcal E_{q_1}\le h_{q_1}$ then $\mathcal E_q\le h_q$, with equality only when all three spectra are
HS spectra; and if $\mathcal E_{q_1}<h_{q_1}$ then $\mathcal E_q<h_q$.
\end{corollary}
\begin{proof}
Put $U(q)=u_q(\lambda)-u_q(\sigma)$ and $K(q)=\ln q+\ln\tanh((q-1)\ln3/2)$, so that $K'=\kappa_q$ and
$G_\lambda=-e^{-K}U/\ln2$. By \eqref{eq:fqe-renyi-single-crossing}, $(e^{-K}U)'=e^{-K}(U'-\kappa_qU)\le0$, with equality
at an order only for HS. For a triple, the average of $G_{\lambda^{(j)}}(q)$ is a positive multiple of
$\mathcal E_q-h_q$. It increases strictly unless all three spectra are HS.
Its sign at $q_1$ therefore gives the two transfers and the equality statement.
\end{proof}

\begin{proof}[Proof of Lemma~\ref{lem:fqe-renyi-single-crossing}]
Fix $q>1$, and put $s=q-1$, $L=\ln3$, $\beta=q\kappa_q-1=qL/\sinh(sL)>0$ and
$\varphi=v_q-\kappa_qu_q$. We maximise this continuous function on the simplex.

\emph{A maximiser has full support.} On the positive coordinates, with $Z=\sum_i\lambda_i^q$,
\[
 \partial_i\varphi=\frac{g(\lambda_i)}Z,\qquad
 g(x)=x^{q-1}(q\ln x+C),\qquad C=1-qv_q-q\kappa_q.
\]
The derivative in a zero coordinate is zero, since $x^{q-1}\ln x\to0$.
At a maximiser on a face, the positive-coordinate derivatives equal a multiplier $\mu$.
But $\sum_i\lambda_i\partial_i\varphi=1-q\kappa_q=-\beta$, so $\mu=-\beta$.
Moving mass from a positive coordinate to a zero one therefore has derivative $\beta>0$,
contradicting maximality.

\emph{Only one coordinate can be larger.} Since
\[
 g'(x)=x^{q-2}\bigl[(q-1)(q\ln x+C)+q\bigr],
\]
the bracket is strictly increasing, so $g$ decreases then increases and takes each value at most twice.
The Lagrange condition thus gives at most two coordinate values, $a>b>0$.
If both occur, $g(a)=g(b)$ implies $g'(a)>0$.
Two coordinates equal to $a$ would admit the trace-preserving variation $(a+t,a-t)$.
Its first variations of $Z$ and $v_q$ vanish, and its second variation of $\varphi$ is
$2g'(a)/Z>0$, again contradicting maximality.
Thus every maximiser is uniform or a permutation of $(a,b,b,b)$.

\emph{The one-large family.} Parametrise it by
$a=(1+3e^{-z})^{-1}$, $b=ae^{-z}$, $z\ge0$, and put
$w=3e^{-qz}/(1+3e^{-qz})$, $p=3e^{-z}/(1+3e^{-z})$.
Direct differentiation gives, for $z>0$,
\[
 \begin{aligned}
 \varphi(z)&=\beta\ln(1+3e^{-z})-\kappa_q\ln(1+3e^{-qz})-zw,\\
 \varphi'(z)&=\beta(w-p)+qzw(1-w)
             =zw(1-w)\bigl(q-\beta R(z)\bigr),
 \end{aligned}
\]
where the elementary hyperbolic identity
\[
 R(z)=\frac{(e^{sz}-1)(1+3e^{-qz})}{z(1+3e^{-z})}
     =\frac{\sinh(sz)}z+
       \tanh\!\left(\frac{z-L}{2}\right)\frac{\cosh(sz)-1}z
 \tag{SR}\label{eq:fqe-renyi-crossing-ratio}
\]
settles the sign. The function $\sinh(sz)/z$ is strictly increasing for $z>0$,
and the second summand has the sign of $z-L$. Consequently
\[
 R(z)<R(L)=\frac{\sinh(sL)}L=\frac q\beta\quad(0<z<L),\qquad
 R(z)>R(L)\quad(z>L).
\]
Hence $\varphi$ strictly increases up to $z=L$ and strictly decreases afterwards.
This also excludes the uniform endpoint $z=0$.
Its unique maximum is $z=L$, namely $\sigma$, proving both the inequality and its equality case.
\end{proof}

\subsection{The universal quadratic variance bound}\label{sec:fqe-quadratic-certificate}
For $0\le r\le1$ the critical normal-form vector $v=(\sqrt{(1+r)/2},\,i\sqrt{(1-r)/2},\,0,\,0)$
(Appendix~\ref{sec:fqe-normal-form}) has cut spectra
\[
\bigl((1+r)/2,(1-r)/2,0,0\bigr),\qquad(1/4)^4,\qquad(1/4)^4,
\tag{QF}\label{eq:fqe-quadratic-family}
\]
since its two nonzero components are in quadrature. This \emph{cluster-to-Bell family} joins the cluster
state ($r=0$) to a product of two Bell pairs ($r=1$).

\begin{theorem}[Sharp quadratic purity variance]
\label{thm:fqe-sharp-quadratic-variance}
Every unit pure four-qubit state satisfies
\[
\boxed{\qquad F^2+3P(P-1)\ge6V_P,
\qquad V_P=\sum_{j=2}^4(p_j-P/3)^2.\qquad}
\tag{QV}\label{eq:fqe-quadratic-variance}
\]
Equality holds along the whole cluster-to-Bell family \eqref{eq:fqe-quadratic-family}, so the
coefficient $3$ cannot be decreased.
\end{theorem}
\noindent[Exact-rational certificate, QVA.]

\begin{corollary}[Quadratic purity variance]
\label{thm:fqe-quadratic-variance}
Every unit pure four-qubit state satisfies
\[
\boxed{\qquad F^2+\frac92(P-1)\ge6V_P.\qquad}
\tag{QK}\label{eq:fqe-quadratic-variance-proved}
\]
Equality holds exactly for HS cut spectra, cluster cut spectra, or products
of two Bell pairs, up to ordering of cuts and eigenvalues.
The coefficient $9/2$ cannot be decreased, and the coefficient $6$ cannot
be increased while retaining $9/2$.
\end{corollary}
\begin{proof}[Proofs in outline]
On critical states Proposition~\ref{prop:fqe-critical-quadratic-variance} of the supplement
proves (QV) by a joint-cut analytic argument, and an elementary box bound on the purities proves it whenever
$F\ge3/2-P$. The certificate proves it for all states at once. Homogenised by $N=\lVert z\rVert^2$, the
defect of (QV) is a qubit-symmetric, conjugation-even local-unitary invariant of bidegree $(6,6)$ that
vanishes at HS and cluster; such invariants form an $83$-dimensional space, and $83$ explicit
Gaussian-integer states determine its elements (Section~\ref{sec:fqe-invariant-dimensions} of the supplement).
Ninety-five six-copy covariants, built from $z$ and its spin flip and vanishing at HS and cluster, have squared
norms in that space. With positive integers $q_i$ and $D$, the certificate reads
$2D\,(F^2+3P(P-1)-6V_P)=\sum_iq_i\lVert\Phi_i\rVert^2$ for unit states; its homogeneous
form is~\eqref{eq:fqe-qv-allstate-sos} of the supplement. That homogeneous identity holds at the $83$ states, hence everywhere.
Every term on the right is nonnegative.
For the corollary, the defect of (QK) exceeds that of (QV) by $3(P-1)(3/2-P)\ge0$ on $1\le P\le3/2$, and the
purity box makes (QK) strict above $3/2$; equality is then analysed at $P=1$ by
Proposition~\ref{prop:fqe-critical-variance} of the supplement and at $P=3/2$ by the purity box, which leaves purities
$(1,\tfrac14,\tfrac14)$ and hence products of two Bell pairs. The family \eqref{eq:fqe-quadratic-family} fixes both coefficients.
The complete proofs are in Section~\ref{sec:sup-qk} of the supplement~\cite{Companion}
(exact rational certificate; checks QVA-1--QVA-3, QG-1, QG-3, SD-4).
\end{proof}

\subsection{Determinant reduction for four-point spectra}
\label{sec:fqe-renyi-determinant}

\begin{lemma}[Determinant monotonicity at fixed second and third moments]
\label{lem:fqe-renyi-determinant}
For $1<q<2$, let $\lambda,\lambda'$ be four-point spectra with the same second moment $p$ and third moment $t$.
If $\prod_i\lambda_i<\prod_i\lambda'_i$, then $H_q(\lambda)<H_q(\lambda')$.
Consequently, at fixed $p$ and $t$, the maximum of $H_q$ is attained at a spectrum of one of the two types
\[
(a,b,c,c)\quad\text{or}\quad(a,a,b,c),
\qquad a\ge b\ge c\ge0,
\tag{DM}\label{eq:fqe-renyi-double-root}
\]
including their degenerations.
\end{lemma}
\begin{proof}[Proof idea]
Write $e_k$ for the elementary symmetric polynomials of the eigenvalues and $d=\prod_i\lambda_i$. Newton's identities fix $e_1,e_2,e_3$, so $\prod_i(\lambda_i+s)$ depends on the spectrum only through $d$; a
Stieltjes representation of $x^q$ then shows that $\sum_i\lambda_i^q$ decreases in $d$, and a spectrum of maximal
determinant has a repeated outer pair. The complete proof is in Section~\ref{sec:sup-determinant} of the supplement~\cite{Companion}.
\end{proof}

\begin{corollary}[Six-variable classical reduction]
\label{cor:fqe-renyi-six-variable}
Let three four-component probability vectors be constrained by any
conditions involving only their total second and third moments $P,T$
and the sum of their squared second moments
$U=\sum_jp_j^2$. For every $1<q<2$, the supremum of their mean
R\'enyi entropy is unchanged if each cut is required to have one
of the spectral forms \eqref{eq:fqe-renyi-double-root}.
Up to permutation of cuts, only four patterns need be considered:
$k=0,1,2,3$ cuts with two largest equal, the others with two
smallest equal. Each cut has two free real spectral parameters.
In particular, this reduction is exact for the classical relaxation
$P\ge1$, $F=17P-12T-12\ge0$, and
$F^2+\tfrac92(P-1)\ge6(U-P^2/3)$.
Theorem~\ref{thm:fqe-renyi-classical} bounds this relaxation at the two anchor orders; with
Corollary~\ref{cor:fqe-renyi-order-transfer} and order monotonicity this gives the envelope on it for $q>1$.
\end{corollary}
\begin{proof}
At fixed $p_j,t_j$ for each cut, replacing that spectrum by one of
maximum determinant preserves $P,T,U$ and thus all the stated
constraints. Lemma~\ref{lem:fqe-renyi-determinant} shows that this
replacement cannot decrease any cut entropy. The two forms in (DM)
have two free coordinates after the trace constraint. The cut
permutation symmetry reduces the eight form choices to four.
\end{proof}

\subsection{A sharp cubic R\'enyi bound from binary grouping}
\label{sec:fqe-renyi-cubic-bound}

Binary grouping gives an independent entropy bound that will control the region of large moment slack. Set
\[
q_b=\frac{4\ln2}{4\ln2-1},\qquad
r_b=q_b-1=\frac1{4\ln2-1},\qquad
\frac32<q_b<\frac{313}{200}<\frac{1587}{1000}=q_s.
\tag{BT}\label{eq:fqe-binary-threshold}
\]
Thus $q_b/(r_b\ln2)=4$ and $1/2<r_b<1$.
The bounds on $q_b$ follow from three terms of the atanh series:
$56/81+2/1215<\ln2<1123/1620<3/4$.

\begin{lemma}[Binary grouping]
\label{lem:fqe-renyi-grouping}
For any finite probability vectors $u,v$ and $0\le a\le1$, set
$w=a^2(3-2a)$. Then
\[
H_{q_b}(au,(1-a)v)
\le4a(1-a)+wH_{q_b}(u)+(1-w)H_{q_b}(v).
\tag{BG}\label{eq:fqe-renyi-grouping}
\]
For $0<a<1$, equality holds exactly when $a=1/2$ and
$H_{q_b}(u)=H_{q_b}(v)$.
In particular $H_q(a,1-a)\le4a(1-a)$ holds for every $a$ if and only if
$q\ge q_b$; its equality cases are $a=0,1/2,1$.
\end{lemma}
\begin{proof}
Write $q=q_b$, $r=r_b$, $L=\ln2$ and
$h(w)=-w\ln w-(1-w)\ln(1-w)$.
The log-sum inequality gives, for positive $A,B$,
$\ln(A+B)\ge w\ln A+(1-w)\ln B+h(w)$.
Apply it with $A=a^q2^{-rH_q(u)}$ and
$B=(1-a)^q2^{-rH_q(v)}$. The required assertion follows if
\[
J(a):=q\{w\ln a+(1-w)\ln(1-a)\}+h(w)+4rL a(1-a)\ge0.
\]
The function $J$ is symmetric about $1/2$, and
$J(1/2)=J(1)=0$, with the latter understood by continuity.
Since $w-a=a(1-a)(2a-1)$ and $4rL=q$, differentiation cancels the
terms outside the derivative of $w$:
\[
J'(a)=6a(1-a)\left(q\ln\frac a{1-a}-\ln\frac w{1-w}\right).
\]
For $a>1/2$, put $t=a/(1-a)>1$. Then
\[
\frac w{1-w}=t\,T_2(t),\qquad T_2(t)=\frac{t(t+3)}{3t+1},
\qquad \operatorname{sgn}J'(a)=\operatorname{sgn}\{r\ln t-\ln T_2(t)\}.
\]
The function $s\mapsto\ln T_2(e^s)$ is strictly convex for $s>0$, since
\[
\left(t\frac d{dt}\right)^2\ln T_2(t)
=\frac{24t(t^2-1)}{(t+3)^2(3t+1)^2}>0.
\]
Its derivative starts at $1/2$ and tends to $1$.
Consequently $\ln T_2(e^s)-rs$ starts at zero with negative derivative,
is strictly convex, tends to $+\infty$, and has exactly one further zero.
Thus $J$ increases and then decreases between its zero endpoints $1/2,1$;
it is strictly positive in between. Symmetry proves $J\ge0$ throughout.
Equality in the log-sum inequality at $a=1/2$ requires $A=B$,
which is precisely the stated entropy condition.

Taking both conditional vectors pure proves the binary bound at $q_b$;
order monotonicity proves it for every $q\ge q_b$ with the same equality
cases. Necessity follows as $a=\epsilon\downarrow0$:
$H_q(\epsilon,1-\epsilon)=q\epsilon/((q-1)\ln2)+o(\epsilon)$,
whereas $4\epsilon(1-\epsilon)=4\epsilon+O(\epsilon^2)$.
Orders $q\le1$ are excluded by monotonicity from any $1<q<q_b$.
\end{proof}

\begin{proposition}[Sharp cubic entropy bound]
\label{prop:fqe-cubic-bound}
For every four-point probability vector $\lambda$ and every $q\ge q_b$,
\[
H_q(\lambda)\le\frac{10}3-6p+\frac83t
=4e_2(\lambda)+8e_3(\lambda),\qquad
p=\sum_i\lambda_i^2,\quad t=\sum_i\lambda_i^3.
\tag{CB}\label{eq:fqe-cubic-bound}
\]
The threshold $q_b$ is necessary for a bound valid on the entire simplex.
Equality occurs exactly at pure, uniform-two and uniform-four spectra.
\end{proposition}
\begin{proof}
By order monotonicity it suffices to take $q=q_b$.
Group the entries as
$\lambda=(au,a(1-u),(1-a)v,(1-a)(1-v))$.
For $0<a<1$, Lemma~\ref{lem:fqe-renyi-grouping} and its binary consequence give
\[
H_q(\lambda)\le4a(1-a)
+4a^2(3-2a)u(1-u)+4(1-a)^2(1+2a)v(1-v).
\]
The right side is exactly $4e_2+8e_3$, by expanding across the two groups.
If $a=0$ or $1$ the assertion is the binary bound.
The identity with $(p,t)$ follows from Newton's identities and $\sum\lambda_i=1$.
For equality with $0<a<1$, grouping forces $a=1/2$ and equal conditional
entropies; binary equality then makes both conditional spectra pure or
both uniform-two. This gives uniform-two or uniform-four. The boundary
case also permits a pure spectrum. Each such spectrum stays an equality
at every order, and all other spectra are strict already at $q_b$.
The binary restriction proves sharpness of the threshold.
\end{proof}

\begin{corollary}[Analytic cubic envelope]
\label{cor:fqe-renyi-high-f}
Every triple of four-point probability vectors satisfies, for every $q\ge q_b$,
\[
\mathcal E_q\le\frac53-\frac{20}{27}(P-1)
-\frac2{27}(F-1/2).
\tag{HC}\label{eq:fqe-renyi-high-f}
\]
In particular $\mathcal E_q\le5/3$ whenever $20(P-1)+2F\ge1$.
Equality in the $5/3$ ceiling holds exactly at a permutation of
$((1/2,1/2,0,0),(1/4)^4,(1/4)^4)$.
No quadratic variance bound or physical compatibility is required.
\end{corollary}
\begin{proof}
Sum \eqref{eq:fqe-cubic-bound} over the three cuts, substitute
$T=(17P-12-F)/12$, and divide by three.
Equality in the ceiling requires $20(P-1)+2F=1$ and equality in all three
scalar bounds. Each cut is then pure, uniform-two or uniform-four, with entropy
$0$, $1$ or $2$ at every order, and a mean of $5/3$ forces exactly one uniform-two and two
uniform-four spectra; these have $P=1$, $F=1/2$ and attain $5/3$ at every order.
\end{proof}

\subsection{An exact Lagrangian dual for the classical relaxation}
\label{sec:fqe-renyi-covering}

We now prove the anchor-order bounds used with Corollary~\ref{cor:fqe-renyi-order-transfer}.
The argument uses a classical relaxation of the cut-spectrum constraints: three four-point probability vectors
$\lambda^{(2)},\lambda^{(3)},\lambda^{(4)}$ with
$p_j=\sum_i(\lambda^{(j)}_i)^2$, $t_j=\sum_i(\lambda^{(j)}_i)^3$, $P=\sum p_j\ge1$,
$F=17P-12T-12\ge0$ ($T=\sum t_j$), and the quadratic bound \eqref{eq:fqe-quadratic-variance-proved}.
The cut spectra of every physical state form such a triple (Lemma~\ref{lem:fqe-purity}, Theorem~\ref{thm:fqe-moment},
Corollary~\ref{thm:fqe-quadratic-variance}). Write $\varepsilon=P-1$, $U=\sum p_j^2$,
$f_j=17p_j-12t_j$, and $\mathcal E_q=\frac13\sum_jH_q(\lambda^{(j)})$ in bits.

\begin{lemma}[Lagrangian dual bound]\label{lem:fqe-dual}
For $\theta=(\alpha,\beta,\gamma)\in\mathbb R^2\times[0,\infty)$ put
\[
\Lambda_\theta(\lambda)=H_q(\lambda)+\alpha(p-\tfrac13)+\beta(f-4)-\gamma(p^2-\tfrac19),
\qquad
\ell_\theta(\varepsilon,F)=\alpha\varepsilon+\beta F-\frac\gamma6\Bigl(F^2+\frac{17}2\varepsilon+2\varepsilon^2\Bigr).
\]
For every triple of the relaxation, $\sum_jH_q(\lambda^{(j)})\le\sum_j\Lambda_\theta(\lambda^{(j)})-\ell_\theta(\varepsilon,F)$.
The function $\ell_\theta$ is concave, so on a rectangle its minimum is attained at a corner.
\end{lemma}
\begin{proof}
$\sum_j(p_j-\frac13)=\varepsilon$, $\sum_j(f_j-4)=F$ and $\sum_j(p_j^2-\frac19)=U-\frac13$, hence
$\sum_jH_q=\sum_j\Lambda_\theta-\alpha\varepsilon-\beta F+\gamma(U-\frac13)$. The quadratic bound reads
$6U\le F^2+\frac92\varepsilon+2P^2$, i.e.\ $U-\frac13\le\frac16(F^2+\frac{17}2\varepsilon+2\varepsilon^2)$, and $\gamma\ge0$.
\end{proof}

\begin{lemma}[Sharp third-moment bound at fixed purity]\label{lem:fqe-third-moment}
If a four-point probability vector has purity $p=\frac14+x$, then $0\le x\le\frac34$ and
$\bigl|t-\frac1{16}-\frac34x\bigr|\le x^{3/2}/\sqrt3$.
The upper bound is attained only at the one-large spectrum
$((1+3s)/4,(1-s)/4,(1-s)/4,(1-s)/4)$, up to permutation,
where $s=\sqrt{4x/3}$. The lower bound is attained precisely at its
one-small counterpart when $x\le1/12$; at $x=0$ both are uniform.
\end{lemma}
\begin{proof}
The vectors with given $p$ form a round two-sphere $\{\lambda=\frac14\mathbf1+w:\ \sum w_i=0,\ |w|^2=x\}$
of radius $\sqrt x$. Optimising on the whole sphere gives a valid bound
even when it extends outside the simplex. On it $t=\frac1{16}+\frac34x+\sum w_i^3$; critical points of the cubic
$\sum w_i^3$ have at most two distinct coordinates, namely the patterns $(3,1)$ and $(2,2)$.
The pattern $(2,2)$ has $\sum w^3=0$, and the patterns $(z,z,z,-3z)$ with $12z^2=x$ give
$\sum w^3=-24z^3=\mp x^{3/2}/\sqrt3$.
The upper extremiser has three entries $(1-s)/4\ge0$ for every $x\le3/4$;
the lower has smallest entry $(1-3s)/4\ge0$ exactly when $x\le1/12$.
\end{proof}

\paragraph{Exact contact identities.}
Write $c_1=(1/2,1/2,0,0)$ and $c_2=(1/4)^4$.
Their $(H_q,p,f,p^2)$ values, independent of $q$, are
$(1,1/2,11/2,1/4)$ and $(2,1/4,7/2,1/16)$.
Consequently, with $\Lambda^\star_i=\Lambda_\theta(c_i)$,
\[
\Lambda^\star_1=1+\frac\alpha6+\frac32\beta-\frac5{36}\gamma,
\qquad \Lambda^\star_2=2-\frac\alpha{12}-\frac\beta2+\frac7{144}\gamma,
\qquad \Lambda^\star_1+2\Lambda^\star_2-\ell_\theta(0,1/2)=5,
\tag{DC}\label{eq:fqe-dual-contact}
\]
since $\ell_\theta(0,1/2)=\beta/2-\gamma/24$.
\begin{theorem}[Two R\'enyi anchors]
\label{thm:fqe-renyi-classical}
Let $q_s=1587/1000$ and $q_u=8/5$. Every triple satisfying $P\ge1$, $F\ge0$ and
\eqref{eq:fqe-quadratic-variance-proved} obeys one of the following.
\begin{enumerate}
\item If $F\ge3/20$ or $20\varepsilon+2F\ge1$, then $\mathcal E_{q_s}\le5/3$, with equality only at a permutation of
$(c_1,c_2,c_2)$.
\item Otherwise $\mathcal E_{q_u}\le h_{q_u}$, with equality only when all three spectra are HS spectra.
\end{enumerate}
\end{theorem}
\begin{proof}[Proof outline]
The cubic envelope first removes $20\varepsilon+2F\ge1$.
On the residual, $\varepsilon<1/20$ and every $p_j<11/20$.
For $F<3/20$ use the HS supporting inequality at $q_u$.
For $3/20\le F\le11/25$ use the strict dual band at $q_s$.
The remaining region lies near the cluster triple: the capped cubic bound
handles cuts with $p_j\le47/100$, and the sliver bound handles the unique
cut above that threshold. Table~\ref{tab:fqe-renyi-coverage} and
Section~\ref{sec:sup-renyi-classical-proof} of the supplement~\cite{Companion}
give the complete partition, summation and equality argument.
\end{proof}

\begin{proof}[Proof of Corollary~\ref{cor:fqe-renyi-gw}]
For $0<q<1$ use Corollary~\ref{cor:fqe-renyi}, and for $q=1$ use Theorem~\ref{thm:fqe-entropy-max}.
For $q>1$, the three cut spectra of $\psi$ form a triple of the relaxation
(Lemma~\ref{lem:fqe-purity}, Theorem~\ref{thm:fqe-moment} and Corollary~\ref{thm:fqe-quadratic-variance}); recall
$q_s<q_*<q_u$ and that $\mathcal E_q$ and $h_q$ decrease in $q$.

In case (1) of Theorem~\ref{thm:fqe-renyi-classical}, $\mathcal E_{q_s}\le5/3<h_{q_s}$. Corollary~\ref{cor:fqe-renyi-order-transfer}
gives $\mathcal E_q<h_q$ for $1<q\le q_s$, and $\mathcal E_q\le\mathcal E_{q_s}\le5/3$ for $q\ge q_s$. So
$\mathcal E_q\le\max\{h_q,5/3\}$, strictly for $q<q_*$; for $q\ge q_*$ equality forces $\mathcal E_{q_s}=5/3$, hence the
cluster-type triple.

In case (2), Corollary~\ref{cor:fqe-renyi-order-transfer} gives $\mathcal E_q\le h_q$ for $1<q\le q_u$, with equality only
for three HS spectra, and $\mathcal E_q\le\mathcal E_{q_u}\le h_{q_u}<5/3$ for $q\ge q_u$. As $h_q<5/3$ for $q>q_*$, equality
occurs only for $q\le q_*$ and three HS spectra.

Three HS spectra imply $P=1,F=0$, so Proposition~\ref{prop:fqe-onecut} gives the two HS LU orbits.
A physical state with triple $(c_1,c_2,c_2)$ has two maximally mixed cut marginals;
Proposition~\ref{prop:fqe-two-flat-rigidity} identifies it with the cluster endpoint of
\eqref{eq:fqe-quadratic-family}.
\end{proof}

\begin{remark}[Scope of the relaxation]\label{rem:fqe-renyi-covering-scope}
The classical relaxation uses only $P\ge1$, $F\ge0$ and (QK); no joint purity bound, spectrum cap or four-tangle information.
It is needed only at the two anchor orders: Corollary~\ref{cor:fqe-renyi-order-transfer} acts cut by cut and needs no
compatibility at all.
The multiplier $\beta$ may have either sign; the dual has no gap on the sliver only after the third-moment link of
Lemma~\ref{lem:fqe-third-moment} (a constant $\theta$ on a fixed $(\varepsilon,F)$ cell fails there).
The equality classification uses the same strict local bounds as the inequality;
it requires no additional covering.
\end{remark}

\subsection{A spectral cap from spin flip}
\label{sec:fqe-spectral-spinflip}

The following bound controls the part of a cut spectrum above $1/2$
directly by two-copy quantities. It holds for every pure state, not only on
the critical locus, and needs neither a four-qubit normal form nor a
polynomial certificate.
Write $x_+=\max\{x,0\}$ and use the Hilbert--Schmidt norm throughout.

\begin{proposition}[Spectral spin-flip bound]
\label{prop:fqe-spectral-spinflip}
For every unit pure four-qubit state and every balanced cut $A|A^c$,
\begin{equation}\label{eq:fqe-spectral-spinflip}
 \boxed{\quad
 (2\lambda_{\max}(\rho_A)-1)_+^2
 \le \tau+\frac12W_1
 =2(P-1)-\frac12W_1\le2(P-1).
 \quad}
\end{equation}
Both coefficients on the first right-hand side are optimal. In particular,
$P=1$ forces every balanced eigenvalue to be at most $1/2$.
\end{proposition}

The matrix inequality behind the proposition holds in every bipartite
dimension. The two vectors below need only have the same Schmidt
coefficients; their Schmidt bases may be unrelated.

\begin{lemma}[Overlap and marginal mismatch]
\label{lem:fqe-isospectral-overlap}
Let $\psi,\phi$ be unit bipartite pure states with common squared Schmidt
coefficients $\lambda_1\ge\cdots\ge\lambda_n\ge0$. Let
$\rho_A,\rho_B$ and $\sigma_A,\sigma_B$ be their marginals. Then
\begin{equation}\label{eq:fqe-isospectral-overlap}
 |\langle\psi,\phi\rangle|^2+
 \frac12\bigl(\|\rho_A-\sigma_A\|_F^2+
                    \|\rho_B-\sigma_B\|_F^2\bigr)
 \ge (2\lambda_1-1)_+^2.
\end{equation}
For each common Schmidt spectrum with $\lambda_1\ge1/2$, the bound is
attained by changing the sign of every Schmidt term except the first.
\end{lemma}

\begin{proof}
The case $\lambda_1\le1/2$ is immediate. Assume $\lambda_1>1/2$.
Pad the coefficient matrices with zeros if the local dimensions differ,
and choose Schmidt bases for $\psi$. Its coefficient matrix is
$D=\operatorname{diag}(\sqrt{\lambda_i})$; write $W$ for that of $\phi$.
Thus $WW^\dagger$ and $W^\dagger W$ have eigenvalues $\lambda_i$.
Put
\[
 \Lambda=D^2,\qquad r=1-\lambda_1,\qquad d=2\lambda_1-1,
 \qquad S=\operatorname{diag}(0,r-\lambda_2,\ldots,r-\lambda_n).
\]
The entries of $S$ are nonnegative and increasing. Trace rearrangement
therefore gives, for $R=WW^\dagger$ and for $R=W^\dagger W$,
\begin{equation}\label{eq:fqe-isospectral-rearrange}
 \operatorname{tr}(SR)\ge
 \sum_{i=2}^n(r-\lambda_i)\lambda_i
 =r^2-\sum_{i=2}^n\lambda_i^2.
\end{equation}
Indeed, diagonalising $R$ expresses its diagonal as a doubly stochastic
average of permutations of the decreasing vector $(\lambda_i)$; pairing
it with the increasing diagonal of $S$ is minimised in the displayed order.

Define the quadratic form
\[
 \mathcal B_D(W)=\|W\|_F^2+|\operatorname{tr}(DW)|^2
       -\operatorname{tr}\bigl(\Lambda(WW^\dagger+W^\dagger W)\bigr).
\]
Its useful lower bound has an explicit positive remainder. Let
$W_{\rm t}=(W_{ij})_{i,j\ge2}$ and
$t=\sum_{i=2}^n\sqrt{\lambda_i}W_{ii}$, so $\operatorname{tr}(DW)=\sqrt{\lambda_1}W_{11}+t$. Since
$\Lambda+S=rI+dE_{11}$, with $E_{11}$ the first matrix unit, and $1-2r=d$, the left side below is
$d\bigl(\|W_{\rm t}\|_F^2-|W_{11}|^2\bigr)+|\sqrt{\lambda_1}W_{11}+t|^2$; for $r>0$, completing the square in
$W_{11}$ with $\lambda_1-d=r$ gives
\begin{equation}\label{eq:fqe-isospectral-calibration}
 \begin{split}
 \mathcal B_D(W)-\operatorname{tr}\bigl(S(WW^\dagger+W^\dagger W)\bigr)
 ={}&d\left(\|W_{\rm t}\|_F^2-\frac{|t|^2}{r}\right)
       +\frac{|rW_{11}+\sqrt{\lambda_1}t|^2}{r}\ge0.
 \end{split}
\end{equation}
Cauchy--Schwarz gives
$|t|^2\le r\sum_{i\ge2}|W_{ii}|^2\le r\|W_{\rm t}\|_F^2$.
This proves positivity without any condition on $W$.
When $r=0$, the remainder is simply $\|W_{\rm t}\|_F^2\ge0$.

Apply~\eqref{eq:fqe-isospectral-rearrange} twice to obtain
\[
 \mathcal B_D(W)\ge
 2\left(r^2-\sum_{i=2}^n\lambda_i^2\right).
\]
Set $c_2=\sum_i\lambda_i^2$. In these coordinates
$\langle\psi,\phi\rangle=\operatorname{tr}(DW)$, $\rho_A=\Lambda$, $\sigma_A=WW^\dagger$,
and $\rho_B,\sigma_B$ are the transposes of $\Lambda,W^\dagger W$. Since
$\operatorname{tr}(WW^\dagger)^2=\operatorname{tr}(W^\dagger W)^2=c_2$, the mismatch terms
are $2c_2-2\operatorname{tr}(\Lambda WW^\dagger)$ and
$2c_2-2\operatorname{tr}(\Lambda W^\dagger W)$. With $\|W\|_F^2=1$, the left-hand
side of~\eqref{eq:fqe-isospectral-overlap} is therefore
$\mathcal B_D(W)-1+2c_2$. Its lower bound is
$2r^2+2\lambda_1^2-1=d^2$, as required. Taking
$W=\operatorname{diag}(\sqrt{\lambda_1},-\sqrt{\lambda_2},\ldots,
-\sqrt{\lambda_n})$ proves sharpness.
\end{proof}

\begin{proof}[Proof of Proposition~\ref{prop:fqe-spectral-spinflip}]
Apply Lemma~\ref{lem:fqe-isospectral-overlap} to
$\phi=\widetilde\psi=\sigma_y^{\otimes4}\overline\psi$.
Spin flip preserves every Schmidt spectrum and
$|\langle\psi,\widetilde\psi\rangle|^2=\tau$.
For a two-qubit marginal on sites $i,j$, Pauli expansion gives
\[
 \rho_{ij}-\widetilde\rho_{ij}
 =\frac12\left(\mathbf r_i\cdot\boldsymbol\sigma\otimes I
              +I\otimes\mathbf r_j\cdot\boldsymbol\sigma\right),
 \qquad
 \|\rho_{ij}-\widetilde\rho_{ij}\|_F^2
 =\|\mathbf r_i\|^2+\|\mathbf r_j\|^2.
\]
Thus the two marginal mismatch terms add to $W_1/2$.
The purity identity gives the remaining equalities and inequality
in~\eqref{eq:fqe-spectral-spinflip}.

For sharpness take
$\psi=(a|00\rangle+b|11\rangle)_{12}\otimes
(|00\rangle+|11\rangle)_{34}/\sqrt2$, with real $a,b\ge0$ and
$a^2+b^2=1$. On $12|34$, the left side is one, while
$\tau=4a^2b^2$ and $W_1=2(a^2-b^2)^2$, so equality holds throughout.
The choices $a=b$ and $b=0$ force respectively the coefficients of
$\tau$ and $W_1$ to be at least $1$ and $1/2$.
\end{proof}

Equation~\eqref{eq:fqe-spectral-spinflip} supplies a fully analytic quantum
compatibility constraint. The separate-cut bounds do not by themselves
replace the quadratic variance certificate: SCB-5 gives exact classical spectra
satisfying all three caps and even $F\ge\tau$ (hence the critical
envelope~\eqref{eq:fqe-critical-envelope}) but violating (QK), so compatibility
between the three cut spectra still matters.

\subsection{Quantitative rigidity of two maximally mixed cuts}
\label{sec:fqe-renyi-two-flat}

\begin{proposition}[Crossed purity costs and two-flat rigidity]
\label{prop:fqe-two-flat-rigidity}
Let $\psi$ be any unit pure four-qubit state. Fix a balanced cut $k$,
write its ordered spectrum as
$\lambda_1\ge\lambda_2\ge\lambda_3\ge\lambda_4\ge0$,
and put
\[
\Delta_k=\sum_{j\ne k}(p_j-1/4),\qquad
p_j=\operatorname{tr}\rho_{1j}^2.
\]
If $\lambda_1\ge1/2$, then
\[
\boxed{\quad
\Delta_k\ge\frac14W_1+
2(\lambda_2\lambda_3+\lambda_2\lambda_4+
\lambda_3\lambda_4)
\ge\frac14W_1+(\lambda_3+\lambda_4)^2.\quad}
\tag{TF}\label{eq:fqe-two-flat}
\]
The first inequality is sharp for every chosen spectrum with
$\lambda_1\ge1/2$. On the critical locus, equality holds exactly when
all nonzero squares of the other normal-form amplitudes have a phase
opposite to the square of the largest amplitude. The noncritical
sharpness family in Proposition~\ref{prop:fqe-spectral-spinflip} also
attains the first inequality.

Consequently, whenever \emph{two} balanced cuts of a pure
four-qubit state have maximally mixed marginals $(1/4)^4$,
the third has Schmidt rank at most two. Up to local
unitaries and a permutation of the qubits, every such
state belongs to the attained cluster-to-Bell family
\eqref{eq:fqe-quadratic-family}. In particular the exact
R\'enyi envelope of Proposition~\ref{prop:fqe-renyi-rank-two} of the supplement
applies for all $q>0$.
\end{proposition}
\begin{proof}
The purity identity and Proposition~\ref{prop:fqe-spectral-spinflip} give
\[
2\Delta_k
=2(P-p_k-1/2)
=W_1+\tau-(2p_k-1)
\ge\frac12W_1+(2\lambda_1-1)^2-(2p_k-1).
\]
Using $\sum_i\lambda_i=1$ gives
\[
(2\lambda_1-1)^2-(2p_k-1)
=4(\lambda_2\lambda_3+\lambda_2\lambda_4+
\lambda_3\lambda_4).
\]
This proves the first bound. Since
$\lambda_2\ge(\lambda_3+\lambda_4)/2$,
the second follows by dropping the nonnegative
$\lambda_3\lambda_4$ term.
On the critical locus, write the chosen spectrum as $\lambda_i=|w_i|^2$
in the normal form, so $\tau=|\sum_iw_i^2|^2$.
The bound is then the reverse triangle inequality, with exactly the stated
opposite-phase equality condition. Choosing those phases realises every
ordered spectrum with $\lambda_1\ge1/2$, proving sharpness.

Suppose now two cuts have purity $1/4$. Already one
maximally mixed two-qubit cut forces all four one-qubit
marginals to be $I/2$ by purity of complementary
reductions. If $k$ is the remaining cut, then
$p_k=P-1/2=1/2+\tau/2\ge1/2$. Since
$p_k=\sum_i\lambda_i^2\le\lambda_1$, the hypothesis
$\lambda_1\ge1/2$ holds. The first bound and
$\Delta_k=0$ force at most one among
$\lambda_2,\lambda_3,\lambda_4$ to be nonzero.
The opposite-phase equality condition means that, up to
overall phase and a signed permutation, $w$ has the form
\[
w=(\sqrt a,\ i\sqrt{1-a},\ 0,\ 0),\qquad 1/2\le a\le1.
\]
The magic-basis action is
$\operatorname{diag}(w)\mapsto O_1\operatorname{diag}(w)O_2^T$, $O_1,O_2\in SO(4)$: the same permutation on both
sides permutes the entries of $w$, a sign on one side changes the sign of one entry, and extra signs on the zero
coordinates make both factors proper rotations. This action is induced by local one-qubit unitaries on
each side of the chosen cut, so this is exactly the
normal form \eqref{eq:fqe-quadratic-family} up to LU and
cut relabelling.
\end{proof}

\section{Seven qubits: copy moments and the five-copy proofs}\label{sec:fqe-seven}

We define the copy hierarchy used in Section~\ref{sec:fqe-seven-entropy}, locate the four-qubit inequalities
within it, and record the ingredients of both seven-qubit
proofs. The complete proofs are in the supplement~\cite{Companion};
Appendix~\ref{sec:ico} constructs the icosahedral family. For a pure state of $n$ qubits put
\[
 \pi_k(\psi)=\binom nk^{-1}\sum_{|S|=k}\operatorname{tr}\rho_S^2.
\]

\paragraph{Pauli expansion.} Let $A_j$ be the sum of the squared expectation values of the $3^j\binom nj$ Pauli strings
of weight $j$ (acting nontrivially on exactly $j$ qubits), so $A_0=1$ and $A_1=\sum_i\|\mathbf r_i\|^2$ (the $W_1$ of
the four-qubit sections). Expanding each marginal in Pauli strings gives
$\pi_k=2^{-k}\binom nk^{-1}\sum_{j\le k}\binom{n-j}{k-j}A_j$~\cite{Rains}; at $n=7$, $k=3$ this is
$280\pi_3=35+15A_1+5A_2+A_3$. An AME state has $A_1=\dots=A_k=0$, which makes it a pure quantum code of distance
$k+1$~\cite{Scott}. For stabilizer states the $A_j$ and the corresponding shadow coefficients $S_w$ (defined in Proposition~\ref{prop:stab7}) are integers; with
$A_1=A_2=0$ the shadow identity $4S_1=7-A_3$ forces $A_3\ge3$, and if $A_1+A_2\ge1$ instead, integrality gives
$15A_1+5A_2+A_3\ge5$. Thus every seven-qubit stabilizer state has $\pi_3\ge19/140$ (Proposition~\ref{prop:stab7}).

\paragraph{Copy moments.} Purities are linear observables of the two-copy density operator: if $\mathrm{SW}_S$ swaps the two
copies of the qubits in $S$, then $\operatorname{tr}\rho_S^2=\operatorname{tr}[\mathrm{SW}_S\,(|\psi\rangle
\langle\psi|)^{\otimes2}]$. A lower bound on $\pi_k$ therefore follows from any convex set of operators
that contains the twirled $m$-copy moments
\[
  X_m=\operatorname{Av}_{\sigma,K}\int\bigl(U|\psi\rangle\langle\psi|U^\dagger\bigr)^{\otimes m}\,dU,
  \qquad m\ge2,
\]
where $U=u_1\otimes\cdots\otimes u_n$ ranges over local unitaries (LU), and
$\operatorname{Av}_{\sigma,K}$ also averages over qubit permutations and complex conjugation (the supplement writes $X_m^+$).
These averages preserve the mean purity objective, and the partial trace over one copy maps $X_{m+1}$ to $X_m$.

Spin flip supplies further positivity constraints. With
$\epsilon=\bigl(\begin{smallmatrix}0&1\\-1&0\end{smallmatrix}\bigr)$ and
$\tilde\psi=\epsilon^{\otimes n}\overline\psi$ the spin-flipped state, the twirl of
$|\psi^{\otimes(m-j)}\tilde\psi^{\otimes j}\rangle\langle\psi^{\otimes(m-j)}\tilde\psi^{\otimes j}|$ is an average of
rank-one projectors, $1\le j\le m/2$; we denote it by $\Lambda_j(X_m)$ when the flipped copies are the last $j$, and set
$\Lambda_0(X_m)=X_m$. Since $\epsilon\overline u\epsilon^{-1}=u$ for $u\in SU(2)$, the flip commutes with the
local twirl; the extra phase for $U(2)$ cancels in projectors. It also commutes with qubit
permutations and complex conjugation. Thus $\Lambda_j$ is a fixed linear image of $X_m$: partial
transpose on the $j$ flipped copies followed by conjugation by the corresponding $\epsilon$ factors.
Minimising $\pi_k$ over trace-one, twirl-invariant operators $X_m$ supported on the symmetric subspace,
subject to $X_m\succeq0$ and $\Lambda_j(X_m)\succeq0$, is a semidefinite programme. Its optimum $L_m$ bounds $\min\pi_k$ from below, $L_m\le L_{m+1}$, and $L_m\to\min\pi_k$; the
level $m=2$ is the shadow-enumerator linear programme of Rains~\cite{Rains}
(Proposition~\ref{prop:rainsppt} and Theorem~\ref{thm:spinflip-ppt-hierarchy} of the supplement~\cite{Companion}).

\paragraph{The four-qubit inequalities in this hierarchy.} A certificate at level $m$ is an identity
\[
 \text{objective}-\text{bound}=\sum_{j}\operatorname{tr}\bigl[\Xi_j\,\Lambda_j(X_m)\bigr],\qquad \Xi_j\succeq0,
\]
valid for all $X_m$; on a genuine state each term is an average of squared norms
$\|\Xi_j^{1/2}(\psi^{\otimes(m-j)}\otimes\tilde\psi^{\otimes j})\|^2$.
At $n=4$ the purity bound $P\ge1$ of Lemma~\ref{lem:fqe-purity} is the level-two linear-programming (LP) minimum
$\pi_2=1/3$ (Theorem~\ref{thm:lptable} of the supplement). Its proof evaluates $\tau=\operatorname{tr}(\rho\tilde\rho)$, the
shadow term. The sextic identity~\eqref{eq:fqe-remainder} is a level-three certificate (it holds as a linear identity in the three-copy moment) with one flipped copy: every
term is $\langle w,Aw\rangle$ with $w=z\otimes z\otimes\tilde z$ and $A\succeq0$
(Appendix~\ref{sec:fqe-sextic}). Its objective $F$ differs from the purity, so it uses the same positivity
conditions with a different objective.

At $n=7$ the levels separate (Table~\ref{tab:fqe-seven-levels}). Five copies are the first level that gives the sharp
purity bound (levels two to four are provably not sharp). With a purity band and an eigenvalue cap as extra constraints, they also give the sharp bound for the entropy majorant of
Section~\ref{sec:fqe-seven-entropy}.
\begin{table}[ht]
\centering\small
\begin{tabular}{@{}cll@{}}
\toprule
$m$ & $L_m$ at $n=7$ & status\\
\midrule
$2$ & $L_2=1/8$ & exact (LP-2; a putative AME weight distribution)\\
$3$ & $L_3=\tfrac{125}{978}=0.1278118\ldots$ & proved (Thm.~\ref{thm:threecopy})\\
$4$ & $\tfrac{211}{1600}\le L_4<\tfrac{6597}{50000}<\tfrac{121}{917}$ & bounds proved (Thm.~\ref{thm:fourcopy}, Prop.~\ref{prop:fourcopy-not-sharp}); $L_4\approx0.131877$ numerical (FS)\\
$5$ & $L_5=\tfrac{121}{917}=0.131952\ldots$ & verified-arithmetic certificate (Thm.~\ref{thm:fqe-seven-purity})\\
\bottomrule
\end{tabular}
\caption{The seven-qubit copy hierarchy for $\pi_3$. Strictness $L_2=1/8<L_3$ is read off the table and $L_3<L_4$ is proved
(Corollary~\ref{cor:strict-copy-hierarchy} of the supplement), and $L_4<L_5$ by the exact feasible pseudo-moment of Proposition~\ref{prop:fourcopy-not-sharp} (exact certificate; checks NS-0--NS-4).}
\label{tab:fqe-seven-levels}
\end{table}

\paragraph{Ingredients of the five-copy proofs.} Section~\ref{sec:fqe-seven-entropy} outlines the proofs of
Theorems~\ref{thm:fqe-seven-purity} and~\ref{thm:fqe-seven-entropy} in three steps (value, equality moment, orbit), taken
at a point $\psi_y$ of the icosahedral family: $y=37/524$ for the purity and $y=x_*$ for the entropy. The complete proofs
are Sections~\ref{sec:fivecopy} and~\ref{sec:sup-seven-entropy} of the supplement. Here we define the objects used in those outlines
and record the sizes of the computations.

\emph{Faces.} By representation theory the positivity conditions at level five decompose into blocks. The
certificates use some of them, $\Gamma_b(X_5)$, each a compression (followed by a fixed congruence) of $\Lambda_{j_b}(X_5)$
with $j_b\in\{0,1,2\}$ flipped copies. Exact matrices $W_b$ (rational for the purity; for the entropy defined by exact
formulas at $x_*$, with enclosures used only to check the identity) have columns in the kernels of the blocks at $\psi_y$,
and the \emph{faces} are $W_b^{\mathsf T}\Gamma_b(X_5)W_b\succeq0$. With positive definite $C_b$, the objective minus its value at
$\psi_y$ (for the entropy, the value minus the objective) equals $\sum_b\operatorname{tr}[C_bW_b^{\mathsf T}\Gamma_b(X_5)W_b]$
plus, for the entropy, further nonnegative terms. For the purity there are twelve faces:
\[
 \pi_3-\frac{121}{917}=\sum_{b=1}^{12}\operatorname{tr}\bigl[C_bW_b^{\mathsf T}\Gamma_b(X_5)W_b\bigr]
\]
for every normalised $X_5$, in the $3285$-dimensional space of conjugation-even five-copy moments
(Proposition~\ref{prop:fivecopy-finite-probes} of the supplement). For the entropy the objective is the
mean of the joint majorant of Section~\ref{sec:fqe-seven-entropy}; there are seventeen faces, and the extra terms are the
purity-band constraints times nonnegative cubic polynomials and matrix squares weighted by the eigenvalue cap.

\emph{Probes and verified arithmetic.} In both cases the target functional and the face entries vanish at $X_5(\psi_y)$ and at
$X_5'=\tfrac{d}{dx}X_5(\psi_x)|_{x=y}$. They therefore lie in the $3283$-dimensional \emph{critical space}
of linear functionals annihilating these two moments. A computer evaluates the faces at $3424$ rational probe
moments, and a rank computation with rigorous error bounds shows that the probes separate the critical space; solving
the probe equations gives each $C_b$ (for the entropy, the first twelve) as a verified small correction of the identity
matrix. In practice numerical linear algebra proposes an approximate inverse, and rigorous rounding-error bounds validate
it; Section~\ref{sec:fivecopy-positive-repair} of the supplement gives the rank test, the error model and the margins. The arithmetic is floating point with rigorous error bounds, so the proof uses a verified-arithmetic certificate
rather than an exact-rational computation. The same rank computation shows that the face entries span the critical space, which
is the input of the equality step. The purity certificate also gives a quantitative bound: $\pi_3-121/917$ is at least the
least eigenvalue of the $C_b$ times the total face energy $\sum_b\operatorname{tr}[W_b^{\mathsf T}\Gamma_b(X_5)W_b]$.

\emph{Equality moment and orbit.} At equality every face vanishes. Since the face entries span the critical space, the
moment then lies on the tangent line $X_5(\psi_y)+tX_5'$; this is Lemma~\ref{lem:fivecopy-tangent-line} of the
supplement, which holds for any number of qubits and copies. The invariant that fixes the tangent parameter is $H=104S_1-25A_2$, with $S_1$
the weight-one shadow coefficient (Proposition~\ref{prop:stab7}) and $A_2$ the weight-two Pauli sum of the Pauli expansion above (the invariant $\mathcal E_2$ of Theorem~\ref{thm:moment-orbit-rigidity}); its family
value is $h=155+96y$. Both $H$ and $H^2$ are linear in $X_5$, so at $X_5(\psi_y)+tX_5'$ they force
$(h+96t)^2=h^2+192ht$, that is $t=0$; equivalently, the centred square $(H-h)^2$ is linear in the moment and lies in the critical
space, so it vanishes at equality, yet on the actual state it equals $(96t)^2$. Hence $X_5(\psi)=X_5(\psi_y)$. The orbit step is Theorem~\ref{thm:moment-orbit-rigidity} of the supplement, valid for every
$y\in(1/32,1/12)$: its thirteen invariants, of bidegree at most $(4,4)$ in $\psi,\bar\psi$, are all determined by the
four-copy marginal of $X_5$, and force a local frame in which $\psi$ is fixed by an antilinear operator built from the spin
flip, that of Proposition~\ref{prop:kramers-conference-shell}; the remaining matrix $Q$ is complex symmetric of order six
with zero diagonal, and the real amplitude is fixed by $y$; the conference-matrix conclusion, unique up to diagonal
unitaries and permutations, is Theorem~\ref{thm:complexsos} (the identity~\eqref{eq:fqe-complex-sos}) with Proposition~\ref{prop:confphase}; and
Corollary~\ref{cor:icosahedral-labelled-orbits} counts the $42$ labelled LU orbits. Steps 2 and 3 together are
Theorem~\ref{thm:fivecopy-family-equality} of the supplement, stated for every $y\in(1/32,1/12)$.

\section{Seven qubits: the icosahedral state}\label{sec:ico}

Label the qubits $s,0,1,\dots,5$. Let $s_{ij}=-1$ for $\{i,j\}$ an edge of the pentagon
$1\!-\!3\!-\!2\!-\!4\!-\!5\!-\!1$ and $s_{ij}=+1$ otherwise. These signs represent the regular two-graph of the six
icosahedral axes: a triangle is negative when its three signs multiply to $-1$, and every pair
lies in exactly two negative triangles. The two-graph has symmetry group $A_5\cong PSL(2,5)$. For $0\le x\le\tfrac12$ put
$\alpha=\sqrt{(1-2x)/30}$, $\beta=\sqrt x$ and
\begin{equation}\label{eq:ico}
\begin{split}
  \psi_x ={}& |0\rangle_s\Bigl(\alpha\sum_{i<j}s_{ij}|e_i+e_j\rangle + \beta|111111\rangle\Bigr)\\
         &+ |1\rangle_s\Bigl(\beta|000000\rangle - \alpha\sum_{i<j}s_{ij}|111111-e_i-e_j\rangle\Bigr).
\end{split}
\end{equation}

\begin{theorem}[the icosahedral family]\label{thm:ico}
\leavevmode\begin{enumerate}
  \item Every one-qubit marginal of every $\psi_x$ is $I/2$.
  \item $\pi_3(\psi_x)$ is a quadratic polynomial in $x$, minimised within the family at $x=37/524$. The
    \emph{icosahedral state} $\Psiico=\psi_{37/524}$ has $\pi_3(\Psiico)=121/917=0.131952\ldots$
  \item Its $20$ three-body marginals avoiding $s$ have purity $4381/34322$ and its $15$ containing $s$
    have purity $2363/17161$. Its pair purities are $8641/34322$ ($15$ pairs avoiding $s$) and
    $8761/34322$ ($6$ pairs containing $s$).
  \item Throughout the family, writing $u=\alpha^2=(1-2x)/30$,
    its triple spectra are $(4u)^6,(3u+x)^2$ on cuts avoiding $s$,
    and $(8u),(4u)^5,(u+x)^2$ on cuts containing $s$.
\end{enumerate}
\end{theorem}

\begin{proof}
Let $C$ be the $6\times6$ zero-diagonal symmetric conference matrix with entries $C_{ij}=s_{ij}$ ($i\ne j$).
Every pair of vertices lies in two negative and two positive triangles, so $C^2=5I_6$.
In~\eqref{eq:ico} the $|0\rangle_s$ and $|1\rangle_s$ branches are supported on ordinary Hamming weights
$\{2,6\}$ and $\{0,4\}$, respectively. Disjointness of these weight sets makes $\rho_s$ diagonal, with each branch of weight $15u+x=1/2$.
Within each branch the ordinary weights have one parity, so every $\rho_i$ is diagonal, and
$P(b_i=1)=5u+x+10u=1/2$. This proves item~(1).

For an ordinary triple $I=\{i,j,k\}$ with complement $J=\{0,\dots,5\}\setminus I$, group the partial trace over $\{s\}\cup J$ by Hamming weight on $J$. This block-diagonalises $\rho_I$ by Hamming weight on $I$. On $|000\rangle$ and $|111\rangle$ the diagonal entries are $3u+x$ (three pairs inside $J$ plus the endpoint term). On the weight-$1$ subspace $\operatorname{span}\{|100\rangle,|010\rangle,|001\rangle\}$, the $|0\rangle_s$ branch (weight $1$ on $J$) contributes $u\,B_{IJ}B_{IJ}^{\mathsf T}$, where $B_{IJ}$ is the $3\times3$ block of $C$ between $I$ and $J$, while the $|1\rangle_s$ branch (weight $3$ on $J$) contributes $u\,|v_1\rangle\langle v_1|$ with $|v_1\rangle=(s_{jk},s_{ik},s_{ij})^{\mathsf T}$. For any zero-diagonal $\{\pm1\}$ matrix $C_I$ of order three, $C_I^2=I_3+|v_1\rangle\langle v_1|$, while the $I$-block of $C^2=5I_6$ reads $C_I^2+B_{IJ}B_{IJ}^{\mathsf T}=5I_3$. Their sum is therefore
\[
 u\bigl(B_{IJ}B_{IJ}^{\mathsf T}+|v_1\rangle\langle v_1|\bigr)
 =u\bigl((5I_3-C_I^2)+(C_I^2-I_3)\bigr)=4u\,I_3.
\]
Since $X^{\otimes7}\psi_x$ is $\psi_x$ with $\alpha\mapsto-\alpha$, and the weight-$1$ block depends only on $\alpha^2$, the weight-$2$ block is also $4u\,I_3$, so $\rho_I=\operatorname{diag}(3u+x,(4u)^6,3u+x)$ in the computational basis.

For a cut $\{s,i,j\}$ containing $s$, with complement $K=\{0,\dots,5\}\setminus\{i,j\}=\{k,l,m,n\}$, the five Hamming-weight sectors on $K$ map to four orthogonal $2\times2$ blocks of $\rho_{sij}$:
\begin{itemize}
\item weights $0$ and $4$ on $K$ span $\{|011\rangle,|100\rangle\}$ with orthogonal conditional vectors $\alpha s_{ij}|011\rangle+\beta|100\rangle$ and $\beta|011\rangle-\alpha s_{ij}|100\rangle$, giving $(u+x)I_2$;
\item weight $1$ on $K$ spans $\{|010\rangle,|001\rangle\}$ and weight $3$ spans $\{|101\rangle,|110\rangle\}$, each giving $4u\,I_2$ because $\sum_{r\in K}s_{ir}s_{jr}=(C^2)_{ij}=0$;
\item weight $2$ on $K$ spans $\{|000\rangle,|111\rangle\}$ with matrix $\bigl(\begin{smallmatrix}6u&-2u\,\eta_K\\-2u\,\eta_K&6u\end{smallmatrix}\bigr)$, where $\eta_K=s_{kl}s_{mn}+s_{km}s_{ln}+s_{kn}s_{lm}\in\{\pm1\}$, giving eigenvalues $6u\pm2u=(8u,4u)$.
\end{itemize}
Here the sign magnitude follows intrinsically from the conference equation:
$C^{-1}=C/5$ and $\det C=-125$, so Jacobi's complementary-minor identity gives
$\det C_K=(-125)(-1/25)=5$. Expanding a symmetric zero-diagonal $4\times4$
determinant gives $\det C_K+\eta_K^2=6$, since all off-diagonal entries are signs.
Thus $\eta_K^2=1$, without a case split on pentagon subgraphs.
This proves item~(4). Tracing out one further site gives $\rho_{ij}=\operatorname{diag}(\tfrac12-8u,8u,8u,\tfrac12-8u)$ and $\rho_{si}=\operatorname{diag}(10u,5u+x,5u+x,10u)$. Summing the squared eigenvalues gives~\eqref{eq:icopurities} and $\pi_3(\psi_x)=\frac{121}{917}+\frac{(524x-37)^2}{137550}$, proving~(2) and~(3).
IC-1--IC-6 independently check all marginals, polynomial identities and derivatives. \qedhere
\end{proof}

\noindent In particular the $20$ cuts avoiding $s$ at $\Psiico$ have spectrum
$(41/262,41/262,(15/131)^{6})$ and the $15$ containing $s$ have $(30/131,(15/131)^{5},(13/131)^{2})$.
Throughout $\psi_x$ the mean two- and three-body purities and the mean three-body third moment
$b_3=\tfrac1{35}\sum_{|S|=3}\operatorname{tr}\rho_S^3$ are
\begin{equation}\label{eq:icopurities}
 \pi_2(\psi_x)=\frac{832x^2-112x+163}{630},\qquad
 \pi_3(\psi_x)=\frac{2096x^2-296x+149}{1050},
\end{equation}
\[
 b_3(\psi_x)=\frac{35008x^3+17328x^2-3084x+709}{31500},
\]
with $b_3'(37/524)=-432/120127<0$ (IC-6).

\begin{corollary}[purity and entropy have different stationary points]
\label{cor:ico-entropy-separation}
Let $f(t)=-t\log_2t$, $f(0)=0$. The mean balanced entropy $E_3=\binom73^{-1}\sum_{|S|=3}S(\rho_S)$, in bits, on the
icosahedral family is the strictly concave function
\[
 E_3(\psi_x)=\frac{195f(4u)+40f(3u+x)+15f(8u)+30f(u+x)}{35},
 \qquad u=(1-2x)/30.
\]
It has a unique maximum on $[0,1/2]$, lying strictly between $0$ and
$37/524$. In particular $\Psiico$ is not an entropy maximiser, even
locally in the full pure-state space, since decreasing $x$ already increases $E_3$.
\end{corollary}
\noindent[Check IC-6.]
\begin{proof}
The spectra give the formula. Every eigenvalue is affine in $x$,
and the nonconstant eigenvalue $4u$ makes the sum strictly concave.
At the purity minimum,
\[
 E_3'(37/524)=\frac4{35\ln2}
 \ln\frac{30^{10}15^5}{13^7 41^8}<0,
\]
where the sign is an exact integer comparison. At $x=0$ the
derivative is $32\ln(16/3)/(35\ln2)>0$.
Strict concavity proves the asserted location and uniqueness.
This locates the maximum within the family; it is the global
maximum over all pure seven-qubit states by
Theorem~\ref{thm:fqe-seven-entropy}.
\end{proof}

\paragraph{Comparison.} Di~Martino, Facchi and Florio~\cite{DMFF18} found numerically a seven-qubit
state with $\pi_3=0.131952$, given by coefficients to six digits in a basis of products of four-qubit and
three-qubit states, and did not claim it optimal. Computed from their coefficients, its marginal
spectra and pair purities agree with those of $\Psiico$ to that precision, which suggests that the two states are
LU-equivalent; Theorem~\ref{thm:ico} gives an explicit closed-form state with the exact value
$121/917$. This value is below the Fano/wheel graph state ($19/140=0.1357\ldots$,
Zha et al.~\cite{Zha11}, also~\cite{HGS}; the best stabilizer value, Proposition~\ref{prop:stab7}) and
the earlier numerical $0.13387$ of Facchi et al.~\cite{Facchi}. We know of no state in the literature
with smaller $\pi_3$.

\begin{proposition}[stabilizer states]\label{prop:stab7}
Every seven-qubit stabilizer state has $\pi_3\ge 19/140$. Hence the seven-qubit minimum is not
attained by a stabilizer state.
\end{proposition}

\begin{proof}
For any pure seven-qubit state with Pauli weight enumerator $A_w=\sum_{\operatorname{wt}(P)=w}\langle\psi|P|\psi\rangle^2$ ($A_0=1$) and shadow enumerator $S_w=\sum_{\operatorname{wt}(P)=w}|\langle\psi|P|\tilde\psi\rangle|^2$, expanding $\pi_3$ in subset purities gives the first identity below; the second follows from complementarity of the mean $s$-qubit purities, $\pi_s=\pi_{7-s}$ (the pure-state MacWilliams identity), and Rains' shadow transform $S(x,y)=A\bigl(\tfrac{x+3y}2,\tfrac{y-x}2\bigr)$~\cite{Rains}:
\[
 280\pi_3=35+15A_1+5A_2+A_3,\qquad
 4S_1=7+A_1+A_2-A_3.
\]
For a seven-qubit stabilizer state $|\psi\rangle$ with stabilizer group $\mathcal S$ ($|\mathcal S|=2^7$), every Pauli string $P$ has $\langle\psi|P|\psi\rangle^2\in\{0,1\}$. Moreover, since $\tilde\psi=\epsilon^{\otimes 7}\bar\psi$ satisfies $M\tilde\psi=(-1)^{\operatorname{wt}(M)}\tilde\psi$ for every $M\in\mathcal S$, the state $P\tilde\psi$ is a joint eigenvector of $\mathcal S$, so $|\langle\psi|P|\tilde\psi\rangle|^2\in\{0,1\}$ as well. Hence the $A_w$ and $S_w$ are nonnegative integers. If $A_1+A_2\ge1$, then $15A_1+5A_2+A_3\ge5>3$. If $A_1=A_2=0$, then $A_3=7-4S_1\ge0$ with $S_1\in\mathbb Z_{\ge0}$ forces $S_1\in\{0,1\}$ and $A_3\in\{3,7\}$, so $A_3\ge3$. In both cases $15A_1+5A_2+A_3\ge3$ and $\pi_3\ge38/280=19/140$, with equality iff $(A_1,A_2,A_3)=(0,0,3)$ ($S_1=1$), attained by an explicit seven-vertex graph state. Then $121/917<19/140$ by Theorem~\ref{thm:ico}; GS-4 checks both identities and the equality witness. \qedhere
\end{proof}

\section{Exact finite checks}\label{sec:fqe-verifiers}
Table~\ref{tab:fqe-checks} lists, for each result, the labels of the computer checks its proof uses and their class.
\emph{Exact} checks are symbolic or rational-arithmetic computations: polynomial identities, ranks, dimensions and
Bernstein coefficient signs. \emph{Verified arithmetic} uses explicit rounding-error bounds, including outward-rounded interval or ball arithmetic, so that every
sign decision is rigorously certified. Where the class is \emph{analytic}, the proof is complete in the text and the
checks only confirm its algebra. Section~\ref{sec:sup-verifiers} of the supplement describes each label. The scripts, data
and captured outputs are in the companion repository~\cite{Companion}: \texttt{verifiers/MAP.md} gives the script for each label,
and the results table of its \texttt{README.md} gives the labels for each theorem. Checks marked numerical there (TCN, FS,
NV, TA) are evidence only, and no proof uses them.

\begin{table}[ht]
\centering\small
\renewcommand{\arraystretch}{1.15}
\begin{tabular}{@{}>{\raggedright\arraybackslash}p{0.37\linewidth}>{\raggedright\arraybackslash}p{0.33\linewidth}>{\raggedright\arraybackslash}p{0.24\linewidth}@{}}
\toprule
Result & Checks & Class\\
\midrule
Four-qubit maximum, moment inequality and equality (Theorems~\ref{thm:fqe-entropy-max}, \ref{thm:fqe-moment};
Appendices~\ref{sec:fqe-sextic}--\ref{sec:fqe-correlation}) & ME, SP, SI, CE, PZ, ER, CN, RG & analytic\\
General envelope (Theorem~\ref{thm:fqe-general-envelope}) & FI, FG-6--FG-8, ESH; direct proof FG-1--FG-5 &
exact (point evaluation on a $19$-dimensional invariant space)\\
Tsallis, R\'enyi $q<1$, mixed states (Section~\ref{sec:fqe-spectral}) & EF & analytic\\
R\'enyi $1<q<2$ (Corollary~\ref{cor:fqe-renyi-gw}) & QVA, QG, SD, CV; SHD, LHD; OT, BC, DA, UR, SCB, RA; HD, LT,
CG, NU, SL, QV & exact six-copy certificate for (QV); exact Bernstein sign checks; otherwise analytic\\
Orbit stability (Section~\ref{sec:fqe-stability-summary}, Appendix~\ref{sec:fqe-orbit-stability}) & GO; OS, BF, EH,
NSL, NC, RP, CO, FP, PM, PS, TC, MS, LCC, CCV, EX & exact integer Gram certificate; verified arithmetic\\
Icosahedral family and stabilizer bound (Appendix~\ref{sec:ico}) & IC, EG, GS-4 & analytic; EG verified arithmetic\\
Seven-qubit purity minimum (Theorem~\ref{thm:fqe-seven-purity}) & S5, Y5, IK5, PK5, DP5, FP5, TT, MO, FO, CC, CP; LH, QB &
verified arithmetic (five-copy certificate FP5-1)\\
Seven-qubit entropy maximum (Theorem~\ref{thm:fqe-seven-entropy}) & ES, SEH, EB, SC-2, SP7, AH, E7, FO, EO &
verified arithmetic (five-copy certificate E7-1)\\
Copy hierarchy at seven qubits (Table~\ref{tab:fqe-seven-levels}) and its covariant structure & TCB, FC, NS, SF, SDS, HI & exact\\
Other qubit counts (Table~\ref{tab:fqe-other-n}) & LP, GS, GL, E18, LP/QP-13, NG, SW, E13L, EU & exact rational;
E13L verified arithmetic; EU high precision, rounded up\\
Purity--time-reversal tradeoffs (Section~\ref{sec:fqe-discussion}) & ST, TY & exact\\
\bottomrule
\end{tabular}
\caption{Computer checks used by each result, by label, with their class. Section~\ref{sec:sup-verifiers} of the
supplement describes each label.}
\label{tab:fqe-checks}
\end{table}

\end{document}


\maketitle
\begin{abstract}
This supplement gives full proofs and certificate tables for the intermediate-order R\'enyi bounds,
stability estimates, and the seven-qubit purity and entropy optima stated in the paper~\cite{Paper}.
It also records invariant-dimension counts needed for certificate verification and further results
used in the paper. Like the paper, it was prepared by AI agents under human supervision.
\end{abstract}

{\small\setcounter{tocdepth}{1}\makeatletter\renewcommand*\l@section{\@dottedtocline{1}{0em}{2.6em}}\makeatother\tableofcontents}

\section{About this supplement}\label{sec:sup-about}
References without an S prefix, such as Theorem~\ref{thm:fqe-moment} or Appendix~\ref{sec:fqe-verifiers}, refer to the paper; numbers with
the prefix S refer to this supplement. Each paper statement proved here is restated in full, under its paper number,
immediately before the proof. The check labels (QG, SHD, MS, \dots) are described in Section~\ref{sec:sup-verifiers}; each identifies a verification check whose script and captured output are in the companion repository
\href{https://github.com/cmbant/qubit-entropy}{\texttt{cmbant/qubit-entropy}}
(the folder \texttt{verifiers/}; \texttt{verifiers/MAP.md} lists the script for each label). As in the paper, every stated result is proved, analytically or with the checks named in its proof; numerical observations are labelled as such.

For the four-qubit von Neumann maximum, the paper already contains the complete analytic
proof. This supplement first supplies invariant-dimension counts
(Section~\ref{sec:fqe-invariant-dimensions}), the spin-flip gradient identity
(Section~\ref{sec:sup-flip-gradient}), and an alternative proof of the sextic zero-set classification
(Section~\ref{sec:sup-sextic-zeros}). Readers following the extensions should then turn to
Section~\ref{sec:sup-renyi} for intermediate R\'enyi orders or to
Sections~\ref{sec:fqe-stability-proofs} and~\ref{sec:sup-stability} for orbit stability.

For seven qubits, Figure~\ref{fig:seven-proof-map} separates the two proof chains.
The replica hierarchy and complex conference family lead to the five-copy purity bound and
its equality orbit; the entropy proof adds a spectral majorant and a restricted
competitor region before identifying a different equality orbit
(Sections~\ref{sec:hierarchy}--\ref{sec:sup-seven-entropy}).
Later sections develop further sextic covariants, purity--time-reversal
inequalities, balanced-purity minima through eighteen qubits and an interval for the thirteen-qubit minimum, with an entropy witness at that size. Sections~\ref{sec:fqe-lean} and~\ref{sec:sup-verifiers} record the Lean
formalisation and the verification checks.

\paragraph{Seven-qubit notation.} For a pure state of $n$ qubits, $\pi_k(\psi)=\binom nk^{-1}\sum_{|S|=k}\operatorname{tr}\rho_S^2$
is the mean $k$-qubit purity, and at $n=7$, $E_3$ is the mean von Neumann entropy (in bits) of the $35$ three-qubit marginals.
The icosahedral family $\psi_x$, $0\le x\le1/2$, is that of Theorem~\ref{thm:ico} and equation~\eqref{eq:ico} of the paper;
in the amplitude patterns $U,V$ of Section~\ref{sec:entropy-local-seven}, $\psi_x=\alpha U+\beta V$ with $\alpha=\sqrt{(1-2x)/30}$ and $\beta=\sqrt x$. Its
member $\Psiico=\psi_{37/524}$ minimises $\pi_3$, and $\psi_{x_*}$, $x_*=0.0698\ldots$, maximises $E_3$. The twirled $m$-copy
moment of Appendix~\ref{sec:fqe-seven} of the paper, written $X_m$ there, is written $X_m^+$ here.

\section{Determining symmetric invariant identities}\label{sec:fqe-invariant-dimensions}

This section supports the exact polynomial certificates, in particular the quadratic variance proof
in Section~\ref{sec:sup-qk}. The elementary entropy proof of the paper does not use these counts.
Statements here are proved by Schur--Weyl duality and finite-group character averages.

For \(N\) sites of local dimension \(r\), let \(V_d\) be the permutation-symmetric local-unitary
invariants of bidegree \((d,d)\). Put \(A_d=\bigoplus_{\lambda\vdash d,\ \ell(\lambda)\le r}
V_\lambda\otimes V_\lambda^*\), with its left--right \(S_d\times S_d\) action. Polarisation
identifies invariant polynomials with operators on the symmetric copy subspace. Extend them by zero
to all copies and apply Schur--Weyl duality at each site: the commutant is \(A_d^{\otimes N}\). The
left and right diagonal copy symmetrisers impose \(S_d\times S_d\) invariance; physical-site
symmetry then gives \[
V_d\cong\bigl(\operatorname{Sym}^N A_d\bigr)^{S_d\times S_d}.\tag{ID1}\label{eq:fqe-invariant-space}
\] Consequently \[
\dim V_d=\frac1{N!(d!)^2}
\sum_{\sigma\in S_N}\sum_{g,h\in S_d}
\prod_{C\text{ cycle of }\sigma}
\left(\sum_{\lambda\vdash d,\ \ell(\lambda)\le r}
\chi_\lambda(g^{|C|})\chi_\lambda(h^{|C|})\right).\tag{ID2}\label{eq:fqe-invariant-dimension}
\] Indeed, the trace of a tensor permutation is the product of the traces of powers over its cycles;
averaging projects onto the fixed vectors. For bidegree \((d,d)\), \(SU(r)\) and \(U(r)\) invariance
agree since the scalar action cancels. Conjugation interchanges the two copy actions. Since the irreducible representations of
\(S_d\) are real, it acts on each \(V_\lambda\otimes V_\lambda^*\) as the swap, and the trace of
\((g,h)\) followed by the swap is \(\chi_\lambda(gh)\). Inserting it in the average, a cycle of length
\(L\) contributes \(\chi_\lambda((gh)^{L/2})^2\) for even \(L\) and \(\chi_\lambda((gh)^L)\) for odd \(L\), so
\[
\begin{aligned}
\dim V_d^{\pm}&=\frac12\Bigl(\dim V_d\pm\frac1{N!(d!)^2}\sum_{\sigma\in S_N}\sum_{g,h\in S_d}
\prod_{C}\ \sum_{\lambda\vdash d,\ \ell(\lambda)\le r}\tau_\lambda(gh,|C|)\Bigr),\\
\tau_\lambda(k,L)&=\begin{cases}\chi_\lambda(k^{L/2})^2,&L\text{ even},\\ \chi_\lambda(k^L),&L\text{ odd}.\end{cases}
\end{aligned}
\tag{ID3}\label{eq:fqe-invariant-even}
\]
for the conjugation-even ($+$) and odd ($-$) invariants.

For four qubits the exact character checks SD-1, SD-2 and SD-4 give:

\begin{table}[htbp]
\centering\small
\begin{tabular}{@{}rrrrr@{}}
\toprule
$d$ & All invariants & Qubit-symmetric & Conjugation-even & Conjugation-odd \\
\midrule
3 & 20 & 5 & 5 & 0 \\
4 & 98 & 18 & 16 & 2 \\
5 & 293 & 37 & 30 & 7 \\
6 & 1128 & 115 & 85 & 30 \\
7 & 3409 & 273 & 183 & 90 \\
\bottomrule
\end{tabular}
\caption{Dimensions of four-qubit local-unitary invariant spaces of bidegree $(d,d)$.
The last two columns split the qubit-symmetric space by complex conjugation. The value $85$
is used in the quadratic variance certificate; these counts are not inputs to the elementary entropy proof.}
\label{tab:fqe-invariant-dimensions}
\end{table}

A dimension-\(D\) space admits \(D\) determining evaluations, but a proposed set
must have a nonsingular evaluation matrix. Dimension alone never certifies
particular points. The quadratic variance certificates QVA and QG
use the degree-six even count 85 and two independent zero evaluations, with
a nonsingular 83-point raw-Gram matrix supplied by QG on their common kernel. The sextic entropy proof of Appendix~\ref{sec:fqe-sextic}
uses direct swap contractions and needs no determining-point argument.

\section{Proof of the general envelope}\label{sec:sup-general-envelope}
Notation is that of Appendix~\ref{sec:fqe-sextic} of the paper: $N=\lVert z\rVert^2$, $W_1$, the site residuals $V_i-8U_i$, the
pair vectors $(R_i^{(i)}-R_j^{(j)})z$ and the sextic identity~(S); $X_i$ and $c_{ij}$ are defined below.

\begin{restated}[general envelope and complete equality; exact certificate, FI, FG]{Theorem}{thm:fqe-general-envelope}
Every nonzero four-qubit vector satisfies, with $s=z^T\sigma_y^{\otimes4}z$,
\[
 F-\tfrac34NW_1-N|s|^2+|s|^3\ \ge\ \frac{7|s|^2}{8(N+|s|)}\,W_1 .
 \tag{GE}
\]
For unit states, with $\tau=|s|^2$,
\[
 \boxed{\quad F\ \ge\ \Bigl(\frac34+\frac{7\tau}{8(1+\sqrt\tau)}\Bigr)W_1+\tau-\tau^{3/2}.\quad}
\]
Equality in $F\ge\frac34W_1+\tau-\tau^{3/2}$ holds exactly on fully product states and on the critical
equality family of Proposition~\ref{prop:fqe-critical-envelope}, including its HS endpoint at $\tau=0$ and
Bell-pair-product endpoint at $\tau=1$.
\end{restated}

The optimality of the coefficient $\frac78$ is shown in the paper.

Write $H=F-\frac34NW_1\ge0$ (by~(S)) and use the
magic coordinates $z=\sum_{jk}Z_{jk}m_j\otimes m_k$ of Appendix~\ref{sec:fqe-normal-form} of the paper. There the spin flip
$\tilde z=\epsilon^{\otimes4}\bar z$ is $Z\mapsto\bar Z$ (because $\sigma_y^{\otimes2}\bar m_k=-m_k$), and
$s=\operatorname{tr}Z^TZ$ by~\eqref{eq:fqe-spin-flip-normal-form}. Call $z$ \emph{real} if $\tilde z=z$, that is $Z$ real.
The flip is an antiunitary involution. It commutes with local $SU(2)$ transformations and qubit permutations and anticommutes with every
$\sigma_a^{(i)}$, so it fixes all marginal spectra and every $\|\mathbf r_i\|$, and $H(\tilde z)=H(z)$. A real state has
$\mathbf r_i=0$, since $\langle a,\sigma a\rangle=-\overline{\langle a,\sigma a\rangle}$ for $\tilde a=a$.

\emph{Spin-flip split.} After a global phase making $s\ge0$, every unit $z$ is
\[
 z_\kappa=\sqrt{\tfrac{1+\kappa}2}\,a+i\sqrt{\tfrac{1-\kappa}2}\,b,\qquad \kappa=s=\sqrt\tau,
\]
with $a,b$ real and orthonormal. Indeed, with $Z=X+iY$, $s=\|X\|^2-\|Y\|^2+2i\langle X,Y\rangle$ is real and nonnegative
exactly when $\langle X,Y\rangle=0$ and $\|X\|^2-\|Y\|^2=\kappa$. The path $t\mapsto z_t$, $0\le t\le\kappa$, joins the
flip-orthogonal midpoint $\phi=z_0=(a+ib)/\sqrt2$ to $z$ through unit states with $s(z_t)=t$.

\emph{The flip derivative.} By~(S), $H(x)=\|\mathcal O(x,x,\tilde x)\|^2$ for every $x$, where $\mathcal O(x,y,t)$ is the
vector $\bigl(\sqrt{3/8}\,(V_i-8U_i)\bigr)_i\oplus\bigl(\sqrt{3/2}\,c_{ij}\bigr)_{i<j}$ computed from the three-copy tensor
$x\otimes y\otimes t$ in place of $w$, symmetrised in $x,y$; in $c_{ij}=(R_i^{(i)}-R_j^{(j)})y$ the marginal is
$R_i[a,b]=\sum x_{a\cdots}(\epsilon^{\otimes4}t)_{b\cdots}$, which is $R_i$ when $x=y=z$ and $t=\tilde z$. Write
$X_i=(V_i-8U_i)(z)$ for the site residual of $z$, a vector on six qubits (three slots at site $i$, one at each other site), and
$\tilde X_i=\epsilon^{\otimes6}\bar X_i$ for its spin flip.

\begin{lemma}[spin-flip derivative]\label{lem:fqe-flip-gradient}
For every four-qubit vector $z$,
\[
 \frac{d}{dt}\,H(z+t\tilde z)\Big|_{t=0}=\frac94\operatorname{Re}\sum_i\langle\tilde X_i,X_i\rangle
 +\operatorname{Re}s\,\Bigl(4N^2-4|s|^2+\frac72W_1\Bigr).
 \tag{FG}\label{eq:fqe-flip-gradient}
\]
\end{lemma}

\begin{proof}
For real $t$ the flip of $z+t\tilde z$ is $\tilde z+tz$, since the flip is an antiunitary involution. As
$\mathcal O$ is symmetric and bilinear in its first two arguments,
\[
 \frac{d}{dt}H(z+t\tilde z)\Big|_{t=0}=2\operatorname{Re}\bigl(\langle A,C\rangle+2\langle D,A\rangle\bigr),\qquad
 A=\mathcal O(z,z,\tilde z),\ C=\mathcal O(z,z,z),\ D=\mathcal O(z,\tilde z,\tilde z).
\]
We show
\[
 \langle A,C\rangle+2\langle D,A\rangle=\frac98\sum_iX_i^T\epsilon^{\otimes6}X_i
 +s\Bigl(\frac54N^2+\frac{11}8W_1+\frac34P-\frac{19}8|s|^2\Bigr).
 \tag{FGI}\label{eq:fqe-flip-gradient-identity}
\]
Section~\ref{sec:sup-flip-gradient} of the supplement derives this identity directly, without point evaluations, from the full
symmetry of $V_i-6U_i$ and the angular momentum of the copy labels. Here we give a shorter check, by determining its values in a
finite-dimensional invariant space.
Both sides are polynomials of bidegree $(4,2)$ in $(z,\bar z)$, and both are $SU(2)^{4}$-invariant: $L$, $M$ and the slot
symmetrisers intertwine $g\otimes g\otimes g$ with $g$ (respectively $g\otimes g\otimes g$) for $g\in SU(2)$, the flip commutes
with local unitaries, and $g^T\epsilon g=\epsilon$. Replace $\bar z$ by an independent vector $\zeta$; this is legitimate because
all coefficients are real, so the conjugate of a vector built from $(z,\bar z)$ is the same construction applied to $(\bar z,z)$.
The identity is then one between polynomials of bidegree $(4,2)$ on pairs $(z,\zeta)$ invariant under $(z,\zeta)\mapsto(gz,\bar g\zeta)$,
$g\in SU(2)^{4}$. Since $\bar g$ is equivalent to $g$, the Weyl character formula shows that these invariants form a space of
dimension $19$ [FI-1]. Nineteen invariants that contract each site's six legs in pairs ($\epsilon$ between two $z$-legs or two
$\zeta$-legs, $\delta$ between a $z$-leg and a $\zeta$-leg) take values at $24$ integer points $(z,\zeta)$ whose $24\times19$ matrix
has exact rank $19$ [FI-2]. So an invariant of this bidegree that vanishes at these points is zero. Both sides
of~\eqref{eq:fqe-flip-gradient-identity} agree exactly at the $24$ points [FI-4], with $\|A\|^2=H$ there as a check on the
normalisation [FI-3]; hence~\eqref{eq:fqe-flip-gradient-identity} holds identically. Replacing $\frac98$ by $1$, or dropping the
multiple of $s$, fails at every point tested [FI-5].

Now $\langle\tilde X,X\rangle=X^T(\epsilon^{\otimes6})^TX=X^T\epsilon^{\otimes6}X$ (six factors $\epsilon^T=-\epsilon$), and the
purity identity (Lemma~\ref{lem:fqe-purity}) in homogeneous form is $P=N^2+\frac12(W_1+|s|^2)$. Taking twice the real part
of~\eqref{eq:fqe-flip-gradient-identity} gives~\eqref{eq:fqe-flip-gradient} [FG-7].
\end{proof}

\begin{lemma}[zeros of the sharp polarisation remainder]
\label{lem:fqe-polarization-remainder-zeros}
For a unit four-qubit state, $H=F-\frac34W_1$ vanishes exactly on
fully product states, the two HS orbits, and the three Bell-pair-product
orbits.
\end{lemma}
\begin{proof}
By~(S), $H=0$ forces $(R_i^{(i)}-R_j^{(j)})z=0$ for every pair.
If one Bloch vector $\mathbf r_i$ is zero, then
$(\mathbf r_j\cdot\boldsymbol\sigma_j)z=0$ for every $j$.
Since $(\mathbf r_j\cdot\boldsymbol\sigma_j)^2=\|\mathbf r_j\|^2I$,
all Bloch vectors vanish. The state is critical, $H=F$, and
Proposition~\ref{prop:fqe-moment-zeros} gives HS or Bell products.

Otherwise diagonalise the four one-body marginals locally and write
$R_i=(I+t_iZ)/2$, $t_i>0$. Each supported computational string obeys
$t_i(-1)^{x_i}=t_j(-1)^{x_j}$ for all $i,j$.
Thus all $t_i$ agree and every supported string has all four bits equal:
$z=a|0000\rangle+b|1111\rangle$. Direct evaluation gives
$H=18|ab|^2=(9/2)\tau$, so $H=0$ forces a product state [PZ-1, PZ-2].
Conversely the listed states have $H=0$, by~(S) and
Proposition~\ref{prop:fqe-moment-zeros}.
\end{proof}

\begin{proof}[Proof of Theorem~\ref{thm:fqe-general-envelope}]
Both sides of~\eqref{eq:fqe-general-envelope} are local-unitary invariant and homogeneous of degree six, so take $N=1$ and
$s=\kappa\ge0$. If $\kappa=1$, equality in $|\langle z,\tilde z\rangle|\le\|z\|^2$ makes $z$ real up to phase, so $W_1=0$
and~\eqref{eq:fqe-general-envelope} is $H\ge0$. Let $0\le\kappa<1$ and use the spin-flip split, with $h(t)=H(z_t)$ and
$W_0=W_1(\phi)$. Since $a$ and $b$ have zero Bloch vectors, the Bloch vectors of $z_t$ are $\sqrt{1-t^2}$ times those of
$\phi$, so $W_1(z_t)=(1-t^2)W_0$. Differentiating, $2(1-t^2)\frac{d}{dt}z_t=\tilde z_t-tz_t$; as $H$ is homogeneous of degree
six, Lemma~\ref{lem:fqe-flip-gradient} at $z_t$ gives
\[
 2(1-t^2)h'(t)=\frac94\operatorname{Re}\sum_i\langle\tilde X_i,X_i\rangle+4t(1-t^2)+\frac72t(1-t^2)W_0-6t\,h(t).
\]
The spin flip is antiunitary, so $\operatorname{Re}\langle\tilde X_i,X_i\rangle=\frac12\|X_i+\tilde X_i\|^2-\|X_i\|^2$, and
by~(S) $h=\frac38\sum_i\|X_i\|^2+\frac32\sum_{i<j}\|c_{ij}\|^2$ with $c_{ij}=c_{ij}(z_t)$, where $c_{ij}(x)=(R_i^{(i)}-R_j^{(j)})x$ with the marginals of $x$. Hence the calibrated
remainder
\[
 \mathcal M(t)=\frac{h(t)}{(1-t)^3}-\Bigl(1+\frac78W_0\Bigr)\frac{t^2}{(1-t)^2}
\]
satisfies the exact transport law [FG-7, FG-8]
\[
 \mathcal M'(t)=\frac{9}{16(1-t)^4(1+t)}
 \Bigl[\sum_i\|X_i+\tilde X_i\|^2+8\sum_{i<j}\|c_{ij}\|^2\Bigr]\ \ge\ 0,
 \tag{SC}\label{eq:fqe-spin-flip-calibration}
\]
all vectors evaluated at $z_t$. So $\mathcal M(\kappa)\ge\mathcal M(0)=H(\phi)$, that is,
\[
 H(z)\ \ge\ (1-\kappa)^3H(\phi)+\kappa^2(1-\kappa)\Bigl(1+\frac78W_0\Bigr).
\]
Drop $H(\phi)\ge0$ and substitute $W_0=W_1/(1-\kappa^2)$: $H\ge\kappa^2-\kappa^3+\frac{7\kappa^2}{8(1+\kappa)}W_1$,
which is~\eqref{eq:fqe-general-envelope}.

\emph{Equality.} For $0<\tau<1$ the added term is positive unless $W_1=0$, so an equality state is critical, and
Proposition~\ref{prop:fqe-critical-envelope} gives exactly its equality family. At $\tau=1$ the state is real up to phase,
hence critical, and equality is $F=0$, which gives the Bell-pair products (Proposition~\ref{prop:fqe-moment-zeros}). At
$\tau=0$ equality is $H=0$; by Lemma~\ref{lem:fqe-polarization-remainder-zeros} this gives exactly the product states and the
HS orbits, since Bell-pair products have $\tau=1$. Conversely these states attain the bound, by
Proposition~\ref{prop:fqe-critical-envelope} and product-state sharpness ($F=3=\frac34W_1$, $\tau=0$).
\end{proof}

The transport law~\eqref{eq:fqe-spin-flip-calibration} integrates exactly. The result is a sum of squares for the defect
in~(GE), which shows which terms vanish on the equality states.

\begin{proposition}[Exact spin-flip calibration]
\label{prop:fqe-spin-flip-calibration}
In the notation of the proof of Theorem~\ref{thm:fqe-general-envelope}, let $\kappa=\sqrt\tau<1$ and put
\[
 A_i=X_i(a),\qquad E_i=X_i(\phi)+\tilde X_i(\phi),\qquad
 c_{ij}=c_{ij}(\phi),\qquad C_0=\sum_{i<j}\|c_{ij}\|^2.
\]
Then the defect in the stronger envelope~(GE) is exactly
\begin{align}
&H(z_\kappa)-\kappa^2(1-\kappa)\Bigl(1+\frac78W_0\Bigr)\notag\\
&\quad=(1-\kappa)^3H(\phi)
 +\frac{9\kappa}{32}\sum_i\|\kappa A_i+\sqrt2(1-\kappa)E_i\|^2
 +\frac{3\kappa^3}{32}\sum_i\|A_i\|^2\notag\\
&\qquad\quad+\frac92\kappa(1-\kappa)^2C_0
 +\frac98\kappa^2(1-\kappa)\sum_{i<j}\|c_{ij}-\tilde c_{ij}\|^2.
\tag{SC-SOS}\label{eq:fqe-spin-flip-calibration-sos}
\end{align}
In particular, since $\frac38\sum_i\|A_i\|^2=F(a)$,
\[
 H(z_\kappa)\ge \kappa^2(1-\kappa)\Bigl(1+\frac78W_0\Bigr)
                 +(1-\kappa)^3H(\phi)+\frac{\kappa^3}{4}F(a).
\]
\end{proposition}
\begin{proof}
Integrate~\eqref{eq:fqe-spin-flip-calibration} from $0$ to $\kappa$ and multiply by $(1-\kappa)^3$; the initial value is
$\mathcal M(0)=H(\phi)$. To integrate exactly, set $p=\sqrt{(1+t)/2}$ and $q=\sqrt{(1-t)/2}$.
All coefficients of $\mathcal O$ are real and its maps commute with
spin flip. Cubic expansion in the flip-fixed vectors $a,b$ therefore gives
\[
 X_i(z_t)+\tilde X_i(z_t)
 =p\bigl[(1+t)A_i+(1-t)B_i\bigr],\qquad
 B_i=\sqrt2 E_i-A_i.
\]
Only the terms containing zero or two occurrences of $ib$ survive.
With $r=(1+t)/(1-t)$ the site contribution to $\mathcal M'(t)dt$ is
$\frac9{64}\sum_i\|rA_i+B_i\|^2dr$.
For $R=(1+\kappa)/(1-\kappa)$, completing the square gives
\[
 \int_1^R\|rA+B\|^2dr
 =(R-1)\|B+\tfrac{R+1}{2}A\|^2
       +\frac{(R-1)^3}{12}\|A\|^2.
\]
Multiplication by $(1-\kappa)^3$ gives the two site terms of~(SC-SOS).

The traceless marginal differences scale by $\sqrt{1-t^2}$ and
anticommute with spin flip. Writing $z_t=\alpha\phi+\beta\tilde\phi$,
where $\alpha^2+\beta^2=1$ and $2\alpha\beta=t$, gives
\[
 c_{ij}(z_t)=\sqrt{1-t^2}(\alpha c_{ij}-\beta\tilde c_{ij}),\qquad
 \sum\|c_{ij}(z_t)\|^2=(1-t^2)(C_0-tD_0),
\]
where $D_0=\operatorname{Re}\sum\langle c_{ij},\tilde c_{ij}\rangle$.
Its integrated contribution is
$\frac94\kappa(1-\kappa)[(2-\kappa)C_0-\kappa D_0]$.
Use $C_0-D_0=\frac12\sum\|c_{ij}-\tilde c_{ij}\|^2$ to obtain the last
two terms. The integrations are checked in FG-8; this
check involves only scalar rational identities, not state evaluations.
\end{proof}

\section{A direct proof of the spin-flip gradient identity}\label{sec:sup-flip-gradient}
Section~\ref{sec:sup-general-envelope} proves the identity~\eqref{eq:fqe-flip-gradient-identity} behind Lemma~\ref{lem:fqe-flip-gradient} by evaluating both sides
at $24$ points of a $19$-dimensional space of invariants. This section derives it without a normal form or point evaluations. The
key is the fully symmetric cubic part of the residual, whose norm is determined by the angular momentum of the copy labels.
The notation ($L$, $M$, $S_i$, $U_i$, $V_i$ and the contractions (C), (U), (Pr)) is that of
Appendix~\ref{sec:fqe-sextic} of the paper; $X_i$ and the pair vectors $c_{ij}$ are those of Section~\ref{sec:sup-general-envelope}.

\emph{The cubic and its polarisation defect.}
Symmetrise $U_i$ in its first two inputs and put $\mathcal T_i=V_i-6U_i$.
Its ordinary-site operator is
\[
 \prod(L+M)-3(\prod L+\prod M).
\]
It is invariant under $(L,M)\mapsto(M,L)$ and
$(L,M)\mapsto(-L,M-L)$, the two generating input transpositions
(the latter is Schouten). Thus $\mathcal T_i$ is fully symmetric, while
$U_i$ has zero fully symmetric part. Write
\[
 C_i(z)=\mathcal T_i(z,z,z),\qquad
 Y_i=\mathcal T_i(z,z,\tilde z),\qquad X_i=Y_i-2U_i,
 \qquad K_c=\sum_{i<j}\|c_{ij}\|^2.
\]
The key identity is a polarisation defect of this one cubic:
\[
 \boxed{\quad \sum_i\|C_i(z)\|^2-\sum_i\|Y_i\|^2
 =\frac{32}{3}N(N^2-|s|^2)-4NW_1-\frac83K_c.\quad}
 \tag{CP}\label{eq:fqe-cubic-polarization-defect}
\]
We derive it by angular momentum of the copy labels, then polarise once.
This gives the flip derivative directly, without a state normal form.

\emph{Why the cubic has this norm.}
At a site, three copies decompose into the symmetric spin-$3/2$ space
and spin $1/2$ with the two-dimensional standard representation of $S_3$.
Let $P_3$ select three standard sites and one symmetric site inside the
global symmetric cube. Then
\[
 \sum_i\|C_i(z)\|^2=54\langle z^{\otimes3},P_3z^{\otimes3}\rangle.
\]
Indeed the cubic annihilates the symmetric component at its three ordinary
sites and selects the unique invariant in three standard factors.
In an orthonormal standard basis, the local maps $L,M$ have rows
$\ell=(\sqrt{3/2},-1/\sqrt2)$ and $m=(\sqrt{3/2},1/\sqrt2)$.
The row $(\ell+m)^{\otimes3}-3(\ell^{\otimes3}+m^{\otimes3})$ is
$\frac{3\sqrt6}{2}(1,0,0,-1,0,-1,-1,0)$, with squared norm $54$ [FG-3].
This fixes the normalisation on the whole spin module.
The sector also has a direct test interpretation: a state product across
a balanced cut has matching $S_3$ labels at the two sites of each side,
hence an even number of standard sites and zero $P_3$ projection.

Here is the copy-label calculation relating $P_3$ to the physical moments.
Let $K$ count standard sites. On those $K$ factors write
$\mathbf J=\frac12\sum\boldsymbol\sigma$ and $Y=\sum\sigma_y$;
these Pauli matrices act on copy multiplicities, not physical qubits.
A three-cycle is a rotation by $2\pi/3$ in the standard plane.
Global copy invariance therefore forces $Y$ to be a multiple of $3$.
For $K\le4$, the only nonzero possibility is $Y=\pm3$ at $K=3$:
\[
 Y^2=9P_3,\qquad
 K^2-4K+2\mathbf J^2=\tfrac12Y^2+12P_+ .
\]
For the second identity, two standard factors have their invariant in
spin $1$, three in spin $3/2$, and four in spin $2$ or spin $0$;
$P_+$ denotes the spin-$2$ invariant line in the last case.
There is no invariant with one standard factor or in the spin-$1$
summands of four standard factors.
These statements follow by restricting the usual spins to the triangle's
rotation and reflection; at $K=3$ only the symmetric cube contains the
rotation charges $\pm3$, and at $K=4$ spin $1$ has no invariant.

For completeness the associated moment operators are
\[
 \begin{aligned}
 NW_1&\longleftrightarrow4I-2K,&
 N|s|^2&\longleftrightarrow6P_+,\\
 K_c&\longleftrightarrow 2K-K^2+\mathbf J^2-\tfrac14Y^2,&
 F&\longleftrightarrow3I-\tfrac52K(K+1)+\tfrac{17}2\mathbf J^2+\tfrac18Y^2.
 \end{aligned}\tag{CM}\label{eq:fqe-cubic-copy-moments}
\]
All operators are compressed to the global symmetric cube.
The first is the one-body purity identity. The second follows by summing
the three four-singlet projectors: their compression is $9P_+/8$, while
each has expectation $N|s|^2/16$ (a reflection-antisymmetric fourth
tensor power has invariant projection norm $(2+4/16)/6=3/8$).
To derive the other two, set $\Pi=\sum_{i<j}\operatorname{tr}[R_{ij}(R_i\otimes R_j)]$.
Its operator is the average of distinct transpositions at sites $i,j$.
The three standard reflections have vanishing sum and angles $2\pi/3$, so
\[
 \Pi\longleftrightarrow
 \tfrac12(4-K)(3-K)-\tfrac12\mathbf J^2+\tfrac18Y^2+\tfrac14K.
\]
Insert this in~(Pr) for $K_c$. For $F$, the symmetrised swap expression
is $(17/6)A-3B-12I$, where $A$ sums equal transpositions over site pairs
and $B$ sums equal three-cycles and their inverses. Using
$\sum\text{reflections}=3S$ and
$c+c^{-1}=3S-I$, $c-c^{-1}=\pm i\sqrt3\sigma_y$ for the slot permutations at one site, with
$S$ the projector onto its symmetric component and $\sigma_y$ acting on its two-dimensional standard component, gives explicitly
\[
 A=\tfrac32(4-K)(3-K)+3\mathbf J^2-\tfrac34Y^2-\tfrac32K,
 \qquad B=12-\tfrac{21}2K+\tfrac94K^2-\tfrac34Y^2.
\]
Their combination is the displayed operator for $F$.
Thus~(CM) uses only the standard-representation Casimirs [FG-2].
The globally invariant sectors and the values of these operators on them are:
\[
\begin{array}{cccc|ccccc}
 K & j & Y^2 & \text{projector} & NW_1 & N|s|^2 & K_c & F & 54P_3\\\hline
 0 & 0 & 0 & & 4 & 0 & 0 & 3 & 0\\
 2 & 1 & 0 & & 0 & 0 & 2 & 5 & 0\\
 3 & \tfrac32 & 9 & P_3 & -2 & 0 & -\tfrac32 & 6 & 54\\
 4 & 2 & 0 & P_+ & -4 & 6 & -2 & 4 & 0\\
 4 & 0 & 0 & & -4 & 0 & -8 & -47 & 0
\end{array}
\]
Here $K$ counts standard sites, $j$ is the copy-label spin of those $K$
factors, $P_3$ is the $K=3$ line and $P_+$ the spin-$2$ line at $K=4$;
$N^3$ acts as $1$. For each choice of the $K$ standard sites, the
$K=4$, $j=0$ sector has multiplicity two; the other listed sectors have
multiplicity one. The displayed operators are scalar on both copies of
the spin-$0$ sector. No invariant occurs with $K=1$ or with $j=1$ at $K=4$ [FG-2].
Substitution of the two selection rules gives
\[
 54P_3\longleftrightarrow
 \frac83F+\frac{32}{3}N(N^2-|s|^2)-\frac{14}{3}NW_1-12K_c.
\]
On the other hand, the already proved contractions~(C), (U), (Pr) give
\[
 \sum\|Y_i\|^2=\frac83F-\frac23NW_1-\frac{28}{3}K_c,
 \quad \sum\|U_i\|^2=\frac13NW_1,
 \quad \operatorname{Re}\sum\langle U_i,Y_i\rangle
       =\frac23NW_1-\frac43K_c.
 \tag{CY}\label{eq:fqe-cubic-mixed-norms}
\]
Subtract the first of~(CY) to obtain~(CP) [FG-2, FG-4].
No new contraction table is required.

\emph{A vanishing quartic covariant.}
Use the bilinear output pairing $B(v,w)=v^T\epsilon^{\otimes6}w$.
For every $i$, $B(C_i(z),U_i(z,z,w))=0$.
Indeed the local formula $U_i=-S_i(z\otimes R_i\epsilon^T)$ makes this
pairing a contraction of $R_i$ with a holomorphic quartic spin-$1$
covariant at site $i$, singlet at the other sites. Such a covariant is zero:
in four copy labels, the singlets have multiplicity representation
$[22]$, on which the Klein four group acts trivially. Spin $1$ has
multiplicity $[31]$, the standard traceless representation on four letters,
which has no Klein-four fixed vector. Hence their tensor product has
no globally symmetric component [FG-3]. This is the only vanishing statement
needed below.

\emph{One polarisation gives the transport identity.}
Now regard $w=\tilde z$ as independent of $z$ and differentiate in the
$w$ direction $z$, denoting this derivative by $\mathcal D$.
With the four-qubit pairing $B_4(x,y)=x^T\epsilon^{\otimes4}y$, write $\bar s=B_4(w,w)$, $n=B_4(z,w)$; physically these are $\bar s,N$.
All Hermitian expressions above extend polynomially by replacing their
conjugate tensors by their spin flips with $z,w$ exchanged.
Put
\[
\begin{gathered}
 u=\sum B(U_i,U_i),\quad q=\sum B(Y_i,U_i),\quad
 p=\sum c_{ij}^T\epsilon^{\otimes4}c_{ij},\\
 y=\sum B(Y_i,Y_i),\quad d=\sum B(C_i(z),\mathcal T_i(z,w,w)).
\end{gathered}
\]
Since the fully symmetric parts of $U$ and $c$ vanish,
$\mathcal D\sum\|U_i\|^2=-u$ and $\mathcal DK_c=-p$.
The vanishing quartic covariant, differentiated in $z$ along $w$, gives
$\sum B(C_i(z),U_i(w,w,z))=3q$.
Consequently $\mathcal D\operatorname{Re}\sum\langle U_i,Y_i\rangle=2q$.
Using $\mathcal Dn=s$, $\mathcal DW_1=0$ and~(CY) yields
\[
 u=-\tfrac13sW_1,\qquad q=\tfrac13sW_1+\tfrac23p.
\]
The derivative of the left side of~(CP) is $2(d-y)$.
Its right side gives
\[
 d-y=\tfrac{16}{3}s(n^2-s\bar s)-2sW_1+\tfrac43p.
\]
Finally $X=Y-2U$ and the same cubic polarisation give
\[
 \begin{aligned}
 \Delta:=\mathcal DH&=\tfrac38(d+2y-8q-4u)-\tfrac32p,\\
 b:=\sum B(X_i,X_i)&=y-4q+4u.
 \end{aligned}
\]
The cancellation is now transparent:
\[
 \Delta-\tfrac98b=\tfrac38(d-y)+\tfrac32q-6u-\tfrac32p
 =s\bigl(2n^2-2s\bar s+\tfrac74W_1\bigr).
\]
The purity identity $P=n^2+(W_1+s\bar s)/2$ turns this precisely into~(FGI) [FG-4, FG-9].
Thus the spin-flip transport law follows from cubic symmetry, the copy-label
Casimirs, and one absent quartic representation, rather than normal-form
block evaluations.

\section{A weighted-variance proof of the sextic zero set}\label{sec:sup-sextic-zeros}
The paper derives the zero set of $F$ (Proposition~\ref{prop:fqe-moment-zeros}) from the critical envelope. This section
gives an independent proof: on the critical normal form, $F$ is an explicit sum of twelve squared moduli, whose
simultaneous vanishing is solved directly. The notation is that of Appendices~\ref{sec:fqe-sextic}
and~\ref{sec:fqe-normal-form} of the paper.

\begin{proof}[Second proof of Proposition~\ref{prop:fqe-moment-zeros}]
The identity \eqref{eq:fqe-remainder} makes every nonzero zero critical. Use the elementary
\hyperref[sec:fqe-normal-form]{normal form} and write the state in its \(G_{abcd}\) coordinates
\(v=(a,b,c,d)\). These differ only by fixed coordinate signs from the magic-basis diagonal
coordinates. Its three cut spectra are the squared moduli of \(v,H_+v,H_-v\), where \[
H_+=\frac12\begin{pmatrix}1&1&1&1\\1&1&-1&-1\\1&-1&1&-1\\1&-1&-1&1\end{pmatrix},\qquad
H_-=\frac12\begin{pmatrix}1&1&1&-1\\1&1&-1&1\\1&-1&1&1\\-1&1&1&1\end{pmatrix}.
\] This follows by the same two-block Hadamard diagonalisation for each cut. Put \(p_i=|v_i|^2\),
\(x_i=v_i^2\), \(s=\sum_i x_i\), and \(B=\sum_i p_ix_i\). The critical-slice identity is \[
\boxed{\quad F(G_{abcd})=
\sum_i\ \sum_{\substack{j<k\\j,k\ne i}}
\left|\bar v_i(v_j^2-v_k^2)+2v_i(p_j-p_k)\right|^2.\quad} \tag{Z}\label{eq:fqe-critical-factor}
\] Here is a direct derivation. The rows of \(2H_+\) and \(2H_-\) together are all four-sign vectors
modulo an overall sign. Expanding the sixth power and cancelling terms with an odd sign multiplicity
gives \[
T=\frac34N^3+\frac98N|s|^2-\frac94N\sum_i p_i^2
+3\sum_i p_i^3-\frac32\operatorname{Re}(\bar sB).
\] Also \(P=N^2+|s|^2/2\): this is the purity identity (Lemma~\ref{lem:fqe-purity}) in homogeneous form for a critical state, where \(P=N^2+\tau/2\) and \(\tau=|s|^2\) by \eqref{eq:fqe-spin-flip-normal-form} (the signs relating \(G_{abcd}\) to the magic-basis coordinates do not affect the squares \(x_i\)). Substitution gives \[
F=27N\sum_i p_i^2-36\sum_i p_i^3-4N^3-5N|s|^2
+18\operatorname{Re}(\bar sB).
\] Expanding the right side of \eqref{eq:fqe-critical-factor} gives this same expression: for each
\(i\) use \(\sum_{j<k}|y_j-y_k|^2=3\sum_j|y_j|^2-|\sum_j y_j|^2\) on the three remaining indices,
and its bilinear version for the cross terms. Thus \eqref{eq:fqe-critical-factor} is an elementary
weighted-variance identity.

We now classify its zeros; rescale to \(N=1\). For \(p_i>0\) let \(u_i=x_i/p_i\), so \(|u_i|=1\).
Vanishing of each summand says \[
x_j-x_k=-2u_i(p_j-p_k),\qquad i,j,k\text{ distinct},\quad p_i>0. \tag{ZE}\label{eq:fqe-zero-equations}
\] If between two and three coordinates are nonzero, choose nonzero \(i,j\) and zero \(k\). Then
\(|x_j|=2p_j\), impossible. A single nonzero coordinate is a Bell-pair product across \(12|34\).

It remains to take all \(p_i>0\). If they are equal, \eqref{eq:fqe-zero-equations} makes all \(x_i\)
equal too. Thus, up to phase, \(v=(\pm1,\pm1,\pm1,\pm1)/2\). One of \(H_+v,H_-v\) has just one
nonzero coordinate: the sign parity selects which. These are Bell-pair products across the other two
cuts.

Otherwise choose \(p_1\ne p_2\). Equation \eqref{eq:fqe-zero-equations} for the pair \(1,2\) with \(i=3,4\) gives
\(u_3=u_4=:u\). If \(p_3\ne p_4\), the same reasoning for the pair \(3,4\) gives \(u_1=u_2=:w\). Now
\eqref{eq:fqe-zero-equations} for the pair \(3,4\) with \(i=1\) reads \(u(p_3-p_4)=-2w(p_3-p_4)\), that is
\(u=-2w\), and for the pair \(1,2\) with \(i=3\) it reads \(w(p_1-p_2)=-2u(p_1-p_2)\), that is \(w=-2u\); but
\(|u|=|w|=1\), which is impossible. Hence
\(p_3=p_4=:t\) and \(x_3=x_4=ut\). Relabel \(1,2\) so \(p_1\ne t\). Applying
\eqref{eq:fqe-zero-equations} to the pair \(1,3\), with \(i=2,4\), gives \(u_2=u\). The equations
for the pairs \(1,2\) and \(1,3\) now give \[
x_1=u(3p_2-2p_1)=u(3t-2p_1).
\] Thus \(p_2=t\), and \(|x_1|=p_1\) forces \(p_1=3t\) and \(u_1=-u\). Normalisation yields, up to
position and common phase, \[
v=(\sqrt3,\pm i,\pm i,\pm i)/\sqrt6. \tag{HS}\label{eq:fqe-hs-family}
\]

These sign families are exactly the two $M_4$ LU orbits, by the explicit normal-form
identification in the proof of Proposition~\ref{prop:fqe-onecut}; the fixed coordinate signs
relating $G_{abcd}$ to $G(v)$ preserve the displayed set of sign choices. Conversely, all the
listed vectors make every term of \eqref{eq:fqe-critical-factor} vanish.
\end{proof}

\section{The R\'enyi bounds: full proofs and further results}\label{sec:sup-renyi}
The intermediate-order R\'enyi proof works with a \emph{relaxation} of the
quantum problem. Take three arbitrary four-point probability vectors
$\lambda^{(j)}$, $j=2,3,4$, and set
$P=\sum_{j,i}(\lambda_i^{(j)})^2$ and
$T=\sum_{j,i}(\lambda_i^{(j)})^3$. Impose the necessary constraints
$P\ge1$, $F=17P-12T-12\ge0$, and the quadratic purity-variance bound
(QK). We do not assume these inequalities characterise physically
realisable cut spectra. An upper bound for this larger set proves the
quantum bound; identifying the equality states requires the separate
rigidity arguments of the paper. The bound (QK)
is deduced below from the sharper six-copy certificate (QV).

The remaining optimisation has two anchor orders. Determinant monotonicity
reduces individual spectra to repeated-root charts; the supporting bounds
then cover an HS region, a strict intermediate band, a capped region and a
narrow cluster region. Single crossing transfers the HS anchor to lower
orders, and monotonicity of R\'enyi entropy transfers the cluster anchor to
higher orders. The final subsections give independent estimates and explain
why local filtering cannot replace the unrestricted compatibility argument.

\subsection*{Proofs of results of the paper}
\subsection{Proof of Theorem~\ref{thm:fqe-sharp-quadratic-variance} and Corollary~\ref{thm:fqe-quadratic-variance}}\label{sec:sup-qk}
\begin{restated}[Sharp quadratic purity variance; exact-rational certificate, QVA]{Theorem}{thm:fqe-sharp-quadratic-variance}
Every unit pure four-qubit state satisfies
\[
\boxed{\qquad F^2+3P(P-1)\ge6V_P,
\qquad V_P=\sum_{j=2}^4(p_j-P/3)^2.\qquad}
\tag{QV}
\]
Equality holds along the whole cluster-to-Bell family \eqref{eq:fqe-quadratic-family}, so the
coefficient $3$ cannot be decreased.
\end{restated}
Write $p_j=\operatorname{tr}\rho_{1j}^2\in[1/4,1]$ and $U=\sum_jp_j^2$. Maximising the convex function $U$ at fixed
$P$ over the box $[1/4,1]^3$, filling one coordinate at a time towards an endpoint, gives the \emph{purity box}
\[
U\le
\begin{cases}
(P-1/2)^2+1/8,&1\le P\le3/2,\\
1+(P-5/4)^2+1/16,&3/2\le P\le9/4,\\
2+(P-2)^2,&9/4\le P\le3.
\end{cases}
\tag{JB}\label{eq:fqe-purity-box}
\]
The family \eqref{eq:fqe-quadratic-family} has $P=1+r^2/2$, $F=(1-r^2)/2$ and $V_P=(1+2r^2)^2/24$, which gives
equality and, for $r>0$, $(6V_P-F^2)/(P(P-1))=3$ exactly (QV-3). Proposition~\ref{prop:fqe-critical-quadratic-variance}
covers critical states, including $P=1$ by Lemma~\ref{lem:fqe-purity}. When $F\ge3/2-P$, the purity box
\eqref{eq:fqe-purity-box} gives the inequality directly by substituting into $F^2+5P^2-3P-6U$ (QV-2).
Thus only noncritical states with $1<P<3/2$ and $F<3/2-P$ need the certificate, which
treats all states at once.

\begin{proof}
For $z\in\mathbb C^{16}$ put $N=\lVert z\rVert^2$, use unnormalised marginals in $P,T,U$, and set
\[
 \mathfrak F=17NP-12T-12N^3,\qquad
 \mathfrak D=\mathfrak F^2+5N^2P^2-3N^4P-6N^2U.
\]
Then $\mathfrak D(z)=N^6\bigl(F^2+3P(P-1)-6V_P\bigr)(z/\sqrt N)$, so it suffices to prove $\mathfrak D\ge0$.
It is a local-unitary invariant of bidegree $(6,6)$, symmetric under qubit permutations and complex
conjugation, and it vanishes at HS and at the cluster state, which are equality cases. We exhibit it as a sum of
squared norms.

Put $\widetilde z=\epsilon^{\otimes4}\bar z$, where $\epsilon=\bigl(\begin{smallmatrix}0&1\\-1&0\end{smallmatrix}\bigr)$.
Since $\epsilon\bar U=U\epsilon$ for $U\in SU(2)$, apply the six-copy spin intertwiners to
$z^{\otimes a}\otimes\widetilde z^{\otimes(6-a)}$, $a\in\{3,4\}$, and stack all $24$ qubit permutations.
For a spin tuple $p=(j_1,\dots,j_4)$, one twice-spin $j_i\in\{0,2,4,6\}$ of the six copies at each qubit, call the resulting map $W_{a,p}$ and equip its magnetic rows with the positive weights
\[
 w_p=\bigotimes_{j\in p}\left(\frac{\operatorname{lcm}_{r}\binom jr}{\binom jk}\right)_{k=0}^{j},
\]
repeated over the permutations. These are the invariant spin inner products: the spin-$j/2$ lowering basis has weights $1/\binom jk$.
The data in the verifier map specifies integer multiplicity vectors $v_{ki}$, integer weights $q_{ki}$
and an integer denominator $D$. Put $\Phi_{ki}=W_{a_k,p_k}v_{ki}$.
Intertwining makes their squared norms local-unitary invariant; summing over permutations makes them
qubit-symmetric, and the real intertwiners and weights make them conjugation-even.

There are three finite checks: the highest-weight identities and vanishing of each $\Phi_{ki}$ at HS and
cluster (QVA-1); positivity of $D$ and every $q_{ki}$ (QVA-2); and the identity
\begin{equation}\label{eq:fqe-qv-allstate-sos}
 2D\,\mathfrak D(z)=\sum_{k,i}q_{ki}\lVert\Phi_{ki}(z)\rVert_{w_{p_k}}^2
\end{equation}
at the $83$ determining states (QVA-3).
Both sides lie in the space $\mathcal V$ of qubit-symmetric, conjugation-even local-unitary
invariants of bidegree $(6,6)$ vanishing at HS and cluster. The ambient space has dimension $85$
(Section~\ref{sec:fqe-invariant-dimensions}, SD-4); HS and cluster evaluations are independent,
since the normalised values of $(N^6,N^3T)$ are $(1,5/12)$ and $(1,3/8)$. Thus $\dim\mathcal V=83$.
QG supplies $83$ Gram entries in $\mathcal V$, built from five-copy covariants multiplied by $N$
or coupled to one further copy (QG-1), whose evaluation matrix at these states
has determinant $627608$ modulo $1000003$ (QG-3). Evaluation is therefore injective on $\mathcal V$,
and \eqref{eq:fqe-qv-allstate-sos} holds identically. Every summand is nonnegative, so $\mathfrak D\ge0$.
\end{proof}

\begin{restated}[Quadratic purity variance]{Corollary}{thm:fqe-quadratic-variance}
Every unit pure four-qubit state satisfies
\[
\boxed{\qquad F^2+\frac92(P-1)\ge6V_P.\qquad}
\tag{QK}
\]
Equality holds exactly for HS cut spectra, cluster cut spectra, or products
of two Bell pairs, up to ordering of cuts and eigenvalues.
The coefficient $9/2$ cannot be decreased, and the coefficient $6$ cannot
be increased while retaining $9/2$.
\end{restated}
\begin{proof}
Write $\mathcal K=F^2+\frac92(P-1)-6V_P$ and $\mathcal V_Q=F^2+3P(P-1)-6V_P$, so that
$\mathcal K-\mathcal V_Q=3(P-1)(3/2-P)$. By Lemma~\ref{lem:fqe-purity}, $P\ge1$. For $1\le P\le3/2$,
Theorem~\ref{thm:fqe-sharp-quadratic-variance} gives $\mathcal K\ge3(P-1)(3/2-P)\ge0$, strictly for $1<P<3/2$.
For $P\ge3/2$ the purity box \eqref{eq:fqe-purity-box} gives
$\mathcal K\ge F^2+4(P-\tfrac32)(\tfrac{27}8-P)$ on $[3/2,9/4]$ and
$\mathcal K\ge F^2+(P-\tfrac94)(\tfrac{39}2-4P)+\tfrac{27}8>0$ on $[9/4,3]$ (QV-2), so $\mathcal K\ge0$, strictly for
$P>3/2$. Equality therefore forces $P=1$ or $P=3/2$.

At $P=3/2$ the box gives $\mathcal K\ge F^2$, so equality forces $F=0$ and $U=9/8$; the purities are a permutation of
$(1,1/4,1/4)$. The marginal of purity one makes the state a product across that cut. On either crossed cut its
marginal is a product of two one-qubit states, and purity $1/4$ forces both to be maximally mixed, so both factors
are Bell pairs. At $P=1$, Proposition~\ref{prop:fqe-critical-variance} gives
$\mathcal K=F^2-6V_P\ge s^4(1-s)(2-s)$ with $0\le s\le1$ (QV-1). Equality requires $s\in\{0,1\}$ and equality in
\eqref{eq:fqe-critical-variance}, which gives exactly HS or cluster spectra. All three types attain equality.
Bell products force the coefficient $9/2$, and cluster spectra force the coefficient $6$.
\end{proof}

The Lean formalisation proves~\eqref{eq:fqe-qv-allstate-sos} for all $z$ directly, without the dimension count
(Section~\ref{sec:fqe-lean}). An older direct certificate for (QK), from eighty squared covariant norms in sixteen modules, is
kept as a regression (QG-2, QG-4).

\subsection{Proof of Lemma~\ref{lem:fqe-renyi-determinant}}\label{sec:sup-determinant}
At fixed trace, second moment and third moment, the determinant is the only
remaining elementary symmetric function of a four-point spectrum. The lemma
shows that increasing this determinant improves R\'enyi entropy for
$1<q<2$, reducing each spectral optimisation to a repeated-root boundary.
\begin{restated}[Determinant monotonicity at fixed second and third moments]{Lemma}{lem:fqe-renyi-determinant}
For $1<q<2$, let $\lambda,\lambda'$ be four-point spectra with the same second moment $p$ and third moment $t$.
If $\prod_i\lambda_i<\prod_i\lambda'_i$, then $H_q(\lambda)<H_q(\lambda')$.
Consequently, at fixed $p$ and $t$, the maximum of $H_q$ is attained at a spectrum of one of the two types
\[
(a,b,c,c)\quad\text{or}\quad(a,a,b,c),
\qquad a\ge b\ge c\ge0,
\tag{DM}
\]
including their degenerations.
\end{restated}
\begin{proof}
Newton's identities fix $e_1=1$, $e_2=(1-p)/2$ and $e_3=(1-3p+2t)/6$ (UR-3). Hence, for $s>0$ and $d=\prod_i\lambda_i$,
\[
 \prod_i(\lambda_i+s)=M(s)+d,\qquad \sum_i\prod_{j\ne i}(\lambda_j+s)=M'(s),\qquad M(s)=s^4+s^3+e_2s^2+e_3s,
\]
where $M$ depends only on $(p,t)$ and both quantities are positive. Since $x^2/(x+s)=x-s+s^2/(x+s)$,
\[
 \sum_i\frac{\lambda_i^2}{\lambda_i+s}=1-4s+\frac{s^2M'(s)}{M(s)+d},
\]
which is strictly decreasing in $d$ for every $s>0$. For $1<q<2$ and $x\ge0$, the substitution $s=xu$ and
$\int_0^\infty u^{q-2}(1+u)^{-1}\,du=\pi/\sin(\pi(q-1))$ give
\[
 x^q=\frac{\sin(\pi(q-1))}{\pi}\int_0^\infty s^{q-2}\,\frac{x^2}{x+s}\,ds .
\]
Each of these integrals converges, so summing over the four eigenvalues gives, for $d<d'$,
\[
 \sum_i\lambda_i^q-\sum_i\lambda_i'^q=\frac{\sin(\pi(q-1))}{\pi}\int_0^\infty s^{q}M'(s)
 \Bigl(\frac1{M(s)+d}-\frac1{M(s)+d'}\Bigr)ds>0,
\]
and $H_q=(1-q)^{-1}\log_2\sum_i\lambda_i^q$ gives $H_q(\lambda)<H_q(\lambda')$.

For (DM), the ordered spectra $\lambda_1\ge\lambda_2\ge\lambda_3\ge\lambda_4\ge0$ with trace one and the given $p,t$
form a compact set, so the determinant attains its maximum on it, and by the first part a spectrum of maximal
determinant maximises $H_q$. Suppose it has $\lambda_1>\lambda_2$ and $\lambda_3>\lambda_4$. Let
$g(z)=\prod_i(z-\lambda_i)$, $m_1=(\lambda_1+\lambda_2)/2$ and $m_2=(\lambda_3+\lambda_4)/2$. Then $g(m_1)<0$,
$g(m_2)<0$ and $g(\lambda_2)=g(\lambda_4)=0$, so for $0<\delta<\min\{-g(m_1),-g(m_2)\}$ the quartic $g+\delta$ changes
sign on each of $(\lambda_4,m_2)$, $(m_2,\lambda_2)$, $(\lambda_2,m_1)$ and $(m_1,\infty)$. Its four roots are therefore
distinct and exceed $\lambda_4\ge0$. They have the same $e_1,e_2,e_3$, hence the same $p,t$, and product $d+\delta$,
contradicting maximality. So $\lambda_1=\lambda_2$ or $\lambda_3=\lambda_4$, which is (DM).
\end{proof}

\subsection{The classical relaxation: multiplier table}\label{sec:sup-renyi-classical}
The proof of Theorem~\ref{thm:fqe-renyi-classical} of the paper uses the multiplier table below and the
supporting lemmas, at the orders $q_s=1587/1000$ and $q_u=8/5$ of that theorem.
The next subsection combines them; the supporting inequalities follow afterwards.

\paragraph{Multiplier table.}
The table specifies the supporting function $\Lambda_\theta$ of Lemma~\ref{lem:fqe-dual}, with
$\theta=(\alpha,\beta,\gamma)$ multiplying the centred cut purity $p-\frac13$, the moment slack $f-4$ and minus
the centred squared purity, $-(p^2-\frac19)$, respectively. Here $\varepsilon=P-1$, $F=17P-12T-12$, and $h_q$ is the mean R\'enyi entropy of $M_4$.
Only two rational multiplier rows are needed: L for the strict band and HT
for the cluster sliver, both at $q_s=1587/1000$. Divide the integer entries
by $10000$. The HS supporting inequality at $q_u=8/5$ has its own explicit
multiplier (Lemma~\ref{lem:fqe-hs-polynomial}); the cubic bounds use none.
The partition in Table~\ref{tab:fqe-renyi-coverage} specifies where each is used.
The L estimate is proved on the slightly larger rectangle
$[0,1/9]\times[3/20,11/25]$ in $(\varepsilon,F)$; the HT estimate uses
$\varepsilon\le1/100$ and distinguishes one cut of purity above $47/100$
from two of purity at most $29/100$.
\begin{table}[htbp]
\centering\small
\begin{tabular}{crrr}
\toprule
Band&$10000\alpha$&$10000\beta$&$10000\gamma$\\
\midrule
L&39700&1500&15900\\
\midrule
HT&35425&1872&11235\\
\bottomrule
\end{tabular}
\caption{Exact supporting multipliers at the lower anchor $q_s$; divide
by $10000$. L supplies a strict common bound, while HT uses separate bounds
for the distinguished cut and the two near-uniform cuts.}
\label{tab:fqe-dual-co}
\end{table}

\subsection{Proof of Theorem~\ref{thm:fqe-renyi-classical}}\label{sec:sup-renyi-classical-proof}
\begin{restated}[Two R\'enyi anchors]{Theorem}{thm:fqe-renyi-classical}
Let $q_s=1587/1000$ and $q_u=8/5$. Every triple satisfying $P\ge1$, $F\ge0$ and
\eqref{eq:fqe-quadratic-variance-proved} obeys one of the following.
\begin{enumerate}
\item If $F\ge3/20$ or $20\varepsilon+2F\ge1$, then $\mathcal E_{q_s}\le5/3$, with equality only at a permutation of
$(c_1,c_2,c_2)$.
\item Otherwise $\mathcal E_{q_u}\le h_{q_u}$, with equality only when all three spectra are HS spectra.
\end{enumerate}
\end{restated}
\begin{proof}
The following partition is exhaustive and disjoint. In the last four rows,
$20\varepsilon+2F<1$; in the last two, $F>11/25$.
\begin{table}[htbp]
\centering\small
\begin{tabular}{@{}p{.36\linewidth}cl@{}}
\toprule
Region (first applicable row) & Order & Bound used \\
\midrule
$20\varepsilon+2F\ge1$ & $q_s$ & Cubic envelope (BC) \\
$F<3/20$ & $q_u$ & HS support (HD) \\
$3/20\le F\le11/25$ & $q_s$ & Strict band L (LT, SHD, LHD) \\
$\max_jp_j\le47/100$ & $q_s$ & Capped cubic bound (CG) \\
$\max_jp_j>47/100$ & $q_s$ & Cluster sliver (NU, SL) \\
\bottomrule
\end{tabular}
\caption{The anchor partition. Rows are applied in order: each triple is handled by the first row whose region
contains it, so every row excludes the regions above it. HD gives HS equality; BC gives cluster equality; the other
three regions are strict. SHD and CG both use the monotonicity of $A_q$ in Lemma~\ref{lem:fqe-binary-quartic}, proved by Cauchy's formula;
SHD adds the quartic checks SHD-3.}
\label{tab:fqe-renyi-coverage}
\end{table}

\emph{Cubic envelope.} Corollary~\ref{cor:fqe-renyi-high-f}, with $q_s>q_b$,
handles the first row and classifies its equality triples.
Henceforth $\varepsilon<1/20$, $F<1/2$, and
\[
 p_j\le P-1/2<11/20<9/16<49/75.
\]
These are the purity caps required by HD and L; they are consequences of
the region, not hypotheses on physical states.

\emph{HS zone.} For $F<3/20$, Lemma~\ref{lem:fqe-hs-polynomial} gives
$\Lambda_\theta\le h_{q_u}$, strictly off HS, and
$\ell_\theta\ge0$ on $[0,1/9]\times[0,3/20]$, which contains $(\varepsilon,F)$.
The dual lemma therefore gives $\mathcal E_{q_u}\le h_{q_u}$,
with equality only at three HS spectra.

\emph{Strict band.} For $3/20\le F\le11/25$,
Lemma~\ref{lem:fqe-rational-dual} gives
$\Lambda_\theta<W-10^{-5}$, where $3W=5+\min\ell_\theta$ on the
L rectangle. Thus
\[
 3\mathcal E_{q_s}\le\sum_j\Lambda_\theta-\ell_\theta
 <5+\min\ell_\theta-\ell_\theta-3\cdot10^{-5}<5.
\]

\emph{Near the cluster triple.} The remaining region has $F>11/25$ and
$\varepsilon<(1-2F)/20<3/500<1/100$.
If every cut has $p_j\le47/100$, Corollary~\ref{cor:fqe-capped-cluster}
gives the strict ceiling. Otherwise label a cut with $p_2>47/100$.
It is unique, since two such cuts would give
$P>47/50+1/4>101/100$. The other two have
$p_j\le P-p_2-1/4<29/100$.
For this last case only, use determinant replacement
(Corollary~\ref{cor:fqe-renyi-six-variable}) to put each cut on an ordered
double-root chart, preserving every $p_j,t_j$ and the constraints.

\emph{Cluster sliver.} Set $X=(p_3-1/4)+(p_4-1/4)=1/2+\varepsilon-p_2\le1/25$.
Writing $x_j=p_j-1/4$, Lemma~\ref{lem:fqe-third-moment} for cuts $3,4$ gives
\[
 F\in I=[F_c-\kappa_aX,F_c+\kappa_aX],\qquad
 F_c=-1+8\varepsilon+9p_2-12t_2,\qquad \kappa_a=4\sqrt3/5,
\]
since the error is at most
$4\sqrt3(x_3^{3/2}+x_4^{3/2})\le4\sqrt3X^{3/2}\le\kappa_aX$.
The actual $F$ lies in $I\cap(11/25,1/2)$, exactly the window needed by SL.
Use the HT row $\theta_s=(3.5425,0.1872,1.1235)$ and put
\[
 \Lambda^\star_i=\Lambda_{\theta_s}(c_i),\qquad
 m=\min_{f\in\partial I}\ell_{\theta_s}(\varepsilon,f),\qquad
 \ell_0=\ell_{\theta_s}(0,1/2).
\]
NU bounds each of the two near-uniform cuts by $\Lambda^\star_2$.
SL bounds the distinguished cut by $\Lambda^\star_1+m-\ell_0$.
Concavity of $\ell_{\theta_s}$ in $F$ gives $\ell_{\theta_s}(\varepsilon,F)\ge m$; hence Lemma~\ref{lem:fqe-dual}
and the contact identity~\eqref{eq:fqe-dual-contact} give
\[
 3\mathcal E_{q_s}
 \le \Lambda^\star_1+2\Lambda^\star_2-\ell_0+m-\ell_{\theta_s}(\varepsilon,F)
 \le \Lambda^\star_1+2\Lambda^\star_2-\ell_0=5.
\]
Here NU and SL are Lemmas~\ref{lem:fqe-near-uniform-dual}
and~\ref{lem:fqe-cluster-sliver}.

\emph{Equality.} HD forces three HS spectra. The strict band and capped region are strict.
In the sliver, equality in SL forces the distinguished cut to be $c_1$ and $\varepsilon=0$.
Then $X=0$, so both other cuts are $c_2$, giving $F=1/2$, excluded in this region.
Thus the replaced triple, and hence the original triple, has a strict bound there.
Only the HS equality of HD and the cluster equality of the cubic envelope remain.
\end{proof}

\subsection{A monotone path from the repeated-largest chart}
\label{sec:fqe-low-tilt-reduction}
\begin{lemma}[Low third-moment tilt; analytic]\label{lem:fqe-low-tilt-reduction}
Let $3/2\le q\le8/5$ and $0\le\omega\le9/5$.
Among spectra $(a,a,b,c)$ with $a\ge b\ge c\ge0$ of a fixed purity,
$H_q-\omega t$ is maximised only by two equal pairs.
Consequently, whenever $0\le\beta\le3/20$, the maximum at fixed purity of the supporting function
$\Lambda_\theta$ of Lemma~\ref{lem:fqe-dual} is attained on the single pattern $(a,b,c,c)$.
\end{lemma}
\begin{proof}
\emph{The path.} Write a repeated-largest spectrum of purity $p$ as
\[
 (1/4+u,\,1/4+u,\,1/4-u+d,\,1/4-u-d),\qquad
 x=p-1/4=4u^2+2d^2,\quad 0\le d\le2u,
\]
the last condition being the ordering. Increase $u$ to $u_*=\sqrt x/2$ at fixed $x$, with
$d=\sqrt{(x-4u^2)/2}$. Then $d'=-2u/d$ and $d\le2u$, so the middle entry $1/4-u+d$ decreases,
stays at most $1/4+u$, and the smallest entry has derivative $2u/d-1\ge0$: it never falls below its
initial value $c\ge0$. The endpoint is two equal pairs. Along the path
\[
 t=\frac1{16}+\frac34x+12u^3-3xu,\qquad
 \frac{dt}{du}=36u^2-3x=6(4u^2-d^2)\ge0,
\]
strictly in the interior unless the path is constant.

\emph{The slope.} Put $r=q-1\in[1/2,3/5]$, $L=\ln2$ and $Z=\sum_i\lambda_i^q$, so that
$H_q=\log_2Z/(1-q)$. The path preserves trace and purity and moves the three distinct entries
$a,b,c$; interpolating $x^r$ quadratically at them gives $dZ=\frac q3[a,b,c]x^r\,dt$, hence
\[
 \frac{dH_q}{dt}=\eta(a,b,c):=-\frac{q[a,b,c]x^r}{3rLZ}.
\]
For distinct $a,b,c\ge0$ with mean $m$ and $0<r<1$,
\[
 -[a,b,c]x^r\ge\frac{r(1-r)}2\,m^{r-2}.
 \tag{DD}\label{eq:fqe-dd-power}
\]
Indeed, let $T$ be the cubic Taylor polynomial of $x^r$ at $m$. Then
$(x^r-T)''=-r(1-r)\{x^{r-2}-m^{r-2}-(r-2)m^{r-3}(x-m)\}\le0$ by the tangent inequality for the convex
function $x^{r-2}$, so $x^r-T$ is concave on $[0,\infty)$ and $[a,b,c](x^r-T)\le0$. As
$[a,b,c](x-m)^2=1$ and $[a,b,c](x-m)^3=a+b+c-3m=0$, $[a,b,c]T=-\frac12r(1-r)m^{r-2}$.
Here $m=(1-a)/3$. With $Z\le a^r\sum_i\lambda_i=a^r$ this gives
\[
 \eta(a,b,c)\ge\frac{q(1-r)3^{2-r}}{6L\,a^r(1-a)^{2-r}},\qquad
 a^r(1-a)^{2-r}=[a(1-a)]^r(1-a)^{2-2r}\le\frac9{16}\left(\frac49\right)^r\le\frac38,
\]
using $1/4\le a\le1/2$ and $r\ge1/2$. With $q(1-r)=q(2-q)\ge16/25$, $L<7/10$ and
$3^{2-r}\ge3^{7/5}>9/2$ (as $3^7>(9/2)^5$),
\[
 \eta(a,b,c)>\frac{64}{35}>\frac95\ge\omega.
\tag{LT}\label{eq:fqe-low-tilt-slope}
\]
Thus $d(H_q-\omega t)/du=(\eta-\omega)\,dt/du>0$ on every nonconstant path, and continuity covers zero or
coalescing entries. This proves the first claim.

For the second, Lemma~\ref{lem:fqe-renyi-determinant} reduces a fixed-purity maximum of $\Lambda_\theta$ to the
patterns $(a,b,c,c)$ and $(a,a,b,c)$. At fixed purity the variable part of $\Lambda_\theta$ is $H_q-12\beta t$ with
$12\beta\le9/5$, and two equal pairs already belong to the first pattern.
\end{proof}

\subsection{An analytic HS supporting inequality}\label{sec:fqe-hs-polynomial}

At the upper anchor $q_u=8/5$, choosing $\beta=1/12$ makes the
fixed-purity optimisation a bound on $H_{8/5}-t$. We first show that its
maximiser has one large eigenvalue, then prove the resulting one-variable
supporting inequality. The outer dual is nonnegative for $F\le3/20$,
which determines the split between the two anchor regions.
Throughout this subsection put $q=8/5$, $r=q-1=3/5$ and $L=\ln2$.

\begin{lemma}[A third-moment tilt at fixed purity; proved]\label{lem:fqe-hs-tilt}
Among four-point probability vectors of a fixed purity $p\le9/16$,
$H_{8/5}(\lambda)-\sum_i\lambda_i^3$ is maximised only by the one-large spectrum
\[
 \lambda(s)=\bigl((1+3s)/4,(1-s)/4,(1-s)/4,(1-s)/4\bigr),
 \qquad p=(1+3s^2)/4,
\]
up to permutation (uniform when $p=1/4$).
\end{lemma}
\begin{proof}
Let $Z=\sum_i\lambda_i^q$ and let $a,b,c$ be three distinct entries.
Their mean is at most $1/3$, so the divided-difference bound~\eqref{eq:fqe-dd-power} gives
$-[a,b,c]x^r\ge\frac12r(1-r)3^{2-r}$. Also
$Z=\sum_i\lambda_i\lambda_i^r\le p^r$ by concavity of $x^r$.
Consequently
\[
 \eta(a,b,c):=-\frac{q[a,b,c]x^r}{3rLZ}
 \ge\frac8{75L}\frac{3^{7/5}}{(9/16)^{3/5}}
 >\frac{16}{105}(3\cdot2^{12})^{1/5}>1.
 \tag{HTS}\label{eq:fqe-hs-tilt-slope}
\]
The last comparisons use $L<7/10$ and
$105^5=12762815625<3\cdot2^{32}=12884901888$.

Lemma~\ref{lem:fqe-low-tilt-reduction}, with $\omega=1$, reduces the maximum
to the pattern $(a,b,c,c)$. Write it as
\[
 (1/4+u+d,1/4+u-d,1/4-u,1/4-u),\qquad
 x=p-1/4=4u^2+2d^2,\quad 0\le d\le2u.
\]
Decrease $u$ to $u_0=\sqrt{x/12}$ at fixed $x$, setting
$d=\sqrt{(x-4u^2)/2}$. Ordering and positivity persist, and $b=c$ at
the endpoint: it is the one-large spectrum.
Along this path
\[
 t=\frac1{16}+\frac34x+3xu-12u^3,\qquad
 \frac{dt}{du}=3x-36u^2\le0.
\]
As in the preceding chart reduction, $dH_q/dt=\eta(a,b,c)>1$.
Thus decreasing $u$ strictly increases $H_q-t$ unless the spectrum
is already one-large. Continuity handles zero or coalescing entries.
At equality the strict paths force every determinant-maximal replacement to be
one-large. That replacement preserves $p,t$, so equality in the sharp third-moment bound
(Lemma~\ref{lem:fqe-third-moment}) forces the original spectrum to be the same.
The uniform case $p=1/4$ is immediate.
\end{proof}

\begin{lemma}[HS supporting inequality; proved]\label{lem:fqe-hs-polynomial}
At $q=q_u=8/5$ put $u=3^{3/5}$ and
\[
 k=\frac{16}{3L}\frac{u-1}{u+1},\qquad
 \beta=\frac1{12},\qquad \gamma=\frac{10}3,\qquad
 \alpha=\frac32k+\frac{65}{36}.
\]
Every four-point probability vector with $p\le9/16$ satisfies
$\Lambda_{(\alpha,\beta,\gamma)}(\lambda)\le h_{8/5}$,
with equality only at HS. For $0\le\varepsilon\le1/9$ and
$0\le F\le3/20$,
\[
 \ell_{(\alpha,\beta,\gamma)}(\varepsilon,F)
 \ge\frac{149}{324}\varepsilon
       +F\left(\frac1{12}-\frac59F\right)\ge0.
\]
\end{lemma}
\begin{proof}
At fixed $p$, the only variable part of $\Lambda$ is $H_q-t$.
Lemma~\ref{lem:fqe-hs-tilt} reduces its maximum to $\lambda(s)$.
Write $f(p)=H_q(\lambda(s))$, $p_* =1/3$, and
\[
 A(s)=\frac92(s-\tfrac13)^2(s+\tfrac16),\qquad
 W(p)=f(p)-\beta A(s)-\gamma(p-p_*)^2.
\]
Direct moment evaluation gives
\[
 \Lambda(\lambda(s))=W(p)+\frac32k(p-p_*),\qquad
 W(p_*)=h_q,\qquad W'(p_*)=-\frac32k,
 \qquad \frac{d^2 A}{dp^2}=\frac6s.
\]
We show that $W$ is first convex and then concave, allowing either
interval to be empty. The needed scalar calculation is explicit below.
Put
\[
 x=\left(\frac{1+3s}{1-s}\right)^{1/5},\quad
 J=\frac{L f''(p)}{2\gamma+6\beta/s},\quad
 h_4(x)=1+x+x^2+x^3+x^4.
\]
Since $p\le9/16$, $s<2/3$, so $1<x<2$ and $0<x^5-1<8$.
Differentiating
\[
 L\frac{dH_q(\lambda(s))}{ds}
 =-\frac{2(x^5+3)(x^3-1)}{x^8+3}
\]
gives the identity
\[
 \frac{dJ}{dx}
 =-\frac{8(x^5+3)^3 Q(x)}
 {5x^3(43x^5-31)^2(x^8+3)^3h_4(x)^3}.
 \tag{HCJ}\label{eq:fqe-hs-curvature-identity}
\]
Here, with $y=x-1$ and $R=x^5-1$,
$Q(x)=\sum_{j=0}^7R^jP_j(y)$, where the five columns below are the
coefficients of $1,y,y^2,y^3,y^4$ in $P_j$.
\[
\begin{array}{c|rrrrr}
 j&1&y&y^2&y^3&y^4\\\hline
0&225600&457600&471040&240000&48640\\
1&582160&1108000&1118304&557584&110864\\
2&690608&1195520&1018512&418768&68128\\
3&756896&1304520&1151162&522393&96703\\
4&516296&613195&462027&220948&46208\\
5&181031&120245&477&-23717&-6797\\
6&-259&2795&1978&774&129\\
7&43&0&0&0&0
\end{array}
\]
This displayed identity proves the sign without a parameter certificate:
$P_0,\ldots,P_4,P_7$ have positive coefficients, and, for $0\le y<1$,
\[
 P_5(y)\ge181031-23717-6797=150517,\qquad P_6(y)\ge-259.
\]
Thus $P_5+RP_6\ge150517-8\cdot259>0$. Hence $Q>0$ and $J$ strictly decreases.
As $W''=(2\gamma+6\beta/s)(J/L-1)$, the asserted curvature pattern follows.

To turn this curvature pattern into a supporting line, subtract the
tangent at $p_*$ and write
$D(p)=W(p)-W(p_*)-W'(p_*)(p-p_*)$.
Then $D(p_*)=D'(p_*)=0$, and $D'$ first increases and then decreases.
If $D(1/4)<0$, the contact $p_*$ must lie in the strictly concave portion:
otherwise $D'<0$ throughout its left side, which would give
$D(1/4)>0$. In the concave portion, the tangent strictly
dominates to the right of $p_*$; to its left, $D$ has no maximum except
at $1/4$ or $p_*$. Thus $D<0$ away from $p_*$.
In the present case the
left-endpoint gap is strictly positive, so $D(1/4)<0$:
\[
 \Delta=-D(1/4)=h_q+\frac k8-2+\frac\beta{12}+\frac\gamma{144}
 =h_q+\frac k8-2+\frac{13}{432}>\frac1{5000}.
 \tag{HUC}\label{eq:fqe-hs-uniform-contact}
\]
For completeness, this small constant has an elementary rational proof.
Let $u_0=19331/10000$, $v=1/(2u_0+1)$ and
\[
 L_-=2\sum_{j=0}^5\frac{1}{(2j+1)3^{2j+1}},\qquad
 B_+=2\sum_{j=0}^4\frac{v^{2j+1}}{2j+1}
             +\frac{2v^{11}}{11(1-v^2)}.
\]
The formula $h_q=8/3-(5/(3L))\ln(1+1/u)$ and the atanh series give $L>L_-$ and
$\ln(1+1/u_0)<B_+$. Since $u_0^5<27$, $u>u_0$; the function
$-5\ln(1+1/u)+2(u-1)/(u+1)$ increases with $u>0$.
The purely rational comparison
\[
 \frac{301}{144}L_- -5B_+ +\frac{2(u_0-1)}{u_0+1}
 >\frac{21}{50000}
\]
therefore gives $3L\Delta>21/50000>3L/5000$, using $L<7/10$.
This proves~\eqref{eq:fqe-hs-uniform-contact}.
The tangent bound is strict away from $p_*$ by the strict curvature
calculation. Thus $\Lambda\le h_q$, with equality only at HS.

Finally $u>19/10$ and $L<7/10$ give
$k>480/203>7/3$. Substitution into the outer dual yields
\[
 \ell=\left(\alpha-\gamma(\tfrac{17}{12}+\tfrac\varepsilon3)\right)\varepsilon
          +F\left(\tfrac1{12}-\tfrac59F\right)
 \ge\frac{149}{324}\varepsilon+F\left(\tfrac1{12}-\tfrac59F\right).
\]
The last term is nonnegative for $F\le3/20$, as required.
All global maximisation and sign arguments above are analytic;
HD checks the displayed identities and rational comparisons.
\end{proof}

\subsection{The strict supporting band}\label{sec:fqe-rational-dual}
The bounded-head and capped-cubic proofs below use the same near-uniform
moment inequality. Readers following dependencies rather than the
case partition can first consult
Lemma~\ref{lem:fqe-near-uniform-moment} and the moment-rectangle argument
in Section~\ref{sec:sup-nu-moment}, then return to these two applications.
\begin{lemma}[Strict dual band; analytic with exact polynomial sign certificates]
\label{lem:fqe-rational-dual}
For band $L$ of Table~\ref{tab:fqe-dual-co}, set
$W=5/3+\min\ell_\theta/3$, with the minimum over its rectangle's corners.
At $q_s=1587/1000$, every four-point spectrum with $p\le49/75$ satisfies
$\Lambda_\theta<W-10^{-5}$.
\end{lemma}
\begin{proof}
At fixed purity the variable part is $H_q-(9/5)t$.
Lemma~\ref{lem:fqe-low-tilt-reduction} reduces the maximum to $(a,b,c,c)$.
Lemma~\ref{lem:fqe-small-head-dual} handles $a\le1/2$ and
Lemma~\ref{lem:fqe-large-head-dual} handles $1/2\le a\le4/5$.
These cover the domain: $p\ge a^2+(1-a)^2/3$ and its value $49/75$ at $a=4/5$
force $a\le4/5$. The corner minimum of the outer dual is $1837/125000$,
so $W-10^{-5}=2507333/1500000$ as required.
\end{proof}

\begin{lemma}[Binary quartic correction; analytic]\label{lem:fqe-binary-quartic}
Let $1<q<2$. Put $v=(1-2u)^2$,
\[
 m_q(v)=\frac{(1+\sqrt v)^q+(1-\sqrt v)^q}{2},\qquad
 D_q(v)=\frac{\ln m_q(v)}{(q-1)\ln2}=1-H_q(u,1-u).
\]
Then $D_q'''(v)>0$ on $0<v<1$, and
\[
 A_q(u)=\frac{4u(1-u)-H_q(u,1-u)}{u(1-u)(1-2u)^2}
\]
is strictly increasing on $0<u<1/2$.
Consequently, a quartic bound
$H_q(u,1-u)\le4u(1-u)-a u(1-u)(1-2u)^2$
at $u=u_0$ holds throughout $u_0\le u\le1/2$.
\end{lemma}
\begin{proof}
Write $a=\tan^2\!\bigl(\tfrac{\pi}{2q}\bigr)$; since $\pi/4<\pi/(2q)<\pi/2$, $a>1$. Put $f=\ln m_q$.
The proof is Cauchy's formula for $\ln m_q$ on the slit plane, after removing the one zero of $m_q$.

\emph{Analyticity.} Let $g(w)=\tfrac12\bigl((1+w)^q+(1-w)^q\bigr)$ with principal powers. It is holomorphic on
$\mathbb C\setminus\bigl((-\infty,-1]\cup[1,\infty)\bigr)$ and even, so $m_q(v)=g(\sqrt v)$ does not depend on the
sign of the root and is holomorphic on $\Omega=\mathbb C\setminus[1,\infty)$; at $v=0$ it is the convergent series
$\sum_k\binom q{2k}v^k$. It is real on the real axis and $m_q(\bar v)=\overline{m_q(v)}$.
For $\operatorname{Im}v\ge0$ take $w=\sqrt v$ in the closed first quadrant. Then $\arg(1-w)=\arg(w-1)-\pi$, so
\begin{equation}
 m_q(v)=\tfrac12\bigl[(1+w)^q+e^{-i\pi q}(w-1)^q\bigr],\qquad \arg(w-1)\in[0,\pi].
 \label{eq:fqe-bq-upper}
\end{equation}
The right side is continuous on the closed first quadrant, so $m_q$ extends continuously to the closed upper
half-plane, and to the closed lower half-plane by conjugation.

\emph{Zeros.} Let $m_q(v)=0$ and $\operatorname{Im}v\ge0$, $w=\sqrt v$. Then $|1+w|^q=|1-w|^q$, so $\operatorname{Re}w=0$ and $v=-y^2$
is real, $w=iy$, $y\ge0$. Now $(1\pm iy)^q=(1+y^2)^{q/2}e^{\pm iq\arctan y}$, so
$m_q(-y^2)=(1+y^2)^{q/2}\cos(q\arctan y)$. As $0<q\arctan y<q\pi/2<\pi$ for $y>0$, this vanishes only at
$q\arctan y=\pi/2$, that is $v=-a$. (For $v\in[0,1)$, $m_q>0$.) The zero is simple:
$g'(w)=\tfrac q2\bigl((1+w)^{q-1}-(1-w)^{q-1}\bigr)$, and at $w=iy_0$, $y_0=\sqrt a$,
\[
 g'(iy_0)=iq(1+y_0^2)^{(q-1)/2}\sin\bigl((q-1)\arctan y_0\bigr)\ne0,
\]
since $0<(q-1)\arctan y_0<\pi/2$; and $m_q'(v)=g'(w)/(2w)$. Hence
$h(v)=m_q(v)/(1+v/a)$ is holomorphic and zero-free on $\Omega$, and continuous and zero-free on the closed upper
half-plane away from the cut; on the cut see the next step.

\emph{Boundary values on the cut.} For $t>1$, \eqref{eq:fqe-bq-upper} gives
$2m_q(t+i0)=(1+\sqrt t)^q+e^{-i\pi q}(\sqrt t-1)^q$ with
$\operatorname{Im}\,2m_q(t+i0)=-\sin(\pi q)(\sqrt t-1)^q>0$, because $\pi<\pi q<2\pi$.
So $m_q(t+i0)\ne0$ and
\[
 \theta_q(t)=\arg\bigl(2m_q(t+i0)\bigr)\in(0,\pi)
\]
is continuous on $[1,\infty)$, with $\theta_q(t)\to0$ as $t\downarrow1$, since $m_q(1)=2^{q-1}>0$.

\emph{The logarithm.} $\Omega$ is simply connected and $h$ is zero-free on it with $h>0$ on $(-\infty,1)$
(there $m_q$ and $1+v/a$ change sign together at $-a$). So $F=\ln h$ with $F(0)=0$ is holomorphic on $\Omega$,
real on $(-\infty,1)$, and $F(\bar v)=\overline{F(v)}$. The closed upper half-plane is simply connected and $h$
is continuous and zero-free on it, so $F$ extends continuously to it. On the cut, $\operatorname{Im}F(t+i0)$ is
continuous and equals $\theta_q(t)$ modulo $2\pi$ (as $1+t/a>0$). Both vanish at $t=1$, so
$\operatorname{Im}F(t+i0)=\theta_q(t)$ for all $t\ge1$ and $F(t+i0)-F(t-i0)=2i\theta_q(t)$.
In particular $F$ is bounded near $v=1$.

\emph{Growth.} For $|w|>1$ in the first quadrant, $(1+w)^q=w^q(1+1/w)^q$ and $(w-1)^q=w^q(1-1/w)^q$ (the
arguments add without wrapping), so by \eqref{eq:fqe-bq-upper}
\[
 m_q(v)=w^q\bigl(c+\varepsilon(1/w)\bigr),\qquad c=\tfrac12\bigl(1+e^{-i\pi q}\bigr)\ne0,
\]
with $\varepsilon$ holomorphic on $|z|<1$ and $\varepsilon(0)=0$; $c\ne0$ because $q$ is not an odd integer.
Hence $F(v)-\bigl(\tfrac q2\mathrm{Log}\,v-\mathrm{Log}(1+v/a)+\mathrm{Log}\,c\bigr)$ has exponential tending to
$1$ as $|v|\to\infty$ in the closed half-plane, which is connected, so it tends to a constant in $2\pi i\mathbb Z$.
Thus $|F(v)|\le C\ln|v|$ for large $|v|$, in both half-planes.

\emph{The integral formula.} Fix $0<v<1$. By Cauchy's formula for $F'''$ on the region
$\{|\zeta|<R\}$ with the strip $\{\operatorname{Re}\zeta\ge1,\ |\operatorname{Im}\zeta|\le\epsilon\}$ removed,
\[
 F'''(v)=\frac{3!}{2\pi i}\oint\frac{F(\zeta)}{(\zeta-v)^4}\,d\zeta .
\]
Let $\epsilon\downarrow0$ and then $R\to\infty$. $F$ is bounded near $1$, so the short segment near the
tip $\zeta=1$ contributes $O(\epsilon)$; the arc contributes $O(R^{-3}\ln R)$; and by the jump of $F$ across the cut
the two edges contribute $2i\int_1^R\theta_q(t)(t-v)^{-4}\,dt$. Therefore
\[
 F'''(v)=\frac6\pi\int_1^\infty\frac{\theta_q(t)}{(t-v)^4}\,dt>0 .
\]
Since $m_q=(1+v/a)e^F$ and $(\ln(1+v/a))'''=2/(a+v)^3$,
\[
 f'''(v)=(\ln m_q)'''(v)=\frac2{(a+v)^3}+\frac6\pi\int_1^\infty\frac{\theta_q(t)}{(t-v)^4}\,dt>0
 \qquad(-a<v<1),
\]
which is $D_q'''>0$ on $0<v<1$.

\emph{Monotonicity of $A_q$.} $D_q(0)=0$ and $D_q(1)=1$, so
\[
 \frac{A_q(u)}4=\frac{D_q(v)-v}{v(1-v)}=-[0,v,1]D_q .
\]
For $0<v<v'<1$, $[0,v,1]D_q-[0,v',1]D_q=(v-v')[0,v,v',1]D_q$, and by the mean value theorem for
divided differences (Rolle's theorem; $D_q$ is continuous on $[0,1]$) the last factor is
$D_q'''(\xi)/6>0$. So $A_q$ decreases in $v=(1-2u)^2$: it is strictly increasing in $u$.
\end{proof}

\subsection{The bounded-head chart}
\label{sec:fqe-small-head-dual}

\begin{lemma}[Supporting bound on the bounded-head chart; analytic; exact rational checks]\label{lem:fqe-small-head-dual}
At $q=1587/1000$ and the multiplier $\theta=(397/100,3/20,159/100)$,
every ordered spectrum $(a,b,c,c)$ with $a\le1/2$ satisfies
\[
 \Lambda_\theta\le\frac{8023}{4800}
 <\frac53+\frac{1837}{375000}-10^{-5}.
\]
So on this chart the band-$L$ dual bound needs only rational arithmetic: rational enclosures of eight node values (SHD-1) for the binary
quartic correction of Lemma~\ref{lem:fqe-binary-quartic} and eight one-variable quartic sign checks (SHD-3).
\end{lemma}
\begin{proof}
Put $r=q-1$, $L=\ln2$, $h(u)=H_q(u,1-u)$, $d=u(1-u)$, and write
\[
 \lambda=\Bigl(\frac{1-u+\delta}2,\frac{1-u-\delta}2,\frac u2,\frac u2\Bigr),\qquad v=\delta^2,
\]
a head pair of total weight $1-u$ and a uniform tail pair of weight $u$.
The head cap and ordering give $0\le u\le1/2$ and $0\le\delta\le\min\{u,1-2u\}$.
We treat $u\le1/4$ by a head-deficit estimate and $u\ge1/4$ by the near-uniform moment bound.

\emph{Small tail, $0\le u\le1/4$.}
For the conditional vectors of the two pairs, Jensen's inequality for the logarithm (as in the proof
of~\eqref{eq:fqe-capped-pairs}) gives
\[
 H_q(\lambda)\le h(u)+(1-w)\bigl(1-D_q(z)\bigr)+w
 =1+h(u)-(1-w)D_q(z),\qquad w=\frac{u^q}{(1-u)^q+u^q},
\]
where $z=v/(1-u)^2\le u^2/(1-u)^2\le1/9$ is the squared imbalance of the head pair and $D_q$ is the
function of Lemma~\ref{lem:fqe-binary-quartic}. That lemma shows that $(D_q(z)-z)/(z(1-z))=A_q/4$ decreases in
$z\in(0,1)$. Put $\kappa=\tfrac{81}8\bigl(D_q(1/9)-\tfrac19\bigr)$, its value at $z=1/9$. Here
$D_q(1/9)=1-H_q(1/3,2/3)\ge249/2000>1/9$, since $H_q(1/3,2/3)<1751/2000$ (power sums and the atanh series, as for the
nodes below), so $\kappa>0$. For $0<z\le1/9$,
\[
 D_q(z)\ge z+\kappa z(1-z)\ge z\bigl(1+\tfrac89\kappa\bigr)=9D_q(1/9)\,z\ge\frac{2241}{2000}z ,
\]
since $9\cdot249/2000=2241/2000$.

Next $1-w=(1-u)^2k_q(u)$ with
\[
 k_q(u)=\frac{(1-u)^{q-2}}{(1-u)^q+u^q}=\frac{(1-u)^{-2}}{1+(u/(1-u))^q},
\]
which increases with $q$ because $u\le1-u$; hence $k_q\ge k_{3/2}$. With $t=\sqrt{u/(1-u)}$,
\[
 \frac{k_{3/2}(u)}{1+2u}=\Bigl[1+\Bigl(\frac{t(1-t)}{1+t^2}\Bigr)^3\Bigr]^{-1}\ge\frac{64}{65},
\]
since $0\le t(1-t)\le1/4$ and $1+t^2\ge1$, so the cube is at most $1/64$.
Combining, $(1-w)D_q(z)\ge\frac{2241}{2000}\cdot\frac{64}{65}(1+2u)v=\frac{8964}{8125}(1+2u)v$, so
\[
 H_q(\lambda)\le1+h(u)-\frac{11}{10}(1+2u)v.                 \tag{SHD}\label{eq:fqe-small-head-deficit}
\]

The binary entropy is controlled by the quartic correction of Lemma~\ref{lem:fqe-binary-quartic}:
$h(u)=4d-A_q(u)\,d(1-4d)$ with $1-4d=(1-2u)^2\ge0$, and $A_q$ is increasing. Take the nodes and values
\[
\begin{array}{c|cccccccc}
u_i&0&\frac1{100}&\frac3{100}&\frac1{20}&\frac2{25}&\frac{13}{100}&\frac{39}{200}&\frac{47}{200}\\[2pt]\hline
a_i&\frac{99}{1000}&\frac{217}{1000}&\frac{289}{1000}&\frac{83}{250}&\frac{377}{1000}&\frac{43}{100}&\frac{12}{25}&\frac{101}{200}
\end{array}
\]
and $u_8=1/4$. They satisfy $A_q(u_i)\ge a_i$: at $u_0=0$ in the sense $A_q(0^+)=4-q/(rL)>99/1000$, and at the other seven
points by interval enclosures of the power sum $u^q+(1-u)^q$ and of the logarithm (atanh series) [exact certificate; SHD-1].
Since $A_q$ is increasing, $A_q\ge a_i$ on $[u_i,u_{i+1}]$, hence $h(u)\le4d-a_id(1-4d)$ there.

Substituting~\eqref{eq:fqe-small-head-deficit} in the definition of $\Lambda_\theta$ gives
\[
 \Lambda_\theta\le\frac{1999}{1200}+h(u)-\frac{179}{50}d
       -\frac{159}{100}d^2
       +\left(\frac3{200}+\frac{37}{50}u-\frac{159}{100}u^2\right)v
       -\frac{159}{400}v^2.
\]
The coefficient of $v$ is positive on $[0,1/4]$. Using $v\le u^2$, dropping the last term and inserting the bound for $h$ leaves, on $[u_i,u_{i+1}]$, the quartic
\[
 P_i(u)=\frac{1999}{1200}+\frac{21}{50}d-a_i\,d(1-4d)-\frac{159}{100}d^2
       +\Bigl(\frac3{200}+\frac{37}{50}u-\frac{159}{100}u^2\Bigr)u^2 .
\]
It remains to show $P_i\le8023/4800$ on $[u_i,u_{i+1}]$. Each $8023/4800-P_i$ is a quartic in $u$; written in the
Bernstein basis of its interval, all five coefficients are positive [exact-rational certificate; SHD-3]. The smallest coefficient is $1/6400$, at $u=1/4$.

\emph{Large tail, $1/4\le u\le1/2$.} Here we show that $\Lambda_\theta$ does not exceed its uniform value.
The coordinate bounds give $p\le11/32$, by maximising a convex
quadratic separately on $[1/4,1/3]$ and $[1/3,1/2]$.
Put $x=p-1/4$, $y=\sum(\lambda_i-1/4)^3$.
Here $y\ge0$, so $v=y/x$ lies in $[0,9/50]$, and $x\le X=3/32$.
With $b=12\beta=9/5$ and $c=\alpha+8\beta-\gamma/2=35/8$, the identity~\eqref{eq:fqe-uniform-dual-difference},
valid for every multiplier, gives $\Lambda_\theta((1/4)^4)-\Lambda_\theta(\lambda)=2-H_q-cx+by+\gamma x^2$.
We apply Lemma~\ref{lem:fqe-moment-rectangle} with the SHD column of Table~\ref{tab:fqe-moment-rectangle}.
On the stated rectangle,
$17/10<V<19/10$, $-3/50<W<3/8$, $1\le N<6/5$ and $|B+2Cv+XD_5|<19/10$,
with $A,B,C,D_5$ those of Lemma~\ref{lem:fqe-near-uniform-moment}.
The bound $V^2-2W>8/3$ from the capped-cubic proof gives $Q=8/3$, and
$8/3-2X(19/10)(3/50)>(7/4)(6/5)^2$ gives $m=7/4>2\rho\gamma$; $C>7/5$ gives
$\Phi_{vv}(X,v)\ge(2C-X(19/10)^2)/(\rho(6/5)^2)>4=\kappa$.
At the single point $v_*=27/200$,
\[
 \Phi(X,v_*)>1/800,\qquad |\Phi_v(X,v_*)|<1/100,
\]
using the eight-term alternating lower sum for
$\ln(1+XV+X^2W)$ and $693147/10^6<L<693148/10^6$.
The lemma gives a dual difference greater than $(1/800-1/80000)x>0$.
Thus $\Lambda_\theta\le\Lambda_\theta((1/4)^4)=8023/4800$.
The uniform case $x=0$ is immediate.

The small-tail and large-tail cases together prove the stated bound.
\end{proof}

\subsection{The large-head chart by two Hermite bounds}
\label{sec:fqe-large-head-dual}
Throughout this subsection $q=1587/1000$, $r=q-1$, $L=\ln2$ and
$\theta=(397/100,3/20,159/100)$. Put
\begin{gather*}
 f(a,z)=\Lambda_\theta\left(a,\frac{(1-a)(1+2z)}3,
              \frac{(1-a)(1-z)}3,\frac{(1-a)(1-z)}3\right),\\
 B=\frac{1+2z^2}3,\qquad C=\frac{1+6z^2+2z^3}9,\qquad
 K=\frac{(1+2z)^q+2(1-z)^q}{3^q}.
\end{gather*}
Here $1/2\le a\le4/5$, $0\le z\le1$, and the purity and third moment are
$p=a^2+B(1-a)^2$, $t=a^3+C(1-a)^3$.
The objective is $f<T=2507333/1500000$. The first lemma supplies the
shape condition $f_{aaa}\le0$. The second then bounds $f$ by a quadratic
in $a$, replaces $K(z)$ by one cubic minorant, and proves positivity of
that quadratic's gap by completing the square. Thus the two-dimensional
chart reduces to univariate polynomial signs.

\begin{lemma}[Largest-entry curvature; exact polynomial sign certificates]\label{lem:fqe-large-head-curvature}
On this whole chart, $f_{aaa}\le0$.
\end{lemma}
\begin{proof}
Write $h(a,K)=-\ln(a^q+K(1-a)^q)/(rL)$. The remaining part of $f$ is a polynomial in $a$
whose only fourth derivative comes from $-\gamma p^2$, so
$f_{aaaa}=h_{aaaa}-24\gamma(1+B)^2$.

\emph{The shape of $h_{aaaa}$ in $K$.}
Set $x=2a-1\in[0,3/5]$, $\rho=(1-a)/a=(1-x)/(1+x)\in[1/4,1]$, $w=K\rho^q$ and
\[
 y=\frac{a^q-K(1-a)^q}{a^q+K(1-a)^q}=\frac{1-w}{1+w}.
\]
Apart from the positive factor $16q/(rL(1-x^2)^4)$, $h_{aaaa}$ is
\begin{align*}
 J(y)={}&6q^3y^4-24q^2xy^3+(-8q^3+36qx^2+8q)y^2\\
 &+(24q^2x-24x^3-24x)y+2q^3-36qx^2-8q+6x^4+36x^2+6.
\end{align*}
On $0\le x\le3/5$,
\[
 J'(y)\big|_{y=x}=8x(q-1)\{3(q-1)^2x^2-2q^2+q+3\}\le0,\qquad
 J'''(y)=144q^2(qy-x)\ge0\quad(y\ge x).
\]
So on $[x,1]$ the derivative $J'$ is convex and starts nonpositive. Hence $J$ decreases and then
increases, and has no interior maximum on any subinterval:
\[
 J(y)\le\max\{J(y_\ell),J(y_h)\}\qquad(x\le y_\ell\le y\le y_h\le1).
\]

\emph{Envelopes for $y$.}
Write $K=\lambda_1^q+2\lambda_2^q$ with $\lambda=\bigl(\frac{1+2z}3,\frac{1-z}3,\frac{1-z}3\bigr)$.
Convexity of $s^q$ gives $K\ge3^{-r}>1/2$, since $3^r<2$.
Concavity of $s^r$ with weights $\lambda_i$ gives $K=\sum\lambda_i\lambda_i^r\le B^r\le1$.
If $B\le1/2$, then $K\le2^{-r}<2/3$, because $2^q>3$.
For $0<\rho\le1$ we have $\rho^2\le\rho^q\le\rho^{3/2}\le\rho(1+\rho)/2$ and $\rho^q\le\rho$.
Hence $w\ge\rho^2/2$, $w\le\rho(1+\rho)/2=(1-x)/(1+x)^2$ always, and $w\le2\rho/3$ when $B\le1/2$.
Since $y=(1-w)/(1+w)$ decreases in $w$, we get $y_\ell\le y\le y_h$ with
\[
 y_h=\frac{(1+x)^2-(1-x)^2/2}{(1+x)^2+(1-x)^2/2},\quad
 y_A=\frac{(1+x)-\frac23(1-x)}{(1+x)+\frac23(1-x)},\quad
 y_B=\frac{(1+x)^2-(1-x)}{(1+x)^2+(1-x)},
\]
and $y_\ell=y_A$ if $B\le1/2$, $y_\ell=y_B$ otherwise.
Moreover $y_A-x$ and $y_B-x$ have the numerators $\frac13(1-x^2)$ and $(1-x)(x+x^2)$ over the
positive denominators, so $x\le y_\ell$, and $y_h\le1$.

\emph{Three polynomial inequalities.}
Let $L_0=0.693147<L$. For $0\le x\le3/5$,
\begin{align*}
 J(x,y_A)&\le\tfrac{1696}{25}\,\frac{rL_0(1-x^2)^4}{16q},\qquad
 J(x,y_h)\le\tfrac{1696}{25}\,\frac{rL_0(1-x^2)^4}{16q},\\
 J(x,y_B)&\le\tfrac{4293}{50}\,\frac{rL_0(1-x^2)^4}{16q}.
\end{align*}
Clearing the denominators ($D_A^4$ with $D_A$ linear, $D_B^4$ and $D_h^4$ with quadratic denominators)
turns each into a polynomial inequality of degree $12$, $16$ and $16$ in $x$.
Each is certified by the Bernstein expansion on $[0,3/5]$, all of whose coefficients are positive (LHD-1).
Thus $h_{aaaa}<1696/25=24\gamma(16/9)$ if $B\le1/2$, and $h_{aaaa}<4293/50=24\gamma(9/4)$ otherwise.
Since $B\ge1/3$, respectively $B\ge1/2$, this is $h_{aaaa}<24\gamma(1+B)^2$, that is, $f_{aaaa}<0$.

\emph{The point $a=1/2$.}
There, with $v=(1-K)/(1+K)$,
\[
 f_{aaa}=\frac{16q(q+1)v-16q^3v^3/r}{L}
             -\frac{54}5(1-C)-\frac{477}{25}(1-B^2).
\]
The first term increases with $v\le1/3$. Since $K^{(4)}>0$ (because $q(q-1)(q-2)(q-3)>0$),
cubic Hermite interpolation at $0,1$ gives, with $k=5247/10000<K(0)=3^{-r}$,
\[
 K\ge K_*(z)=k+(3-3k-2q/3)z^2+(2q/3-2+2k)z^3.
\]
Replacing $K$ by $K_*$ and $L$ by $L_0$, and clearing
$r(1+K_*)^3$, reduces $-f_{aaa}(1/2,z)\ge0$ to $(1-z)\sum_{j=0}^{12}c_jz^j\ge0$, where every $c_j$ is
positive (LHD-2). Lowering $K(0)$ to $k$ preserves the minorant because its coefficient is
$1-3z^2+2z^3=(1-z)^2(1+2z)\ge0$. Together with $f_{aaaa}<0$ this proves the lemma.
\end{proof}

\begin{lemma}[The whole large-head chart; exact polynomial sign certificates]\label{lem:fqe-large-head-dual}
Every spectrum on the above chart satisfies
$\Lambda_\theta<T=2507333/1500000$.
\end{lemma}
\begin{proof}
Put $a_0=127/200$, $S=1-a_0=73/200$, $d=a_0-1/2=27/200$ and $\xi=(a-a_0)/d\in[-1,11/9]$.
By Lemma~\ref{lem:fqe-large-head-curvature}, the quadratic Hermite interpolant in $a$
at the single node $1/2$ and the double node $a_0$ is a global upper bound,
since its remainder has the sign of $f_{aaa}(\eta,z)(a-1/2)(a-a_0)^2\le0$:
\[
 f(a,z)\le Q=f_0(1-\xi^2)+f_l\xi^2+dD(\xi+\xi^2),\quad
 f_0=f(a_0,z),\ f_l=f(\tfrac12,z),\ D=f_a(a_0,z).
\]
Write $f=h(a,K)+P$, where $P$ is a polynomial in $(a,B,C)$ and does not depend on $K$.
Then only the three values $h(a_0,K)$, $h(\tfrac12,K)$, $h_a(a_0,K)$ in $Q$ depend on $K$.

\emph{$Q$ decreases in $K$ on $[.52,1]$.}
Let $R=(a_0/S)^q\in[12/5,5/2]$, $c=rL$ and $\lambda=dq/(a_0S)=42849/46355$.
Differentiating,
\[
 h_K(a_0,K)=\frac{-1}{c(R+K)},\quad h_K(\tfrac12,K)=\frac{-1}{c(1+K)},\quad
 h_{aK}(a_0,K)=\frac{qR}{c\,a_0S(R+K)^2},
\]
and therefore
\[
 -c(R+K)^2(1+K)\,Q_K=(1+K)\,(R+K-\lambda R\xi)+\xi^2\{(R+K)(R-1)-\lambda R(1+K)\}.
\]
For $-1\le\xi\le11/9$ both terms are nonnegative, the first strictly positive. The first is linear in $\xi$, positive at both ends:
$\frac{11}9\lambda<1.13$ and $K>.13R$. The brace increases in $R$ and decreases in $K$, and is
$\frac{17}5\cdot\frac75-\frac{24}5\lambda>0$ at $R=12/5$, $K=1$ (LHD-3).

\emph{A cubic minorant of $K$.}
Because $K^{(4)}>0$, cubic Hermite interpolation at the double nodes $3/20,17/20$ gives
\[
 K(z)\ge\kappa(z)=.5238033+.0142244z+.4204281z^2+.0384307z^3,
 \qquad .52<\kappa<1.                                      \tag{LH1}
\]
Each coefficient of $\kappa$ is a rational lower rounding of the interpolant's coefficient (LHD-3).
Since $.52<\kappa\le K\le1$, we have $Q(K)\le Q(\kappa)$.

\emph{Rational bounds for the logarithms.} Set
\[
 \rho=2.40792594,\quad u=2.45774282,\quad H_0=1.10489073,\quad H_l=1.32818582.
\]
They approximate $R$, $1/(rL)$, $h(a_0,3/4)$, $h(\tfrac12,3/4)$ respectively with error less than $2\cdot10^{-8}$.
For $v\in[.52,1]$ define
\[
 \widehat h_0=H_0-2u\frac{v-3/4}{2\rho+v+3/4},\quad
 \widehat h_l=H_l-2u\frac{v-3/4}{v+11/4},\quad
 \widehat h_a=-\frac{2qu}{1-(27/100)^2}\Bigl(\frac{\rho-v}{\rho+v}-\frac{27}{100}\Bigr).
\]
The identity $\ln((1+y)/(1-y))=2y+e_y$ has $|e_y|\le2|y|^3/(3(1-y^2))$.
Here $h(a_0,v)-h(a_0,\tfrac34)=-\frac1{rL}\ln\frac{R+v}{R+3/4}$ and
$h(\tfrac12,v)-h(\tfrac12,\tfrac34)=-\frac1{rL}\ln\frac{1+v}{7/4}$. The logarithms equal $\ln\frac{1+y}{1-y}$ with
$y=(v-\frac34)/(2R+v+\frac34)$, respectively $y=(v-\frac34)/(v+\frac{11}4)$, and $|y|<.0383$, respectively $|y|<.0704$.
Rounding the constants and the remainder give
\[
 |h(a_0,v)-\widehat h_0|<e_0=10^{-4},\quad
 |h(\tfrac12,v)-\widehat h_l|<e_l=6\cdot10^{-4},\quad
 |h_a(a_0,v)-\widehat h_a|<e_a=10^{-6}.                         \tag{LH2}
\]
These include the rounding errors; LHD-3 checks the scalar arithmetic.
On $[-1,11/9]$ we have $|1-\xi^2|\le1$ and $|d(\xi+\xi^2)|\le11/30<1$, so with
$P_0=P(a_0)$, $P_l=P(\tfrac12)$ and $P_a=\partial_aP(a_0)$,
\[
 Q(\kappa)\le F=(\widehat h_0+P_0)(1-\xi^2)+(\widehat h_l+P_l)\xi^2+d(\widehat h_a+P_a)(\xi+\xi^2)
   +e_0+e_a+e_l\xi^2,\qquad v=\kappa(z).
\]

\emph{The completed square.}
Put $E=(v+2\rho+3/4)(v+11/4)(v+\rho)>0$. Then $E\,(T-F)=c_0+c_1\xi+c_2\xi^2$, where after
substituting $v=\kappa(z)$ and the polynomials $B,C$ the $c_i$ are polynomials in $z$.
On $[0,1]$,
\[
 c_2\ge\tfrac1{10},\qquad 4c_0c_2-c_1^2\ge\tfrac3{100},
\]
because all Bernstein coefficients of $c_2$ (degree $13$) and of $4c_0c_2-c_1^2$ (degree $26$) are at least
these values (LHD-4). Thus the quadratic in $\xi$ is positive for every real $\xi$, so $f\le Q\le F<T$.
\end{proof}

\subsection{A strengthened cubic bound below purity $47/100$}
\label{sec:fqe-capped-cubic}
We now fix $q=1587/1000$, $r=q-1$, $L=\ln2$ and $x=p-1/4$.
\begin{lemma}[Capped cubic slack; analytic with explicit scalar arithmetic]
\label{lem:fqe-capped-cubic}
Every four-point probability vector with $p\le47/100$ satisfies
\[
H_q(\lambda)\le \frac{10}3-6p+\frac83t
-\frac{11}{10}(p-1/4)(1/2-p).
\tag{CG}\label{eq:fqe-capped-cubic}
\]
Equality holds only at the uniform-four spectrum.
\end{lemma}
\begin{proof}
In centred moments $y=\sum_i(\lambda_i-1/4)^3$, the assertion is
\[
2-H_q\ge \frac{171}{40}x-\frac83y-\frac{11}{10}x^2.
\]
We first prove it for $p\le21/50$, then use binary grouping for the rest.

\emph{The moment argument, $x\le17/100$.}
The claim follows from~\eqref{eq:fqe-moment-rectangle} of Lemma~\ref{lem:fqe-moment-rectangle} with $b=8/3$,
$c=171/40$, $\gamma=11/10$ and $\mu>0$; we apply that lemma with the CG column of
Table~\ref{tab:fqe-moment-rectangle}. Here $N=1+xV+x^2W$, $V=A+Bv+Cv^2$, $W=C/4+D_5v$, $v=y/x$, and
$A,B,C,D_5$ are those of Lemma~\ref{lem:fqe-near-uniform-moment}; the rectangle lemma does not require
its smaller purity domain. The third-moment bound gives $|v|<6/25$.
On $0\le x\le X=17/100$, $|v|\le6/25$, direct rational bounds give
\[
5/3<V<9/4,\quad -21/100<W<1,\quad 1\le N<3/2,\qquad
V^2-2W\ge\frac{1706}{625}-\frac9{10}|v|+5v^2>\frac83,
\]
and $|B+2Cv+XD_5|<213/100$. Then $(8/3-2X(9/4)(21/100))/(3/2)^2>1=m$, so $(\ln N)_{xx}<-1$, and
$m>(11/5)\rho=2\rho\gamma$; $C>144/100$ gives $\Phi_{vv}(X,v)\ge(2C-X(213/100)^2)/(\rho(3/2)^2)>2=\kappa$.
At $v_*=0$ put
$z=XA+X^2C/4$, use the eight-term lower sum
$\ln(1+z)\ge\sum_{j=1}^8(-1)^{j+1}z^j/j$, and use
$693147/10^6<L<693148/10^6$: then $\Phi(X,0)>1/250$ and $|\Phi_v(X,0)|<1/20$.
The lemma gives $\mu\ge1/250-(1/20)^2/4>0$. This proves the claim strictly
when $x>0$; $x=0$ is the uniform spectrum.

\emph{Two binary bounds from endpoint values.}
Write $h(u)=H_q(u,1-u)$ and $d_0=u(1-u)$.
The monotonicity of $A_q$ in Lemma~\ref{lem:fqe-binary-quartic} (proved there by Cauchy's formula) gives
\[
 h(u)\le4d_0-\frac{11}{40}d_0(1-4d_0)\quad(3/100\le u\le1/2),
\qquad
 h(u)\le4d_0-\frac38d_0(1-4d_0)\quad(1/12\le u\le1/2).
\tag{BQ}\label{eq:fqe-capped-binary}
\]
Only the two lower endpoints require checking. Explicit sufficient bounds are
\[
\begin{array}{c|c|c|c}
u_0&u_0^q>&(1-u_0)^q>&h(u_0)<\\\hline
3/100&19/5000&1191/1250&1093/10000\\
1/12&193/10000&871/1000&5713/20000
\end{array}
\]
The power bounds follow by raising to the thousandth power. The power sums are then at least
$S=4783/5000$ and $S=8903/10000$, and the entropy caps follow from
$-\ln(1-z)\le z+z^2/2+z^3/(3(1-z))$ with $z=1-S$; the two quartics exceed them (CG-2).

It follows that $4u(1-u)-h(u)>(11/10)d(1/4-d)$ whenever
$3/100\le u\le4/15$ and $3/100\le d\le\min\{2/25,d_0\}$.
For $u\le1/12$, use the first quartic and the fact that
$d(1/4-d)$ increases on $[0,1/8]$.
For $u\ge1/12$, use the second quartic: the concave function $d_0(1-4d_0)$
has its minimum at an endpoint of $11/144\le d_0\le44/225$, and
\[
\frac38\min\left\{\frac{11}{144}\frac{25}{36},
\frac{44}{225}\frac{49}{225}\right\}
>\frac{187}{12500}
\ge\frac{11}{10}d(1/4-d).
\]

\emph{The two pairs, $21/50\le p\le47/100$.}
Determinant monotonicity preserves $p,t$ and increases $H_q$, so it suffices
to use the two ordered charts in \eqref{eq:fqe-sliver-coordinates}.
Put $v=\delta^2$, $d=1/2-p=u-u^2-v/2\in[3/100,2/25]$.
Ordering gives $3/100<u<4/15$ on S and $3/100<u<3/32$ on B.
For S use $p\le1-3u+3u^2$; for B use
$p\le1/2-u+3u^2/2$, $u\le1/2$, and convexity of this quadratic on
$[3/32,1/2]$. Also on S,
\[
\frac{v}{(1-u)^2}\le\frac{2(u-u^2-3/100)}{(1-u)^2}<\frac23.
\]
For two conditional probability vectors, Jensen's inequality for the logarithm gives
\[
H_q((1-u)a,ub)\le h(u)+(1-w)H_q(a)+wH_q(b),\qquad
w=\frac{u^q}{(1-u)^q+u^q}.
\]
One conditional vector in each chart is uniform-two. We claim
\[
H_q(\lambda_S)\le1+h(u)-(1+2u)v,\qquad
H_q(\lambda_B)\le1+h(u)-(3-2u)v.
\tag{CG3}\label{eq:fqe-capped-pairs}
\]
For S, the conditional squared imbalance $z=v/(1-u)^2$ is less than $2/3$.
Its smaller probability is therefore greater than $1/12$; the second bound in
\eqref{eq:fqe-capped-binary} gives a deficit
$z+(3/32)z(1-z)\ge(33/32)z$.
Its weight times $z$ is $v/[(1-u)^2(1+\rho^q)]$ with $\rho=u/(1-u)<1$, and
$\rho^q\le\rho^{3/2}\le\rho(\rho+1/4)$ because $\sqrt\rho\le\rho+1/4$. Hence
$(1-u)^2(1+\rho^q)\le(1-u)^2+u^2+u(1-u)/4$, and the S inequality follows from
\[
 32(1+2u)\Bigl((1-u)^2+u^2+\frac{u(1-u)}4\Bigr)=32+8u(1-7u+14u^2)\le33 .
\]
Indeed $1-7u+14u^2=14(u-1/4)^2+1/8$ is at most $1$ for $u\le1/8$ and at most $11/32$ for
$1/8\le u\le4/15$, so $8u(1-7u+14u^2)\le\max\{1,11/15\}$.

For B the binary cubic bound gives deficit at least $z$ at order $q$.
Here $M=(1-u)^q+u^q\le1-u$, because $q\ge3/2$, $\sqrt u<1/3$, and
$(1-u)^{3/2}\le(1-u)(1-u/2)$.
Thus $u^{q-2}/M\ge u^{-2/5}/(1-u)\ge3-2u$ on $u\le3/32$, as $u^2((1-u)(3-2u))^5\le1$ there:
its logarithmic derivative has the sign of $24u^2-35u+6>0$, and at $u=3/32$ the inequality is
$(3/32)^2(1305/512)^5<1$.
This proves \eqref{eq:fqe-capped-pairs}.

In these coordinates the cubic bound is
$1+4u-4u^2-(1+2u)v$ on S and
$1+4u-4u^2-(3-2u)v$ on B, and the correction is $(11/10)d(1/4-d)$.
By~\eqref{eq:fqe-capped-pairs} the slack is at least $4u(1-u)-h(u)$, which exceeds the correction
by the binary estimate above, since $d\le u-u^2=d_0$. This completes the proof, with strictness.
\end{proof}

\begin{corollary}[The capped cluster region; analytic]
\label{cor:fqe-capped-cluster}
For three four-point spectra with $1\le P\le101/100$, $F\ge11/25$ and every
$p_j\le47/100$, one has $\mathcal E_{q_s}<5/3$.
No quadratic variance bound is required.
\end{corollary}
\begin{proof}
Put $\varepsilon=P-1$. Convexity gives
$U\le(47/100)^2+(7/25+\varepsilon)^2+1/16$.
The sum of the corrections in \eqref{eq:fqe-capped-cubic} is therefore at least
$(11/10)(33/2500+19\varepsilon/100-\varepsilon^2)$.
The cubic envelope gives the uncorrected bound
$3\mathcal E_{q_s}\le5+(1-20\varepsilon-2F)/9$.
Subtracting the correction leaves at most $5-89/75000<5$:
the maximum is at $F=11/25$, $\varepsilon=0$.
\end{proof}

\subsection{An analytic moment bound near the uniform spectrum}\label{sec:sup-nu-moment}
The following argument bounds the near-uniform cuts in the cluster sliver. It applies to all four-point
spectra, without a chart reduction or physical compatibility assumption. Its rectangle form,
Lemma~\ref{lem:fqe-moment-rectangle}, is shared with the bounded-head chart and the capped region.

\begin{lemma}[Near-uniform moment minorant; analytic]
\label{lem:fqe-near-uniform-moment}
Let $\lambda$ be a four-point probability vector. Put
$w_i=\lambda_i-1/4$, $x=\sum w_i^2\le1/25$, and $y=\sum w_i^3$.
For $q=q_s=1587/1000$, write $r=q-1$ and
\[
A=2qr,\quad B=\frac83qr(q-2),\quad
C=\frac83qr(q-2)(q-3),\quad D_5=\frac23(q-4)C.
\]
If $x>0$, set $v=y/x$, $V=A+Bv+Cv^2$, $W=C/4+D_5v$.
The notation $D_5$ distinguishes this Taylor coefficient from the
global third-moment sum $T$.
Then $|v|<3/25$, $V,W>0$, and
\[
2-H_q(\lambda)\ge
\frac{\ln(1+xV+x^2W)}{r\ln2}.
\tag{NU}\label{eq:fqe-near-uniform-moment}
\]
\end{lemma}
\begin{proof}
The sixth derivative of $z^q$ is positive on $z>0$. Taylor's formula
through degree five at $z=1/4$, summed over the four entries, therefore gives
\[
4^{q-1}\sum_i\lambda_i^q
\ge1+Ax+By+C\sum_iw_i^4+D_5xy.
\]
Here the trace-zero Newton identity is
$\sum w_i^5=5xy/6$. The estimate extends to zero entries by continuity.
The elementary square identity
\[
\sum_i\left(w_i^2-\frac x4-\frac yx w_i\right)^2
=\sum_iw_i^4-\frac{x^2}4-\frac{y^2}x\ge0
\]
proves the required moment lower bound, since $C>0$.
Lemma~\ref{lem:fqe-third-moment} gives $|y|\le x^{3/2}/\sqrt3$,
hence $|v|\le1/\sqrt{75}<3/25$. The logarithm is increasing,
and its division by $r\ln2$ gives \eqref{eq:fqe-near-uniform-moment}.
For completeness, the rational bounds at $q_s$
\[
 A\ge\tfrac95,\quad |B|\le\tfrac{21}{20},\quad
 C\ge\tfrac75,\quad -\tfrac{12}5\le D_5<0
\]
give $V\ge\tfrac{837}{500}$ and
$W\ge\tfrac7{20}-\tfrac{12}5\cdot\tfrac3{25}
=\tfrac{31}{500}>0$ for $|v|<3/25$.
This proves the asserted positivity without relying on a later case split.
\end{proof}

The bound~\eqref{eq:fqe-near-uniform-moment} is used three times in the same way: on a rectangle
in $(x,v)$ it is concave enough in $x$ to reduce to the far edge $x=X$, and convex enough in $v$ there
to reduce to one point.
\begin{lemma}[Moment rectangle; analytic]
\label{lem:fqe-moment-rectangle}
Keep $q=q_s$ and the notation of Lemma~\ref{lem:fqe-near-uniform-moment}, put $L=\ln2$,
$N(x,v)=1+xV(v)+x^2W(v)$ and $\rho=21/50=(3/5)(7/10)$, so that $rL<\rho$. For constants $b,c$ and
$\gamma\ge0$ let
\[
\Phi(x,v)=\frac{\ln N(x,v)}{rLx}+bv-c+\gamma x\qquad(x>0).
\]
Let $X>0$ and $v_*\in\mathcal I=[v_{\min},v_{\max}]$, and suppose that on $[0,X]\times\mathcal I$
\[
1\le N\le N_+,\quad V\le V_+,\quad W\ge-W_-,\quad V^2-2W\ge Q,\quad
|B+2Cv+XD_5|\le S,
\tag{MR1}\label{eq:fqe-rect-bounds}
\]
with $W_-\ge0$, and that the constants $m,\kappa,f_0,\eta$ satisfy
\[
Q-2XV_+W_-\ge mN_+^2,\quad m>2\rho\gamma,\quad
2C-XS^2\ge\kappa\rho N_+^2,\quad \kappa>0,
\tag{MR2}\label{eq:fqe-rect-curvature}
\]
\[
\Phi(X,v_*)>f_0,\qquad|\Phi_v(X,v_*)|<\eta.
\tag{MR3}\label{eq:fqe-rect-point}
\]
Then $\Phi$ is strictly decreasing in $x$ and $\Phi>\mu:=f_0-\eta^2/(2\kappa)$ on
$(0,X]\times\mathcal I$. Consequently every four-point probability vector with $0<x\le X$ and
$v=y/x\in\mathcal I$ satisfies
\[
2-H_q(\lambda)-cx+by+\gamma x^2>\mu x.
\tag{MR}\label{eq:fqe-moment-rectangle}
\]
\end{lemma}
\begin{proof}
Fix $v\in\mathcal I$ and put $\ell(x)=\ln N$. Then
\[
\ell''(x)=\frac{2W-V^2-2xW(V+xW)}{N^2}.
\]
Since $N\ge1$, $V+xW\ge0$ for $x>0$. The last term is therefore nonpositive when $W\ge0$; when $W<0$ it is
at most $2XW_-V_+$, as $V+xW\le V$. The numerator is at most $-Q+2XV_+W_-\le-mN_+^2<0$, so
$\ell''\le-m$. Put $\mathcal F(x)=rLx\Phi=\ell(x)-rLx(c-bv-\gamma x)$. Then $\mathcal F(0)=0$ and
$\mathcal F''=\ell''+2rL\gamma\le-m+2\rho\gamma<0$, so $\mathcal F(x)/x$, and with it $\Phi$, is strictly
decreasing in $x$ on $(0,X]$.

At $x=X$, since $1\le N\le N_+$, $C>0$ and $2C-XS^2>0$,
\[
\Phi_{vv}=\frac{2CN-X(B+2Cv+XD_5)^2}{rLN^2}\ge\frac{2C-XS^2}{\rho N_+^2}\ge\kappa .
\]
Completing the tangent square at $v_*$ gives
$\Phi(X,v)\ge\Phi(X,v_*)-\Phi_v(X,v_*)^2/(2\kappa)>\mu$, and so $\Phi(x,v)\ge\Phi(X,v)>\mu$.

The Taylor and projection inequality in the proof of Lemma~\ref{lem:fqe-near-uniform-moment} holds for
every four-point vector; $x\le1/25$ entered only the bound on $|v|$. As $N\ge1>0$, it gives
$2-H_q(\lambda)\ge\ln N/(rL)=x\Phi+cx-by-\gamma x^2$, which is~\eqref{eq:fqe-moment-rectangle}.
\end{proof}

In each application~\eqref{eq:fqe-rect-point} is checked at one point by rational arithmetic. With
$z=XV(v_*)+X^2W(v_*)\in(0,1)$ and an even $k$, $\ln(1+z)\ge\sum_{j=1}^k(-1)^{j+1}z^j/j$; if this sum is
positive, dividing it by $rL_+X$ bounds $\ln N/(rLX)$ below whenever $L<L_+$. The slope
$\Phi_v(X,v_*)=(B+2Cv_*+XD_5)/(rLN(X,v_*))+b$ is monotone in $L$, so it is bounded by substituting the two
rational endpoints of an enclosure of $L$. Table~\ref{tab:fqe-moment-rectangle} lists the three
instances; each application derives~\eqref{eq:fqe-rect-bounds} from its own coefficient bounds.
\begin{table}[htbp]
\centering\small
\begin{tabular}{@{}lccc@{}}
\toprule
 & NU & SHD (large tail) & CG (moment part)\\
\midrule
target & $\Lambda_\theta(c_2)-\Lambda_\theta(\lambda)\ge dx$ & $\Lambda_\theta(\lambda)\le\Lambda_\theta((1/4)^4)$ & \eqref{eq:fqe-capped-cubic}, $p\le21/50$\\
$X$; $\mathcal I$; $v_*$ & $1/25$; $[-3/25,3/25]$; $1/20$ & $3/32$; $[0,9/50]$; $27/200$ & $17/100$; $[-6/25,6/25]$; $0$\\
$b$; $c$; $\gamma$ & $12\beta$; $\alpha+8\beta-\gamma/2$; $\gamma$ (HT) & $9/5$; $35/8$; $159/100$ & $8/3$; $171/40$; $11/10$\\
$N_+$; $V_+$; $W_-$ & $13/12$; $41/20$; $0$ & $6/5$; $19/10$; $3/50$ & $3/2$; $9/4$; $21/100$\\
$Q$; $m$ & $(837/500)^2-4/3$; $6/5$ & $8/3$; $7/4$ & $8/3$; $1$\\
$S$; $\kappa$ & $753/500$; $5$ & $19/10$; $4$ & $213/100$; $2$\\
$f_0$; $\eta$ & $21/2500$; $1/50$ & $1/800$; $1/100$ & $1/250$; $1/20$\\
$k$; $(L_-,L_+)$ & $4$; $(6931/10^4,1733/2500)$ & $8$; $(693147,693148)/10^6$ & $8$; $(693147,693148)/10^6$\\
$\mu$ & $>d=1/125$ & $=1/800-1/80000$ & $=1/250-1/1600$\\
\bottomrule
\end{tabular}
\caption{The three instances of Lemma~\ref{lem:fqe-moment-rectangle}, all at $q_s$ with $\rho=21/50$.
For NU and SHD, $(\alpha,\beta,\gamma)$ is the multiplier of the lemma and the target is the dual
difference $2-H_q-cx+by+\gamma x^2$; for CG the target is the centred form of~\eqref{eq:fqe-capped-cubic}.
The hypotheses are checked by NU-3, SHD-4 and CG-1 through one shared function.}
\label{tab:fqe-moment-rectangle}
\end{table}

\subsection{An analytic uniform-cut dual bound}\label{sec:sup-nu-dual}
\begin{lemma}[Analytic uniform-cut dual bound; NU]
\label{lem:fqe-near-uniform-dual}
At the order $q_s=1587/1000$, with the HT multiplier row, every four-point spectrum
with $p\le29/100$ satisfies
\[
\Lambda_\theta(c_2)-\Lambda_\theta(\lambda)
\ge d\,(p-1/4),\qquad c_2=(1/4)^4.
\tag{ND}\label{eq:fqe-near-uniform-dual}
\]
HT labels the cluster-sliver multiplier row. The order and multipliers are exactly those of the row HT of
Table~\ref{tab:fqe-dual-co}, and $d=1/125$. Equality occurs only at $c_2$.
\end{lemma}
The proof uses two further constants, checked by NU-3: $f_0=21/2500$ and $\eta=1/50$.
\begin{proof}
Put $b=12\beta$, $c=\alpha+8\beta-\gamma/2$ and $x=p-1/4$.
The uniform moment identities give
\[
\Lambda_\theta(c_2)-\Lambda_\theta(\lambda)
=2-H_q(\lambda)-cx+by+\gamma x^2 .
\tag{UD}\label{eq:fqe-uniform-dual-difference}
\]
By Lemma~\ref{lem:fqe-near-uniform-moment}, $x\le X=1/25$ and $|v|<D=3/25$ when $x>0$. We apply
Lemma~\ref{lem:fqe-moment-rectangle} with the NU column of Table~\ref{tab:fqe-moment-rectangle}.
At $q=q_s$ the coefficients satisfy
\[
\frac95\le A\le\frac{19}{10},\quad
-\frac{21}{20}\le B<0,\quad
\frac75\le C\le\frac32,\quad -\frac{12}5\le D_5<0.
\]
These are rational comparisons at the single order $q_s$ (NU-2). Consequently
\[
\frac{837}{500}\le V\le\frac{41}{20},\quad
0<W<\frac23,\quad 1\le N<\frac{13}{12},\quad
|B+2Cv+XD_5|\le\frac{21}{20}+3D+\frac{12}5X=\frac{753}{500},
\]
and $V^2-2W\ge Q=(837/500)^2-4/3$. This is~\eqref{eq:fqe-rect-bounds} with $W_-=0$.
For~\eqref{eq:fqe-rect-curvature}, $Q>(6/5)(13/12)^2$, $2\rho\gamma\le(21/25)(6/5)<6/5$ as
$0\le\gamma\le6/5$, and $C\ge7/5$ gives
$14/5-(1/25)(753/500)^2>5\rho(13/12)^2$.
For~\eqref{eq:fqe-rect-point} take $k=4$ and
\[
\frac{6931}{10000}<L<\frac{1733}{2500}.
\]
The latter follows by summing ten terms of
$L=2\sum_{k\ge0}((2k+1)3^{2k+1})^{-1}$ and bounding the geometric tail.
Direct rational substitution of the multiplier row gives $\Phi(X,v_*)>f_0$ and
$|\Phi_v(X,v_*)|<\eta$ at $v_*=1/20$. Since $\mu=f_0-\eta^2/10>d$,
\eqref{eq:fqe-moment-rectangle} and~\eqref{eq:fqe-uniform-dual-difference} prove
\eqref{eq:fqe-near-uniform-dual}.
At $x=0$ the spectrum is $c_2$; otherwise $d>0$ makes it strict.
\end{proof}

\subsection{An analytic bound on the distinguished cluster cut}\label{sec:sup-sliver}
We now treat the distinguished cut in the cluster sliver. Elementary entropy estimates reduce the
required bound to finitely many polynomial signs, with no interval covering or derivative grid.
Write $q_s=1587/1000$ and let $\theta_s$ be the
HT row of Table~\ref{tab:fqe-dual-co}. Set
\[
\kappa=4\sqrt3/5,\quad X=\tfrac12+\varepsilon-p,\quad
F_c=-1+8\varepsilon+9p-12t.
\]
\begin{lemma}[Analytic cluster sliver; SL]
\label{lem:fqe-cluster-sliver}
At the order $q_s=1587/1000$,
on either ordered chart, suppose
\[
p\ge47/100,\quad X\ge0,\quad 0\le\varepsilon\le1/100,\quad
[F_c-\kappa X,F_c+\kappa X]\cap[11/25,1/2]\ne\varnothing.
\]
For either sign, put
\[
G_{\pm}=\Lambda_{\theta_s}(c_1)-\Lambda_{\theta_s}(\lambda;q_s)
+\ell_{\theta_s}(\varepsilon,F_c\pm\kappa X)
-\ell_{\theta_s}(0,1/2).
\]
Then $G_{\pm}\ge0$, with equality only at
$\lambda=c_1$, $\varepsilon=0$.
\end{lemma}
\begin{proof}
The proof first bounds the two chart coordinates, then replaces the entropy by an explicit
minorant of its deficit. The resulting four polynomial bounds are reduced to one-variable
endpoint signs by removing $\varepsilon$ and using concavity in $v$.

\emph{Coordinates and domains.}
Put $v=\delta^2$. In the two charts use
\[
\begin{aligned}
\lambda_S&=((1-u+\delta)/2,(1-u-\delta)/2,u/2,u/2),\\
\lambda_B&=((1-u)/2,(1-u)/2,(u+\delta)/2,(u-\delta)/2),\\
p&=\tfrac12-u+u^2+v/2,\\
t_S&=\tfrac14-\tfrac34u+\tfrac34u^2+\tfrac34(1-u)v,\qquad
t_B=\tfrac14-\tfrac34u+\tfrac34u^2+\tfrac34uv.
\end{aligned}
\tag{SV}\label{eq:fqe-sliver-coordinates}
\]
In both charts $0\le u<1/16$, $0\le v<1/20$.
Here are explicit domain reductions. Ordering in $S$ gives $u\le1/2$,
$v\le(1-2u)^2$ and $p\le1-3u+3u^2$; hence $u<1/4$.
Since $X\le1/25$ and $\kappa<7/5$, the window condition gives
$(9/2)(1-2u)v\le49/250$.
Successively this gives $v<9/100$, $u<1/10$, $v<11/200$,
$u<1/16$ and $v<1/20$, using $u-u^2\le3/100+v/2$ each time
(verifier-checked, SL-3).
In $B$, ordering gives $u\le1/2$, $v\le u^2$ and
$p\le1/2-u+3u^2/2$; thus already $u<1/30$.

\emph{Entropy estimates.}
Set $(K,c_s)=(7801/2000,5/2)$.
On $0\le u\le1/16$, the binary entropy obeys
\[
H_{q_s}(u,1-u)\le Ku-c_su^{8/5}+\frac{12}5u^2.
\tag{SE}\label{eq:fqe-sliver-binary}
\]
Indeed $(1-u)^q\ge1-qu+q(q-1)u^2/2$. If
$z=qu-q(q-1)u^2/2-u^q$, then $0\le z\le qu$ and
$\ln(1-z)\ge-z-z^2/(2(1-z))$ give the sharper bound
\[
H_q(u,1-u)\le\frac q{(q-1)\ln2}u-\frac{u^q}{(q-1)\ln2}
+\frac{q^2/(2(1-q/16))-q(q-1)/2}{(q-1)\ln2}u^2.
\]
The last coefficient is less than $12/5$. The linear coefficient is
less than $K$, while $16^{8/5-q_s}>51/50$ exceeds $c_s(q_s-1)\ln2$.
These are rational comparisons, integer-power comparisons and the
atanh bounds $693147/10^6<\ln2<693148/10^6$.

Separating the head and tail pairs in the power sum gives
\[
H_{q_s}(\lambda)\le1+Ku-c_su^{8/5}+\frac{12}5u^2
-\begin{cases}
\frac{113}{100}v+\frac32uv-\frac38v^2,&S,\\
3v,&B.
\end{cases}
\tag{SH}\label{eq:fqe-sliver-head-tail}
\]
To see the head estimate, let $M=(1-u)^q+u^q$.
The even Taylor expansion of the head power gives an increase of at least
$(q(q-1)/2)(1-u)^{q-2}v$ over $M$. The factor
\[
\frac{(1-u)^{q-2}}M
=\frac{(1-u)^{-2}}{1+(u/(1-u))^q}\ge1+\frac32u
\]
follows from $(u/(1-u))^q\le7u/25$ and $(1-u)^{-2}\ge1+2u$.
The coefficient $\frac{q(q-1)}2(1-u)^{q-2}/M$ of $v$ in the resulting relative increase is at most
$\frac{q(q-1)}2(1-u)^{-2}\le\frac{q(q-1)}2\bigl(\frac{16}{15}\bigr)^2<27/50$. Applying $\ln(1+z)\ge z-z^2/2$ proves the $S$ estimate,
since $q/(2\ln2)>113/100$ and
$(27/50)^2/(2(q-1)\ln2)<3/8$.
In $B$ the same argument acts on the tail: its factor is
$u^{q-2}/M\ge3$, and the logarithm increment is at most $1/100$
when $v\le u^2$. The resulting coefficient exceeds $3$.
All estimates extend to $u=0$ by continuity.

\emph{Polynomial lower bounds.}
Inserting \eqref{eq:fqe-sliver-head-tail} in $G_{\pm}$
gives the following lower bound, throughout
$u\le1/16$, $v\le1/20$, $\varepsilon\le1/100$:
\[
P(u,v)+\varepsilon(A+Bv),\qquad
P=au+bv+cu^2+duv+fv^2+gu^2v+c_su^{8/5}.
\tag{SP}\label{eq:fqe-sliver-polynomial}
\]
The coefficients are exact rationals. In each row label, the first letter is the chart ($S$ or $B$)
and the second is the sign in $G_\pm$:
\[
\begin{array}{c|rrrrrr|rr}
&a&b&c&d&f&g&A&B\\\hline
S-&2/125&7/500&-571/100&-33/10&-281/100&6&185/100&8\\
S+&2/125&7/500&-6&69/50&-103/20&0&147/100&16\\
B-&2/125&19/100&-6&3&-5&0&1/2&0\\
B+&2/125&19/100&-571/100&-3/2&-5/2&0&1/2&0
\end{array}
\tag{ST}\label{eq:fqe-sliver-coefficients}
\]
For clarity, this table involves polynomial algebra only. Expand the
formula for $G$ using \eqref{eq:fqe-sliver-coordinates}, enclose $\kappa$
between $13856/10000$ and $13857/10000$, and bound negative terms of
higher degree by $u\le1/16$, $v\le1/20$, $\varepsilon\le1/100$.
At $\varepsilon=0$, charge them to $u^2$, $uv$ or $v^2$; in the
coefficient of $\varepsilon$, retain positive terms proportional to $v$.
The difference from every row in \eqref{eq:fqe-sliver-coefficients}
then has nonnegative coefficients (verifier-checked, SL-2). No transcendental sign is left.

\emph{Removing $\varepsilon$ (verifier-checked, SL-3).}
In $S$ the window and $X\ge0$ imply
\[
v\le\min\{2(u-u^2)+2\varepsilon,\ V(u)+21\varepsilon/10\},\qquad
V(u)=\frac{3/50+(7/5)(u-u^2)}{26/5-9u}.
\]
Put $v_0=\min\{v,2(u-u^2),V(u)\}$. Then $v-v_0\le21\varepsilon/10$.
From the two $S$ rows, $P_v\ge-C_0+2fv$, where
$C_0=\max\{0,-b+\max(0,-d)/16\}$. Direct substitution gives
\[
A-\tfrac{21}{10}C_0+\tfrac1{20}\min\{0,B+\tfrac{21}5f\}>9/10.
\]
Thus $G_\pm\ge P(u,v_0)+9\varepsilon/10$. Hence it remains to prove $P\ge0$ with
$\varepsilon=0$, $v\le\min\{2(u-u^2),V(u)\}$ and $u\le1/16$.
In chart $B$ simply discard the positive $\varepsilon/2$ term;
there $u<1/30$ and $v\le u^2$.

\emph{One-variable signs.}
The function $P$ is concave in $v$, so its minimum is at an endpoint.
At $v=0$ the claim follows from $u^{8/5}\ge3u^2$ on $0\le u\le1/16$. In chart $B$,
$b+du+fv\ge3/40$, so this also proves the $v=u^2$ case.

For the $S$ endpoint $v=2(u-u^2)$, that endpoint is the smaller one
only when $u<1/140$. Dividing $P$ by $u>0$ and dropping positive terms gives
\[
P/u\ge a+2b+c_su^{3/5}-Du,\quad
D=2b-c-2\min\{d,(139/140)d\}-4f.
\]
This is a concave function of $u$. It is positive at both $0$ and $1/140$;
for the latter check use $(1/140)^{3/5}>51/1000$.

Finally $v=V(u)$ is the smaller endpoint only for $u>1/150$.
Use the following quadratic minorant of $u^{8/5}$:
for $Z=23/40$, $U=Z^5$, $z=2/5$, $u_0=z^5$, put
\[
Q_{Z,z}(u)=z^8+\frac85z^3(u-u_0)
+\frac{Z^8-z^8-(8/5)z^3(U-u_0)}{(U-u_0)^2}(u-u_0)^2.
\tag{SQ}\label{eq:fqe-sliver-quadratic}
\]
Since the third derivative of $u^{8/5}$ is negative,
Hermite's remainder gives $u^{8/5}\ge Q_{Z,z}(u)$ for $0\le u\le U$.
After replacing $u^{8/5}$ by this minorant, multiply $P(u,V(u))$
by $(26/5-9u)^2$. The resulting polynomial $R$ has degree at most five.
For $L=1/150$, $U=Z^5$, write
\[
R(L+(U-L)x)=\sum_{i=0}^n B_i\binom ni x^i(1-x)^{n-i}.
\]
Every $B_i$ is positive. Explicit integer lower bounds on $10^6 B_i$ are
\[
\begin{array}{c|r|rrrrrr}
&n&\multicolumn{6}{c}{\text{coefficient lower bounds}}\\\hline
S-&5&793&27467&39030&38325&27813&9606\\
S+&4&238&51693&86240&106860&116069&
\end{array}
\tag{SB}\label{eq:fqe-sliver-bernstein}
\]
These are two polynomial expansions on single intervals, with no
subdivision. They prove the last endpoint signs. Together with the
strict $v=0$ and $\varepsilon$ bounds, this proves the asserted equality case.
\end{proof}

\subsection*{Further results}
\noindent The proof of Corollary~\ref{cor:fqe-renyi-gw} in the paper uses none of the results of
Sections~\ref{sec:fqe-renyi-above-one}--\ref{sec:fqe-renyi-filter-limit}. They give an independent near-one defect estimate
with its quantitative form, the sharp purity-variance profile on the critical locus with an analytic proof of (QV) on critical
states (used for the equality case of (QK) at $P=1$), and a shorter proof of the envelope for states with a rank-two cut.
\subsection{An independent near-one defect estimate}\label{sec:fqe-renyi-above-one}

Use $H_q$, $h_q$ and $\mathcal E_q$ from Section~\ref{sec:fqe-spectral}, and set
$m_q=2^{-q}+3\,6^{-q}$. Entropies are in bits throughout this section; the proof below uses
$H_q^{\mathrm{nat}}=H_q\ln2$ when working in nats.
The Tsallis result alone does not give an upper R\'enyi bound when $q>1$.
The anchor-order argument proves the full HS--cluster envelope of Corollary~\ref{cor:fqe-renyi-gw} for all $q>1$.
Here we give an independent near-one proof using the nonnegative Hermite interpolation error to control
the logarithm's curvature; it also yields the quantitative defect estimate~\eqref{eq:fqe-renyi-defect}.

\begin{theorem}[R\'enyi interval above one]
\label{thm:fqe-renyi-above-one}
For every pure four-qubit state and every $1<q\le5/4$,
\[
\boxed{\quad\mathcal E_q(\psi)\le\frac{\log_2 m_q}{1-q}.\quad}
\]
Equality holds exactly on the two $M_4$, $\overline{M_4}$ LU orbits.
\end{theorem}

The argument rests on a scalar slack estimate. Set $p=1/6$, $s=1/2$, and
\[
f_q(x)=\frac{x^q-x}{1-q},\qquad f_1(x)=-x\ln x,\qquad f_q(0)=0.
\]
Let $\ell_q$ be the secant of $f_q$ through $p,s$, and $\widehat h_q$ its cubic
Hermite interpolant there, as in Corollary~\ref{cor:fqe-tsallis}.

\begin{lemma}[Scalar slack control]
\label{lem:fqe-renyi-slack}
For $1\le q\le5/4$ and $0\le x\le1$,
\[
 \bigl(f_q(x)-\ell_q(x)\bigr)^2
 \le\frac{141}{100}\bigl(\widehat h_q(x)-f_q(x)\bigr).
 \tag{RA}\label{eq:fqe-renyi-scalar}
\]
Consequently, for any four-component probability vector $\lambda$, put
$S_q=\sum_i f_q(\lambda_i)$, $S_q^*=f_q(s)+3f_q(p)$,
and $D_q=\sum_i(\widehat h_q-f_q)(\lambda_i)$. Then
\[
 (S_q-S_q^*)^2\le\frac{141}{25}D_q.
 \tag{RB}\label{eq:fqe-renyi-sum}
\]
\end{lemma}
The proof (Section~\ref{sec:sup-renyi-slack}; check RA-1) puts
$G=\widehat h_q-f_q-\frac{100}{141}(f_q-\ell_q)^2$, which vanishes with its derivative at $p$ and $s$, and shows
$G^{(4)}>0$ on $(0,1]$ from an elementary bound $\Phi_q<141/100$ on the factor its fourth derivative contains;
the Hermite remainder then gives $G\ge0$, and Cauchy--Schwarz gives the summed bound.

\begin{proof}[Proof of Theorem~\ref{thm:fqe-renyi-above-one}]
Put $r=q-1$, $L=4^{-r}$, $M=\sum_i\lambda_i^q=1-rS_q$,
and $m=m_q=1-rS_q^*$. Convexity of $x^q$ gives $M\ge L$ and $m\ge L$.
For $u>0$, integration of $(t-1)/t$ gives
\[
u-1-\ln u\le\frac{(u-1)^2}{2\min(u,1)}.
\]
Apply this with $u=M/m$; since $\min(u,1)\ge L/m$, it follows that
\[
H_q^{\mathrm{nat}}(\lambda)-H_q^{\mathrm{nat}}(s,p,p,p)
\le\frac{S_q-S_q^*}{m}+
\frac{r(S_q-S_q^*)^2}{2mL}.
\]
By \eqref{eq:fqe-renyi-sum} and $S_q=\sum_i \widehat h_q(\lambda_i)-D_q$, this is at most
\[
\frac{\sum_i \widehat h_q(\lambda_i)-S_q^*}{m}
-\frac{1-141r/(50L)}{m}D_q.
\tag{RD}\label{eq:fqe-renyi-cut}
\]
For $r\le1/4$, the last coefficient is positive, because
\[
\frac{141r}{50L}\le\frac{141}{200}\sqrt2<1
\qquad\left(\left(\tfrac{141}{200}\sqrt2\right)^2=\tfrac{19881}{20000}\right).
\]
Write $a_q,b_q$ for the cubic and quadratic coefficients of $\widehat h_q$ and
$\alpha_q=-b_q-17a_q/12$. Corollary~\ref{cor:fqe-tsallis} proves
$a_q,\alpha_q>0$. Sum \eqref{eq:fqe-renyi-cut} over the three cuts and
use $T=(17P-12-F)/12$. With $\mathcal D_q=\sum_jD_q(\rho_{1j})$ and
$\kappa=1-141\sqrt2/200>0$, the result is the defect estimate
\[
3m_q\ln2\bigl(\mathcal E_q(M_4)-\mathcal E_q(\psi)\bigr)
\ge\alpha_q(P-1)+\frac{a_q}{12}F+\kappa\mathcal D_q\ge0.
\tag{RE}\label{eq:fqe-renyi-defect}
\]
Equality forces $P=1,F=0$, so Proposition~\ref{prop:fqe-moment-zeros} gives the two HS
orbits; both attain the bound.
\end{proof}

\subsection{Proof of Lemma~\ref{lem:fqe-renyi-slack}}\label{sec:sup-renyi-slack}
Notation is that of Section~\ref{sec:fqe-renyi-above-one}: $p=1/6$, $s=1/2$,
$f_q(x)=(x^q-x)/(1-q)$, $\ell_q$ the secant of $f_q$ through $p,s$, and $\widehat h_q$ its cubic Hermite
interpolant there.

The proof rests on an elementary bound for the function $\Phi_q$ that appears in the fourth derivative.

\begin{lemma}[An elementary fourth-derivative bound; proved, RA-1]
\label{lem:fqe-slack-fourth-derivative}
Write $\ell_q(x)=\mu_qx+\nu_q$ and, for $1<q\le5/4$, put
\[
 A(q)=\frac{2(q+1)}{3-q}\left(\frac1{q-1}-\mu_q\right),\qquad
 D(q)=\frac{4(2q-1)(3-2q)}{(q-1)(3-q)(2-q)},
\]
and $\Phi_q(x)=2\nu_q+A(q)x-D(q)x^q$.
Define $\Phi_1$ by continuity. Then $\Phi_q(x)<141/100$ for
$1\le q\le5/4$ and $0\le x\le1$.
\end{lemma}
\begin{proof}
Set $r=q-1$, $p=1/6$, $s=1/2$, and $\delta=r(2-r)(1-r)$.
The function $\Phi_q$ is concave in $x$. Its derivative at $p$ is
$p^rN_p/\delta$, where
\[
 N_p=(2+r)(1-r)(3\cdot3^r-1)-4(1+r)(1-4r^2).
\]
Since $\ln3>1$, $3^r\ge1+r$, and substitution gives
$N_p\ge11r^2+13r^3\ge0$. Thus its maximum on $[0,1]$ is attained
in $[p,1]$.

We show that $\Phi_q(x)$ increases with $q$ on this latter interval.
Three endpoint derivatives suffice. Define
\[
 \begin{split}
 N_0&=6(1-r)3^r-6+5r+7r^2,\\
 N_1&=(2+r)(1-r)(3\cdot2^{-r}-6^{-r})-4(1+r)(1-4r^2),\\
 J_0&=(N_0'-(\ln6)N_0)\delta-N_0\delta',\\
 J_p&=(N_p'-(\ln6)N_p)\delta-N_p\delta',\qquad
 J_1=N_1'\delta-N_1\delta'.
 \end{split}
\]
Primes here mean $r$ derivatives. Direct substitution into $\Phi_q$ gives
\[
 \partial_q\Phi_q(p)=\frac{p^rJ_0}{3\delta^2},\qquad
 \partial_q\partial_x\Phi_q(p)=\frac{p^rJ_p}{\delta^2},\qquad
 \partial_q\partial_x\Phi_q(1)=\frac{J_1}{\delta^2}.
\]
All three $J$ are positive for $0<r\le1/4$. Here are coarse bounds
that permit this assertion to be checked with small rational numbers.
Write $J=\sum_{k=2}^6c_kr^k+R$; the constant and linear coefficients
are zero. The five coefficients have the following lower bounds,
and $|R|\le Mr^7$:
\[
\begin{array}{c|rrrrr|r}
 &c_2&c_3&c_4&c_5&c_6&M\\\hline
J_0&4&-6&15&-14&0&3\\
J_p&27&-2&-33&52&-29&3\\
J_1&4&86&-75&2&0&14\\
\end{array}
\]
For completeness, these bounds require only
$69/100<\ln2<7/10$ and $109/100<\ln3<11/10$.
Both follow from the first five terms of
$\ln u=2\sum_{j\ge0}z^{2j+1}/(2j+1)$, $z=(u-1)/(u+1)$,
with tail at most $2z^{11}/[11(1-z^2)]$.
Expand the displayed $N$ and $J$ through degree six and substitute these
two logarithm intervals to get the five columns. To check the remainder,
each $J$ is a polynomial of degree at most six plus terms
$Q(r)e^{\alpha r}$ with $\deg Q\le5$.
For $Q(r)=\sum_jQ_jr^j$, its tail after degree six has absolute value at most
\[
 r^7 e^{|\alpha|/4}\sum_j
       \frac{|Q_j|\,|\alpha|^{7-j}}{(7-j)!}.
\]
Use $3^{1/4}<4/3$, $6^{1/4}<8/5$ and $2^{1/4}<6/5$ for the
three possible exponential factors. The resulting sums are respectively
less than $3,3,14$, giving the last column. In particular the table yields
$J_0>2r^2$, $J_p>24r^2$ and $J_1>3r^2$ on $0<r\le1/4$;
for the last row use $86r-75r^2\ge0$.

Now let $W(x)=\partial_q\partial_x\Phi_q(x)$ and
$k(q)=q(q-1)D(q)>0$. Differentiating $\partial_x^2\Phi_q=-k(q)x^{q-2}$ gives
\[
 W'(x)=-k(q)x^{q-2}\bigl(\partial_q\ln k(q)+\ln x\bigr).
\]
The bracket increases with $x$, so $W$ has no interior minimum.
Its endpoint values at $p,1$ are positive by the table. Thus $W\ge0$,
and $\partial_q\Phi_q(x)\ge\partial_q\Phi_q(p)>0$ on $[p,1]$.
It remains only to bound $\Phi_{5/4}$.

Put $a=6^{-1/4}$, $b=2^{-1/4}$, and $t=x^{1/4}$. Then
\[
 \Phi_{5/4}(x)=2(b-a)+\frac{36}{7}(3b-a)t^4-\frac{64}{7}t^5
 \le2(b-a)+\frac{36\cdot9^4}{35\cdot20^4}(3b-a)^5,
\]
by maximising the last two terms over $t\ge0$; their maximum is at
$t=9(3b-a)/20$. Integer fourth powers give $a>319/500$ and $b<841/1000$.
Consequently
\[
 \Phi_{5/4}(x)
 <\frac{203}{500}+\frac{36\cdot9^4}{35\cdot20^4}
                         \left(\frac{377}{200}\right)^5
 <\frac{141}{100}.
\]
Concavity in $x$ and monotonicity in $q$ give the assertion everywhere;
the endpoint $q=1$ follows by continuity.
\end{proof}

\begin{proof}[Proof of Lemma~\ref{lem:fqe-renyi-slack}]
Let $g=f_q-\ell_q$ and
$G=\widehat h_q-f_q-(100/141)g^2$. Both $G$ and $G'$ vanish at $p,s$.
Since $f_q^{(4)}=-q(2-q)(3-q)x^{q-4}<0$, differentiation gives
\[
 G^{(4)}(x)=-f_q^{(4)}(x)\left(1-\frac{100}{141}\Phi_q(x)\right)>0
 \qquad(x>0).
\]
The cubic Hermite remainder, with repeated nodes $p,p,s,s$, gives
$G(x)=G^{(4)}(\xi)(x-p)^2(x-s)^2/24\ge0$ for $x>0$;
continuity gives the result at zero. No derivative at zero is needed.
Finally $\sum_i\ell_q(\lambda_i)=S_q^*$, since $\ell_q$ is affine and
$\sum_i\lambda_i=s+3p=1$. Cauchy--Schwarz gives the summed bound.
\end{proof}

\subsection{Critical-locus variance refinements}\label{sec:fqe-renyi-compatibility}
The universal bound (QK) suffices for the anchor argument. We now give its sharper
critical-locus refinements: the fixed-variance profile on $P=1$, used to classify equality
in (QK), and the analytic proof of the sharp bound~\eqref{eq:fqe-quadratic-variance} on every critical state.
Put $p_j=\operatorname{tr}\rho_{1j}^2$ and $U=\sum_jp_j^2$.

\begin{proposition}[Sharp purity-variance tradeoff at minimal purity]
\label{prop:fqe-critical-variance}
Suppose a pure four-qubit state has $P=1$. Set
\[
V=\sum_{j=2}^4(p_j-1/3)^2,\qquad s=(24V)^{1/4}.
\]
Then $0\le s\le1$ and the sharp bound at each fixed $V$ is
\[
\boxed{\quad F\ge\frac32s^2-s^3.\quad}
\tag{RF}\label{eq:fqe-critical-variance}
\]
For $a=(1+2s)/3$, $b=(1-s)/3$, equality holds exactly when, up to ordering of
cuts and eigenvalues, the three spectra are
\[
\begin{split}
&\left(\frac12,\frac a2,\frac b2,\frac b2\right),\\
&\left(\frac{1+b+2\sqrt{ab}}4,\frac{1+b-2\sqrt{ab}}4,
\frac{1-b}4,\frac{1-b}4\right)\quad\text{on each of the other two cuts}.
\end{split}
\tag{RF-eq}\label{eq:fqe-variance-spectra}
\]
These spectra are attained for every $s\in[0,1]$.
In particular, $T+U\le3/4$, equivalently $F\ge12V$; equality in this weaker
bound occurs only at the HS spectra ($s=0$) or cluster spectra ($s=1$).
\end{proposition}
\noindent[Proved under the hypothesis $P=1$.]
\begin{proof}
By Lemma~\ref{lem:fqe-purity}, the state is critical with $\tau=0$. As in the proof of
Proposition~\ref{prop:fqe-critical-envelope}, its Bloch correlation matrix is then $\Theta=\mathbf m\mathbf n^T$ with unit
$\mathbf m,\mathbf n\in\mathbb R^3$ (Lemma~\ref{lem:fqe-critical-correlation}(2) at $\kappa=0$), and the cut $c\in\mathbb Z/3$ has the
four eigenvalues $(1+\sigma\cdot g^{(c)})/4$, $g^{(c)}_a=m_{a+c}n_a$, as $\sigma$ runs over the sign triples with
$\sigma_1\sigma_2\sigma_3=-1$ (Lemma~\ref{lem:fqe-p1-cuts}). Put $A_a=m_a^2$, $B_a=n_a^2$ and $k=-m_1m_2m_3n_1n_2n_3$.
The three vectors $g^{(c)}$ have squared norms
\[
u_c=\sum_aA_{a+c}B_a,\qquad \sum_cu_c=1,
\]
and the same coordinate product $-k$. The sign sums $\sum_\sigma\sigma\cdot g=0$, $\sum_\sigma(\sigma\cdot g)^2=4|g|^2$ and
$\sum_\sigma(\sigma\cdot g)^3=-24g_1g_2g_3$ give, for the corresponding cut,
\[
p_j=\frac{1+u_c}{4},\qquad
\operatorname{tr}\rho_{1j}^3=\frac{1+3u_c+6k}{16}.
\]
Set $X=1-3e_2(A)$ and $Y=1-3e_2(B)$, where $e_2(A)=\sum_{i<j}A_iA_j$.
Then $0\le X,Y\le1$, and expanding the three cyclic sums gives
\[
\sum_c u_c^2=\frac{1+2XY}{3}.
\]
We sharpen the elementary Schur estimate by keeping $X$ and $Y$.
For any nonnegative triple $A$ of sum one and $X=1-3e_2(A)$,
\[
27e_3(A)\le h(X),\qquad
h(X)=1-3X+2X^{3/2}=(1-\sqrt X)^2(1+2\sqrt X).
\tag{RP}\label{eq:fqe-triple-product}
\]
This is the cubic discriminant identity
\[
 4X^3-\bigl(27e_3(A)-1+3X\bigr)^2
   =27\prod_{i<j}(A_i-A_j)^2\ge0.
\]
Taking the upper square-root bound proves~\eqref{eq:fqe-triple-product}.
Equality requires two equal coordinates, and the upper sign selects
$((1+2r)/3,(1-r)/3,(1-r)/3)$, $r=\sqrt X$, up to permutation.
This includes the uniform triple and coordinate-vector endpoints.

The transfer from $X,Y$ to their geometric mean is algebraic as well.
Set $g(t)=(1-t)^2(1+2t)$, $\xi=X^{1/4}$, $\zeta=Y^{1/4}$,
$q=\xi\zeta$, $d=(\xi-\zeta)^2$, and
$w=(1-\xi^2)(1-\zeta^2)$. Direct factorisation gives
\begin{equation}\label{eq:fqe-product-transfer-remainder}
 \begin{split}
 h(\sqrt{XY})^2-h(X)h(Y)
 &=g(\xi\zeta)^2-g(\xi^2)g(\zeta^2)\\
 &=d\bigl\{w[6q(q+2)+2d]+d(1+2q)^2\bigr\}\ge0.
 \end{split}
\end{equation}
Every factor is nonnegative for $0\le X,Y\le1$, and the difference
is strictly positive when $X\ne Y$. Consequently
\[
27k\le27\sqrt{e_3(A)e_3(B)}
\le\sqrt{h(X)h(Y)}\le h(\sqrt{XY}).
\]
The moment formulas above give
\[
V=\frac{XY}{24},\qquad F=\frac12-\frac{27}{2}k.
\]
Thus $s=(XY)^{1/4}$ and $F\ge(1-h(s^2))/2=3s^2/2-s^3$.
For $0<s<1$, equality requires $X=Y=s^2$, $k>0$, and both $A,B$ to be
permutations of $(a,b,b)$. The shift $c$ that aligns their distinguished
coordinates gives $|g^{(c)}_a|=(a,b,b)$ up to order; the other two shifts give
$(\sqrt{ab},\sqrt{ab},b)$. As every $g^{(c)}$ has negative coordinate product,
all sign choices give the same spectral multisets, and the eigenvalue formula
proves \eqref{eq:fqe-variance-spectra}.
At $s=0$, equality forces $k=1/27$ and hence both triples uniform; at $s=1$
both are coordinate vectors. These give the stated endpoint spectra.

For attainability, take $\mathbf m=-\mathbf n=(\sqrt a,\sqrt b,\sqrt b)$, so $A=B=(a,b,b)$ and $k=ab^2=e_3(A)$.
The product auxiliary vector with Bloch vectors $\mathbf m,\mathbf n$ has $\Theta=\mathbf m\mathbf n^T$ and $\tau=0$, and it is
$\sum_kv_km_k$ for a critical $G(v)$, as at the end of the proof of Proposition~\ref{prop:fqe-critical-envelope}.
Finally,
\[
\frac32s^2-s^3-12V=\frac{s^2}{2}(1-s)(3+s)\ge0,
\]
with equality exactly at $s=0,1$. This proves the weaker bound and its
exhaustive equality statement as a corollary.
\end{proof}

\subsection{A joint-cut proof on the critical locus}
\label{sec:fqe-critical-quadratic-variance}

\begin{proposition}[Sharp quadratic variance on critical states]
\label{prop:fqe-critical-quadratic-variance}
Every critical unit four-qubit state satisfies
\[
 F^2+3P(P-1)\ge6V_P.
 \tag{CQ}\label{eq:fqe-critical-quadratic-variance}
\]
Equality holds at HS spectra and along the full cluster-to-Bell family
\[
 \bigl((1+\kappa)/2,(1-\kappa)/2,0,0\bigr),\qquad
 (1/4)^4,\qquad(1/4)^4,\qquad 0\le\kappa\le1,
\]
up to ordering of cuts and eigenvalues, and only there.
In particular, the defect in \eqref{eq:fqe-quadratic-variance-proved} is at least
$3\tau(1-\tau)/4$ on the critical locus. The proof uses no variance certificate. At $P=1$ the sharper
profile~\eqref{eq:fqe-critical-variance} of Proposition~\ref{prop:fqe-critical-variance} holds.
\end{proposition}
\begin{proof}
Use the auxiliary correlation matrix of Proposition~\ref{prop:fqe-critical-envelope}:
\[
 \Theta=U\operatorname{diag}(\kappa,-\kappa,1)V^T,\qquad U,V\in SO(3),
 \qquad \tau=\kappa^2,\qquad P=1+\kappa^2/2.
\]
Write $r=Ue_3$, $s=Ve_3$, $A_i=r_i^2$, $B_i=s_i^2$, and
$X=1-3e_2(A)$, $Y=1-3e_2(B)$. Thus $0\le X,Y\le1$.
We retain these two directional parameters in both the permanent and purity estimates.

\emph{The permanent bound with directions retained.}
Put $z=\sqrt{XY}$ and
\[
 b_\kappa(x)=\kappa^2+\frac23\kappa(1-\kappa)^2(1-x)
             +\frac29(1-\kappa)^3(1-3x+2x^{3/2}).
\]
The polar-decomposition Cauchy--Schwarz bound from
Proposition~\ref{prop:fqe-critical-envelope}, now using
\eqref{eq:fqe-triple-product} instead of discarding $X,Y$, gives
\[
 |\operatorname{perm}\Theta|
 \le\sqrt{b_\kappa(X)b_\kappa(Y)}\le b_\kappa(z).
\]
For completeness, the last inequality is geometric concavity.
Write $b_\kappa(x)=a-bx+cx^{3/2}$, where
\[
 a=(4\kappa^3+3\kappa^2+2)/9,\quad
 b=2(1-\kappa)^2/3,\quad c=4(1-\kappa)^3/9.
\]
The sign of $d^2\log b_\kappa(e^v)/dv^2$ is the negative of
$4ab-9ac t+bc t^3$, $t=\sqrt x$.
This polynomial decreases on $[0,1]$, since $b\le3a$, and its value at one is
$4\kappa^2(1-\kappa)^2(4\kappa^2+\kappa+1)/9\ge0$.
The cases with zero arguments follow by continuity.
Consequently the exact formula for $F$ gives
\[
 F\ge\kappa^2(1-\kappa)+\frac32(1-\kappa)^2z
                  -(1-\kappa)^3z^{3/2}
 \ge f_\kappa(z),
\]
where, using $\sqrt z\le(1+z)/2$,
\[
 f_\kappa(z)=(1-\kappa)\kappa^2
       +\frac12(1-\kappa)^2(1+2\kappa)z
       +\frac12(1-\kappa)^3z(1-z)\ge0.
\]

\emph{The three purities as one Fourier coefficient.}
Let $\omega=e^{2\pi i/3}$, $D=\operatorname{diag}(1,\omega,\omega^2)$, and
\[
 u_c=\sum_{j=0}^2\Theta_{j+c,j}^2,\qquad
 Z=\sum_c\omega^c u_c
   =\operatorname{tr}(D\Theta\overline D\Theta^T).
\]
Indices are cyclic. The cut formula $p_c=(1+u_c)/4$ implies
$6V_P=|Z|^2/4$.
The symmetric unitary matrices $M=U^TDU$ and $N=V^T\overline D V$
have $|M_{33}|^2=X$, $|N_{33}|^2=Y$.
Their last-row off-diagonal squared norms are $1-X,1-Y$;
their leading $2\times2$ blocks have squared Frobenius norms $1+X,1+Y$.
Expanding $Z$ in these blocks and applying Cauchy--Schwarz gives
\[
 |Z|\le z+2\kappa\sqrt{(1-X)(1-Y)}
                +\kappa^2\sqrt{(1+X)(1+Y)}
       \le B_\kappa(z):=(1-\kappa)^2z+2\kappa+\kappa^2.
\]
To see the last step, fix $z$ and put $S=X+Y\ge2z$.
The last two terms become
$2\kappa\sqrt{1-S+z^2}+\kappa^2\sqrt{1+S+z^2}$,
which decreases with $S$ for $0\le\kappa\le1$; evaluate at $S=2z$.

\emph{An elementary scalar comparison.}
For $0\le\kappa,z\le1$,
\[
 4f_\kappa(z)^2+9\kappa^2-B_\kappa(z)^2
       -3\kappa^2(1-\kappa^2)
   =(1-\kappa)^2(1-z)R_\kappa(z),
\]
where, putting $d=1-\kappa$,
\[
 R_\kappa(z)=-d^4z^3+3d^3(1+\kappa)z^2
                    -4\kappa d^2z+2\kappa^2(1+2\kappa^2).
\]
Since $z^3\le z^2$,
\[
 R_\kappa(z)\ge
 2d^3(1+2\kappa)z^2-4\kappa d^2z+2\kappa^2(1+2\kappa^2)\ge0.
\]
The final quadratic is nonnegative: its discriminant is
$-16\kappa^3d^3(3+2\kappa+4\kappa^2)\le0$ and its leading and constant
coefficients are nonnegative (the endpoints follow directly).
Combining the estimates proves
\[
 4F^2+9\tau-24V_P\ge3\tau(1-\tau),
\]
which is exactly~\eqref{eq:fqe-critical-quadratic-variance}.

For $0<\kappa<1$ the scalar comparison is strict unless $z=1$.
Then $X=Y=1$, and equality in the Fourier triangle bound forces all nonzero
entries of $\Theta$ onto one cyclic diagonal. Their magnitudes are
$1,\kappa,\kappa$, with negative product since $\det\Theta=-\kappa^2$.
The cut eigenvalue formula gives precisely the displayed family.
At $\kappa=0$, equality requires $z=0$ or $z=1$; the latter is its cluster
endpoint, while the former requires $F=0$ and hence HS by the critical
envelope equality case. At $\kappa=1$, equality requires $F=0$ and
$|Z|=3$, forcing the Bell endpoint.
Conversely the family spectra are attained by the correlation matrices
$\operatorname{diag}(\kappa,-\kappa,1)$, and direct substitution gives equality.
\end{proof}

\subsection{Independent arguments on restricted domains}\label{sec:fqe-renyi-restricted}
The proof of Corollary~\ref{cor:fqe-renyi-gw} in the paper needs no eigenvalue cap and no rank condition. The following result gives a shorter,
independent proof of the envelope for states with a rank-two cut, at every order (for $q\ge2$ see
Section~\ref{sec:fqe-renyi-high-order}).

\begin{proposition}[Sharp fixed-spectrum rank-two cut envelope]
\label{prop:fqe-renyi-rank-two}
Suppose one of the three balanced cuts of a pure four-qubit state has
Schmidt rank at most two and nonzero eigenvalues $a,1-a$,
where $1/2\le a\le1$. Then for every $q>0$,
\[
\boxed{\qquad
\mathcal E_q(\psi)\le
\frac43+\frac{1}{3}H_q(a,1-a)
\le\frac53.\qquad}
\tag{Rk2}\label{eq:fqe-renyi-rank-two}
\]
For each fixed $a$, the first bound is sharp; equality holds precisely
when the other two balanced cut spectra are both $(1/4)^4$.
The attained cluster-to-Bell family \eqref{eq:fqe-quadratic-family}
realises equality for every $a\in[1/2,1]$ and every order $q>0$.
In particular the global HS--cluster R\'enyi envelope holds on
this entire rank-two boundary without any purity or other cap hypothesis.
\end{proposition}
\begin{proof}
A density matrix on a four-dimensional cut has R\'enyi entropy
at most $\log_2 4=2$, with equality exactly for the uniform-four
spectrum. The stipulated cut has entropy $H_q(a,1-a)$, so bounding
each of the other two cut entropies by $2$ proves the first
inequality and its equality classification. A binary probability
vector has entropy at most $1$, proving the second inequality.
The physical states in \eqref{eq:fqe-quadratic-family} have the
stipulated binary spectrum and both crossed cuts uniform-four
for all $a=(1+r)/2\in[1/2,1]$; they therefore attain the bound
at every $q>0$. Since the HS entropy $h_q$ is at least $5/3$
up to $q_*$, and the cluster ceiling is $5/3$ thereafter,
the final assertion follows.
\end{proof}

\subsection{A cap-free purity-controlled R\'enyi bound}\label{sec:fqe-renyi-purity-bound}
The defect estimate of Theorem~\ref{thm:fqe-renyi-above-one} also gives a ceiling at every order $q\ge5/4$ that
improves linearly with the purity sum. It is not used in the proof of Corollary~\ref{cor:fqe-renyi-gw}.

\begin{corollary}[Cap-free purity-controlled R\'enyi bound; exact-rational check, RA-5]
\label{cor:fqe-renyi-purity}
For every unit pure four-qubit state and every $q\ge5/4$,
\[
\boxed{\qquad \mathcal E_q(\psi)\le h_{5/4}-\frac{13}{20}(P-1),
\qquad h_{5/4}=\frac{\log_2m_{5/4}}{1-5/4}.\qquad}
\tag{RH}\label{eq:fqe-renyi-high-purity}
\]
In particular, $P\ge10/9$ implies $\mathcal E_q(\psi)<5/3$
for $q\ge5/4$. The bound \eqref{eq:fqe-renyi-high-purity} holds for every triple of four-point spectra with
$P\ge1$ and $F=17P-12T-12\ge0$, since the proof uses only these and the per-cut estimate
\eqref{eq:fqe-renyi-cut}.
\end{corollary}
\begin{proof}
At $q_0=5/4$, the defect estimate \eqref{eq:fqe-renyi-defect}
and $F,\mathcal D_{q_0}\ge0$ give
\[
\mathcal E_{q_0}(\psi)\le h_{q_0}-c(P-1),
\qquad c=\frac{\alpha_{q_0}}{3m_{q_0}\ln2}.
\]
The exact-rational enclosures RA-5 establish
\[
20\alpha_{q_0}-39m_{q_0}\ln2>0,\qquad
4\ln m_{q_0}+\frac{313}{180}\ln2>0.
\]
Equivalently $c>13/20$ and $h_{q_0}<313/180$.
R\'enyi entropy is nonincreasing in its order, so
$\mathcal E_q(\psi)\le\mathcal E_{q_0}(\psi)$ for $q\ge q_0$.
Since $P\ge1$, the first assertion follows. If $P\ge10/9$,
\[
\mathcal E_q(\psi)<\frac{313}{180}
-\frac{13}{20}\left(\frac{10}{9}-1\right)=\frac53,
\]
as asserted.
\end{proof}

\subsection{The unrestricted joint purity bound}\label{sec:fqe-joint-purity}

\begin{theorem}[Joint purity compatibility; consequence of QK]
\label{thm:fqe-joint-purity}
Every pure four-qubit state satisfies
\[
\boxed{\qquad T+\sum_{j=2}^4p_j^2\le3P-\frac94.\qquad}
\tag{JV}\label{eq:fqe-joint-purity}
\]
Equality holds exactly for HS cut spectra, cluster cut spectra, or products
of two Bell pairs, up to ordering of the cuts and eigenvalues.
In the HS case the states belong to the two HS LU orbits.
\end{theorem}
\begin{proof}
This is an elementary consequence of the stronger quadratic variance bound
\eqref{eq:fqe-quadratic-variance-proved}, together with $F\ge0$ and
$p_j\in[1/4,1]$. Set
\[
U=\sum_jp_j^2,\qquad G=3P-\frac94-T-U,\qquad
Q=F^2+\frac92(P-1)-6\bigl(U-P^2/3\bigr)\ge0.
\]
Since $T=(17P-12-F)/12$, direct algebra gives the useful identity
\[
12G=F(1-2F)+2(P-1)(3-2P)+2Q.
\tag{JI}\label{eq:fqe-joint-implication}
\]
It proves $G\ge0$ whenever $1\le P\le3/2$ and $0\le F\le1/2$.

For the remaining cases, use the purity box \eqref{eq:fqe-purity-box} of Section~\ref{sec:sup-qk}, the
maximum of the convex function $U$ at fixed $P$ over $[1/4,1]^3$.
Using $12G=19P-15+F-12U$, the first piece, for
$1\le P\le3/2$ and $F\ge1/2$, yields
\[
12G\ge F-\frac12+(P-1)(19-12P)\ge0.
\]
For $3/2\le P\le9/4$ the second piece yields
\[
12G\ge F+\frac12(2P-3)(31-12P)\ge0,
\]
and for $9/4\le P\le3$ the last piece yields
\[
12G\ge F-12P^2+67P-87\ge F+3>0.
\]
The last inequality follows from concavity of the quadratic and its
values $3,6$ at the two endpoints.

Equality in the first case forces $Q=0$, so the equality classification of
Corollary~\ref{thm:fqe-quadratic-variance} gives HS, cluster or Bell-pair
spectra. In the $F\ge1/2$ case equality requires
$P=1$, $F=1/2$, $U=3/8$, again giving $Q=0$ and the cluster
spectra. In the second box range equality requires
$P=3/2$, $F=0$ and $U=9/8$; its purity triple is
$(1,1/4,1/4)$, and the state is a product of two Bell pairs.
The final range is strict. All three listed types attain equality.
\end{proof}

\subsection{What local filtering can and cannot reduce}
\label{sec:fqe-renyi-filter-limit}

\begin{lemma}[One-site filtering majorises a locally maximally mixed pair]
\label{lem:fqe-single-filter}
Let $\rho$ be a two-qubit density matrix with both one-qubit
marginals equal to $I/2$. Let $K$ be any nonzero filter on
one of its qubits, and normalise
$\rho_K=(K\otimes I)\rho(K^\dagger\otimes I)$ by its trace
(with the factors interchanged if the filter acts on the second qubit).
Then
\[
\operatorname{spec}(\rho)\prec\operatorname{spec}(\rho_K).
\tag{FM}\label{eq:fqe-filter-majorization}
\]
In particular, filtering a single qubit of any pure critical four-qubit
state cannot increase the R\'enyi entropy of \emph{any} of the three
balanced cuts, at any $q>0$.
\end{lemma}
\begin{proof}
The positive factor in the polar decomposition of $K$ suffices,
since the unitary factor only conjugates $\rho_K$. By a positive
rescaling assume $\operatorname{tr}K^\dagger K=2$; the given
single-qubit marginal makes the filtered state already normalised.
Then $K^\dagger K=I+s\,\mathbf n\cdot\boldsymbol\sigma$,
$0\le s\le1$, where $\mathbf n$ is a real unit vector. Put
$N=(\mathbf n\cdot\boldsymbol\sigma)\otimes I$.

Let $\Theta=(\sigma_y\otimes\sigma_y)\mathcal K$ be the
two-qubit spin-flip antiunitary, where $\mathcal K$ is
complex conjugation in the computational basis.
Expanding $\rho$ in Pauli matrices shows that
$\Theta\rho\Theta^{-1}=\rho$, because the two one-site
Bloch vectors vanish and spin flip changes the sign of
each individual Pauli factor. Also
$\Theta N\Theta^{-1}=-N$.
The matrices
\[
B_\pm=\rho^{1/2}(I\pm sN)\rho^{1/2}
\]
are positive, have the same spectrum by conjugation with
$\Theta$, and satisfy $(B_++B_-)/2=\rho$.
The nonzero eigenvalues of $B_+$ coincide with those of
$(K\otimes I)\rho(K^\dagger\otimes I)$, because these
matrices have the form $X^\dagger X$ and $XX^\dagger$.
Convexity of every Ky Fan sum under averaging therefore gives
$\operatorname{spec}(\rho)\prec\operatorname{spec}(B_+)
=\operatorname{spec}(\rho_K)$.

For a critical four-qubit state, each qubit marginal is $I/2$.
On each balanced cut choose the two-qubit side containing the
filtered site. Its reduced state has both one-qubit marginals $I/2$;
the two sides of a pure bipartition have the same nonzero spectrum.
Apply the result above and Schur concavity of R\'enyi entropy.
\end{proof}

\paragraph{Two filters need not preserve this ordering.}
The single-site conclusion is \emph{not} an orbitwise
monotonicity theorem. For an exact counterexample let
\[
z=(2+2i)(|0011\rangle+|1100\rangle)
 +|0101\rangle-|0110\rangle-|1001\rangle+|1010\rangle.
\]
It has squared norm $20$ and all four one-qubit marginals
maximally mixed. Apply $\operatorname{diag}(2,1)$ at
\emph{both} sites 2 and 3, and normalise the resulting
unnormalised vector $z'$ of squared norm $89$.
The three cut spectra of $z/\sqrt{20}$ are, respectively,
\[
(2/5,2/5,1/5,0),\qquad
(13/20,1/4,1/20,1/20),\qquad
(13/20,1/4,1/20,1/20),
\]
whereas those of $z'/\sqrt{89}$ are
\[
(32,32,25,0)/89,\quad
(65,16,4,4)/89,\quad
(52,20,16,1)/89.
\]
Consequently, their products of cut purities are
\[
\prod_jp_j(z/\sqrt{20})=\frac{21609}{250000}
\;>\;
\frac{40544579889}{496981290961}
=\prod_jp_j(z'/\sqrt{89}).
\tag{FN}\label{eq:fqe-filter-nogo}
\]
This proves that the \emph{mean} R\'enyi-2 entropy increases
under two local filters. The same is true at $q=3/2$, inside
the intermediate-order interval: if $M(\lambda)=\sum_i\lambda_i^{3/2}$,
the original moments obey $M_1>59/100$ and
$M_2=M_3>67/100$, while the filtered moments obey
$M'_1<581/1000$, $M'_2<720/1000$, and $M'_3<631/1000$.
Each strict bound follows by squaring rational bounds on the
square roots of the displayed rational eigenvalues. Thus
\[
\prod_jM_j>\frac{264851}{10^6}>
\frac{3299499}{12500000}>\prod_jM'_j,
\]
so $\mathcal E_{3/2}(z'/\sqrt{89})
>\mathcal E_{3/2}(z/\sqrt{20})$.
Both states obey the global HS--cluster envelope of Corollary~\ref{cor:fqe-renyi-gw}.
The valid single-filter lemma therefore cannot be iterated to
replace all noncritical states by critical representatives.

\section{Proofs of the stability theorems}\label{sec:fqe-stability-proofs}
This section proves the global and local stability results of Appendix~\ref{sec:fqe-orbit-stability} of the paper.
We first write the surrogate $Q$ intrinsically in the magic matrix and prove the sharp global overlap bound. We then compute
the transverse quadratic forms and obtain explicit local radii. The
two sharp critical-locus theorems of Appendix~\ref{sec:fqe-critical-orbit-stability} are proved in Section~\ref{sec:sup-stability}.

\begin{restated}[Explicit global LU-orbit stability; from Theorem~\ref{thm:fqe-global-overlap}]{Theorem}{thm:fqe-orbit-stability}
For every unit pure four-qubit state,
\[
 d(\psi)^2\le\frac{12}{6+\sqrt6}\,Q(\psi),\qquad
 \boxed{\ \delta(\psi)\ge k\,\frac{6+\sqrt6}{12}\,d(\psi)^2\ },\qquad k\,\frac{6+\sqrt6}{12}>\frac1{20}.
\]
For every critical state the optimal bound is $\delta\ge C_*d^2$, with
$C_*\approx0.19058$ (Theorem~\ref{thm:fqe-p1-sharp}).
On an explicit neighbourhood of the two orbits the coefficient improves:
\[
 d(\psi)\le1/20\quad\Longrightarrow\quad
 \delta(\psi)\ge\frac3{10}d(\psi)^2.
\]
Its sharp limiting coefficient is
$c_*=(2-\ln3)/(2\ln2)$, in the precise sense of
\eqref{eq:fqe-entropy-local-sharp}. The power two is optimal.
No optimal global coefficient is claimed; the equality states of
Theorem~\ref{thm:fqe-p1-sharp} show that it is at most $C_*$.
\end{restated}
\noindent The global bounds are proved at the end of Section~\ref{sec:fqe-global-overlap}, the explicit radius and the
limiting coefficient in Section~\ref{sec:fqe-entropy-orbit-local}; the critical coefficient $C_*$ is
Theorem~\ref{thm:fqe-p1-sharp}, proved in Section~\ref{sec:sup-stability}.

\subsection{The intrinsic moment identity}
\label{sec:fqe-bell-extraction}

Here is a convenient arbitrary-state identity. For a state with magic matrix $Z$ (Appendix~\ref{sec:fqe-normal-form} of the paper) put $s=\operatorname{tr}(Z^TZ)$,
$\mathcal N=ZZ^TZ-sZ/2$, and $C=\sum_j e_3(\rho_{1j})$, where $e_3$ is the third elementary symmetric
polynomial of the eigenvalues. Then
\begin{equation}\label{eq:fqe-bell-cubic}
 C=\frac32\|\operatorname{adj}Z\|_F^2+\frac12\|\mathcal N\|_F^2,
 \qquad Q=5-36C\quad(\|Z\|_F=1).
\end{equation}
For a direct derivation let $D_{12},D_{13},D_{14}$ be the three flattening determinants
in the computational basis, and set
$J=\operatorname{tr}((Z^TZ)^2)/4-s^2/8$. Expanding the four-by-four
determinants in the stated magic basis gives
$D_{12}=\det Z$, $D_{13}=(\det Z+J)/2$,
$D_{14}=(-\det Z+J)/2$. If $M_{1j}$ is the flattening matrix with $\rho_{1j}=M_{1j}M_{1j}^\dagger$, then
$e_3(\rho_{1j})$ is the sum of the $3\times3$ principal minors of $M_{1j}M_{1j}^\dagger$, which by Cauchy--Binet is
$\|\operatorname{adj}M_{1j}\|_F^2$; the holomorphic gradient of $\det M_{1j}$ is the transposed adjugate, and the flattening
and the magic change of basis are unitary reshapings, so $e_3(\rho_{1j})=\|\nabla D_{1j}\|^2$. Here
$\nabla D_{12}=\nabla\det Z$ has norm $\|\operatorname{adj}Z\|_F$, and $\nabla J=\mathcal N$. By the parallelogram law
$\|\nabla D_{13}\|^2+\|\nabla D_{14}\|^2=\tfrac12(\|\operatorname{adj}Z\|_F^2+\|\mathcal N\|_F^2)$, and adding
$\|\nabla D_{12}\|^2$ proves the first identity in \eqref{eq:fqe-bell-cubic}. The second is Newton's identity
$6e_3(\rho)=1-3\operatorname{tr}\rho^2+2\operatorname{tr}\rho^3$ for $\operatorname{tr}\rho=1$, summed over $j=2,3,4$:
$36C=18-18P+12T$, while $Q=18P-12T-13$ for unit states.

Section~\ref{sec:sup-bellframe} also gives a constructive distance certificate from three chosen Bell-pair products.

\subsection{Closest-orbit geometry and the sharp overlap bound}
\label{sec:fqe-global-overlap}
We prove the global overlap comparison by a rational polynomial certificate on a normal slice.
The first step shows that this slice contains a closest representative of every state; the second identifies
the constraints that make the certificate nonnegative there.

\begin{restated}[Sharp HS overlap stability; exact rational certificate, GO-1--GO-3]{Theorem}{thm:fqe-global-overlap}
For every unit pure four-qubit state,
\begin{equation*}
 Q(\psi)\ge1-\Phi(\psi)
 \ge\frac{6+\sqrt{6(1+\sqrt{\tau(\psi)})}}{12}\,d(\psi)^2
 \ge\frac{6+\sqrt6}{12}\,d(\psi)^2,\qquad \frac{6+\sqrt6}{12}>\frac7{10}.
\end{equation*}
Equality in $Q\ge1-\Phi$ holds on $\mathcal M$ ($Q=0$, $\Phi=1$) and on the three Bell-pair-product LU orbits
($Q=\Phi=1/2$), so the coefficient one is optimal.
\end{restated}

\paragraph{Coverage by the normal slice.}
Take a closest HS representative and transform it to
$Z_*=\operatorname{diag}(\sqrt3,i,i,i)/\sqrt6$ (here and in the rest of the supplement, magic-matrix indices run over $0,\dots,3$, index $i$ labelling the magic
vector $m_{i+1}$ of Appendix~\ref{sec:fqe-normal-form} of the paper); choose phase so
$h=\langle Z_*,Z\rangle\ge0$ is real. Up to global phase, the union
$\mathcal M$ is the $O(4)\times O(4)$ orbit of $Z_*$: the proof of Proposition~\ref{prop:fqe-onecut} of the paper
puts every HS state in the $SO(4)\times SO(4)$ orbit of $Z_*$ or of $\overline{Z_*}$, and $\operatorname{diag}(1,-1,-1,-1)$
on one side maps $Z_*$ to $\overline{Z_*}$.
The arbitrary-state identity \eqref{eq:fqe-bell-cubic} gives
\begin{equation}\label{eq:fqe-global-intrinsic}
 \begin{split}
 Q(Z)&=27N\operatorname{tr}[(ZZ^\dagger)^2]
      -18\operatorname{tr}[(ZZ^\dagger)^3]
      -18\|ZZ^TZ-sZ/2\|_F^2-4N^3,\\
 s&=\operatorname{tr}(Z^TZ),\\
 W_1(Z)&=\|ZZ^\dagger-\overline{ZZ^\dagger}\|_F^2
          +\|Z^\dagger Z-\overline{Z^\dagger Z}\|_F^2.
 \end{split}
\end{equation}
The $W_1$ formula follows because the six one-qubit Pauli matrices
on either magic pair form an orthogonal basis of imaginary skew matrices,
each with Frobenius norm squared four (BF-2).
These formulas show directly that $N$, $Q$, $W_1$ and $s$ are invariant under
both real orthogonal factors. Thus the transformation preserves every
quantity needed below.

A closest representative maximises the real height
$\sqrt6\operatorname{Re}\langle Z_*,Z\rangle=3X_{00}+\sum_jY_{jj}$ over phase and both orthogonal
factors. Write $Z=\sqrt3X+iY$ with $X,Y$ real. The first-order conditions for the six left
and six right generators and for the phase have rank ten (simultaneous lower rotations fix
$Z_*$), and are exactly
\begin{equation}\label{eq:fqe-slice-stationary}
 Y_{00}=\sum_{j}X_{jj},\qquad X_{j0}=Y_{0j}/3,\qquad Y_{j0}=3X_{0j},\qquad Y_{jk}=Y_{kj}
\end{equation}
($1\le j<k\le3$ in the last relation). Simultaneous lower rotations preserve
\eqref{eq:fqe-slice-stationary} and conjugate the symmetric matrix
$\operatorname{Sym}(X_{jk})+2(Y_{jk})$, so one of them makes it diagonal: three further linear
conditions (NSL-1--NSL-3).
Consequently every closest-orbit problem has a representative
$Z=\sqrt3X+iY$ with exactly the following 19 real coordinates:
\begin{align*}
 X_{00}&=t,&Y_{00}&=u_1+u_2+u_3,\\
 X_{jj}&=u_j,&Y_{jj}&=t+a_j,\\
 X_{0j}&=p_j,&Y_{0j}&=q_j,\\
 X_{j0}&=q_j/3,&Y_{j0}&=3p_j,\\
 X_{jk}&=e_{jk}+b_{jk},&X_{kj}&=e_{jk}-b_{jk},\\
 Y_{jk}&=Y_{kj}=-e_{jk}/2\qquad(1\le j<k\le3).
\end{align*}
Here and below $1\le j\le3$ for cross-block and diagonal entries.
Set
\[
 h_0=6t+a_1+a_2+a_3=\sqrt6h,\quad
 N=3\|X\|_F^2+\|Y\|_F^2,\quad e=N-h_0^2/6.
\]
For a normalised state, $e=1-\Phi$.
This coverage does not assume criticality or distinct singular values.

\paragraph{Positive constraints at a closest point.}
For any unit HS ray $\eta$ put
\[
 g_\eta=h^2-|\langle\eta,Z\rangle|^2\ge0.
\]
Write $Y_3=(Y_{jk})_{j,k=1}^3$.

Fixing either orthogonal factor and reflecting the other across an arbitrary
real hyperplane gives the stronger one-factor Procrustes constraints
\[
 K_0=\begin{pmatrix}3t&q^T\\q&Y_3\end{pmatrix}\succeq0,\qquad
 K_1=\begin{pmatrix}3t&3p^T\\3p&Y_3\end{pmatrix}\succeq0.
\]
Indeed these are $\sqrt6\operatorname{Re}(ZZ_*^\dagger)$ and
$\sqrt6\operatorname{Re}(Z_*^\dagger Z)$, respectively, and both are
symmetric on the normal slice. A reflection $I-2vv^T$ changes the real
height by $-2v^TK_i v$, which cannot be positive at a closest
point. Reflections are admissible because both HS orbits are included.
Their principal blocks also give $t\ge0$ and $Y_3\succeq0$.
These constraints retain the correlation between each cross block and
the positive lower block.

Let $G_{ij}$ be the real skew matrix with entries $(i,j)=1$,
$(j,i)=-1$. Use the following ten infinitesimal actions on $Z$:
left $Z\mapsto G_{0j}Z$ and right $Z\mapsto-ZG_{0j}$, left $G_{jk}$ for $1\le j<k\le3$, and
phase multiplication by $i\sqrt3$. Denote them by $T_\ell$.
The height Hessian satisfies
\begin{equation}\label{eq:fqe-height-matrix}
 H_{\ell m}(Z)=-\sqrt6\operatorname{Re}
  \left\langle Z,\frac{T_\ell T_m+T_mT_\ell}{2}Z_*\right\rangle,
 \qquad H(Z)\succeq0.
\end{equation}
Indeed the chosen representative maximises real height along each group
curve. Every entry of $H$ is rational linear in the 19 coordinates.

A final family tests a continuous set of lower-block axes. Put
$S=\operatorname{Sym}(X_{jk})_{j,k=1}^3$. For
$\sigma_0,\sigma\in\{\pm1\}$ and
$(c,s)\in\{(1,0),(1/2,1/2),(-1/2,1/2),(-1,0),
(-1/2,-1/2),(1/2,-1/2)\}$, define
\begin{align*}
 r_0&=\sigma_0Y_{00}+\sigma\operatorname{tr}Y_3,&
 i_0&=-\sigma_0X_{00}-\sigma\operatorname{tr}S,\\
 L_{\sigma_0,\sigma,c,s}
   &=(h_0-cr_0-3si_0)I-c(3S-\sigma Y_3)-3s(Y_3+\sigma S).
\end{align*}
Then $L_{\sigma_0,\sigma,c,s}\succeq0$. To prove this, for each real
unit $v\in\mathbb R^3$ use the HS matrix with top component
$\sigma_0i/\sqrt6$ and lower block
$v v^T/\sqrt2+\sigma i(I-vv^T)/\sqrt6$.
The closest-height inequality at phase $\phi$, where
$\cos\phi=c$ and $\sin\phi=\sqrt3s$, is exactly $v^TLv\ge0$.

The fidelity floor in Lemma~\ref{lem:fqe-hs-fidelity-floor} also gives a
positive quadratic matrix. Put
\[
 s_R=3\|X\|_F^2-\|Y\|_F^2,\quad
 s_I=2\sum_{ij}X_{ij}Y_{ij},\quad \ell=h_0^2-N,\qquad
 B_0=\begin{pmatrix}\ell+s_R&s_I\\s_I&(\ell-s_R)/3\end{pmatrix}\succeq0.
\]
Indeed $s=s_R+i\sqrt3s_I$ and the homogeneous fidelity floor gives
$\ell\ge|s|$. Both diagonal entries of $B_0$ are nonnegative and
$3\det B_0=\ell^2-|s|^2\ge0$.

\paragraph{The positive module.}
Below, $n=N=3\|X\|_F^2+\|Y\|_F^2$ denotes the homogeneous norm squared, so every block has total degree six.
Let $\mathcal A$ average over simultaneous lower signed permutations.
They preserve $Z_*$, the normal slice and all closest-point inequalities.
The target polynomial is invariant under this averaging.
For a positive matrix $B$ and a positive rational Gram matrix $G$, write
\[
 \mathcal T_G(B;v)=\sum_{a,b}G_{ab}v_a^TBv_b\ge0.
\]
A factorisation of $G$ makes this a sum of nonnegative quadratic forms;
no exact factorisation is required in the certificate checker.
The weighted matrices used below are
\begin{equation}\label{eq:fqe-global-module}
 \begin{split}
 \mathcal C_0=\bigg\{&1,\frac{g_\eta}6,\frac{h_0t}6,\frac{B_0}{36},
       \frac{nh_0H}{216},\frac{nh_0Y_3}{36},\\
 &\frac{h_0H}{216},\frac{h_0Y_3}{36},
       \frac{h_0K_0}{216},\frac{h_0K_1}{216}\bigg\}.
 \end{split}
\end{equation}
Here $\eta$ ranges over the supplied HS rays. All members are positive at a closest point.
The supplied HS rays are checked independently by $N=1,W_1=\tau=Q=0$,
so the established equality classification proves that every one is HS.

\begin{lemma}[An HS fidelity floor from real singular values]\label{lem:fqe-hs-fidelity-floor}
For every unit pure four-qubit state,
\[
 \Phi\ge\frac{1+\sqrt\tau}{6}\ge\frac16,\qquad
 d^2\le2\left(1-\sqrt{\frac{1+\sqrt\tau}{6}}\right)
      \le2(1-1/\sqrt6).
\]
The fidelity inequality is attained by the GHZ state, for which
$\tau=1$ and $\Phi=1/3$.
\end{lemma}
\begin{proof}
Choose phase so that $s=\operatorname{tr}(Z^TZ)$ is real and nonnegative.
Then $\|\operatorname{Re}Z\|_F^2=(1+|s|)/2$.
Apply a real singular-value decomposition to $\operatorname{Re}Z$, writing
its nonnegative diagonal entries as $m\ge x\ge y\ge z$.
The transformed imaginary part can be arbitrary. Two HS representatives,
with large real component $1/\sqrt2$ in the $m$ position and the other
three components all $+i/\sqrt6$ or all $-i/\sqrt6$, give
\[
 \Phi\ge\tfrac12m^2+\tfrac16(x+y+z)^2
       \ge\tfrac13(m^2+x^2+y^2+z^2).
\]
The first inequality follows by averaging their squared overlaps; the
imaginary diagonal entries contribute further nonnegative squares.
For the second, six times its difference is
\[
 m^2+(x+y+z)^2-2(x^2+y^2+z^2)
 =(m^2-x^2)+y(2x-y)+z(2x-z)+2yz\ge0.
\]
Both real orthogonal transformations preserve the phase-inclusive union
of the two HS orbits, so these comparisons are admissible. Finally
$|s|=\sqrt\tau$, since $\tau=|\operatorname{tr}Z^TZ|^2$ in the magic basis. For GHZ the magic
diagonal is $(1,0,0,-1)/\sqrt2$, signed-permutation equivalent to $(1,1,0,0)/\sqrt2$.
Formula~\eqref{eq:fqe-signed-diagonal-overlap}
gives maximum squared overlap $1/3$ over the 32 HS representatives
(RP-3), proving the equality claim.
\end{proof}

\begin{proof}[Proof of Theorem~\ref{thm:fqe-global-overlap}]
Place a closest HS representative at
$Z_*=\operatorname{diag}(\sqrt3,i,i,i)/\sqrt6$ so that the state has a
representative $Z=\sqrt3X+iY$ on the $19$-real-coordinate stationary normal
slice \eqref{eq:fqe-slice-stationary},
with $h_0=6t+a_1+a_2+a_3=\sqrt6h\ge0$, $n=3\|X\|_F^2+\|Y\|_F^2$, and
$e=n-h_0^2/6$. For a unit state, $n=1$ and $e=1-\Phi(\psi)\ge0$ by
Cauchy--Schwarz. The homogeneous sextic target $(Q-n^2e)/36$ is a polynomial in the
$19$ slice coordinates.

Enlarge the positive module $\mathcal C_0$ of \eqref{eq:fqe-global-module} to
\begin{equation}\label{eq:fqe-global-overlap-module}
 \mathcal C_Q
 =\mathcal C_0\cup\left\{\frac{e}{6},\;\frac{h_0L_{\sigma_0,\sigma,c,s}}{216}\right\},
\end{equation}
where $\eta$ ranges over $50$ supplied unit HS rays, checked independently in GO-1 by
$n=1$ and $Q=W_1=\tau=0$. Only the four lower-frame choices
\[\begin{gathered}
 (\sigma_0,\sigma,c,s)\in\{(-1,-1,-1/2,1/2),\ (-1,1,1/2,1/2),\\
 (1,-1,-1,0),\ (1,1,1,0)\}
\end{gathered}\]
are needed in this module. The additional scalar constraint $e/6\ge0$ is
multiplied by quadratic polynomial vectors in the slice coordinates, and the
lower-frame matrices $h_0L_{\sigma_0,\sigma,c,s}/216\succeq0$ are paired with
quadratic matrix multipliers; on the $B_0/36$ and lower-frame blocks where the
simultaneous signed lower-permutation representation splits into conjugate
two-dimensional $S_3$ characters, both paired character components are
retained in the polynomial-vector bases $p_B$.

The certificate supplies $277$ blocks $B\in\mathcal C_Q$ with explicit rational
polynomial-vector bases $p_B$ and positive rational Gram matrices $G_B$ such
that, with the two explicit nonnegative terms below, the polynomial identity
\begin{equation}\label{eq:fqe-global-overlap-sos}
 \frac{Q-n^2e}{36}
 =R_{\rm sq}+R_{\rm ax}+\mathcal A\sum_{B\in\mathcal C_Q}\mathcal T_{G_B}(B;p_B)
\end{equation}
holds identically in $\mathbb Q[x_1,\dots,x_{19}]$ across all $3{,}277$
degree-$\le6$ signed-permutation monomial orbits (including the six dependent
core equations as well as the $3{,}271$ independent ones). GO-2 checks every
Gram matrix $G_B\succ0$ by the compact integer test below, expands every block, applies the signed Reynolds average
$\mathcal A$, and verifies that all $3{,}277$ orbit coefficients match
$(Q-n^2e)/36$ identically over $\mathbb Q$. Evaluating
\eqref{eq:fqe-global-overlap-sos} at a closest representative of a unit state
gives $Q(\psi)\ge e=1-\Phi(\psi)$.

\paragraph{Analytic replacement of the linear lower-frame blocks.}
Put $\gamma=10^{-7}$, $\omega=794424/10^{12}$ and define
\begin{align*}
 J={}&5\left[\sum_j(p_j^2+q_j^2)+\sum_{j<k}(e_{jk}^2+b_{jk}^2)\right]
       +2\sum_{j<k}\left[(u_j-u_k)^2+(a_j-a_k)^2\right],\\
 z={}&(b_{23},-b_{13},b_{12})^T,\qquad L_*=L_{-1,-1,-1/2,1/2},\\
 R_{\rm sq}={}&\frac{\gamma}{27}nh_0^2J,\qquad
 R_{\rm ax}=\frac{\omega}{216}nh_0z^TL_*z.
\end{align*}
The first term is manifestly nonnegative, and the second is nonnegative because $n,h_0\ge0$ and $L_*\succeq0$.
Together they replace all $96$ linear lower-frame Gram blocks, so none of those blocks is supplied numerically.

The reduction follows from linearity in $L$. For each of four common polynomial-vector bases, the $24$ lower-frame
matrices have the same Gram contribution $\gamma I$, and
\[
 \sum_{\sigma_0,\sigma,\,(c,s)}L_{\sigma_0,\sigma,c,s}=24h_0I,
\]
where the sum is over both independent signs and the six phase pairs used in the lower-frame construction.
The signed-permutation average of the sum of squared components of those four bases is $J/3$:
coefficients $(2,2,1)$ on each of the $p,q,e,b$ triples average to $5/3$, and coefficients $(1,1,0)$ on each
of the $u,a$ difference triples average to $2/3$. Their contribution is
$(\gamma nh_0/216)(24h_0)(J/3)=R_{\rm sq}$.
The only additional Gram contribution is $\omega vv^T$, with $v=(0,1,0,-1,0,1)$ in the six-vector basis
whose selected components are $b_{23},b_{13},b_{12}$. Its contribution is $R_{\rm ax}$.
Both remainders are invariant under the signed lower permutations: $z$ is the axial vector of the real
antisymmetric lower block, so it transforms by $(\det O)O$, while $L_*$ transforms by conjugation.
GO-2 independently checks their invariance and the complete reduced polynomial identity.
The lower-frame conditions used by the finite certificate are consequently reduced from $24$ to $4$.

\paragraph{Compact integer Gram data.}
The remaining Gram matrices and their positivity witnesses are stored compactly.
Write $g=10^{12}$, $c$ for the supplied correction denominator, and $s=2^{46}$.
For each block of size $m$, the data consist of a rectangular integer matrix $A$, symmetric integer matrices
$\Delta,C$ (the latter sparse), and the reconstruction rule
\[
 H_{ij}=\left\lfloor\frac{g(AA^T)_{ij}}{s^2}+\frac12\right\rfloor,
 \qquad G_{ij}=\frac{H_{ij}+\Delta_{ij}+C_{ij}/c}{g}.
\]
Put $D_{ij}=c\Delta_{ij}+C_{ij}$. The only positivity test is
\begin{equation}\label{eq:fqe-compact-gram-test}
 2\left(D_{ii}-\sum_{j\ne i}|D_{ij}|\right)>mc\qquad\text{for every }i.
\end{equation}
Indeed each rounding error $H_{ij}-g(AA^T)_{ij}/s^2$ has absolute value at most $1/2$.
Thus \eqref{eq:fqe-compact-gram-test} makes $G-AA^T/s^2$ strictly diagonally dominant with positive diagonal.
Its positive definiteness follows, for example, by applying $2|v_iv_j|\le v_i^2+v_j^2$ to its quadratic form;
adding $AA^T/s^2\succeq0$ proves $G\succ0$.
No square roots, eigenvalues, divisions in an ordered field, or matrix factorisation are needed in this finite check:
rounding is integer quotient and the final tests are integer inequalities.
The $277$ numerical blocks use $2{,}222$ factor columns and $178{,}221$ factor entries; $11$ need no factor.
The exact polynomial identity uses $50$ HS comparisons and the two explicit remainders (GO-1--GO-2).

\paragraph{Integer identity check.}
All basis coefficients have denominators dividing $6$ and all constraint coefficients become integral after
multiplication by the supplied map denominator, so GO-2 compares the identity at a common integer scale, keeping the
integer Gram parts $H+\Delta$ and the sparse corrections $C$ in separate integer accumulators; the correction
denominator $c$ enters only the final coefficient comparisons.

Finally, since $d(\psi)^2=2(1-\sqrt{\Phi(\psi)})$, we have
\[
 1-\Phi(\psi)=\frac{1+\sqrt{\Phi(\psi)}}{2}\,d(\psi)^2.
\]
Lemma~\ref{lem:fqe-hs-fidelity-floor} gives
$\Phi(\psi)\ge(1+\sqrt{\tau(\psi)})/6\ge1/6$, which yields the two middle
inequalities in \eqref{eq:fqe-global-overlap-sharp}. Because
$(2/5)^2=4/25<1/6$, we have $1/\sqrt6>2/5$ and
$(6+\sqrt6)/12=(1+1/\sqrt6)/2>(1+2/5)/2=7/10$ (GO-3). Attainment of
$Q=1-\Phi$ at the HS orbits and the Bell-pair-product orbits is established in
Theorem~\ref{thm:fqe-critical-orbit-sharp}.
\end{proof}

\begin{proof}[Global part of Theorem~\ref{thm:fqe-orbit-stability}]
Theorem~\ref{thm:fqe-global-overlap} gives $d^2\le12Q/(6+\sqrt6)$, and \eqref{eq:fqe-orbit-surrogate} gives
$\delta\ge kQ\ge k\frac{6+\sqrt6}{12}d^2$. With $k>16/225$ and $\sqrt6>2449/1000$,
$k(6+\sqrt6)/12>\frac{16}{225}\cdot\frac{8449}{12000}>\frac1{20}$ (GO-4).
\end{proof}

\subsection{The transverse quadratic form and an explicit local radius}
\label{sec:fqe-orbit-local}
Use the magic representative $Z_* =\operatorname{diag}(a,ib,ib,ib)$,
$a=1/\sqrt2$, $b=1/\sqrt6$, which belongs to one of the HS orbits.
For a perturbation write
\[
 H=\begin{pmatrix}c&u^T\\v&D\end{pmatrix},\quad
 S=(D+D^T)/2,\quad A=(D-D^T)/2.
\]
The coefficient $Q_2(H)$ of $t^2$ in $Q(Z_*+tH)$ is
\begin{align}
 Q_2(H)={}&6\bigl(|c|^2+\|S\|_F^2-|ac-ib\operatorname{tr}S|^2\bigr)
       +20\|\operatorname{Re}A\|_F^2\notag\\
 &+30\sum_{j=1}^3\bigl[(a\operatorname{Im}u_j+b\operatorname{Re}v_j)^2
            +(b\operatorname{Re}u_j+a\operatorname{Im}v_j)^2\bigr].
 \label{eq:fqe-orbit-quadratic}
\end{align}
This is an explicit orthogonal decomposition. On the complex symmetric
block together with $c$, it is six times the norm squared perpendicular
to the complex line through $Z_*$. The real antisymmetric block has
coefficient twenty; the imaginary antisymmetric block is tangent to the
orbit. For each cross-block pair the two displayed linear forms are
orthogonal and have norm squared $a^2+b^2=2/3$, giving coefficient twenty
on the two normal directions. The other two directions are the left and
right real-rotation tangents. Thus the real quadratic-form operator has
eigenvalues $0,6,20$ with multiplicities $11,12,9$. Its kernel is exactly
the radial direction and the tangent space of $\mathcal M$ (including phase).
The actual Hessian eigenvalues are twice these quadratic coefficients.

For clarity, the formula can be derived with only quadratic matrix
expansions. On each cut let $C$ and $B$ be the flattenings of $Z_*$ and
$H$ in the computational basis, and put
$R=CC^\dagger$, $L=CB^\dagger+BC^\dagger$, $V=BB^\dagger$.
Then
\[
 P_2=\sum_{\rm cuts}\operatorname{tr}(L^2+2RV),\quad
 T_2=3\sum_{\rm cuts}\operatorname{tr}(R^2V+RL^2),\quad
 Q_2=18P_2-12T_2-21\|H\|_F^2-3n_1^2,
\]
where $n_1=2\operatorname{Re}\langle Z_*,H\rangle$; substitution gives
\eqref{eq:fqe-orbit-quadratic}. This also specifies every matrix entry
of the quadratic form without a numerical eigendecomposition.

To bound all higher terms, represent $Q(z)$ as a matrix element on
$z^{\otimes3}$. Each $N\operatorname{tr}R_j^2$ is a two-copy swap
matrix element with an identity third copy, and each
$\operatorname{tr}R_j^3$ is a three-cycle matrix element. Taking Hermitian
parts if necessary gives an operator of norm at most
$18\cdot3+12\cdot3+13=103$. Consequently, for $d=\|H\|\le r$,
\begin{equation}\label{eq:fqe-orbit-taylor}
 |Q(Z_*+H)-Q_2(H)|
 \le103d^3(20+15r+6r^2+r^3).
\end{equation}
The constant and linear terms vanish because $Q$ is nonnegative and
zero at $Z_*$. The binomial coefficients count the six tensor slots.

Take a closest $\chi\in\mathcal M$ to $\psi$ and $H=\psi-\chi$.
Stationarity makes $H$ real-orthogonal to the orbit tangent, which contains the phase direction $i\chi$, so
$\langle\chi,H\rangle$ is real, while
$\operatorname{Re}\langle\chi,H\rangle=-d^2/2$. Therefore
$Q_2(H)\ge6(d^2-d^4/4)$. The same bound applies at either orbit by LU
invariance and conjugation. Inserting $r=2^{-10}$ in
\eqref{eq:fqe-orbit-taylor} proves $Q\ge3d^2$ for $d\le2^{-10}$, used for the local
spectral-majorant transfer of Corollary~\ref{cor:fqe-orbit-spectral} of the paper.

\subsection{The sharp entropy Hessian and a larger local neighbourhood}
\label{sec:fqe-entropy-orbit-local}
Let
\[
 c_* =\frac{2-\ln3}{2\ln2}.
\]
The transverse quadratic coefficients of the entropy gap are
\begin{equation}\label{eq:fqe-entropy-hessian-spectrum}
 \begin{array}{c|ccc}
 \text{coefficient}&c_*&1/\ln2&(2+3\ln3)/(3\ln2)\\\hline
 \text{real multiplicity}&10&2&9
 \end{array}
\end{equation}
(the Hessian eigenvalues are twice these coefficients). The ten orbit
tangents are its kernel on the unit sphere. In particular
\begin{equation}\label{eq:fqe-entropy-local-sharp}
 \lim_{r\downarrow0}\ \inf_{0<d(\psi)\le r}
                  \frac{\delta(\psi)}{d(\psi)^2}=c_*.
\end{equation}
The Hessian determines the limiting coefficient. To obtain a bound on a whole neighbourhood, we also
control the higher-order terms through the following radius estimate.
\begin{lemma}[Explicit local radius]\label{lem:fqe-entropy-local-radius}
For $0<r\le1/20$ put
\[
 \epsilon_r=2\sqrt6\,r+6r^2,\qquad
 c(r)=\frac{1-r^2/4}{3\ln2}
 \left[(3-\epsilon_r)
       \left(1-\sqrt3\,r-\frac{r^2}2\right)^2
       -\frac32\ln3\right].
\]
Then
\begin{equation}\label{eq:fqe-entropy-local-radius}
 d(\psi)\le r\quad\Longrightarrow\quad
 \delta(\psi)\ge c(r)d(\psi)^2,\qquad c(1/20)>3/10.
\end{equation}
The bracket is positive and decreasing on this interval, and
$c(r)\to c_*$.
\end{lemma}

\begin{proof}
Fix a closest $\chi\in\mathcal M$, and write
$\psi=h\chi+s\nu$, where $h=1-d^2/2$, $s^2=d^2-d^4/4$,
$\|\nu\|=1$. The vector $\nu$ is complex-orthogonal to $\chi$
and real-orthogonal to its orbit tangents. On cut $j$ write $C_j,B_j$
for the flattenings of $\chi,\nu$, and set
\[
 \sigma_j=C_jC_j^\dagger=I/6+\Pi_j/3,\quad
 L_j=C_jB_j^\dagger+B_jC_j^\dagger,\quad
 V_j=B_jB_j^\dagger,\quad
 K=\sum_j\Pi_j\otimes I-\frac32 I.
\]
The sum is over the three cuts, with $\Pi_j$ acting on the corresponding
pair containing qubit one. Direct substitution at $Z_*$ gives $K\chi=0$.
The relative entropy identity, with natural logarithms, is exact:
\begin{equation}\label{eq:fqe-entropy-relative}
 3\ln2\,\delta(\psi)=\ln3\,\langle\psi,K\psi\rangle
                         +\sum_jD(\rho_j\|\sigma_j).
\end{equation}
Let $\mathcal L_j$ be the derivative of matrix logarithm at $\sigma_j$,
and put $U(\nu)=\sum_j\operatorname{tr}(L_j\mathcal L_j(L_j))$.
The quadratic entropy term is most simply diagonalised before estimating
$U$. Put $\Pi=\Pi_j$, $\Pi^\perp=I-\Pi$ on each cut and define the rational forms
\[
 \mathfrak a=\sum_j\bigl(\operatorname{tr}[(\Pi L_j\Pi)^2]
                         +3\operatorname{tr}[(\Pi^\perp L_j\Pi^\perp)^2]\bigr),\qquad
 \mathfrak b=\langle\nu,K\nu\rangle
                         +3\sum_j\|\Pi L_j\Pi^\perp\|_F^2.
\]
The divided differences of $\ln$ at $1/2,1/6$ give
\[
 U=2\bigl[\mathfrak a+\ln3\,(\mathfrak b-\langle\nu,K\nu\rangle)\bigr].
\]
Thus the coefficient of $s^2$ in \eqref{eq:fqe-entropy-relative} is
$(\mathfrak a+\ln3\,\mathfrak b)/(3\ln2)$: the $K$ terms cancel.

Here are explicit coordinates for the full real normal space at $Z_*$:
\begin{equation}\label{eq:fqe-entropy-normal-blocks}
 H=\begin{pmatrix}
 i\sqrt3\,z&(b\xi+ia\eta)^T\\
 b\eta+ia\xi&zI_3+E+A
 \end{pmatrix},\qquad a=1/\sqrt2,\quad b=1/\sqrt6,
\end{equation}
where $z\in\mathbb C$, $E$ is complex symmetric and traceless,
$A$ is real antisymmetric, and $\xi,\eta\in\mathbb R^3$.
Indeed complex orthogonality to $Z_*$ imposes $H_{00}=i\sqrt3\,z$;
orthogonality to lower rotation tangents makes the antisymmetric part
real; the two cross-block tangents impose the displayed relations.
These mutually orthogonal blocks have real dimensions $10,2,9$ and
norm squares
\[
 N_0=\|E\|_F^2,\qquad N_1=6|z|^2,\qquad
 N_2=\|A\|_F^2+\tfrac23(|\xi|^2+|\eta|^2).
\]
Substitution in the cut formulas gives
\begin{equation}\label{eq:fqe-entropy-normal-forms}
 \mathfrak a=3N_0+3N_1+2N_2,\qquad
 \mathfrak b=-\tfrac32N_0+3N_2.
\end{equation}
This proves the three coefficients and multiplicities in
\eqref{eq:fqe-entropy-hessian-spectrum} without spectral projectors.
For reproducibility, use the two unit traceless diagonal matrices and
three unit symmetric off-diagonal matrices in $E$, and their imaginary
multiples; use $z=1/\sqrt6,i/\sqrt6$; use the three unit antisymmetric
matrices in $A$; finally use each coordinate axis for $\xi,\eta$, scaled
by $\sqrt{3/2}$. EH-1 substitutes this orthonormal basis in all cut
forms and checks every coefficient of \eqref{eq:fqe-entropy-normal-forms}.

Direct substitution gives $K=-3/2$ on the $E$ block and $K=0$ on the
$z$ block. Hermiticity makes their orthogonal complement invariant.
On that complement $K\le3/2$, since it is the sum of three projectors
minus $3I/2$. Consequently $U=6$ per unit norm on the first two blocks
($\mathfrak b-\langle\nu,K\nu\rangle=0$ there),
and $U\ge4+3\ln3>6$ per unit norm on the last block. By \eqref{eq:fqe-entropy-normal-forms}, $\mathfrak a$ and $\mathfrak b$
have no cross terms between the three blocks, and neither has $\langle\nu,K\nu\rangle$, since the first two blocks are
eigenspaces of $K$; so $U\ge6\|\nu\|^2$ on all normals,
the least value being on the $E$ and $z$ blocks. The radius estimate below needs only the
bound $K\ge-3/2$, which holds on the whole space by the definition of $K$.

For the radius bound use the norm induced by $\mathcal L_j$.
Since $\|\mathcal L_j\|\le6$, $\operatorname{tr}\sigma_j^2=1/3$
and $\operatorname{tr}(V_j\sigma_j)\ge1/6$,
\[
 \sum_j\|V_j-\sigma_j\|_{\mathcal L_j}^2
 \le6\sum_j\operatorname{tr}[(V_j-\sigma_j)^2]\le18.
\]
Also $\rho_j\le(1+\epsilon_r)\sigma_j$: write the cut matrix as
$C_j+E_j$ with $\|E_j\|\le d$. The least singular value of $C_j$ is
$1/\sqrt6$, so $\|C_j^{-1}E_j\|\le\sqrt6\,d$, and
$\rho_j=C_j(I+C_j^{-1}E_j)(I+C_j^{-1}E_j)^\dagger C_j^\dagger
\le(1+\sqrt6\,r)^2\sigma_j=(1+\epsilon_r)\sigma_j$.
The resolvent formula for the logarithmic derivative implies
$\|X\|_{\mathcal L_A}^2\ge a^{-1}\|X\|_{\mathcal L_\sigma}^2$
when $A\le a\sigma$. To see the order directly, inverse resolvents
are ordered, their tensor squares are ordered, and integration with
a rescaled variable gives the factor $a^{-1}$.
Twice integrating the entropy Hessian along the segment therefore gives
\[
 D(\rho_j\|\sigma_j)\ge
 \left(\int_0^1\frac{1-v}{1+v\epsilon_r}\,dv\right)
           \|\rho_j-\sigma_j\|_{\mathcal L_j}^2
 \ge\left(\frac12-\frac{\epsilon_r}{6}\right)
           \|\rho_j-\sigma_j\|_{\mathcal L_j}^2.
\]
Now $\rho_j-\sigma_j=hsL_j+s^2(V_j-\sigma_j)$. The reverse
triangle inequality in the sum of the three induced norms gives,
since $h\sqrt6\ge s\sqrt{18}$ for $r\le1/20$,
\[
 \sum_j\|\rho_j-\sigma_j\|_{\mathcal L_j}^2
 \ge s^2(h\sqrt6-s\sqrt{18})^2.
\]
Insert this into \eqref{eq:fqe-entropy-relative}, use
$\langle\psi,K\psi\rangle=s^2\langle\nu,K\nu\rangle\ge-3s^2/2$,
and substitute $h\ge1-r^2/2$, $s\le r$ and
$s^2\ge d^2(1-r^2/4)$; since $(h\sqrt6-s\sqrt{18})^2=6(h-\sqrt3\,s)^2$ and
$6(\frac12-\frac{\epsilon_r}6)=3-\epsilon_r$, this proves \eqref{eq:fqe-entropy-local-radius}.
The rational bounds $\ln2<7/10$, $\ln3<1099/1000$,
$\sqrt6<49/20$ and $\sqrt3<1733/1000$
certify $c(1/20)>3/10$ (EH-2).

Finally the unit normal $H=\operatorname{diag}(0,1,-1,0)/\sqrt2$
belongs to the ten-dimensional least-coefficient space. Along
$(Z_*+tH)/\|Z_*+tH\|$, the distance to $\mathcal M$ is
$|t|+O(t^2)$ and the gap is $c_*t^2+O(t^3)$. This proves
\eqref{eq:fqe-entropy-local-sharp}, and also optimality of the power two.
\end{proof}

\section{Orbit stability: further proofs and results}\label{sec:sup-stability}
We supplement the stability results of Appendix~\ref{sec:fqe-orbit-stability} of the paper and Section~\ref{sec:fqe-stability-proofs}
with a constructive distance certificate, the proofs of the two sharp critical-locus theorems,
a one-parameter refinement on $P=1$, and the proof of the sharp entropy coefficient on the critical locus.
Throughout, $\mathcal M$ is the union of the two HS local-unitary orbits,
including global phase, and
\[
 d(\psi)=\min_{\chi\in\mathcal M}\|\psi-\chi\|,\qquad
 \Phi(\psi)=\max_{\chi\in\mathcal M}|\langle\psi,\chi\rangle|^2,\qquad
 d^2=2(1-\sqrt{\Phi}).
\]
For unit states, $Q=F+(P-1)$ is the sextic moment defect and
$\delta=E(M_4)-E$ is the entropy deficit, as in the paper.

\subsection{A direct certificate from three Bell-pair products}\label{sec:sup-bellframe}
Let $\tilde w=\epsilon^{\otimes4}\bar w$ be the spin flip, as in the paper; in the magic matrix coordinates it is ordinary
conjugation (Section~\ref{sec:sup-general-envelope}). Hence a unit state $w$ with
$\langle\tilde w,w\rangle=0$ has the form $(x+iy)/\sqrt2$ for orthonormal real vectors $x,y$ in the magic coordinates; put
$V=\operatorname{span}_{\mathbb R}(x,y)$. A \emph{Bell-pair product} is a product of two maximally entangled Bell states with arbitrary local
unitaries on the four qubits, with its pairing specified; it is real up to phase in these coordinates.

\begin{proposition}[Three-Bell-frame certificate]\label{prop:fqe-bell-frame}
Let $w$ be unit with $\langle\tilde w,w\rangle=0$. Choose one unit
Bell-pair product $b_j$ in each of the three pairings, and put
\[
 \varepsilon_j=1-2|\langle b_j,w\rangle|^2,\qquad
 E_B=\varepsilon_1+\varepsilon_2+\varepsilon_3.
\]
Then $\varepsilon_j\ge0$ and
\[
 \boxed{\quad d(w)^2\le\frac92 E_B.\quad}
\]
The vectors $b_j$ need not maximise their overlaps. The proof constructs
at most six candidate HS states, two for each unordered pair of pairings.
\end{proposition}
\begin{proof}
Choose phases so the $b_j$ are real. Their squared distances to $V$ are
exactly $\varepsilon_j$. Bell-pair products in distinct pairings have
$|\langle b_i,b_j\rangle|\le1/2$: in the magic matrix coordinates for
the first pairing they are a real rank-one matrix and $O/2$, $O\in O(4)$.
The proper and improper components describe the two other pairings.
This description follows by expanding the standard Bell products and
using $SU(2)\otimes SU(2)\to SO(4)$.

Let $G$ be the $3\times3$ real Gram matrix of the $b_j$, and let
$\lambda_3$ be its smallest eigenvalue. The rank-two projection on $V$
gives $\lambda_3\le E_B$. Since $\operatorname{tr}G=3$,
\[
 \operatorname{tr}G^2\ge\frac{(3-\lambda_3)^2}{2}+\lambda_3^2
                     \ge\frac92-3E_B.
\]
Consequently, for $\Delta_{ij}=1/4-|\langle b_i,b_j\rangle|^2$,
\begin{equation}\label{eq:fqe-bell-gram}
 \Delta_{ij}\ge0,\qquad \sum_{i<j}\Delta_{ij}\le\frac32 E_B.
\end{equation}

Fix a pair $(i,j)$ and use real rotations to make $b_i=E_{00}$
and $b_j=O/2$. Adjust its real sign so
$c=O_{00}/2=|\langle b_i,b_j\rangle|\in[0,1/2]$.
The lower $3\times3$ block of $O$ has singular values $1,1,2c$:
its product with its transpose is $I-zz^T$, with $\|z\|^2=1-4c^2$.
Choose its polar factor $R\in O(3)$, including a factor on any nullspace,
and put $O'=\operatorname{diag}(1,R)$, $b'_j=O'/2$.
The real plane $S=\operatorname{span}(b_i,b'_j)$ is spanned by
$E_{00}$ and $\operatorname{diag}(0,R)/\sqrt3$. Its two normalised
complex-null rays are precisely HS states, since their magic matrices
are $\operatorname{diag}(1/\sqrt2,\,\pm iR/\sqrt6)$.

Put $S_0=\operatorname{span}(b_i,b_j)$. Its Gram matrix has
smallest eigenvalue $1-c$.
Consequently its leakage from $V$ satisfies
\[
 \ell_{ij}:=\operatorname{tr}(P_{S_0}(I-P_V))
 \le\frac{\varepsilon_i+\varepsilon_j}{1-c}
 \le2(\varepsilon_i+\varepsilon_j).
\]
The planes $S_0,S$ share $b_i$. Their other principal-angle cosine is
\[
 r_c=\frac{1+c}{\sqrt{3(1-c^2)}}
     =\sqrt{\frac{1+c}{3(1-c)}}.
\]
Indeed the polar-aligned lower block of $O$ has trace $2+2c$.
The distance squared between the nearer normalised null rays of these
planes is $1-r_c$. Moreover
\[
 g(c):=1-r_c\le\frac74(1/4-c^2)=\frac74\Delta_{ij}.
\]
To check the last inequality, square its nonnegative right-hand
comparison with $r_c$. After multiplication by $3(1-c)$ the difference
is
$(1-2c)[13+c(525-462c)+588c^3(1-2c)]/256\ge0$.

For any two real two-planes with principal-angle cosines $s_1,s_2$,
choose orthonormal bases in which their overlap matrix has singular values
$s_1,s_2$. Reversing an orientation replaces a null ray
$(x+iy)/\sqrt2$ by its conjugate. Thus the largest overlap between
their normalised null rays is $(s_1+s_2)/2$, and their minimum
squared distance (after a phase choice) is
$2-s_1-s_2\le2-s_1^2-s_2^2$, the projection leakage.
The triangle inequality therefore gives
$d(w)\le\sqrt{\ell_{ij}}+\sqrt{g(c)}$.
Average its square over the three unordered pairs, use
\eqref{eq:fqe-bell-gram}, and apply Cauchy--Schwarz to the cross term:
\[
 d(w)^2\le\left(\sqrt{\frac43}+\sqrt{\frac78}\right)^2E_B
          \le\frac92 E_B,
\]
since $(\sqrt{4/3}+\sqrt{7/8})^2<9/2$. The last inequality is strict when $E_B>0$.
\end{proof}

For a general unit state choose phase and real orthonormal $x,y$ so
$\psi=\sqrt{(1+t)/2}\,x+i\sqrt{(1-t)/2}\,y$,
where $t=|\langle\tilde\psi,\psi\rangle|$, and set $w=(x+iy)/\sqrt2$.
At $t=1$, choose any real unit $y$ orthogonal to $x$.
The same constructive certificate gives
\[
 d(\psi)\le\sqrt{2-\sqrt{1+t}-\sqrt{1-t}}+\sqrt{9E_B(w)/2}.
\]
This individual-state certificate requires only chosen Bell products and
linear algebra; the entropy-only bound is proved in
Section~\ref{sec:fqe-global-overlap}.

\subsection{The critical locus: signed Birkhoff reduction and real-density model}
\label{sec:fqe-critical-orbit-proof}
A state is critical when all four one-qubit marginals are $I/2$. On this locus, both the overlap comparison and its Hilbert-space distance
version have optimal constants.

\begin{restated}[Sharp critical-state orbit stability; from Theorem~\ref{thm:fqe-global-overlap}, CO-1--CO-3]{Theorem}{thm:fqe-critical-orbit-sharp}
For every critical unit pure four-qubit state, with $\Phi,d,Q,k$
as in Appendix~\ref{sec:fqe-orbit-stability},
\begin{align*}
 Q(\psi)&\ge\max\{1-\Phi(\psi),\;3/2-2\Phi(\psi)\},\\
 Q(\psi)&\ge\frac{2+\sqrt2}{4}d(\psi)^2,\\
 \delta(\psi)&\ge k\bigl(\max\{1-\Phi,3/2-2\Phi\}+5\tau/2\bigr)
       \ge k\,\frac{2+\sqrt2}{4}\,d(\psi)^2.
\end{align*}
The coefficient one in $Q\ge1-\Phi$, and the coefficient
$(2+\sqrt2)/4$ in \eqref{eq:fqe-critical-distance}, are both optimal:
Bell-pair products attain them. The comparison $Q\ge1-\Phi$ holds for all states
(Theorem~\ref{thm:fqe-global-overlap}).
If the critical state also has $\tau=0$, equivalently if $P=1$, then
\begin{equation*}
 \Phi\ge\frac{2+\sqrt3}{6},\qquad
 Q\ge\frac3{4-\sqrt3}(1-\Phi)\ge\frac{3+\sqrt3}{4}d^2.
\end{equation*}
These coefficients and the fidelity floor are optimal, attained by cluster states.
\end{restated}
Section~\ref{sec:sup-nullsharp} below proves a one-parameter sharpening of
\eqref{eq:fqe-null-critical-sharp} in the Bloch coordinates of \eqref{eq:fqe-null-critical-bloch}. The optimal entropy
coefficient on the whole critical locus is larger: $\delta\ge C_*d^2$ with $C_*=0.19057894701\ldots$ \eqref{eq:fqe-p1-cstar}
(Theorem~\ref{thm:fqe-p1-sharp}), optimal by its equality states; $k(2+\sqrt2)/4$ in \eqref{eq:fqe-critical-entropy}
is the weaker coefficient of the moment route.

We first remove the continuous orbit optimisation exactly.
For complex diagonal matrices $D_v,D_w$ of any common size,
\begin{equation}\label{eq:fqe-signed-diagonal-overlap}
 \max_{A,B\in O(m)}|\operatorname{tr}(D_v^\dagger A D_wB^T)|
 =\max_{\pi,\,\varepsilon_i=\pm1}
                  \left|\sum_i\bar v_i\varepsilon_iw_{\pi(i)}\right|.
\end{equation}
Indeed $|A_{ij}B_{ij}|$ is doubly substochastic, by Cauchy--Schwarz
on each row and column. Complete it to a doubly stochastic matrix;
for any common phase the real overlap is bounded by the largest
assignment sum of the absolute real coefficient entries. Birkhoff's
permutation decomposition gives this bound, and signed permutation
matrices attain it. Maximising the phase proves
\eqref{eq:fqe-signed-diagonal-overlap}.

The magic normal form makes a critical state $D_v$. Here and below the coordinates of $v\in\mathbb C^4$, and the
rows and columns of $4\times4$ matrices, are indexed $0,\dots,3$: $v_i$ is the coefficient of $m_{i+1}\otimes m_{i+1}$ in
$G(v)$ of Appendix~\ref{sec:fqe-normal-form} of the paper, so $D_v=\operatorname{diag}(v)$.
Up to global phase, the union $\mathcal M$ is the $O(4)\times O(4)$ orbit
of $Z_*$: a common orientation reversal in both factors fixes $Z_*$
up to proper rotations, whereas a single reversal gives its conjugate
up to proper rotations. Thus its overlap with $D_v$ is exactly the
maximum over the following 32 diagonal representatives:
\[
 \mathcal H=\left\{\eta:\ \eta_i=1/\sqrt2,\quad
         \eta_j=\varepsilon_j i/\sqrt6\ (j\ne i),\quad i=0,1,2,3\right\}.
\]
Its exact spin-flip-dependent covering profile is proved in
Lemma~\ref{lem:fqe-critical-real-density} below.

The real cut transforms are $H_0=I$ and
\[
 H_1=\frac12\begin{pmatrix}
 1&-1&1&-1\\-1&1&1&-1\\1&1&1&1\\-1&-1&1&1
 \end{pmatrix},\qquad
 H_2=\frac12\begin{pmatrix}
 1&-1&-1&-1\\-1&1&-1&-1\\1&1&-1&1\\-1&-1&-1&1
 \end{pmatrix}.
\]
With $\Delta=\operatorname{diag}(1,-1,1,-1)$, so that $\Delta v$ are the coordinates $(a,b,c,d)$ of $G_{abcd}$, these are
the transforms $H_\pm$ of Section~\ref{sec:sup-sextic-zeros} conjugated by $\Delta$: $H_1=\Delta H_+\Delta$, and $H_2$
is $\Delta H_-\Delta$ with its rows permuted and re-signed, which only permutes the squared moduli $|e_i^TH_2v|^2$ of the cut spectrum.

\begin{proof}[Proof of Theorem~\ref{thm:fqe-critical-orbit-sharp}]
The comparison $Q(\psi)\ge1-\Phi(\psi)$ holds for all states by
Theorem~\ref{thm:fqe-global-overlap}. When $\Phi(\psi)\ge\tfrac12$,
$1-\Phi(\psi)\ge\tfrac32-2\Phi(\psi)$, so \eqref{eq:fqe-critical-fidelity}
holds. Assume henceforth that $\Phi(v)\le\tfrac12$ for a unit critical
normal vector $v\in\mathbb C^4$.

The three real orthogonal cut transforms $H_0=I,H_1,H_2$ displayed above
preserve $N=\|v\|^2=1$, $\tau=|\sum_i v_i^2|^2$ and $Q(v)$, and permute
the $32$ representatives $\mathcal H$ up to the sixth-root-of-unity
phases $(\pm1\pm i\sqrt3)/2$: for any $\eta\in\mathcal H$, one coordinate
is $1/\sqrt2$ and three are $\pm i/\sqrt6$, so under $H_1$ or $H_2$ the
row whose relative signs on those three coordinates match $\eta$ has inner
product $\frac{\pm1\pm i\sqrt3}{2\sqrt2}$ and the other three rows have
$\frac{\pm1\pm i\sqrt3}{2}(\mp i/\sqrt6)$ (CO-1). Thus $H_0,H_1,H_2$,
coordinate permutations, global phase and coordinate sign flips preserve
both $Q(v)$ and $\Phi(v)$. Replace $v$ by $H_mv$ ($m\in\{0,1,2\}$) and
permute coordinates so that
$p_0:=|v_0|^2=\max_{m,i}|e_i^TH_mv|^2$ is the largest of all twelve cut
probabilities. Choose global phase so that $v_0=r_0>0$, and flip signs
$v_j\mapsto\pm v_j$ ($1\le j\le3$) so that $y_j:=\operatorname{Im}v_j\ge0$,
writing $v_j=x_j+iy_j$ and $p_j=x_j^2+y_j^2\le p_0=r_0^2$. Put
$X_2:=\sum_{j=1}^3x_j^2$, $Y_2:=\sum_{j=1}^3y_j^2=1-p_0-X_2$,
$S_x:=\sum_{j=1}^3|x_j|$ and $S_y:=\sum_{j=1}^3y_j$.

Testing $\Phi(v)$ against
$Z_*=(1/\sqrt2,i/\sqrt6,i/\sqrt6,i/\sqrt6)^T\in\mathcal H$ gives
\[
 \Phi(v)\ge\Phi_{\rm im}(v):=\left(\frac{r_0}{\sqrt2}+\frac{S_y}{\sqrt6}\right)^2
 =\frac12-\frac12X_2+\frac13\Bigl(\sqrt3\,r_0S_y-Y_2+\sum_{1\le k<\ell\le3}y_ky_\ell\Bigr).
\]
Since $\Phi_{\rm im}(v)\le\Phi(v)\le\tfrac12$, we have
$S_y\le\sqrt3(1-r_0)$, hence $Y_2\le S_y^2\le3(1-r_0)^2$ and
$X_2=1-r_0^2-Y_2\ge2(1-r_0)(2r_0-1)$. Moreover, since
$0\le y_j\le\sqrt{p_j}\le r_0$ and $\sqrt3<\tfrac{26}{15}<\tfrac74$
($26^2=676>675$), we also have
$Y_2\le r_0S_y\le\tfrac{26}{15}r_0(1-r_0)\le\tfrac74r_0(1-r_0)$ and
$X_2\ge\tfrac1{15}(1-r_0)(15-11r_0)$.

Up to sign, the eight rows of $H_1$ and $H_2$ are
$\tfrac12(1,\varepsilon_1,\varepsilon_2,\varepsilon_3)$ for
$\varepsilon\in\{\pm1\}^3$, so the eight cut probabilities in $H_1v$ and
$H_2v$ are
\[
 \lambda_\varepsilon
 =\frac14\left[\left(r_0+\sum_{j=1}^3\varepsilon_jx_j\right)^2
   +\left(\sum_{j=1}^3\varepsilon_jy_j\right)^2\right]
 =\frac14\left(1+2\sum_{j=1}^3\varepsilon_jc_{0j}
   +2\sum_{1\le k<\ell\le3}\varepsilon_k\varepsilon_\ell c_{k\ell}\right)\le p_0,
\]
where $c_{ij}:=\operatorname{Re}(\bar v_iv_j)=x_ix_j+y_iy_j$
($c_{0j}=r_0x_j$). Choosing $\varepsilon_j=\operatorname{sgn}(x_j)$ gives
$(r_0+S_x)^2\le4p_0$, so $S_x\le r_0$. For each $j\in\{1,2,3\}$, averaging
$\lambda_\varepsilon\le p_0$ over the four sign vectors with
$\varepsilon_j=\operatorname{sgn}(x_j)$ cancels the other terms and yields
$\tfrac14(1+2r_0|x_j|)\le r_0^2$, so $|x_j|\le\frac{4r_0^2-1}{2r_0}$ and
$X_2\le(\max_j|x_j|)S_x\le2r_0^2-\tfrac12$. Combining this with
$Y_2\le\tfrac74r_0(1-r_0)$ gives
\[
 1=r_0^2+X_2+Y_2\le3r_0^2-\frac12+\frac74r_0(1-r_0),\qquad
 0\le\frac14(5r_0-3)(r_0+2),
\]
hence $r_0\ge\tfrac35$. Furthermore, for $\varepsilon_j=\operatorname{sgn}(x_j)$
we have $(\sum_{j=1}^3\varepsilon_jy_j)^2\ge Y_2-2\sum_{k<\ell}y_ky_\ell
=2Y_2-S_y^2\ge2(1-p_0-X_2)-3(1-r_0)^2$, so
\[
 (r_0+S_x)^2\le4p_0-2(1-p_0-X_2)+3(1-r_0)^2=(3r_0-1)^2+2X_2
 \le\left(3r_0-1+\frac{X_2}{3r_0-1}\right)^2,
\]
and therefore $S_x\le\min\bigl(r_0,\,2r_0-1+\frac{X_2}{3r_0-1}\bigr)$. For
$s_2:=\sum_{1\le k<\ell\le3}|x_k||x_\ell|=\tfrac12(S_x^2-X_2)$, if
$X_2\le(1-r_0)(3r_0-1)$ then
\[
 2s_2\le\left(2r_0-1+\frac{X_2}{3r_0-1}\right)^2-X_2
 =(2r_0-1)^2-\frac{X_2\bigl((1-r_0)(3r_0-1)-X_2\bigr)}{(3r_0-1)^2}
 \le(2r_0-1)^2,
\]
while if $X_2\ge(1-r_0)(3r_0-1)$ then
$2s_2\le r_0^2-X_2\le(2r_0-1)^2$. Thus $2s_2\le(2r_0-1)^2$ and
$q_4:=\sum_{1\le k<\ell\le3}x_k^2x_\ell^2\le s_2^2\le\tfrac14(2r_0-1)^4$
for all $X_2\ge0$.

On the critical locus, expanding $Q(v)$ in terms of $p_i=|v_i|^2$ and
$c_{ij}=\operatorname{Re}(\bar v_iv_j)$ gives the exact identity (CO-1)
\begin{equation}\label{eq:fqe-critical-three-term}
 Q(v)=\frac12\Bigl(\sum_{i=0}^3p_i\Bigr)^3
      +18\sum_{j=1}^3D_j\bigl(c_{0j}^2-c_{k\ell}^2\bigr),
\end{equation}
where for each $j\in\{1,2,3\}$ and $\{k,\ell\}=\{1,2,3\}\setminus\{j\}$ we
set $D_j:=p_0+p_j-p_k-p_\ell=2(p_0+p_j)-1=1-2(p_k+p_\ell)\le
D_{\max}:=\min(1,4p_0-1)$. Using $c_{0j}^2=p_0x_j^2$,
$c_{k\ell}^2\le2(x_k^2x_\ell^2+y_k^2y_\ell^2)$ when $D_j\ge0$, and
$c_{k\ell}^2\ge0$ when $D_j<0$, we obtain with $D_j^+:=\max(D_j,0)$ the
decoupled lower bound
\[
 Q(v)+2\Phi(v)-\frac32\ge Q(v)+2\Phi_{\rm im}(v)-\frac32
 \ge\mathcal Y(r_0,y)+\mathcal X(p;x^2),
\]
where
\begin{align*}
 \mathcal Y(r_0,y)&:=\frac23\Bigl(\sqrt3\,r_0S_y-Y_2
   +\sum_{1\le k<\ell\le3}y_ky_\ell\Bigr)-36\sum_{j=1}^3D_j^+y_k^2y_\ell^2,\\
 \mathcal X(p;x^2)&:=\sum_{j=1}^3(18p_0D_j-1)x_j^2
   -36\sum_{j=1}^3D_j^+x_k^2x_\ell^2.
\end{align*}

We show that both $\mathcal Y(r_0,y)$ and $\mathcal X(p;x^2)$ are
nonnegative (CO-2). For $\mathcal Y(r_0,y)$, the four-term AM--GM
inequality on $(2r_0+1)+3(2r_0-1)+4(1-r_0)+4(1-r_0)=6$ gives
$48(4r_0^2-1)(1-r_0)^2\le(3/2)^4=81/16$, so
$9D_{\max}S_y^2\le27(4r_0^2-1)(1-r_0)^2\le729/256<3$. Since
$Y_2=S_y^2-2\sum_{k<\ell}y_ky_\ell$ and
$\sum_{k<\ell}y_k^2y_\ell^2\le\tfrac14S_y^2\sum_{k<\ell}y_ky_\ell$
(as $y_ky_\ell\le\tfrac14(y_k+y_\ell)^2\le\tfrac14S_y^2$), we have
\[
 \mathcal Y(r_0,y)\ge\frac23S_y(\sqrt3\,r_0-S_y)
   +\bigl(2-9D_{\max}S_y^2\bigr)\sum_{1\le k<\ell\le3}y_ky_\ell
 \ge\frac23S_y(\sqrt3\,r_0-S_y)-\sum_{1\le k<\ell\le3}y_ky_\ell.
\]
Using $\sum_{k<\ell}y_ky_\ell\le\tfrac13S_y^2$ and
$S_y\le\sqrt3(1-r_0)$ gives
$\mathcal Y(r_0,y)\ge\tfrac13S_y(2\sqrt3\,r_0-3S_y)
\ge\tfrac{\sqrt3}{3}S_y(5r_0-3)\ge0$.

For $\mathcal X(p;x^2)$, substituting
$D_j=(2p_0-1)+2x_j^2+2y_j^2\ge(2p_0-1)+2x_j^2$ and
$\sum_{j=1}^3x_j^4=X_2^2-2q_4$ into the first sum yields
\[
 \mathcal X(p;x^2)\ge\bigl(18p_0(2p_0-1)-1\bigr)X_2+36p_0(X_2^2-2q_4)
   -36\sum_{j=1}^3D_j^+x_k^2x_\ell^2.
\]
If $X_2=0$, then $\mathcal X=0$; assume $X_2>0$. On
$r_0\in[\tfrac{19}{25},1]$, we have $p_0\ge\tfrac{361}{625}>\tfrac12$, so
$D_j=1-2(p_k+p_\ell)\ge0$ and $D_j\le1-2(x_k^2+x_\ell^2)$. Hence
$\sum_{j=1}^3D_j^+x_k^2x_\ell^2\le q_4-2(X_2q_4-3x_1^2x_2^2x_3^2)
\le(1-\tfrac43X_2)q_4$ by $3x_1^2x_2^2x_3^2\le\tfrac13X_2q_4$, and since
$2p_0+1-\tfrac43X_2>0$ and $q_4\le\tfrac13X_2^2$,
\[
\begin{aligned}
 \frac{\mathcal X(p;x^2)}{X_2}&\ge18p_0(2p_0-1)-1-12(1-p_0)X_2+16X_2^2\\
 &=\left(4X_2-\frac32(1-p_0)\right)^2+\frac{135p_0^2-54p_0-13}{4}
 \ge\frac{16573}{78125}>0.
\end{aligned}
\]
On $r_0\in[\tfrac35,\tfrac{19}{25}]$, using $D_j^+\le D_{\max}$ and
$q_4\le\tfrac14(2r_0-1)^4$ gives
\[
 \frac{\mathcal X(p;x^2)}{X_2}\ge18p_0(2p_0-1)-1+36p_0X_2
   -\frac{9(2p_0+D_{\max})(2r_0-1)^4}{X_2},
\]
which is strictly increasing in $X_2>0$. On $[\tfrac35,\tfrac{16}{25}]$,
setting $D_{\max}=4p_0-1$ and
$X_2\ge X_{2,1}(r_0):=\tfrac1{15}(1-r_0)(15-11r_0)$ and multiplying by
$15X_{2,1}(r_0)>0$ gives the degree-six polynomial
$P_1(r_0)=15\bigl((36r_0^4-18r_0^2-1)X_{2,1}(r_0)+36r_0^2X_{2,1}(r_0)^2
-9(6r_0^2-1)(2r_0-1)^4\bigr)$, which is strictly concave on
$[\tfrac35,1]$ ($P_1''(\tfrac35+u)=-368208u^4-\tfrac{2057376}{5}u^3
-\tfrac{2990232}{25}u^2-\tfrac{1591704}{125}u-\tfrac{226078}{625}<0$ for all
$u\ge0$) with $P_1(\tfrac35)=\tfrac{3693}{78125}>0$ and
$P_1(\tfrac{16}{25})=\tfrac{216609912}{1220703125}>0$. On
$[\tfrac{16}{25},\tfrac{19}{25}]$, setting $D_{\max}=1$ and
$X_2\ge X_{2,2}(r_0):=2(1-r_0)(2r_0-1)$ and multiplying by
$\frac{X_{2,2}(r_0)}{2r_0-1}=2(1-r_0)>0$ gives the degree-five polynomial
$P_2(r_0)=72r_0^5-432r_0^4+432r_0^3-54r_0^2-52r_0+7$, which is strictly
concave on $[\tfrac35,1]$
($P_2''(\tfrac35+u)=-36\bigl(8u^2(9-5u)+\tfrac{288}{5}u+3\bigr)<0$ for
$u\in[0,\tfrac25]$) with $P_2(\tfrac{16}{25})=\tfrac{988047}{9765625}>0$
and $P_2(\tfrac{19}{25})=\tfrac{570453}{9765625}>0$. Thus
$\mathcal X(p;x^2)\ge0$ in all cases, proving \eqref{eq:fqe-critical-fidelity}.

Let $x=\sqrt{\Phi}$ and $c=(2+\sqrt2)/4$. The $32$ overlaps
$q_\eta=|\langle\eta,v\rangle|^2$ sum to $8$, since the frame operator is $8I$ (CO-1), so
$\Phi=\max_\eta q_\eta\ge1/4$ and $x\ge1/2$.
For $x\ge1/\sqrt2$, the first bound gives
$1-x^2\ge2c(1-x)$. For $1/2\le x\le1/\sqrt2$, the second gives
\[
 3/2-2x^2-2c(1-x)
 =2(1/\sqrt2-x)\bigl(x-(2-\sqrt2)/4\bigr)\ge0.
\]
Since $d^2=2(1-x)$, this proves \eqref{eq:fqe-critical-distance}.
Equation~\eqref{eq:fqe-orbit-surrogate} gives \eqref{eq:fqe-critical-entropy}, with
$\mathcal D=Q+5\tau/2$ on critical states and the bound just proved as its last step. Bell products have
$Q=\Phi=1/2$, hence $d^2=2-\sqrt2$; both constants are sharp.
The null-critical refinement is proved in the next paragraph.
\end{proof}

\paragraph{Two auxiliary Bloch spheres on the null-critical locus.}
Let $M=(m_1,m_2,m_3,m_4)$ be the unitary $4\times4$ matrix whose columns are the magic basis vectors of
Appendix~\ref{sec:fqe-normal-form} of the paper, written in the basis
$|00\rangle,|01\rangle,|10\rangle,|11\rangle$, so that $Mv=\sum_kv_km_{k+1}$ for $v\in\mathbb C^4$. For the critical
normal form $D_v$, regard $Mv$ as an auxiliary two-qubit vector.
Its coefficient matrix has determinant $\sum_jv_j^2/2$.
Thus $\tau=0$ is equivalent to $Mv=a\otimes b$ for unit auxiliary
qubits $a,b$. These are coordinates on the critical normal form, not
physical tensor factors of the original four-qubit state.
Let $r,s\in S^2$ be their Bloch vectors. Direct Bell probabilities give
\begin{equation}\label{eq:fqe-null-critical-bloch}
 Q=\frac12+\frac{27}{2}r_1r_2r_3s_1s_2s_3,\qquad
 \Phi=\frac14\max_{\substack{\varepsilon,\eta\in\{\pm1\}^3\\
                         \prod_j\varepsilon_j\eta_j=-1}}
 \left(1+\frac{\varepsilon\cdot r}{\sqrt3}\right)
 \left(1+\frac{\eta\cdot s}{\sqrt3}\right).
\end{equation}
Indeed, the four probabilities $|v_j|^2$ are
$(1+\sum_\alpha\chi_{j\alpha}r_\alpha s_\alpha)/4$, where
\[
 (\chi_{j\alpha})=
 \begin{pmatrix}1&-1&1\\1&1&-1\\-1&-1&-1\\-1&1&1\end{pmatrix}.
\]
Their second and third power sums in
\eqref{eq:fqe-global-intrinsic}, with $Z=D_v$ and $\operatorname{tr}(Z^TZ)=0$, give
the first identity. The 32 HS representatives become exactly the
auxiliary product rays whose Bloch vectors are cube vertices
$\varepsilon/\sqrt3,\eta/\sqrt3$ of opposite parity. The product-state
overlap formula and \eqref{eq:fqe-signed-diagonal-overlap} give the second
identity. These finite identities are checked independently by NC-1--NC-3.

Here is an elementary proof of the geometric claims in
\eqref{eq:fqe-null-critical-sharp}; NC-4 checks its scalar identities and constants.
Write $A=1-1/\sqrt3$, $B=1+1/\sqrt3$, and
$f_0=B^2/4=(2+\sqrt3)/6$.
First we prove $\Phi\ge f_0$. If the signs of the coordinates of
$r,s$ have opposite parity, align both cube vertices with those signs;
$\|r\|_1,\|s\|_1\ge1$ gives the claim. A zero coordinate allows either
parity. In the remaining case put $L=\|r\|_1$, $m=\min_j|r_j|$ and
define $f_r=1+L/\sqrt3$, $g_r=1+(L-2m)/\sqrt3$, with analogous
quantities for $s$. Flipping the sign at a smallest coordinate gives
\[
 4\Phi\ge\max(f_rg_s,g_rf_s)
                  \ge\sqrt{f_rg_rf_sg_s}.
\]
For the ordered coordinate moduli $x\ge y\ge z=m$ of either unit vector,
\[
 f_rg_r-B^2=\frac2{\sqrt3}(x+y-1)+\frac23(xy-z^2)\ge0.
\]
Here $xy\ge z^2$ and $(x+y)^2=1-z^2+2xy\ge1$.
This proves the fidelity floor in all cases.

If $r_1r_2r_3s_1s_2s_3\ge0$, then $Q\ge1/2$, and with $\Phi\ge f_0$ this gives
$Q\ge(1-\Phi)/(2(1-f_0))=\frac3{4-\sqrt3}(1-\Phi)$.
Otherwise put $\mu=3\sqrt3|r_1r_2r_3|$ and
$\nu=3\sqrt3|s_1s_2s_3|$, so $0\le\mu,\nu\le1$ and $2Q=1-\mu\nu$.
For a nonnegative unit triple $x,y,z$,
\[
 x+y+z-1
 =\frac{2(xy+yz+zx)}{x+y+z+1}
 \ge(9-3\sqrt3)xyz.
\]
For positive coordinates this follows from
$1/x+1/y+1/z\ge3\sqrt3$ and $x+y+z\le\sqrt3$;
zero coordinates follow by continuity.
Thus $\|r\|_1\ge1+(\sqrt3-1)\mu$ and similarly for $s$.
Aligning opposite-parity signs in \eqref{eq:fqe-null-critical-bloch} gives
$4\Phi\ge \Lambda:=(B+A\mu)(B+A\nu)$. The identity
\[
 \Lambda-B^2-(4-B^2)\mu\nu=AB\bigl(\mu(1-\nu)+\nu(1-\mu)\bigr)\ge0
\]
gives $1-\Phi\le(1-f_0)(1-\mu\nu)=2(1-f_0)Q$, the infidelity bound. In both cases the distance bound follows
pointwise: $1-\Phi=\frac{1+\sqrt\Phi}2\,d^2\ge\frac{1+\sqrt{f_0}}2\,d^2$, and
$\frac3{4-\sqrt3}\cdot\frac{1+\sqrt{f_0}}2=\frac{3+\sqrt3}4$ because $\sqrt{f_0}=B/2$.
The critical cluster representative $v=(1,i,0,0)/\sqrt2$ has
$Q=1/2$ and $\sqrt{\Phi}=B/2$, attaining all three geometric
bounds.

\begin{lemma}[Real-density geometry and sharp critical covering profile; exact certificate, RP, FP]
\label{lem:fqe-critical-real-density}
Write a unit critical normal vector as $v=x+iy$, with $x,y\in\mathbb R^4$,
and put $A=xx^T+yy^T$. Then $A$ ranges over exactly the real positive
semidefinite matrices of trace one and rank at most two, and
\begin{equation}\label{eq:fqe-critical-density-measurements}
 \lambda_{ji}=e_i^TH_jAH_j^Te_i,\qquad
 \tau=2\operatorname{tr}A^2-1.
\end{equation}
Here $H_0,H_1,H_2$ are the three real orthogonal cut transforms displayed
above. Their bases are mutually unbiased. For their twelve unit vectors
$h_{ji}=H_j^Te_i$ and rank-one projectors $B_{ji}=h_{ji}h_{ji}^T$,
\begin{equation}\label{eq:fqe-critical-real-tomography}
 A=\sum_{j,i}\lambda_{ji}B_{ji}-\tfrac12I,\qquad
 \sum_{j,i}\lambda_{ji}^2=\operatorname{tr}A^2+\tfrac12.
\end{equation}
For each coordinate axis $e_k$ and each of the four lines
$u=(0,\pm1,\pm1,\pm1)/\sqrt3$ in its orthogonal coordinate three-space,
put $P=e_ke_k^T+uu^T$ and $A_P=P/2$. There are sixteen such planes.
With $F_{\rm B}(A,B)=(\operatorname{tr}\sqrt{\sqrt B A\sqrt B})^2$,
\begin{equation}\label{eq:fqe-critical-bures-orbit}
 \Phi(v)=\max_P F_{\rm B}(A,A_P),\qquad
 F_{\rm B}(A,A_P)=\frac12\left[\operatorname{tr}(PA)
                    +2\sqrt{\det\bigl((e_k,u)^TA(e_k,u)\bigr)}\right].
\end{equation}
Thus $d(v)$ is exactly the minimum Bures distance from $A$ to these
sixteen real density matrices. The entire critical locus obeys the sharp bound
\begin{equation}\label{eq:fqe-critical-fidelity-floor}
 \Phi\ge f_{\rm cov}(\tau):=
       \frac13+\frac{\sqrt{1-\tau}}{2\sqrt3},\qquad
 \max_{\substack{\psi\ {\rm critical}\\\tau(\psi)=t}}d(\psi)^2
       =2-2\sqrt{f_{\rm cov}(t)}\quad(0\le t\le1).
\end{equation}
The bound is attained for every $t$ by the cluster-to-GHZ family below.
In particular every critical state has $d^2\le2-2/\sqrt3$, with equality for GHZ.
For any one such plane, put
$\lambda^P_{ji}=\operatorname{tr}(B_{ji}A_P)$ and
$D_P=\sum_{j,i}\lambda_{ji}\ln(\lambda_{ji}/\lambda^P_{ji})$.
The entropy gap has the exact form
\begin{equation}\label{eq:fqe-critical-relative-entropy}
 3\ln2\,\delta=D_P-\tfrac32\ln3\,J_P,\qquad
 J_P=1-\operatorname{tr}(PA)
    =\|A-A_P\|_F^2-\tau/2.
\end{equation}
\end{lemma}

\begin{proof}
The three physical cut vectors are $H_jv$, so their squared moduli are
the stated real measurements. Conversely, for any real $A$ of the
specified rank, its spectral decomposition gives
$v=\sqrt{a_1}e_1+i\sqrt{a_2}e_2$; rank one is included.
The identity $|v^Tv|^2=2\operatorname{tr}A^2-1$ proves the spin-flip claim.
The twelve real lines are the coordinate lines and the half-sign lines
of the regular $24$-cell. Their tight frame identity, valid for every
real symmetric matrix $C$, is
\[
 \sum_{j,i}\operatorname{tr}(B_{ji}C)B_{ji}
                    =C+\tfrac12\operatorname{tr}(C)I.
\]
Taking its trace pairing with $C$ proves both tomography formulas (RP-1).

The two HS rays associated with $P$ are $(e_k\pm iu)/\sqrt2$.
Their squared overlaps with $v$ have sum $\operatorname{tr}(PA)$ and
half their difference has squared value
$\det((e_k,u)^TA(e_k,u))$. Maximising over this pair therefore gives
\eqref{eq:fqe-critical-bures-orbit}, by the two-dimensional formula
$(\operatorname{tr}\sqrt C)^2=\operatorname{tr}C+2\sqrt{\det C}$.
The signed-Birkhoff reduction already proved shows that maximising
these sixteen pairwise maxima is the full physical HS fidelity.

For $q_\eta=|\langle\eta,v\rangle|^2$, the finite HS frame satisfies
\[
 \sum_{\eta\in\mathcal H}q_\eta=8,\qquad
 \sum_{\eta\in\mathcal H}q_\eta^2=\tfrac89(4-\tau).
\]
Both follow by expanding and cancelling odd sign powers (RP-3).
The stronger sharp profile \eqref{eq:fqe-critical-fidelity-floor} follows
from the area-weighted calibration proved below; the moment identity
alone does not give that profile.

Each reference distribution has one entry $1/2$ in each cut and nine
entries $1/6$ in total. The sum of the three projectors selecting its
large entries is $3P/2$ (RP-3). Expanding $D_P$ then gives
\eqref{eq:fqe-critical-relative-entropy}. Finally
$\operatorname{tr}A_P^2=1/2$ and
$2\operatorname{tr}(AA_P)=\operatorname{tr}(AP)$ give the leakage formula.
All matrix and compression identities here are independently checked by RP;
the covering calibration uses FP.
\end{proof}

\paragraph{Area-weighted covering calibration.}
We prove the profile in \eqref{eq:fqe-critical-fidelity-floor} using an
explicit probability measure on the sixteen HS planes. Its only
polynomial positivity input is the following finite exact certificate.

\begin{lemma}[Four-point cubic inequality; exact rational certificate]
\label{lem:fqe-four-point-cubic}
If $X,Y\in\mathbb R^4$ satisfy
$\sum_iX_i=\sum_iY_i=0$ and $\sum_iX_i^2=\sum_iY_i^2=4$, then
\begin{equation}\label{eq:fqe-four-point-cubic}
 \sum_{i,j=1}^4|X_i-Y_j|^3\ge64.
\end{equation}
Equality holds at $X=Y=(-1,-1,1,1)$.
\end{lemma}

\begin{proof}
Put $X=Er$, $Y=Es$, where
\[
 E=\begin{pmatrix}1&1&1\\1&-1&-1\\-1&1&-1\\-1&-1&1\end{pmatrix},
 \qquad E^TE=4I,
\]
so that $\|r\|=\|s\|=1$. Every zero-sum four-vector is represented.
Sort $X$ and $Y$ separately and interleave their entries in increasing
order. There are $\binom84=70$ words with four letters of each kind.
Swapping $X,Y$ and simultaneously negating both vectors reduce these
to exactly twenty-three classes (FP-1). Ties may use either order.
For each word let $z_1,\ldots,z_8$ be its ordered entries and
$g_i=z_{i+1}-z_i\ge0$. The absolute values then give a cubic polynomial
$p=\sum_{ij}|X_i-Y_j|^3/16-4$.
For each of the twenty-three words the supplied rational data give
the identity
\begin{equation}\label{eq:fqe-four-point-cubic-sos}
 p=B_0^TG_0B_0+\sum_{i=1}^7g_i B_i^TG_iB_i
       +(\|r\|^2-1)\ell_r+(\|s\|^2-1)\ell_s.
\end{equation}
The $\ell$ are quadratic polynomials and all eight $G_i$ are positive
definite. FP-2 rebuilds all $210$ coefficients through degree four
in each class and proves positivity of all $184$ Grams using integer
arithmetic. Thus the right side is nonnegative on each ordering cone
and the two spheres, proving the full inequality. The bases and exact
matrix representation are specified in Section~\ref{sec:sup-verifiers}.
The stated equality follows by substitution.
\end{proof}

Let $U=(U_1,U_2)$ be a real orthonormal $4\times2$ frame with plane projector
$\Pi=UU^T$. For each HS plane $P$ use its orthonormal frame
$V_P=(e_k,u)$ and put $c_P=|\det(U^TV_P)|$. The sixteen planes obey
\begin{equation}\label{eq:fqe-area-weighted-fusion}
 \sum_Pc_P^2=\frac83,\qquad
 \sum_Pc_P^2P=\frac89(I+\Pi).
\end{equation}
Indeed, the bivector $\Delta=U_1U_2^T-U_2U_1^T$ satisfies $\Delta\Delta^T=\Delta^T\Delta=\Pi$, and
$\det(U^TV_P)=e_k^T\Delta u$. Fix $k$, put $g=\Delta^Te_k$ (so $g_k=0$,
$|g|^2=\Pi_{kk}$) and sum over the four lines $u$ for this $k$, using $u_j=\pm1/\sqrt3$ ($j\ne k$):
$\sum_u(g\cdot u)^2=\frac43|g|^2$ and
$\sum_u(g\cdot u)^2uu^T=\frac49|g|^2(I-e_ke_k^T)+\frac89(gg^T-\operatorname{diag}(g_j^2))$.
Summing over $k$ with $\sum_kgg^T=\Pi$ and $\sum_k\operatorname{diag}(g_j^2)=\operatorname{diag}(\Pi)$ gives
both identities. FP-3 also checks them without
assuming orthonormality first, using the homogeneous squared area and
its associated matrix. Therefore the explicit weights
$w_P=3c_P^2/8$ satisfy
\begin{equation}\label{eq:fqe-plane-calibration}
 \sum_Pw_P=1,\qquad \sum_Pw_PU^TPU=\frac23I_2,\qquad
 \sum_Pw_Pc_P\ge\frac1{\sqrt3}.
\end{equation}

To prove the last inequality, let $M$ be the magic-basis matrix defined before
\eqref{eq:fqe-null-critical-bloch}. Form the null critical vector
$z=(U_1+iU_2)/\sqrt2$. Its auxiliary product representation
$Mz=a\otimes b$ has unit Bloch vectors $r,s$, as in
\eqref{eq:fqe-null-critical-bloch}.
For the two HS rays $\eta_P=(e_k+iu)/\sqrt2$ and $\bar\eta_P$,
expansion gives
\[
 c_P=\bigl||\langle\eta_P,z\rangle|^2
                 -|\langle\bar\eta_P,z\rangle|^2\bigr|.
\]
In the cube-pair description of the HS rays given after \eqref{eq:fqe-null-critical-bloch}, conjugation reverses both cube vertices.
Indeed $M=\epsilon^{\otimes2}\bar M$, where
$\epsilon=\left(\begin{smallmatrix}0&1\\-1&0\end{smallmatrix}\right)$
(FP-3), and the qubit spin flip reverses the Bloch vector.
Choose the four even-parity cube vertices $e_i$, the rows of $E$.
The sixteen conjugate pairs are represented once by
$(\varepsilon,\eta)=(e_i,-e_j)$, and hence
\[
 c_P=\frac{|e_i\cdot r-e_j\cdot s|}{2\sqrt3}.
\]
The four numbers $X_i=e_i\cdot r$, and likewise $Y_j=e_j\cdot s$,
have sum zero and sum of squares four. Lemma~\ref{lem:fqe-four-point-cubic}
therefore gives
\[
 \sum_Pc_P^3=\frac{\sum_{ij}|X_i-Y_j|^3}{24\sqrt3}
       \ge\frac8{3\sqrt3},
\]
which proves the last calibration inequality.

Now write the real density matrix of any critical state as
$A=U\Lambda U^T$, with $\operatorname{tr}\Lambda=1$ and
$\det\Lambda=(1-\tau)/4$. Rank one is included by completing its
support to any plane. The exact Bures formula gives
\begin{align*}
 \Phi
 &\ge\sum_Pw_P F_{\rm B}(A,P/2)\\
 &=\frac12\operatorname{tr}\!\left(\Lambda\sum_Pw_PU^TPU\right)
                  +\sqrt{\det\Lambda}\sum_Pw_Pc_P\\
 &\ge\frac13+\frac{\sqrt{1-\tau}}{2\sqrt3}.
\end{align*}
This proves the profile for the entire critical locus.

Finally, for $a,b\ge0$, $a+b>0$, the critical normal family
\begin{equation}\label{eq:fqe-covering-equality-family}
 v=\frac{(a+ib,-a+ib,0,0)}{\sqrt{2(a^2+b^2)}},\qquad
 \tau=\frac{(a^2-b^2)^2}{(a^2+b^2)^2},\qquad
 \Phi=\frac13+\frac{ab}{\sqrt3(a^2+b^2)}
\end{equation}
attains equality. FP-4 checks every one of the $32$ ray overlaps against
this maximum. The parameters cover $0\le\tau\le1$, proving pointwise
optimality and the conditional covering radius. The endpoint $a=b$
is a cluster state; $b=0$ is GHZ up to LU.

This model expresses both the entropy and the optimised orbit distance exactly.
In particular, a bound on the classical relative entropy $D_P$ that
controls the explicit plane leakage $J_P$ can improve the entropy
constant without first improving a moment surrogate. Its domain is a
real rank-two density matrix; replacing it by all real density matrices
would admit $I/4$, whose entropy exceeds the critical-state maximum.

\subsection{A one-parameter family of sharp geometric inequalities on the slice $P=1$}\label{sec:sup-nullsharp}
\begin{theorem}[Sharper geometric family on $P=1$; MS]\label{thm:sup-null-family}
Let $\psi$ be a critical unit pure four-qubit state with $\tau=0$, equivalently $P=1$, and let $\Phi,d,Q$ be as in
Theorem~\ref{thm:fqe-critical-orbit-sharp}. Use the two auxiliary Bloch vectors of \eqref{eq:fqe-null-critical-bloch} to put
$X=3\sum r_j^4-1$ and $Y=3\sum s_j^4-1$. Then
\begin{equation}\label{eq:fqe-null-critical-family}
 \begin{gathered}
 \Phi\ge\frac{2+\sqrt3}{6},\qquad 0\le\theta\le\frac1{24},\\
 Q+\theta XY\ge(1+8\theta)\frac3{4-\sqrt3}(1-\Phi)
       \ge(1+8\theta)\frac{3+\sqrt3}{4}d^2.
 \end{gathered}
\end{equation}
For every $\theta$, both geometric coefficients and the fidelity floor are optimal, attained by cluster states.
\end{theorem}
The case $\theta=0$ is the inequality \eqref{eq:fqe-null-critical-sharp} of the paper. The endpoint $\theta=\tfrac1{24}$ is proved by
an elementary multiaffine vertex check in the opposite-parity sector and a univariate polynomial minorant in the same-parity sector.
\paragraph{Sharp mixed-moment geometry.}
With the auxiliary Bloch vectors $r,s$ of
\eqref{eq:fqe-null-critical-bloch}, put
$X=3\sum r_j^4-1$ and $Y=3\sum s_j^4-1$.
We prove the endpoint $\theta=1/24$ in
\eqref{eq:fqe-null-critical-family} by elementary polynomial inequalities in both parity sectors.
Interpolation with the already proved $\theta=0$ bound, which is the inequality
\eqref{eq:fqe-null-critical-sharp} of the paper, then covers
$0\le\theta\le1/24$.

First, every nonnegative unit triple $a,b,c$ obeys the strengthened norm
inequality
\begin{equation}\label{eq:fqe-strengthened-sphere-norm}
 L:=a+b+c\ge1+(\sqrt3-1)u+\tfrac14(2-X-2u),\quad
 u=3\sqrt3abc,\quad X=3(a^4+b^4+c^4)-1.
\end{equation}
To see this, put $B=9-9\sqrt3/2$ and use
$a^4+b^4+c^4=1-(L^2-1)^2/2+4Labc$.
The desired residual is
\[
 L-1-\tfrac38(L^2-1)^2+(3L-B)abc.
\]
For $1\le L\le\sqrt2$, its first three terms equal
$(L-1)[1-3(L-1)(L+1)^2/8]\ge0$, since
$(\sqrt2-1)(\sqrt2+1)^2=1+\sqrt2<8/3$; also $3L>B$.
For $\sqrt2\le L\le\sqrt3$, Schur gives
$abc\ge L(L^2-2)/9$. The resulting lower bound is
\[
 f(L)=-L^4/24+(\sqrt3/2-1)L^3+L^2/12+(3-\sqrt3)L-11/8.
\]
It is concave on that interval, with
$f(\sqrt2)=\sqrt2-11/8>0$ and $f(\sqrt3)=0$, proving the claim.
All these finite identities are checked by MS-1.

Suppose first that $\prod_jr_js_j\le0$.
Put $u=3\sqrt3|r_1r_2r_3|$, $v=3\sqrt3|s_1s_2s_3|$.
For a unit triple $r$, put $t_i=r_i^2$ and $q_r=\sum_{i<j}t_it_j$. The identity
\[
 q_r^2-3t_1t_2t_3
 =\tfrac12\bigl[(t_1t_2-t_2t_3)^2+(t_2t_3-t_3t_1)^2
                         +(t_3t_1-t_1t_2)^2\bigr]\ge0
\]
gives the elementary bound
\begin{equation}\label{eq:fqe-bloch-triple-product}
 0\le u\le3q_r=1-X/2,\qquad 0\le v\le1-Y/2.
\end{equation}
Write $p=1-u$, $q=1-v$, $X=2p(1-\alpha)$ and
$Y=2q(1-\beta)$, with $p,q,\alpha,\beta\in[0,1]$; a zero
$p$ or $q$ is handled by continuity.
Aligning the two cube signs is admissible in this parity sector, so
\eqref{eq:fqe-strengthened-sphere-norm} gives
\[
 4\Phi\ge f_rf_s,\qquad
 f_r=2-(1-1/\sqrt3)p+\frac{p\alpha}{2\sqrt3},\quad
 f_s=2-(1-1/\sqrt3)q+\frac{q\beta}{2\sqrt3}.
\]
Set $a_{24}=4/(4-\sqrt3)$. The polynomial
\[
 \frac{p+q-pq}{2}+\frac{pq(1-\alpha)(1-\beta)}6
                   -a_{24}(1-f_rf_s/4)
\]
is affine in each of its four variables separately.
At the sixteen cube vertices its distinct values are
\[
 0,\quad \frac{3+4\sqrt3}{26},\quad
 \frac{-6+5\sqrt3}{78},\quad
 \frac{2+7\sqrt3}{52},\quad \frac{3+4\sqrt3}{78},
\]
all nonnegative (MS-1). This proves
$Q+XY/24\ge a_{24}(1-\Phi)$ in this sector.

For the remaining parity sector ($\prod_{j=1}^3 r_j s_j\ge0$), sort the
coordinate moduli of $r$ as $a_r\ge b_r\ge c_r\ge0$ and those of $s$ as
$a_s\ge b_s\ge c_s\ge0$. Swap $r$ and $s$ if necessary so that
$c_r\ge c_s$, and set
\[
 x=\sqrt3\,c_r\in[0,1],\quad y=\sqrt3\,c_s\in[0,x],\qquad
 \tilde w_r=3a_rb_r,\quad \tilde w_s=3a_sb_s.
\]
For a sorted unit triple $a\ge b\ge c\ge0$ with $x=\sqrt3\,c$ and
$\tilde w=3ab$, one has $a+b=\sqrt{1-c^2+2ab}=\sqrt{1+(2\tilde w-x^2)/3}$,
$3\sqrt3\,abc=x\tilde w$, and
$\tilde w\in\bigl[x\sqrt{3-2x^2},\,(3-x^2)/2\bigr]$, where the lower and
upper bounds correspond to $b=c$ and $a=b$. Moreover (MS-2),
\[
 X=2-2x^2\Bigl(1-\frac{x^2}{3}\Bigr)-\frac23\tilde w^2
  \ge X_{\rm lin}(x,\tilde w)
  :=2-2x^2\Bigl(1-\frac{x^2}{3}\Bigr)-\Bigl(1-\frac{x^2}{3}\Bigr)\tilde w\ge0,
\]
since $X-X_{\rm lin}=\frac23\tilde w\bigl(\frac{3-x^2}{2}-\tilde w\bigr)=3ab(a-b)^2\ge0$
and $X_{\rm lin}(x,\tilde w)=\frac12(1-x^2)^2+\bigl(1-\frac{x^2}{3}\bigr)\bigl(\frac{3-x^2}{2}-\tilde w\bigr)\ge0$.
Using the branch $(1+(a_r+b_r+c_r)/\sqrt3)(1+(a_s+b_s-c_s)/\sqrt3)$ of $4\Phi$ gives
$4\Phi\ge f_r(x,\tilde w_r)\,g_s(y,\tilde w_s)$ with
\[
 f_r(x,\tilde w_r)=1+\frac1{\sqrt3}\sqrt{1+\frac{2\tilde w_r-x^2}{3}}+\frac x3,\qquad
 g_s(y,\tilde w_s)=1+\frac1{\sqrt3}\sqrt{1+\frac{2\tilde w_s-y^2}{3}}-\frac y3.
\]
Hence $Q+XY/24-a_{24}(1-\Phi)$ is bounded below by
\[
 \mathcal F(x,y,\tilde w_r,\tilde w_s)
 :=\frac{1+xy\tilde w_r\tilde w_s}{2}
   +\frac{X_{\rm lin}(x,\tilde w_r)\,Y_{\rm lin}(y,\tilde w_s)}{24}
   +\frac{a_{24}}{4}f_r(x,\tilde w_r)\,g_s(y,\tilde w_s)-a_{24}.
\]
Since $(3-2x^2)-\bigl(\frac{5-2x}{3}\bigr)^2=\frac{2(1-x)(1+11x)}{9}\ge0$,
we have $\tilde w_r\ge x\sqrt{3-2x^2}\ge\tilde w_{\rm low}(x):=\frac{x(5-2x)}{3}\ge x$
and $\tilde w_s\ge\tilde w_{\rm low}(y)\ge y$. Consequently $2\tilde w_s-y^2\ge\frac{y(10-7y)}{3}\ge0$, so
\[
 Y_{\rm lin}(y,\tilde w_s)\le2-\frac23y
 \le\frac{2\sqrt3}{\sqrt3+1}\Bigl(1+\frac1{\sqrt3}-\frac y3\Bigr)
 \le\frac{2\sqrt3}{\sqrt3+1}\,g_s(y,\tilde w_s),
\]
and likewise $X_{\rm lin}(x,\tilde w_r)\le2\le\frac{2\sqrt3}{\sqrt3+1}\,f_r(x,\tilde w_r)$.
Since $1+(2\tilde w_r-x^2)/3\le2$, the partial derivative with respect to $\tilde w_r$ satisfies
\[
 \frac{\partial\mathcal F}{\partial\tilde w_r}
 \ge Y_{\rm lin}(y,\tilde w_s)\Bigl(\frac{\sqrt3+1}{18\sqrt2\,(4-\sqrt3)}-\frac1{24}\Bigr)\ge0,
\]
and the same calculation gives $\partial\mathcal F/\partial\tilde w_s\ge0$ (MS-2). Monotonicity in $(\tilde w_r,\tilde w_s)$
therefore reduces the inequality to $\tilde w_r=\tilde w_{\rm low}(x)$ and
$\tilde w_s=\tilde w_{\rm low}(y)$.

At $\tilde w=\tilde w_{\rm low}(x)$ one has $x\tilde w_{\rm low}(x)=\frac{x^2(5-2x)}{3}\ge x^2$,
\[
 X_{\rm lin}\bigl(x,\tilde w_{\rm low}(x)\bigr)-\frac49x^2(1-x)^2
 =\frac19(1-x)(18+3x-13x^2)=:X_b(x)\ge0,
\]
and $\sqrt{1+(2\tilde w_{\rm low}(x)-x^2)/3}=\frac13\sqrt{9+10x-7x^2}\ge S_b(x):=1+\frac12x-\frac7{20}x^2$,
since $(9+10x-7x^2)-9S_b(x)^2=\frac{x(10-7x)(63x^2-90x+40)}{400}\ge0$ on $[0,1]$ (MS-2).
Putting $\hat f(x)=1+S_b(x)/\sqrt3+x/3$ and $\hat g(y)=1+S_b(y)/\sqrt3-y/3$, we obtain
\begin{equation}\label{eq:fqe-null-mixed-certificate}
 Q+\frac{XY}{24}-a_{24}(1-\Phi)
 \ge P(x,y):=\frac{1+x^2y^2}{2}+\frac{X_b(x)X_b(y)}{24}+\frac{a_{24}}{4}\hat f(x)\hat g(y)-a_{24}
\end{equation}
for $0\le y\le x\le1$. Write $B=1+1/\sqrt3$. Three univariate polynomials on $[0,1]$ close the bound (MS-3):
\begin{enumerate}
\item $\hat g(y)-\frac B2 X_b(y)=\frac y2 q_2(y)\ge0$, where
 $q_2(y)=\bigl(1+\frac{8\sqrt3}{9}\bigr)+\bigl(\frac{16}{9}+\frac{97\sqrt3}{270}\bigr)y-\bigl(\frac{13}{9}+\frac{13\sqrt3}{27}\bigr)y^2$
 is concave on $[0,1]$ with $q_2(0)>0$ and $q_2(1)=\frac43+\frac{23\sqrt3}{30}>0$.
\item Consequently, for $0\le y\le x\le1$,
 \[
  \frac{\partial P}{\partial x}(x,y)
  =xy^2+\frac{a_{24}}{4}\hat f'(x)\hat g(y)+\frac{X_b'(x)}{24}X_b(y)
  \ge\frac{X_b(y)}{4-\sqrt3}\,q_1(x)\ge0,
 \]
 where $q_1(x):=(4-\sqrt3)\bigl(\frac{a_{24}B}{8}\hat f'(x)+\frac{X_b'(x)}{24}\bigr)
 =\bigl(\frac{11}{20}-\frac35x\bigr)^2+\bigl(\frac{5\sqrt3}{24}-\frac{1189}{3600}\bigr)
 +\bigl(\frac{17\sqrt3}{540}-\frac{133}{2700}\bigr)x+\bigl(\frac{163}{450}-\frac{13\sqrt3}{72}\bigr)x^2>0$ on $[0,1]$.
\item Hence $P(x,y)\ge P(y,y)=yq_3(y)\ge0$ on $[0,1]$, where
 \begin{align*}
  (4-\sqrt3)\,q_3(y)
  &=\Bigl(\frac{13}{25}-\frac{26}{25}y-\frac1{10}y^2\Bigr)^2
   +\Bigl(\frac{11\sqrt3}{18}-\frac{5896}{5625}\Bigr)
   +\Bigl(\frac{737}{7500}-\frac{19\sqrt3}{360}\Bigr)y\\
  &\quad+\Bigl(\frac{173411}{202500}-\frac{79\sqrt3}{162}\Bigr)y^2
   +\Bigl(\frac{756757}{486000}-\frac{419\sqrt3}{972}\Bigr)y^3\\
  &\quad-\Bigl(\frac{21043}{24300}-\frac{52\sqrt3}{243}\Bigr)y^4
   +\frac{169(4-\sqrt3)}{1944}y^5>0,
 \end{align*}
 since every coefficient through $y^3$ and at $y^5$ is strictly positive while the $y^3$ coefficient ($>4/5$)
 exceeds the $y^4$ coefficient ($<1/2$) on $0\le y\le1$. Since $P(0,0)=0$, this proves $P(x,y)\ge0$ on the entire triangle $0\le y\le x\le1$.
\end{enumerate}

Finally,
\[
 Q+\theta XY=(1-24\theta)Q+24\theta(Q+XY/24)
 \ge(1+8\theta)\frac3{4-\sqrt3}(1-\Phi).
\]
The fidelity floor already proved converts this to the displayed distance
bound. Clusters have $Q=1/2$, $X=Y=2$ and
$\Phi=(2+\sqrt3)/6$, attaining both coefficients for every
$\theta$ in the stated interval (MS-4). Thus the full geometric family
is sharp.

\subsection{The minimum-purity slice}
\label{sec:fqe-p1-sharp}
Statements are proved; the finite verifications are interval-certified (outward-rounded ball
arithmetic, every sign decision a certain comparison) or exact, and are listed in
Section~\ref{sec:sup-verifiers} (PM, PS).
This section gives the coordinates, symmetries and equality geometry on the slice $P=1$ ($\tau=0$) that
Section~\ref{sec:fqe-critical-sharp} uses for the whole critical locus.

\paragraph{Geometry of the equality states.}
At an equality point $\delta$ is stationary in every direction, because one Bloch vector is an axis
 and $\delta$ is even in the vanishing coordinate of the other, while four
admissible cube pairs tie (PM-4) and, computationally, their $d^2$-gradients form a regular tetrahedron around $0$ (explanatory; not used). The distance $d^2$
therefore has a conical corner and $\delta-C_*d^2$ has a conical minimum in the three transverse directions;
the one smooth direction, which rotates the second vector within its coordinate plane, is the wall of
Lemma~\ref{lem:fqe-p1-wall}, where the wall function $G_w$ is stationary.  This picture is explanatory; the proof uses the
exact identities of Lemma~\ref{lem:fqe-critical-local-cone}.

\subsubsection*{Coordinates and inputs}
Let $r,s$ be the unit auxiliary Bloch vectors of \eqref{eq:fqe-null-critical-bloch}; indices are taken mod $3$.
\begin{lemma}[Cut spectra; PM-1]\label{lem:fqe-p1-cuts}
For $k\in\mathbb Z/3$ put $u^{(k)}_a=r_{a+k}s_a$ and let
\[
 \mathcal S=\{(-,-,-),(-,+,+),(+,-,+),(+,+,-)\}.
\]
The three physical cuts have the spectra $\{(1+\sigma\cdot u^{(k)})/4:\sigma\in\mathcal S\}$, $k=0,1,2$ in some order.
Hence $E=\frac13\sum_kg(u^{(k)})$, $g(u)=\sum_{\sigma\in\mathcal S}\eta\bigl((1+\sigma\cdot u)/4\bigr)$,
$\eta(x)=-x\log_2x$, $\delta=1+\tfrac12\log_23-E$, and $\sum_k|u^{(k)}|^2=|r|^2|s|^2=1$.
\end{lemma}
\begin{proof}
Let $h_{ji}=H_j^Te_i$ be the unit vectors of Lemma~\ref{lem:fqe-critical-real-density}. Since $h_{ji}$ is real,
$Mh_{ji}$ is a Bell ray; write
$C^{ji}_{\alpha\beta}=\langle Mh_{ji}|\sigma_\alpha\otimes\sigma_\beta|Mh_{ji}\rangle$ for its correlation matrix, which is real
orthogonal with determinant $-1$. Its projector is $\tfrac14(I+\sum_{\alpha\beta}C^{ji}_{\alpha\beta}\sigma_\alpha\otimes\sigma_\beta)$, so
the probability $\lambda_{ji}=|h_{ji}^Tv|^2=|\langle Mh_{ji},Mv\rangle|^2$ is $(1+r^TC^{ji}s)/4$, as $Mv=a\otimes b$.
\emph{Cut $0$.} From the definitions of the $m_k$, $(\sigma_\alpha\otimes\sigma_\alpha)m_k=s^\alpha_km_k$ with
$s^x=(+,+,-,-)$, $s^y=(-,+,-,+)$, $s^z=(+,-,-,+)$. So $C^{0i}$ has diagonal $(s^x_i,s^y_i,s^z_i)$, entries $\pm1$, and being
orthogonal it is diagonal; the four diagonals $(+,-,+),(+,+,-),(-,-,-),(-,+,+)$ are the elements of $\mathcal S$.
\emph{Cuts $1,2$.} Let $u=\tfrac12\bigl(-I+i(\sigma_z-\sigma_x-\sigma_y)\bigr)$, a unitary with $u\sigma_\beta u^\dagger=\sum_\alpha
R_{\alpha\beta}\sigma_\alpha$, where $Re_1=e_2$, $Re_2=-e_3$, $Re_3=-e_1$. Direct multiplication gives
$MH_2M^\dagger=u\otimes u^\dagger$ and $MH_1M^\dagger=(u\otimes u^\dagger)\,\mathrm{SWAP}$, where SWAP exchanges the two auxiliary
qubits and transposes correlation matrices. Since $Mh_{ji}=(MH_jM^\dagger)^\dagger m_i$, this gives
$C^{2i}=R^TC^{0i}R^T$ and $C^{1i}=RC^{0i}R$. As $R_{\alpha\beta}\ne0$ only for $\alpha=\beta+1$, these are the signed
permutations $C^{ji}_{\alpha\beta}=\sigma_\beta[\alpha=\beta+k_j]$ with $(k_j)=(0,2,1)$; the four column-sign triples $\sigma$ of
each cut are distinct and have product $\det C^{ji}=-1$, so they run over $\mathcal S$. Hence
$r^TC^{ji}s=\sigma\cdot u^{(k_j)}$. PM-1 recomputes all twelve $C^{ji}$ exactly from the physical magic basis.
\end{proof}
\begin{lemma}[Symmetries; PM-2]\label{lem:fqe-p1-sym}
$\delta$ and $d^2$ are invariant under: even sign changes of the six coordinates; cyclic shifts of $r$ alone and of $s$
alone; and the exchange $r\leftrightarrow s$.  (The stated operations permute both the admissible cube pairs and the twelve
probabilities. Fidelity alone is also invariant under arbitrary coordinate permutations
of either vector.)
\end{lemma}

\subsubsection*{The wall}
Let $h(t)=H_{\rm b}((1+t)/2)$, with $H_{\rm b}$ the binary entropy defined in Appendix~\ref{sec:fqe-critical-orbit-stability}.
For $r=e_1$ and $s=(\cos\theta,0,\sin\theta)$ (a point of the wall) write
$\delta_w(\theta)=\frac12\log_23-\frac13(1+h(\cos\theta)+h(\sin\theta))$,
$d_w^2(\theta)=2-\sqrt{B(1+(\cos\theta+\sin\theta)/\sqrt3)}$, $B=1+1/\sqrt3$, and $G_w=\delta_w-C_*d_w^2$.
\begin{lemma}[Wall]\label{lem:fqe-p1-wall}
$G_w(\pi/4)=0$ by \eqref{eq:fqe-p1-cstar}, and $G_w(\theta)=G_w(\pi/2-\theta)$; hence $G_w'(\pi/4)=0$.
\end{lemma}
\begin{proof}
The reflection $\theta\mapsto\pi/2-\theta$ exchanges $\cos\theta$ and $\sin\theta$, so it exchanges $h(\cos\theta)$ and
$h(\sin\theta)$ and fixes $\cos\theta+\sin\theta$; thus $\delta_w$ and $d_w^2$ are both invariant.  $G_w$ is smooth near
$\pi/4$ (there $\cos\theta,\sin\theta\in(0,1)$, where $h$ is analytic), and an even function of $\theta-\pi/4$ has zero
derivative at $\pi/4$.
\end{proof}

\subsection{The sharp entropy coefficient on the whole critical locus}
\label{sec:fqe-critical-sharp}
Statements are proved; the finite verifications are interval-certified (outward-rounded ball arithmetic, every sign
decision a certain comparison) or exact, and are listed in Section~\ref{sec:sup-verifiers} (TC, EX, LCC, CCV).
This section proves Theorem~\ref{thm:fqe-p1-sharp} for all critical states.
Section~\ref{sec:fqe-p1-sharp} supplies the coordinates, symmetries, wall function and equality geometry on the slice
$\tau=0$ ($P=1$) that are used here.

The constant $C_*$ is defined in \eqref{eq:fqe-p1-cstar}; the equality states attain it exactly (PM-4), and PS-1 encloses its decimal value.

\begin{restated}[Sharp entropy stability on the critical locus; interval-certified]{Theorem}{thm:fqe-p1-sharp}
Every critical unit pure four-qubit state satisfies
\[
 \delta(\psi)\ \ge\ C_*\,d(\psi)^2 ,
\]
and equality with $d>0$ holds exactly for states whose critical normal form has
one of the following $144$ Bloch-coordinate pairs, all with $P=1$. In the auxiliary Bloch coordinates $(r,s)\in S^2\times S^2$ of the null-critical normal form they are: one of $r,s$ is a signed
coordinate axis and the other is $(\pm e_b\pm e_c)/\sqrt2$, $b\ne c$ ($6\cdot12$ pairs, twice for the swap).
These pairs attain the bound, so $C_*$ is the optimal coefficient on the critical locus. Consequently every universal coefficient in
$\delta\ge Cd^2$ is at most $C_*$, and the question left open is whether $C_*$ is optimal over all states.
\end{restated}

\paragraph{Strategy.}
On the critical locus the twelve cut probabilities are linear in a $3\times3$ correlation matrix $T$, and
$\delta=\delta(T)$ is convex. The fidelity is the maximum of sixteen explicit plane functions. A critical state is
determined by $(r,s,w)$ with unit vectors $r,s$ (the $P=1$ coordinates) and a disc coordinate $w\in\mathbb R^2$,
$|w|^2=\tau$; the equality states sit at $w=0$. Analytic bounds settle $\tau\ge0.36$ (their crossing is at $\tau\approx0.35879$, TC-6); for $\tau\le0.37$ a
six-dimensional interval covering, together with a local cone lemma at the equality points, closes the proof.

\subsubsection*{Critical states as correlation matrices}
Let $M$ be the magic matrix, $v$ a unit critical normal vector, $A=\operatorname{Re}(vv^\dagger)$, and
$\psi=Mv$ the auxiliary two-qubit vector.  Put $T_{\alpha\beta}=\langle\psi|\sigma_\alpha\otimes\sigma_\beta|\psi\rangle$,
$t_{ji}=4\lambda_{ji}-1$ with $\lambda_{ji}$ the cut probabilities of Lemma~\ref{lem:fqe-critical-real-density}.
\begin{lemma}[Correlation-matrix model; analytic, exact check TC-1]\label{lem:fqe-critical-correlation}
\begin{enumerate}
\item The twelve rays $Mh_{ji}$ are Bell rays without local terms; their correlation matrices $C^{ji}$ are signed
permutation matrices (those of Lemma~\ref{lem:fqe-p1-cuts}), and $t_{ji}=\langle C^{ji},T\rangle$.
Consequently, by \eqref{eq:fqe-critical-real-tomography},
$A=\tfrac14I+\tfrac14\sum_{j,i}t_{ji}B_{ji}$: $A$ is an affine function of $T$.
\item $T=R_a\operatorname{diag}(\kappa,-\kappa,1)R_b^{T}$ with $R_a,R_b\in SO(3)$ and $\kappa=\sqrt\tau$.  Writing
$r=R_ae_3$, $s=R_be_3$ and $x_a,y_a$ ($x_b,y_b$) for the first two columns of $R_a$ ($R_b$),
\[
 T=rs^T+\kappa\bigl(x_ax_b^T-y_ay_b^T\bigr).
\]
Rotating the frames about $r$ and $s$ by $\theta_a,\theta_b$ turns $\kappa(x_ax_b^T-y_ay_b^T)$ into
$w_1(x_ax_b^T-y_ay_b^T)+w_2(x_ay_b^T+y_ax_b^T)$ with $w=\kappa(\cos(\theta_a+\theta_b),\sin(\theta_a+\theta_b))$.
\item For a plane $\Pi=VV^T$ of Lemma~\ref{lem:fqe-critical-real-density} the matrix $m_\Pi=V^TAV$ is affine in $T$ and
$\Phi=\max_\Pi\Phi_\Pi$, $\Phi_\Pi=\tfrac12\operatorname{tr}m_\Pi+\sqrt{\det m_\Pi}$.  Equivalently
$\Phi=\max\tfrac14\bigl(1+\sqrt{1-\tau}\,(\varepsilon\!\cdot\! r+\eta\!\cdot\! s)/\sqrt3+\varepsilon^TT\eta/3\bigr)$ over the
$32$ cube pairs of \eqref{eq:fqe-null-critical-bloch}.
\end{enumerate}
\end{lemma}
\begin{proof}
(1) The projector onto a maximally entangled ray is $\tfrac14(I+\sum C_{\alpha\beta}\sigma_\alpha\otimes\sigma_\beta)$,
with no local terms; the proof of Lemma~\ref{lem:fqe-p1-cuts} derives the twelve $C^{ji}$, and $t_{ji}=\langle C^{ji},T\rangle$
because $\lambda_{ji}$ is the expectation of that projector in $\psi$ (PM-1 and TC-1 recompute them, TC-1 together
with the identity $\lambda_{ji}=(\|v\|^2+\langle C^{ji},T\rangle)/4$ exactly for a random rational $v$).
(2) A pure two-qubit vector is $(U\otimes V)(\cos\varphi|00\rangle+\sin\varphi|11\rangle)$ with $0\le\varphi\le\pi/4$; its
correlation matrix is $\operatorname{diag}(\sin2\varphi,-\sin2\varphi,1)$ in the frames rotated by $U,V$, and the concurrence
$\sin2\varphi=|\sum_jv_j^2|=\sqrt\tau$ (the coefficient matrix of $Mv$ has determinant $\sum_jv_j^2/2$).  The local Bloch
vectors are $\cos2\varphi\,r$ and $\cos2\varphi\,s$, of length $\sqrt{1-\tau}$.
(3) $m_\Pi$ is affine in $A$.  The first formula is \eqref{eq:fqe-critical-bures-orbit}.  The second is the overlap of
$\psi$ with a product ray with Bloch vectors $\varepsilon/\sqrt3,\eta/\sqrt3$:
$\langle\psi|\tfrac12(1+\hat\varepsilon\cdot\sigma)\otimes\tfrac12(1+\hat\eta\cdot\sigma)|\psi\rangle
=\tfrac14(1+\hat\varepsilon\cdot r_A+\hat\eta\cdot r_B+\hat\varepsilon^TT\hat\eta)$.
TC-2 and TC-3 check the affine representation of $A$, the equality of the two maxima for 200 random states, and the singular
values $(1,\kappa,\kappa)$, $\det T=-\kappa^2$.
\end{proof}
Thus $\tau=|w|^2$, and a critical state is determined (up to the discrete choices) by $(r,s,w)$.  At $\tau=0$ one recovers the
auxiliary Bloch vectors of \eqref{eq:fqe-null-critical-bloch}.  In these variables
\[
 \begin{gathered}
 \delta=\mathrm{EMAX}-\tfrac13\sum_{j,i}\operatorname{ent}\bigl(\tfrac{1+t_{ji}}4\bigr),\qquad
 d^2=2-2\sqrt{\max_\Pi\Phi_\Pi},\\
 G:=\delta-C_*d^2=\max_\Pi G_\Pi,\qquad G_\Pi=\delta-C_*(2-2\sqrt{\Phi_\Pi}),
 \end{gathered}
\]
with $\mathrm{EMAX}=1+\frac12\log_23$ and $\operatorname{ent}(x)=-x\log_2x$.  Each $G_\Pi$ is analytic wherever all $\lambda_{ji}>0$ and
$\det m_\Pi>0$, and $\delta$ is a convex function of $T$.

\subsubsection*{The chart}
For $\zeta_1,\zeta_2,\zeta_1',\zeta_2'\in[0,1]$ and a sign $\mathrm{sg}=\pm1$ put
\begin{align*}
 r&=\tfrac1N(1,\zeta_1,\zeta_2),\ x_a=\tfrac1n(\zeta_1,-1,0),\ y_a=r\times x_a,\\
 s&=\tfrac1{N'}(\mathrm{sg},\zeta_1',\zeta_2'),\ x_b=\tfrac1{n'}(\zeta_1',-\mathrm{sg},0),\ y_b=s\times x_b,
\end{align*}
$N=\sqrt{1+\zeta_1^2+\zeta_2^2}$, $n=\sqrt{1+\zeta_1^2}$ (and primed), and for $w\in\mathbb R^2$, $|w|\le1$,
\begin{equation}\label{eq:fqe-critical-chart}
 T(y)=rs^T+w_1(x_ax_b^T-y_ay_b^T)+w_2(x_ay_b^T+y_ax_b^T),\qquad y=(\zeta_1,\zeta_2,\zeta_1',\zeta_2',w_1,w_2).
\end{equation}
\begin{lemma}[Reduction to the chart; analytic]\label{lem:fqe-critical-chart}
$\delta$ and $d$ are invariant under $T\mapsto S_aTS_b^T$, where $S_a=D_a\Pi_a$, $S_b=D_b\Pi_b$ with
$\Pi_a,\Pi_b$ cyclic permutation matrices and $D_a,D_b$ diagonal sign matrices, $\det D_a\det D_b=1$ (an even total number of
sign changes), and under the transposition $T\mapsto T^T$.  Every critical state is equivalent under such maps to one
with $T=T(y)$, $y$ in the chart $[0,1]^4\times\{|w|\le1\}$ and $\mathrm{sg}=\pm1$, and then $\tau=|w|^2$.
In the chart the transposition is $(\zeta_1,\zeta_2,\zeta_1',\zeta_2',w_1,w_2)\mapsto(\zeta_1',\zeta_2',\zeta_1,\zeta_2,w_1,\mathrm{sg}\,w_2)$, so it suffices to treat $\zeta_1\le\zeta_1'$.
\end{lemma}
\begin{proof}
The $t_{ji}$ are permuted: this is a linear identity in $T$, and holds on the rank-one matrices $rs^T$ by
Lemma~\ref{lem:fqe-p1-sym}, which span all matrices.  The set of $32$ cube pairs is invariant (PM-2: the parity
$\prod\varepsilon_j\eta_j$ is preserved because $\det D_a\det D_b=1$), the local vectors transform with $S_a,S_b$ up to a
common sign which is absorbed by $(\varepsilon,\eta)\mapsto-(\varepsilon,\eta)$, and $\varepsilon^TT\eta$ is invariant;
so $\Phi$ is invariant by Lemma~\ref{lem:fqe-critical-correlation}(3).  Given $T$, take $r,s$ as in
Lemma~\ref{lem:fqe-critical-correlation}(2); cyclic shifts of $r$ alone and of $s$ alone make $|r_1|,|s_1|$ maximal, and
flips of an even number of the six coordinate signs make $r_1,r_2,r_3,s_2,s_3\ge0$ (the sign of $s_1$ is then determined and
is $\mathrm{sg}$).  The sign patterns of $r$ and of $s$ may have $\det D_a=\det D_b=-1$; then $D_a=-D_a'$, $D_b=-D_b'$ with
$\det D_a'=\det D_b'=1$ and $D_aTD_b^T=D_a'TD_b'^T$, i.e.\ the common sign flip of $(r,s)$, which leaves $rs^T$ fixed, is absorbed.
$(r,s)$ is determined by $T$ up to this common sign when $\tau<1$; at $\tau=1$ the top singular value is degenerate and $r,s$ are
not unique, but $\tau=1$ lies in the analytic range and the chart is used only for $\tau\le0.37$.  Then $r=(1,\zeta_1,\zeta_2)/N$, $s=(\mathrm{sg},\zeta_1',\zeta_2')/N'$ with $\zeta_1,\zeta_2,\zeta_1',\zeta_2'\in[0,1]$, and $T-rs^T$ lies in the
span of the two matrices of \eqref{eq:fqe-critical-chart} by Lemma~\ref{lem:fqe-critical-correlation}(2), with $|w|=\kappa$.
Transposition permutes the twelve $t_{ji}$ (again a linear identity, on rank-one matrices Lemma~\ref{lem:fqe-p1-sym} gives the swap
$r\leftrightarrow s$) and exchanges the roles of the two cube vertices in $\Phi$. Its action on the chart coordinates follows by
direct substitution: $T(y)^T=DT(y')D$, with $y'$ the image coordinates displayed in the lemma and $D=I$ for $\mathrm{sg}=1$,
$D=\operatorname{diag}(1,-1,-1)$ for $\mathrm{sg}=-1$, an admissible sign change (TC-4 also checks it numerically).
\end{proof}

\subsubsection*{Large $\tau$}
\begin{lemma}[Analytic range; TC-6]\label{lem:fqe-critical-large-tau}
If $\tau\ge0.36$ then $\delta\ge C_*d^2$, strictly.
\end{lemma}
\begin{proof}
Put $k=(\ln3-1)/(2\ln2)$.  By \eqref{eq:fqe-critical-entropy} and \eqref{eq:fqe-critical-fidelity},
$\delta\ge k(1-\Phi+5\tau/2)$, and $\Phi\ge f_{\rm cov}(\tau)$ by \eqref{eq:fqe-critical-fidelity-floor}.  With
$x=\sqrt{\Phi}$ the function $k(1-x^2)+5k\tau/2-2C_*(1-x)$ is concave in $x$, hence on $[\sqrt{f_{\rm cov}},1]$ it
is at least its smaller endpoint value; at $x=1$ it is $5k\tau/2>0$.  At $x=\sqrt{f_{\rm cov}(\tau)}$, writing
$\tau=1-u^2$, $f=f_{\rm cov}=\frac13+\frac u{2\sqrt3}$, it is
$\psi(u)=k(1-f)+\frac52k(1-u^2)-2C_*(1-\sqrt f)$, which is concave on $0\le u\le4/5$: $k(1-f)$ is affine,
$-\frac52ku^2$ is concave, and $\sqrt f$ is the square root of a positive affine function, hence concave, with coefficient
$2C_*>0$.  So $\psi$ is positive on $[0,4/5]$ once it is positive at the two endpoints, which TC-6 encloses in ball
arithmetic ($\psi(0)>0.064$, $\psi(4/5)>1.7\cdot10^{-4}$).
\end{proof}

\subsubsection*{The local cone lemma}
At $w=0$ the equality points of Theorem~\ref{thm:fqe-p1-sharp} lie in the chart at the eight points
$(\zeta_1,\zeta_2,\zeta_1',\zeta_2')\in\{(0,0,0,1),(0,0,1,0),(0,1,0,0),(1,0,0,0)\}$, one set for each sign $\mathrm{sg}$.
Let $z_i=|y_i-y^{\rm eq}_i|$ for the four chart coordinates and $z_{5,6}=w_{1,2}$; the box $\mathcal B_\rho$ is
$|z|_\infty\le\rho$, $\rho=1/256$.
\begin{lemma}[Local cone; LCC]\label{lem:fqe-critical-local-cone}
At each equality point exactly four planes $\Pi$ are tied: $G_\Pi=0$ and the others are below $-0.05$.  Their gradients (in $z$)
sum to $0$ and have rank three.  In $\mathcal B_\rho$ one has $G>0$ except at the point itself.
\end{lemma}
\begin{proof}
By Taylor's formula with the Hessian enclosed over the box, $G_\Pi(t\omega)=t\,g_\Pi\!\cdot\!\omega+\frac{t^2}2\omega^T H_\Pi(\xi)\omega$
for $\omega$ on the unit sphere of $|\cdot|_\infty$ (in the orthant of the four chart coordinates) and $0<t\le\rho$, with
$H_\Pi(\xi)$ in an interval matrix $[H_\Pi]$, and $G\ge\sum_\Pi\mu_\Pi G_\Pi$ for weights $\mu_\Pi\ge0$ summing to one on the tied planes.
Write $\mu=\frac14+\nu$.  Because $\sum g_\Pi=0$, the first-order term of the average depends on $\nu$ only.  The tie and this identity are exact, and
are proved symbolically (EX, no floating point or enclosures) at all eight points: exactly four planes have $\Phi_\Pi=\Phi_*$, where
$4\Phi_*=(1+1/\sqrt3)(1+\sqrt{2/3})$ and all other $\Phi_\Pi$ are smaller; the twelve probabilities are $(2\pm\sqrt2)/8$ and $\frac14$, so
$\delta=\frac12\log_23-\frac13(1+2H_{\rm b}(p))$ and $\delta=C_*(2-2\sqrt{\Phi_*})$, i.e.\ $G_\Pi=0$ at the point for the four tied planes, which
is the definition of $C_*$ in \eqref{eq:fqe-p1-cstar}; and $\sum_{\Pi\ \rm tied}\partial_kG_\Pi=0$ for all six $k$, as an identity in
$C_*,\ln p,\ln(1-p),\ln2$.  The reason is the sign symmetry
$z\mapsto(s_1z_1,s_2z_2,s_1s_2z_3,z_4,s_2z_5,s_1z_6)$ at $(0,0,0,1)$, which EX also checks exactly: $T(\sigma y)=DT(y)D'$ with
$D=\operatorname{diag}(1,s_1,s_2)$, $D'=\operatorname{diag}(1,s_1s_2,1)$ permutes the twelve $t_{ji}$ inside each cut and the sixteen
planes, so $G_\Pi\circ\sigma=G_{\pi(\Pi)}$ and only the $z_4$ component of $\sum_\Pi g_\Pi$ can survive; the $z_4$ component vanishes in EX
(it is the derivative of the wall function $G_w$ at $\pi/4$, zero by the reflection symmetry of Lemma~\ref{lem:fqe-p1-wall}).  By the sign symmetry the tied
gradients have the sign form $(s_1a,s_2a,-s_1s_2c,0,-s_2d,-s_1d)$ of the remark below, so their rank is exactly three
(the characters $s_1,s_2,s_1s_2$ carry independent vectors, as the enclosures give $a,c\ne0$); the bound $-0.05$ for the other planes
and the Hessian enclosures are ball arithmetic (LCC).  A piece $\Omega$ of the sphere is
discharged if, for some $\nu$, $\ell=\sum\nu_\Pi g_\Pi\!\cdot\!\omega\ge0$ and $\ell+\rho q>0$ (or $q>0$) with
$q=\frac12\omega^T[H_\mu]\omega$, evaluated in ball arithmetic. This suffices: on $\Omega$ the average is at least
$t\ell+t^2q=t(\ell+tq)$, and $\ell+tq$ is positive for $0<t\le\rho$, being affine in $t$ with $\ell\ge0$ at $t=0$ and
$\ell+\rho q>0$ at $t=\rho$ (or increasing if $q>0$).  LCC finds such weights on $9012$ cells for the eight points.
\end{proof}

\subsubsection*{The covering}
\begin{proposition}[Interval covering; CCV]\label{prop:fqe-critical-covering}
Let $\mathcal R$ be the part of the chart with $|w|^2\le37/100$, both signs $\mathrm{sg}$.  Every box of the bisection tree is
discharged by one of: (G) a certified positive lower bound of $\delta-C_*d^2$; (HS) $d^2\le1/400$ (then
$\delta\ge\frac3{10}d^2$, strictly greater than $C_*d^2$ when $d>0$, by Theorem~\ref{thm:fqe-orbit-stability}); (core) it lies in a box $\mathcal B_\rho$ of
Lemma~\ref{lem:fqe-critical-local-cone}.
\end{proposition}
\begin{proof}
The starting closed boxes cover $[0,1]^4\times[-5/8,5/8]^2$ in each sign sector, hence the required disc.
A split replaces one box by two closed children whose union is the parent; their common boundary is retained.
Boxes crossing the disc or ordering boundary are retained unless they are wholly outside the domain reduced by symmetry.
Thus complete coverage means that every retained branch ends in one of the three certified alternatives in the statement.
A failed or inconclusive sign test only triggers subdivision; it never certifies a box.

There is no need to choose a unique maximising fidelity plane. Since $G=\max_\Pi G_\Pi$,
any convex combination $G_\mu$ is a lower bound for $G$, including on a tie.
Near equality, the exact contacts and gradient cancellation of the local cone lemma replace a strict sign test at the apex.

On a box with centre $c$ and half-widths $h$ (in $y$) use the weights $\mu$ uniform on the best one to four planes.
(i) $\delta$ is convex in the twelve numbers $t_{ji}$, hence $\delta(y)\ge\delta(c)+\sum_iD_i(t_i(y)-t_i(c))$ with
$D_i=(1+\ln\lambda_i(c))/(12\ln2)$, valid for every point of the box since every chart point is a state, so that $t_i(y)\ge-1$
(the tangent of a convex function holds on its whole domain); a ray whose probability at $c$ is not certified positive is left
out of the tangent and bounded by the maximum of $-x\log_2x$ over its probability box ($-x\log_2x$ increases on $[0,1/e]$).
(ii) $\Phi_\Pi$ is concave in the three entries $m$ of $m_\Pi$, so $d_\Pi^2=2-2\sqrt{\Phi_\Pi}$ is convex there:
$d_\Pi^2(m)\le d_\Pi^2(m_c)+a_\Pi\!\cdot\!(m-m_c)+B$, where the Bregman remainder $B$ is convex in $m-m_c$ and so at most its maximum at the
corners of the box $|m-m_c|\le r$; $r$ is a bound for $|m(y)-m(c)|$.
(iii) $t(y)-t(c)$ and $m(y)-m(c)$ are expanded in $y-c$ to second order, with the exact gradient at $c$ and the interval second
derivatives of the nine entries $T_{\alpha\beta}(y)$ over the box (closed forms, ball arithmetic).
With $g_k=\partial_kG_\mu(c)$ this gives $G\ge G_\mu(c)-\sum_k|g_k|h_k-R$, where $R$ is the sum of the second-order terms of (iii)
inserted into the tangent of (i) and, with factor $C_*$, into the linear term of (ii), and of $C_*\max B$, the first-order terms being exact.
Bisection of the widest side (the floating-point routine \texttt{critical\_float.py} only orders the attempts and certifies nothing)
discharges the chart region $\zeta_1\le\zeta_1'$, $|w|^2\le37/100$ with $208953$ boxes (G) in the sector $\mathrm{sg}=+1$, and
$246044$ (G) and $436$ (HS) for $\mathrm{sg}=-1$, besides $42$ boxes inside the local cone boxes in each sector;
the ordering comparison $\zeta_1>\zeta_1'$ is exact on the dyadic endpoints, while the outside-disc screen has a conservative
positive margin exceeding its floating-point roundoff, so it discards only boxes strictly outside $|w|^2\le37/100$.
The orbit point of the chart ($d=0$, for
$\mathrm{sg}=-1$) is covered by (HS).
\end{proof}

\begin{proof}[Proof of Theorem~\ref{thm:fqe-p1-sharp} for critical states]
If $\tau\ge0.36$, apply Lemma~\ref{lem:fqe-critical-large-tau}.  Otherwise Lemma~\ref{lem:fqe-critical-chart} reduces to
the chart with $|w|^2\le37/100$, which Proposition~\ref{prop:fqe-critical-covering} and Lemma~\ref{lem:fqe-critical-local-cone}
cover: $G>0$, or $d>0$ and $\delta\ge\frac3{10}d^2>C_*d^2$, or $d=0$, except at the equality points $y^{\rm eq}$ where $G=0$ by
\eqref{eq:fqe-p1-cstar}.  The equality points have $w=0$ and are the $144$ points of Theorem~\ref{thm:fqe-p1-sharp}.
\end{proof}

\begin{remark}[Structure at the equality points; numerical picture]\label{rem:fqe-critical-structure}
In the coordinates $z=(\zeta_1,\zeta_2,\zeta_1',1-\zeta_2',w_1,w_2)$ at the equality point with $(\zeta_1,\zeta_2,\zeta_1',\zeta_2')=(0,0,0,1)$, the four tied
plane functions have the gradients $g=(s_1a,\,s_2a,\,-s_1s_2c,\,0,\,-s_2d,\,-s_1d)$, $s_1,s_2=\pm1$, with
$a=0.059039$, $c=0.036251$, $d=0.018765$ (ball enclosures, LCC).  They sum to zero and have rank three, with singular values
$0.1239,0.1239,0.0725$: $G$ has a conical minimum in three directions and is smooth in the other three.  The average of the
four tied Hessians is positive definite on the three-dimensional kernel of the gradients, with eigenvalues
$0.0725$ (the wall direction $\theta$) and $0.7285,0.7285$ (two combinations of the chart coordinates with the $w$-plane).
In particular the dependence on $w$ is first order, in the shape of a square $(\pm1,\pm1)$, so that the loss of
fidelity under moving off $\tau=0$ is compensated at first order by the cone and costs nothing along the smooth directions.
Numerically (not a proof; numerical evidence, NV-1) the minimum of $\delta-C_*d^2$ over critical states with fixed
$\tau$ is about $0.31\,\tau$ for every $\tau\in(0,1]$ (about $0.30\tau$ at $\tau=0.4$ and $0.5$), attained near the equality points for small $\tau$ and,
at $\tau=1$, at a real state with $d^2=2-\sqrt2$.
\end{remark}

\begin{remark}[Non-critical states; numerical evidence only]\label{rem:fqe-critical-noncritical}
The theorem is for critical states; whether $C_*$ is optimal over all states is open.  Evidence (not a proof):
at the equality state, random positive local filters of size $\varepsilon\le0.2$ (the twelve non-critical directions of the
$18$-dimensional LU quotient; $12$ directions per size) give $\delta-C_*d^2\ge3\varepsilon^2$ (NV-2); on states obtained by filtering critical states within $0.4$ of the equality state
(normal-form coordinates) with filters of size $0.05$ to $0.4$ the minimum of $\delta/d^2$ is $0.27$.
These sampled tests found no value below $C_*$. Reduction of a non-critical state to its critical representative does not preserve
$\delta$ (by Section~\ref{sec:fqe-renyi-filter-limit}, one filter on one qubit of a critical state cannot raise any
balanced-cut entropy, but filters on two qubits can raise the mean R\'enyi entropy of order $3/2$), and the general coefficient $k(6+\sqrt6)/12\approx0.05$ is a factor $3.8$ below
$C_*$, so a proof needs a quantitative treatment of the filter directions: near the critical locus a Taylor bound in the filter
variables over this six-parameter chart (a covering of the same size), far from it a stronger distance--entropy comparison.
\end{remark}

\section{A common replica hierarchy}\label{sec:hierarchy}

Section~\ref{sec:fourcopy} bounds the mean three-qubit purity from below using four copies of the state. This section places
that bound in a hierarchy of semidefinite relaxations indexed by the number $m$ of copies. It is a bosonic-extension
hierarchy with PPT constraints of the Doherty--Parrilo--Spedalieri type~\cite{DPS}, applied to the pure-state moment problem
rather than to separability; what is new here are the twirled reductions, the sharp certificates and the orbit recovery. Level $m=2$ is exactly the linear
programme of Rains (Proposition~\ref{prop:rainsppt}); the spin-flip blocks give all higher levels, which increase to the true
minimum (Theorem~\ref{thm:spinflip-ppt-hierarchy}); and levels $m=3,4$ are the three- and four-copy relaxations. The three-copy level
has exact optimum $125/978=0.12781\ldots<121/917$, and four copies are provably
stronger (Section~\ref{sec:threecopy}). The optimal three-copy pseudo-moment also rules out cubic sum-of-squares proofs
of the sharp bound. Five copies are sharp: Section~\ref{sec:fivecopy} gives the certificate and the equality case.

\begin{proposition}[two-copy PPT formulation of the Rains LP]\label{prop:rainsppt}
After local-unitary and site-permutation twirling, the Rains LP
of~Rains~\cite{Rains} (stated in Section~\ref{sec:lp} in the notation $a_s$, $A_j$, $S_j$ used here) is exactly the two-copy positive-partial-transpose
relaxation with support on the globally symmetric two-copy subspace: a twirled two-copy operator $X$
and an LP point $(a_k)$ correspond bijectively through $a_k=\tr(XG_{n,k})$ (with $G_{n,k}$ as in
Theorem~\ref{thm:spinflip-ppt-hierarchy}), and $X$ is feasible exactly when the point is.
\end{proposition}
\begin{proof}
Twirling is without loss of generality: the PPT constraint and the objective are invariant under
$U\otimes U$ and site permutations, so averaging a feasible operator keeps it feasible with the same value.
Let $F_i$ swap the two copies at qubit $i$, and $Q_i=2F_i-I$.
If $R_r$ is the projector onto the sum of sectors with exactly $r$
locally antisymmetric sites, then the local-unitary and site-twirled
algebra is the commutative span of the $R_r$. The generating
polynomial for the shadow coefficients gives
$S_{n-r}=2^n\tr(XR_r)$ for a twirled operator $X$, as an algebraic identity in which the enumerator of a
general $X$ is defined by $A_j=\sum_{|T|=j}\tr(XQ_T)$, $Q_T=\prod_{i\in T}Q_i$, so
$S_j\geq0$ is equivalent to $X\succeq0$.
Since $F_{\rm all}=\prod_iF_i=2^{-n}\prod_i(I+Q_i)$, the equality
$a_n=a_0=1$ says $\tr(XF_{\rm all})=1$, forcing globally
symmetric support for positive normalised $X$.
Let $\mathcal T$ be partial transpose on the second copy followed by
the local spin flip $\epsilon^{\otimes n}$, acting on the second copy only. Since
$\mathcal T(F_i)=I-F_i$, and $\mathcal T$ is its own Hilbert--Schmidt adjoint on the span of the $F_S$
(so $\tr[\mathcal T(X)R_r]=\tr[X\mathcal T(R_r)]$),
\[
 F_{\rm all}\sum_{r=0}^n\mathcal T(R_r)t^r
 =2^{-n}\prod_{i=1}^n(tI+Q_i).
\]
On the globally symmetric support this yields
$\tr[\mathcal T(X)R_r]=2^{-n}A_{n-r}$, so $A_j\geq0$ is
equivalent to $\mathcal T(X)\succeq0$. Conversely, the shadow
moments reconstruct $X$ uniquely in the twirled algebra; global
bosonic support gives $a_s=a_{n-s}$, and PPT gives $A_j\geq0$.
\end{proof}

\begin{theorem}[the spin-flip blocks are the bosonic PPT hierarchy]
\label{thm:spinflip-ppt-hierarchy}
For general $n,k$, the two-copy operator
$G_{n,k}=\binom nk^{-1}\sum_{|S|=k}\prod_{i\in S}F_i$
satisfies $\pi_k(z)=\tr[G_{n,k}(|z\rangle\langle z|)^{\otimes2}]$.
For $m\geq2$, minimise the same expectation over normalised positive
$m$-copy operators supported on $\mathrm{Sym}^m((\mathbb C^2)^{\otimes n})$
and positive under partial transposition on every collection of
\emph{global copies}. By Proposition~\ref{prop:rainsppt}, $L_2$ is exactly the Rains LP.
The lower bounds $L_m$ satisfy
$L_m\leq L_{m+1}\leq\min_z\pi_k(z)$.
Their limit equals $\min_z\pi_k(z)$ by compactness and the quantum
de~Finetti representation for infinitely extendible bosonic states.
Local-unitary and site-permutation twirling reduce each finite SDP.
For qubits, let $\mathcal A_m$ be the local copy commutant and,
for $1\le s\le\lfloor m/2\rfloor$, define
\[
 \mathcal T_s(X)=E_sX^{T_{[s]}}E_s^\dagger,
 \qquad E_s=(\epsilon^{\otimes n})^{\otimes s}\otimes I.
\]
The operator $X$ is feasible exactly when it is positive, normalised,
globally bosonic, and every $\mathcal T_s(X)$ is positive.
After twirling these are operators in $\mathcal A_m^{\otimes n}$.
Thus the \emph{complete} Fourier $\omega$- and flipped $\gamma$-blocks,
including all site types, give precisely this hierarchy.
In particular the three- and four-copy SDPs in
\S\ref{sec:threecopy}--\ref{sec:fourcopy} are its levels $m=3,4$
in conjugation-even coordinates. Restricted focused-contraction
cones in \S\ref{sec:sextic-general} are smaller objects and are not being
identified with the complete hierarchy.
\end{theorem}
\begin{proof}
Partial trace maps feasible level $m+1$ operators to feasible level
$m$ operators and preserves the objective, proving monotonicity.
The product $(|z\rangle\langle z|)^{\otimes m}$ is feasible, so $L_m\le\min_z\pi_k(z)$.
For the limit, let $X_m$ attain $L_m$ (the feasible set is compact). For each fixed $j$, the $j$-copy
marginals of $X_m$, $m\ge j$, lie in a compact set; a diagonal subsequence gives limits $X^{(j)}$, each the
partial trace of $X^{(j+1)}$ and supported on $\mathrm{Sym}^j((\mathbb C^2)^{\otimes n})$. By the quantum
de~Finetti theorem for such an infinitely exchangeable family~\cite{HudsonMoody}, $X^{(j)}=\int\sigma^{\otimes j}\,d\mu(\sigma)$ for a
probability measure $\mu$ on states. Bosonic support gives
$1=\operatorname{tr}(FX^{(2)})=\int\operatorname{tr}\sigma^2\,d\mu(\sigma)$ for the swap $F$, so $\mu$ is carried by
pure states $\sigma=|z\rangle\langle z|$. Hence
$\lim_mL_m=\operatorname{tr}[G_{n,k}X^{(2)}]=\int\pi_k(z)\,d\mu\ge\min_z\pi_k(z)$; the PPT constraints are not
needed for the limit.

Unitary conjugation by $E_s$ preserves positivity, so
$\mathcal T_s(X)\succeq0$ is exactly the usual global-copy PPT
constraint. Global copy permutations equate all subsets of the same
size; full transposition equates a subset with its complement. These
are all PPT constraints. The identity
$\epsilon\overline U\epsilon^{-1}=U$ for $U\in SU(2)$ shows that
$\mathcal T_s$ preserves the local commutant. Bosonic support of $X$
implies left and right invariance of $\mathcal T_s(X)$ under
$S_s\times S_{m-s}$, so its other blocks vanish identically.
Positivity on every remaining Fourier block is equivalent to
positivity of the full operator; invertible basis changes preserve it.
For $m=3$ there is just the $1|2$ flip; for $m=4$ there are the
$1|3$ and $2|2$ flips used in the construction.
Finally $X\mapsto\overline X$ preserves all constraints and the real
purity objective. Averaging $X$ with $\overline X$ loses no objective
value and justifies the conjugation-even coordinate reduction.
\end{proof}

\paragraph{Bosonic realignment at four copies.}
One natural convex strengthening of PPT is the realignment
(or computable cross-norm) condition. The copy-$2|2$
version, however, is \emph{already implied} by the
bosonic PPT constraints of
Theorem~\ref{thm:spinflip-ppt-hierarchy}.
For an operator $X$ on $V\otimes V$ write its matrix
elements $X_{ij,kl}$ and define its realignment by
$\mathcal R(X)_{ik,jl}=X_{ij,kl}$.
Let $F$ swap the two $V$ factors. If $XF=X$, then
\[
   \mathcal R(X)_{ik,jl}
     =X_{ij,lk}
     =(X^{T_2})_{ik,lj},
   \qquad\text{i.e.}\quad
   \boxed{\mathcal R(X)=X^{T_2}F}.
\]
Consequently for trace-one $X$ with $X^{T_2}\succeq0$,
$\|\mathcal R(X)\|_1=\|X^{T_2}\|_1=\tr X=1$.
At four copies set $V=H^{\otimes2}$ and group
$(1,2)|(3,4)$. Bosonic support on
$\operatorname{Sym}^4(H)$ implies $XF=X$
for the group swap $(13)(24)$; the $Y_2$
constraint of Section~\ref{sec:fourcopy} is precisely PPT on copies $(3,4)$,
up to its unitary spin-flip congruence.
Thus a $2|2$ CCNR inequality adds no restriction
to the four-copy SDP.

For completeness, the \emph{unequal} copy split
$1|3$ supplies the distinct, valid necessary
condition
\[
                   \|\mathcal R_{1|3}(X_4)\|_1\le1
\]
for any actual (possibly twirled) fourth-power
moment, or any convex mixture of fourth powers:
each unaveraged pure tensor power is a product
across that copy partition, its realignment has
trace norm one, and convexity survives twirling.
No claim is made that this $1|3$ condition is
independent of the existing bosonic PPT constraints
or that it improves the numerical seven-qubit bound;
both questions remain untested.

\subsection{The three-copy spin-flip relaxation}\label{sec:threecopy}

The sharp four-qubit inequality of Theorem~\ref{thm:fqe-moment} is proved by an identity among squared norms of local contractions
of $z\otimes z\otimes\tilde z$, $\tilde z=\epsilon^{\otimes n}\bar z$, $\epsilon=\bigl(\begin{smallmatrix}0&1\\-1&0\end{smallmatrix}\bigr)$
(Appendix~\ref{sec:fqe-sextic}). The same positivity construction gives a semidefinite relaxation for general numbers of qubits.

\paragraph{Moments.} For a unit $z\in(\mathbb C^2)^{\otimes n}$ and $m$ copies, the per-site
twirl of $(zz^*)^{\otimes m}$ by $U^{\otimes m}$, $U\in U(2)$ (which has the same commutant as $SU(2)$), lies in $\mathcal A_m^{\otimes n}$, where
$\mathcal A_m\subset\operatorname{End}((\mathbb C^2)^{\otimes m})$ is the image of
$\mathbb C[S_m]$, and it is positive. For $m=3$, $\mathcal A_3\cong\mathbb C\oplus M_2$ (trivial and
standard $S_3$ blocks; the sign block vanishes on qubits). The twirl is determined by the word values
$\mu(w)=\langle z^{\otimes3}|\bigotimes_i\pi_i|z^{\otimes3}\rangle$, $w=(\pi_1,\dots,\pi_n)\in S_3^n$.
Since $\pi_3$ is invariant under site permutations, global copy permutations and $z\mapsto\bar z$, we
may average $\mu$ over these; the averaged values $\bar\mu$ are constant on orbits of $S_n\times
S_3\times S_3$ and inversion. Two families of positive matrices are linear in $\bar\mu$:
\begin{itemize}
  \item \emph{$\omega$-blocks}: the Fourier blocks $\sum_w\bar\mu(w)\bigotimes_i\rho(\pi_i^{-1})$ of
    the twirl of $(zz^*)^{\otimes3}$, compressed to the copy-symmetric subspace;
  \item \emph{$\gamma$-blocks}: the same for the twirl of $|w\rangle\langle w|$,
    $w=z\otimes z\otimes\tilde z$. Because $\epsilon\bar U\epsilon^{-1}=U$ on $SU(2)$, this twirl is the
    image of the first under the per-site spin-flip map
    $t(\pi)=(I\otimes I\otimes\epsilon)^T\pi^{T_3}(I\otimes I\otimes\epsilon)$, which is expanded exactly
    in permutations (TCB-0).
\end{itemize}
Every one-site antisymmetriser of three qubit copies vanishes, which gives linear relations among the
$\bar\mu$. For $n=7$ there are $37$ orbit values, $23$ relations, $7$ $\omega$-blocks and $8$
$\gamma$-blocks, built exactly in integer bases (TCB-0).

\begin{theorem}[exact certificate; TCB-0--TCB-4]\label{thm:threecopy}
Every pure seven-qubit state satisfies $\pi_3(\psi)\ge 125/978 = 0.12781186\ldots$,
and the exact optimum of the $n=7$ three-copy spin-flip relaxation is $L_3 = 125/978$.
\end{theorem}

\begin{proof}
At the boundary rational pseudo-moment $m^*/70416$ (given by $37$ four-digit integers with
$m^*_{\mathrm{norm}}=70416$ and $m^*_{\mathrm{obj}}=9000$), all $23$ antisymmetriser relations vanish,
nine of the $15$ integer block matrices $B_b(m^*)$ are strictly positive definite, and the remaining six
active blocks $b\in\mathcal A=\{\omega_6,\omega_7,\gamma_4,\gamma_5,\gamma_6,\gamma_7\}$ have integer
kernels $V_b=\ker B_b(m^*)$ of dimensions $k_b=1,14,1,5,22,56$ whose complementary principal pivot
submatrices are strictly positive definite by fraction-free Bareiss elimination (TCB-2, TCB-3), proving
$B_b(m^*)\succeq0$ and hence $L_3\le 125/978$. On each active block $b\in\mathcal A$, let
$G_{\mathrm{site},b}\succ0$ be the integer
site-invariant Gram matrix, $M_{b,0}=V_b^{\mathsf T}G_{\mathrm{site},b}V_b\succ0$, and
$K_{b,q}=V_b^{\mathsf T}B_b(R_q)V_b$ for the $14$ integer RREF basis vectors $R_q$ of $\ker(\mathrm{rel})$.
The $25$ canonically normalised $S_m$-invariant kernel matrices
$\widehat Y_{b,p}=M_{b,0}^{-1}K_{b,p}M_{b,0}^{-1}/\operatorname{tr}(M_{b,0}^{-1}K_{b,p}M_{b,0}^{-1}K_{b,p})$
have basis-independent trace pairings $\operatorname{tr}(\widehat Y_{b,p}K_{b,q})$. Solving the $14$
reduced dual equations at $\lambda=125/978$ with $12$ rational free parameters yields symmetric rational
multipliers $Y_b=\sum_p w_{b,p}\widehat Y_{b,p}$ on the six active kernels that TCB-3 proves strictly
positive definite by Bareiss elimination, together with rational relation multipliers $\nu\in\mathbb Q^{23}$
satisfying the exact $37$-orbit identity (TCB-2)
\[
  \bar\mu_{\mathrm{obj}}-\frac{125}{978}\bar\mu_{\mathrm{norm}}
  =\sum_{b\in\mathcal A}\bigl\langle Y_b,V_b^{\mathsf T}B_b(\bar\mu)V_b\bigr\rangle+\nu\cdot\mathrm{rel}(\bar\mu).
\]
For a genuine state $\bar\mu_{\mathrm{norm}}=1$, $\mathrm{rel}(\bar\mu)=0$, $B_b(\bar\mu)\succeq0$ and
$\bar\mu_{\mathrm{obj}}=\pi_3$, since each $B_b(\bar\mu)$ is obtained from a multiplicity block of a positive
twirled moment by compression to the copy-invariant subspace followed by congruence with a fixed invertible
integer matrix. The identity uses only normalisation, the relations and block positivity, so every feasible
pseudo-moment satisfies $\bar\mu_{\mathrm{obj}}\ge 125/978$, giving $L_3=125/978$.
The four-copy bound of Theorem~\ref{thm:fourcopy} supersedes this value; the three-copy certificate is kept
for the strict separation of Corollary~\ref{cor:strict-copy-hierarchy} and the cubic-SOS obstruction. \qedhere
\end{proof}

\noindent Since $L_3=125/978<121/917$, the three-copy positivity and linear relations cannot certify the sharp
seven-qubit bound. The boundary point $m^*/70416$ is a pseudo-moment, not a quantum state, so it is no physical
counterexample. Through Theorem~\ref{thm:spinflip-ppt-hierarchy}, which identifies block positivity with operator
positivity on every local representation sector, it also excludes every homogeneous real cubic SOS proof of the sharp
bound (Theorem~\ref{thm:sextic-sos-obstruction}).

\begin{corollary}[quantitative strictness of the copy hierarchy]
\label{cor:strict-copy-hierarchy}
Let $L_3$ and $L_4$ denote the optima of the specific
three- and four-copy spin-flip SDPs defined in this section.
Then the exact certificates imply
\[
 L_4-L_3\ \ge\
 \frac{211}{1600}-\frac{125}{978}
 =\frac{3179}{782400}>0.
\]
Thus four copies yield a provably strictly stronger
seven-qubit SDP bound, by more than $0.00406$.
Neither level attains $121/917$: Proposition~\ref{prop:fourcopy-not-sharp} gives $L_3<L_4<6597/50000<121/917$.
\end{corollary}
\begin{proof}
Theorem~\ref{thm:fourcopy} gives $L_4\ge211/1600$ and Theorem~\ref{thm:threecopy} gives $L_3=125/978$.
\end{proof}

\noindent\emph{Numerical} (TCN-1): the relaxation reproduces the four-qubit value $1/3$ at $n=4$ and the AME
values $1/4$, $1/8$ at $n=5,6$. At $n=7$ the tight-tolerance solve (TCN-1b) gives $0.1278118609\ldots$, which
agrees with $125/978$ to $3\times10^{-13}$, so the three-copy bound stays below $121/917$. Rico et
al.~\cite{Rico} report $0.127(8)$, without a certificate, by strengthening the linear programme with the
quantum-code SDP of Angl\`es Munn\'e, Nemec and Huber~\cite{MNH}.

\subsection{The four-copy spin-flip relaxation}\label{sec:fourcopy}

\paragraph{Construction.} At four copies, the per-site algebra is
$\mathcal A_4\cong\mathbb C\oplus M_3\oplus M_2$: the $S_4$ irreps $[4],[3,1],[2,2]$, with $U(2)$
multiplicity spaces of dimension $5,3,1$ (the others vanish on qubits). The twirled moment $X$ of
$(|\psi\rangle\langle\psi|)^{\otimes4}$ lies in $\mathcal A_4^{\otimes n}$; after averaging over site permutations and
$\psi\mapsto\overline\psi$ it is real, $S_n$-symmetric and fixed by global copy permutations on either side, so
each Fourier block of a site-type pattern $(a,b,c)$ ($a$ sites in $[3,1]$, $b$ in $[2,2]$, $c$ in $[4]$)
is supported on the $S_4$-invariant vectors of $[3,1]^{\otimes a}\otimes[2,2]^{\otimes b}$. Splitting
these by the $S_a\times S_b$ isotypic components leaves blocks $\omega_{(a,b),\lambda\mu}$ whose entries
are the free parameters: $167$ at $n=7$. Two flipped moments are linear in $X$:
\[
 \begin{aligned}
  Y_1&=\text{twirl of }|\psi\psi\psi\tilde\psi\rangle\langle\psi\psi\psi\tilde\psi|=T_1^{\otimes n}X,\\
  Y_2&=\text{twirl of }|\psi\psi\tilde\psi\tilde\psi\rangle\langle\psi\psi\tilde\psi\tilde\psi|=T_2^{\otimes n}X,
 \end{aligned}
\]
where $T_1$, $T_2$ are the per-site maps $a\mapsto E\,a^{T_{\mathrm{flip}}}E^{T}$ (partial transpose on
the flipped copies, $E=\epsilon$ on those copies), exact rational $14\times14$ matrices with $12T$
integral. $Y_1$ is supported on $S_3$-invariant and $Y_2$ on $S_2\times S_2$-invariant vectors of the
copies, and both are positive. The relaxation minimises the mean purity $\tr(X\,\mathrm{SWAP}_{12}^{S})$
(averaged over $|S|=k$) subject to $\tr X=1$, $\omega\succeq0$ and the $\gamma$-blocks of $Y_1,Y_2$
$\succeq0$. Tracing out a copy maps these constraints onto the corresponding three-copy ones, so the
four-copy bound is at least the three-copy bound. The numerical construction uses orthonormal irreps;
the certificate uses integer irreps and integer symmetry-adapted bases.

\noindent\emph{Numerical} (FS-0--FS-2): the values at $n=4,5,6$ are $1/3$, $1/4$, $1/8$ (to
$10^{-9}$), the true minima. At $n=7$:
\begin{center}
\begin{tabular}{@{}lcccc@{}}
\toprule
constraints & $\omega$ & $\omega+Y_1$ & $\omega+Y_2$ & $\omega+Y_1+Y_2$\\
\midrule
bound on $\pi_3$ & $-1/7$ & $0.130888$ & $0.131821$ & $0.131877$\\
\bottomrule
\end{tabular}
\end{center}
The combined numerical value is about $7.5\times10^{-5}$ below $121/917$;
$Y_2$ alone nearly reaches the combined value. To improve the conditioning of the numerical solve, every block is first congruence-centred at the Haar average of the moment,
proportional to $\sum_{\pi\in S_4}\pi^{\otimes n}$, on the retained nonzero blocks.
The two largest $n=7$ solves are flagged inaccurate by the solver, so these values are
numerical evidence. The certified bounds are Theorem~\ref{thm:fourcopy} and, from above,
Proposition~\ref{prop:fourcopy-not-sharp}: $211/1600\le L_4<6597/50000<121/917$.

\begin{theorem}[exact certificate; FC-0--FC-4]\label{thm:fourcopy}
Every pure seven-qubit state satisfies $\pi_3(\psi)\ge 211/1600 = 0.131875$.
\end{theorem}

\begin{proof}
Write $x$ for the $167$ real coordinates of the symmetrised moment, with
$\mathrm{on}(x)=\tr X=1$ and $\mathrm{oo}(x)=\pi_3(\psi)$.
The trace weights each site type $(a,b,c)$ by $\frac{7!}{a!b!c!}3^a5^c$.
In the certificate's convention each multiplicity matrix has
$\omega_{kk}=x_{kk}$ and $\omega_{kl}=x_{kl}/2$ for $k\ne l$.
These matrices are positive semidefinite for genuine states. Indeed each nontrivial
Fourier block is
$2R(\omega\otimes\operatorname{adj}K)R^TH$, with $H$ the positive Gram form,
$R$ an integer basis of the allowed isotypic support and $K$ a positive definite Gram matrix.
Conjugation by $H^{1/2}$ makes this a congruence of
$\omega\otimes\operatorname{adj}K$; the scalar trivial block is treated separately.
For a Fourier block $B_j$ of $Y_j$, $j=1,2$, the flip blocks are
$\gamma_g(x)=Q_g^THB_jQ_g\succeq0$, where $Q_g$ spans the corresponding
row-invariant isotypic component. These are Gram compressions of the positive $Y_j$ blocks.

Put $\lambda=211/1600$. The certificate supplies a positive integer $Q_z$ and
integer matrices $Z_g\succ0$ such that
\[
 d_p=\mathrm{oo}_p-\lambda\,\mathrm{on}_p
       -\frac1{Q_z}\sum_g\langle Z_g,\gamma_g(e_p)\rangle
\]
forms a positive definite matrix $D_b=[d_{kl}]$ on every $\omega$-block $b$.
Here $\langle A,B\rangle=\tr(A^TB)$. FC-3 checks positive definiteness by exact
fraction-free elimination. It computes the integer numerator
$12^7\sum_g\langle Z_g,\gamma_g(e_p)\rangle$ using $12T_1,12T_2$,
by adjoint contraction modulo $35$ primes and Chinese remaindering under an integer
triangle-inequality bound. Thus, with the off-diagonal convention above,
\[
 \pi_3(\psi)-\lambda
 =\sum_b\langle D_b,\omega_b(x)\rangle
   +\frac1{Q_z}\sum_g\langle Z_g,\gamma_g(x)\rangle\ge0.
\]
The rational basis construction uses modular pivots only to prove independence.
For the copy-group invariant space, the exact trace of its averaging projector
gives the matching upper bound on dimension. In the subsequent Young-isotypic
decomposition the row subgroup has a one-dimensional fixed line in the matching Specht module,
so each selected row-invariant vector accounts for one multiplicity;
the sum of its irrep dimensions equals that same exact rank. Thus no component
is omitted. A floating rank estimate is not used for completeness. \qedhere
\end{proof}

\begin{remark}[regressions; numerical]
Two numerical regressions accompany the exact proof and are not used in it. FC-1 checks the parametrisation and
the identities $\mathrm{on}=\tr X$, $\mathrm{oo}=$ mean purity, in the same exact coordinates, on random
three-qubit states; FC-2 compares the CRT-recovered adjoint sum with a direct forward computation at $n=3,4$ for random integer $Z$.
\end{remark}

\begin{proposition}[the four-copy relaxation is not sharp; exact certificate, NS-0--NS-4]\label{prop:fourcopy-not-sharp}
Let $L_4$ be the optimum of the $n=7$ four-copy relaxation above ($k=3$; constraints $\tr X=1$, all $78$
$\omega$-blocks and all $271$ flip blocks of $Y_1,Y_2$ positive semidefinite; these are all of its constraints,
with no marginal or further flip equalities). Then
\[
 \frac{211}{1600}\ \le\ L_4\ <\ \frac{6597}{50000}=0.13194\ <\ \frac{121}{917}=0.131952\ldots,
 \qquad \frac{121}{917}-\frac{6597}{50000}=\frac{551}{45850000}.
\]
The feasible point is a pseudo-moment, not a seven-qubit state; it excludes only proofs that use this
relaxation unchanged.
\end{proposition}
\begin{proof}
The lower bound is Theorem~\ref{thm:fourcopy}. For the upper bound, the data file supplies an integer
vector $x$ of the $167$ coordinates of the proof of Theorem~\ref{thm:fourcopy}. Apart from $\tr X=1$
every constraint is homogeneous, so $x/\mathrm{on}(x)$ is feasible once every block of $x$ is positive
semidefinite, and its objective is $\mathrm{oo}(x)/\mathrm{on}(x)$. NS-1 computes $\mathrm{on}(x)>0$
and this ratio exactly; it is $0.1319308\ldots<6597/50000$. NS-2 checks that all $78$ matrices
$2\omega_b(x)$ (integer) are positive definite by exact fraction-free elimination. NS-3 forms
the moment $X(x)$ and $(12T_j)^{\otimes7}X(x)$ modulo each of $34$ primes, recovers every
$12^7\gamma_g(x)=Q_g^TH\,B_j\,Q_g$ as an integer matrix by the Chinese remainder theorem under an
integer triangle-inequality bound on its entries (the bound covers the block sums of the moment, the seven
flip contractions and the compression; the modulus exceeds twice it), and checks that all $271$ are exactly symmetric and
positive definite by exact fraction-free elimination. Hence $L_4\le\mathrm{oo}(x)/\mathrm{on}(x)<6597/50000$.
\end{proof}

\begin{remark}[provenance of the pseudo-moment; numerical]
The point $x$ is a floating-point optimum of the Haar-centred relaxation
($0.1318766470$), mixed with the Haar point at weight $10^{-3}$, moved to the exact coordinates by a per-type
least-squares fit and rounded to a dyadic grid; its least centred block eigenvalue is about $10^{-3}$.
None of this is used in the proof, which checks the stored integers only. NS-0 is a regression of the
modular forward map against a direct floating-point computation at $n=3,4$.
\end{remark}

\section{The complex conference family}\label{sec:sup-conference}
The preceding replica hierarchy shows how the lower-copy relaxations fall
short of the sharp seven-qubit purity bound. An explicit candidate is
provided by the complex conference family, whose states are parametrised
by a complex symmetric $6\times6$ matrix. We prove a sum-of-squares
purity identity \emph{within this family}, classify its equality cases,
and derive its spin-flip equation. This explains the icosahedral geometry,
but does not establish a bound for arbitrary states: the five-copy
certificate in Section~\ref{sec:fivecopy} supplies globality, and the
invariant reconstruction there supplies equality rigidity.
Figure~\ref{fig:seven-proof-map} separates these steps from the later
entropy proof.
\begin{figure}[tbp]
\centering
\begin{tikzpicture}[
  box/.style={draw,rounded corners=2pt,align=center,font=\footnotesize,minimum width=6.4cm,inner sep=3pt,minimum height=1.05cm},
  res/.style={box,very thick},
  arr/.style={-{Stealth[length=2mm]},semithick},
  node distance=0.42cm]
\node[font=\small\bfseries] (P) {Purity minimum};
\node[box,below=0.25cm of P] (p1) {Five-copy kernels and finite probes\\(Props.~\ref{prop:ico-fivecopy-g34-kernels}, \ref{prop:fivecopy-finite-probes})};
\node[box,below=of p1] (p2) {Sharp five-copy certificate:\\$\pi_3-121/917$ is a sum of faces\\(Thm.~\ref{thm:fivecopy-sharp-certificate}, FP5-1)};
\node[box,below=of p2] (p3) {Tangent line and invariant square give\\$X_5^+=X_5^+(\Psiico)$\\(Lemma~\ref{lem:fivecopy-tangent-line}, Thm.~\ref{thm:fivecopy-family-equality})};
\node[box,below=of p3] (p4) {Degree-four invariants fix the orbit\\through the conference family\\(Thm.~\ref{thm:moment-orbit-rigidity}, Thm.~\ref{thm:complexsos})};
\node[res,below=of p4] (p5) {$\pi_3\ge121/917$, with equality only\\on the orbit of $\Psiico$\\(Thm.~\ref{thm:fqe-seven-purity})};
\draw[arr] (p1)--(p2); \draw[arr] (p2)--(p3); \draw[arr] (p3)--(p4); \draw[arr] (p4)--(p5);

\node[font=\small\bfseries,right=4.6cm of P] (E) {Entropy maximum};
\node[box,below=0.25cm of E] (e1) {The family maximiser $\psi_{x_*}$\\is stationary on the whole sphere\\(Prop.~\ref{prop:entropy-seven-stationarity})};
\node[box,below=of e1] (e2) {Contender region: purity band\\and eigenvalues at most $497/1000$\\(Prop.~\ref{prop:seven-competitor-band}, Cor.~\ref{cor:seven-entropy-eigen-cap-497})};
\node[box,below=of e2] (e3) {Joint scalar majorant (Lemma~\ref{lem:seven-joint-scalar-majorant}, AH-6)\\and localised identity (Thm.~\ref{thm:seven-entropy-localized-identity}, E7-1)};
\node[box,below=of e3] (e4) {Invariant and square fix\\the five-copy moment\\(Thm.~\ref{thm:fivecopy-family-equality}, Prop.~\ref{prop:seven-entropy-equality-moment})};
\node[res,below=of e4] (e5) {$E_3\le E_3(\psi_{x_*})$, with equality only\\on the orbit of $\psi_{x_*}$\\(Thm.~\ref{thm:fqe-seven-entropy})};
\draw[arr] (e1)--(e2); \draw[arr] (e2)--(e3); \draw[arr] (e3)--(e4); \draw[arr] (e4)--(e5);

\draw[arr,dashed] (p5.east) to[out=0,in=180] (e2.west);
\draw[arr,dashed] (p4.east)--(e4.west);
\end{tikzpicture}
\caption{How the seven-qubit optima are proved. Solid arrows are the steps of each proof; dashed arrows mark what the
entropy proof takes from the purity proof: the purity bound, which confines the contenders, and the orbit step. Section~\ref{sec:sup-conference} treats the conference family,
Section~\ref{sec:fivecopy} the purity minimum and Section~\ref{sec:sup-seven-entropy} the entropy maximum.}
\label{fig:seven-proof-map}
\end{figure}

\subsection{A sharp complex conference family}\label{sec:conference-phase}

The icosahedral family has the two-branch form $\Omega_{u,v}$ of Theorem~\ref{thm:spinflip-double}, with branch supports far apart in
Hamming distance. This separation makes the low-weight odd Pauli correlations vanish, as the first result shows. We then
apply that cancellation to the family with complex pair amplitudes, which contains the icosahedral state.
In Corollary~\ref{cor:ordinary-triples} through Theorem~\ref{thm:complexsos} below, the distinguished qubit is labelled $0$ and the
ordinary qubits $1,\ldots,6$; in Proposition~\ref{prop:confphase} matrix indices are the ordinary labels $0,\ldots,5$
of~\eqref{eq:ico}.

\begin{theorem}[spin-flip-doubled odd-weight selection]\label{thm:spinflip-double}
Let $u,v$ be vectors on $m$ qubits whose computational-basis
supports have minimum Hamming distance $D>0$. Let
$J_m=\epsilon^{\otimes m}K$, where $K$ is complex conjugation, and set
\[
 |\Omega_{u,v}\rangle=
 \frac{|0\rangle(u+v)+|1\rangle J_m(v-u)}
 {\sqrt{2(\|u\|^2+\|v\|^2)}}.
\]
Every Pauli expectation supported entirely on the last $m$ qubits
of odd weight $q<D$ vanishes. If $u$ is supported at Hamming weight
$w$ and $v=b|1^m\rangle$ with $0<w\le m-2$ and $2w\ne m$,
then \emph{all} $m+1$ one-qubit marginals of $\Omega_{u,v}$
are maximally mixed.
\end{theorem}
\begin{proof}
A Pauli word of weight $q<D$ cannot connect the support of $u$
to that of $v$. Thus its expectations in $u+v$ and in $v-u$
are equal. Spin flip reverses the sign of every odd-weight
Hermitian Pauli expectation:
$\langle J_m a|P|J_m a\rangle
=(-1)^{\operatorname{wt}P}\langle a|P|a\rangle$.
The two branches therefore cancel. For the last assertion,
$D=m-w\ge2$ gives maximal mixing on each of the ordinary
sites. The two branches have equal norm, while their supports
are disjoint when $w\notin\{0,m/2,m\}$, so the distinguished
qubit also has marginal $I/2$.
\end{proof}

\begin{corollary}[ordinary triple reductions in the complex conference
ansatz; SDS-1--SDS-3]\label{cor:ordinary-triples}
The seven-qubit pair/complement family of
Theorem~\ref{thm:complexsos}, which contains the
icosahedral family of Theorem~\ref{thm:ico}, is $\Omega_{u,v}$ with
$m=6$, $w=2$, $u=\sum_{i<j}t_{ij}|ij\rangle$ and
$v=b|1^6\rangle$. For every three ordinary sites
$I=\{i,j,k\}\subseteq\{1,\ldots,6\}$:
\[
 \langle P_I\rangle=0
 \quad\text{for every Pauli word of exact weight }3,
 \qquad \Phi_{I,0}=0,
\]
and
\begin{equation}\label{eq:ordinary-triple-purity}
 \tr\rho_{ijk}^2=
 \frac12\bigl(\tr\rho_{ij}^2+\tr\rho_{ik}^2+
                 \tr\rho_{jk}^2\bigr)-\frac14.
\end{equation}
This holds for all complex pair amplitudes, not only for
conference matrices or the optimal parameter.
\end{corollary}
\begin{proof}
The Hamming separation is $D=4$, so all ordinary-site
Pauli weights one and three vanish by
Theorem~\ref{thm:spinflip-double}. Lemma~\ref{lem:focus-pauli}, where the focused covariants $\Phi_{I,r}$ are
defined (Section~\ref{sec:sextic-general}), then proves the focused-covariant zero.
The Pauli expansion of the normalised triple purity is
$8\tr\rho_{ijk}^2=1+\sum_{\{a,b\}\subseteq I}
(4\tr\rho_{ab}^2-1)$, giving~\eqref{eq:ordinary-triple-purity}.
SDS-1--SDS-3 check Pauli cancellation and
all $20$ triple-purity relations using Gaussian integer amplitudes.
\end{proof}

\begin{theorem}[proved; CC-1]\label{thm:complexsos}
Let $T=(t_{ij})$ be any complex symmetric $6\times6$ matrix with zero
diagonal, and let $b\geq0$ be real. Define the normalised seven-qubit state
\[
 z_{T,b}=|0\rangle\left(\sum_{i<j}t_{ij}|ij\rangle+b|1^6\rangle\right)
  +|1\rangle\left(b|0^6\rangle-
       \sum_{i<j}\overline{t_{ij}}|\overline{ij}\rangle\right),
 \quad 2\left(b^2+\sum_{i<j}|t_{ij}|^2\right)=1.
\]
Here $|ij\rangle$ is the six-bit weight-two vector supported on $i,j$,
and $|\overline{ij}\rangle$ its complement. Put
$x=b^2$, $\sigma=\sum_{i<j}|t_{ij}|^2$, $p_i=\sum_{j\ne i}|t_{ij}|^2$. Then
\begin{equation}\label{eq:complexsos}
\begin{split}
 \pi_3(z_{T,b})-\frac{121}{917}
 ={}&\frac{(524x-37)^2}{137550}
 +\frac4{35}\sum_i(p_i-\sigma/3)^2\\
 &+\frac{48}{35}\sum_{i<j}(|t_{ij}|^2-\sigma/15)^2
 +\frac{24}{35}\sum_{i<j}|(TT^\dagger)_{ij}|^2.
\end{split}
\end{equation}
Consequently the minimum on this entire complex-amplitude family is
$121/917$. Equality requires $x=37/524$, uniform magnitudes $|t_{ij}|$
and $TT^\dagger=(\sigma/3)I$: $T$ is a scaled complex symmetric conference
matrix. The icosahedral state is one equality case.
\end{theorem}
\begin{proof}
Fix an ordinary pair $\{i,j\}\subset\{1,\ldots,6\}$ and let $K=\{1,\ldots,6\}\setminus\{i,j\}=\{k,l,m,n\}$.
When $K$ is traced out to form $\rho_{0ij}$, the five Hamming-weight sectors on $K$ map to four orthogonal two-dimensional subspaces of $\mathcal H_{0ij}$:
\begin{itemize}
\item weights $0$ and $4$ on $K$ give orthogonal kets $t_{ij}|011\rangle+b|100\rangle$ and $b|011\rangle-\overline{t_{ij}}|100\rangle$, contributing $(|t_{ij}|^2+b^2)I_2$ on $\operatorname{span}\{|011\rangle,|100\rangle\}$;
\item weight $1$ on $K$ (at $r\in K$) gives $|0\rangle(t_{ir}|10\rangle+t_{jr}|01\rangle)$, contributing $M_{ij}=\bigl(\begin{smallmatrix}p_i-|t_{ij}|^2&(TT^\dagger)_{ij}\\\overline{(TT^\dagger)_{ij}}&p_j-|t_{ij}|^2\end{smallmatrix}\bigr)$ on $\operatorname{span}\{|010\rangle,|001\rangle\}$, and weight $3$ on $K$ gives $\overline{M_{ij}}$ in the ordered basis $(|101\rangle,|110\rangle)$;
\item weight $2$ on $K$ gives $\bigl(\begin{smallmatrix}Q_K&-2\operatorname{Hf}(T_K)\\-2\overline{\operatorname{Hf}(T_K)}&Q_K\end{smallmatrix}\bigr)$ on $\operatorname{span}\{|000\rangle,|111\rangle\}$, where $Q_K=\sum_{\{u,v\}\subset K}|t_{uv}|^2=\sigma-p_i-p_j+|t_{ij}|^2$ and $\operatorname{Hf}(T_K)=t_{kl}t_{mn}+t_{km}t_{ln}+t_{kn}t_{lm}$.
\end{itemize}
Hence
\[
 \tr\rho_{0ij}^2=2(|t_{ij}|^2+b^2)^2+2(p_i-|t_{ij}|^2)^2+2(p_j-|t_{ij}|^2)^2+4|(TT^\dagger)_{ij}|^2+2Q_K^2+8|\operatorname{Hf}(T_K)|^2.
\]
Tracing out qubit $0$ from $\rho_{0ij}$ leaves $\rho_{ij}$ diagonal on $\{|00\rangle,|11\rangle\}$ with entry $\tfrac12-d_{ij}$ ($d_{ij}=p_i+p_j-2|t_{ij}|^2$, using $\sigma+b^2=\tfrac12$) and equal to $\bigl(\begin{smallmatrix}d_{ij}&2(TT^\dagger)_{ij}\\2\overline{(TT^\dagger)_{ij}}&d_{ij}\end{smallmatrix}\bigr)$ on $\operatorname{span}\{|10\rangle,|01\rangle\}$, so $\tr\rho_{ij}^2=\tfrac12-2d_{ij}+4d_{ij}^2+8|(TT^\dagger)_{ij}|^2$. Summing~\eqref{eq:ordinary-triple-purity} over the $20$ ordinary triples gives $\sum_{i<j<k}\tr\rho_{ijk}^2=2\sum_{i<j}\tr\rho_{ij}^2-5$.

To combine the marginal purities, expand $|\operatorname{Hf}(T_K)|^2$ and $|(TT^\dagger)_{ij}|^2$. Both contain the same $4$-cycle cross terms $2\operatorname{Re}(t_{kl}\overline{t_{km}}t_{mn}\overline{t_{ln}})$, while the squared moduli partition edge pairs of $K_6$ into disjoint and adjacent pairs:
\[
 2\sum_{|K|=4}|\operatorname{Hf}(T_K)|^2-\sum_{i<j}|(TT^\dagger)_{ij}|^2
 =\sigma^2-\frac32\sum_{i=1}^6p_i^2+2\sum_{i<j}|t_{ij}|^4.
\]
Adding the $20$ ordinary triples and the $15$ cuts $\{0,i,j\}$, with $\sum_ip_i=2\sigma$ and $\sigma=\tfrac12-x$, gives
\[
 35\,\pi_3(z_{T,b})=\frac72-4x+64x^2+4\sum_ip_i^2+48\sum_{i<j}|t_{ij}|^4+24\sum_{i<j}|(TT^\dagger)_{ij}|^2.
\]
Since $\sum_i(p_i-\sigma/3)^2=\sum_ip_i^2-2\sigma^2/3$ and $\sum_{i<j}(|t_{ij}|^2-\sigma/15)^2=\sum_{i<j}|t_{ij}|^4-\sigma^2/15$,
\eqref{eq:complexsos} reduces to the scalar identity
$\frac72-4x+64x^2=\frac{35\cdot121}{917}+\frac{7(524x-37)^2}{27510}-\frac{88}{15}\bigl(\tfrac12-x\bigr)^2$, checked by expansion.
CC-1 independently checks all $226$ formal quartic coefficients. Each term on the right of~\eqref{eq:complexsos} is nonnegative, and equality holds iff $x=37/524$, $|t_{ij}|^2=\sigma/15$, $p_i=\sigma/3$, and $(TT^\dagger)_{ij}=0$ ($i\ne j$).
\end{proof}

The classification of order-six complex symmetric conference matrices is due to
Et-Taoui and Makhlouf~\cite{EtTaouiMakhlouf}, who classify the complex symmetric, skew-symmetric and Hermitian
conference matrices of order six. We give the diagonal-congruence form needed here,
with a self-contained proof, and then identify its action on the seven-qubit state.
The application is an equality classification within the pair/complement ansatz.

\begin{proposition}[order-six symmetric conference matrices and restricted-ansatz equality;
proved, CP-4]\label{prop:confphase}
Let $C$ be the real icosahedral $6\times6$ conference matrix ($C_{ij}=s_{ij}$ for $i\ne j$) of Theorem~\ref{thm:ico}.
Every complex symmetric matrix $\widetilde C$ of order six with $\widetilde C_{ii}=0$,
$|\widetilde C_{ij}|=1$ for $i\ne j$, and $\widetilde C\widetilde C^\dagger=5I$ has the form
\[
                 \boxed{\ \widetilde C=DPCP^{\mathsf T}D\ },
\]
where $D$ is diagonal unitary and $P$ is a permutation matrix.
Consequently, \emph{every} state attaining $\pi_3=121/917$ within
the complete complex pair/complement ansatz of
Theorem~\ref{thm:complexsos} is related to $\Psiico$ by
single-qubit diagonal unitaries and a permutation of the six
ordinary qubits. This is a complete \emph{restricted-ansatz} equality
classification; the unrestricted equality statement is
Corollary~\ref{cor:seven-global-from-fivecopy}, proved independently by
moment-orbit rigidity.
\end{proposition}
\begin{proof}
A diagonal unitary congruence first makes the five off-diagonal
entries in row and column $0$ all equal to $1$. Write the remaining
$5\times5$ symmetric core as $B$, with zero diagonal. Because $B^{\mathsf T}=B$, the conference equation $\widetilde C\widetilde C^\dagger=5I_6$ on the core reads
\[
 B\mathbf 1=0,\qquad B\overline B=5I_5-J_5,
\]
where $J_5=\mathbf 1\mathbf 1^{\mathsf T}$. Four unit complex numbers summing to zero form two opposite pairs: for any three unit complex numbers $z_1,z_2,z_3$, the identity $|z_1+z_2+z_3|^2-1=(z_1+z_2)(z_2+z_3)(z_3+z_1)/(z_1z_2z_3)$ shows that $|z_1+z_2+z_3|=1$ iff two of them sum to zero and the third equals the sum.

After removing the common unit phase $B_{12}$ from $B$ and relabelling vertices $\{2,3,4,5\}$, row~$1$ of $B$ is $(0,1,-1,t,-t)$ with $|t|=1$. The off-diagonal equation $(B\overline B)_{12}=-1$ reads
\[
 -\overline{B_{23}}+t\overline{B_{24}}-t\overline{B_{25}}=-1.
\]
The three unit summands on the left sum to $-1$, so one equals $-1$ and the other two sum to zero.
The relabelling $(2\,3)(4\,5)$ followed by $B\mapsto-B$ fixes row~$1$ and sends $B_{23}$ to $-B_{23}$, so
every configuration with $B_{23}=-1$ is carried to one with $B_{23}=1$. Hence the excluded subcases $B_{23}=-1$ of the first two cases below reduce to the third case, $B_{23}=1$, and need no
separate argument.
\begin{itemize}
\item If $t\overline{B_{24}}=-1$ and $-\overline{B_{23}}-t\overline{B_{25}}=0$, then $B_{24}=-t$ and $B_{25}=-B_{23}t$, so the sum of row~$2$ is $1+B_{23}-t-B_{23}t=(1+B_{23})(1-t)=0$, forcing $t=1$ unless $B_{23}=-1$.
\item If $-t\overline{B_{25}}=-1$ and $-\overline{B_{23}}+t\overline{B_{24}}=0$, then $B_{25}=t$ and $B_{24}=B_{23}t$, so the sum of row~$2$ is $1+B_{23}+B_{23}t+t=(1+B_{23})(1+t)=0$, forcing $t=-1$ unless $B_{23}=-1$.
\item Otherwise $-\overline{B_{23}}=-1$, so $B_{23}=1$. Then $(B\overline B)_{12}=-1$ gives $B_{24}=B_{25}$, and zero row-$2$ sum gives $B_{24}=B_{25}=-1$; meanwhile $(B\overline B)_{13}=1+t\overline{B_{34}}-t\overline{B_{35}}=-1$ gives $t\overline{B_{34}}-t\overline{B_{35}}=-2$, forcing $B_{34}=-t$ and $B_{35}=t$. Zero row-$4$ sum ($t-1-t+B_{45}=0$) then gives $B_{45}=1$, and $(B\overline B)_{14}=-1+\overline t-t=-1$ forces $t=\overline t=\pm1$.
\end{itemize}
Hence $t=\pm1$ in all cases. Applying the same argument to each row $i$ shows that row $i$ is a unit phase $\omega_i$ times a real sign vector; since $B_{ij}=\pm\omega_i=\pm\omega_j$, all $\omega_i$ agree up to sign, so $B$ is a single unit phase times a real $\{\pm1\}$ matrix (also checked by CP-4 across all $3^5$ row-matching patterns).

A diagonal congruence removes this common core phase while preserving
row $0$. The now-real core has exactly two negative and two positive
entries per row. Therefore its negative-edge graph is a simple
$2$-regular graph on five vertices, necessarily a pentagon.
All such pentagons differ by a permutation, so the real conference
matrix is $PCP^{\mathsf T}$. Restoring the removed diagonal phases
proves the boxed classification.

Finally, if $D=\operatorname{diag}(e^{i\phi_i})$ and
$\Sigma=\sum_i\phi_i$, applying
$\operatorname{diag}(e^{i\Sigma/2},1)$ to the distinguished qubit,
and $\operatorname{diag}(1,e^{i(\phi_i-\Sigma/4)})$
to ordinary qubit $i$, realises $C\mapsto DCD$ on the
pair/complement state without changing either endpoint amplitude.
Ordinary-qubit permutations implement $P$.
\end{proof}

\subsection{A spin-flip equation for the conference family}\label{sec:sup-conference-shell}
\begin{proposition}[an exact spin-flip equation forces the conference shell]
\label{prop:kramers-conference-shell}
Let $J=J_7$. Choose a distinguished qubit $s$
and local $Z$ axes, and define
\[
 H_s=-\frac12Z_s-\frac14\sum_{i\ne s}Z_i,
 \qquad {\cal A}_s=H_sJ.
\]
A normalised pure seven-qubit state satisfying
$z={\cal A}_sz$ lies in the \emph{complex
pair/complement family} of
Theorem~\ref{thm:complexsos},
up to a special-qubit diagonal $SU(2)$ phase.
Conversely all members of that family
satisfy this equation in that phase frame.
Thus every such $z$ has $\pi_3(z)\ge121/917$, with equality
only on the icosahedral LU/permutation orbit.
In its displayed conference frame the
icosahedral state satisfies
$\Psiico=H_sJ\Psiico$.
\end{proposition}
\begin{proof}
Since $J^2=-I$ and $JH_sJ^{-1}=-H_s$,
the real-linear operator $\mathcal A_s=H_sJ$
obeys $\mathcal A_s^2=H_s^2$.
For a computational string with
special-site bit $x_s$ and ordinary
Hamming weight $w$, the eigenvalue
of $H_s$ is
\[
 h(x)=\begin{cases}
 -2+w/2,&x_s=0,\\
 -1+w/2,&x_s=1.
 \end{cases}
\]
Hence $z=\mathcal A_sz$ implies $H_s^2z=z$,
restricting support to $h(x)^2=1$:
exactly $15+1+1+15=32$ strings, those of
ordinary weights $2,6$ at $x_s=0$ and $0,4$ at $x_s=1$.

On this shell the equation identifies the
weight-four amplitudes with the
\emph{negative complex conjugates}
of the complementary weight-two amplitudes,
and makes the two endpoint amplitudes complex conjugates.
A diagonal $SU(2)$ phase at the special site
makes both endpoint amplitudes real and
nonnegative. This is exactly the complex
pair/complement ansatz, and the converse
follows by the same spin-flip sign calculation.
Theorem~\ref{thm:complexsos} and
Proposition~\ref{prop:confphase} give the purity
assertions. The icosahedral relation follows
coefficientwise from~\eqref{eq:ico}.
\end{proof}

\section{The five-copy level at seven qubits}\label{sec:fivecopy}
The five-copy certificate proves that $\Psi_{\rm ico}$ minimises $\pi_3$ over all pure
seven-qubit states, uniquely up to local unitaries and qubit permutations
(Corollary~\ref{cor:seven-global-from-fivecopy}). The proof has three
parts.
\begin{enumerate}
\item \emph{Value.} A rational five-copy dual supported on twelve exact
  physical icosahedral kernels, with every Gram positive definite
  (Theorem~\ref{thm:fivecopy-sharp-certificate}). The identity is determined by
  $3424$ direct probes (Section~\ref{sec:fivecopy-direct-probes}) and a
  verified right-inverse repair of an identity-Gram baseline, using $3754$
  independent symmetric coordinates and at most seven columns per frame.
\item \emph{Equality moment.} The repair directions span the whole
  critical space, so at equality the five-copy moment lies on the
  tangent line of the icosahedral family
  (Lemma~\ref{lem:fivecopy-tangent-line}); an invariant and its square
  fix the tangent parameter at zero
  (Theorem~\ref{thm:fivecopy-family-equality}).
\item \emph{Orbit.} Thirteen invariants of bidegree at most $(4,4)$
  already determine the icosahedral orbit
  (Theorem~\ref{thm:moment-orbit-rigidity}).
\end{enumerate}
Steps 2 and 3 hold at every family point with $1/32<x<1/12$ and combine into
Theorem~\ref{thm:fivecopy-family-equality}, which the entropy proof of
Section~\ref{sec:seven-entropy-equality} uses as well.
Levels two, three and four of the hierarchy are provably not sharp
(\S\S\ref{sec:hierarchy}--\ref{sec:fourcopy}; for level four,
Proposition~\ref{prop:fourcopy-not-sharp}), so five copies is the first
sharp level.

\subsection{Site algebra and the icosahedral kernels}\label{sec:fivecopy-kernels}

\begin{lemma}[Exact five-copy site algebra; S5-1]\label{lem:fivecopy-site-algebra}
On one qubit's five-copy space, the commutant of collective $\mathrm{SU}(2)$ is
$\mathcal A_5\simeq\mathbb C\oplus M_4(\mathbb C)\oplus M_5(\mathbb C)$.
Choose rational bases of the highest-weight spaces in Hamming weights
$m=0,1,2$ by taking the kernels of the qubit raising operator.  In the
resulting matrix-unit coordinates, the spin-flip partial-transpose maps
on copy sets $\{5\}$ and $\{4,5\}$ are rational $42\times42$ involutions;
every entry has denominator dividing $30$.
\end{lemma}
\begin{proof}
The highest-weight kernels have dimensions $1,4,5$; the corresponding
$\mathrm{SU}(2)$ irreducibles have dimensions $6,4,2$, giving $1\cdot6+4\cdot4+5\cdot2=32$.
Schur--Weyl duality gives the stated commutant.  For a weight-$m$ integer
representation $\rho_m$ of $S_5$, its matrix unit is
$E^{(m)}_{ij}=\frac{d_m}{120}\sum_{\sigma\in S_5}
\rho_m(\sigma^{-1})_{ji}U_\sigma$.
Apply $M\mapsto (\epsilon_J^{\mathsf T}M\epsilon_J)^{T_J}$ to these units,
then read the transformed highest-weight blocks.  The exact integer
calculation S5-1 checks all $42$ matrix units against their physical
$32\times32$ operators, reconstructs each transformed operator, and checks
both involutions and the denominator claim.
\end{proof}

\begin{lemma}[Rational coordinates for the leading two-flip blocks; Y5-1]
\label{lem:fivecopy-young-blocks}
Let $F\cong\mathbb Q^4$ and $V\cong\mathbb Q^5$ carry the $S_5$
representations $[4,1]$ and $[3,2]$ from Lemma~\ref{lem:fivecopy-site-algebra}.
For the seven-site type with one $F$ sector and six $V$ sectors, the
$\mathrm{GL}(V)$ modules of shapes $(5,1)$ and $(4,2)$ both have dimension
$420$.  Under the two-flip copy subgroup $H=S_3\times S_2$, their
$F$-tensored invariant spaces have dimensions $137$ and $148$,
respectively.  Both invariant spaces have explicit rational bases.
\end{lemma}
\begin{proof}
Pieri decomposition identifies the two modules as
\[
 \ker\bigl(\operatorname{Sym}^5V\otimes V\to\operatorname{Sym}^6V\bigr),
 \qquad
 \ker\bigl(\operatorname{Sym}^4V\otimes\operatorname{Sym}^2V
       \xrightarrow{\delta}\operatorname{Sym}^5V\otimes V\bigr),
\]
where the first map is multiplication and $\delta$ polarises one factor
of the second symmetric power.  Both are surjective, so their kernel
dimensions are $5\binom94-\binom{10}4=420$ and
$\binom84\binom62-5\binom94=420$.
Y5-1 gives sparse integer generators annihilated by each map, checks
rank $420$ over a prime field, and hence obtains rational bases of the
full kernels.  Averaging their tensor products with $F$ over the twelve
elements of $H$ gives invariant bases.  The character averages are
$137$ and $148$, and Y5-1 checks that many independent averaged columns
and their invariance under every element of $H$ exactly.
\end{proof}

\begin{proposition}[Exact leading icosahedral five-copy kernels; IK5-1--IK5-4, PK5-1--PK5-3]
\label{prop:ico-fivecopy-g34-kernels}
At $\Psi_{\rm ico}$, twelve five-copy moment blocks have the
following ranks and kernel dimensions in their $H$-invariant
multiplicity spaces, where $j$ is the number of flipped copies:
\[
\begin{array}{cc|c|cc}
 j&\text{site type}&\mathrm{GL}(5)\text{ shape} & \operatorname{rank} & \dim\ker\\ \hline
0&(0,0,7)&(7)&2&2\\
0&(0,0,7)&(6,1)&3&3\\
0&(0,0,7)&(5,2)&3&6\\
1&(0,0,7)&(7)&2&14\\
1&(0,0,7)&(6,1)&3&27\\
1&(0,0,7)&(5,2)&3&36\\
2&(0,0,7)&(7)&2&31\\
2&(0,0,7)&(6,1)&3&57\\
2&(0,0,7)&(5,2)&3&72\\
2&(0,1,6)&(6)&8&69\\
2&(0,1,6)&(5,1)&12&125\\
2&(0,1,6)&(4,2)&20&128
\end{array}
\]
For the four original blocks (the $j=2$ rows of shapes $(7)$, $(6)$, $(5,1)$ and $(4,2)$), each physical kernel is $F^{-\mathsf T}K$ for the stored rational
feature-annihilator basis $K$ and the rational Gram transport $F$
of the proof. PK5-1 constructs compact integer physical bases for all four
blocks.
PK5-2 directly constructs physical integer bases for the two additional
all-$V$ blocks, with maximum coefficient $3$ in both cases.
PK5-3 constructs exact physical kernel bases for the six $j=0,1$ blocks.
The stored feature-annihilator bases are sparse; the two one-$F$ Young blocks have at most
$13$ and $21$ nonzero coordinates per vector, respectively.
\end{proposition}
\begin{proof}
In the explicit gauge of Theorem~\ref{thm:ico}, write
$\Psi_{\rm ico}=(\sqrt{15}\,u+\sqrt{37}\,v)/\sqrt{524}$,
where $u$ has thirty signed unit entries and $v$ has two unit entries.
The flipped rank-one moment is generated by
$\Psi_{\rm ico}^{\otimes3}\otimes(J_7\Psi_{\rm ico})^{\otimes2}$.
The pentagon-site cycle preserves $u$ and $v$.  For the three blocks
with one $[4,1]$ site it permutes the six ordinary-site tensor slots,
acting only on Young and spin-multiplicity \emph{columns}.  The
$\mathrm{GL}(5)$ and $H$-invariant row coordinates stay fixed.  Hence
the seven site placements have the same joint row span as three
representatives.  The fully symmetric block has one placement.  The
scripts IK5-1--IK5-4 check the appropriate pentagon-cycle covariance entrywise
in integer tensors.

Using the exact local maps of Lemma~\ref{lem:fivecopy-site-algebra}, the
four scripts contract those representatives at six integer amplitude ratios
and interpolate the resulting degree-five polynomial matrix.  Every
binary64 tensor contraction is integer-exact: an absolute-sum bound is
$32^5 5^5\cdot24\cdot3^6<2^{53}$.  The subsequent binary64 matrix
products are chunked below $2^{52}$, and integer accumulations are
checked against their type bounds.  The four feature matrices have
sizes $33\times128$, $77\times768$, $137\times3840$ and
$148\times11520$, respectively.  For the last, the integral projector numerator
$40I-14FE+(FE)^2$ selects the kernel of the polarisation map of
Lemma~\ref{lem:fivecopy-young-blocks}.

At the physical ratio each matrix is, up to a nonzero scalar,
$15A+\sqrt{555}\,B$ with integral $A,B$ collecting the even and the odd
powers of $t$ in the degree-five polynomial matrix $C(t)=\sum_{d=0}^5 t^d C_d$.
The key simplification is that these two parts have disjoint column supports. To see this, note that the diagonal local unitary
\[
 D=\operatorname{diag}(-i,-1)_s\otimes\operatorname{diag}(1,e^{i\pi/4})^{\otimes6}
\]
multiplies $|s,\mathbf a\rangle$ by $e^{i\pi(6-2s+|\mathbf a|)/4}$, which is $1$ on the
pair and complement strings and $-1$ on the two endpoints, so $Du=u$ and $Dv=-v$.
For the spin flip $J_7=\epsilon^{\otimes7}K$, with
$\epsilon=\bigl(\begin{smallmatrix}0&1\\-1&0\end{smallmatrix}\bigr)$ and $K$ complex
conjugation, every diagonal unitary $G$ satisfies $\epsilon\bar G\epsilon^{-1}=\overline{\det G}\,G$,
so $J_7D=cDJ_7$ with $|c|=1$. Hence $D$ maps
$(u+tv)^{\otimes(5-j)}\otimes(J_7(u+tv))^{\otimes j}$ to a constant multiple of the
same tensor at $-t$. On each site, $D^{\otimes5}$ commutes with the copy permutations
and multiplies every weight vector of the highest-weight bases of
Lemma~\ref{lem:fivecopy-site-algebra} by a scalar, so it fixes the row coordinates
and acts on the columns by a diagonal unimodular matrix $\Delta$. Thus
$C(-t)=C(t)\Delta$. Since $C(t)$ is real, each column that is not identically zero is
either even or odd in $t$. In particular
\[
 C_{\mathrm{even}}(t)\,C_{\mathrm{odd}}(s)^{\mathsf T}=0\quad(\forall t,s),
 \qquad A B^{\mathsf T}=0,
\]
for the Euclidean column pairing and for any diagonal column weighting.
For any $k=x+\sqrt{555}\,y\in\mathbb Q(\sqrt{555})^n$, the relation
$k^{\mathsf T}(15A+\sqrt{555}\,B)=(15x^{\mathsf T}A+555y^{\mathsf T}B)+\sqrt{555}(x^{\mathsf T}B+15y^{\mathsf T}A)=0$
forces both summands, which are supported on disjoint columns, to vanish in each component, so
$x^{\mathsf T}A=x^{\mathsf T}B=y^{\mathsf T}A=y^{\mathsf T}B=0$.
Every left kernel over $\mathbb Q(\sqrt{555})$ is therefore the scalar extension of the rational
intersection $\ker A^{\mathsf T}\cap\ker B^{\mathsf T}$ (explaining why all
$\sqrt{555}$ coefficients vanish across all twelve blocks), and
$\operatorname{rank}_{\mathbb Q(\sqrt{555})}(15A+\sqrt{555}\,B)=\operatorname{rank}_{\mathbb Q}[A\ B]$.
IK5-1 checks that the integer
realification for shape $(5,1)$ has rational rank $24$.  For shape $(4,2)$, IK5-2 checks a nonzero $20\times20$
minor over $\mathbb Q(\sqrt{555})$ and $128$ independent annihilators of
all $11520$ columns.  IK5-3 and IK5-4 check rational realification
ranks $4$ and $16$ for shapes $(7)$ and $(6)$.  All four scripts verify their sparse annihilators
against every column in both field components.  To identify the physical kernels,
let $U$ be the integer invariant basis in the appropriate Young module,
$H$ its rational positive-definite physical Gram, and $Y$ a column of
the contracted moment.  The scripts store $A_{\rm feat}=U^{\mathsf T}Y$.
Since $Y\in\operatorname{im}U$, its physical covariant column is
\[
 U^{\mathsf T}HY=F A_{\rm feat},\qquad
 F=(U^{\mathsf T}HU)(U^{\mathsf T}U)^{-1}\in\operatorname{GL}_d(\mathbb Q).
\]
Thus, if $K$ is the stored rational basis satisfying
$K^{\mathsf T}A_{\rm feat}=0$, then $F^{-\mathsf T}K$ is a rational
basis of the physical moment kernel.  The invertible change preserves all
four ranks and kernel dimensions.  This distinction matters: the stored $K$ itself need not be a
physical kernel.

PK5-1 avoids forming the dense rational transport. Select $r$ independent
columns $Q$ from the common feature image, whose exact dimension $r$ is
already certified above. Reconstruct their ambient columns as
$Y_Q=U(U^{\mathsf T}U)^{-1}Q$, then form the physical column basis
$C_Q=U^{\mathsf T}H Y_Q$. The reconstruction is integral and
$\ker C_Q^{\mathsf T}$ is the physical moment kernel. Integer Hermite
normal form and lattice reduction give bases whose largest absolute
coefficients are respectively $3,8,14,28$ in the four original blocks' order.
The compact artifact stores the selected columns, their exact physical
images and the four kernel bases. PK5-1 reconstructs those images and
checks both ranks and exact annihilation. These data do not assert any
dual equality or positive multiplier.

For the two additional all-$V$ shapes, use the complete Young models
\[
 \mathcal Y_{61}=\ker(\operatorname{Sym}^6V\otimes V\to\operatorname{Sym}^7V),
 \qquad
 \mathcal Y_{52}=\ker(\operatorname{Sym}^5V\otimes\operatorname{Sym}^2V
       \xrightarrow{\delta}\operatorname{Sym}^6V\otimes V).
\]
The maps are surjective and the model dimensions are $720,840$.
Their characters are $h_6h_1-h_7$ and $h_5h_2-h_6h_1$; averaging over
$H$ gives invariant dimensions $60,75$. PK5-2 constructs integral invariant
bases $U$, checks these characters exactly, and verifies the spanning
ranks by nonzero modular minors and matching dimension upper bounds.

Contract the all-$V$ physical tensor and row-symmetrise its six-plus-one
or five-plus-two multiplicity indices. The seven choices of singleton
have three pentagon-cycle orbits, and the twenty-one choices of pair
have five. Taking one representative of each therefore captures every
Young and spin column. Projection against $U$ with the \emph{physical}
Gram gives covariant features $U^{\mathsf T}H Y$ directly; other Pieri
summands are orthogonal and need no explicit projector. The six integer
amplitude ratios and interpolation above give physical feature pairs
$15A+\sqrt{555}B$ of sizes $60\times384$ and $75\times640$, with $A B^{\mathsf T}=0$
by the same column-parity argument.
In each case PK5-2 checks the exact rational rank
$\operatorname{rank}[A\ B]=3$ (which already equals the rank over $\mathbb Q(\sqrt{555})$
by row orthogonality $A B^{\mathsf T}=0$, and is also checked by a nonzero
three-dimensional minor after mapping $\sqrt{555}$ into a finite prime field).
Integer HNF and
lattice reduction then give $57$ and $72$ independent rational physical
annihilators, checked against every column of both $A$ and $B$.

For $j=0,1$, use the same three all-$V$ Young modules with the invariant
subgroups $S_5$ and $S_4\times S_1$. Projection of their two-flip invariant
spaces onto these subgroups gives complete rational bases: their exact
character dimensions are respectively $(4,6,9)$ and $(16,30,39)$.
PK5-3 repeats the physical contraction with
$\psi^{\otimes(5-j)}\otimes(J_7\psi)^{\otimes j}$.
The common feature ranks are $(2,3,3)$ in both cases, determined by
$\operatorname{rank}[A\ B]$ via $A B^{\mathsf T}=0$ and checked against
matching finite-field lower bounds and full rational annihilators.

\end{proof}

A compact rational baseline uses all twelve kernels. Its Gram matrices are
identities in the supplied dyadic bases; PK5-4 checks their exact support
and nonsingularity. Thus its multipliers are strictly positive on every
listed kernel. The baseline is repaired to an exact sharp dual by the verified
bounds of Section~\ref{sec:fivecopy-positive-repair}.

\subsection{Direct probes for a sharp five-copy certificate}
\label{sec:fivecopy-direct-probes}

Let $W=(\mathbb C^2)^{\otimes5}$, let $U_\sigma$ permute its five
copies, and let $G$ be any element of the rational local commutant
$\mathcal A_5$. On seven sites put
\[
 P=\frac1{120}\sum_{\sigma\in S_5}U_\sigma^{\otimes7},\qquad
 X(G)=PG^{\otimes7}P.
\]
These linear probes determine identities on the moment space; they need not themselves be positive moments. For $j=0,1,2$, let
$H_j=S_{5-j}\times S_j$, let $\phi_j$ be the local $j$-copy spin-flip transpose,
and put $\Phi_j=\phi_j^{\otimes7}$.

\begin{lemma}[direct contracted dual probes; DP5-1]
\label{lem:fivecopy-direct-probes}
Suppose $Z$ is a site-permutation-invariant commutant operator supported
on the global $H_j$-invariant subspace. Choose left-coset representatives
$a\in H_j\backslash S_5$ and right-coset representatives
$b\in S_5/H_j$. Then
\[
 \operatorname{tr}[Z\Phi_j(X(G))]
 =\binom5j^{-2}\sum_{a,b}
       \operatorname{tr}\left[Z\,\phi_j(U_aGU_b)^{\otimes7}\right].
\]
There are $1,25,100$ terms for $j=0,1,2$, respectively.
The traces on the right are evaluated directly in the chosen Schur models;
no full gamma map is needed. Write
$a_\sigma=\operatorname{tr}(U_\sigma G)$, and let $\mathcal T$ be the
ten transpositions in $S_5$. The sharp target on this probe is
\[
 q(G)=\frac1{1200}\sum_{\sigma\in S_5,\,t\in\mathcal T}
       a_{t\sigma}^3a_\sigma^4
       -\frac{121}{917\cdot120}\sum_{\sigma\in S_5}a_\sigma^7.
\]
Equality of this polynomial with the sum of the contracted dual probes
for every $G\in\mathcal A_5$ is equivalent to the sharp dual identity
on the real, site-averaged bosonic moment space.
\end{lemma}
\begin{proof}
The spin-flip transpose is self-adjoint for the physical trace pairing.
Expanding both copies of $P$ gives $120^2$ terms. Both $Z$ and $\Phi_j(Z)$
have left and right $H_j$-invariant support: partial transpose exchanges
left and right permutations of the flipped copies and preserves the
unflipped ones. The terms therefore group into $\binom5j^2$ equal-weight
coset pairs. Moving $\Phi_j$ back to the probe gives the first formula.
The same proof applies with any numbers of copies and sites in place of $5$ and $7$.

On the bosonic subspace, the purity operator may be averaged over all
ten choices of the swapped global-copy pair. This average commutes with
$P$, so cyclicity removes one of the two projectors in its trace against
$X(G)$. All $35$ choices of the three swapped sites have the same trace
against an identical-site product $G^{\otimes7}$, yielding the formula
for $q$. Finally the span of the powers $G^{\otimes7}$ is the symmetric
seventh tensor power of $\mathcal A_5$, by polarisation. The trace pairing
on this site-symmetric algebra is nondegenerate. This proves the last claim.
Probes over $\mathbb Q(i)$ are also allowed: real and imaginary parts of
$G=A+iB$ evaluations are rational combinations of the eight real probes
$A+tB$, $t=0,\ldots,7$, by degree-seven interpolation.
\end{proof}

\begin{proposition}[the critical space and finite probes; DP5-1]
\label{prop:fivecopy-finite-probes}
Let $\mathcal M_5$ be the real span of the normalised five-copy moments
of pure seven-qubit states after local-unitary twirling, site averaging
and averaging with the complex conjugate state (the conjugation-even
part). Then $\dim\mathcal M_5=3285$. The \emph{critical space}
$\mathcal C\subset\mathcal M_5^*$ of functionals vanishing at the
icosahedral moment $X_0$ and at its derivative $X_0'$ along the
normalised family $\psi_x$ of~\eqref{eq:ico} at $x_0=37/524$ has
dimension $3283$. If functionals in $\mathcal C$ have an evaluation
matrix of rank $3283$ on some finite set of probes, they span
$\mathcal C$, and evaluation on those probes is injective on
$\mathcal C$.
\end{proposition}
\begin{proof}
By the last part of Lemma~\ref{lem:fivecopy-direct-probes}, $\mathcal M_5$
is dual to the site-symmetric probe polynomials, that is, to the permutation-symmetric local-unitary
invariants of bidegree $(5,5)$ for $N=7$ sites and local dimension $r=2$. Their Schur--Weyl character sum
(\eqref{eq:fqe-invariant-dimension} and~\eqref{eq:fqe-invariant-even} of
Section~\ref{sec:fqe-invariant-dimensions}) gives dimension $5844$, and the trace of the conjugation swap is
$726$ (DP5-1), so the conjugation-even part has dimension
$(5844+726)/2=3285$. This is the dimension of the linear span $\mathcal M_5$; the normalised moments
lie in its affine hyperplane $\nu=1$. The normalisation functional $\nu$ has
$\nu(X_0)=1$ and $\nu(X_0')=0$, and DP5-1 checks that the weight-two Pauli enumerator
$A_2$ has derivative $96/131\ne0$ there, so $X_0'\ne0$ and $X_0,X_0'$ are independent. Hence
$\dim\mathcal C=3283$. A rank-$3283$ evaluation matrix of elements of
$\mathcal C$ gives both remaining assertions.
\end{proof}

The criterion is applied to the twelve-face baseline in
Section~\ref{sec:fivecopy-positive-repair}, where rank and repair bounds
are verified for the rational data.

\subsection{A positive repair lemma}
\label{sec:positive-repair}

The seven-qubit purity and entropy certificates
(Theorems~\ref{thm:fivecopy-sharp-certificate}
and~\ref{thm:seven-entropy-localized-identity}) both correct a strictly positive
baseline, which fails the target identity, into an exact positive identity. The
correction is found from finitely many evaluations, and its existence and
positivity follow from two verified norm bounds. The common step is the following
finite-dimensional real linear algebra; each application supplies its own spaces
and verifies the bounds.

\begin{lemma}[positive repair]
\label{lem:positive-repair}
Let $\mathcal V$ be a finite-dimensional real vector space, $\mathcal C\subseteq\mathcal V$
a subspace of dimension $d$, and $\mathcal E:\mathcal V\to\mathbb R^m$ a linear map. Let
$\Phi:\mathbb R^N\to\mathcal V$ be linear with $\Phi(\mathbb R^N)\subseteq\mathcal C$, let
$\rho\in\mathcal C$, and put $A=\mathcal E\Phi\in\mathbb R^{m\times N}$ and
$r=\mathcal E\rho\in\mathbb R^m$. Suppose $L\in\mathbb R^{d\times m}$ and
$R\in\mathbb R^{N\times d}$ satisfy
\begin{equation}\label{eq:positive-repair-inverse}
 \|I_d-LAR\|_2\le\delta<1 .
\end{equation}
\begin{enumerate}
\item[(a)] $A$ has rank $d$; $\Phi$ maps onto $\mathcal C$, and $\mathcal E$ is injective
  on $\mathcal C$.
\item[(b)] $z=R(LAR)^{-1}Lr$ satisfies $\Phi(z)=\rho$ and
  $\|z\|_2\le\|R\|_2\|Lr\|_2/(1-\delta)$. If the entries of $A,r,L,R$ lie in a
  subfield $K\subseteq\mathbb R$, then $z\in K^N$.
\item[(c)] Let $\Delta_b:\mathbb R^N\to\operatorname{Sym}_{k_b}(\mathbb R)$, $b=1,\ldots,B$, be
  linear maps with $\|\Delta_b(y)\|_2\le c\|y\|_2$ for every $y$. If
  \begin{equation}\label{eq:positive-repair-margin}
   \frac{c\,\|R\|_2\|Lr\|_2}{1-\delta}<1,
  \end{equation}
  then $I+\Delta_b(z)\succ0$ for every $b$.
\end{enumerate}
\end{lemma}
\begin{proof}
By~\eqref{eq:positive-repair-inverse} the Neumann series for $(LAR)^{-1}$ converges, so
$LAR$ is invertible with $\|(LAR)^{-1}\|_2\le1/(1-\delta)$. Hence $\operatorname{rank}A\ge d$.
Since $A=\mathcal E\Phi$ factors through the $d$-dimensional space $\mathcal C$, also
$\operatorname{rank}A\le d$. Equality forces $\Phi(\mathbb R^N)=\mathcal C$ and
$\ker\mathcal E\cap\mathcal C=0$, proving (a).

Since $AR$ has rank $d$, its image is $A(\mathbb R^N)=\mathcal E(\mathcal C)$, which
contains $r$. Write $r=ARy$; applying $L$ gives $y=(LAR)^{-1}Lr$, so $Az=r$. Thus
$\mathcal E(\Phi(z)-\rho)=0$ with $\Phi(z)-\rho\in\mathcal C$, and (a) gives $\Phi(z)=\rho$.
The norm bound follows from the bound on $\|(LAR)^{-1}\|_2$, and $z\in K^N$ because
$(LAR)^{-1}$ has entries in $K$. For (c),
$\|\Delta_b(z)\|_2\le c\|z\|_2<1$ by (b) and~\eqref{eq:positive-repair-margin}, so every
eigenvalue of $I+\Delta_b(z)$ is positive.
\end{proof}
No singular value enters the conclusion: $\delta$ and the margin bound the exact matrices,
and $L,R$ are arbitrary candidates (in both applications, binary64 singular-vector data read
as exact dyadic rationals).

\subsection{A positive twelve-face certificate}
\label{sec:fivecopy-positive-repair}

We construct the certificate by correcting a positive baseline on twelve exact physical
kernels. In each supplied rational kernel basis, the identity Gram is
strictly positive. A sufficiently small correction can therefore enforce the polynomial
identity while preserving positivity. The same rank computation also gives the equality case
(Corollary~\ref{cor:seven-global-from-fivecopy}).

\begin{theorem}[sharp five-copy certificate; verified-arithmetic certificate, FP5-1]
\label{thm:fivecopy-sharp-certificate}
There is a rational sharp five-copy dual supported on the twelve
physical kernels of Proposition~\ref{prop:ico-fivecopy-g34-kernels}:
rational $C_b\succ0$ with
\[
 q(G)=\sum_b\operatorname{tr}\bigl[C_b\,W_b^{\mathsf T}\Gamma_b(G)W_b\bigr]
 \qquad\text{for every }G\in\mathcal A_5 .
\]
Consequently the five-copy relaxation has value $121/917$. Moreover the
functionals $X\mapsto\operatorname{tr}[W_bSW_b^{\mathsf T}\Gamma_b(X)]$,
$S$ symmetric, span the $3283$-dimensional critical space of
Proposition~\ref{prop:fivecopy-finite-probes}.
\end{theorem}
The $C_b$ solve a rational linear system whose solvability is certified with rigorous floating-point bounds;
they are not computed in exact arithmetic, which is why the theorem is a verified-arithmetic certificate.
The rational coefficients therefore exist but are not printed: their existence and the strict positivity of every
$C_b$ follow from the verified inverse test~\eqref{eq:fivecopy-verified-repair} in the proof.
\begin{proof}
Write $W_b=U_bK_bB_b$ for the supplied rational basis of the $b$th
physical kernel (PK5-4 checks that it annihilates the icosahedral
moment block), and let $\Gamma_b(G)$ be the contracted probe matrix,
including its positive multiplicity weight. The physical inner product
is used on the ambient Young model, so an identity Gram in these
coordinates is a positive physical multiplier even though the columns of
$W_b$ need not be orthonormal.

Let $Q_b$ be the rational orthogonal matrix given by the supplied
Householder reflections, and put $r_b=\min(7,k_b)$, where
$k_b=\operatorname{cols}W_b$. Restrict repairs to
\[
 C_b=I+Q_b\begin{pmatrix}D_b&E_b\\E_b^{\mathsf T}&0\end{pmatrix}Q_b^{\mathsf T},
 \qquad D_b=D_b^{\mathsf T}\in\mathbb R^{r_b\times r_b}.
\]
Each reflection is $I-2vv^{\mathsf T}/(v^{\mathsf T}v)$; orthogonality is
checked with integers. Let $z_b$ list the upper-triangular entries of $D_b$
and all entries of $E_b$. Then
$\|C_b-I\|_2\le\sqrt2\|z_b\|_2\le\tfrac32\|z_b\|_2$.
There are $\sum_b[r_bk_b-r_b(r_b-1)/2]=3754$ independent coordinates.
Width six would give only $3249$, too few to span the critical space
(FP5-4), so seven is the smallest possible uniform width for this argument.

For every vector $v$, the smooth function
$v^{\mathsf T}W_b^{\mathsf T}\Gamma_b(X_5^+(\psi_x))W_bv$ is
nonnegative and vanishes at $x=37/524$, so its derivative vanishes there.
Thus every repair functional is critical. The target $q$ is critical by
Theorem~\ref{thm:ico}.
We apply Lemma~\ref{lem:positive-repair} with $\mathcal V=\mathcal M_5^*$,
$\mathcal C$ the critical space ($d=3283$, Proposition~\ref{prop:fivecopy-finite-probes}),
$N=3754$, $\Phi(z)=\sum_b\operatorname{tr}[(C_b-I)W_b^{\mathsf T}\Gamma_bW_b]$,
$\Delta_b(z)=C_b-I$ with $c=\tfrac32$, and
\[
 \rho=q-\sum_b\operatorname{tr}[W_b^{\mathsf T}\Gamma_bW_b].
\]
By the preceding paragraph $\Phi(\mathbb R^N)\subseteq\mathcal C$ and $\rho\in\mathcal C$.
Let $\mathcal E$ evaluate at the $3424$ supplied rational probes,
with the supplied nonzero dyadic row weights, so that $A$ is the rational evaluation
matrix of the repair directions and $r=\mathcal E\rho$.
The verifier takes the leading $3283$ singular triples of a floating
evaluation of $A$, reads their binary64 entries as exact dyadic
rationals, and forms from them rational matrices $L$ and $R$ of sizes
$3283\times3424$ and $3754\times3283$. With the error bounds below it
establishes
\begin{equation}\label{eq:fivecopy-verified-repair}
 \|I-LAR\|_2\le\delta<1,\qquad
 \frac{3\|R\|_2\|Lr\|_2}{2(1-\delta)}<1 ;
\end{equation}
the logged upper bounds are $\delta<0.070$ and $0.324$. These are
\eqref{eq:positive-repair-inverse} and~\eqref{eq:positive-repair-margin}
for the exact rational matrices. FP5-3 reimplements the probe evaluator independently and
reproduces all face traces and repair coefficients on non-scalar probes.

Lemma~\ref{lem:positive-repair}(a) gives rank $3283$, so the repair
functionals span the critical space (the rank criterion of
Proposition~\ref{prop:fivecopy-finite-probes}). Part~(b), with $K=\mathbb Q$,
gives rational $z$ with $\Phi(z)=\rho$, which is the asserted identity, and
part~(c), using $\|C_b-I\|_2\le\tfrac32\|z_b\|_2\le\tfrac32\|z\|_2$, gives every
$C_b\succ0$. Lemma~\ref{lem:fivecopy-direct-probes} turns
the identity into an exact rational feasible sharp dual, and the
icosahedral moment is feasible with value $121/917$.
\end{proof}

\begin{lemma}[equality moments lie on the tangent line]
\label{lem:fivecopy-tangent-line}
Let $\mathcal M$ be a finite-dimensional real moment space with normalisation
functional $\nu$. Fix $X_0,X_0'\in\mathcal M$ with
$\nu(X_0)=1$, $\nu(X_0')=0$ and $X_0'\ne0$, and let
$\mathcal C=\{f\in\mathcal M^*:f(X_0)=f(X_0')=0\}$. Let $\Gamma_b$ be
linear maps to symmetric matrices, and $W_b$ matrices such that the
functionals $X\mapsto\operatorname{tr}[W_bSW_b^{\mathsf T}\Gamma_b(X)]$
span $\mathcal C$. If $\nu(X)=1$ and
$W_b^{\mathsf T}\Gamma_b(X)W_b=0$ for every $b$, then
$X=X_0+tX_0'$ for some $t\in\mathbb R$.
\end{lemma}
\begin{proof}
Every functional of the spanning set, hence every $f\in\mathcal C$,
vanishes at $X$. Since
$\nu(X_0)=1$ and $\nu(X_0')=0$, the vectors $X_0,X_0'$ are independent,
$\mathcal C$ has codimension two, and its annihilator is their span.
Normalisation fixes the coefficient of $X_0$ at one.
The lemma uses no property of the number of copies, the number of qubits or the target.
\end{proof}

\paragraph{Rational models and exact target values.}
The verification uses only the exact certificate data. Each baseline factor
$B_b$ is lower triangular with positive dyadic diagonal; its invertibility
is therefore checked directly over the rationals. There are $24502$
potentially nonzero factor entries, rather than $48434$ entries in dense
factors. Each is stored as an odd signed integer of at most $53$ bits times
an integral power of two, and integer shifts reconstruct it exactly.
The twelve frames are regenerated from
fixed PCG64 seeds; FP5-4 fingerprints all $36{,}480$ original Householder
integers before checking their exact orthogonality. The probe archive uses
lossless compression, with signed $16$-bit probe numerators in column order.
The baseline, frames and probes occupy $441612$ bytes. PK5-4 fingerprints
the triangular factors and checks their full physical kernel support;
the probe fingerprint preserves every probe value and row weight.
In the rational
one-site model write its physical Gram as $G_d=L_dD_dL_d^{\mathsf T}$,
with rational $L_d$ and positive diagonal rational $D_d$.
Changing coordinates by $L_d^{\mathsf T}$ makes the physical Gram diagonal
without square roots. In a symmetric power the monomial of exponent $m$
has squared norm
\[
 \frac{\prod_i m_i!}{(\sum_i m_i)!}\prod_i(D_d)_{ii}^{m_i}.
\]
Thus both the kernel columns and their physical dual columns are rational.
They, and their products with the rational repair frames, are formed
exactly before conversion to binary64. The conversion is correctly
rounded; exact zero counts and exponent guards exclude underflow.
The local spin-flip and copy-permutation maps are formed exactly as
rational $42\times42$ matrices.

Each source probe has integer coordinates divided by $8192$. Consequently
all $a_\sigma$ in Lemma~\ref{lem:fivecopy-direct-probes} are rational with
this denominator. If their integer numerators are $a'_{\sigma}$, the target
is evaluated first as the exact fraction
\[
 \frac{917\sum_{\sigma,t}(a'_{t\sigma})^3(a'_\sigma)^4
                  -1210\sum_\sigma(a'_\sigma)^7}
 {917\cdot1200\cdot8192^7}.
\]
For probes commuting with the copy transposition $(12)$, the paired cosets
are also identified under $(a,b)\mapsto(a(12),(12)b)$.
The commutation is checked exactly before using this reduction.
An independent exact scalar control (FP5-2) takes $G$ to be the identity
on the local spin-$5/2$ block and zero elsewhere. Then
$q(G)=(796/917)6^7$, and only the two-flip all-$V$ shape $(7)$ and
one-$F$ shape $(6)$ survive. Their forms are respectively
$6^7d_7d_7^{\mathsf T}$ and $(7/3)6^7d_6d_6^{\mathsf T}$, where the right
functionals on the ambient polynomial models are $e_5^7$ and
$(e_3+e_4)\otimes e_5^6$. This follows because, for this $G$, every
local two-flip block has rank one, so each contracted gamma form is a
symmetric power of a rank-one local form.
FP5-2 checks every thin coefficient and the baseline trace against these
exact rational forms, including the ten zero faces.

\paragraph{Forward-error bounds for the probes.}
These bounds, and those of the entropy certificate E7-1, hold under one arithmetic model:
IEEE-754 binary64 with round-to-nearest, so that every floating-point addition, multiplication, division or square root
(including any fused multiply--add inside the single-threaded BLAS calls) returns the exact result times $1+\delta$
with $|\delta|\le u$, away from underflow and overflow, which are treated separately below.
Put $u=2^{-53}$ and $\gamma_s=su/(1-su)$.
For a large face the baseline trace is evaluated as
$\operatorname{tr}(TC)$, where $C=V_RV_L^{\mathsf T}$ is formed once.
Its tensor indices are contracted in decreasing factor dimension, with
inner products blocked at $128$ terms. The first repair term is evaluated
as $(T^{\mathsf T}V_{LU})^{\mathsf T}V_R$, so both repair actions use
only $r_b\le7$ columns. The smaller faces use the factored baseline trace.
These are cyclic trace and transpose identities; no additional
certificate data are retained. Forming $C$ costs error depth $k+2$,
including its two column conversions.
For every signed contraction the verifier evaluates a positive companion.
For a column matrix $V$, write $n(V)_i=\|V_{i,:}\|_2$. If $T_+$ is the
nonnegative tensor action obtained by replacing the local matrices by
their absolute values, Minkowski and Cauchy give
\[
 n(TV)\le T_+n(V),\qquad
 \|A^{\mathsf T}TV\|_F\le\bigl\||A|^{\mathsf T}T_+n(V)\bigr\|_2,
 \qquad
 |\operatorname{tr}(A^{\mathsf T}TV)|\le n(A)^{\mathsf T}T_+n(V).
\]
Thus single-column tensor actions on $n(V_R)$ and $n(V_{RU})$ bound the
two coefficient terms and the trace. The positive coefficient vectors
are summed over coset terms before taking their norms; the two term
bounds are added, and the face bounds are combined in Euclidean norm.
The resulting scalar radius bounds the Frobenius norm of the entire
coefficient error. Row norms are rounded upwards with $1+2^{-20}$;
their squared inputs are normal by the conversion guards below.
The polynomial functionals being evaluated are unchanged.
A degree-$n$ symmetric-power recurrence has at most five terms per entry
at each step. Including conversion of the local matrix gives error depth
at most $6n$. A matrix product with inner dimension $d$, split into
chunks of length $c$ and summed pairwise, has depth at most
\[
 \min(d,c)+\left\lceil\log_2\lceil d/c\rceil\right\rceil.
\]
If the ambient model has factor dimensions $d_i$, total dimension $D$,
kernel dimension $k$, and $t$ coset terms, define
\begin{align*}
 a&=\sum_i\left(\min(d_i,128)+\lceil\log_2\lceil d_i/128\rceil\rceil\right),\\
 f&=\min(D,256)+\lceil\log_2\lceil D/256\rceil\rceil,\\
 h&=\min(D,128)+\lceil\log_2\lceil D/128\rceil\rceil+\lceil\log_2 k\rceil+1,\\
 s&=8+\lceil t/8\rceil+16.
\end{align*}
With $g\le7$ the local degree of the model, the coefficient and trace error depths are respectively $a+f+6g+s$
and $a+h+6g+s$ for a factored trace. For a multiplier trace its depth is
\[
 k+2+6g+s+
 \sum_i\left[\min(d_i^2,128)+\lceil\log_2\lceil d_i^2/128\rceil\rceil\right].
\]
The final $16$ covers the two rational column conversions,
the addition of the two thin repair terms, multiplicity scaling,
and summation over all twelve models. A common error budget for the
positive companion is
\[
 b=Dk+\sum_i d_i+6g+2\binom5j^2+64.
\]
If its computed value is $p$, the corresponding error radius is bounded
by $\gamma_{a+f+6g+s}p/(1-\gamma_b)$, or with the corresponding trace
depth for the trace. An additional factor $1+2^{-20}$ bounds rounding in the positive
bounds themselves. Every scalar accumulation and Frobenius norm used in these bounds has
fewer than $2^{29}$ terms. The input guards give $|h_{ij}|<2^{10}$ for
every exact local matrix, $D<2^{13}$, $k<2^8$, multiplicity below $2^{15}$
and model-column entries below $2^{20}$. Each degree-$n$ symmetric-power
entry is bounded by $(5\cdot2^{10})^n<2^{13n}$. Since the total degree is
seven, signed tensor actions are bounded by $2^{13+91+20}=2^{124}$ and
norm companions by $2^{128}$; subsequent
column products, traces, multiplicities and at most $100$ coset terms keep
all mathematical intermediates below $2^{200}$. These bounds also hold
with a factor of two for the rounded computations, so overflow is
excluded without repeatedly scanning the large arrays. An underflow
error is at most $2^{-1074}$ per operation. Even allowing $2^{100}$
operations and amplification by $2^{200}$ gives less than $2^{-774}$;
the verifier adds $2^{-500}$ to each coefficient and trace radius.
The initial row-norm squares are normal. Norms of the accumulated
positive coefficient vectors are taken after the multiplicity and coset
scales: loss of subnormal squares changes such a norm by less than
$2^{-520}$, and its contribution to a radius is multiplied by
$\gamma_{a+f+6g+s}<2^{-30}$. The same $2^{-500}$ allowance covers it.
FP5-2 checks row scales between $1$ and $2^{32}$, and model conversion
guards keep every nonzero input column above $2^{-400}$.
In the inverse validation, possible dot-product underflows, even after
multiplication by a guarded $d_i$ and taking a Frobenius norm, contribute
less than $2^{-600}$. The separate norm allowances and these terms sum
to less than $2^{-500}$, far below the coarse rational margins.

\paragraph{Validation of the rational inverse.}
The probe evaluator first gives rectangular coefficients in the first
$r_b$ frame columns. To obtain the independent coordinates, multiply
each coefficient array $M_b$ by the full $Q_b$, halve the diagonal,
average the two entries on each off-diagonal of the leading square,
and retain its upper triangle and the remaining rectangular block.
These selections and averages contract the Euclidean error norm.
If the initial row radius is $\epsilon$, a radius after this change is
\[
 \epsilon+\sum_b\left[
 \left(\gamma_{k_b}+\frac{u}{1-u}\right)\|M_b\|_F\|\widetilde Q_b\|_F
 +u\|\operatorname{fl}(M_b\widetilde Q_b)\|_F\right],
\]
with the same upward norm allowance. Here $\widetilde Q_b$ is the
correctly rounded rational orthogonal matrix. FP5-4 checks this packing
against independent exact matrix contractions on every supplied frame.

Let $\widetilde A,\widetilde r$ be the computed centres, and let $e_j$
bound the Euclidean norm of row $j$ of $A-\widetilde A$ and $\tau_j$
bound $|r_j-\widetilde r_j|$. The preceding radii provide these bounds.
Take the first $3283$ left and right singular vectors of
$\widetilde A$ merely as candidate data, and regard their binary64
entries, and the computed reciprocal singular values $d_i$, as exact
dyadic rationals. Write these matrices as $U,V,d$ and put
\[
 L=\operatorname{diag}(d)U^{\mathsf T},\qquad R=V^{\mathsf T}.
\]
An independent Gram-product error bound verifies
$\|R\|_2\le v=1+2^{-12}$.
For the computed centres
$H=\operatorname{fl}(\operatorname{diag}(d)\operatorname{fl}(U^{\mathsf T}\widetilde A))$
and $M=\operatorname{fl}(HV^{\mathsf T})$, the inverse defect is at most
\begin{align*}
 \|I-M\|_F
 &+(\gamma_m+u+u\gamma_m)d_{\max}\|U\|_F\|\widetilde A\|_Fv
   +\gamma_n\|H\|_F\|V\|_F\\
 &+v\|\operatorname{diag}(d)|U|^{\mathsf T}e\|_2,
 \qquad m=3424,\quad n=3754.
\end{align*}
Likewise, if $p=\operatorname{fl}(\operatorname{diag}(d)
\operatorname{fl}(U^{\mathsf T}\widetilde r))$, then
\[
 \|Lr\|_2\le
 \left\||p|+\operatorname{diag}(d)|U|^{\mathsf T}
       (\tau+\gamma_m|\widetilde r|)+\frac{u}{1-u}|p|\right\|_2.
\]
All norms and positive bounds are rounded upwards using the same
$1+2^{-20}$ allowance. Substituting these bounds into
$3v\|Lr\|_2/[2(1-\delta)]$ verifies
\eqref{eq:fivecopy-verified-repair}. Thus neither a numerical rank assumption nor a dense exact inverse is
needed.

\subsection{Degree-four invariants fix the orbit}\label{sec:fivecopy-orbit}
The moment certificate fixes invariant values; we must now recover the state. The invariants needed for this
step have degree at most four and are linear in the four-copy marginal of the five-copy moment. Combined with the tangent-line lemma this gives one
equality theorem for every family point with $1/32<x<1/12$ (Theorem~\ref{thm:fivecopy-family-equality}), applied below to the
purity and in Section~\ref{sec:seven-entropy-equality} to the entropy.

\begin{theorem}[Degree-four invariants identify the icosahedral family; proved, FO-1]
\label{thm:moment-orbit-rigidity}
Fix $y\in(1/32,1/12)$. Let $z$ be a unit pure seven-qubit state whose thirteen
product-$\mathrm{SU}(2)$, site-permutation and complex-conjugation invariant polynomials
\[
 A_1,\ C,\ D,\ \mathcal E_2,\ \mathcal E_3,\ \mathcal E_4,\ B_1,\ S_1,\ Q,\ R,\ F,\ L,\ \pi_3,
\]
all of bidegree at most $(4,4)$ and defined below, take their values at $\psi_y$.
Then $z$ is local-unitarily and site-permutation equivalent to $\psi_y$.
In particular, relabelling the distinguished site as $0$ and the ordinary sites as $1,\ldots,6$,
in a suitable local frame its one-body marginals are $I/2$ and
\[
 T_{0j}=v e_ze_z^{\mathsf T},\qquad T_{ij}=w e_ze_z^{\mathsf T}\quad(i,j>0),
 \qquad v=\frac{1-8y}{3},\quad w=\frac{32y-1}{15}.
\]
The interval contains both $37/524$ and the entropy parameter $x_*$.
\end{theorem}

\begin{proof}
Write $r_i^a=\langle\sigma_a^{(i)}\rangle$ and
$T_{ij}^{ab}=\langle\sigma_a^{(i)}\sigma_b^{(j)}\rangle$.
At site $i$, concatenate $T_{ij}$ for $j>i$ and $T_{ji}^{\mathsf T}$
for $j<i$ into the real $3\times18$ matrix $M_i$.  The incident-axis
sum of squares
\[
 C=\sum_i\sum_{\substack{a<b\\c<d}}
    (M_i^{ac}M_i^{bd}-M_i^{ad}M_i^{bc})^2
  =\frac12\sum_i\bigl[(\operatorname{tr}M_iM_i^{\mathsf T})^2
  -\operatorname{tr}(M_iM_i^{\mathsf T})^2\bigr]
\]
has bidegree $(4,4)$ in $z,\bar z$.  Also sum, over ordered quadruples
of distinct sites, the nonnegative tetrad squares
\[
 D_{ijkl}=\sum_{a,b,c,d}
 (T_{ij}^{ab}T_{kl}^{cd}-T_{ik}^{ac}T_{jl}^{bd})^2.
\]
The site-sum $D$ likewise has bidegree $(4,4)$.
Local $SU(2)$ actions rotate each Pauli index orthogonally; each
displayed squared tensor norm is unchanged.  Site permutations
relabel summands.  Complex conjugation reflects the $Y$ Pauli axis,
another orthogonal action.  Hence $C,D$ and
$A_1=\sum_{i,a}(r_i^a)^2$ are conjugation-even invariant polynomials.

By Theorem~\ref{thm:ico} and its proof, at $\psi_y$ every one-body marginal is $I/2$ and the two-body computational-basis marginals are $\rho_{si}=\operatorname{diag}(10u,5u+y,5u+y,10u)$ and $\rho_{ij}=\operatorname{diag}(\tfrac12-8u,8u,8u,\tfrac12-8u)$ with $u=(1-2y)/30$. In this displayed frame the axial pair tensors are $T_{0j}=ve_ze_z^{\mathsf T}$ and $T_{ij}=we_ze_z^{\mathsf T}$ with $v=2(10u)-2(5u+y)=(1-8y)/3$ and $w=2(\tfrac12-8u)-2(8u)=(32y-1)/15$, so $A_1=C=D=0$ at $\psi_y$ (checked independently by FO-1). On the stated interval $v>w>0$.
The hypothesis forces $A_1(z)=C(z)=D(z)=0$.
Thus all one-body Bloch vectors vanish and $\operatorname{rank}M_i\le1$
for every $i$.  At each site choose a unit vector $n_i$ spanning the
column space of $M_i$, or any unit vector if $M_i=0$.
Every $T_{ij}$ then has the form $c_{ij}n_i n_j^{\mathsf T}$.
Product local rotations send all $n_i$ to $e_z$, leaving a real
symmetric scalar edge array $(c_{ij})$.  The tetrad equalities become
\begin{equation}\label{eq:fivecopy-pair-tetrads}
 c_{ij}c_{kl}=c_{ik}c_{jl}=c_{il}c_{jk}
 \qquad(i,j,k,l\text{ distinct}).
\end{equation}

The following four scalar graph invariants are again conjugation-even,
site-symmetric and of bidegree at most $(4,4)$:
\[
 \mathcal E_2=\sum_{i<j}\|T_{ij}\|_F^2,\qquad
 \mathcal E_3=\sum_{i<j<k}\operatorname{tr}(T_{ij}T_{jk}T_{ki}),
\]
\[
 \mathcal E_4=\frac13\sum_{i<j<k<l}
     \sum_{\text{three cycles on }\{i,j,k,l\}}
       \operatorname{tr}(T_{ij}T_{jk}T_{kl}T_{li}),
 \qquad B_1=\sum_{i<j}\|T_{ij}\|_F^4.
\]
In the $\mathcal E_4$ expression the indices inside the trace follow each listed cycle.
Their family values are
\[
 \mathcal E_2=6v^2+15w^2,\quad \mathcal E_3=15v^2w+20w^3,\quad
 \mathcal E_4=20v^2w^2+15w^4,\quad B_1=6v^4+15w^4.
\]
In particular $\mathcal E_3>0$, so the edge array has a nonzero triangle.
By~\eqref{eq:fivecopy-pair-tetrads}, a vertex outside this triangle
is either connected to all its vertices or to none; in the latter
case it has no edge to any other vertex.  Two vertices connected to
the triangle have a nonzero mutual edge.  Thus the nonzero edges form
a complete graph on one subset, with the other vertices isolated.
For the scalar edge reconstruction, temporarily index the seven sites $1,\dots,7$, with a nonzero triangle on $1,2,3$, and put
$\tau=c_{12}c_{13}/c_{23}\ne0$ and
$\varepsilon=\operatorname{sgn}\tau$.
Set $t_1=\sqrt{|\tau|}$ and
$t_j=c_{1j}/(\varepsilon t_1)$ on the active subset, and $t_j=0$
elsewhere.  The tetrads give $c_{ij}=\varepsilon t_it_j$ for every pair.
Every nonzero triangle product therefore has sign $\varepsilon$;
since $\mathcal E_3>0$, $\varepsilon=1$.

Put $x_i=t_i^2\ge0$, $e_k=e_k(x_1,\ldots,x_7)$, and
$p_k=\sum_i x_i^k$.  The graph invariants above become
$\mathcal E_2=e_2$, $\mathcal E_3=e_3$, $\mathcal E_4=e_4$, $B_1=e_2(x_1^2,\ldots,x_7^2)$.
Newton's identities give
\[
 B_1=e_2^2-2e_1e_3+2e_4,\qquad
 e_1=\frac{\mathcal E_2^2+2\mathcal E_4-B_1}{2\mathcal E_3}=r+6\ell,
 \qquad r=\frac{v^2}{w},\quad\ell=w.
\]
They also determine $p_2=e_1^2-2e_2$ and $p_3=e_1^3-3e_1e_2+3e_3$.
For $\mu=e_1/7$ and $S_k=\sum_i(x_i-\mu)^k$, evaluating at the family profile $(r,\ell,\dots,\ell)$ with deviations $\frac67(r-\ell)$ and $-\frac17(r-\ell)$ (six times) gives (FO-1)
\[
 S_2=\frac67(r-\ell)^2>0,\qquad
 S_3=\frac{30}{49}(r-\ell)^3>0,\qquad
 S_3^2=\frac{25}{42}S_2^3.
\]
The sharp seven-variable skewness inequality is
$S_3\le(5/\sqrt{42})S_2^{3/2}$, with equality only for one
high value and six equal low values.  Indeed, maximising $S_3$ at
fixed zero mean and $S_2$ by Lagrange multipliers yields at most two
values; if the higher value occurs $m$ times, the ratio is
$(7-2m)/\sqrt{7m(7-m)}$, maximal only at $m=1$.
The recovered mean and variance give one value $r=v^2/w$ and six values
$\ell=w$. All are positive. Flipping the local axis directions makes all
$t_i$ positive, so there is one distinguished vertex, with incident weight
$\sqrt{r\ell}=v$, and every ordinary edge has weight $w$.

The pair correlations have now fixed the local axes and edge weights. Return to labels $0,\ldots,6$,
with $0$ distinguished. We next use the remaining invariants to recover the state in that frame.
The symbols $Q,R,F,L$ below denote seven-qubit invariants local to this proof. Let
$J_7=\epsilon^{\otimes7}K$ be the spin flip, with $K$ complex conjugation,
and form the $21$ one-body spin-flip vectors $V_{ia}=\sigma_a^{(i)}J_7z$
($0\le i\le6$, $a\in\{x,y,z\}$), their Gram matrix $G=V^\dagger V$,
the overlaps $h=V^\dagger z$ and $q_i=\sum_a|h_{ia}|^2$.
Put $d_i=\sum_{j\ne i}\|T_{ij}\|_F^2$ and define
\[
 S_1=\sum_iq_i,\qquad Q=\sum_iq_i^2,\qquad
 R=\sum_id_iq_i,
\]
\[
 F=\sum_{i\ne j}\bigl(\|T_{ij}\|_F^2q_i
                -\|T_{ij}^{\mathsf T}h_i\|^2\bigr),\qquad
 L=\sum_{i<j}\operatorname{Re}(h_i^\dagger T_{ij}h_j),
\]
where $T_{ij}=T_{ji}^{\mathsf T}$ for $i>j$.
Each summand of $F$ is nonnegative because the operator norm of a
matrix is at most its Frobenius norm.  The respective bidegrees in
$(z,\bar z)$ are $(2,2)$ for $S_1$, $(4,4)$ for $Q,R,F$, and $(3,3)$
for $L$.  All five quantities are invariant under product $SU(2)$
and site permutations: $J_7U=UJ_7$, so the Pauli indices of $h_i$
and $T_{ij}$ rotate together.  Under complex conjugation, with
$\Delta=\operatorname{diag}(1,-1,1)$, they transform as
$h_i(\bar z)=\Delta\overline{h_i(z)}$ and
$T_{ij}(\bar z)=\Delta T_{ij}(z)\Delta$; the real part makes $L$ invariant.
A global phase multiplies all $h_i$ by one phase.  Their values equal those of $\psi_y$ by hypothesis.

In the displayed frame of~\eqref{eq:ico}, $J_7|b\rangle=(-1)^{7-|b|}|\bar b\rangle$. The pair and complement strings have total weights $2$ and $5$, and the endpoint strings have weights $6$ and $1$, so their spin-flip signs must be retained separately. Comparing complementary amplitudes, the antilinear map $\sigma_z^{(s)}J_7$ acts on each string of the pair/complement pattern with sign $-1$ and on the endpoint pattern with sign $+1$. At an ordinary site $i$ the map is $\sigma_z^{(i)}J_7=(-1)^{b_s+b_i}\sigma_z^{(s)}J_7$, which reverses the sign on both endpoints and on the $10$ pair/complement strings with $b_i\ne b_s$; these $10$ strings, now with sign $+1$, and the other $20$, with sign $-1$, contribute $-10u$, while the endpoints contribute $-2y$. The transverse components of every $h_i$ vanish: $J_7\psi_y$ has the same support as $\psi_y$ ($s=0$ with ordinary weights $\{2,6\}$, $s=1$ with $\{0,4\}$), $X_i$ or $Y_i$ at an ordinary site changes the parity of the ordinary weight, and $X_s$ or $Y_s$ moves a string into the other branch, whose weight set is disjoint. This yields the family frame values (checked by FO-1)
\[
 h_s^*=-1+4y,\qquad h_o^*=-\frac{1+4y}{3},\qquad
 q_0^*=(h_s^*)^2,\quad q_o^*=(h_o^*)^2.
\]
Both $h_s^*$ and $h_o^*$ are negative. The pair profile gives
$d_0=6v^2$ and $d_i=v^2+5w^2$ for $i>0$, so $d_0>d_i$.
Since $F(\psi_y)=0$, the hypothesis gives $F(z)=0$; every pair weight
is positive, so each $h_i=\zeta_i e_z$ is axial.

The equalities for $S_1$ and $R$, with $d_0\ne d_i$, determine
$q_0=q_0^*$ and $\sum_{i=1}^6q_i=6q_o^*$.
The equality for $Q$ then gives
$\sum_{i=1}^6q_i^2=6(q_o^*)^2$.
Equality in Cauchy--Schwarz forces $q_i=q_o^*$ for every $i>0$.
Thus all $|\zeta_i|$ equal their nonzero family values.
Moreover
\[
 L(z)=\sum_{i<j}a_{ij}\operatorname{Re}(\bar\zeta_i\zeta_j)
 \le\sum_{i<j}a_{ij}|\zeta_i||\zeta_j|=L(\psi_y),
\]
where all $a_{ij}$ are positive.  The hypothesis on $L$ makes this an
equality, so every pair of phases agrees.  A global phase of $z$
can make $h_{0z}=h_s^*$ and $h_{iz}=h_o^*$ exactly.

The one-site Bloch vectors vanish; the Gram $G$ therefore has
identity diagonal $3\times3$ blocks and off-diagonal blocks $T_{ij}$.
Let $c$ have axial entries $c_{0z}=-1/2$, $c_{iz}=-1/4$ for $i>0$,
and zero transverse entries.  Since $T_{0j}=ve_ze_z^{\mathsf T}$ and $T_{ij}=we_ze_z^{\mathsf T}$, only axial entries of $Gc$ survive, with
$(Gc)_{0z}=-\tfrac12-\tfrac32v=h_s^*$ and $(Gc)_{iz}=-\tfrac14-\tfrac v2-\tfrac54w=h_o^*$; so $Gc=h$, and
$c^\dagger h=-\tfrac12h_s^*-\tfrac32h_o^*=1$.  Consequently
\[
 \|z-Vc\|^2=1-2\operatorname{Re}(c^\dagger h)
                   +c^\dagger Gc=0.
\]
Thus $z=(-Z_0/2-\sum_{i=1}^6Z_i/4)J_7z$.
Proposition~\ref{prop:kramers-conference-shell} therefore puts $z$ in the
complex pair/complement shell $z_{T,b}$, where $T$ is the shell's symmetric amplitude matrix (not the
correlation blocks $T_{ij}$ above), with entries $t_{ij}$. On that shell
$\langle Z_0Z_j\rangle=1-4b^2-4p_j$ and $\langle Z_iZ_j\rangle=1-4(p_i+p_j)+8|t_{ij}|^2$, with $\sigma$ and $p_j$ as
in~\eqref{eq:complexsos}. Equating these with $v$ and $w$, and using $\sum_jp_j=2\sigma=1-2b^2$, gives
$b^2=y$, $p_j=(1-2y)/6=\sigma/3$ and $|t_{ij}|^2=(1-2y)/30=\sigma/15$, so the second and third squares
of~\eqref{eq:complexsos} vanish and its first is $\pi_3(\psi_y)-121/917$. Since $z$ has the purity of $\psi_y$,
\eqref{eq:complexsos} gives $\sum_{i<j}|(TT^\dagger)_{ij}|^2=0$: $T$ is a scaled symmetric conference matrix.
Proposition~\ref{prop:confphase} classifies it by diagonal congruence and
ordinary-site permutations. Its displayed local unitaries preserve the
endpoint amplitudes for every $b$, so $z$ is equivalent to $\psi_y$.
At $y=37/524$ this recovers the purity orbit and the pair values
$19/131$ and $11/131$ checked independently by MO-1--MO-2.
\end{proof}

\begin{corollary}[Fourth-moment orbit rigidity]\label{cor:fourth-moment-orbit-rigidity}
Let $X_r^+(z)$ denote the normalised $r$-copy moment of a pure seven-qubit
state after product-local-unitary and site-permutation twirling and the
complex-conjugation average. Fix $y\in(1/32,1/12)$. If
$X_r^+(z)=X_r^+(\psi_y)$ for some $r\ge4$, then $z$ is local-unitarily and
site-permutation equivalent to $\psi_y$. In particular this applies to
both the purity and entropy optimisers.
\end{corollary}
\begin{proof}
A twirl-invariant polynomial of bidegree $(k,k)$ with $k\le r$ is a
linear functional of $X_r^+$, after multiplying by
$\operatorname{tr}(zz^*)^{r-k}=1$.  The thirteen invariants of
Theorem~\ref{thm:moment-orbit-rigidity} are of this kind with $k\le4$;
the $r$-copy moment determines the four-copy moment by partial trace.
\end{proof}

Lemma~\ref{lem:fivecopy-tangent-line} and Corollary~\ref{cor:fourth-moment-orbit-rigidity}
combine into the equality step shared by the purity and entropy theorems.

\begin{theorem}[Equality from a sharp five-copy identity on the family; proved]
\label{thm:fivecopy-family-equality}
Fix $y\in(1/32,1/12)$, and put $X_y=X_5^+(\psi_y)$ and
$X_y'=\frac{d}{dx}X_5^+(\psi_x)|_{x=y}$. Let $\Gamma_b$ be linear maps to
symmetric matrices, and $W_b$ matrices such that
the functionals $X\mapsto\operatorname{tr}[W_bSW_b^{\mathsf T}\Gamma_b(X)]$,
  with $b$ varying and $S$ ranging over symmetric matrices, span the critical space of
  linear functionals $f$ on the span of the twirled five-copy moments with
  $f(X_y)=f(X_y')=0$.
If a normalised pure seven-qubit state $z$ has
$W_b^{\mathsf T}\Gamma_b(X_5^+(z))W_b=0$ for every $b$, then
$X_5^+(z)=X_y$, and $z$ is local-unitarily and site-permutation
equivalent to $\psi_y$.
\end{theorem}
\begin{proof}
The family values above give $H(\psi_x)=155+96x$ for the invariant
$H=104S_1-25\mathcal E_2$, so $X_y'\ne0$.
Lemma~\ref{lem:fivecopy-tangent-line} gives $X_5^+(z)=X_y+tX_y'$.
Writing $h=H(\psi_y)$, both $H$ and $(H-h)^2$ are linear functionals
of the normalised five-copy moment, since their bidegrees are at most $(4,4)$.
The latter is critical, hence vanishes at $z$. But
$H(z)-h=96t$, so $t=0$. Thus $X_5^+(z)=X_y$, and
Corollary~\ref{cor:fourth-moment-orbit-rigidity} with $r=5$ gives the orbit.
\end{proof}

\begin{corollary}[seven-qubit purity minimum and rigidity; verified-arithmetic certificate, FP5-1]
\label{cor:seven-global-from-fivecopy}
Every normalised pure seven-qubit state satisfies $\pi_3\ge121/917$.
Equality holds precisely on the icosahedral local-unitary/permutation
orbit, and there the whole normalised five-copy moment equals that of
$\Psi_{\rm ico}$. Quantitatively, with $c>0$ the least eigenvalue of the
$C_b$,
\[
 \pi_3(z)-\frac{121}{917}\ \ge\ c\sum_b\operatorname{tr}
 \bigl[W_b^{\mathsf T}\Gamma_b(X_5^+(z))W_b\bigr].
\]
\end{corollary}
\begin{proof}
For a pure state the flipped five-copy moments are averages of rank-one
projectors, so every $\Gamma_b(X_5^+(z))\succeq0$. The identity of
Theorem~\ref{thm:fivecopy-sharp-certificate} gives the bound and the
displayed inequality. At equality every
$W_b^{\mathsf T}\Gamma_b(X_5^+(z))W_b$ vanishes, and
Theorem~\ref{thm:fivecopy-family-equality} with $y=37/524$ gives the
moment and orbit, using the spanning assertion of
Theorem~\ref{thm:fivecopy-sharp-certificate}. Conversely,
Theorem~\ref{thm:ico} gives equality on that orbit.
\end{proof}

\begin{corollary}[Labelled LU orbits in the family; proved, EO-1]
\label{cor:icosahedral-labelled-orbits}
For each fixed $y\in(1/32,1/12)$, the states obtained by permuting the
qubits of $\psi_y$ form exactly $42$ distinct orbits under local unitaries
and global phase. They form one orbit when qubit permutations are also
allowed. Representatives are indexed by a choice of distinguished qubit
and a permutation of the last three of the six ordered ordinary labels,
leaving the first three ordinary labels fixed.
\end{corollary}
\begin{proof}
The incident pair weights are $d_s=6v^2$ at the distinguished site and
$d_o=v^2+5w^2$ elsewhere, with $d_s>d_o$. These labelled quantities are
LU invariants, so different choices of distinguished site give seven
different LU classes of configurations.

Fix that site. Every canonical permuted state has the same nonzero
positive axial pair profile. A local unitary mapping two such states
must send each local $Z$ axis to itself or its negative. Positivity of
every pair coefficient forces the seven signs to agree. Thus all local
unitaries are diagonal, or all are anti-diagonal. A simultaneous bit flip
sends the real conference matrix $C$ to $-C$, which is already realised
by diagonal congruence $D=iI$ and the local unitaries of
Proposition~\ref{prop:confphase}.

For diagonal unitaries, equality of the two nonzero endpoint amplitudes
shows that the pair matrices are related by $C'=DCD$ with $D$ diagonal
unitary. Conversely every such congruence is implemented by the local
unitaries displayed in that proposition, preserving the endpoints.
When $C,C'$ have real sign entries, $d_id_j\in\{1,-1\}$ for every
$i\ne j$. Using three distinct indices gives
$D=\eta\operatorname{diag}(\epsilon_i)$, where
$\eta\in\{1,i\}$ and $\epsilon_i\in\{1,-1\}$, up to an irrelevant
common sign. The exact LU criterion for ordinary permutations is therefore
\[
 C'=\tau\operatorname{diag}(\epsilon_i)
                C\operatorname{diag}(\epsilon_i),\qquad \tau\in\{1,-1\}.
\]
Normalise the first ordinary row to all positive signs. Orthogonality to
that row says that every row of the five-vertex core has two negative
entries. Its negative-edge graph is therefore a pentagon. There are
$(5-1)!/2=12$ labelled pentagons, all attained by ordinary permutations
of $C$. In the normalised frame the remaining equivalence above is
identity or complementation: after choosing $\epsilon_0=1$, the first
row forces $\epsilon_i=\tau$ for every $i>0$. Complementation pairs
the twelve pentagons without fixed points, leaving six classes.

The six permutations fixing ordinary labels $0,1,2$ and permuting
$3,4,5$ represent all six classes. In the base pentagon
$1\!-\!3\!-\!2\!-\!4\!-\!5$ the vertices $1,2$ are not adjacent, and a
permutation fixing them keeps them non-adjacent, whereas they are adjacent
in every complement; and two such images coincide only if the permutations
differ by a dihedral symmetry fixing two vertices, that is, the identity.
EO-1 independently checks all $720$ ordinary permutations: the switching
and switching-complement automorphism counts are $60$ and $60$, and
the six classes each contain $120$ permutations. It also verifies the
explicit diagonal local gauges on both amplitude-pattern basis vectors. Combining these with the seven
distinguished-site choices gives $7\cdot6=42$ LU orbits.
\end{proof}

\subsection{Proof of Theorem~\ref{thm:fqe-seven-purity}}\label{sec:sup-seven-purity}
\begin{restated}[the seven-qubit purity minimum; verified-arithmetic certificate FP5-1, exact FO-1]{Theorem}{thm:fqe-seven-purity}
Every pure seven-qubit state satisfies $\pi_3(\psi)\ge121/917$. Equality holds if and only if $\psi$ is
equivalent to $\Psiico$ under local unitaries and permutations of the qubits.
\end{restated}
\begin{proof}
Corollary~\ref{cor:seven-global-from-fivecopy} combines the positive
twelve-face identity of Theorem~\ref{thm:fivecopy-sharp-certificate}
with the equality-moment and orbit reconstruction of
Theorem~\ref{thm:fivecopy-family-equality}.
Theorem~\ref{thm:ico} computes the attaining value at $\Psiico$.
\end{proof}
\subsection{Nondegeneracy of the purity minimiser}\label{sec:seven-purity-nondegenerate}
We now quantify the purity increase near the minimising orbit of Corollary~\ref{cor:seven-global-from-fivecopy}.
The Lagrange Hessian is positive transverse to the local-unitary orbit, giving an explicit quadratic basin.

\begin{theorem}[exact full-space Hessian spectrum; LH-1--LH-4]
\label{thm:icolocal}
The state $\Psiico$ is a strict local minimiser of $\pi_3$ on
$\mathbb CP^{127}$ modulo the action of local unitaries. The real
$256\times256$ Lagrange-Hessian of the purity functional at $\Psiico$ is
positive semidefinite, with kernel exactly the $22$-dimensional span
of $i\Psiico$ and the $21$ one-site Pauli rotations
$iX_j\Psiico,iY_j\Psiico,iZ_j\Psiico$, $j\in\{s,0,\ldots,5\}$ (qubits labelled as in~\eqref{eq:ico}).
For the constrained quadratic form $Q$ described below and every
nonzero real vector $v$ orthogonal to those rotations,
\[
             Q(v)\ge\frac{350-6\sqrt{2005}}{4585}\|v\|^2>\frac{78}{4585}\|v\|^2,
\]
with equality in the first bound on the $18$-dimensional eigenspace inside the special-qubit spin-flip sector described below.
Global minimality is Corollary~\ref{cor:seven-global-from-fivecopy}; this theorem adds nondegeneracy.
\end{theorem}
\begin{proof}
For each $|S|=3$, flatten $z$ into $M_S$ and set
$R_S=M_SM_S^\dagger$. Direct differentiation of the homogeneous
quartic objective gives the stationary equation
\[
   \frac1{35}\sum_{|S|=3}(R_S\otimes I_{S^c})z=\pi_3(z)z,
\]
where the operator notation includes the canonical reshuffling of tensor
factors. The family quadratic in Theorem~\ref{thm:ico} is stationary at
$x=37/524$, so Lemma~\ref{lem:seven-family-symmetric-criticality} (Section~\ref{sec:entropy-local-seven}, which uses only the symmetry of the family) proves
full-space stationarity. LH-1 independently checks the displayed equation
for $z=\Psiico=\sqrt{15/524}\,U+\sqrt{37/524}\,V$ (where $U$ and $V$ are the parts of~\eqref{eq:ico} multiplying $\alpha$ and $\beta$), with the signed
integer support tensors from~\eqref{eq:ico}.

Write $v$ for a variation and $N_S$ for its flattening. The Lagrange
second-variation form (half the usual real Hessian) is
\[
 Q(v)=\frac1{35}\sum_{|S|=3}
 \left(\tr\bigl[(N_SM_S^\dagger+M_SN_S^\dagger)^2\bigr]+
 2\tr(R_SN_SN_S^\dagger)\right)-2\pi_3(\Psiico)\|v\|^2.
\]
LH-2 produces a real symmetric integer-matrix representation
$H=(A+\sqrt{555}\,B)/18340$ of this form on all $256$ real amplitude
directions, and LH-3 verifies that the $22$ displayed symmetry tangent
vectors are annihilated in $\mathbb Q(\sqrt{555})$.

To analyse $H$ and the entropy Hessian along the family~\eqref{eq:ico}, we use a common symmetry decomposition defined over $\mathbb Z$.
First observe that the local diagonal unitary
$T=-Z_s\otimes\operatorname{diag}(1,i)^{\otimes6}$ of Lemma~\ref{lem:seven-family-symmetric-criticality}
acts on $|s,b_0\cdots b_5\rangle$ with eigenvalue $-(-1)^s i^w$, where $w=\sum_{j=0}^5 b_j$.
Hence $T$ has eigenvalue $+1$ on the $32$-dimensional support subspace
$\mathcal H_{\mathrm{supp}}$ ($(s,w)\in\{(0,2),(1,4),(0,6),(1,0)\}$), eigenvalue $-1$ on the
$32$-dimensional special-qubit flip subspace $\mathcal H_{\mathrm{flip}}=X_s\mathcal H_{\mathrm{supp}}$
($(s,w)\in\{(1,2),(0,4),(1,6),(0,0)\}$), and eigenvalues $\pm i$ on the $64$-dimensional odd-weight
subspace $\mathcal H_{\mathrm{odd}}$ ($w\in\{1,3,5\}$). Both objectives are invariant under local unitaries and complex conjugation. Since $T\psi_x=\psi_x$ and
$\psi_x$ is real, their Hessians commute with these two symmetries, decoupling into the six computational-basis blocks
\begin{equation}\label{eq:ico-hessian-six-blocks}
 \mathbb R\mathcal H_{\mathrm{supp}}\oplus i\mathbb R\mathcal H_{\mathrm{supp}}\oplus
 \mathbb R\mathcal H_{\mathrm{flip}}\oplus i\mathbb R\mathcal H_{\mathrm{flip}}\oplus
 \mathbb R\mathcal H_{\mathrm{odd}}\oplus i\mathbb R\mathcal H_{\mathrm{odd}}
\end{equation}
of dimensions $32,32,32,32,64,64$. $T$ maps $\mathbb R\mathcal H_{\mathrm{odd}}$ isometrically onto
$i\mathbb R\mathcal H_{\mathrm{odd}}$, and on $\mathcal H_{\mathrm{flip}}$ a signed bit-complement $DX^{\otimes7}$
($D=+1$ on $(0,4),(1,6)$, $-1$ on $(1,2),(0,0)$) conjugates the real block into the imaginary one.

Two signed incidence structures of the conference matrix finish the reduction (LH-4 checks every identity below).
On the $15$ ordinary pairs, the disjoint-pair matching matrix $(L_{\mathrm{disj}})_{p,q}=s_ps_qs_{(p\cup q)^c}$
($p\cap q=\emptyset$) satisfies $(L_{\mathrm{disj}}-2)(L_{\mathrm{disj}}-3)(L_{\mathrm{disj}}+1)(L_{\mathrm{disj}}+3)=0$,
with integer eigenprojectors $P_2,P_3,P_{-1},P_{-3}$ of ranks $1,5,5,4$. Pairing each pair string with its
bit-complement and adding the two endpoints reduces $\mathcal H_{\mathrm{supp}}$ and $\mathcal H_{\mathrm{flip}}$ to a
$4\times4$ block on $P_2$ and the endpoints and $2\times2$ blocks on $P_3,P_{-1},P_{-3}$ (superscripts $\pm$ below
denote the bit-complement sign). On the $20$ ordinary triples, the vertex-incidence matrices
$(E_{\mathrm{in}})_{T,i}=s_{kl}$ ($T=\{i,k,l\}$) and $(E_{\mathrm{out}})_{T,i}=(E_{\mathrm{in}})_{T^c,i}$ satisfy
$E_{\mathrm{in}}^{\mathsf T}E_{\mathrm{in}}=E_{\mathrm{out}}^{\mathsf T}E_{\mathrm{out}}=10I_6$ and
$E_{\mathrm{in}}^{\mathsf T}E_{\mathrm{out}}=2C$ by $C^2=5I_6$. The orthogonal columns of $F_1=E_{\mathrm{in}}$ and
$F_2=E_{\mathrm{out}}C-E_{\mathrm{in}}$ and their rank-$8$ complement $P_{8,40}$ reduce $\mathcal H_{\mathrm{odd}}$
to scalars on $P_{8,40}^\pm$ and two $4\times4$ blocks tensored with $I_6$.

In every block of $18340Q=A+\sqrt{555}\,B$, $A$ preserves and $B$ exchanges the $U$- and $V$-type subspaces. Changing the sign on one of
them is orthogonal and sends $\sqrt{555}\mapsto-\sqrt{555}$, so each characteristic polynomial is Galois-invariant
and lies in $\mathbb Z[t]$ (LH-4):
\begin{itemize}
\item on $\mathbb R\mathcal H_{\mathrm{supp}}$ (dimension $32$), $18340Q$ is strictly positive definite with eigenvalues
$1440^4$ (on $P_{-3}^+$), $1648^9$ (on $P_3^-\oplus P_{-3}^-$), $3600^5$ (on $P_3^+$), $4080^5$ (on $P_{-1}^+$),
$5488^5$ (on $P_{-1}^-$), $4440^1$ and $9680^1$ (on $\operatorname{span}\{U,V\}$),
and $6864\pm16\sqrt{10726}$ (on the complementary $P_2$/endpoint plane);
\item on $i\mathbb R\mathcal H_{\mathrm{supp}}$ (dimension $32$), the eigenvalues are $0^8$ (spanned by $i\Psiico$ and
$iZ_s\Psiico,iZ_0\Psiico,\ldots,iZ_5\Psiico$: three in the $P_2$/endpoint block and five on $P_{-1}^-$), $688^5$ (on $P_{-1}^+$),
$1048^1$ (in the $P_2$/endpoint block), $1440^5$ (on $P_3^-$), $1648^9$ (on $P_3^+\oplus P_{-3}^+$), and
$3600^4$ (on $P_{-3}^-$);
\item on $\mathbb R\mathcal H_{\mathrm{flip}}\cong i\mathbb R\mathcal H_{\mathrm{flip}}$ ($32$ dimensions each),
they are $1400\pm24\sqrt{2005}$ on $P_3\oplus P_{-3}$ (multiplicity $9$ per sector, $18$ total),
$1400\pm24\sqrt{1195}$ on $P_{-1}$ (multiplicity $5$ per sector, $10$ total), $0^1$ ($iY_s\Psiico$ and
$iX_s\Psiico$), and the three roots of $t^3-13600t^2+55582720t-63383040000=0$ (all $>1900$) on the $P_2$/endpoint
$4\times4$ block;
\item on $\mathbb R\mathcal H_{\mathrm{odd}}\cong i\mathbb R\mathcal H_{\mathrm{odd}}$ ($64$ dimensions each),
they are the scalars $2248^8$ and $840^8$ on $P_{8,40}^\pm$ (multiplicity $16$ each across real and imaginary
parts), $0^6$ ($iY_0\Psiico,\ldots,iY_5\Psiico$ and $iX_0\Psiico,\ldots,iX_5\Psiico$), $2248^6$, and
$2484\pm4\sqrt{286081}$ (multiplicity $6$ each per sector) on the first $4\times4$ block, and the
four roots of $t^4-10400t^3+32376000t^2-37446912000t+14433914880000=0$ (multiplicity $6$ per sector, $12$ total,
all $>900$) on the second $4\times4$ block.
\end{itemize}
The kernel therefore has dimension $0+8+2+12=22$, spanned by $i\Psiico$ and the $21$ one-site Pauli rotations. Since
$44^2=1936<2005<2025=45^2$, the smallest positive eigenvalue of $18340Q$ is
$1400-24\sqrt{2005}>1400-24\cdot45=320>312=4\cdot78$, giving the exact transverse minimum
$(1400-24\sqrt{2005})/18340=(350-6\sqrt{2005})/4585>312/18340=78/4585$.
The usual nondegenerate transverse-Hessian argument gives strict local
minimality modulo the compact local-unitary group.
\end{proof}

\begin{lemma}[quantitative orbit basin for quartic swap observables]
\label{lem:quarticbasin}
Let $F(z)=\langle z^{\otimes2},Gz^{\otimes2}\rangle$ for a
Hermitian $G$ with $\|G\|_{\mathrm{op}}\le1$. Suppose $F$ is invariant
under a compact group of unitary transformations, and $z_0$ is a
unit stationary point of $F$ on the unit sphere. Let
$\mathcal O$ be the orbit of $z_0$ including global phases.
If the Lagrange second-variation form satisfies
$Q(v)\ge h\|v\|^2$ on the real orthogonal complement of the
orbit tangent space at $z_0$, then every unit vector $z$
satisfies, with $\delta=\min_{\phi\in\mathcal O}\|z-\phi\|$,
\[
             F(z)-F(z_0)\ge(h-4\delta-\delta^2)\delta^2.
\]
\end{lemma}
\begin{proof}
The compact orbit contains a closest point $\phi$ to $z$.
Put $v=z-\phi$. Differentiating the squared-distance
function along the orbit shows that $v$ is real orthogonal to
every orbit tangent at $\phi$. Because the group preserves
$F$, the same transverse Hessian bound holds at $\phi$.
Write $p=F(\phi)$ and expand the quartic polynomial
$F(\phi+v)-2p\|\phi+v\|^2+p$. Its constant and linear
terms vanish by normalisation and stationarity, and its
quadratic term is exactly $Q(v)$. The remaining terms have
degrees three and four in $v$. In the two-copy matrix-element
expression, each multilinear term is bounded by the product
of its input vector norms because $\|G\|_{\rm op}\le1$.
There are $\binom43=4$ cubic terms and one quartic term, so
their total absolute value is at most
$4\|v\|^3+\|v\|^4$.
Finally $\|\phi+v\|=\|z\|=1$, so the polynomial on the
left equals $F(z)-p$. Combining the bounds proves the claim.
\end{proof}

\begin{theorem}[exact quantitative basin; QB-1--QB-2]
\label{thm:icoquantbasin}
Let $\mathcal O_{\mathrm{ico}}$ be the full local-unitary orbit
of $\Psiico$ including global phase, and for any normalised pure
seven-qubit state $z$ put
\[
 \delta(z)=\min_{\phi\in\mathcal O_{\mathrm{ico}}}\|z-\phi\|_2.
\]
If $\delta(z)\le1/250$, then
\[
 \boxed{\quad \pi_3(z)\ge\frac{121}{917}
             +\frac{\delta(z)^2}{1024}.\quad}
\]
Here $\delta(z)^2=2-2\max_{U\in\mathrm{LU}}|\langle z|U\Psiico\rangle|$.
\end{theorem}
\begin{proof}
Take $G=\binom73^{-1}\sum_{|S|=3}F_S$ where $F_S$
is the unitary two-copy subsystem swap. Thus
$\|G\|_{\mathrm{op}}\le1$, and its quartic expectation
is the unnormalised mean three-body purity.
The exact full-space Hessian certificate
Theorem~\ref{thm:icolocal} gives $h=78/4585$ in
Lemma~\ref{lem:quarticbasin}. QB-1 checks the rational
strict inequality
\[
 \frac{78}{4585}-\frac4{250}-\frac1{250^2}
  -\frac1{1024}=\frac{285123}{14672000000}>0.
\]
Since $4\delta+\delta^2$ increases with $\delta\ge0$,
the displayed quadratic basin follows. For unit vectors,
minimising the distance over global phase gives the overlap form of $\delta$ (QB-2).
\end{proof}

\section{The sharp mean entropy at seven qubits}\label{sec:sup-seven-entropy}
The family entropy maximiser $\psi_{x_*}$ has, over all its triple
reductions, four distinct positive marginal eigenvalues. A one-variable polynomial majorant of the entropy
touching all four would require degree at least seven
(Proposition~\ref{prop:seven-entropy-majorant-degree}).
The proof instead pairs each marginal with its spin flip to obtain
a degree-five joint majorant. An entropy--purity slope and an eigenvalue
cap first restrict the states that could improve on $\psi_{x_*}$;
a localised five-copy identity then proves the bound on that region.
The equality-moment reconstruction used for purity applies again at
$x_*$, but selects a different orbit. The final subsections establish
nondegeneracy, exclude mixed-state maximisers, and show that every local
entropy maximum has full-rank marginals.
\subsection{The family maximiser}
\label{sec:entropy-local-seven}

Write \(E_3(z)=35^{-1}\sum_{|S|=3}S(\rho_S)\) for a unit
seven-qubit vector, and retain the two real amplitude patterns
\[
 U=\sum_{i<j}s_{ij}\bigl(|0,e_i+e_j\rangle
       -|1,\mathbf1-e_i-e_j\rangle\bigr),\qquad
 V=|0,\mathbf1\rangle+|1,\mathbf0\rangle,
\]
so \(\psi_x=\alpha U+\beta V\), \(\alpha=\sqrt{(1-2x)/30}\), \(\beta=\sqrt x\),
\(\langle U,V\rangle=0\), \(\|U\|^2=30\), and \(\|V\|^2=2\).

\begin{lemma}[Symmetric criticality of the conference family; proved, ES-1]
\label{lem:seven-family-symmetric-criticality}
Let $\mathcal F$ be a real function on the unit sphere, invariant under
local unitaries, qubit permutations and complex conjugation, and
differentiable at $\psi_x$ for some $0<x<1/2$. If $x$ is a stationary point
of $x\mapsto\mathcal F(\psi_x)$, then $\psi_x$ is stationary on the whole
unit sphere. This applies to $\pi_3$ (a polynomial) and to $E_3$, whose
marginals have full rank for $0<x<1/2$ by Theorem~\ref{thm:ico}.
\end{lemma}
\begin{proof}
We exhibit a common symmetry group of $U,V$ whose real fixed space is
exactly $\operatorname{span}_{\mathbb R}\{U,V\}$. Ordinary labels are $0,\ldots,5$,
with the distinguished qubit denoted by $s$.
The diagonal unitary
\[
 T=-Z_s\otimes\operatorname{diag}(1,i)^{\otimes6}
\]
fixes $U,V$. Its eigenvalue on $|s,b\rangle$ is $-(-1)^s i^{|b|}$,
so its fixed space consists of the pair and complement sectors
$(s,|b|)=(0,2),(1,4)$ and the two endpoints $(0,6),(1,0)$.

Put $g=(1\,3\,2\,4\,5)$, $p=(0\,1)(2\,4)$ and
$\epsilon=(1,1,1,-1,1,-1)$. The conference signs satisfy
$s_{g(i)g(j)}=s_{ij}$ and
$\epsilon_i\epsilon_j s_{p(i)p(j)}=s_{ij}$.
If $P_\pi$ permutes the ordinary qubits by $\pi$, then $P_g$ and
\[
 W=\left(-Z_s\otimes\bigotimes_{i=0}^5
           \operatorname{diag}(1,-i\epsilon_i)\right)P_p
\]
fix $U,V$ and both endpoints. On either pair sector, dividing the
coefficients by $s_{ij}$ turns their actions into ordinary permutations
of the fifteen pairs. These permutations act transitively: $g$ has
three pair orbits (pairs through $0$, pentagon edges, and pentagon
diagonals), while $p$ sends $\{0,3\}$ to $\{1,3\}$ and
$\{0,2\}$ to $\{1,4\}$, connecting all three.
Thus a real vector fixed by $T,P_g,W$ has one coefficient on each
signed pair sector and one on each endpoint.

Finally the local unitary
\[
 Q=\left(\operatorname{diag}(i,1)_s\otimes
       \operatorname{diag}(1,e^{i\pi/4})^{\otimes6}\right)X^{\otimes7}
\]
fixes $U,V$. On the displayed support it exchanges complementary
pair strings with a minus sign and exchanges the endpoints with a
plus sign. Its fixed space within the preceding four-dimensional
space is therefore exactly $\operatorname{span}_{\mathbb R}\{U,V\}$.

At a real $\psi_x$, conjugation invariance makes the spherical gradient
real. The four displayed symmetries act orthogonally on
$\mathbb C^{128}\cong\mathbb R^{256}$, preserve $\mathcal F$ and fix
$\psi_x$, so they fix that gradient, which therefore lies in this plane. Orthogonality to $\psi_x$ leaves only the
one-dimensional family tangent. Stationarity within the family makes that last component vanish,
proving full-space stationarity.
\end{proof}

\begin{proposition}[Full-space stationarity; proved, ES-1]
\label{prop:entropy-seven-stationarity}
The unique maximiser \(\psi_{x_*}\) of \(E_3\) within the icosahedral
family is a stationary point on the whole projective space
\(\mathbb CP^{127}\). Its parameter is the unique root in \((0,1/2)\) of
\begin{equation}\label{eq:entropy-seven-star}
 2^{32}(1-2x_*)^{15}=3^8(1+8x_*)^8(1+28x_*)^7,
 \qquad x_*=0.069828\ldots.
\end{equation}
In particular \(E_3(\psi_{x_*})=2.96535602947767\ldots\) bits, and
\(1/32<x_*<37/524\), so \(x_*/u_*\in(1,37/15)\) with \(u_*=(1-2x_*)/30\).
\end{proposition}
\begin{proof}
By Lemma~\ref{lem:seven-family-symmetric-criticality}, it suffices to
find the family stationary point. The marginal spectra of
Theorem~\ref{thm:ico} give
\[
 E_3'(x)=\frac4{35\ln2}\ln
  \frac{(4u)^{13}(8u)^2}{(3u+x)^8(u+x)^7},
 \qquad u=(1-2x)/30.
\]
Strict concavity and the endpoint derivative signs in
Corollary~\ref{cor:ico-entropy-separation} give a unique interior
zero, equivalent to~\eqref{eq:entropy-seven-star}. At this zero the symmetry
lemma gives full-projective stationarity. At \(x=1/32\) one has \(u=x\) and
\(E_3'(1/32)=4\ln512/(35\ln2)>0\), so \(1/32<x_*<37/524\) by concavity
and Corollary~\ref{cor:ico-entropy-separation}; since \(x/u\) increases
with \(x\) and equals \(37/15\) at \(x=37/524\), \(x_*/u_*\in(1,37/15)\).
\end{proof}

Throughout the entropy sections, \(\phi=\psi_{x_*}\), \(u_*=(1-2x_*)/30\) and
\(E_*=E_3(\phi)\); a \emph{competitor} is a pure state \(z\) with \(E_3(z)\ge E_*\).

\subsection{The fixed-purity entropy envelope}
\label{sec:seven-entropy-envelope}

For \(1/8\le p\le1\), let
\begin{equation}\label{eq:seven-entropy-envelope}
 t(p)=\frac{1+\sqrt{7(8p-1)}}8,\qquad
 g_8(p)=-t(p)\log_2t(p)
 -(1-t(p))\log_2\frac{1-t(p)}7.
\end{equation}

To bound entropy at fixed purity, we use a quadratic majorant of \(f(x)=-x\ln x\), in the spirit of the
Hermite majorants of the four-qubit sections.

\begin{lemma}[Quadratic majorant of \(-x\ln x\); proved]
\label{lem:seven-entropy-quadratic-majorant}
Let \(0<b<a\), let \(T(x)=f(b)+f'(b)(x-b)\) be the tangent of \(f\) at \(b\), and put
\(c=(f(a)-T(a))/(a-b)^2\) and \(Q(x)=T(x)+c(x-b)^2\). Then \(f(x)\le Q(x)\) for
\(0\le x\le a\), with equality only at \(x\in\{b,a\}\).
\end{lemma}
\begin{proof}
Put \(R=f-T\), so \(R(b)=R'(b)=0\) and \(R'(y)=-\ln(y/b)\).
The function \(u(y)=R(y)+(y-b)^2/(2b)\) has \(u(b)=0\) and
\(u'(y)=y/b-1-\ln(y/b)\ge0\), with equality only at \(y=b\). Hence
\(R(x)<-(x-b)^2/(2b)\) for \(0\le x<b\), and \(R(a)>-(a-b)^2/(2b)\), that is
\(c>-1/(2b)\); together, \(R(x)<c(x-b)^2\) on \([0,b)\). On \((b,\infty)\) the
divided difference \(\varrho(y)=R(y)/(y-b)^2\) has
\(\varrho'(y)=N(y)/(y-b)^3\) with \(N(y)=(y+b)\ln(y/b)-2(y-b)\). Since \(N(b)=0\)
and \(N'(y)=\ln(y/b)+b/y-1>0\) for \(y\ne b\), \(\varrho\) is strictly increasing, so
\(R(x)<\varrho(a)(x-b)^2=c(x-b)^2\) for \(b<x<a\).
\end{proof}

\begin{lemma}[Entropy envelope at fixed purity; all dimensions]
\label{lem:seven-fixed-purity-envelope}
For every $d\ge2$ and every $d$-dimensional density matrix $\rho$,
\[
 S(\rho)\le g_d(\operatorname{tr}\rho^2),\qquad
 t_d(p)=\frac{1+\sqrt{(d-1)(dp-1)}}d,
\]
where $g_d(p)$ is the entropy of
$(t_d(p),((1-t_d(p))/(d-1))^{d-1})$.
Equality holds exactly at that spectrum, including the uniform
and pure endpoints. For $d=8$ this is~\eqref{eq:seven-entropy-envelope}.
\end{lemma}
\begin{proof}
The cases $p=1/d$ and $p=1$ are immediate. Otherwise put
$a=t_d(p)$, $b=(1-a)/(d-1)$, so $a^2+(d-1)b^2=p$.
Cauchy--Schwarz on all eigenvalues except $\lambda_i$ gives
$p\ge\lambda_i^2+(1-\lambda_i)^2/(d-1)$, hence $\lambda_i\le a$.
The preceding dimension-independent quadratic majorant gives
\[
 S(\rho)\ln2\le\sum_iQ(\lambda_i)=g_d(p)\ln2,
\]
since the sum depends only on the trace and purity and can
be evaluated on $(a,b^{d-1})$. Equality requires every
$\lambda_i\in\{a,b\}$; if $j$ eigenvalues equal $a$, the trace
forces $(j-1)(da-1)=0$, hence $j=1$.
\end{proof}

\subsection{Spin-flip spectral pairing and the pure-state obstruction}
\label{sec:seven-spinflip-spectral}

We first bound opposite-pair sums in the marginal spectrum, then combine these bounds across overlapping cuts.
For a normalised three-qubit state \(\rho\), write its
eigenvalues as \(\lambda_1\ge\cdots\ge\lambda_8\).
Put
\[
 e(\rho)=8\operatorname{tr}\rho^2-1,\quad
 b(\rho)=\sum_{1\le i<j\le3}\sum_{a,b=1}^3
    \bigl[\operatorname{tr}(\rho\,\sigma_a^{(i)}
                       \sigma_b^{(j)})\bigr]^2,\quad
 v(\rho)=\sum_{i=1}^4
   \left(\lambda_i+\lambda_{9-i}-\frac14\right)^2.
\]
The convention for \(b\) sums precisely the
weight-two Pauli squares on these three sites;
one- and three-body terms do not enter.

\begin{proposition}[Spin-flip spectral pairing; SP7-1]
\label{prop:seven-spinflip-spectral-pair}
Every three-qubit density matrix satisfies the
\emph{sharp spectral inequality}
\begin{equation}\label{eq:seven-spinflip-pair}
 \boxed{\qquad b(\rho)\ge4v(\rho).\qquad}
\end{equation}
Equality holds, in particular, for every
\[
 \rho_r=(1-r)\frac{I_8}{8}+r\,|g\rangle\langle g|,
 \qquad 0\le r\le1,\quad |g\rangle\in(\mathbb C^2)^{\otimes3}
 \text{ pure},
\]
where \(e(\rho_r)=7r^2\),
\(b(\rho_r)=3r^2\), \(v(\rho_r)=3r^2/4\).
For a pure seven-qubit state, letting \(e_S=e(\rho_S)\)
and \(R=\sum_{|T|=3}e_T\), there is the stronger
\emph{joint spectral compatibility} inequality
\begin{equation}\label{eq:seven-spinflip-cross-cut}
 \boxed{\qquad e_S+16v(\rho_S)\le R
       \qquad\text{for all }|S|=3.\qquad}
\end{equation}
The latter also holds for mixed global states.
\end{proposition}
\begin{proof}
Let \(\widetilde\rho=Y^{\otimes3}\rho^T Y^{\otimes3}\)
be the spin-flipped marginal. This is positive,
has the same eigenvalues as \(\rho\), and flips the
sign of each odd Pauli weight. Pauli orthogonality gives
\[
 \operatorname{tr}\rho^2=\frac{1+e}{8},\qquad
 \operatorname{tr}(\rho\widetilde\rho)
      =\frac{1-e+2b}{8},
\]
hence
\(\operatorname{tr}\rho^2+
  \operatorname{tr}(\rho\widetilde\rho)=(1+b)/4\).
The von Neumann trace rearrangement inequality implies
\[
 \operatorname{tr}(\rho\widetilde\rho)
   \ge\sum_{i=1}^8\lambda_i\lambda_{9-i}.
\]
Consequently
\[
 \frac{1+b}{4}\ge
 \sum_{i=1}^4(\lambda_i+\lambda_{9-i})^2
 =\frac14+v(\rho),
\]
proving~\eqref{eq:seven-spinflip-pair}.

For every pure three-qubit vector \(g\), complementarity
of one- and two-qubit marginal purities gives
\(\sum_{i<j}\operatorname{tr}\rho_{ij}(g)^2
 =\sum_i\operatorname{tr}\rho_i(g)^2\).
Its Pauli expansion shows that the total weight-two
Pauli sum is exactly \(3\), independently of \(g\).
Depolarisation by \(r\) multiplies all nontrivial
Pauli expectations by \(r\), proving \(b=3r^2\)
and \(e=7r^2\). The spectrum is
\(((1+7r)/8,((1-r)/8)^7)\), whose four opposite
pair sums give \(v=3r^2/4\); equality follows.

For a seven-qubit global state, let
\(b_{ij}\) denote the sum of weight-two Pauli
squares on the physical pair \(ij\). Then
\(b(\rho_S)=\sum_{\{i,j\}\subset S}b_{ij}\)
and each such pair occurs in \emph{four other}
three-site subsets. Since all weights in every
\(e_T\) are nonnegative,
\[
 R-e_S\ge4b(\rho_S)\ge16v(\rho_S).
\]
This proves~\eqref{eq:seven-spinflip-cross-cut}.
\end{proof}

\begin{lemma}[Sharp entropy--purity slope; proved, SP7-5]\label{lem:seven-entropy-linear-purity}
Every $d$-dimensional density matrix, $d\ge3$, satisfies
\begin{equation}\label{eq:entropy-purity-general-slope}
 \log_2d-S(\rho)\ge c_d\bigl(d\operatorname{tr}\rho^2-1\bigr),
 \qquad c_d=\frac{(d-1)\log_2(d-1)}{d(d-2)}.
\end{equation}
The constant is sharp: equality holds exactly at the uniform spectrum and at
$\bigl((d-1)/d,(1/[d(d-1)])^{d-1}\bigr)$. In particular, for $d=8$,
\begin{equation}\label{eq:seven-entropy-linear-purity}
 3-S(\rho)\ge\frac{7\log_27}{48}(8\operatorname{tr}\rho^2-1)
 \ge\frac{49}{120}(8\operatorname{tr}\rho^2-1)
 \ge\frac25(8\operatorname{tr}\rho^2-1),
\end{equation}
and the last two comparisons are strict unless $\rho=I_8/8$.
These bounds include rank-deficient matrices.
\end{lemma}
\begin{proof}
Put $m=d-1\ge2$. By Lemma~\ref{lem:seven-fixed-purity-envelope} it suffices to treat the one-spike spectrum
$((1+mr)/d,((1-r)/d)^m)$, $0\le r\le1$. Its normalised purity excess is $d\operatorname{tr}\rho^2-1=mr^2$ and its
entropy loss is $\log_2d-S=F_m(r)/(d\ln2)$, where
\[
 F_m(r)=(1+mr)\ln(1+mr)+m(1-r)\ln(1-r).
\]
The derivative of $F_m(r)/r^2$ has numerator $N=rF_m'-2F_m$, with $N(0)=N'(0)=0$ and
\[
 N''=rF_m'''=\frac{m(m+1)r\bigl(2mr-(m-1)\bigr)}{(1-r)^2(1+mr)^2}.
\]
So $N''$ changes sign once, from negative to positive: $N'$ is negative and then increasing, and $N$ has at most one
interior zero. At $r_*=(m-1)/m$ one has $1+mr_*=m$ and $1-r_*=1/m$, so
\[
 F_m(r_*)=(m-1)\ln m,\qquad F_m'(r_*)=2m\ln m,\qquad N(r_*)=0.
\]
Hence $F_m/r^2$ strictly decreases on $(0,r_*]$ and strictly increases on $[r_*,1]$, with minimum $m^2\ln m/(m-1)$.
Dividing by $dm\ln2$ gives~\eqref{eq:entropy-purity-general-slope}, with equality only at $r\in\{0,r_*\}$.
For $d=8$, $c_8=7\log_27/48$; since $7^5=16807>16384=2^{14}$, $\log_27>14/5$ and
$7\log_27/48>49/120>2/5$. SP7-5 checks these identities and constants.
\end{proof}

\begin{corollary}[Sharpened universal high-entropy spectral
restriction; proved, SC-2, SP7-6]
\label{cor:seven-entropy-eigen-cap-497}
For any (pure or mixed) seven-qubit state \(\omega\),
\[
 E_3(\omega)\ge\frac{59307}{20000}
 \quad\Longrightarrow\quad
 \boxed{\ \lambda_{\max}(\omega_S)
      <\frac{497}{1000}=0.497
      \quad\text{for every }|S|=3.\ }
\]
Every competitor satisfies the hypothesis, since
\(E_*\ge E_3(\psi_{7/100})>59307/20000\) (SC-2).
\end{corollary}
\begin{proof}
Fix a cut \(S\) with eigenvalues in descending
order. Let \(t=\lambda_1\), \(a=\lambda_8\)
and \(b=(1-t-a)/6\), so \(0\le a\le(1-t)/7\).
By concavity of entropy,
\[
 S(\omega_S)\le
 -t\log_2t-a\log_2a-6b\log_2b.
\]
For the four opposite-pair sums of eigenvalues
in \(v(\omega_S)\), one sum is \(t+a\), and
the other three sum to \(1-t-a\).
Cauchy--Schwarz thus yields
\[
 v(\omega_S)\ge\frac43
      \left(t+a-\frac14\right)^2.
\]
The spin-flip compatibility inequality
\eqref{eq:seven-spinflip-cross-cut} and
Lemma~\ref{lem:seven-entropy-linear-purity}
(with the rational slope \(49/120\))
applied to the other 34 marginals yield
\[
\begin{split}
 35(3-E_3(\omega))
 &\ge 3-S(\omega_S)
      +\frac{49}{120}\sum_{T\ne S}e_T\\
 &\ge 3-S(\omega_S)+\frac{98}{15}v(\omega_S)
 \ge J(t,a),
\end{split}
\]
where, continuously at \(a=0\),
\[
 J(t,a)=3+t\log_2t+a\log_2a+6b\log_2b
      +\frac{392}{45}
           \left(t+a-\frac14\right)^2.
\]
Apply the tangent inequality \(p\ln(p/q)\ge p-q\)
(\(p\ge0,q>0\)) to the three probability weights
\((t,6b,a)\) with comparison weights
\(q_1=1/2\), \(6q_2=1/2\) (\(q_2=1/12,\;8q_2=2/3\)), and
\(q_3=1/192\) (\(8q_3=1/24\)), whose sum is
\(q_1+6q_2+q_3=1+1/192\). Since
\(3+t\log_2t+a\log_2a+6b\log_2b
 =\bigl[t\ln(8t)+6b\ln(8b)+a\ln(8a)\bigr]/\ln2\)
and \(\ln(8q_1)=\ln4=2\ln2\),
\(\ln(8q_2)=\ln(2/3)=\ln2-\ln3\),
\(\ln(8q_3)=\ln(1/24)=-3\ln2-\ln3\), this gives
\[
\begin{aligned}
 3+t\log_2t+a\log_2a+6b\log_2b
 &\ge 2t+(1-t-a)(1-\log_23)-a(3+\log_23)
      -\frac1{192\ln2}\\
 &=1+t-(1-t)\log_23-4a-\frac1{192\ln2}.
\end{aligned}
\]
Completing the square in \(a\) gives
\[
 -4a+\frac{392}{45}\left(t+a-\frac14\right)^2
 =4t-1-\frac{45}{98}
  +\frac{392}{45}\left(t+a-\frac{47}{98}\right)^2
 \ge 4t-1-\frac{45}{98}.
\]
The resulting lower bound \(5t-\frac{45}{98}-(1-t)\log_23-\frac1{192\ln2}\) increases in \(t\)
(slope \(5+\log_23>0\)). For \(t\ge t_0=497/1000\), using only
\(3^5=243<256=2^8\) (\(\log_23<8/5\)) and
\(\ln2=2\operatorname{arctanh}(1/3)>2/3\)
(\(1/(192\ln2)<1/128\)), we obtain
\[
 J(t,a)
 >5t_0-\frac{45}{98}-\frac85(1-t_0)-\frac1{128}
 =\frac{4755759}{3920000}
 >\frac{4851}{4000}
 =35\left(3-\frac{59307}{20000}\right),
\]
with exact rational margin \(1779/3920000>0\) (SP7-6),
contradicting \(E_3(\omega)\ge59307/20000\).
Therefore \(t<497/1000\) on all 35 cuts.
\end{proof}

\begin{proposition}[Competitor purity band; AH-1]
\label{prop:seven-competitor-band}
Every pure seven-qubit state with $E_3(z)\ge E_*$ obeys
\begin{equation}\label{eq:seven-competitor-band}
 \frac{121}{917}\le\pi_3(z)<p_c:=\frac{8693}{64000}.
\end{equation}
\end{proposition}
\begin{proof}
The lower bound is Corollary~\ref{cor:seven-global-from-fivecopy}.
Lemma~\ref{lem:seven-entropy-linear-purity} gives
$3-E_3\ge(2/5)(8\pi_3-1)$. Since
$E_*>59307/20000$ (SC-2), rearrangement gives
$\pi_3<(1+(5/2)(3-59307/20000))/8=8693/64000$.
\end{proof}

\begin{lemma}[Quadratic joint spin-flip relations; proved, AH-5]
\label{lem:seven-joint-spinflip}
For any triple marginal $R$ of the real family $\psi_x$, put
$\widetilde R=\epsilon^{\otimes3}R^{\mathsf T}
(\epsilon^{\otimes3})^{\mathsf T}$ (with $\epsilon=i\sigma_y$ this is the spin flip
$Y^{\otimes3}R^{\mathsf T}Y^{\otimes3}$ of Section~\ref{sec:seven-spinflip-spectral}),
$E=(R+\widetilde R)/2$, $O=(R-\widetilde R)/2$, and $u=(1-2x)/30$.
If the triple avoids the distinguished site, then
\[
 O=0,\qquad (E-4uI)(E-(3u+x)I)=0.
\]
If it contains that site, then
\begin{equation}\label{eq:seven-joint-spinflip}
 [E,O]=0,\qquad O(E-6uI)=0,\qquad
 2u(E-4uI)(E-(u+x)I)-(5u-x)O^2=0.
\end{equation}
\end{lemma}
\begin{proof}
In the computational basis $\{|b_1b_2b_3\rangle\}$, $\epsilon^{\otimes3}$ maps $|b_1b_2b_3\rangle$ to $(-1)^{3-(b_1+b_2+b_3)}|\bar b_1\bar b_2\bar b_3\rangle$.
By the block diagonalisation in the proof of Theorem~\ref{thm:ico}, on any cut $I=\{i,j,k\}$ avoiding $s$ the marginal $R=\operatorname{diag}(3u+x,(4u)^6,3u+x)$ is diagonal with bit-complement-symmetric entries, so $\widetilde R=R$, $O=0$, and $(E-4uI)(E-(3u+x)I)=0$.
On any cut $\{s,i,j\}$ containing $s$, $R$ is diagonal with bit-complement-symmetric entries on the six basis states outside $\operatorname{span}\{|000\rangle,|111\rangle\}$ (two eigenvalues $u+x$ and four eigenvalues $4u$, where $O=0$), while on $\operatorname{span}\{|000\rangle,|111\rangle\}$ one has $R=\bigl(\begin{smallmatrix}6u&\mp2u\\\mp2u&6u\end{smallmatrix}\bigr)$ and $\epsilon^{\otimes3}$ maps $|000\rangle\mapsto-|111\rangle$ and $|111\rangle\mapsto|000\rangle$, flipping the off-diagonal sign: $\widetilde R=\bigl(\begin{smallmatrix}6u&\pm2u\\\pm2u&6u\end{smallmatrix}\bigr)$. Thus $E=6u\,I_2$ and $O=\bigl(\begin{smallmatrix}0&\mp2u\\\mp2u&0\end{smallmatrix}\bigr)$ on $\operatorname{span}\{|000\rangle,|111\rangle\}$, where $O^2=4u^2I_2$ and $2u(6u-4u)(6u-u-x)-(5u-x)(4u^2)=0$. This proves~\eqref{eq:seven-joint-spinflip} across all eight basis directions; AH-5 checks all $35$ cuts directly.
\end{proof}

\subsection{The degree of a sharp scalar entropy majorant}
\label{sec:seven-entropy-majorant-degree}

\begin{proposition}[Minimum replica degree for a sharp scalar
entropy majorant]\label{prop:seven-entropy-majorant-degree}
Let \(f(t)=-t\log_2t\) on \([0,1]\), with \(f(0)=0\).
For \(k\ge2\) distinct interior nodes \(0<a_1<\dots<a_k<1\),
any polynomial \(q\) satisfying \(q(t)\ge f(t)\) on \([0,1]\)
and \(q(a_j)=f(a_j)\) for all \(j\) has degree at least
\(2k-1\). This threshold is attained: the degree-\((2k-1)\)
Hermite interpolant matching both \(f\) and \(f'\) at every node
majorises \(f\) on \([0,1]\).

At the seven-qubit entropy maximiser \(\psi_{x_*}\), a
\emph{single} scalar entropy majorant sharp on every triple
reduction therefore requires degree at least \(7\), since the
combined spectra have four distinct positive eigenvalues.
A joint majorant in a marginal and its spin flip avoids this:
Lemma~\ref{lem:seven-joint-scalar-majorant} has degree five.
\end{proposition}
\begin{proof}
If \(h=q-f\ge0\) and \(h(a_j)=0\) at an interior point,
then \(h'(a_j)=0\). Thus \(h\) has at least \(2k\)
zeros counted with multiplicity, so iterated Rolle's theorem
forces \(h^{(2k-1)}\) to vanish somewhere in the interval
spanned by the nodes. If \(\deg q\le2k-2\), however,
\[
 h^{(2k-1)}(t)=-f^{(2k-1)}(t)
 =-\frac{(2k-3)!}{t^{2k-2}\ln2}<0
 \qquad (t>0),
\]
a contradiction. For the converse, let \(H\) denote the
Hermite polynomial matching \(f,f'\) at all \(k\) nodes.
For \(0<t<1\), the Hermite remainder formula gives
\[
 f(t)-H(t)
 =\frac{f^{(2k)}(\xi)}{(2k)!}
   \prod_{j=1}^k(t-a_j)^2
 \le0,
 \qquad
 f^{(2k)}(\xi)
 =-\frac{(2k-2)!}{\xi^{2k-1}\ln2}<0.
\]
Continuity extends the inequality to \(t=0\).
By Theorem~\ref{thm:ico} the cut eigenvalues of \(\psi_x\) are
\(4u,3u+x,u+x,8u\) with \(u=(1-2x)/30\); two of them coincide
only when \(x/u\in\{1,3,5,7\}\). By
Proposition~\ref{prop:entropy-seven-stationarity},
\(x_*/u_*\in(1,37/15)\), so the cuts of \(\psi_{x_*}\) have four
distinct nodes. The asserted degree bound follows.
\end{proof}

\subsection{A joint degree-five entropy majorant}
\label{sec:seven-joint-entropy-majorant}

The joint spin-flip relations let us build a majorant with the same contacts for all seven choices of
distinguished site. Its moment functional is therefore invariant under qubit permutations.
Throughout this subsection $f(t)=-t\log_2t$, with $f(0)=0$, and
$u=u_*$, $c=4u$, $a=3u+x_*$, $d=u+x_*$, $b=8u$.
Let $P$ be the degree-five Hermite interpolant to $X\mapsto f(X/8)$
at $8c,8a,8d$, including the first derivatives. Put
\[
 e_0=48u,\quad v_0=256u^2,\quad
 h=\frac{f(b)+f(c)}2,\quad
 h_e=\frac{f'(b)+f'(c)}{16},\quad
 h_v=\frac{f'(b)-f'(c)}{512u}.
\]
Choose the following exact rational coefficients:
\[
 \alpha_2=-\frac{115001089107}{10^{12}},\qquad
 \alpha_3=\frac{18166193766}{10^{12}},\qquad
 \beta_1=\frac{14438812336}{10^{12}},
\]
and define the remaining coefficients by
\begin{align*}
 \beta_0&=\frac{v_0h_v-h+P(e_0)}{v_0^2}-\beta_1e_0,\\
 \alpha_1&=\frac{h_e-P'(e_0)-v_0^2\beta_1}{v_0}
             -2\alpha_2e_0-3\alpha_3e_0^2,\\
 \alpha_0&=h_v-2v_0(\beta_0+\beta_1e_0)
             -\alpha_1e_0-\alpha_2e_0^2-\alpha_3e_0^3.
\end{align*}
For $e=4(\lambda+\mu)$ and $v=16(\lambda-\mu)^2$, set
\begin{equation}\label{eq:seven-joint-H}
 H(\lambda,\mu)=P(e)+v\sum_{j=0}^3\alpha_je^j
                          +v^2(\beta_0+\beta_1e)
                 =\sum_{k+l\le5}h_{kl}\lambda^k\mu^l.
\end{equation}
It is a symmetric polynomial of total degree five.

\begin{lemma}[Joint scalar majorant; exact-rational coefficient certificate AH-6]
\label{lem:seven-joint-scalar-majorant}
For all $\lambda,\mu\ge0$,
\[
 H(\lambda,\mu)\ge\frac{f(\lambda)+f(\mu)}2.
\]
Equality holds exactly at
$(c,c),(a,a),(d,d),(b,c),(c,b)$, with first-order contact at each pair.
\end{lemma}
\begin{proof}
The diagonal contacts follow from the definition of $P$. Substitution
in the three formulas for $\beta_0,\alpha_1,\alpha_0$ gives the value
and both first derivatives at $(b,c)$; symmetry gives $(c,b)$.

Section~\ref{sec:seven-joint-analytic-proof} proves the inequality by a
positive Stieltjes-kernel bound and a cubic Hermite minorant for the binary
entropy correction. A single univariate discriminant polynomial remains;
its fixed positive-coefficient certificate is checked in exact rational
arithmetic (AH-6). That proof also excludes any further equality points.
\end{proof}

\begin{corollary}[Degree five is minimal]\label{cor:seven-joint-degree-five}
Let $G(\lambda,\mu)$ be a polynomial of total degree $m$ with $G(\lambda,\mu)\ge\tfrac12[f(\lambda)+f(\mu)]$ on
$[0,497/1000]^2$ and $G(t,t)=f(t)$ for $t\in\{c,a,d\}$. Then $m\ge5$. Every joint majorant with equality at the
joint eigenvalue pairs of the marginals of $\psi_{x_*}$ satisfies these hypotheses, so $H$ has the least possible degree.
\end{corollary}
\begin{proof}
The points $c,a,d$ are distinct (Proposition~\ref{prop:seven-entropy-majorant-degree}) and lie in $(0,497/1000)$.
The restriction $g(t)=G(t,t)$ has degree at most $m$, satisfies $g\ge f$ on $[0,497/1000]$ and touches $f$ at three
interior points. The proof of Proposition~\ref{prop:seven-entropy-majorant-degree} uses $g\ge f$ only on an interval
containing the nodes in its interior, so it gives $m\ge\deg g\ge5$. For the second statement, Lemma~\ref{lem:seven-joint-spinflip}
shows that a marginal avoiding the distinguished site equals its spin flip, with eigenvalues $c$ and $a$, and that on the
eigenspace of $E$ for the eigenvalue $d$ of a marginal containing it the odd part $O$ vanishes, because $O(E-6uI)=0$
and $d\ne6u$ (as $x_*/u_*<37/15$); so $(c,c),(a,a),(d,d)$ are joint
eigenvalue pairs. This concerns majorants in a marginal and its spin flip only; it is not a lower bound on the copy
hierarchy.
\end{proof}

\begin{proposition}[The five-copy entropy target]
\label{prop:seven-joint-entropy-target}
For a positive semidefinite $8\times8$ matrix $R$, let $\widetilde R$ denote its three-qubit
spin flip. Then
\begin{equation}\label{eq:seven-joint-matrix-majorant}
 \operatorname{tr}f(R)\le
 \mathcal P(R):=\sum_{k+l\le5}h_{kl}
                         \operatorname{tr}(R^k\widetilde R^{\,l}).
\end{equation}
Commutation of $R$ and $\widetilde R$ is not required.
For normalised seven-qubit states put
\[
 \mathcal P_5(z)=\frac1{35}\sum_{|S|=3}\mathcal P(\rho_S(z)).
\]
Every normalised pure seven-qubit state satisfies $E_3(z)\le\mathcal P_5(z)$,
and $\mathcal P_5(\phi)=E_*$ for every choice of the distinguished qubit of $\phi$.
After homogenisation, $\mathcal P_5$ is a permutation-invariant
linear functional on five-copy moments.
\end{proposition}
\begin{proof}
Take orthonormal eigenbases $u_i,v_j$ of $R,\widetilde R$, with
eigenvalues $\lambda_i,\mu_j$. The matrix
$w_{ij}=|\langle u_i,v_j\rangle|^2$ is doubly stochastic, so
\[
 \mathcal P(R)=\sum_{ij}w_{ij}H(\lambda_i,\mu_j)
 \ge\frac{\operatorname{tr}f(R)+
                \operatorname{tr}f(\widetilde R)}2
 =\operatorname{tr}f(R).
\]
Lemma~\ref{lem:seven-joint-spinflip} shows that every
marginal of $\phi$ commutes with its spin flip and has only the five
joint pairs listed in Lemma~\ref{lem:seven-joint-scalar-majorant}.
This proves equality. Multiplication of the term of degree $k+l$ in unnormalised
marginals by $\nu^{5-k-l}$, where $\nu=\|z\|^2$, gives the homogeneous
five-copy functional.
\end{proof}

Theorem~\ref{thm:seven-entropy-localized-identity} proves the matching
quantum inequality for every competitor, completing the entropy bound.

\subsection{A positive-kernel proof of the joint entropy majorant}
\label{sec:seven-joint-analytic-proof}

We now prove the scalar-majorant inequality of Lemma~\ref{lem:seven-joint-scalar-majorant}
for the polynomial $H$ defined there. The proof separates the diagonal Hermite remainder from
a binary entropy correction, then reduces nonnegativity of the combined gap to the
sign of one univariate polynomial. Only the fixed coefficient bounds in
Table~\ref{tab:seven-joint-discriminant} require arithmetic verification
(AH-6, exact-rational interval enclosures); no subdivision of the entropy gap is used.

Retain $A(e)=\sum_{j=0}^3\alpha_je^j$, $B(e)=\beta_0+\beta_1e$,
$e_0=48u$ and $v_0=e_0^2/9$ from the definition of $H$ before~\eqref{eq:seven-joint-H}, and put
\[
 q(e)=(e-8a)(e-8c)(e-8d),\qquad
 K(e)=\frac{P(e)-f(e/8)}{q(e)^2}.
\]
At the three apparent singularities take the continuous extensions.
The family bound $x_*/u_*<37/15$ places $e_0$ strictly above these three nodes.
The Stieltjes remainder formula for quintic Hermite interpolation gives
\begin{equation}\label{eq:seven-joint-stieltjes}
 K(e)=\frac1{8\ln2}\int_0^\infty
 \frac{s\,ds}{(s+e)(s+8a)^2(s+8c)^2(s+8d)^2}>0.
\end{equation}
Indeed, $-t\ln t=\int_0^\infty[t/(t+s)-t/(1+s)]\,ds$, and only the term
$-s/(t+s)$ is not affine in $t$; the affine difference between $f(e/8)$ and
$f(e)/8$ contributes no interpolation error. The Hermite remainder of
$e\mapsto1/(s+e)$ at the double nodes $8a,8c,8d$ is
$q(e)^2/[(s+e)(s+8a)^2(s+8c)^2(s+8d)^2]$, which gives the identity.

Set
\[
 k_0=K(e_0),\qquad \gamma=-K'(e_0)/k_0,
 \qquad L(e)=1+\gamma(e-e_0).
\]
The kernel has the rational lower bound
\begin{equation}\label{eq:seven-joint-kernel-jensen}
 K(e)\ge\frac{k_0}{L(e)},\qquad e\ge0,
 \qquad k_0>0,\quad0<\gamma<1/e_0.
\end{equation}
To see this, normalise the positive integrand of $K(e_0)$ to a probability
measure. Then $K(e)/k_0$ is the mean of $\phi(y)=[1+(e-e_0)y]^{-1}$ at
$y=1/(s+e_0)$. Since $1+(e-e_0)y=(s+e)/(s+e_0)>0$ for $e\ge0$, $\phi$ is
convex on the support, and Jensen's inequality gives
\eqref{eq:seven-joint-kernel-jensen}, because the mean of $y$ is
$\gamma$. Its support also gives $0<\gamma<1/e_0$, so $L(e)>0$ for
all $e\ge0$. Both value and first derivative agree at $e_0$.

For the binary correction define
\[
 \Phi(z)=\frac{(1+\sqrt z)\ln(1+\sqrt z)
                    +(1-\sqrt z)\ln(1-\sqrt z)}2,
 \qquad 0\le z\le1.
\]
Its cubic Hermite interpolant, with double contacts at $0$ and $1/9$, is
\begin{equation}\label{eq:seven-joint-binary-minorant}
 \Phi(z)\ge\frac z2+\eta_2z^2+\eta_3z^3,
 \quad
 \eta_2=\frac{1593}4\ln2-243\ln3-9,
 \quad
 \eta_3=\frac{81}2-\frac{9477}4\ln2+1458\ln3>0.
\end{equation}
The interpolation data at $1/9$ are
$\Phi(1/9)=\tfrac53\ln2-\ln3$ and
$\Phi'(1/9)=\tfrac34\ln2$. The power series
$\Phi(z)=\sum_{j\ge1}z^j/[2j(2j-1)]$ gives
$\Phi^{(4)}>0$ on $[0,1)$, so the Hermite error has the sign of
$z^2(z-1/9)^2$. Continuity includes $z=1$.
The positive third divided difference also gives $\eta_3>0$.

For $\lambda,\mu\ge0$, put $e=4(\lambda+\mu)$,
$v=16(\lambda-\mu)^2$. If $e>0$ then $0\le v/e^2\le1$ and
\[
 \frac{f(\lambda)+f(\mu)}2
 =f(e/8)-\frac{e}{8\ln2}\Phi(v/e^2).
\]
Consequently, for $e>0$, the entropy gap is bounded below by a cubic in $v$:
\begin{equation}\label{eq:seven-joint-cubic-reduction}
 H(\lambda,\mu)-\frac{f(\lambda)+f(\mu)}2
 \ge\frac{F(e,v)}{8\ln2\,e^5L(e)},
\end{equation}
where
\begin{align*}
 U(e)&=8\ln2\,k_0q(e)^2,\\
 V(e)&=\tfrac12+8\ln2\,eA(e),\\
 W(e)&=\eta_2+8\ln2\,e^3B(e),\\
 F(e,v)&=e^5U(e)+e^4L(e)V(e)v
                 +e^2L(e)W(e)v^2+\eta_3L(e)v^3.
\end{align*}
This follows by substituting the two lower bounds above into
$q(e)^2K(e)+vA(e)+v^2B(e)+e\Phi(v/e^2)/(8\ln2)$.

\paragraph{Reduction to one polynomial sign.}
The discriminant of $F(e,\cdot)$ is
\begin{equation}\label{eq:seven-joint-discriminant}
 \operatorname{disc}_v F(e,v)
    =-e^{10}L(e)^2(e-e_0)^2\Pi(e),
\end{equation}
where $\Pi$ is the degree-eighteen polynomial obtained by dividing
\begin{equation}\label{eq:seven-joint-discriminant-polynomial}
\begin{split}
 R(e)={}&27\eta_3^2U^2+4eLUW^3+4\eta_3e^2L^2V^3\\
       &-e^2L^2V^2W^2-18\eta_3eLUVW
\end{split}
\end{equation}
by $(e-e_0)^2$; $U,V,W,L$ in this expression are evaluated at $e$.
The factor is exact. The contact $(\lambda,\mu)=(b,c)$ of the Lemma maps
to $(e_0,v_0)$, and since $b\ne c$ (so $v_0>0$) the map
$(\lambda,\mu)\mapsto(e,v)$ is a local diffeomorphism there. The entropy gap
has value and gradient zero at this contact, and both minorant defects vanish
to second order there: $q^2(K-k_0/L)$ at $e_0$, and the binary defect at
$v/e^2=1/9$. Hence $F=F_v=F_e=0$ at $(e_0,v_0)$. The cubic $F(e_0,\cdot)$
has a double root at $v_0$, so its discriminant vanishes; and the
differential of the discriminant at a cubic with double root $r$ is a
multiple of $\delta p\mapsto\delta p(r)$, so its $e$-derivative is a
multiple of $F_e(e_0,v_0)=0$. Dividing out $e^{10}L(e)^2$ preserves this
double zero (AH-6b checks the cubic identity symbolically).

The sign needed is $\Pi(e)>0$ for every $e\ge0$. On each finite interval $[l,h]$ of
Table~\ref{tab:seven-joint-discriminant}, expand $\Pi(l+(h-l)t)$ in the degree-eighteen Bernstein
basis; on the final half-line expand $\Pi(5/4+t)$ in powers of $t$. Every coefficient is strictly positive.

\begin{table}[htbp]
\centering
\begin{tabular}{ccc}
\toprule
$e$ range & basis & lower bound on every coefficient\\
\midrule
$[0,3/4]$ & Bernstein & $2\cdot10^{-7}$\\
$[3/4,1]$ & Bernstein & $5\cdot10^{-10}$\\
$[1,5/4]$ & Bernstein & $9\cdot10^{-11}$\\
$[5/4,\infty)$ & powers of $e-5/4$ & $2\cdot10^{-10}$\\
\bottomrule
\end{tabular}
\caption{Exact rational coefficient certificate AH-6c for $\Pi>0$.
Each row checks nineteen coefficients.}
\label{tab:seven-joint-discriminant}
\end{table}

AH-6 computes outward dyadic rational enclosures of the coefficients of $\Pi$ from the exact root
equation for $x_*$ and atanh series with rational tails, converts them to the bases of the table, and
verifies the bounds by integer comparisons. These are coefficient signs of a fixed polynomial, not sampled
values of the entropy inequality.

\paragraph{Completion and equality.}
For $e>0$, $e\ne e_0$, the discriminant is strictly negative. Thus the
cubic $F(e,\cdot)$ has exactly one real root. Its leading coefficient
is positive, and $F(e,0)=e^5U(e)\ge0$, so that root is nonpositive
and $F(e,v)\ge0$ for $v\ge0$. Equality requires $v=0$ and
$e\in\{8a,8c,8d\}$.
At $e=e_0$, the double contact gives
\[
 F(e_0,v)=(v-v_0)^2(\eta_3v+m),\qquad
 m=e_0^5U(e_0)/v_0^2>0,
\]
so equality requires $v=v_0$. Equation~\eqref{eq:seven-joint-cubic-reduction}
therefore proves the majorant, with exactly the five asserted contacts.
For $e=0$, $\lambda=\mu=0$ and the gap is $P(0)=q(0)^2K(0)>0$
by~\eqref{eq:seven-joint-stieltjes}.
This completes the proof on the full nonnegative quadrant.

\subsection{The localised five-copy inequality}
\label{sec:seven-entropy-certificate}

The scalar majorant bounds entropy by a five-copy functional. We now prove the matching bound on that
functional using a quantum compatibility identity.
Write $\nu=\|z\|^2$, $R_S=\operatorname{tr}_{S^c}|z\rangle\langle z|$,
and $\Pi(z)=35^{-1}\sum_{|S|=3}\operatorname{tr}R_S^2$.
Let $\mathcal H(z)$ be the homogeneous lift of
$\mathcal P_5$: its term $h_{kl}\operatorname{tr}(R_S^k\widetilde R_S^l)$
is multiplied by $\nu^{5-k-l}$. Thus
$\mathcal H(z)=\mathcal P_5(z)$ when $\nu=1$.
Put $p_0=121/917$, $p_c=8693/64000$, and $\kappa=497/1000$.
For a vector that need not be normalised, $X_5^+(z)$ denotes the unnormalised twirled moment
$\nu^5X_5^+(z/\|z\|)$, with $X_5^+(0)=0$; it is a real polynomial of degree ten in $z,\bar z$, so both sides
of the identity below are homogeneous of degree ten.

\begin{theorem}[Localised entropy identity; verified-arithmetic certificate E7-1]
\label{thm:seven-entropy-localized-identity}
There are seventeen positive definite real Gram matrices $C_b$, two
nonnegative cubic-moment polynomials $K_+,K_-$, and quadratic marginal
matrix covariants $A_j,B_j$, such that the following polynomial identity
holds for every complex seven-qubit vector $z$:
\begin{align}
 E_*\nu^5-\mathcal H(z)
 &=\sum_{b=1}^{17}\operatorname{tr}[C_bG_b(z)]
   +(p_c\nu^2-\Pi(z))K_+(z)
   +(\Pi(z)-p_0\nu^2)K_-(z)\notag\\
 &\quad+\frac1{35}\sum_{|S|=3}\left[
       \nu\sum_j\operatorname{tr} A_j(R_S)^\dagger A_j(R_S)
       +\sum_j\operatorname{tr}B_j(R_S)^\dagger
                   (\kappa\nu I-R_S)B_j(R_S)\right].
 \label{eq:seven-entropy-localized-identity}
\end{align}
Here $G_b(z)=W_b^{\mathsf T}\Gamma_b(X_5^+(z))W_b\succeq0$ are Gram
matrices of five-copy features: $\Gamma_b$ is a block of the twirled
five-copy moment with zero, one or two spin-flipped copies, as in
Section~\ref{sec:fivecopy}, and the columns of $W_b$ lie in its kernel at
$\phi=\psi_{x_*}$. Every term on the right vanishes at $\phi$, and the
functionals $X\mapsto\operatorname{tr}[W_bSW_b^{\mathsf T}\Gamma_b(X)]$
($b\le12$, $S$ symmetric) span the $3283$-dimensional space of linear
functionals vanishing on $X_5^+(\phi)$ and on its family derivative. In
particular,
\begin{equation}\label{eq:seven-entropy-compatible-bound}
 p_0\le\pi_3(z)\le p_c,\qquad R_S\preceq\kappa I\quad(|S|=3),
 \qquad \nu=1
 \quad\Longrightarrow\quad \mathcal P_5(z)\le E_*.
\end{equation}
\end{theorem}

\paragraph{The positive sources.}
The construction uses the one-site multiplicity modules
$F=\mathbb R^4$ and $V=\mathbb R^5$ of the five-copy qubit Schur algebra
from Section~\ref{sec:fivecopy-kernels}. In the following table
$[a,b]V$ denotes the two-row Young module inside
$\operatorname{Sym}^aV\otimes\operatorname{Sym}^bV$.
The middle column is the number of spin-flipped copies; the last column
is the number of kernel features.
\begin{center}\small
\begin{tabular}{lrr@{\qquad}lrr}
\toprule
module & flips & features & module & flips & features\\
\midrule
$\operatorname{Sym}^7V$ &2&31&
$F\otimes\operatorname{Sym}^6V$ &2&69\\
$F\otimes[5,1]V$ &2&125&$F\otimes[4,2]V$ &2&128\\
$[6,1]V$ &2&57&$[5,2]V$ &2&72\\
$\operatorname{Sym}^7V$ &0&2&$[6,1]V$ &0&3\\
$[5,2]V$ &0&6&$\operatorname{Sym}^7V$ &1&14\\
$[6,1]V$ &1&27&$[5,2]V$ &1&36\\
$\operatorname{Sym}^2F\otimes\operatorname{Sym}^5V$ &2&94&
$\bigwedge^2F\otimes\operatorname{Sym}^5V$ &2&46\\
$[4,3]V$ &0&2&$[4,3]V$ &1&20\\
$[4,3]V$ &2&44&&&\\
\bottomrule
\end{tabular}
\end{center}
Each feature is projected onto the exact physical kernel at $\phi$;
$G_b$ includes the positive multiplicity weight and the average over
local unitaries and site permutations. Positivity holds for every
complex pure state. The last five modules supply directions absent
from the twelve-face purity certificate.

For three copies the corresponding one-site multiplicity module has
dimension two. The modules $\operatorname{Sym}^7\mathbb R^2$,
$[6,1]\mathbb R^2$, and the one-spin-$3/2$
$\operatorname{Sym}^6\mathbb R^2$ module have respectively two, one and
one exact kernel features at $\phi$. Positive sums of their squared
norms give $K_+$ and $K_-$. In the invariant bases $U_b$ of E7-K3, the kernel
features in $[6,1]\mathbb R^2$ and one-spin-$3/2$ $\operatorname{Sym}^6\mathbb R^2$
are constant rational vectors annihilating the entire family $\psi_x$,
while the two kernel features in $\operatorname{Sym}^7\mathbb R^2$ depend on $x$
and are taken at $x=x_*$. E7-K3 checks their physical coordinate forms directly.
Section~\ref{sec:sextic-general} gives a structural interpretation of these features:
the spin-$3/2$ feature is the one-site focused covariant $\Phi_{i,0}$ of
Lemma~\ref{lem:focus-pauli}, and the $\operatorname{Sym}^7\mathbb R^2$ pair
are the one-body and purity-gradient Wronskian
relations~\eqref{eq:zero-focus-family} of Proposition~\ref{prop:zero-focus}.
That proposition also proves their completeness analytically: the entire
symmetric scalar cubic Gram has rank two on $0<x<1/2$, with the
positive two-source determinant $72x(1-2x)$ in
\eqref{eq:zero-focus-gradient-gram}. Thus its kernel is exactly this pair.

To define the marginal sources put, for normalised $R$,
\[
 E=(R+\widetilde R)/2,\quad O=(R-\widetilde R)/2,\quad
 p=\operatorname{tr}R^2,\quad q=\operatorname{tr}R\widetilde R,
\]
\[
 p_o=6c^2+2a^2,\quad p_i=b^2+5c^2+2d^2,\quad
 \alpha=(p_i-p_o)/c^2,
\]
and set
\[
 Q_0=((p-p_o)-\alpha(p-q))I,\quad
 Q_1=O(E-6u_*I),\quad Q_2=[E,O].
\]
The $A_j$ and $B_j$ are fixed real linear combinations of these three
covariants. Homogenisation replaces $p_o$ by $p_o\nu^2$ and $6u_*$ by
$6u_*\nu$. Lemma~\ref{lem:seven-joint-spinflip} gives
$Q_0=Q_1=Q_2=0$ for every marginal of $\phi$: in the out-class
$p=q=p_o$, while in the in-class $p=p_i$ and $p-q=c^2$.
Consequently both the ordinary and cap-weighted Gram matrices of the
$Q_j$ are tight at $\phi$.

\begin{proof}
We describe the finite verification and the exact identity it establishes.
All choices of bases and fixed coefficients are supplied as integers,
rationals or binary rationals. The only other constants are $x_*$,
its positive square root ratios, and the logarithms in $H$.

Let $U_b$ be an integer basis fixed by the copy subgroup, $D_b$ its
physical Gram matrix, and $Y_b(t)$ the degree-five physical image of
$U+tV$. With $S_b=U_b^{\mathsf T}D_bU_b$ and
$X_b(t)=U_b^{\mathsf T}D_bY_b(t)$, select independent columns $X_b^0$.
The physical kernel projector in invariant coordinates is
\begin{equation}\label{eq:seven-entropy-kernel-projector}
 P_b=I-S_b^{-1}X_b^0
       \bigl((X_b^0)^{\mathsf T}S_b^{-1}X_b^0\bigr)^{-1}
       (X_b^0)^{\mathsf T},\qquad
 t=\sqrt{x_*/u_*}.
\end{equation}
The fixed target factors $B_b$ are lower-trapezoidal dyadic matrices
in invariant coordinates; the exact features are $U_bP_bB_b$, with the
rational diagonal-Gram transport used in the purity certificate.
These baselines need no justification of their own: the complete repair
check below establishes the identity for them. Repeated image columns are
stored once and expanded exactly, and no least squares or QR is performed
during verification.
E7-K0 checks the added two-row Young modules. E7-K1 verifies the image polynomials by
integer contractions at six determining homogeneous amplitude pairs,
and checks subgroup fixedness. E7-K2 proves the asserted generic image
rank $r_b$: all $(r_b+1)$-minors have degree at most $5(r_b+1)$,
and exact rank factorisations at $5(r_b+1)+1$ integer points make those
minors identically zero. Modular elimination only selects pivots;
the factorisations themselves use integers. Arb verifies invertibility
of the displayed $r_b$-column Gram matrix at $x_*$. The symmetric cubic
kernel instead follows from the closed Gram and its two analytic
relations in Proposition~\ref{prop:zero-focus}. E7-K3 checks their
physical coordinates, the other cubic kernels and the exact generating
polynomials. These steps prove
kernel membership exactly, rather than inferring it from small singular values.

Let $\mathcal C_*$ be the real, conjugation-even, permutation-invariant
five-copy polynomial space annihilating the moment of $\phi$ and its
family tangent. The invariant space has dimension $3285$, by
Proposition~\ref{prop:fivecopy-finite-probes}. Evaluation and family
differentiation are independent: normalisation has value one and zero
derivative, while $\pi_3'(x_*)\ne0$ by the explicit family quadratic.
Hence $\dim\mathcal C_*=3283$. Every source above belongs to
$\mathcal C_*$, since its kernel feature vanishes at $\phi$.
The target belongs to $\mathcal C_*$ by first-order scalar contact and
the entropy stationarity equation for $x_*$.

Start with the identity Gram in each projected five-copy basis, and
with the fixed positive cubic and marginal sources. Repair only the
first twelve Grams, using the same rational orthogonal frames $Q_b$ as in
Theorem~\ref{thm:fivecopy-sharp-certificate}:
\[
 C_b=I+Q_b\begin{pmatrix}J_b&E_b\\E_b^{\mathsf T}&0\end{pmatrix}Q_b^{\mathsf T}
 \quad(b\le12),\qquad C_b=I\quad(b>12),
\]
where $J_b=J_b^{\mathsf T}$ has size $r_b=\min(7,k_b)$.
Let $z_b$ list the upper triangle of $J_b$ and all entries of $E_b$.
Then $\|C_b-I\|_2\le\sqrt2\|z_b\|_2\le\tfrac32\|z\|_2$.
There are $3754$ independent real repair coordinates $z=(z_b)_{b\le12}$,
with the same exact frame and coordinate audit as FP5-4.
The columns of the evaluation matrix are the repair functionals
$X\mapsto\operatorname{tr}[W_bSW_b^{\mathsf T}\Gamma_b(X)]$, $b\le12$, which
lie in $\mathcal C_*$: each face $W_b^{\mathsf T}\Gamma_b(X_5^+(\psi_x))W_b$ is
positive semidefinite along the family and vanishes at $\phi$, so its
derivative vanishes there too.
We apply Lemma~\ref{lem:positive-repair} over $K=\mathbb R$ with
$\mathcal C=\mathcal C_*$ ($d=3283$) inside the $3285$-dimensional invariant space, $N=3754$,
$\Phi(z)=\sum_{b\le12}\operatorname{tr}[(C_b-I)G_b]$,
$\Delta_b(z)=C_b-I$ with $c=3/2$, and $\rho$ the left side
of~\eqref{eq:seven-entropy-localized-identity} minus the baseline right side
(identity Grams and the fixed cubic and marginal sources). Then
$\rho\in\mathcal C_*$, since the target, every source and each identity Gram $\operatorname{tr}G_b$ (the $S=I$ repair functionals above) lie in $\mathcal C_*$.
Let $\mathcal E$ evaluate, with fixed dyadic row weights, at the $3424$ rational direct probes of
Proposition~\ref{prop:fivecopy-finite-probes}. E7-1 encloses every
entry of the exact evaluation matrix $A$ and residual $r$ by explicit
forward rounding bounds and Arb input enclosures. Its proposed binary
matrices $L,R$ satisfy
\[
 \|I-LAR\|_2\le\delta<1,\qquad
 \frac{3\|R\|_2\|Lr\|_2}{2(1-\delta)}<1,
\]
which are \eqref{eq:positive-repair-inverse} and~\eqref{eq:positive-repair-margin}.
The complete cold log gives verified upper bounds $\delta<0.0311$ and
a Gram correction below $0.1354$.
Lemma~\ref{lem:positive-repair} shows that these repair functionals span
$\mathcal C_*$, that evaluation is injective there, and that
$z=R(LAR)^{-1}Lr$ yields the exact identity with $C_b\succ0$ for every repaired
block. The unmodified Grams are identities. This proves
\eqref{eq:seven-entropy-localized-identity}; every term is nonnegative
under the hypotheses of~\eqref{eq:seven-entropy-compatible-bound}.
\end{proof}

\paragraph{Arithmetic bounds in E7-1.}
E7-1 uses the IEEE-754 arithmetic model and the row-norm companions
of the purity certificate (the paragraphs following
Theorem~\ref{thm:fivecopy-sharp-certificate}). Large-face traces use the
multiplier $V_RV_L^{\mathsf T}$ formed once, while both repair actions
use at most seven columns. The symmetric and exterior $F^2$ actions
are quadratic products of the local $4\times4$ matrix; replacing their
signs by plus gives their positive companions, and their depth is bounded
by the same $6g$ degree allowance. In particular the exterior action
retains its alternating sign in the signed contraction.
The coefficient and factored-trace depths are $a+f+6g+s$ and
$a+h+6g+s$, with chunks $(128,256,128)$; multiplier traces use the
corresponding squared-factor dimensions, as in FP5-1.
Here $g=7$, and the positive-companion budget is
$b=Dk+\sum_i d_i+42+2\binom5j^2+64$.
The extra summation allowance covers all seventeen faces. Every signed
path has depth at least $64$, asserted by the verifier.

The entropy certificate also has Arb input radii $E_W$ on the feature
bases. Replace each $|W|$ by $|W|+E_W/(64u)$, rounded upwards,
\emph{before} forming its row norms or left companions.
Since $\gamma_m\ge64u$, the bilinear terms containing $E_W$ bound
input errors, and the $|W|$ term bounds rounding errors. Minkowski and
Cauchy therefore give the same coefficient Frobenius and trace bounds
for the exact Arb-enclosed features. The input guards and explicit
$2^{-500}$ allowances for propagated subnormal errors are those of
FP5-1. The rectangular coefficients are then rotated and packed using
the exact rational frames and the rigorous transformation radius in
that certificate; this yields one row radius for the $3754$ independent
coordinates. The validated inverse uses these row radii, the source
radii and the unchanged $3283$-dimensional critical space.
The joint trace products have local degree seven and at
most $120^2$ copy-coset terms, with integer dot products below $2^{53}$ before rational division. The two
cubic localisers are combined into one degree-seven polynomial in ten local variables ($5560$ monomials),
built from rational local maps that E7-K3 checks. These bound the exact functions in the identity.

\begin{theorem}[Sharp seven-qubit mean entropy; verified-arithmetic certificate E7-1]
\label{thm:seven-entropy-global-maximum}
Every pure seven-qubit state satisfies
\begin{equation}\label{eq:seven-entropy-global-maximum}
 \boxed{\quad E_3(z)\le E_3(\psi_{x_*})
 =\frac{39f(c)+8f(a)+6f(d)+3f(b)}7
 =2.965356029477675899\ldots\quad}
\end{equation}
where $x_*$ is the root specified in
Proposition~\ref{prop:entropy-seven-stationarity}. Equality holds if and only if $z$ is local-unitarily and site-permutation
equivalent to $\psi_{x_*}$. With the qubit labels fixed, this equality set
consists of exactly $42$ LU orbits.
\end{theorem}
\begin{proof}
It suffices to consider $E_3(z)\ge E_*$. Proposition~\ref{prop:seven-competitor-band}
and Corollary~\ref{cor:seven-entropy-eigen-cap-497} put every such state
in the region of~\eqref{eq:seven-entropy-compatible-bound}. Therefore
Proposition~\ref{prop:seven-joint-entropy-target} and
Theorem~\ref{thm:seven-entropy-localized-identity} give
$E_3(z)\le\mathcal P_5(z)\le E_*$, with equality at $\phi$. Proposition~\ref{prop:seven-entropy-equality-moment}
and Corollary~\ref{cor:icosahedral-labelled-orbits} prove the equality assertions.
\end{proof}

\subsection{The entropy equality orbit}\label{sec:seven-entropy-equality}

The positive identity and the invariant-square equality theorem shared with
purity fix the five-copy moment, after which the degree-four invariants
recover the orbit.

\begin{proposition}[Entropy equality fixes the five-copy moment]
\label{prop:seven-entropy-equality-moment}
If a normalised pure seven-qubit state $z$ satisfies $E_3(z)=E_*$, then
\[
                     X_5^+(z)=X_5^+(\psi_{x_*}).
\]
Consequently it is local-unitarily and site-permutation equivalent to
$\psi_{x_*}$.
\end{proposition}
\begin{proof}
Proposition~\ref{prop:seven-competitor-band} and Corollary~\ref{cor:seven-entropy-eigen-cap-497} apply to $z$, so
Proposition~\ref{prop:seven-joint-entropy-target} and
Theorem~\ref{thm:seven-entropy-localized-identity} give
$E_*\le\mathcal P_5(z)\le E_*$. Every term of the positive identity
therefore vanishes; since each $C_b$ is positive definite, so does every face
$W_b^{\mathsf T}\Gamma_b(X_5^+(z))W_b$. Apply
Theorem~\ref{thm:fivecopy-family-equality} with $y=x_*$, which lies in its
interval by Proposition~\ref{prop:entropy-seven-stationarity}: the spanning
hypothesis is the last assertion of
Theorem~\ref{thm:seven-entropy-localized-identity}. The converse follows from
local-unitary and permutation invariance and $E_3(\phi)=E_*$.
\end{proof}

\begin{restated}[seven-qubit mean-entropy maximum and equality; verified-arithmetic E7-1; exact AH-6, FO-1, EO-1; uses Theorem~\ref{thm:fqe-seven-purity}]{Theorem}{thm:fqe-seven-entropy}
Every pure seven-qubit state satisfies
\[
 E_3(\psi)\le E_3(\psi_{x_*})=2.965356029477675899\ldots\ \text{bits}.
\]
Equality holds if and only if $\psi$ is local-unitarily and qubit-permutation equivalent to $\psi_{x_*}$.
With fixed qubit labels, the equality set consists of exactly $42$ LU orbits.
\end{restated}
\begin{proof}
Theorem~\ref{thm:seven-entropy-global-maximum} combines the
competitor restrictions
(Proposition~\ref{prop:seven-competitor-band} and
Corollary~\ref{cor:seven-entropy-eigen-cap-497}),
the joint majorant (Proposition~\ref{prop:seven-joint-entropy-target}),
and the localised five-copy identity
(Theorem~\ref{thm:seven-entropy-localized-identity}).
Proposition~\ref{prop:seven-entropy-equality-moment} fixes the equality
moment; Corollary~\ref{cor:fourth-moment-orbit-rigidity}
identifies the orbit, and Corollary~\ref{cor:icosahedral-labelled-orbits}
counts its $42$ labelled LU orbits.
\end{proof}
\subsection{Nondegeneracy of the maximiser}\label{sec:seven-entropy-nondegenerate}
Having identified the global maximum and its equality orbit, we now quantify the entropy loss near that orbit.
The constrained Hessian at \(\phi\) is negative definite transverse to the symmetry orbit, giving
an explicit quadratic basin.

\begin{theorem}[Strict full-space local entropy maximum; exact certificate SEH-1--SEH-3 with an analytic perturbation bound]
\label{thm:entropy-seven-local}
The state \(\psi_{x_*}\) is a \emph{strict local maximiser} of
\(E_3\) modulo local unitaries and global phase. More precisely,
for every nonzero real unit-sphere tangent vector \(v\) orthogonal to the
local-unitary and phase tangent space at \(\psi_{x_*}\),
\begin{equation}\label{eq:entropy-seven-hessian-gap}
   D^2(E_3|_{\|z\|=1})_{\psi_{x_*}}[v,v]
      <-\frac{239}{1000}\,\|v\|^2.
\end{equation}
The \(22\) phase and local-unitary symmetry infinitesimals have
zero constrained Hessian.
\end{theorem}
\begin{proof}
Extend \(E_3(z)\) off the unit sphere by the same formula
\(-35^{-1}\sum_{|S|=3}\operatorname{tr}(R_S\log_2 R_S)\), with \(R_S\) not renormalised:
for unit \(z\) and \(r>0\), \(E_3(rz)=r^2E_3(z)-r^2\log_2r^2\). Hence \(E_3\) is the mean entropy only on the
unit sphere, and its radial derivative \(\frac12\frac{d}{dr}E_3(rz)|_{r=1}=E_3(z)-1/\ln2\) is the multiplier \(\lambda\) below.
For a three-qubit cut flatten \(z\) to an \(8\times16\) matrix \(M\)
and put \(R=MM^\dagger\). For a variation \(v\) with flattening \(\dot M\),
write \(\dot R=\dot MM^\dagger+M\dot M^\dagger\), and denote the Fréchet
derivative of the matrix logarithm by
\[
 {\cal L}_R(H)=\int_0^\infty(R+tI)^{-1}H(R+tI)^{-1}\,dt.
\]
The unnormalised second variation of the mean entropy is
\begin{equation}\label{eq:entropy-seven-H-form}
 D^2 E_3(z)[v,v]
 =-\frac1{35\ln2}\sum_{|S|=3}
 \left[2\operatorname{tr}\dot M^\dagger(\ln R+I)\dot M
 +\operatorname{tr}\dot R\,{\cal L}_R(\dot R)\right].
\end{equation}
At the stationary unit vector put
\(\lambda=\langle z,\partial_{\bar z} E_3(z)\rangle
=E_3(z)-1/\ln2\).
The constrained Hessian is \(H=D^2E_3-2\lambda I_{256}\);
it splits into a real and imaginary \(128\times128\) block
because \(z\) is real. Its phase/LU symmetry generators, given
by \(iz\) and \(i\sigma_a^{(j)}z\), span \(G\).

For the certificate take the nearby \emph{rational} vector
\(z_0=a_0U+b_0V\), with \(a_0=16934604/10^8\),
\(b_0=26425139/10^8\). Exact integer signs in
\eqref{eq:entropy-seven-star}, evaluated at
\(698287976009/10^{13}\) and \(698287976010/10^{13}\), bracket
\(x_*\). Squaring rational endpoints proves
\(|a-a_0|,|b-b_0|<5\cdot10^{-9}\) and hence
\(\|\psi_{x_*}-z_0\|<3\cdot10^{-8}\) (SEH-1).
Every reduced eigenvalue along this short segment is at least
\(1/12\), and each difference between two \emph{distinct}
reduced eigenvalues at \(z_0\) exceeds \(1/64\); both are
exact rational checks.

At \(z_0\) the spectral projectors are rational: if \(\ell\)
ranges over the two or three distinct eigenvalues of \(R\), then
\[
 P_\ell=\prod_{h\ne\ell}\frac{R-hI}{\ell-h},\qquad
 \ln R=\sum_\ell(\ln\ell)P_\ell,\qquad
 {\cal L}_R(H)=\sum_{\ell,h}
 \frac{\ln\ell-\ln h}{\ell-h}P_\ell H P_h,
\]
with diagonal coefficient \(1/\ell\). SEH-2 uses rational proxies
for the four logarithms and \(1/\ln2\), each with error less than
\(10^{-13}\), bounded by exact rational \(\operatorname{arctanh}\)
series. It forms a rational fixed-point matrix \(\widehat B\)
approximating \(B=-H+4GG^T\), including the symmetry penalty,
with entrywise rounding checked by integer arithmetic.

The exact Hessian at \(z_0=a_0U+b_0V\) respects the symmetry-adapted orthogonal integer decomposition of Theorem~\ref{thm:icolocal}.
Rounding could disrupt this structure in \(\widehat B\), so SEH-2 checks the decomposition exactly for \(\widehat B\). The decomposition consists of the six computational-basis sectors~\eqref{eq:ico-hessian-six-blocks}, the disjoint-pair projectors \(P_2,P_3,P_{-1},P_{-3}\) on \(\mathcal H_{\mathrm{supp}}\) and \(\mathcal H_{\mathrm{flip}}\), and the triple-incidence projectors \(F_1,F_2,P_{8,40}\) on \(\mathcal H_{\mathrm{odd}}\). In both \(128\times128\) blocks, SEH-2 verifies by exact integer multiplication that the \(16\) isotypic blocks (sizes \(1\times1\), \(2\times2\), and \(4\times4\), with diagonal integer Gram matrices \(D_{\mathrm{blk}}\)) span \(\mathbb R^{128}\) and are exact invariant subspaces of \(\widehat B\), and that every \(1\times1\), \(2\times2\), and \(4\times4\) symmetric integer matrix \(V_{\mathrm{blk}}^{\mathsf T}(\widehat B-0.239395\,I_{128})V_{\mathrm{blk}}\) has strictly positive leading principal minors. Hence \(\lambda_{\min}(\widehat B)>0.239395\) with no floating-point basis or Gram conditioning step.

It remains to bound \(B(\psi_{x_*})-\widehat B\) analytically. Write \(\varepsilon=\|z_0-\psi_{x_*}\|<3\cdot10^{-8}\).
Using \(R\succeq I/12\) along the segment,
\[
 \|R(z_0)-R(z_*)\|_{\rm op}\le3\varepsilon,\quad
 \|\ln R(z_0)-\ln R(z_*)\|_{\rm op}\le36\varepsilon,\quad
 \|{\cal L}_{R(z_0)}-{\cal L}_{R(z_*)}\|_{F\to F}
 \le432\varepsilon.
\]
Also \(\|\dot R(v)\|_F\le3\|v\|\) and
\(\|\dot R_{z_0}(v)-\dot R_{z_*}(v)\|_F
\le2\varepsilon\|v\|\). Inserting these estimates in
\eqref{eq:entropy-seven-H-form} with \(\|\dot M\|_F=\|v\|=1\) and
\(\|{\cal L}_R\|\le12\), each cut changes by at most
\(2\cdot36\varepsilon+2\cdot(2\varepsilon)\cdot12\cdot3+3\cdot432\varepsilon\cdot3
=4104\varepsilon\), hence the mean by at most
\(4104\varepsilon/\ln2<5.93\cdot10^3\varepsilon\). Parametrise the segment from $z_*$ to $z_0$ over $[0,1]$;
then $\|\dot R\|_F\le3\varepsilon$ along it. Hence
\(|\operatorname{tr}\dot R(\ln R+I)|\le\|\dot R\|_F\|\ln R+I\|_F<3\sqrt8\cdot\frac32\,\varepsilon\) per cut and
\(|\|z_0\|^2-1|<3\varepsilon\), so \(\lambda(z)=E_3(z)-\|z\|^2/\ln2\) obeys \(2|\lambda(z_0)-\lambda(z_*)|<50\varepsilon\); the \(22\) columns \(iz,i\sigma_a^{(j)}z\)
of \(G\) have norm \(\|z\|\), so \(4\|G(z_0)G(z_0)^T-G(z_*)G(z_*)^T\|_{\rm op}\le8\cdot22(1+\varepsilon)\varepsilon<180\varepsilon\).
These add less than \(4\cdot10^2\varepsilon\). The operator-norm change is therefore below
\(10^4\varepsilon\). This budget is analytic; SEH-3 checks only the resulting margin.
The four rational logarithm enclosures, their divided differences,
and the explicit \(10^{-10}\)-grid rounding used to assemble
\(\widehat B\) add less than \(10^{-5}\) in operator norm.
Consequently
\[
 \|B(\psi_{x_*})-\widehat B\|_{\rm op}
 <10^4(3\cdot10^{-8})+10^{-5}=0.00031.
\]
The exact block positivity bound now implies \(B(\psi_{x_*})\succ0\), with
\(\lambda_{\min}(B)>0.239395-0.00031=0.239085>0.239\) (SEH-3).
By Proposition~\ref{prop:entropy-seven-stationarity} and LU/phase
invariance, \(H\) annihilates every vector in \(G\). On \(G^\perp\)
the penalty vanishes, so \(-H=B\succ0\) and
\eqref{eq:entropy-seven-hessian-gap} follows. A local slice
transverse to the compact unitary-group orbit gives the claimed
strict local maximum modulo symmetry.
\end{proof}

\begin{theorem}[quadratic stability near the maximiser; proved from SEH-1--SEH-3, rational constants EB-0--EB-3]
\label{thm:entropy-seven-quantbasin}
Let $\mathcal O_E$ be the local-unitary, global-phase
and qubit-permutation orbit of $\phi=\psi_{x_*}$, the equality set of
Theorem~\ref{thm:seven-entropy-global-maximum}.
For any normalised pure seven-qubit state $z$ write
$\delta(z)=\min_{\phi\in\mathcal O_E}\|z-\phi\|_2$.
If $\delta(z)\le1/20000$, then
\begin{equation}\label{eq:entropy-seven-quantbasin}
\boxed{\qquad
 E_3(z)\le E_3(\psi_{x_*})-\frac{\delta(z)^2}{20}.
 \qquad}
\end{equation}
\end{theorem}
\begin{proof}
Let $\phi'\in\mathcal O_E$ be a closest point to $z$; by symmetry we may take $\phi'=\phi$.
After the allowed global phase its overlap with $z$ is real
and nonnegative. Thus
\[
 z=\cos\theta\,\phi+\sin\theta\,w,\qquad
 \|w\|=1,\quad\langle\phi,w\rangle=0,\quad
 \delta(z)=2\sin(\theta/2).
\]
Here $\theta=0$ is trivial. The nearest-point
condition implies that $w$ is real orthogonal
to every local-unitary and phase tangent vector
at $\phi$. The full-space constrained Hessian
certificate of Theorem~\ref{thm:entropy-seven-local}
then gives, for $e(t)=E_3(\cos t\,\phi+\sin t\,w)$,
\[
 e'(0)=0,\qquad e''(0)<-\frac{239}{1000}.
\]

We obtain a uniform third-derivative bound with no
numerical spectral estimates. For each $|S|=3$,
let $M(t)$ be the $8\times16$ flattening of the
unit vector $\gamma(t)=\cos t\,\phi+\sin t\,w$,
and let $N=M'$, $R=MM^\dagger$. Along this
unit-sphere geodesic,
$M''=-M$, $\|M\|_F=\|N\|_F=1$, so
\[
 \|R'\|_F\le2,\quad \|R''\|_F\le4,\quad
 R'''=-4R',\quad\|R'''\|_F\le8.
\]
At $t=0$ all triple marginals have smallest
eigenvalue at least $1/12$ by the exact
eigenvalue enclosures in SEH-1. As
$\|R'(t)\|_{\rm op}\le2$, Weyl's inequality
gives, for $0\le t\le1/10000$,
\[
 \lambda_{\min}(R(t))
 \ge\frac1{12}-2t
 \ge\frac1{12}-\frac1{5000}
 >\frac1{16}=:m.
\]
The eigenvalues are at most $1$ because $\gamma(t)$
is normalised. Hence
$\|\ln R+I\|_F<6$.

Write
$\mathcal L_R(K)=\int_0^\infty
 (R+sI)^{-1}K(R+sI)^{-1}\,ds$
for the Fr\'echet derivative of $\ln R$.
The resolvent integral bounds imply
\[
 \|\mathcal L_R(K)\|_F\le m^{-1}\|K\|_F,\qquad
 \|D\mathcal L_R[H](K)\|_F
   \le m^{-2}\|H\|_F\|K\|_F.
\]
Differentiating the entropy of one cut three times
(with natural logarithms internally) yields
\[
 \frac{d^3}{dt^3}[-\operatorname{tr}R\log_2 R]
 =-\frac1{\ln2}
 \left[
 \operatorname{tr}R'''(\ln R+I)
 +3\operatorname{tr}R''\mathcal L_R(R')
 +\operatorname{tr}R'\,
       D\mathcal L_R[R'](R')
 \right].
\]
Therefore its absolute value is bounded by
\[
 \frac{8\cdot6+3\cdot4\cdot2\cdot16
                     +2^3\cdot16^2}{\ln2}
 =\frac{2480}{\ln2}<3720,
\]
using $\ln2>2/3$. Averaging the $35$ cuts
gives the same bound $|e'''(t)|<3720$.
Every step holds uniformly along the geodesic.

For $\delta\le1/20000$,
$\theta=2\arcsin(\delta/2)\le2\delta\le1/10000$.
Taylor's theorem and the negative Hessian give
\[
 e(\theta)-e(0)
 \le-\frac{239}{2000}\theta^2
       +\frac{3720}{6}\theta^3
 \le-\left(\frac{239}{2000}-\frac{620}{10000}\right)
       \theta^2
 =-\frac{23}{400}\theta^2
 \le-\frac1{20}\delta^2,
\]
as $\theta\ge\delta$.
EB-0--EB-3 check all needed rational constant
inequalities independently; the spectral and Hessian
inputs are exactly SEH-1--SEH-3. This proves
\eqref{eq:entropy-seven-quantbasin}.
\end{proof}

\begin{corollary}[Global quadratic stability at seven qubits; proved from the certified theorems cited in the proof]\label{cor:seven-global-quadratic}
Let $\delta_p(z)$ be the distance from a normalised pure seven-qubit state $z$ to the orbit of $\Psiico$ under local
unitaries, global phase and qubit permutations, and $\delta(z)$ the distance to $\mathcal O_E$ as in
Theorem~\ref{thm:entropy-seven-quantbasin}. There are constants $c_p,c_e>0$ such that every such $z$ satisfies
\[
 \pi_3(z)-\frac{121}{917}\ge c_p\,\delta_p(z)^2,\qquad
 E_3(\psi_{x_*})-E_3(z)\ge c_e\,\delta(z)^2 .
\]
\end{corollary}
\begin{proof}
Both distances are at most $\sqrt2$, since $\min_\theta\|z-e^{i\theta}\phi\|^2=2-2|\langle z,\phi\rangle|$. If
$\delta_p(z)\le1/250$, a qubit permutation $\sigma$ carries a nearest point of the orbit into the local-unitary and
phase orbit of $\Psiico$; $\pi_3$ is permutation invariant, so Theorem~\ref{thm:icoquantbasin} applied to $\sigma z$
gives the bound with $1/1024$. The set $\{\delta_p\ge1/250\}$ of unit vectors is compact and $\pi_3-121/917$ is
continuous on it. It is positive there, because equality holds only on the orbit
(Corollary~\ref{cor:seven-global-from-fivecopy}), so it has a positive minimum $m_p$. Hence
$c_p=\min\{1/1024,m_p/2\}$ works. The entropy bound follows in the same way from
Theorem~\ref{thm:entropy-seven-quantbasin} for $\delta\le1/20000$, continuity of $E_3$, and the equality case of
Theorem~\ref{thm:seven-entropy-global-maximum}. The constants come from compactness and are not explicit.
\end{proof}

\subsection{Mixed states attain neither optimum}\label{sec:seven-mixed-roofs}
For a seven-qubit density matrix $\rho$ and $|S|=3$ let
\[
 E_S(\rho)=\inf_{\rho=\sum_\ell p_\ell|\psi_\ell\rangle\langle\psi_\ell|}\sum_\ell p_\ell\,S\bigl(\rho_S^{(\ell)}\bigr),\qquad
 L_S(\rho)=\inf_{\rho=\sum_\ell p_\ell|\psi_\ell\rangle\langle\psi_\ell|}\sum_\ell p_\ell\bigl(1-\operatorname{tr}(\rho_S^{(\ell)})^2\bigr)
\]
be the convex roofs of the entropy (in bits) and of the linear entropy of the cut $S|S^c$, where $\rho_S^{(\ell)}$ is the
marginal of $\psi_\ell$.

\begin{lemma}[no flat extremal planes]\label{lem:seven-flat-extremal-planes}
Let $f$ be a real function on the unit sphere of $(\mathbb C^2)^{\otimes7}$, invariant under phases and local unitaries,
and let $\phi$ be a global extremum of $f$ such that $f$ is $C^2$ near $\phi$, every one-qubit marginal of $\phi$ is $I/2$,
and the constrained Hessian $H$ of $f$ at $\phi$ satisfies $H[w,w]\ne0$ for every nonzero real tangent vector
$w$ ($\operatorname{Re}\langle\phi,w\rangle=0$) orthogonal to
\[
 G=\operatorname{span}_{\mathbb R}\{i\phi,\ i\sigma_a^{(j)}\phi\}.
\]
Then no complex subspace of dimension at least two containing $\phi$ has all its unit vectors global extrema of the
same kind.
\end{lemma}
\begin{proof}
Suppose $V\ni\phi$ is such a subspace with $\dim V\ge2$. Pick a unit vector $\psi\in V$ with $\langle\phi,\psi\rangle=0$, and let
$w$ be $\psi$ or $i\psi$. The great circle $t\mapsto\cos t\,\phi+\sin t\,w$ lies in the unit sphere of $V$, so $f$ is
constant on it, and at the critical point $\phi$ its second derivative along this geodesic is $H[w,w]$; hence $H[w,w]=0$.
The vectors spanning $G$ are tangent, as $\langle\phi,\sigma_a^{(j)}\phi\rangle$ is real, and the orbit of $\phi$ consists of
critical points, so differentiating the vanishing constrained gradient along it gives $Hg=0$ for $g\in G$. Writing $w=g+w'$
with $g\in G$ and $w'$ tangent and orthogonal to $G$, $0=H[w,w]=H[w',w']$, so $w'=0$: both $\psi$ and $i\psi$ lie in $G$.

But $G\cap iG=\{0\}$. Let $\mathfrak k$ be the real span of $iI$ and the $i\sigma_a^{(j)}$, and suppose $g=A\phi=iA'\phi$ lies in $G\cap iG$, with
$A,A'\in\mathfrak k$. Put $B=-A'$, so $A\phi=-iB\phi$. Since $A,B$ are anti-Hermitian,
\[
 \langle\phi,[A,B]\phi\rangle=\langle B\phi,A\phi\rangle-\langle A\phi,B\phi\rangle=-2i\|B\phi\|^2 .
\]
The commutator $[A,B]$ is a sum of traceless one-site operators, and every one-qubit marginal of $\phi$ is $I/2$, so the left
side vanishes. Hence $B\phi=0$ and $A\phi=0$. Thus $\psi=0$, a contradiction.
\end{proof}

\begin{proposition}[mixed-state roofs; proved from the certified theorems cited in the proof]\label{prop:seven-mixed-roofs}
Every seven-qubit density matrix $\rho$ satisfies
\[
 \frac1{35}\sum_{|S|=3}E_S(\rho)\le E_3(\psi_{x_*}),\qquad \frac1{35}\sum_{|S|=3}L_S(\rho)\le1-\frac{121}{917}.
\]
Equality holds in the first bound only for pure states local-unitarily and permutation equivalent to $\psi_{x_*}$, and in
the second only for pure states equivalent to $\Psiico$.
\end{proposition}
\begin{proof}
Using one ensemble of $\rho$ for all $35$ infima bounds the left sides by $\sum_\ell p_\ell E_3(\psi_\ell)$ and
$\sum_\ell p_\ell(1-\pi_3(\psi_\ell))$, so Theorem~\ref{thm:seven-entropy-global-maximum} and
Corollary~\ref{cor:seven-global-from-fivecopy} give the bounds and the pure equality cases.

Suppose the entropy bound is attained. The left side is at most the ensemble mean of $E_3(\psi_\ell)$, and each term is
at most $E_3(\psi_{x_*})$, so every member of every ensemble of $\rho$ is a pure maximiser. Every unit
vector $\chi$ in the support of $\rho$ belongs to some ensemble, since $\rho-p|\chi\rangle\langle\chi|\succeq0$ for
$p=\langle\chi,\rho^+\chi\rangle^{-1}$, with $\rho^+$ the inverse of $\rho$ on its support; so every unit vector of the
support is a maximiser. Pick a unit vector $\phi$ in the support; after a local unitary, a qubit permutation and a phase,
applied to $\rho$, which change neither side, $\phi=\psi_{x_*}$. Lemma~\ref{lem:seven-flat-extremal-planes} applies to
$f=E_3$ at $\phi$: near $\phi$ all triple marginals have full rank, so $E_3$ is smooth there, the one-qubit marginals are
$I/2$ (Theorem~\ref{thm:ico}), and Theorem~\ref{thm:entropy-seven-local} gives $H[w,w]<-\frac{239}{1000}\|w\|^2$ for tangent $w$
orthogonal to $G$. So the support has dimension one and $\rho$ is pure.

The linear-entropy case is the same with $f=\pi_3$, a global minimum and $\Psiico$ in place of $\psi_{x_*}$: the Lagrange form
$Q$ of Theorem~\ref{thm:icolocal} is the constrained Hessian, positive semidefinite with kernel exactly $G$.
\end{proof}
Thus mixed-state exclusion follows from two properties of the pure optimiser, maximally mixed one-qubit marginals
and a Hessian nondegenerate transverse to the local-unitary orbit (Lemma~\ref{lem:seven-flat-extremal-planes}); the
four-qubit proof of
Corollary~\ref{cor:fqe-convex-roof} instead uses the square flattenings, which the $3|4$ cuts do not have.

\subsection{Full-rank necessity at every local entropy maximum}
\label{sec:seven-entropy-fullrank}
At a rank-deficient cut, a perturbation can create a small positive eigenvalue. The resulting entropy gain
dominates the smooth losses on the other cuts, ruling out a local maximum.

\begin{proposition}[No rank-deficient local maxima of mean balanced entropy]
\label{prop:seven-entropy-full-rank}
Let \(z\) be a normalised pure state of seven qubits. If \(z\)
is a local maximiser of the mean three-body von Neumann entropy
\(E_3(z)\) on the full pure-state space, then
\[
                  \rho_S(z)\succ0
                  \qquad\text{for every }|S|=3.
\]
More generally, for any finite multipartite pure-state system,
a positive weighted sum of subsystem entropies cannot have a
local maximum at a state with a rank-deficient included marginal,
provided the included subsystem has dimension no greater than
its complement. This holds without assuming permutation or
local-unitary symmetry.
\end{proposition}
\begin{proof}
Suppose the reduction at a cut \(A|B\), with dimensions
\(d_A\le d_B\), has rank \(r<d_A\). Flatten the normalised
state \(z\) into a \(d_A\times d_B\) matrix \(M\). Choose unit
vectors \(u\in\ker M^\dagger\) and \(w\in\ker M\), and let
\(v\) have flattening \(u w^\dagger\). Then
\(\langle z,v\rangle=0\) and \(\|v\|=1\).
The normalised perturbation
\[
 z_t=\frac{z+t v}{\sqrt{1+t^2}},\qquad
 p=\frac{t^2}{1+t^2}
\]
has, at the selected cut, the exact reduced state
\[
 \rho_A(z_t)=(1-p)\rho_A(z)+p\,uu^\dagger,
\]
with orthogonal supports. Hence
\[
 S(\rho_A(z_t))-S(\rho_A(z))
   =h_2(p)-pS(\rho_A(z))
   =\frac{2t^2|\ln|t||}{\ln2}+O(t^2)
   \qquad(t\longrightarrow0),
\]
where \(h_2\) is the binary entropy.

For every other included subsystem \(T\), its reduced state
\(\rho_T(z_t)\) is a real-analytic positive semidefinite
Hermitian matrix in \(t\). Its eigenvalues separated away from
zero at \(t=0\) contribute
\[
 S_T^{\rm positive}(t)-S_T^{\rm positive}(0)
       =a_T t+O(t^2)
\]
by analytic functional calculus on the positive spectral
cluster. The eigenvalues emanating from zero contribute
\(-\sum_{\lambda_i(t)\to0}\lambda_i(t)\log_2\lambda_i(t)
\ge0\) for sufficiently small \(t\), since all such
eigenvalues lie in \([0,1]\). Thus
\[
 S(\rho_T(z_t))-S(\rho_T(z))
       \ge a_Tt-C_Tt^2
\]
for small positive and negative \(t\); this also holds for
the selected cut with its strictly positive
\(t^2|\ln|t||\) improvement kept separately.

Summing with strictly positive weights, including the
selected cut, gives
\[
 E_w(z_t)-E_w(z)
 \ge at+c\,t^2|\ln|t||-Ct^2,\qquad c>0,
\]
for both signs of sufficiently small \(t\).
Choose the sign of \(t\) so that \(at\ge0\).
Then the right side is positive for small nonzero \(t\),
contradicting local maximality.
For seven qubits all 35 balanced cuts have dimensions
\(8\le16\) and positive equal weights, proving the claim.
\end{proof}

\begin{corollary}[Stabilizer exclusion at every non-AME size]
\label{cor:stabilizer-entropy-noname}
For $n\ge2$ and $k=\lfloor n/2\rfloor$, a pure stabilizer state is a local
maximiser of mean $k$-body von Neumann entropy if and only if it is
AME$(n,2)$. Consequently no stabilizer state is an unrestricted
entropy maximiser at $n=4$ or at any $n\ge7$; in particular, neither are the graph-state minimisers of $\pi_k$ in
Table~\ref{tab:minima}.
\end{corollary}
\begin{proof}
An AME state attains the cut-dimension upper bound $k$ bits on every cut.
Conversely, every marginal of a stabilizer state is a normalised projector, $\rho_S=2^{-r_S}\Pi_S$ with
$\operatorname{rank}\Pi_S=2^{r_S}$ (Hein--Eisert--Briegel~\cite{HEB}). If every $k$-qubit marginal had full rank $2^k$,
each would be maximally mixed and the state would be AME$(n,2)$. So a non-AME stabilizer state has a rank-deficient
$k$-qubit marginal, and $k\le n-k$, so Proposition~\ref{prop:seven-entropy-full-rank} applies with equal weights.
The AME qubit sizes are $2,3,5,6$ (Section~\ref{sec:lp}).
\end{proof}

\paragraph{The remaining sections.} The four- and seven-qubit proofs are now complete. The remaining sections collect
independent consequences and generalisations, with one exception: Lemma~\ref{lem:focus-pauli} and
Proposition~\ref{prop:zero-focus} of Section~\ref{sec:sextic-general} supply the focused-covariant zero of the
conference family and the symmetric cubic kernel of the seven-qubit entropy certificate. Apart from these two
statements, none of the remaining sections is needed for the preceding global optima.

\section{Focused covariants of the four-qubit sextic}\label{sec:sextic-general}

Throughout this section $n$ is the number of qubits and $N=\|z\|^2$ (in the four-qubit
sections the norm squared is denoted by $N$, and in the seven-qubit entropy certificate by $\nu$), and $W_1=\sum_i\|\mathbf r_i\|^2$ with
$\mathbf r_i^a=\langle z|\sigma_a^{(i)}|z\rangle$.
The four-qubit moment identity \eqref{eq:fqe-remainder} is not an ordinary positive
three-copy replica operator: the latter has a negative sector on the
copy-symmetric cube (Appendix~\ref{sec:fqe-correlation}).
The successful proof uses the spin-flipped mixed tensor
$w=z_x\otimes z_y\otimes\widetilde z_t$,
$\widetilde z=\epsilon^{\otimes n}\bar z$, with
$\epsilon=\left(\begin{smallmatrix}0&1\\-1&0\end{smallmatrix}\right)$.
A complementary Pauli-coordinate description of third moments is derived in
\eqref{eq:pauli-triple} below.
At one site write
\[
 L|xyt\rangle=\epsilon_{yt}|x\rangle,\qquad
 M|xyt\rangle=\epsilon_{xt}|y\rangle,
\]
and let $S_i$ project the local three-copy tensor at site $i$ onto
$\operatorname{Sym}^3(\mathbb C^2)$.

\begin{proposition}[focused-site sextic covariants]\label{prop:focus}
For a subset $I\subseteq[n]$ and $0\le r\le n-|I|$, define
\[
 \Phi_{I,r}(z)=\left(\prod_{i\in I}S_i\right)
 \sum_{\substack{A\subseteq I^c\\|A|=r}}
 \left(\prod_{j\in A}M_j\right)
 \left(\prod_{j\in I^c\setminus A}L_j\right)
 (z_x\otimes z_y\otimes\widetilde z_t).
\]
Each $\Phi_{I,r}$ is a local-$SU(2)$ covariant of bidegree $(2,1)$ (quadratic in $z$, linear in $\bar z$), and
\begin{equation}\label{eq:focus-reflection}
 \Phi_{I,r}=\Phi_{I,n-|I|-r}.
\end{equation}
Its Gram matrix $G_I(z)=(\langle\Phi_{I,r},\Phi_{I,s}\rangle)_{r,s}$
is positive semidefinite. After~\eqref{eq:focus-reflection}, at most
$\lfloor(n-|I|)/2\rfloor+1$ distinct vectors remain.

For $n=4$ and $I=\{i\}$ put $\mathcal R_i=4\Phi_{i,1}-15\Phi_{i,0}$, a residual different from the
$V_i-8U_i$ of \eqref{eq:fqe-remainder}. The
homogeneous sextic identity then has the reduced form
\begin{equation}\label{eq:focus-four}
32(17NP-12T-12N^3)
=3\sum_{i=1}^4\left(
 15\|\Phi_{i,0}\|^2+
 \|4\Phi_{i,1}-15\Phi_{i,0}\|^2\right),
\end{equation}
where $N=\|z\|^2$, $P=\sum_{j=2}^4\tr R_{1j}^2$ and
$T=\sum_{j=2}^4\tr R_{1j}^3$.
\end{proposition}
\begin{proof}
Interchanging the two identical factors $z_x,z_y$ swaps $L$ with
$M$ at every contracted site, while the symmetric projectors at the
focused sites are unchanged. This gives~\eqref{eq:focus-reflection}.
Gram positivity follows from the Hilbert-space inner product.
In the notation of Appendix~\ref{sec:fqe-sextic}, $U_i=\Phi_{i,0}$ and
$V_i=\sum_{r=0}^3\Phi_{i,r}=2(\Phi_{i,0}+\Phi_{i,1})$.
Thus $\mathcal R_i=2V_i-19U_i=2Y_i-7U_i$ with $Y_i=V_i-6U_i$. Writing $F=17NP-12T-12N^3$ for the
left side of~\eqref{eq:focus-four} divided by $32$, the contractions
\eqref{eq:fqe-cubic-mixed-norms} give
\[
 \sum_i\|\mathcal R_i\|^2=4\sum_i\|Y_i\|^2-28\operatorname{Re}\sum_i\langle U_i,Y_i\rangle+49\sum_i\|U_i\|^2
 =\tfrac{32}3F-5NW_1 .
\]
Hence $32F=15NW_1+3\sum_i\|\mathcal R_i\|^2$, and $\sum_i\|U_i\|^2=NW_1/3$ turns this
into~\eqref{eq:focus-four}. SF-1 checks the identity on one integer four-qubit tensor by exact rational arithmetic.
\end{proof}

\begin{lemma}[the exact-support Pauli selection rule]\label{lem:focus-pauli}
Let $R_I$ be the unnormalised reduction of an arbitrary pure $n$-qubit
state to the focused sites $I$. Then, up to the fixed sign
$(-1)^{n-|I|}$,
\[
 \Phi_{I,0}= \left(\prod_{i\in I}S_i\right)
 \left(z_x\otimes (R_I\epsilon_I^T)_{y_I,t_I}\right),
 \qquad \epsilon_I=\epsilon^{\otimes|I|},
\]
with the $x$-slots on $I^c$ left intact. In particular, if
$\langle z|P|z\rangle=0$ for every Pauli word $P$ of exact support
$I$, then $\Phi_{I,0}(z)=0$.
\end{lemma}
\begin{proof}
At a contracted site,
$\sum_t\epsilon_{yt}\epsilon_{tu}=-\delta_{yu}$,
so the two remaining replicas form $R_I$. Expand
$R_I=2^{-|I|}\sum_{P\in\{I,X,Y,Z\}^{I}}
\langle z|P|z\rangle P$.
If $P$ has an identity factor at some $i\in I$, then
$\operatorname{vec}(I\epsilon^T)$ is a local singlet in the $y,t$
slots. Its tensor product with any $x$-qubit is killed by $S_i$.
Only terms supported on every site survive.
\end{proof}

\begin{proposition}[complete scalar cubic gradients; analytic, SF-5 controls]
\label{prop:zero-focus}
For any $n$, let $Q_r(z)=\sum_{|S|=r}\tr R_S^2$, $N=\|z\|^2$,
and $V_r(z)=\sum_{|S|=r}(R_S\otimes I_{S^c})z$.
Then
\[
 \Phi_{\varnothing,r}(z)=(-1)^nV_r(z),\qquad
 \frac{\partial Q_r}{\partial\bar z}=2V_r(z),\qquad
 \langle z,V_r(z)\rangle=Q_r(z).
\]
More generally, let $v$ be a local-unitary scalar covariant of bidegree
$(2,1)$ with real coefficients in the $L,M$ tensor basis. Its closure
$p_v=\langle z,v(z)\rangle$ is real and
$v=\tfrac12\partial_{\bar z}p_v$.
The vectors $V_S=(R_S\otimes I)z$, with one $S$ from each complementary
pair $\{S,S^c\}$, form a basis of this real space, of dimension $2^{n-1}$.
Its permutation-invariant subspace has basis
$V_0,\ldots,V_{\lfloor n/2\rfloor}$ and dimension
$\lfloor n/2\rfloor+1$. Thus the zero-focus weight sums exhaust
the symmetric scalar cubic source.
Consequently, for any fixed real $\lambda$,
\begin{equation}\label{eq:gradient-sextic}
 \mathcal G_{r,\lambda}(z):=
 \|\Phi_{\varnothing,r}(z)-\lambda\Phi_{\varnothing,0}(z)\|^2
 =\|V_r(z)-\lambda Nz\|^2\ge0
\end{equation}
is a universal homogeneous sextic. It vanishes exactly when $z$
is a projective stationary point of $Q_r/N^2$ with
$Q_r/N^2=\lambda$, for nonzero $z$. Moreover,
\begin{equation}\label{eq:gradient-octic}
 \begin{split}
 N\mathcal G_{r,\lambda}(z)
 &=\bigl(N\|V_r\|^2-Q_r^2\bigr)+(Q_r-\lambda N^2)^2,\\
 N\|V_r\|^2-Q_r^2
 &=\sum_{a<b}|z_a(V_r)_b-z_b(V_r)_a|^2\ge0.
 \end{split}
\end{equation}
Along the seven-qubit icosahedral family $\psi_x$ ($0\le x\le1/2$),
\begin{equation}\label{eq:zero-focus-family}
\begin{aligned}
 &\Phi_{\varnothing,1}(\psi_x)=\tfrac72\Phi_{\varnothing,0}(\psi_x),\\
 &4(104x-7)\Phi_{\varnothing,3}(\psi_x)-2(524x-37)\Phi_{\varnothing,2}(\psi_x)
   +(96x^2+3628x-263)\Phi_{\varnothing,0}(\psi_x)=0.
\end{aligned}
\end{equation}
At $\Psiico=\psi_{37/524}$ the second relation simplifies to
$\Phi_{\varnothing,3}=\frac{605}{131}\Phi_{\varnothing,0}$.
(Both relations are proved below; SF-5 checks them exactly.)
In fact the entire zero-focus Gram on this family is
\begin{equation}\label{eq:zero-focus-gradient-gram}
 G(x)=q(x)q(x)^T+\frac{x(1-2x)}8q'(x)q'(x)^T,
 \qquad q=(1,\tfrac72,21\pi_2,35\pi_3)^T.
\end{equation}
It has rank two for $0<x<1/2$ and rank one at both endpoints.
For the two source combinations $V_0$ and $131V_2-52V_3$, its
$2\times2$ Gram determinant is exactly $72x(1-2x)$.
Thus $\mathcal G_{3,605/131}$ is a nontrivial
sextic SOS vanishing at $\Psiico$, while at $x=x_*$ the two relations
in~\eqref{eq:zero-focus-family} give the $\operatorname{Sym}^7\mathbb R^2$
three-copy kernel features of Theorem~\ref{thm:seven-entropy-localized-identity}.
Neither gives a lower bound on $\pi_3$; Theorem~\ref{thm:sextic-sos-obstruction}
excludes a sharp fixed-coefficient cubic SOS certificate even after including this form.
\end{proposition}
\begin{proof}
At an $L$ site, the first replica index survives and the other
two are identified after spin flip; at an $M$ site the second
replica index survives and the first is identified with the
conjugate. For a set $S$ of $M$ sites, the surviving tensor is
$(-1)^n(R_S\otimes I_{S^c})z$.
Differentiating $\tr R_S^2$ by the product rule yields
$2(R_S\otimes I)z$. Euler's homogeneous identity gives the
inner product.

To prove that these scalar cubics exhaust the possibilities, contract the output of a general scalar cubic with
$\bar z$. Locally its two-copy invariant operator is a real combination
of identity and swap. Their tensor products are Hermitian and commute
with the simultaneous copy exchange. Since the two holomorphic inputs
are already symmetric, the two antiholomorphic inputs are symmetric
as well; hence $p_v$ is real and $\partial_{\bar z}p_v=2v$.
All quartic local-unitary invariants of bidegree $(2,2)$ are linear combinations of the
swap expectations $\tr R_S^2$. On the symmetric square, complementary
swap words agree. These are their only relations: the local
antisymmetric-site sets $A$ have even cardinality, and the words act
on them by the characters $(-1)^{|S\cap A|}$. Characters of this
even-subset group are indexed exactly by $S$ modulo complementation
and are linearly independent. Each such sector occurs on qubits.
Expectations on $z^{\otimes2}$ determine a Hermitian operator on
the symmetric square, by its polynomial coefficients.
This proves the full basis statement. After site averaging, only
$|S|$ remains. The weight-$r$ swap sum has value
$[t^r](1+t)^{n-|A|}(1-t)^{|A|}$, a polynomial of degree $r$ in $|A|$
with leading coefficient $(-2)^r/r!$. For
$0\le r\le\lfloor n/2\rfloor$ these polynomials are independent on
the equally many distinct even values of $|A|$, proving the symmetric
basis statement.

Projective stationarity of $Q_r/N^2$ is
$V_r=(Q_r/N)z$, proving the equality criterion.
The octic equation is the orthogonal decomposition of
$V_r$ into its component along $z$ and its perpendicular
component; its final equality is the usual complex
Lagrange identity. Along the icosahedral family $\psi_x$,
one-body maximal mixing gives the first relation in~\eqref{eq:zero-focus-family}.
For $r\in\{2,3\}$, the fixed-space argument of
Lemma~\ref{lem:seven-family-symmetric-criticality} places the real gradient
$4V_r(\psi_x)$ in the plane
$\operatorname{span}\{U,V\}=\operatorname{span}\{\psi_x,\psi_x'\}$, where
$\psi_x'\perp\psi_x$, $\langle\psi_x,V_r(\psi_x)\rangle=Q_r(\psi_x)$ and, by the gradient formula,
$\langle\psi_x',V_r(\psi_x)\rangle=\tfrac14 Q_r'(\psi_x)$.
Hence $V_r=Q_r\psi_x+\tfrac14Q_r'\psi_x'/\|\psi_x'\|^2$.
Since $30(a')^2+2(b')^2=1/[2x(1-2x)]$, taking inner products gives
\eqref{eq:zero-focus-gradient-gram}. The explicit purity polynomials give
$131q_2'-52q_3'=24$. The displayed Gram determinant is therefore
$24^2x(1-2x)/8=72x(1-2x)$, proving rank two throughout the open
interval; continuity gives rank one at its endpoints.
Eliminating $\psi_x'$ between $r=2$ and $r=3$ gives
$Q_2'\Phi_{\varnothing,3}-Q_3'\Phi_{\varnothing,2}-(Q_2'Q_3-Q_3'Q_2)\Phi_{\varnothing,0}=0$;
substituting $Q_2=21\pi_2$ and $Q_3=35\pi_3$ from Theorem~\ref{thm:ico} yields the second
relation in~\eqref{eq:zero-focus-family}.
Thus the two displayed relations span its kernel for every interior $x$:
their source rows are independent, since $104x-7$ and $524x-37$
cannot vanish together. SF-5 compares the closed Gram formula with every
coefficient of an independent exact tensor contraction, and checks its
specialisation at $x=37/524$ with null vectors
$(-7,2,0,0)$ and $(-605,0,0,131)$ in the
$\Phi_{\varnothing,0},\ldots,\Phi_{\varnothing,3}$ basis.
\end{proof}

\begin{remark}[scope of cubic-contraction searches]
Theorem~\ref{thm:sextic-sos-obstruction} rules out every fixed-coefficient
cubic SOS certificate of the sharp seven-qubit purity bound,
including all nonempty-focus forms, zero-focus gradients and their
cross-sector couplings. The
covariant identities above remain useful algebraic descriptions;
they do not evade the complete three-copy obstruction.
\end{remark}

\begin{remark}[joint quadratic/cubic Pauli enumerators]
For a trace-one density matrix $\rho$ define $r_P=\tr(\rho P)$ for Hermitian
Pauli strings $P$. The ordinary weight enumerator
$A_j=\sum_{\operatorname{wt}P=j}r_P^2$ controls mean subsystem
purity. There is a natural signed \emph{triple} enumerator:
\[
 C_j=\!\!\sum_{\substack{P,Q:[P,Q]=0\\
           |\operatorname{supp}P\cup\operatorname{supp}Q|=j}}
 \eta(P,Q)\,r_Pr_Qr_{R(P,Q)},
 \qquad PQ=\eta(P,Q)R(P,Q),\ \eta(P,Q)\in\{\pm1\},
\]
The sum is over ordered pairs of Pauli words, including the identity;
$R(P,Q)$ is the canonical Hermitian Pauli word.
For $b_k=\binom nk^{-1}\sum_{|S|=k}\tr\rho_S^3$,
\begin{equation}\label{eq:pauli-triple}
 b_k=\frac1{4^k\binom nk}
 \sum_{j=0}^k\binom{n-j}{k-j}C_j.
\end{equation}
Expand $\rho_S^3$ in the Pauli basis: products of
anticommuting $P,Q$ cancel between orders and every commuting
product is $\pm R(P,Q)$. Group by the union of supports to obtain
the binomial coefficient. Thus $A_j,C_j$ are natural joint
second/third-moment coordinates. For pure states the third moments are sextic in the amplitudes;
unlike $A_j$, the individual $C_j$ need not be positive. This coordinate identity is independent
of the covariant certificates and is not used in the cubic-SOS obstruction below.
\end{remark}

\section{The complete cubic-SOS obstruction}
\label{sec:sextic-all-sos}

For seven qubits let $N(z)=\|z\|^2$, let $\mathcal F_3(z)$ be the homogeneous
quartic mean three-body purity ($\mathcal F_3=N^2\pi_3$), and put
\[
 F_\lambda(z)=N(z)\bigl(\mathcal F_3(z)-\lambda N(z)^2\bigr),\qquad
 \lambda_{\mathrm{SOS},6}
 =\max\{\lambda:F_\lambda\text{ is a sum of squares of real cubics}\}.
\]
The polynomials are in all $256$ real amplitude coordinates. No
restriction to the conference family or to covariant subsets is imposed.

\begin{theorem}[Complete real cubic-SOS obstruction; TCB]
\label{thm:sextic-sos-obstruction}
The maximal homogeneous degree-six real SOS lower bound satisfies
\[
 \boxed{\quad
 \lambda_{\mathrm{SOS},6}=L_3=\frac{125}{978}<\frac{121}{917}.\quad}
\]
In particular, for every $\lambda>125/978$ there
is no identity
\[
 F_\lambda(z)=\sum_a|p_a(z)|^2+\sum_b|q_b(z,\bar z)|^2
\]
with fixed-coefficient holomorphic cubics $p_a$ and bidegree-$(2,1)$
cubics $q_b$, nor a real homogeneous cubic SOS.
This includes arbitrary local $[2,1]$ sectors and cross-focused
couplings. It excludes neither higher-degree SOS nor non-SOS
inequalities or state-dependent reductions.
\end{theorem}
\begin{proof}
Every squared holomorphic cubic is a positive linear functional of
$|z\rangle\langle z|^{\otimes3}$. Every squared mixed cubic is a
positive linear functional of
$|zz\widetilde z\rangle\langle zz\widetilde z|$, since spin flip
is invertible on the conjugate factor. Average any proposed real
cubic SOS over global phase: only opposite phase weights pair, so
its terms become squared moduli of their $(3,0)$ and $(2,1)$
components, since $(\operatorname{Re}p)^2+(\operatorname{Im}p)^2=|p|^2$. Twirling their positive Gram matrices over the local
unitaries, sites and conjugation preserves positivity and $F_\lambda$; an average of sums of squares is
again a sum of squares.
By Theorem~\ref{thm:spinflip-ppt-hierarchy}, the complete three-copy
Fourier blocks contain both kinds of Gram forms, with no omitted
local representation sector.

Polynomial equality on every $z$ implies equality of the compressed
Hermitian coefficient operators on $\operatorname{Sym}^3H$: the
expectations in all $z^{\otimes3}$ determine such an operator by
its polynomial coefficients. Thus the twirled identity holds on
any feasible averaged moment, not only on physical pure powers.
The argument uses three properties of the moment: its twirled operator $X$ is positive semidefinite on every
$\omega$-block of the $\operatorname{Sym}^3H$ sector, the flipped image $\mathcal T(X)$ is positive semidefinite on
every $\gamma$-block of the $\operatorname{Sym}^2H\otimes H$ sector, and its word values satisfy the one-site
antisymmetriser relations, so that the pairing is well defined. In the isotypic decomposition of the local
three-copy algebra the compressed blocks $\omega_1,\dots,\omega_7$ and $\gamma_0,\dots,\gamma_7$ have sizes $(1,1,1,3,5,11,21)$ and
$(1,1,2,4,8,16,32,64)$.
The boundary pseudo-moment $m^*/70416$ of Theorem~\ref{thm:threecopy} has positive semidefinite blocks,
satisfies the relations and has objective value $125/978$, so it evaluates the left side as
\[
 F_\lambda(m^*/70416)=\frac{125}{978}-\lambda,
\]
while every positive Gram form pairs nonnegatively with it. An SOS representation therefore forces
$\lambda\le125/978$.
Conversely, the TCB dual certificate gives an SOS at $125/978$: its positive
$\omega$ and $\gamma$ Gram forms factor into cubic squared moduli,
and its antisymmetriser relations vanish identically.
\end{proof}

\noindent\emph{Remark.} The relations have exact rank $23$ (TCB); the $37-23=14$ free coordinates match
the independent invariant count from~\eqref{eq:fqe-invariant-dimension} (HI). This count is not
used in the proof.

\section{Exact minima of the balanced purity}\label{sec:lp}

For a pure state $\psi$ of $n$ qubits and $k=\lfloor n/2\rfloor$ write
$\pi_k(\psi)=\binom nk^{-1}\sum_{|S|=k}\tr\rho_S^2$ for the mean balanced-cut purity. Since
$\tr\rho_S^2\ge2^{-k}$, $\pi_k\ge2^{-k}$, with equality iff $\psi$ is absolutely maximally entangled
(AME$(n,2)$), which happens only for $n=2,3,5,6$ (Rains~\cite{Rains} and Scott~\cite{Scott} for $n\ge8$;
Huber--G\"uhne--Siewert~\cite{HGS} for $n=7$). This section shows that for $n\le18$, $n\ne7,13$,
the minimum of $\pi_k$ is the value of the Rains linear programme and is attained by a graph state
(Table~\ref{tab:minima}; $n=18$ is Theorem~\ref{thm:eighteen}).

\begin{table}[htbp]
\centering
\begin{tabular}{@{}cccl@{}}
\toprule
$n$ & $2^{-k}$ & $\min\pi_k$ (the Rains LP value) & attained by \\
\midrule
4 & $1/4$ & $1/3$ & cluster ($P_4$) or Higuchi--Sudbery state\\
5 & $1/4$ & $1/4$ & AME(5,2), e.g.\ $C_5(1)$\\
6 & $1/8$ & $1/8$ & AME(6,2), e.g.\ $C_6(2,3)$\\
8 & $1/16$ & $3/35$ & graph state (Table~\ref{tab:graphs})\\
9 & $1/16$ & $1/14$ & $C_9(3,4)$\\
10 & $1/32$ & $1/21$ & $C_{10}(1,2,4)$ (Remark~\ref{rem:ten})\\
11 & $1/32$ & $1/28$ & graph state (Table~\ref{tab:graphs})\\
12 & $1/64$ & $5/224$ & $C_{12}(3,5,6)$\\
14 & $1/128$ & $331/27456$ & graph state (Theorem~\ref{thm:fourteenfifteen})\\
15 & $1/128$ & $1/104$ & graph state (Theorem~\ref{thm:fourteenfifteen})\\
16 & $1/256$ & $1/156$ & induced subgraph of $C_{17}(1,3,4,5)$\\
17 & $1/256$ & $1/208$ & $C_{17}(1,3,4,5)$\\
18 & $1/512$ & $7/2288$ & graph state (Theorem~\ref{thm:eighteen})\\
\bottomrule
\end{tabular}
\caption{Minimum of $\pi_k$ for $n\le18$ qubits, $n\ne7,13$ (for $n=7$ the minimum $121/917$ is
Corollary~\ref{cor:seven-global-from-fivecopy}). $C_n(D)$ is the circulant graph joining $i$ and $i\pm d$ for $d\in D$.}
\label{tab:minima}
\end{table}

Write $A_j=\sum_{\mathrm{wt}(P)=j}\langle\psi|P|\psi\rangle^2$ for the weight distribution over the
$4^n$ Pauli strings $P$ ($A_0=1$, $\sum_jA_j=2^n$). Expanding $\rho_S$ in Pauli strings gives
\begin{equation}\label{eq:as}
  a_s := \binom ns^{-1}\sum_{|S|=s}\tr\rho_S^2 = \frac{1}{2^s\binom ns}\sum_{j\le s}\binom{n-j}{s-j}A_j,
\end{equation}
and for pure states $a_s=a_{n-s}$. Rains' shadow inequalities~\cite{Rains} state that the coefficients
$S_j$ of $S(x,y)=\sum_jA_j\bigl(\tfrac{x+3y}{2}\bigr)^{n-j}\bigl(\tfrac{y-x}{2}\bigr)^{j}$ are
non-negative. The \emph{Rains LP} minimises $a_k$ over $A\in\mathbb R^{n+1}_{\ge0}$ with $A_0=1$,
$a_s=a_{n-s}$ and $S_j\ge0$.

\begin{theorem}[exact certificate; LP-1, GS-1]\label{thm:lptable}
\begin{enumerate}
  \item For $n\in\{4,5,6,8,9,10,11,12\}$, $\pi_k(\psi)$ is at least the value in Table~\ref{tab:minima} for every pure $n$-qubit state.
  \item For $n\in\{4,5,6,8,9,10,11,12\}$ the graph state of Table~\ref{tab:graphs} has $\pi_k$ equal to
    the LP bound. Hence $\min\pi_k$ equals the LP bound for these $n$.
\end{enumerate}
\end{theorem}

\begin{proof}
(1) For each $n$, LP-1 exhibits rational multipliers $e_s\in\mathbb Q$, $y_i\ge 0$ such that
$a_k - L A_0 - \sum_s e_s(a_s-a_{n-s}) - \sum_i y_iS_i$, as a linear form in $A$, has non-negative
coefficients. On any feasible $A$ the right-hand terms vanish or are non-negative, so $a_k\ge L$.
(Part~(2) shows that $L$ is the LP value.)
(2) For the graph state $|G\rangle$ and a cut $S$, $\tr\rho_S^2 = 2^{-\operatorname{rk}_2\Gamma[S,S^c]}$,
with $\Gamma[S,S^c]$ the biadjacency matrix over $\mathbb F_2$ (Hein--Eisert--Briegel~\cite{HEB}).
GS-1 computes all ranks exactly. \qedhere
\end{proof}

\begin{table}[htbp]
\centering
\begin{tabular}{@{}c>{\raggedright\arraybackslash}p{8cm}l@{}}
\toprule
$n$ & graph & ranks over all $k$-sets $S$\\
\midrule
4 & path $P_4$ (linear cluster) & $1^{2}\,2^{4}$ \\
5 & $C_5(1)$ & $2^{10}$ \\
6 & $C_6(2,3)$ & $3^{20}$ \\
8 & edges 20 21 31 32 41 43 50 51 52 54 60 61 63 64 70 71 76 & $2^6\,3^8\,4^{56}$ \\
9 & $C_9(3,4)$ & $3^{18}\,4^{108}$ \\
10 & $C_{10}(1,2,4)$ & $4^{132}\,5^{120}$ \\
11 & edges 20 21 31 40 41 42 51 53 54 60 61 63 64 70 71 76 81 83 84 85 86 87 92 93 96 98 $\{10\}$0--6 $\{10\}$9 & $4^{66}\,5^{396}$\\
12 & $C_{12}(3,5,6)$ & $4^{6}\,5^{378}\,6^{540}$ \\
\bottomrule
\end{tabular}
\caption{Graph states attaining the LP bound ($r^m$: $m$ cuts of rank $r$). For $n=11$,
``$\{10\}$0--6'' means vertex 10 is joined to $0,\dots,6$.}
\label{tab:graphs}
\end{table}

\begin{remark}[ten qubits]\label{rem:ten}
The $C_{10}(1,2,4)$ graph state is $3$-uniform, with weight distribution
$A=(1,0,0,0,10,72,160,240,285,200,56)$ (GS-2): a $[[10,0,4]]$ stabilizer state. By \eqref{eq:as},
$\pi_5$ depends on $A_4$ and $A_5$ as well as the distance, so not every $[[10,0,4]]$ state attains
$1/21$: the Petersen graph state has $\pi_5=3/56$ (GS-3). Scott's Table~I~\cite{Scott} lists only
the Petersen weight distribution for $n=10$ (value $3/56$, with a caption noting that other
distributions exist); Zha et al.~\cite{Zha14} give $13/336$ as a heuristic ``theoretical
limitation''. The distribution of $C_{10}(1,2,4)$ is the extremal Type~I enumerator $W_{10,1}$ of
Danielsen--Parker~\cite{DP}, so the value $1/21$ is implicit in published data. We have found no earlier
statement that $\min\pi_5=1/21$.
\end{remark}

\begin{theorem}[exact finite certificates; LP-14, LP-15, GS-14, GS-15]
\label{thm:fourteenfifteen}
The minimum mean balanced-cut purity over all pure qubit states is
\[
 \min_{\psi\in(\mathbb C^2)^{\otimes14}}\pi_7(\psi)=\frac{331}{27456},
 \qquad
 \min_{\psi\in(\mathbb C^2)^{\otimes15}}\pi_7(\psi)=\frac1{104}.
\]
Both minima are attained by graph states.
\end{theorem}
\begin{proof}
For each $n=14,15$, the checks LP-14 and LP-15
verify, coefficient by
coefficient in all $n+1$ Pauli-weight coordinates, the exact rational
Rains-LP dual identity
\[
 a_{\lfloor n/2\rfloor}(A)-L A_0
  =\sum_{s< n/2}e_s(a_s-a_{n-s})
      +\sum_{i=0}^n y_i S_i(A)+\sum_{j=0}^n w_j A_j,
       \quad y_i,w_j\ge0,
\]
where the finite lists of rational coefficients are supplied literally
in the script. This proves the universal lower bounds.

For the graph witnesses with symmetric binary adjacency matrices
specified by the hexadecimal row strings in that script, exact
enumeration of the graph stabilizer gives, respectively,
\[
 \begin{array}{c|c|l}
 n & (A_0,\ldots,A_7) & \text{balanced-cut binary rank histogram}\\ \hline
 14 &(1,0,0,0,0,0,161,576)&5^{60}\,6^{1684}\,7^{1688}\\
 15 &(1,0,0,0,0,0,105,540)&5^{35}\,6^{1380}\,7^{5020}.
 \end{array}
\]
The 14- and 15-vertex graphs have 47 and 56 edges. An independent
enumeration of every $7|(n-7)$ binary cut matrix computes its rank
$r_S$, yielding $\operatorname{tr}\rho_S^2=2^{-r_S}$. Averaging the rank
histograms gives exactly $331/27456$ and $1/104$, matching the
certified lower bounds.
\end{proof}

\begin{theorem}[exact finite certificates; GL-1, GL-2]
\label{thm:sixteenseventeen}
The minimum mean balanced purity is $1/156$ for $n=16$ and $1/208$
for $n=17$. It is attained by the graph state $C_{17}(1,3,4,5)$ at
$n=17$, and by its induced graph on vertices $0,\dots,15$ at $n=16$.
\end{theorem}
\begin{proof}
GL-1 verifies exact Rains-LP dual identities with nonnegative rational
shadow and Pauli-weight coefficients, proving the stated universal
lower bounds. GL-2 computes every balanced cut rank over $\mathbb F_2$.
For $n=16$ the ranks are $5^{10}6^{500}7^{6680}8^{5680}$;
for $n=17$ they are $6^{510}7^{4080}8^{19720}$.
Inserting $2^{-r}$ for each cut of rank $r$ yields $1/156$ and
$1/208$ respectively, matching the lower bounds.
\end{proof}

\section{An exact eighteen-qubit balanced-purity minimum}
\label{sec:eighteen}

\begin{theorem}[exact Rains dual and graph certificate; E18-1--E18-4]
\label{thm:eighteen}
For every normalised pure $18$-qubit state,
\[
 \boxed{\quad\pi_9(\psi)\ge\frac7{2288},\qquad
 \min_\psi\pi_9(\psi)=\frac7{2288}.\quad}
\]
Equality is attained by the graph state whose eighteen
adjacency rows (vertex $0$ is the lowest bit, in hexadecimal)
are
\[
\begin{gathered}
\texttt{36e72 0fb7d 1b80a 0cad6 25e0b 30583}\\
\texttt{1c10b 2b228 1a262 3999b 27031 3221f}\\
\texttt{2a696 1dd87 2a45b 273ce 22b65 1deb1}.
\end{gathered}
\]
Its entire graph-state Pauli weight distribution is
\[
 A_0=1,\quad
 (A_8,A_{10},A_{12},A_{14},A_{16},A_{18})
 =(2754,18360,77112,110160,50949,2808)
\]
and all other weights vanish. Its $9|9$ cut-rank
distribution is $7^{9180}9^{39440}$.
The graph is of Type II (all vertex degrees odd) and has
global spin-flip overlap $\tau_{18}=1$.
\end{theorem}
\begin{proof}
E18-1 verifies the rational dual identity for the
18-qubit Rains LP, with constant $7/2288$.
Specifically, using $a_s=a_s(A)$ and shadow coefficients
$S_i(A)$, it verifies coefficientwise
\[
 a_9-\tfrac7{2288}A_0
 =\sum_{s=0}^{8}e_s(a_s-a_{18-s})
 +\tfrac1{1166880}S_4
 +\tfrac1{1633632}S_6
 +\tfrac1{2498496}S_{11}
 +\sum_{j=0}^{18}w_jA_j
\]
with $w_j\ge0$ and the rational multipliers
\[
 (e_0,\ldots,e_8)=
 \left(\tfrac5{4004},-\tfrac{291}{2002},\tfrac{81}{364},
 \tfrac{278}{1001},-\tfrac{333}{1001},0,
 -\tfrac{101}{143},-\tfrac{54}{143},\tfrac{51}{26}\right).
\]
For a pure state the complementarity terms vanish and
every remaining term is nonnegative.

E18-2 checks the displayed binary graph is simple,
symmetric and has odd degrees; exhaustive enumeration of
its $2^{18}$ stabilizers gives the displayed weight
distribution, corresponding to an extremal Type II
self-dual additive $GF(4)$ code of minimum weight eight.
E18-3 computes every one of the $\binom{18}{9}=48620$
binary cut ranks, obtaining $9180$ cuts of rank $7$
and $39440$ of rank $9$. Thus
\[
 \pi_9=
 \frac{9180/128+39440/512}{48620}=\frac7{2288}.
\]
E18-4 independently checks the same purity from the
Pauli weight formula. The matching lower and upper
bounds prove the theorem.
\end{proof}

\begin{remark}[relationship to the known quaternary $S_{18}$ code]
An extremal Hermitian self-dual quaternary
$[18,9,8]$ code is classical and well documented
(see Cheng--Sloane and Harada--Munemasa's
\href{https://www.math.is.tohoku.ac.jp/~mharada/research/codes/hsd4.htm}
{code database}).
The graph above is a local-Clifford graph form of a
generator in that database. E18-1--E18-4 make the
purity/cut-rank attainment claim entirely reproducible
without downloading the external generator.
We do not claim novelty for the underlying code or
the existence of an extremal $[18,9,8]$ code.
The equality between its purity and the Rains-LP
lower bound is the conclusion here.
\end{remark}

\begin{corollary}[further sharp time-reversal-symmetric minima;
TY-1--TY-2]\label{cor:typeII-sixteen-twenty}
Inside the full spin-flip-invariant marginal class,
\[
 \boxed{\quad
 \min_{\tau_{16}=1}\pi_8=\frac1{143},\quad
 \min_{\tau_{18}=1}\pi_9=\frac7{2288},\quad
 \min_{\tau_{20}=1}\pi_{10}=\frac{131}{77792}.
 \quad}
\]
For $n=16$ and $20$ the respective attaining graph
states are the circulants $C_{16}(2,3,4,8)$ and
$C_{20}(3,4,5,10)$. Their balanced-cut rank
distributions are
\[
\begin{array}{c|l|c}
n & \text{ranks} & \pi_{n/2}\\\hline
16&6^{1096}7^{6882}8^{4892}&1/143\\
20&8^{12312}9^{96900}10^{75544}&131/77792.
\end{array}
\]
In particular the positive spin-flip coefficients
$c_{16}=1/1716$ and $c_{20}=103/1478048$
in Theorem~\ref{thm:purity-spinflip-tradeoff} are
\emph{sharp with their stated constant terms}.
The $n=20$ constant $1193/739024$ remains only a
global Rains lower bound, not a proven global minimum.
\end{corollary}
\begin{proof}
Theorem~\ref{thm:purity-spinflip-tradeoff} at
$\tau_n=1$ gives the lower values $1/143$
and $131/77792$ for $n=16,20$. TY-1--TY-2
check all $2^n$ graph stabilizers and all
balanced-cut ranks, with exact fraction arithmetic.
Each circulant has odd degree at every vertex, hence
$\tau_n=1$. Theorem~\ref{thm:eighteen} handles
$n=18$ and also exhibits a $\tau_{18}=1$ minimiser.
A stronger coefficient at $16$ or $20$ would fail
on the respective circulant witness, proving its
sharpness.
\end{proof}

\section{Sharp purity--time-reversal tradeoffs}\label{sec:spinflip-tradeoffs}

At four qubits, $\pi_2=\tfrac13+\tfrac16(W_1+\tau_4)$ (Lemma~\ref{lem:fqe-purity}): a large time-reversal overlap $\tau_4$ costs purity.
This section proves the corresponding inequality for larger even $n$, using the exact minima of Section~\ref{sec:lp}. The paper's Discussion explains the motivation. For odd $n$, $\tau_n$ vanishes identically; Section~\ref{sec:odd-spinflip} gives the odd-$n$ tradeoff.

For a normalised pure even-$n$ qubit vector $\psi$ let
\[
 \tau_n(\psi)=|\langle\psi,J_n\psi\rangle|^2=|\psi^T\sigma_y^{\otimes n}\psi|^2,
 \qquad J_n=\epsilon^{\otimes n}K,
\]
so that $\tau_4$ is the four-tangle $\tau$ of Lemma~\ref{lem:fqe-purity}, and let
$\pi_{n/2}$ be the mean purity of the $\binom n{n/2}$ balanced marginals. The Pauli weight distribution $A_j$, the mean
$s$-site purities $a_s$ (so $\pi_{n/2}=a_{n/2}$, and $a_s=a_{n-s}$ for pure states) and Rains' non-negative shadow coefficients $S_i$
are as in Section~\ref{sec:lp}. The new ingredient is the identity
$\tau_n=2^{-n}\sum_j(-1)^jA_j$, since $\sum_j(-1)^jA_j=2^n\operatorname{tr}(\rho\,J_n\rho J_n^{-1})$. The inequalities below hold on the
\emph{unrestricted} pure state space: no spin-flip symmetry is imposed.

\begin{theorem}[exact certificates; ST-1--ST-6]
\label{thm:purity-spinflip-tradeoff}
For each $n$ in the table below, every normalised pure $n$-qubit state satisfies
\[
 \boxed{\qquad
 \pi_{n/2}(\psi)\ge L_n+c_n\tau_n(\psi),\qquad}
\]
with
\[
\begin{array}{c|cccccc}
n&4&8&10&14&16&20\\\hline
L_n&1/3&3/35&1/21&331/27456&1/156&1193/739024\\
c_n&1/6&1/70&1/168&5/6864&1/1716&103/1478048
\end{array}
\]
For each $n$, $c_n$ is the largest coefficient compatible with the constant $L_n$: a graph state with
every vertex of odd degree has $\tau_n=1$ and $\pi_{n/2}=L_n+c_n$.
\end{theorem}
\begin{proof}
The checks ST-1--ST-6 supply rational multipliers $e_s$,
$y_i\ge0$ and residual coefficients $w_j\ge0$ satisfying the exact coefficientwise identity
\begin{equation}\label{eq:tradeoff-dual}
 a_{n/2}(A)-L_nA_0-
 c_n\,2^{-n}\sum_{j=0}^n(-1)^jA_j
 =\sum_{s< n/2}e_s(a_s-a_{n-s})
  +\sum_{i=0}^n y_iS_i+\sum_{j=0}^n w_jA_j
\end{equation}
of linear forms in $A$ (the $s=0$ term encodes the normalisation $a_0=A_0=1=2^{-n}\sum_jA_j=a_n$).
For a pure state $A_0=1$, $a_s=a_{n-s}$ and $A_j,S_i\ge0$, and the left side is
$\pi_{n/2}-L_n-c_n\tau_n$; this gives the inequalities. ST-1--ST-6 check every coefficient and sign in
rational arithmetic, with no numerical solver.

For sharpness, let $G$ be a graph whose vertices all have odd degree. The product of the stabilizer
generators $X_vZ_{N(v)}$ of $|G\rangle$ is, up to a phase, $X^{\otimes n}Z^{\otimes n}$, a phase times
$\sigma_y^{\otimes n}$, so $|G\rangle$ is an eigenvector of $\sigma_y^{\otimes n}$; as $|G\rangle$ is
real, $\tau_n=|\langle G|\sigma_y^{\otimes n}|G\rangle|^2=1$. ST-1--ST-6 compute $\pi_{n/2}$ exactly
for such a graph at each $n$ and find $L_n+c_n$: at $n=4,8,10,14$ the spin-flip extension of a graph on
$n-1$ vertices, at $n=16,20$ the circulants $C_{16}(2,3,4,8)$ and
$C_{20}(3,4,5,10)$.
\end{proof}

For $n\le16$ the constant $L_n$ is the minimum of $\pi_{n/2}$ over pure states, attained by a graph
state (Theorems~\ref{thm:lptable}, \ref{thm:fourteenfifteen}
and~\ref{thm:sixteenseventeen}); at $n=20$ it is an exact
lower bound whose attainment is open.

\begin{corollary}[time-reversal breaking and stability]
\label{cor:spinflip-rigidity}
For $n\in\{4,8,10,14,16\}$ every minimiser of $\pi_{n/2}$ has $\langle\psi,J_n\psi\rangle=0$, and for
$n\in\{4,8,10,14,16,20\}$ every pure state with $\pi_{n/2}(\psi)\le L_n+\delta$ has
$\tau_n(\psi)\le\delta/c_n$.
\end{corollary}
\begin{proof}
Each $c_n>0$; subtract $L_n$ in Theorem~\ref{thm:purity-spinflip-tradeoff}.
\end{proof}

\begin{remark}[four qubits and the parity-preserving sizes]
At $n=4$ the inequality is the identity $\pi_2=\tfrac13+\tfrac16\bigl(\sum_i\|\mathbf r_i\|^2+\tau_4\bigr)$
of Lemma~\ref{lem:fqe-purity}, with the Bloch-vector term dropped. For larger $n$ the certificates
come from Rains' shadow inequalities, not from an analogue of the four-qubit sextic remainder, and they
exclude time-reversal-symmetric minimisers without any graph or conference-matrix ansatz. At $n=6$
and $n=12$ the known balanced-purity minima are attained by states with $\tau_n=1$, so no positive
coefficient can accompany the sharp constant there.
\end{remark}

\subsection{Odd $n$: the time-reversal overlap on single-qubit complements}\label{sec:odd-spinflip}

For odd $n$ every pure state has $\tau_n=0$, but the time-reversal overlap survives on reduced states.
For a $k$-qubit state $\rho$ put $\tilde\rho=\sigma_y^{\otimes k}\bar\rho\,\sigma_y^{\otimes k}$ and $R_\rho=\operatorname{tr}(\rho\tilde\rho)$, the spin-flip overlap of
Serrano-Ens\'astiga, Giraud and Martin~\cite{SGM} (Lemma~\ref{lem:odd-sector} below is the odd-sector companion of their even-sector identity $\operatorname{tr}\rho^2+R_\rho=2^{1-k}\sum_{\mathrm{wt\ even}}\langle P\rangle^2$).
For a pure state $\psi$ of $n$ qubits write $\rho_{\bar s}$ for the $(n-1)$-qubit reduction without qubit $s$, and
\[
 O(\psi)=\frac1n\sum_{s=1}^n\tfrac12\bigl(\operatorname{tr}\rho_{\bar s}^2-R_{\rho_{\bar s}}\bigr),
\]
the mean \emph{T-odd purity} of the single-qubit complements. As before $\pi_m$ is the mean purity of the $\binom nm$ marginals on
$m$ qubits, and $A_j$ and $a_s$ are as above.

\begin{lemma}[sector formulas; proved]\label{lem:odd-sector}
For a state $\rho_S$ on a set $S$ of $k$ qubits with Pauli expectations $\langle P\rangle$,
$\operatorname{tr}\rho_S^2=2^{-k}\sum_{P\subseteq S}\langle P\rangle^2$ and
$\tfrac12(\operatorname{tr}\rho_S^2-R_{\rho_S})=2^{-k}\sum_{P\subseteq S,\ \mathrm{wt}P\ \mathrm{odd}}\langle P\rangle^2$.
Hence $a_s=\sum_jA_j\binom{n-j}{s-j}/(2^s\binom ns)$ and, for odd $n$,
$O=\sum_{j\ \mathrm{odd}}\dfrac{n-j}{n\,2^{n-1}}A_j$.
\end{lemma}
\begin{proof}
The first formula is the Pauli expansion $\rho_S=2^{-k}\sum_P\langle P\rangle P$. For each of $\sigma=X,Y,Z$ one has $\sigma_y\bar\sigma\sigma_y=-\sigma$, so
$\sigma_y^{\otimes k}\bar P\sigma_y^{\otimes k}=(-1)^{\operatorname{wt}P}P$ and
$\tilde\rho_S=2^{-k}\sum_P(-1)^{\operatorname{wt}P}\langle P\rangle P$; therefore
$R_{\rho_S}=2^{-k}\sum_P(-1)^{\operatorname{wt}P}\langle P\rangle^2$. A weight-$j$ Pauli word lies in $n-j$ of the $n$ complements
$\bar s$ and in $\binom{n-j}{s-j}$ of the $s$-subsets.
\end{proof}

\begin{theorem}[odd-$n$ tradeoff; proved, exact checks ST-7, ST-8]\label{thm:odd-spinflip-tradeoff}
Let $n=2m+1\ge5$ be odd, and put $c_5=1$ and $c_n=(n+1)/4$ for $n\ge7$. Every normalised pure $n$-qubit state satisfies the identity
\begin{equation}\label{eq:odd-tradeoff}
 \pi_m(\psi)+c_n\,O(\psi)=\frac12+\sum_{j\ \mathrm{odd},\,j<n}w_j^{(n)}A_j(\psi),
\end{equation}
where, with $\binom ab=0$ for $b<0$ or $b>a$,
\begin{equation}\label{eq:odd-tradeoff-weights}
 w_j^{(n)}=\frac{2^m\bigl[\binom m{j-1}+3\binom mj\bigr]+(-1)^{(n-j)/2}\binom m{(n-j)/2}+2c_n\binom{n-1}j-\binom nj}{2^n\binom nj}.
\end{equation}
For example $\pi_2+O=\tfrac12+\tfrac15A_1$ at $n=5$,
$\pi_3+2O=\tfrac12+\tfrac3{28}A_1+\tfrac3{140}A_3$ at $n=7$, and
$\pi_4+\tfrac52O=\tfrac12+\tfrac5{96}A_1+\tfrac5{448}A_3+\tfrac{11}{4032}A_5$ at $n=9$. The weights are positive on the odd $j\le n-4$
and zero for $j=n-2,n$. Consequently
\[
 \pi_m(\psi)+c_n\,O(\psi)\ \ge\ \frac12 ,
\]
with equality if and only if $A_j(\psi)=0$ for every odd $j\le n-4$. The bound is attained by the GHZ state for every
odd $n$. In particular, at $n=5$ equality holds exactly when all one-qubit marginals are maximally mixed.
\end{theorem}
\begin{proof}
Put $B_j=A_j/\binom nj$. By Lemma~\ref{lem:odd-sector}, $a_s=2^{-s}\sum_j\binom sjB_j$ and $O=2^{1-n}\sum_{j\ \mathrm{odd}}\binom{n-1}jB_j$.
For a pure state $A_0=1$ and $a_s=a_{n-s}$, so it suffices to find $e_s$ with
\[
 a_m+c\,O-\tfrac12A_0-\sum_{s=2}^me_s(a_s-a_{n-s})=\sum_{j\ \mathrm{odd}}w_jA_j
\]
as an identity of linear forms in $B$, for $c=c_n$. Multiply the coefficient of $B_j$ by $y^j$, sum over $j$, and put $x=(1+y)/2$. Since
$\sum_j2^{-s}\binom sjy^j=x^s$ and $2^{1-n}\sum_{j\ \mathrm{odd}}\binom{n-1}jy^j=\tfrac12\bigl(x^{n-1}-(1-x)^{n-1}\bigr)$, the left side becomes
$x^m+\tfrac c2(x^{n-1}-(1-x)^{n-1})-\tfrac12-\sum_se_s(x^s-x^{n-s})$. On the right, the four parts of the numerator
of~\eqref{eq:odd-tradeoff-weights} sum over odd $j$ to the odd parts in $y$ of $2^m(3+y)(1+y)^m$, $y(y^2-1)^m$, $2c(1+y)^{n-1}$ and
$-(1+y)^n$. In $x$, after division by $2^n$, these are $\tfrac12[(1+x)x^m-(2-x)(1-x)^m]$, $\tfrac12(-1)^m(2x-1)x^m(1-x)^m$,
$\tfrac c2(x^{n-1}-(1-x)^{n-1})$ and $-\tfrac12(x^n-(1-x)^n)$. The terms in $c$ cancel, for every $c$, and the identity becomes
$2\sum_{s=2}^me_s(x^s-x^{n-s})=Q(x)$, where
\[
 Q(x)=x^m-x^{m+1}-1+x^n+(2-x)(1-x)^m-(1-x)^n-(-1)^m(2x-1)x^m(1-x)^m .
\]
Each of the pairs $\{x^m,-x^{m+1}\}$ and $\{-1,x^n\}$ and the remaining three terms together change sign under
$Q(x)\mapsto x^nQ(1/x)$ (for the last group use $(x-1)^n=-(1-x)^n$). Hence $Q=\sum_{s\le m}q_s(x^s-x^{n-s})$ with $q_s=[x^s]Q$, and
reading off coefficients for $s\le m$,
\[
 q_s=(-1)^s\Bigl[2\tbinom ms+\tbinom m{s-1}-\tbinom ns\Bigr]-\delta_{s0}+\bigl(1+(-1)^m\bigr)\delta_{sm},
\]
so $q_0=q_1=0$ (as $n=2m+1$) and $e_s=q_s/2$ proves~\eqref{eq:odd-tradeoff}.

Let $N_j=2^n\binom njw_j$ and $k=(n-j)/2$. At $j=n$, $N_n=0+1+0-1=0$. At $j=n-2$, $N_3=4-2+8-10=0$ for $n=5$, and for
$m\ge3$ both binomials in the first term vanish and $2c_n=m+1$, so $N_{n-2}=-m+(m+1)2m-m(2m+1)=0$.
For $j\le n-4$ ($k\ge2$, so $n\ge7$), $(m+1)\binom{n-1}j-\binom nj=\tfrac{2k(m+1)-n}n\binom nj$, which exceeds $\binom nj$ because
$2k(m+1)-n\ge4(m+1)-n=2m+3>n$; and
$\binom nj=\binom n{2k}\ge\binom mk$ (send a $k$-subset $T$ of $\{1,\dots,m\}$ to $T\cup(T+m)$). So $N_j>0$, and for $n=5$, $j=1$ directly
$N_1=28+1+8-5=32$. The equality statement follows because $A_j\ge0$. For GHZ$_n=(|0\rangle^{\otimes n}+|1\rangle^{\otimes n})/\sqrt2$, a Pauli word with an $X$ or
$Y$ on some but not all sites has zero expectation, and a $Z$-word of weight $j<n$ has $\langle P\rangle\ne0$ only if $j$ is even, so
$A_j=0$ for odd $j<n$. ST-7 checks~\eqref{eq:odd-tradeoff-weights} and the multipliers $e_s$ against an independent exact
elimination for odd $n\le41$; ST-8 evaluates both sides of \eqref{eq:odd-tradeoff} exactly on Gaussian-integer states at $n=5,7,9$
(three random states, GHZ, the ring graph state and $W$) from the reduced density matrices and the Pauli expectations, without
Lemma~\ref{lem:odd-sector}.
\end{proof}

\begin{remark}[interpretation]
The inequality is the odd-$n$ counterpart of
Theorem~\ref{thm:purity-spinflip-tradeoff}: a small balanced purity forces a large mean T-odd purity, $O\ge(\tfrac12-\pi_m)/c_n$, with $1/c_n=4/(n+1)$ for $n\ge7$.
The coefficient $c_n$ is the least for which~\eqref{eq:odd-tradeoff}, with the same multipliers $e_s$, has nonnegative weights:
$O$ carries the weight $(n-j)/(n2^{n-1})>0$ on $A_j$, so lowering $c_n$ makes $w_{n-2}^{(n)}=0$ negative. It is also optimal over all pure states. For
$\Psi_n=\mathrm{GHZ}_{n-2}\otimes\Phi^+$ the only odd Pauli weights are $A_{n-2}=2^{n-3}$ (the full-support $X/Y$ words of
the GHZ factor) and $A_n=3\cdot2^{n-3}$ (those words times the three Bell correlations), so the residual
of~\eqref{eq:odd-tradeoff} vanishes, while $O=1/(2n)>0$ by Lemma~\ref{lem:odd-sector}. Any smaller coefficient therefore
fails at $\Psi_n$ (ST-9).
For example, at $n=7$ the hypothesis $\pi_3<19/140$, the purity of the best stabilizer states (Proposition~\ref{prop:stab7}), gives
$O>51/280$.
\end{remark}

\section{Thirteen qubits: purity interval and an entropy witness}\label{sec:sup-thirteen}
At thirteen qubits the Rains linear programme (level two of the hierarchy of Section~\ref{sec:hierarchy}) has value
$25/1408$, but a state attaining it would be a pure $((13,1,6))_2$ code, which does not exist. The three-copy level, in the
Schur--Weyl coordinates of Section~\ref{sec:threecopy-sw}, gives a lower bound strictly above the LP value; measuring one
qubit of the fourteen-qubit graph state of Theorem~\ref{thm:fourteenfifteen} gives an upper bound
(Proposition~\ref{prop:thirteeninterval}). The exact minimum remains open. A graph-code superposition built on the same
witness gives $\max E_6>5.8336$ (Corollary~\ref{cor:thirteen-entropy-witness}). The paper uses only these
thirteen-qubit statements and Theorem~\ref{thm:threecopy-sw}(a); the other sizes are recorded with the same certificates.
\subsection{The three-copy level for all $n$: Schur--Weyl form}\label{sec:threecopy-sw}

The orbit coordinates of \S\ref{sec:threecopy} grow exponentially in $n$. Schur--Weyl duality for the
site symmetry reduces level $m=3$ of Theorem~\ref{thm:spinflip-ppt-hierarchy} to $O(n^3)$ real
parameters and blocks of size $O(n)$, with every step exact over $\mathbb Q$
(\texttt{three\_copy\_schur\_weyl.py}).

\paragraph{Reduction.} Per site, $\mathcal A_3=\operatorname{End}(S^{[3]})\oplus\operatorname{End}(S^{[21]})
\cong\mathbb C\oplus M_2$, with $S^{[3]}=\langle|000\rangle\rangle$ and
$S^{[21]}=\langle b_1=|100\rangle-|010\rangle,\ b_2=|010\rangle-|001\rangle\rangle$ (Gram matrix
$G_0=\bigl(\begin{smallmatrix}2&-1\\-1&2\end{smallmatrix}\bigr)$), on which global copy permutations act
by the standard representation $\rho$ of $S_3$. Sort the sites by type: $a$ sites in the $[21]$ sector.
The site-symmetrised twirl $\Omega$ of $(zz^*)^{\otimes3}$ lies in
$\bigoplus_a\bigl(M_2^{\otimes a}\bigr)^{S_a}$, and by Schur--Weyl duality for $GL_2$,
$\bigl(M_2^{\otimes a}\bigr)^{S_a}=\bigoplus_{\nu}\operatorname{End}(U_\nu)$ over two-row partitions
$\nu=(a-q,q)$, with $U_\nu=\operatorname{Sym}^{a-2q}\otimes{\det}^{q}$ and multiplicity
$f_\nu=\binom aq-\binom a{q-1}$. Global copy symmetry confines the block on $U_\nu$ to the
$\rho(S_3)$-invariant vectors $w_1,\dots,w_r$ of $U_\nu$, so $\Omega$ is the family
$X_{a,\nu}=\sum_{k,l}M_{kl}|w_k\rangle\langle w_l|$ with real symmetric $M=M_{a,\nu}\succeq0$
(the $\omega$-blocks; their entries are the parameters). For a test element $g=(g_0,g_1)\in\mathbb C\oplus M_2$
of $\mathcal A_3$,
\[
 P(g)=\tr\bigl[\Omega\,g^{\otimes n}\bigr]
 =\sum_{a=0}^n\binom na g_0^{\,n-a}\sum_{\nu}f_\nu\,\tr\bigl[X_{a,\nu}\,\rho_\nu(g_1)\bigr],
\]
and the normalisation and every mean purity $\pi_k$ are fixed linear functionals of the parameters:
$\sum_k\binom nk\pi_k\,t^k$ is $P$ evaluated on the identity plus $t$ times a copy transposition.
The flipped moment of $z\otimes z\otimes\tilde z$ has generating polynomial $P(T^*g)$, where $T^*$ is the
per-site spin-flip map on test elements (a fixed rational $5\times5$ matrix), and the same block form
with $S_2$-invariant vectors (copies $1,2$): the $\gamma$-blocks $\Gamma_{a,\nu}$. Matching the two
polynomial forms determines each $\Gamma_{a,\nu}$ as an exact rational linear function of the parameters
(the \emph{$\gamma$-map}); it is computed by isolating each type $a$ on rational points of the unit
circle and solving one square rational system per type. Level $m=3$ is then
\[
 L_3(n,k)=\min\Bigl\{\pi_k(M):\ \text{normalised},\ M_{a,\nu}\succeq0,\ \Gamma_{a,\nu}(M)\succeq0\Bigr\}.
\]
The sizes are small: $(14,12,18)$ parameters, $\omega$- and $\gamma$-blocks at $n=7$, $(68,41,53)$
at $n=13$, $(209,87,105)$ at $n=19$ and $(502,152,176)$ at $n=25$, with blocks of size at most $13$.
SW-1 checks the construction against genuine states: for a random integer state at $n=4,5$ the exact
generating polynomial lies in the parameter family, its flipped values equal those of the $\gamma$-map, and
its $\gamma$-blocks are positive semidefinite to floating tolerance (a regression, not a proof input); the Haar moment gives the Haar purity $(2^k+2^{n-k})/(2^n+1)$ exactly.

\begin{theorem}[three-copy bounds for all $n$; exact certificates SW-0--SW-4]\label{thm:threecopy-sw}
For every pure state of $n$ qubits:
\begin{enumerate}
  \item[(a)] $n=13$: \ $\pi_6\ \ge\ 18171249/10^9=0.018171249>25/1408$; \quad
             $n=19$: \ $\pi_9\ \ge\ 23947609/10^{10}=0.0023947609>93/38896$;
  \item[(b)] $n=13$: \ $\pi_5\ \ge\ 15699039/(5\cdot10^8)>1/32$; \quad
             $n=19$: \ $\pi_7\ \ge\ 78133177/10^{10}>1/128$;
  \item[(c)] $n=25$: \ $\pi_9\ \ge\ 4882879/(25\cdot10^8)=0.0019531516>1/512$.
\end{enumerate}
Consequently no pure $13$-qubit state is $5$-uniform, no pure $19$-qubit state is $7$-uniform and no
pure $25$-qubit state is $9$-uniform, that is, there is no pure $((13,1,6))_2$, $((19,1,8))_2$ or
$((25,1,10))_2$ quantum code.
\end{theorem}

\begin{proof}
For each case \texttt{data/three\_copy\_sw\_certificates.json} gives $\lambda$ and rational matrices
$Z_{a,\nu}$, which SW-2 proves positive definite. SW-3 forms the exact residual
$d=\pi_k-\lambda\cdot\mathrm{norm}-\sum\langle Z_{a,\nu},\Gamma_{a,\nu}(\cdot)\rangle$, a linear functional
of the parameters, and proves each of its $\omega$-blocks positive definite. For a genuine state the
parameters are normalised with $M_{a,\nu}\succeq0$ and $\Gamma_{a,\nu}\succeq0$, so
$\pi_k-\lambda=\sum\langle D_{a,\nu},M_{a,\nu}\rangle+\sum\langle Z_{a,\nu},\Gamma_{a,\nu}\rangle\ge0$.
Since $\tr\rho_S^2\ge2^{-k}$ for $|S|=k$, with equality only for a maximally mixed marginal, a
$k$-uniform state has $\pi_k=2^{-k}$; and a pure $((n,1,d))_2$ code is exactly a $(d-1)$-uniform
state. \qedhere
\end{proof}

\noindent\emph{Code-theoretic priority.}
The nonexistence of pure $((13,1,6))_2$ and $((25,1,10))_2$
codes is \emph{not new}. Huber's 2017 dissertation
\cite[\S5.11.4]{HuberThesis17} explicitly excludes those code
parameters, and $((31,1,12))_2$, by applying Rains's
shortening theorem for pure quantum codes and a linear
programming nonexistence bound for the shortened codes.
The later Terwilliger-algebra SDP certificates of
Angl\`es Munn\'e and Huber~\cite{AMH26}
independently exclude $((13,1,6))_2$ and
$((19,1,8))_2$.
Theorem~\ref{thm:threecopy-sw}(b,c) supplies a
\emph{different certificate family} and, importantly, the
explicit \emph{quantitative} inequalities
$\pi_5(13)>1/32$, $\pi_7(19)>1/128$, and
$\pi_9(25)\ge4882879/(25\cdot10^8)
=1/512+133/(5\cdot10^9)$.
These quantitative lower bounds concern \emph{all}
pure states, not just their exact uniformity status.
The three-copy SDP is tighter than the Rains LP for
the $25$-qubit $9$-body problem, but that fact does
not imply a new code-nonexistence theorem.
The seven-qubit three-copy numerical value is
$125/978>1/8$ (TCN-1, TA-1), compatible with the
known nonexistence of AME$(7,2)$.
The quantitative bounds in part (a) do not
require the previous code-nonexistence results.

\paragraph{Where three copies beat the Rains LP} (TA-1; numerical). The Rains LP is level
$L_2$ (Proposition~\ref{prop:rainsppt}). For $4\le n\le20$ and $k=\lfloor n/2\rfloor$, $L_3$ agrees with
$L_2$ to $10^{-12}$ except at $n=7$, where $L_3=0.1278118609407\ldots$ (agreeing with $125/978$ to
$10^{-16}$), at $n=13$, where $L_3=0.0181712542184\ldots$, and at $n=19$, where
$L_3=0.0023947628089\ldots$. For $n\le18$ every other $L_2$ is attained by a graph state
(Table~\ref{tab:minima} and Theorem~\ref{thm:eighteen}), so $L_3=L_2$ is forced there. The three exceptions are exactly the cases
$n\le20$ in which the LP optimum would require a nonexistent code: $((7,1,4))_2$, $((13,1,6))_2$ and
$((19,1,8))_2$ (\cite{HGS} and Theorem~\ref{thm:strictLP1319}). At $n=20$ level three gives no obstruction to the
LP equality profile, consistent with its open status; nor (numerically) at $n=23,24$ (at the uniformity levels $k=8,9$ of the putative codes, not $\lfloor n/2\rfloor$), while at $n=25$, $k=9$
it excludes the code (Theorem~\ref{thm:threecopy-sw}(c)): so far it obstructs exactly at $n\equiv1\pmod6$. The double-precision solver loses accuracy from
$n=15$; TA-1 uses an exactly re-centred multiprecision solve (\texttt{search\_three\_copy\_sw\_certificate.py}).

\subsection{The thirteen-qubit interval}\label{sec:sup-thirteen-interval}
\begin{proposition}[Product Pauli postselection convexity for stabilizer states]
\label{prop:pauliprojection}
Let $|\Omega\rangle$ be a pure $n$-qubit stabilizer state, and let
$M$ be a set of $t$ sites with $\rho_M=I/2^t$. On each $i\in M$ choose
an arbitrary pure vector $|u_i\rangle$ with Bloch coordinates
$r_i=(r_{i,x},r_{i,y},r_{i,z})$, and set
$|\psi_u\rangle=2^{t/2}(\bigotimes_{i\in M}\langle u_i|)\Omega\rangle$.
For each $a\in\{x,y,z\}^M$, let $\psi_a$ be the normalized state
obtained by projecting every $i\in M$ onto either eigenstate of
$\sigma_{a_i}$. Then, for every subset $T$ of the unmeasured sites,
\[
 \operatorname{tr}\rho_T(\psi_u)^2
  =\sum_{a\in\{x,y,z\}^M}
      \left(\prod_{i\in M}r_{i,a_i}^2\right)
      \operatorname{tr}\rho_T(\psi_a)^2.
\]
Thus the entire subsystem-purity profile after arbitrary product
postselection on $M$ is in the convex hull of the $3^t$ product-Pauli
profiles. No such product measurement can improve the minimum of any
fixed linear combination of these purities beyond its best product-Pauli
measurement. The one-site statement is the case $t=1$.
\end{proposition}
\begin{proof}
Expand the stabilizer density operator as
$|\Omega\rangle\langle\Omega|=2^{-n}\sum_{g\in\mathcal S}g$.
The normalized postselected density is
\[
 \rho_u=2^{-(n-t)}\sum_{g\in\mathcal S}
       \left(\prod_{i\in M}\langle u_i|g_i|u_i\rangle\right)g_{M^c}.
\]
The condition $\rho_M=I/2^t$ says that no nonidentity stabilizer is
supported inside $M$. Hence distinct stabilizers have distinct Pauli
restrictions to $M^c$, up to their fixed signs. In the Pauli expansion
of $\rho_T$, each nonzero coefficient is therefore a single product
$\pm\prod_{i\in M}r_{i,b_i}$, with $r_{i,I}=1$ and
$b_i\in\{I,x,y,z\}$. Purity is the sum of the squares of these
coefficients divided by $2^{|T|}$. Since
$\sum_{a_i=x,y,z}r_{i,a_i}^2=1$ for each measured site, expanding the
identity factors gives the displayed convex combination. Changing a
Pauli outcome only changes coefficient signs, so every $\psi_a$ has
the same purity profile for its $2^t$ possible outcomes.
\end{proof}

{\sloppy %
\begin{proposition}[exact thirteen-qubit interval; LP-13, QP-13, SW-2--SW-4]
\label{prop:thirteeninterval}
For all pure thirteen-qubit states,
\[
 \frac{18171249}{10^9}\ \le\
 \min_\psi\pi_6(\psi)\ \le\ \frac{85}{4576}.
\]
The upper bound is attained by a stabilizer state obtained by
measuring $X$ on vertex $4$ (indexed from $0$) of the 14-vertex graph
in Theorem~\ref{thm:fourteenfifteen}. The Rains LP gives only $25/1408$
(Theorem~\ref{thm:strictLP1319}).
\end{proposition}
\begin{proof}
The lower bound is Theorem~\ref{thm:threecopy-sw}(a). LP-13 checks an
exact rational Rains dual for $25/1408$ as in
Theorem~\ref{thm:fourteenfifteen}. For the $X$-projection witness,
QP-13 counts the restricted graph stabilizers exactly. Its
Pauli-weight distribution has $A_0=1$, $A_1=\cdots=A_4=0$,
$A_5=15$, $A_6=204$. Formula~\eqref{eq:as} gives
\[
 \pi_6=\frac{\binom{13}{6}+8A_5+A_6}
                 {64\binom{13}{6}}=\frac{85}{4576}.
\]
QP-13 also checks all $14\times3$ choices of distinguished qubit
and Pauli measurement and, by Proposition~\ref{prop:pauliprojection},
establishes that this is the best projection from this particular
14-qubit stabilizer witness in \emph{any} one-qubit pure basis.
No global thirteen-qubit optimality is asserted.
\end{proof}

\begin{theorem}[quantitative Rains-LP gaps at 13 and 19 qubits; LP-13, LP-19,
SW-2--SW-4]\label{thm:strictLP1319}
The unrestricted balanced-purity minima satisfy
\[
 \boxed{\frac{25}{1408}<\frac{18171249}{10^9}\le\min_{\|\psi\|=1}\pi_6(\psi)
      \le\frac{85}{4576}\quad(n=13),}
\]
\[
 \boxed{\frac{93}{38896}<\frac{23947609}{10^{10}}\le\min_{\|\psi\|=1}\pi_9(\psi)
      \le\frac{15201}{5912192}\quad(n=19).}
\]
The values $25/1408$ and $93/38896$ are the exact optima of the Rains
\emph{linear programmes}. The improved nineteen-qubit upper bound is attained by the explicit graph
in \texttt{code/verify\_nineteen\_graph\_improvement.py} (NG-1--NG-3).
\end{theorem}
\begin{proof}
The middle inequalities are Theorem~\ref{thm:threecopy-sw}(a); the upper
bounds are Proposition~\ref{prop:thirteeninterval} and the nineteen-qubit graph of NG-1--NG-3.
The exact LP-13 dual of Proposition~\ref{prop:thirteeninterval}
has strictly positive Pauli-weight slack coefficients for $j=1,\ldots,5$.
Thus equality in the LP would force $A_1=\cdots=A_5=0$, and hence every
five-body reduced state would be maximally mixed. Such a vector would
be a pure one-dimensional quantum code with parameters
$((13,1,6))_2$. A second, qualitative proof that the LP value is not
attained is the nonexistence of $((13,1,6))_2$ codes, proved by
Angl\`es Munn\'e and Huber~\cite{AMH26} and independently by
Theorem~\ref{thm:threecopy-sw}(b), together with compactness.

For $n=19$, the script LP-19 checks the exact rational dual identity
\[
 a_9-\frac{93}{38896}A_0
 =\sum_{s\in\{0,1,2,7\}}e_s(a_s-a_{19-s})
  +\sum_{i\in\{0,1,2,3,4,8,12,16\}}y_iS_i
  +\sum_{j=1}^{7}w_j A_j,
 \quad y_i>0,\ w_j>0.
\]
All rational multipliers are literal in the script, and every
coefficient of $A_0,\ldots,A_{19}$ is checked exactly.
Starting from the graph witness specified by nineteen hexadecimal adjacency
row masks in GS-19, toggle precisely the three edges $(2,6)$, $(2,16)$, and
$(2,18)$ (vertices indexed from zero). NG-1--NG-3 check the resulting
symmetric binary adjacency matrix, enumerate all $2^{19}$ stabilizer elements,
and obtain the improved purity numerator $121608$, giving
$\pi_9=15201/5912192$. An independent enumeration of all
$\binom{19}{9}=92378$ cut ranks yields
$6^2\,7^{840}\,8^{26696}\,9^{64840}$ and the same purity.
The improvement over the earlier GS-19 upper bound $449/173888$ is exactly
$5/454784$. Neither graph is claimed globally optimal.

Any state attaining the LP bound at $n=19$ would have
$A_1=\cdots=A_7=0$, hence would define a pure
$((19,1,8))_2$ code, which~\cite{AMH26} and
Theorem~\ref{thm:threecopy-sw}(b) exclude: the same qualitative
second proof.
\end{proof}

\par}

\subsection{A thirteen-qubit entropy witness}
\label{sec:thirteen-entropy-witness}

Let $G_{13}$ be the graph obtained from the fourteen-vertex graph of
Theorem~\ref{thm:fourteenfifteen} by pivoting on edge $(4,1)$ and
deleting vertex $4$, with the remaining vertices relabelled
$0,\ldots,12$ in order. This is the graph rule for the $X$ measurement
of vertex $4$, so up to local Clifford gates $G_{13}$ is the witness
of Proposition~\ref{prop:thirteeninterval}; as
$\pi_6(G_{13})=85/4576>2^{-6}$, some of its balanced cuts are rank
deficient. A superposition of
$256$ graph-basis vectors of $G_{13}$, all fixed by five graph
stabilizers, removes every rank deficiency and raises the attained
mean entropy above $5.8336$ bits. The five stabilizers make each
balanced marginal unitarily block diagonal with blocks of size at most
eight, so the entropy is enclosed from small exact blocks.

\begin{lemma}[Pauli-code compression of a marginal; proved]
\label{lem:pauli-code-compression}
Let $\psi\in(\mathbb C^2)^{\otimes k}\otimes(\mathbb C^2)^{\otimes m}$
be a joint eigenvector of commuting Pauli operators
$P^{(j)}=P_A^{(j)}\otimes P_B^{(j)}$, $j=1,\ldots,r$, and let
$\rho_A=\tr_B|\psi\rangle\langle\psi|$. Then $\rho_A$ commutes
with every $P_A^{(j)}$. Suppose the group generated by the
$P_A^{(j)}$ modulo phases is also generated by Paulis
$\bar X_1,\bar Z_1,\ldots,\bar X_e,\bar Z_e,\bar Z_{e+1},\ldots,\bar Z_{e+c}$,
which commute except that $\bar X_i$ and $\bar Z_i$ anticommute for
$i\le e$. Extend this set to a full canonical set
$\bar X_i,\bar Z_i$, $1\le i\le k$, and let $V$ be a matrix with
$V^\dagger V=dI$, $d>0$, such that $V^\dagger\bar Z_iV/d$ and
$V^\dagger\bar X_iV/d$ are the standard $Z_i$ and $X_i$. Then
\[
 \frac{V^\dagger\rho_AV}{d}
 =I_{2^e}\otimes\bigoplus_{u\in\mathbb F_2^c}D_u,
 \qquad \dim D_u=2^{k-e-c},
\]
where the first factor is qubits $1,\ldots,e$ and $u$ labels the
computational values of qubits $e+1,\ldots,e+c$.
\end{lemma}
\begin{proof}
If $P^{(j)}\psi=\lambda\psi$, then $|\lambda|=1$ and
$(P_A^{(j)}\otimes I)\psi=\lambda(I\otimes P_B^{(j)\dagger})\psi$, so
\[
 P_A^{(j)}\rho_AP_A^{(j)\dagger}
 =\tr_B\bigl[(I\otimes P_B^{(j)\dagger})|\psi\rangle\langle\psi|
             (I\otimes P_B^{(j)})\bigr]=\rho_A.
\]
Commuting with a Pauli does not depend on its phase, so $\rho_A$
commutes with the generated group. After conjugation by $V/\sqrt d$
it commutes with $X_i,Z_i$ for $i\le e$, hence is the identity on
those qubits, and with $Z_i$ for $e<i\le e+c$, hence is block
diagonal in their computational values.
\end{proof}

\begin{corollary}[Thirteen-qubit entropy witness; verified arithmetic E13L-1--E13L-2]
\label{cor:thirteen-entropy-witness}
Let $g$ be the graph state of $G_{13}$ and $g_v=Z^vg$,
$v\in\mathbb F_2^{13}$, its orthonormal graph basis, with qubit $i$
as bit $i$ of $v$. Let $a_v,b_v$ be the integer coefficients in
\texttt{data/thirteen\_entropy\_local.json}, ordered by increasing
labels $v$ in the binary annihilator of
\[
                     3987,\quad3302,\quad3039,\quad2939,\quad968.
\]
The normalized state
\[
 \zeta=\frac{\sum_v(a_v+ib_v)g_v}{\sqrt{\sum_v(a_v^2+b_v^2)}}
\]
has every balanced marginal positive definite, with all eigenvalues
above $1/2100^2$, and
\[
       5.83360583836270199<E_6(\zeta)<5.83360583836270200.
\]
Hence the thirteen-qubit balanced-entropy maximum satisfies
$\max_{\|z\|=1}E_6(z)>5.8336$.
\end{corollary}
\begin{proof}
Write $K_j=X_jZ^{N(j)}$ for the graph stabilizers and
$K^w=\prod_jK_j^{w_j}$. Since $K_jZ^v=(-1)^{v_j}Z^vK_j$, one has
$K^wg_v=(-1)^{w\cdot v}g_v$. Every occupied label annihilates the five
displayed words, so $\zeta$ is fixed by their five graph
stabilizers, which are $\pm X^wZ^{\sum_jw_jN(j)}$.
Orthonormality of the graph basis makes the denominator exact.
The computational amplitudes are
$\langle x|\zeta\rangle=(-1)^{q(x)}\sum_v(a_v+ib_v)(-1)^{v\cdot x}
/\sqrt{2^{13}\sum_v(a_v^2+b_v^2)}$, with $q$ the edge form of $G_{13}$;
E13L forms the integer numerators by a Walsh transform and checks
the exact norm identity.

For each of the $1716$ balanced cuts $A|\bar A$, E13L restricts the
five stabilizers to $A$, checks that the restrictions are independent,
and computes a canonical set by symplectic elimination whose span
equals theirs. It builds a Gaussian-integer $V_A$ with
$V_A^\dagger V_A=d_AI$, $d_A$ a power of two, and checks exactly that
$V_A$ conjugates every canonical Pauli to its standard form.
Lemma~\ref{lem:pauli-code-compression} then gives
\[
 \rho_A\simeq I_{2^e}\otimes\bigoplus_{u\in\mathbb F_2^c}D_u,
 \qquad 2^e\sum_u\tr D_u=1,
\]
with $(e,c)=(1,3)$ at $312$ cuts and $(e,c)=(2,1)$ at $1404$ cuts.
The blocks $D_u$ are therefore $4\times4$ or $8\times8$, and
$S(\rho_A)=2^e\sum_uS(D_u)$.
E13L encloses the entries of each $D_u$ in Arb from the exact
normalized amplitudes, isolates its eigenvalues rigorously and checks
that each exceeds $1/2100^2$. This proves positive definiteness.
Summing the enclosed $-\lambda\log_2\lambda$ over all blocks and
cuts gives
$E_6(\zeta)\in[5.8336058383627019939735\pm10^{-40}]$, which lies
strictly inside the stated interval.
\end{proof}

\begin{remark}[Scope]
The state $\zeta$ is not claimed stationary, and no global
thirteen-qubit entropy bound is asserted here; the global maximum
remains open. The script E13L also checks Hessian and roundoff
data (E13L-0, E13L-3--E13L-5) and prints a conclusion about a nearby
local maximum; that statement is not part of this supplement, and only
E13L-1--E13L-2 are used here.
\end{remark}

\section{Lean formalisation}\label{sec:fqe-lean}
The formalisation is the \texttt{lean/} folder of the companion repository \href{https://github.com/cmbant/qubit-entropy}{\texttt{cmbant/qubit-entropy}}, a standalone Lake project
(\texttt{lake build}) using Lean~4 and Mathlib, whose theorem docstrings give the TeX labels of the paper and this
supplement. Table~\ref{tab:fqe-lean} lists what it proves, with the main Lean declarations; the
companion README gives the build commands. The default \texttt{lake build} omits the final overlap certificate and
the unconditional quadratic-variance certificate; these have separate targets described below.
The formalised results use only Lean's standard three axioms
(\texttt{propext}, \texttt{Classical.choice}, \texttt{Quot.sound}), except for the finite compiled checks behind
Theorems~\ref{thm:fqe-global-overlap} and~\ref{thm:fqe-sharp-quadratic-variance} and any results that depend on them.
For the first, three finite
checks of the certificate data (the identity~\eqref{eq:fqe-global-overlap-sos} with the Gram test~\eqref{eq:fqe-compact-gram-test}, the
target polynomial, and the constraint matrices) are evaluated by compiled code (\texttt{native\_decide}, which adds one axiom per
check); their soundness and the whole reduction to the normal slice are kernel-checked. The 32-variable
polynomial identities behind~\eqref{eq:fqe-remainder} and~\eqref{eq:fqe-flip-gradient-identity} are split into 7 and 22
pieces, each checked by kernel evaluation in its own declaration.

For Theorem~\ref{thm:fqe-sharp-quadratic-variance} the formalisation proves the identity~\eqref{eq:fqe-qv-allstate-sos} for
all $z$ without the dimension count of Section~\ref{sec:sup-qk}. The difference of its two sides is written as a
sesquilinear form $\overline{z^{\otimes6}}^{\mathsf T}Kz^{\otimes6}$ with an integer matrix $K$ assembled from the
covariants $\Phi_{ki}$ and from copy permutations; the form vanishes once $PKP=0$ for the symmetriser $P$ of the six copies.
The integer spin intertwiner of $(\mathbb C^2)^{\otimes6}$ block-diagonalises each qubit factor into sectors of twice-spin $0,2,4,6$, and
$PKP=0$ becomes one integer matrix identity in each of the $164$ spin sectors that carry symmetrised vectors, with bases
of their images certified by a character average and pivot certificates. These finite checks run as one compiled checker
(\texttt{native\_decide}); a second compiled check identifies its intertwiner array with the mathematical definition.
The remaining table facts are kernel-evaluated. The soundness of the checker, the block calculus, and the
deduction of (QV), (QK) and the R\'enyi envelope above one are kernel-checked.

To include these results, build \nolinkurl{FourQubitEntropy.QVEqualityMain} separately (it imports
\nolinkurl{FourQubitEntropy.QVCheck.Unconditional} and adds the equality and optimality statements of (QV) and (QK)), and, after generating
the overlap data as directed in the companion README, \nolinkurl{FourQubitEntropy.GOGlue.Final}.
Their consequences inherit the same compiled checks: in particular, the unconditional R\'enyi envelope inherits
the quadratic-variance checks, and the unconditional global orbit-stability bounds inherit the overlap checks.

The analytic arguments and certificates of the paper do not depend on the Lean
development.

{\small
\begin{longtable}{@{}>{\raggedright\arraybackslash}p{0.14\linewidth}>{\raggedright\arraybackslash}p{0.49\linewidth}>{\raggedright\arraybackslash}p{0.31\linewidth}@{}}
\caption{Scope of the Lean formalisation.}\label{tab:fqe-lean}\\
\toprule
Topic & Formalised statements & Main declarations \\
\midrule
\endfirsthead
\toprule
Topic & Formalised statements & Main declarations \\
\midrule
\endhead
Four-qubit maximum &
Theorem~\ref{thm:fqe-entropy-max}: the bound, attainment by $M_4$ and the equality case; one-cut rigidity on every cut
(Proposition~\ref{prop:fqe-onecut}); the critical normal form (Theorem~\ref{thm:fqe-critical-normal-form}). &
\texttt{meanEntropy\_le}, \texttt{meanEntropy\_eq\_max\_iff}, \texttt{one\_cut\_rigidity\_12}, \texttt{critical\_normal\_form}, \texttt{critical\_iff\_LU\_Gv}\\
Moment inequality &
Theorem~\ref{thm:fqe-moment}: the sharp $F\ge\frac34NW_1$ from~\eqref{eq:fqe-remainder}, optimality of $\frac34$ at
product states, the zero set of $F$ (Proposition~\ref{prop:fqe-moment-zeros}) and the sharp third-moment lemma. &
\texttt{sexticF\_ge}, \texttt{sexticF\_ge\_sharp}, \texttt{moment\_zeros}, \texttt{third\_moment}\\
Envelopes &
The spin-flip derivative identity~\eqref{eq:fqe-flip-gradient-identity} (Lemma~\ref{lem:fqe-flip-gradient}); the
general envelope of Theorem~\ref{thm:fqe-general-envelope} in the homogeneous form~\eqref{eq:fqe-general-envelope}, with
optimality of its coefficient $\frac78$ and its complete equality classification; the critical envelope
(Proposition~\ref{prop:fqe-critical-envelope}) in full: the closed form, its equality classification and the critical moment
boundary. &
\texttt{flipGradient}, \texttt{general\_envelope\_homog}, \texttt{general\_envelope\_coeff\_iff}, \texttt{envelope\_equality\_iff},
\texttt{critical\_closed\_form}, \texttt{critical\_envelope\_eq\_iff}\\
Orbit distance &
The distance $d$ to the $M_4$ orbits as a minimum, with $d^2=2-2\sqrt\Phi$; the fidelity floor
(Lemma~\ref{lem:fqe-hs-fidelity-floor}) with its attainment at GHZ; hence the global estimates of Theorem~\ref{thm:fqe-orbit-stability} and
Corollary~\ref{cor:fqe-orbit-spectral} (the local-radius clauses and the limiting coefficient are not formalised). &
\texttt{orbitDist\_sq}, \texttt{fidelity\_floor}, \texttt{fidelity\_floor\_ghz\_attained}, \texttt{entropy\_gap\_orbit\_\allowbreak stability\_dist\_all}\\
HS overlap stability &
Theorem~\ref{thm:fqe-global-overlap}, $Q\ge1-\Phi$ for every unit state: a closest representative on the $19$-coordinate normal
slice with all its constraints, $Q$ on the slice, and the certificate~\eqref{eq:fqe-global-overlap-sos} on data generated from
the certificate file. &
\texttt{globalOverlap\_all}, \texttt{globalOverlap\_of\_frame\_cert}, \texttt{go\_certificate\_sound}\\
Critical orbit stability &
Theorem~\ref{thm:fqe-critical-orbit-sharp} with its sharp constants (the case $\Phi\ge\frac12$ from
Theorem~\ref{thm:fqe-global-overlap}) and its null-critical refinement~\eqref{eq:fqe-null-critical-sharp}, attained at
the cluster state. &
\texttt{critical\_orbit\_sharp\_\allowbreak dist\_all}, \texttt{Phi\_eq\_max}, \texttt{bell\_product\_sharp}, \texttt{null\_critical\_sharp}\\
Spectral extensions &
The cubic-majorant criterion with its equality clause (Proposition~\ref{prop:fqe-majorant}); Tsallis $0<q<2$ and R\'enyi $0<q<1$ with equality
cases; failure of $M_4$ for Tsallis $q>2$; the quadratic deficit bounds in $W_1$ and $\tau$; the rank-two and $q\ge2$
R\'enyi bounds; the HS--cluster crossing $1587/1000<q_*<1589/1000$; Lemma~\ref{lem:fqe-renyi-grouping} and
Proposition~\ref{prop:fqe-cubic-bound} with their equality cases, the necessity of $q_b$ and $\tfrac32<q_b<\tfrac{313}{200}$; the single crossing
(Lemma~\ref{lem:fqe-renyi-single-crossing}) with its equality case and the order transfer
(Corollary~\ref{cor:fqe-renyi-order-transfer}), also for the three cut spectra of a state. &
\texttt{majorant\_le}, \texttt{majorant\_eq\_iff\_of\_pos}, \texttt{tsallis\_le}, \texttt{renyi\_le}, \texttt{tsallis\_cluster4\_gt}, \texttt{entropy\_stability},
\texttt{renyi\_crossing}, \texttt{renyi\_grouping\_eq\_iff}, \texttt{renyi\_cubic\_eq\_iff}, \texttt{renyi\_cubic\_iff},
\texttt{single\_crossing\_statement}, \texttt{meanRenyi\_order\_transfer}\\
R\'enyi above one &
The envelope $\mathcal E_q\le\max\{h_q,5/3\}$ of Corollary~\ref{cor:fqe-renyi-gw} for every $q>1$, with its equality cases (HS
equality as the two $M_4$ orbits, cluster equality spectrally), from (QK); within it: the band-$L$
bound (Lemma~\ref{lem:fqe-rational-dual}, from the bounded-head bound of Lemma~\ref{lem:fqe-small-head-dual} and the large-head
Lemmas~\ref{lem:fqe-large-head-curvature} and~\ref{lem:fqe-large-head-dual}), the capped cubic bound
(Lemma~\ref{lem:fqe-capped-cubic}), the binary quartic lemma at $q_s$ (Lemma~\ref{lem:fqe-binary-quartic}, by a real-variable route that differs from the
Cauchy-formula proof in the text), the HS supporting inequality (Lemma~\ref{lem:fqe-hs-polynomial}, with the low-tilt
reduction), the cluster sliver (Lemma~\ref{lem:fqe-cluster-sliver}), the dual bound (Lemma~\ref{lem:fqe-dual}), determinant
monotonicity (Lemma~\ref{lem:fqe-renyi-determinant}), the near-uniform dual bound (Lemma~\ref{lem:fqe-near-uniform-dual}), the
large-moment-slack regime (Corollary~\ref{cor:fqe-renyi-high-f}) and the case analysis of
Theorem~\ref{thm:fqe-renyi-classical}. &
\texttt{renyi\_envelope\_above\_\allowbreak one\_unconditional}, \texttt{renyi\_envelope\_above\_one\_QK}, \texttt{largeHeadDual}, \texttt{smallHeadDual}, \texttt{det\_reduction}, \texttt{nearUniformDual},
\texttt{meanRenyi\_high\_F}\\
Quadratic variance &
The sharp bound (QV) of Theorem~\ref{thm:fqe-sharp-quadratic-variance} through the identity~\eqref{eq:fqe-qv-allstate-sos}
for every $z$, with the certificate data committed as Lean literals generated from the certificate file and the identity
checked by compiled code, its equality along the cluster-to-Bell family
and optimality of the coefficient $3$; (QK) of Corollary~\ref{thm:fqe-quadratic-variance} with its equality classification
and both coefficients sharp; the critical-state bound of Proposition~\ref{prop:fqe-critical-variance} with its equality case
and attainment. &
\texttt{qvCertIdentity}, \texttt{sharpQuadraticVarianceBound}, \texttt{quadraticVarianceBound}, \texttt{checkAll\_true},
\texttt{qv\_coefficient\_iff}, \texttt{qk\_equality\_\allowbreak classification\_product}, \texttt{qk\_coefficients\_sharp}, \texttt{critical\_rf}\\
Convex roofs &
The von Neumann, Tsallis ($0<q<2$) and R\'enyi ($0<q<1$) roof ceilings, attained only by pure states in the two $M_4$
orbits (Corollary~\ref{cor:fqe-convex-roof}). &
\texttt{convexRoof\_vn\_eq\_iff} and its Tsallis and R\'enyi analogues\\
Spin-flip bounds &
Both inequalities of~\eqref{eq:fqe-two-flat}, their sharpness, the equality case on the critical locus and the
Schmidt-rank-two consequence and the identification of states with two maximally mixed cuts with the cluster-to-Bell
family, with their R\'enyi and von Neumann values (Proposition~\ref{prop:fqe-two-flat-rigidity}); the spectral spin-flip bound
(Proposition~\ref{prop:fqe-spectral-spinflip}) and the overlap inequality behind it
(Lemma~\ref{lem:fqe-isospectral-overlap}, with attainment); Lemma~\ref{lem:fqe-real-gram}. &
\texttt{two\_flat\_TF\_12}, \texttt{two\_flat\_eq\_iff\_12}, \texttt{two\_flat\_QF}, \texttt{two\_flat\_renyi}, \texttt{spectral\_spinflip\_12}, \texttt{isospectral\_overlap\_rect},
\texttt{real\_gram\_rect}\\
Icosahedral family &
Normalisation, maximally mixed one-qubit marginals, and the purity and spectrum of every pair and three-qubit marginal;
hence $\min_x\pi_3(\psi_x)=121/917$ at $x=37/524$, and $b_3'(37/524)=-432/120127$. The family entropy $E_3(x)$ is strictly
concave with unique maximiser $x_*$, the root in $(0,\tfrac12)$ of $2^{32}(1-2x)^{15}=3^8(1+8x)^8(1+28x)^7$, with
$0.0698287<x_*<0.0698288$ and $2.9653560294776<E_3(x_*)<2.9653560294777$ bits. &
\texttt{icosahedralState\_eq\_ico}, \texttt{pairCut\_spectrum}, \texttt{star\_value}\\
Seven-qubit reductions &
The global joint scalar majorant with its equality set (Lemma~\ref{lem:seven-joint-scalar-majorant}, AH-6); the spectral pair
(Proposition~\ref{prop:seven-spinflip-spectral-pair}); the joint spin-flip identity on the family (Lemma~\ref{lem:seven-joint-spinflip});
the minimal majorant degree (Proposition~\ref{prop:seven-entropy-majorant-degree}); the eigenvalue cap of
Corollary~\ref{cor:seven-entropy-eigen-cap-497} for pure states. Assuming $\pi_3\ge121/917$: the competitor band
(Proposition~\ref{prop:seven-competitor-band}), so every pure state with $E_3$ above the SC-2 threshold has purity in the band and
eigenvalues at most $497/1000$, and $E_3$ is bounded by the joint functional; with the localised bound E7-1 as a further hypothesis this
gives $E_3\le E_3(\psi_{x_*})$ for every state. &
\texttt{joint\_scalar\_majorant}, \texttt{joint\_scalar\_majorant\_\allowbreak eq\_iff}, \texttt{seven\_entropy\_max\_\allowbreak reduction\_joint},
\texttt{spinflip\_spectral\_pair}, \texttt{tripleCut\_\allowbreak spinflip\_family}, \texttt{majorant\_degree\_isLeast},
\texttt{eigenCap\_497\_pure}, \texttt{competitor\_\allowbreak purity\_band}\\
Spin-flip doubling and five-copy steps &
Theorem~\ref{thm:spinflip-double} and Corollary~\ref{cor:ordinary-triples}; the tangent-line lemma
(Lemma~\ref{lem:fivecopy-tangent-line}) in abstract form, for any copy number and number of qubits; the moment-level step of
Theorem~\ref{thm:fivecopy-family-equality}; the family relations of Proposition~\ref{prop:zero-focus}. &
\texttt{spinflip\_double\_odd}, \texttt{ordinary\_\allowbreak triple\_purity}, \texttt{tangent\_line}, \texttt{equality\_\allowbreak moment\_centred},
\texttt{zeroFocus\_\allowbreak wronskian\_\allowbreak unconditional}\\
Balanced-purity minima &
For $4\le n\le18$, $n\ne7,13$: $\min\pi_{\lfloor n/2\rfloor}$ equals the value of Table~\ref{tab:minima}
(Theorems~\ref{thm:lptable}, \ref{thm:fourteenfifteen}, \ref{thm:sixteenseventeen} and~\ref{thm:eighteen}), with
Rains' shadow inequalities and the graph-state purity $2^{-\mathrm{rk}_2}$ proved rather than quoted. &
\texttt{minPurity}$n$, \texttt{shadow\_nonneg}, \texttt{purity\_gsn}\\
Entropy--purity slope &
For every $d\ge3$ the sharp slope \eqref{eq:entropy-purity-general-slope} of Lemma~\ref{lem:seven-entropy-linear-purity} and,
for every $d\ge2$, the fixed-purity envelope of Lemma~\ref{lem:seven-fixed-purity-envelope}, both with their equality
cases; hence the entropy ceilings of Table~\ref{tab:fqe-other-n} for $8\le n\le18$, $n\ne13$, from the formalised purity minima. &
\texttt{entropy\_purity\_slope\_bits}, \texttt{entropy\_le\_gEnv}, \texttt{meanEntropy\_le}, \texttt{entropyCeiling}$n$\\
Odd-$n$ tradeoff &
The identity of Theorem~\ref{thm:odd-spinflip-tradeoff} for every odd $n\ge5$, with the coefficients $c_n$ and its
equality case. &
\texttt{odd\_tradeoff\_sharp}, \texttt{odd\_tradeoff\_sharp\_eq\_iff}\\
Even-$n$ tradeoff &
The bound of Theorem~\ref{thm:purity-spinflip-tradeoff} for $n\in\{8,10,14,16,20\}$, with its sharpness, the optimality of
$c_n$ for $L_n$ and Corollary~\ref{cor:spinflip-rigidity}. &
\texttt{purity\_spinflip\_\allowbreak tradeoff\_even}, \texttt{purity\_spinflip\_\allowbreak tradeoff\_even\_sharp}\\
\midrule
Not formalised: four qubits &
The local-radius and limiting-coefficient clauses of Theorem~\ref{thm:fqe-orbit-stability} and
Corollary~\ref{cor:fqe-orbit-spectral} (Lemma~\ref{lem:fqe-entropy-local-radius}); Theorem~\ref{thm:fqe-p1-sharp} and its covering
proof (Proposition~\ref{prop:fqe-critical-covering} and its lemmas); the critical fidelity profile~\eqref{eq:fqe-critical-fidelity-floor}
(Lemmas~\ref{lem:fqe-critical-real-density} and~\ref{lem:fqe-four-point-cubic}); Theorem~\ref{thm:sup-null-family} for $\theta>0$;
Proposition~\ref{prop:fqe-bell-frame}; the near-one proof of Theorem~\ref{thm:fqe-renyi-above-one} (Lemma~\ref{lem:fqe-renyi-slack},
the defect estimate~\eqref{eq:fqe-renyi-defect} and the R\'enyi clause of Corollary~\ref{cor:fqe-orbit-spectral}) and
Corollary~\ref{cor:fqe-renyi-purity}, although the bound and equality case of Theorem~\ref{thm:fqe-renyi-above-one} follow from the
formalised envelope; Proposition~\ref{prop:fqe-spin-flip-calibration}; the equality and defect clauses of
Proposition~\ref{prop:fqe-critical-quadratic-variance} (its inequality is (QV)) and the equality clause of
Theorem~\ref{thm:fqe-joint-purity}; Lemma~\ref{lem:fqe-single-filter}; the R\'enyi roofs for $1<q<q_*$ in
Corollary~\ref{cor:fqe-convex-roof}. & \\
Not formalised: seven qubits &
Theorems~\ref{thm:fqe-seven-purity} and~\ref{thm:fqe-seven-entropy} (FP5-1, E7-1, attainment and the longer decimal; only the family
values and the reductions under explicit hypotheses above are formalised), with the five-copy chain
(Lemmas~\ref{lem:fivecopy-site-algebra}, \ref{lem:fivecopy-young-blocks}, \ref{lem:fivecopy-direct-probes}, \ref{lem:positive-repair},
Propositions~\ref{prop:ico-fivecopy-g34-kernels}, \ref{prop:fivecopy-finite-probes}, Theorem~\ref{thm:fivecopy-sharp-certificate},
Corollary~\ref{cor:seven-global-from-fivecopy}), orbit rigidity (Theorem~\ref{thm:moment-orbit-rigidity},
Corollaries~\ref{cor:fourth-moment-orbit-rigidity} and~\ref{cor:icosahedral-labelled-orbits}; of
Theorem~\ref{thm:fivecopy-family-equality} only the moment-level step in abstract form: the tangent line and the elimination of $t$
by the centred square, given $H(X_y')\ne0$; the family value $H(\psi_y)=155+96y$ (FO-1) and the passage to the orbit are not formalised) and the entropy certificates
(Theorems~\ref{thm:seven-entropy-localized-identity}, \ref{thm:seven-entropy-global-maximum},
Proposition~\ref{prop:seven-entropy-equality-moment}, the five-copy part of Proposition~\ref{prop:seven-joint-entropy-target},
Corollary~\ref{cor:seven-joint-degree-five}, the mixed-state clause of Corollary~\ref{cor:seven-entropy-eigen-cap-497});
full-space stationarity (Lemma~\ref{lem:seven-family-symmetric-criticality}); the Hessian, basin and stability results
(Theorems~\ref{thm:icolocal}, \ref{thm:icoquantbasin}, \ref{thm:entropy-seven-local}, \ref{thm:entropy-seven-quantbasin},
Lemmas~\ref{lem:quarticbasin} and~\ref{lem:seven-flat-extremal-planes}, Corollary~\ref{cor:seven-global-quadratic},
Proposition~\ref{prop:seven-mixed-roofs}); Propositions~\ref{prop:stab7} and~\ref{prop:seven-entropy-full-rank},
Corollary~\ref{cor:stabilizer-entropy-noname}; Theorem~\ref{thm:complexsos} and Propositions~\ref{prop:confphase}
and~\ref{prop:kramers-conference-shell}. & \\
Not formalised: replica hierarchy and covariants &
Proposition~\ref{prop:rainsppt} and Theorem~\ref{thm:spinflip-ppt-hierarchy}; Theorems~\ref{thm:threecopy} and~\ref{thm:fourcopy},
Corollary~\ref{cor:strict-copy-hierarchy}, Proposition~\ref{prop:fourcopy-not-sharp}; Proposition~\ref{prop:focus},
Lemma~\ref{lem:focus-pauli}, Proposition~\ref{prop:zero-focus} beyond its family relations, Theorem~\ref{thm:sextic-sos-obstruction};
the Pauli weight distribution and Type~II claim of Theorem~\ref{thm:eighteen} and the $n=18$ entry of
Corollary~\ref{cor:typeII-sixteen-twenty}. & \\
\bottomrule
\end{longtable}
}

\section{Verification checks and certificate status}\label{sec:sup-verifiers}
The labels below identify the finite checks used in the paper and this supplement. They have
different roles: \emph{exact symbolic or rational checks} establish algebraic
identities, ranks and coefficient signs; \emph{verified-arithmetic checks}
prove bounds using explicit rounding-error control; and
\emph{exploratory numerical checks} are diagnostics, not proof inputs.
In particular, the five-copy identities for purity (FP5-1, rational) and for
entropy (E7-1, whose coefficients involve the algebraic parameter $x_*$ and
logarithms) are verified-arithmetic certificates.
The surrounding reductions and equality arguments are given in the
proofs, not delegated to a table of scripts.

Table~\ref{tab:fqe-checks} of the paper lists which results use which checks. The companion's
\texttt{verifiers/MAP.md} identifies the script, supporting data and
committed output for each label. The groups NV, TCN, FS and TA are
numerical evidence only, and no proof uses them.

\subsection{Checks of the paper's appendices}\label{sec:sup-paper-checks}
Table~\ref{tab:fqe-checks} of the paper lists which results use the checks below. The four-qubit von Neumann, Tsallis and
single-crossing proofs do not assume a computational certificate; their checks confirm algebra. Finite decimals used as
bounds specify exact rationals.
\begin{description}
\item[ME-1--ME-6] Exact polynomial arithmetic in the sixteen amplitudes and their conjugates: two
independent sum-of-squares forms of $F$, coefficient by
coefficient (ME-1, ME-2); the purity identity of Lemma~\ref{lem:fqe-purity} (ME-3); the Hermite
coefficients $a,b$ and the entropy-gap identity of Section~\ref{sec:fqe-entropy_maximum-entropy-bound} (ME-4);
the values of $(P,T,F)$ at $M_4$ and at a product of two Bell pairs (ME-5); and, on the critical slice,
the cut spectra, the spin-flip amplitude $\sum_kv_k^2$, Clifford realisations of the coordinate
symmetries and the equality solutions (ME-6).
\item[SP-1--SP-5] The elementary swap derivation of Appendix~\ref{sec:fqe-sextic}: the $\epsilon$-maps,
spin-flip transpose and antisymmetry relations behind \eqref{eq:fqe-kernels} (SP-1); the generic
two-qubit swap, cycle and partial-transpose contractions with the marginal trace identities and the pair norms
\eqref{eq:fqe-pair-norms} (SP-2);
the eleven type coefficients of the table and the two class reductions (SP-3); the three contractions
\eqref{eq:fqe-contractions} and the identity \eqref{eq:fqe-remainder} (SP-4); and the local polarisation norm
\eqref{eq:fqe-local-polarization}, including degenerate Schmidt values (SP-5).
\item[SI-1--SI-3] An independent check of \eqref{eq:fqe-remainder}: the tensors $U_i,V_i$ and the pair vectors built
from their definitions (SI-1), the identity compared coefficient by coefficient in the sixteen amplitudes and their
conjugates (SI-2), and the product-state value $F=\tfrac34NW_1$ (SI-3).
\item[CE-1--CE-5] The critical envelope \eqref{eq:fqe-critical-envelope}: on $G(v)$, with $v,\bar v$ independent symbols,
$F=\tfrac12N^3+\tfrac74N\tau+\tfrac94\operatorname{perm}\Theta$, $P=N^2+\tau/2$, $\det\Theta=-N\tau$ and the eigenvalues
$N^2,\tau,\tau$ of $\Theta^T\Theta$ (CE-1); the sign-triple and cyclic-diagonal sums (CE-2); the expansion of
$\operatorname{perm}(\kappa I+(1-\kappa)uu^T)$ and its Maclaurin value (CE-3); the twelve equality matrices $O$ among the
signed permutations, and the bound attained on all of them (CE-4). CE-5 is a numerical sanity check, not used.
\item[PZ-1--PZ-2] Bloch-operator invertibility, the positive-Bloch support exclusion, and the generic GHZ identity
$H=(9/2)N\tau$ of Lemma~\ref{lem:fqe-polarization-remainder-zeros}.
\item[FI-1--FI-5] The spin-flip gradient identity~\eqref{eq:fqe-flip-gradient-identity} of Lemma~\ref{lem:fqe-flip-gradient}:
the dimension $19$ of the $SU(2)^4$-invariants of bidegree $(4,2)$ (and $8$ for $(2,2)$) as an exact character constant term
(FI-1); $19$ site-wise $\epsilon/\delta$ contraction invariants with an exact rank-$19$ evaluation matrix at $24$ integer points,
with $\bar z$ replaced by an independent vector (FI-2); (S) in the normalisation used, $\|A\|^2=F-\frac34NW_1$ (FI-3); the identity
at all $24$ points in exact integer arithmetic (FI-4); negative regressions with the coefficient $\frac98$ replaced by $1$ or the
multiple of $s$ dropped (FI-5).
\item[FG-6--FG-8] The integration proving Theorem~\ref{thm:fqe-general-envelope}: the passage
from~\eqref{eq:fqe-flip-gradient-identity} to~\eqref{eq:fqe-flip-gradient}, the pencil derivative, the scaling of $W_1$, the
integrating factor and the final substitution (FG-7); the scalar rational identities giving the exact transport
law~\eqref{eq:fqe-spin-flip-calibration} and its positive-norm integral~\eqref{eq:fqe-spin-flip-calibration-sos} (FG-8). FG-6 is a
finite-difference sanity check of~\eqref{eq:fqe-flip-gradient} and of the envelope at random states, not used in the proof. FG-1--FG-5
and FG-9, in the same script, check the direct proof of~\eqref{eq:fqe-flip-gradient-identity} in Section~\ref{sec:sup-flip-gradient}.
\item[ESH-1--ESH-3] Optimality of the coefficient $7/8$ in Theorem~\ref{thm:fqe-general-envelope}: on the GHZ family
$c|0000\rangle+s|1111\rangle$, $\tau$, $W_1$ and $F$ computed exactly from the tensor definitions, and the slack with $7/8$
replaced by $C$ equals $(7/2-4C)\tau+(1+4C)\tau^{3/2}$ ; exact symbolic.
\item[ER-1--ER-5] The equality case: the one-cut spectrum and spin-flip amplitude on $G(v)$ with the
Pauli actions that move its coordinates (ER-1); the weighted triangle-deficit identity (ER-2); the
entropy-gap identity and the fact that $(\tfrac12,\tfrac16,\tfrac16,\tfrac16)$ is the only spectrum in
$\{\tfrac16,\tfrac12\}$ with trace one (ER-3); the critical-slice factorisation \eqref{eq:fqe-critical-factor}'s second proof of Proposition~\ref{prop:fqe-moment-zeros}
against direct partial traces (ER-4); and the vanishing of $F$ on every listed zero representative together
with the determinant ratio that separates $M_4$ from $\overline{M_4}$ (ER-5).
\item[CN-1--CN-4] The normal form of Appendix~\ref{sec:fqe-normal-form}: the magic-basis generators and
moment maps (CN-1); the amplitude formula $G(v)=G_{abcd}$ (CN-2); the block-elimination identities of the
real singular-value lemma (CN-3); and rank-deficient and repeated-block rectangular examples (CN-4).
\item[EF-1--EF-7] The spectral corollaries of Section~\ref{sec:fqe-spectral}: the general Hermite
coefficient identity \eqref{eq:fqe-tsallis-coefficient} (EF-1); the cluster-state cut spectra (EF-2); the
tie at $q=2$, the Tsallis advantage $1/144$ of the cluster state at $q=3$ and the all-$q$ moment identity (EF-3);
the von Neumann and $q=2$ limits and the stability constant $4.57862$ (EF-4); the derivative and
logarithmic-derivative identities, endpoint values and rational logarithm bounds of the Tsallis sign proof (EF-5);
the strengthened polarisation and spin-flip coefficients, including the small-$\tau$ values $0.454$ and $0.602$ with and without the envelope term (EF-6); and the strengthened defect identity with the
leading coefficient of the determinant pencil used in the mixed-state exclusion (EF-7).
\item[RA-1--RA-5] The fourth-derivative identity, three endpoint derivative numerators,
fifteen printed coarse Taylor-coefficient bounds and their tails in
Lemma~\ref{lem:fqe-slack-fourth-derivative} (RA-1), logarithm absorption,
moment-defect algebra, the HS--cluster crossing brackets of Proposition~\ref{prop:fqe-renyi-crossing}: the integer comparisons giving $3/2<q_*<8/5$ and the exact enclosures giving $1587/1000<q_*<1589/1000$ (RA-4), and the strict
rational constants in the cap-free purity bound (RA-5). The concavity,
derivative-shape and Hermite-remainder proof is analytic; no parameter grid
is used.
\item[SCB-1--SCB-5] Generic complex-matrix calibration algebra in dimensions
two through six, the trace-rearrangement gap identity, exact unitary
regressions, the generic two-qubit Pauli spin-flip identity, the
physical sharpness family, and a classical counterexample to deriving QK
from the separate-cut caps. Positivity and the arbitrary-dimensional
overlap bound in Section~\ref{sec:fqe-spectral-spinflip} are analytic.
\item[UR-4] The quantitative crossed-purity algebra, including the unrestricted
$W_1/4$ term and the critical rank-two equality case, for
Proposition~\ref{prop:fqe-two-flat-rigidity}; the rigidity argument itself is analytic.

\item[OT-1--OT-4] Exact symbolic algebra in the analytic single-crossing proof
(Lemma~\ref{lem:fqe-renyi-single-crossing}, every $q>1$): the family formula, its derivative and the gradient profile (OT-1);
the Euler identity and the second variation splitting equal coordinates (OT-2);
the hyperbolic identity~\eqref{eq:fqe-renyi-crossing-ratio}, its contact value, the derivative of $\sinh(sz)/z$
and the integrating factor (OT-3); and a negative control where increasing $\kappa_q$ by $1/1000$ destroys HS
stationarity (OT-4). Full support, exclusion of repeated larger entries and all sign and equality arguments are
proved analytically; these checks are not proof inputs.
\item[SHD-1--SHD-4] The interval enclosures of the eight node values of the binary
quartic correction and $H_q(1/3,2/3)$; the grouping
constants and dual identities; the exact Bernstein coefficients of the eight one-variable quartics (SHD-3, with a negative
control); and the large-tail coefficient bounds and the SHD column of
Table~\ref{tab:fqe-moment-rectangle} (Lemma~\ref{lem:fqe-moment-rectangle}) in
Lemma~\ref{lem:fqe-small-head-dual}. The monotonicity of $A_q$ (Lemma~\ref{lem:fqe-binary-quartic}, by Cauchy's
formula), the grouping
and global derivative arguments are analytic.
\item[LHD-1--LHD-4] Lemmas~\ref{lem:fqe-large-head-curvature} and~\ref{lem:fqe-large-head-dual}.
LHD-1: the identities for the biased binary fourth derivative $J(x,y)$, the range of $K$, the three envelopes
of $y$, and the Bernstein expansions of degree $12$, $16$, $16$ on $[0,3/5]$ with all coefficients positive (with a negative control).
LHD-2: the polynomial with positive coefficients at $a=1/2$.
LHD-3: the monotonicity constants in $K$, the cubic minorant $\kappa$, the logarithm constants and error budgets.
LHD-4: the two Bernstein expansions of degree $13$ and $26$ behind the completed square, each with all coefficients
at least the stated bound. The Hermite remainder signs and the quasi-convexity of $J$ are proved in the text; no
subdivision or external coefficient file is used.
\item[BC-1--BC-3] Exact grouping derivative cancellation, geometric-convexity
and cubic grouping identities, equality contacts, lower-order negative control,
and the high-$F$ transfer. Log-sum, the single-crossing inference and the global
entropy bound are analytic in Section~\ref{sec:fqe-renyi-cubic-bound}.
This covers the analytic high-$F$ case $20\varepsilon+2F\ge1$ of Theorem~\ref{thm:fqe-renyi-classical} without
interval arithmetic.
\item[DA-1--DA-5] Exact-symbolic HS tangency, cluster contacts, evenness,
dual and sliver algebra; exact comparison of every printed multiplier and band with the
executable data; closed-band coverage and finite crossing brackets; exact wrong-multiplier
and shifted-contact negative controls.
\item[CV-1--CV-5] An independent quaternion parametrisation of the cut spectra on $P=1$ (CV-1, CV-2), cyclic purity and moment
identities, the cubic discriminant and nonnegative geometric-mean transfer remainder, and the full equality-family data for Proposition~\ref{prop:fqe-critical-variance}.
The cut-spectrum formula and equality inference are analytic.
\item[CJ-1--CJ-5] Directional permanent and geometric-concavity identities,
the Fourier variance factor and shared-unitary block expansion, the scalar
factorisation and quadratic discriminant, the sharp cluster-to-Bell family,
and a scalar-only negative control for
Proposition~\ref{prop:fqe-critical-quadratic-variance}.
The Cauchy--Schwarz bounds, global sign and equality arguments are analytic;
these checks use no coefficient certificate.
\item[QVA-1--QVA-4] The compact universal certificate for Theorem~\ref{thm:fqe-sharp-quadratic-variance}:
exact six-copy $\mathrm{SU}(2)$ intertwiners and vanishing at HS and cluster (QVA-1);
positivity of the denominator and $95$ integer norm weights (QVA-2); and the exact identity
\eqref{eq:fqe-qv-allstate-sos} at the $83$ determining Gaussian-integer states of QG (QVA-3).
There are $95$ squared norms in $14$ spin sectors, with denominator $D=10235456181520$ and positive integer
weights of at most $14$ decimal digits. Their sparse multiplicity vectors have $716$ nonzero integer entries,
all of absolute value at most $22$. There are no Gram matrices or matrix positivity checks.
QVA-4 retains the invariance regression for every squared norm under conjugation, qubit permutations and four
Gaussian-integer local maps per qubit. The replay uses integer arithmetic throughout.
The data encodes each highest-weight row by its negative and positive supports in the zero-based
six-bit computational basis. For $j=0,2,4,6$ these give $5,9,5,1$ rows $H_j$, respectively;
each is supported at Hamming weight $(6-j)/2$ and killed by the raising matrix $E$.
Concatenate $H_jE^k/k!$ in increasing $j$, then $k=0,\ldots,j$, to obtain $C_6$.
For each sector, \texttt{a} and \texttt{spins} specify $W_{a,p}$; each entry of \texttt{vectors} is a flat
list $[j_1,c_1,\ldots,j_s,c_s]$ specifying $\sum_tc_te_{j_t}$ with zero-based multiplicity indices,
and the corresponding entry of \texttt{weights} is its positive coefficient $q_{ki}$.
The proof needs no equality-curve jets, auxiliary rank computation or coefficient solve.
\item[EU-1, EU-2] The sharp entropy--purity slope of Lemma~\ref{lem:seven-entropy-linear-purity} for
every dimension: the one-spike identities for $N''$, $F_m(r_*)$, $F_m'(r_*)$ and $c_d$ (EU-1); and the entropy ceilings
of Table~\ref{tab:fqe-other-n}, evaluated at 60 digits and rounded up (EU-2).
\item[QG-1--QG-5] Mixed-sextic recoupling, every permuted HS/cluster zero, and the nonsingular 83-point evaluation matrix of
83 raw Gram entries built from five-copy blocks, multiplied by $N$ or coupled to one further copy (QG-1, QG-3):
these make the 83 states determining for the invariant space of
Section~\ref{sec:sup-qk}. The same script checks an independent direct proof of (QK): 80 squared covariant norms in 16
modules with strictly positive rational coefficients from an exact $80\times80$ solve, and a signed norm identity forcing
$P\in\{1,3/2\}$ at equality (QG-2, QG-4). QG-5 checks invariance of every raw Gram entry under complex conjugation, qubit
transpositions and five Gaussian-integer multiples of local unitaries on full-support states (a regression of the
construction). The ambient dimension 85 follows from \eqref{eq:fqe-invariant-even} and SD-4.

\item[RG-1--RG-3] The correlation geometry of Appendix~\ref{sec:fqe-correlation}: the Pauli trace formulas
for $\operatorname{tr}R^2$ and $\operatorname{tr}R^3$ of an arbitrary two-qubit matrix (RG-1); the determinant
identity \eqref{eq:fqe-correlation-volume} (RG-2); and the generalised eigenvalues $4,-47,-47$ of the
replica witness $K_{\rm rep}$ on the three pairings (RG-3).
\item[IC-1--IC-6] The icosahedral family of Appendix~\ref{sec:ico}: one-qubit marginals, the
quadratic $\pi_3(\psi_x)$ and its exact minimiser $x=37/524$, the marginal and pair purities, the
computational-basis block diagonalisation of the 35 triple and 21 pair marginals (proved analytically from $C^2=5I_6$ in Theorem~\ref{thm:ico}),
and the exact signs of both entropy derivatives in Corollary~\ref{cor:ico-entropy-separation}.
\item[EG-1] Interval-arithmetic enclosures of the gap $E_3(\psi_{x_*})-E_3(\Psiico)$ and of $\pi_3(\psi_{x_*})-121/917$ on
the icosahedral family, from its explicit spectra.
\item[GS-4] The MacWilliams and shadow-integrality identity for seven-qubit stabilizer states and an explicit seven-vertex graph witness attaining $\pi_3=19/140$; the lower bound is proved analytically in Proposition~\ref{prop:stab7}.

\item[SD-1, SD-2, SD-4] Dimensions of the spaces of local-unitary invariant polynomials of four qubits
for copy numbers up to $8$, by a character average and independently by a multiplicity count (SD-1, SD-2),
and the qubit-symmetric conjugation-even and odd counts of \eqref{eq:fqe-invariant-even} (SD-4). The degree-six
even count $85$ is a proof input for QVA and QG.

\end{description}

\subsubsection*{Orbit-stability certificates}\label{sec:fqe-stability-verifiers}
The entropy-maximum proof is analytic; the quantitative stability comparisons use exact or
outward-rounded interval certificates. The global overlap identity is checked by GO-2 in exact integer arithmetic. The local quadratic forms (OS, EH), the
Bell-frame witness (BF) and the closest-point normal slice (NSL) check proofs in this supplement and are listed below. The two sharp critical-locus
theorems are proved in this supplement, and their checks, listed below, are the analytic
critical-moment comparison supported by the exact symbolic checks CO-1--CO-3 and the covering profile FP-2 over $\mathbb Q$, the analytic mixed null-critical proof supported by the exact symbolic checks MS-1--MS-4, and for the sharp entropy coefficient the interval certificates PS, TC, LCC and CCV on the
critical locus, with the exact checks PM (the slice $P=1$) and EX. Normal-form coverage, constraint positivity,
chart reductions and entropy-transfer arguments are analytic. Interval certificates use outward-rounded
ball arithmetic, and every sign decision is a certain comparison of balls; no floating tolerance decides
anything. Certificate data files give every Gram matrix as integers over a stated denominator, with an
integer Cholesky approximation whose residual is checked to be strictly diagonally dominant with positive
diagonal, using integer arithmetic; no search solver is a verification input.
\begin{description}
\item[GO-1--GO-4] The sharp overlap certificate for Theorem~\ref{thm:fqe-global-overlap}:
independent verification of all $50$ used unit HS rays ($46$ baseline plus four rational two-factor
skew-rotation rays) by $n=1$ and $Q=W_1=\tau=0$ (GO-1); exact rounded low-rank reconstruction and the integer dominance test
\eqref{eq:fqe-compact-gram-test} for all $277$ remaining rational Gram blocks in $\mathcal C_Q$;
invariance of the two explicit analytic remainders; and exact equality
of \eqref{eq:fqe-global-overlap-sos} for $(Q-n^2e)/36$ across all $3{,}277$ degree-$\le6$ signed
lower-permutation monomial orbits on the $19$-variable closest-point normal slice (GO-2); the
ordered-singular-value remainder of Lemma~\ref{lem:fqe-hs-fidelity-floor} and the rational chain giving
$Q\ge1-\Phi\ge\frac{6+\sqrt6}{12}d^2>\frac7{10}d^2$ (GO-3); rational series bounds giving $k>16/225$, and the
margin $k(6+\sqrt6)/12>1/20$ of Theorem~\ref{thm:fqe-orbit-stability} (GO-4). Normal-slice coverage and the
positivity of the closest-point constraints are analytic (Section~\ref{sec:fqe-global-overlap}).
\end{description}

\subsection{Further four-qubit and stability checks}
\begin{description}
\item[FG-1--FG-5, FG-9] The direct proof of the spin-flip gradient identity (Section~\ref{sec:sup-flip-gradient}): Schouten symmetry
of $V-6U$ (FG-1); the copy-label Casimir selection rules, the moment operators~\eqref{eq:fqe-cubic-copy-moments}, and the
polarisation defect~\eqref{eq:fqe-cubic-polarization-defect} on the full invariant multiplicities (FG-2); the cubic intertwiner's
normalisation $54$ and the Klein-four rule excluding the quartic spin-one covariant (FG-3); the mixed-norm
algebra~\eqref{eq:fqe-cubic-mixed-norms} and its polarisation giving~\eqref{eq:fqe-flip-gradient-identity} (FG-4); nonzero coefficients
for the wrong factor or omitted correction (FG-5). FG-9 evaluates every intermediate identity of the proof from the tensor
definitions, in exact arithmetic at Gaussian-integer states, as a regression check. The checks FG-6--FG-8 of the same script
are listed in Section~\ref{sec:sup-paper-checks}.
\item[CO-1--CO-3] The critical variance and three-term moment identity \eqref{eq:fqe-critical-three-term},
the 32-ray tight frame and its permutation up to sixth-root-of-unity phases under $H_1,H_2$ (CO-1);
the exact polynomial identities and rational concavity bounds $P_1,P_2$ for $Q\ge\tfrac32-2\Phi$ on
$\Phi\le\tfrac12$ (CO-2); and the sharp distance constant $(2+\sqrt2)/4$ and Bell equality (CO-3).
The signed-Birkhoff reduction and case reductions are analytic
(Section~\ref{sec:fqe-critical-orbit-proof}).
\item[OS-1--OS-3] The full local quadratic form from direct flattenings, its exact spectrum and orbit
kernel, the local sextic radius, and HS, Bell and cluster regressions
(Section~\ref{sec:fqe-orbit-local}).
\item[BF-1--BF-4] All coefficients of the quartic determinant and cubic gradient identities behind \eqref{eq:fqe-bell-cubic}, the magic
Pauli normalisation of the $W_1$ formula, and the Bell-pair Gram algebra,
compatible-plane geometry and HS/Bell representatives of Proposition~\ref{prop:fqe-bell-frame}, whose
constructive plane argument is analytic.
\item[EH-1--EH-3] The full entropy quadratic forms from all cut flattenings, explicit orthonormal normal
blocks, exact coefficient multiplicities, the rational radius bound and a least-coefficient normal; the
resolvent estimate and the sharp limiting argument are analytic
(Section~\ref{sec:fqe-entropy-orbit-local}).
\item[NSL-1--NSL-3] The closest-point normal slice: the thirteen first-order conditions have rank ten and are
exactly \eqref{eq:fqe-slice-stationary}; lower rotations conjugate the gauge matrix; the stationary and gauge
conditions cut out exactly the span of the nineteen slice coordinates. That a closest point exists and
maximises the height is analytic.
\item[NC-1--NC-4] The two auxiliary Bloch spheres on the null-critical locus
(Section~\ref{sec:fqe-critical-orbit-proof}): the auxiliary magic transform is unitary with determinant
$\sum_jv_j^2/2$ (NC-1); the 32 HS rays are exactly the opposite-parity cube pairs (NC-2); the Bell
projectors and the moment formula \eqref{eq:fqe-null-critical-bloch} (NC-3); and the scalar remainder
identities, sharp constants and cluster equality of the elementary proof of
\eqref{eq:fqe-null-critical-sharp} of the paper (NC-4). The unit-sphere inequalities and the overlap maximisation are
analytic. NC-5 and NC-6 of the same script are further checks not used here.

\item[RP-1--RP-4] The three real mutually unbiased bases and the tight tomography identity, all sixteen
HS planes, the paired overlap/Bures compression formulas, purity, the fourth frame moment, sharp GHZ
covering equality and entropy-gradient matrices; the rank-two coverage and the relative-entropy identity
are analytic (Lemma~\ref{lem:fqe-critical-real-density}).
\item[FP-1--FP-4] All $70$ interleavings and their $23$ symmetry classes (FP-1), every coefficient and positive
Gram of \eqref{eq:fqe-four-point-cubic-sos} ($184$ Grams, all $4830$ coefficients through degree four
compared exactly; FP-2), the homogeneous area-weighted fusion identity and the magic-basis spin flip (FP-3), and every
overlap of the matching family \eqref{eq:fqe-covering-equality-family} (FP-4); ordering coverage, the Bloch
bridge and the area-weighted Bures argument are analytic.
\item[PM-1, PM-2, PM-4, PS-1] The slice $P=1$ (Section~\ref{sec:fqe-p1-sharp}): exact cut correlation
matrices from the physical magic basis, symmetry generators, and the exact equality
spectra, fidelities and full $144$-pair symmetry orbit (PM); $C_*$ to eighteen digits in interval arithmetic (PS-1);
the symmetry arguments and the wall lemma are analytic.
\item[NV-1--NV-2] Numerical evidence only, no certificate: the minimum of $\delta-C_*d^2$ over critical states at fixed
$\tau$ (NV-1) and random positive local filters of the equality state and of nearby critical states (NV-2), quoted in the
remarks after Theorem~\ref{thm:fqe-p1-sharp}.

\item[TC-1--TC-8] The correlation-matrix chart of the critical locus
(Section~\ref{sec:fqe-critical-sharp}): exact Bell correlation matrices of the twelve rays and
$\lambda_{ji}=(\|v\|^2+\langle C^{ji},T\rangle)/4$ for a rational $v$ (TC-1); the analytic range
$\tau\ge0.36$ in ball arithmetic (TC-6) and the sign symmetry at the equality point (TC-7). TC-2--TC-5 are
numerical regressions of the analytic Lemmas~\ref{lem:fqe-critical-correlation}
and~\ref{lem:fqe-critical-chart} and of the ball evaluator against floating point and finite
differences; TC-8 is a sampled audit of the cell bounds and of the HS rule, a regression and not a
proof input.
\item[EX] Exact symbolic statements behind the local cone lemma at all eight chart equality points: the
sign symmetry permutes the rays within each cut and the planes, preserving the trace and determinant of
each plane matrix; exactly four planes tie, at $4\Phi_*=(1+1/\sqrt3)(1+\sqrt{2/3})$, with $G_\Pi=0$ there by the
definition of $C_*$; and the tied gradients sum to zero identically.
\item[LCC] The four tied planes, their gradients and interval Hessians at the eight chart equality points,
and the cone cells of Lemma~\ref{lem:fqe-critical-local-cone}.
\item[CCV] The covering of Proposition~\ref{prop:fqe-critical-covering}; a floating-point twin only orders
the attempts and certifies nothing.
\item[UR-3] The fixed-trace Newton identities $e_2=(1-p)/2$ and $e_3=(1-3p+2t)/6$ used in the proof of
Lemma~\ref{lem:fqe-renyi-determinant} (Section~\ref{sec:sup-determinant}); the rest of that proof is analytic.
The same verifier checks the quantitative crossed-purity bound UR-4 in
Section~\ref{sec:sup-paper-checks}; its UR-1 and UR-2 are further checks not used here.
\item[HD-1--HD-5] Algebra in the analytic HS supporting proof of
Lemma~\ref{lem:fqe-hs-polynomial} at $q_u=8/5$: the fixed-purity path to the one-large spectrum (HD-1),
the slope constant $105^5<3\cdot2^{32}$ (HD-2), the one-large moment formula and the displayed scalar
curvature identity (HD-3), the rational uniform-contact gap and the outer dual (HD-4), and the
lower-band target and wrong-slope/excessive-tilt controls (HD-5). The maximisation, divided-difference and curvature
sign arguments and all equality inferences are proved in the text.
\item[LT-1--LT-3] The generic identities in the low-tilt chart reduction
(Lemma~\ref{lem:fqe-low-tilt-reduction}): the divided-difference identities behind the bound
\eqref{eq:fqe-dd-power} and the slope constant $64/35>9/5$ (LT-1),
fixed-purity path and derivative ratio (LT-2), and negative controls for extending
$q$ to $2$ or dropping the tilt bound (LT-3). The feasible monotone path is proved
in the text.
\item[CG-1--CG-5] The explicit arithmetic in the capped strengthened cubic bound
(Lemma~\ref{lem:fqe-capped-cubic} and Corollary~\ref{cor:fqe-capped-cluster}): the coefficient bounds and the CG column of
Table~\ref{tab:fqe-moment-rectangle} (CG-1), the divided-difference algebra and two binary endpoints, with the kernel algebra of the Cauchy-integral
proof of monotonicity in Lemma~\ref{lem:fqe-binary-quartic}, which is all this lemma uses of
that lemma (CG-2),
and the quartic endpoint bounds (CG-3), the pair identities, the two weight polynomials and the capped-cluster assembly (CG-4), and
negative controls for an excessive correction or an omitted purity cap (CG-5). All maximisation and
sign arguments are in the proof; CG adds no Bernstein sign test or parameter covering.
\item[NU-1--NU-3] Exact-symbolic fifth-order Newton and projection identities,
rational coefficient and logarithm bounds at the order $q_s$, and the NU column of
Table~\ref{tab:fqe-moment-rectangle} in Lemma~\ref{lem:fqe-near-uniform-dual}.
NU-3, SHD-4 and CG-1 check their columns of that table, the hypotheses of
Lemma~\ref{lem:fqe-moment-rectangle}, through one shared function. This is algebra for a written analytic
proof, with no subdivision, Hessian grid or interval sign test.
\item[SL-1--SL-4] Exact finite power/logarithm constants for the head/tail entropy estimates,
four polynomial lower bounds, rational domain and $\varepsilon$-removal inequalities,
and comparison of all 11 printed Bernstein coefficients with exact rationals.
The distinguished HT cut is proved analytically in Lemma~\ref{lem:fqe-cluster-sliver};
there is no subdivision or interval derivative grid. The verifier checks both printed tables
against the symbolic formulas.
\item[QV-1--QV-3] The residual identity and the minimum-purity reduction (QV-1), the purity-box bounds for (QV) and (QK)
(QV-2), and the cluster-to-Bell calibration
\eqref{eq:fqe-quadratic-family} of the paper with both forced coefficients (QV-3), used in Section~\ref{sec:sup-qk}; the box
extremum and the equality inferences are analytic.
\item[JI-1--JI-3] Exact symbolic verification of the implication identity
\eqref{eq:fqe-joint-implication}, all three purity-box residual formulas
\eqref{eq:fqe-purity-box}, and rational data at HS, cluster and Bell products.
These check algebra only; the fixed-sum box maximum and equality inference
are analytic.
\item[MS-1--MS-4] The strengthened sphere norm identities and all sixteen multiaffine vertices for the
opposite-parity reduction (MS-1), the exact same-parity parametrisation, monotonicity in $\tilde w$ and
polynomial minorants (MS-2), the three univariate polynomial decompositions closing the same-parity triangle
on $[0,1]$ (MS-3), and the sharp cluster interpolation and equality family (MS-4); the sphere, monotonicity
and coverage arguments are analytic.
\end{description}
\subsection{Seven-qubit and higher-qubit checks}
\begin{description}
\item[ES-1] Exact phase-support and conference-switching identities, and the common real fixed space
$\operatorname{span}_{\mathbb R}\{U,V\}$. Lemma~\ref{lem:seven-family-symmetric-criticality} gives the analytic
stationarity argument for every differentiable LU/permutation/conjugation-invariant functional;
Proposition~\ref{prop:entropy-seven-stationarity} applies it to entropy.
\item[SEH-1--SEH-3] The root bracket for $x_*$ and the spectral floor $1/12$ (SEH-1); exact invariance and
Sylvester positive definiteness of all $16$ canonical symmetry-adapted integer blocks (sizes at most $4\times4$)
of the two $128\times128$ Hessian blocks above $0.239395$, with no floating-point basis (SEH-2); and the
perturbation bound giving $-239/1000$ (SEH-3), for
Theorem~\ref{thm:entropy-seven-local}.
\item[EB-0--EB-3] The rational constants of the third-derivative and Taylor bounds in
Theorem~\ref{thm:entropy-seven-quantbasin}.
\item[SC-2] The rational logarithm lower witness $E_3(\psi_{7/100})>59307/20000$ used for the competitor band and the cap.
\item[SP7-1, SP7-5, SP7-6] Spin-flip spectral pairing and overlapping-cut compatibility,
the sharp entropy--purity slope at $d=8$ (proved analytically in Lemma~\ref{lem:seven-entropy-linear-purity} from $N(r_*)=0$ and $7^5>2^{14}$), and the competitor eigenvalue cap $497/1000$ (proved analytically by a relative-entropy tangent lower bound and square completion in Corollary~\ref{cor:seven-entropy-eigen-cap-497}).
\item[AH-1, AH-5] The exact competitor purity band and candidate joint spin-flip identities (proved analytically in Lemma~\ref{lem:seven-joint-spinflip}).
\item[AH-6] Exact rational coefficient certificate for the global joint scalar majorant
(Section~\ref{sec:seven-joint-analytic-proof}): rational root/logarithm enclosures and the positive kernel constants (AH-6a),
the symbolic cubic discriminant and double factor (AH-6b), and positive Bernstein and shifted-power coefficients (four rows) of the
degree-eighteen quotient (AH-6c). The positive-kernel and binary Hermite reductions are analytic.
\item[E7-K0--E7-K3] Exact integer family contractions for seventeen five-copy sources, exact generic
rank bounds by determining-point minor identities, and exact three-copy kernel source polynomials.
\item[E7-1] Verified-arithmetic certificate of Theorems~\ref{thm:seven-entropy-localized-identity}
and~\ref{thm:seven-entropy-global-maximum}: exact physical kernel projection with Arb, evaluation of all
$3424$ direct probes with forward error bounds, a validated rank-$3283$ inverse, and a strictly positive
Gram repair in $3754$ independent symmetric coordinates.
The cubic purity multipliers and marginal spectral-cap sources are included in the identity.
\item[TCB-0--TCB-4] The three-copy certificate of Theorem~\ref{thm:threecopy}: integer block kernels and
antisymmetriser relations, primal feasibility of the boundary pseudo-moment $m^*/70416$, the exact dual identity
at $L_3=125/978$ in all $37$ orbit coordinates from the six active-block kernels, and positive definiteness
of the primal pivot submatrices and rational kernel multipliers by fraction-free Bareiss elimination.
\item[TCN-1--TCN-3] Numerical only: floating-point solves of the three-copy relaxation for $n=4,\ldots,7$ (TCN-1),
with a tight-tolerance solve at $n=7$ (TCN-1b), a strictly feasible dual at $n=7$ (TCN-2) and its rounding to
integer data (TCN-3).
\item[FC-0--FC-4] The four-copy certificate of Theorem~\ref{thm:fourcopy}: exact integer $S_4$ irreps,
matrix units and the flip maps $T_1,T_2$ (FC-0); numerical regressions of the parametrisation and of the
modular adjoint contraction (FC-1, FC-2, not used in the proof); positive definiteness of the $271$ integer
flip multipliers and of every residual $\omega$-block, with the residual sum recovered exactly by the
Chinese remainder theorem over $35$ primes under an integer magnitude bound (FC-3); and the conclusion $\pi_3\ge211/1600$ (FC-4). The
integer certificate is a data file found by a floating-point SDP search; only its exact check is used.
The reported smallest eigenvalue is diagnostic; exact Sylvester tests decide positivity.
Runtime about $12$ minutes.
\item[NS-0--NS-4] The exact four-copy pseudo-moment of Proposition~\ref{prop:fourcopy-not-sharp}: a regression of
the modular forward map against a direct floating-point computation at $n=3,4$ (NS-0, not used in the proof); the
exact objective $\mathrm{oo}(x)/\mathrm{on}(x)<6597/50000$ (NS-1); positive definiteness of all $78$ $\omega$-blocks
(NS-2) and of all $271$ flip blocks, recovered exactly by the Chinese remainder theorem over $34$ primes under an
integer magnitude bound (NS-3), by fraction-free elimination; and the conclusion $L_4<6597/50000<121/917$ (NS-4).
The integer point comes from a floating-point solve; only its exact check is used. Runtime about $13$ minutes.
\item[FS-0--FS-2] Numerical only: floating-point solves of the four-copy relaxation at $n=4,5,6,7$, the
source of the table in Appendix~\ref{sec:fqe-seven}; the two largest $n=7$ solves
are flagged inaccurate by the solver and are not certified bounds.
\item[S5-1, Y5-1] The exact $S_5$ matrix units and physical spin-flip reconstruction of Lemma~\ref{lem:fivecopy-site-algebra}, and the rational two-flip Young models of Lemma~\ref{lem:fivecopy-young-blocks}.
\item[IK5-1--IK5-4, PK5-1--PK5-4] The exact icosahedral feature ranks and annihilators, the twelve exact physical kernels, and the support and invertibility of the supplied kernel bases (Proposition~\ref{prop:ico-fivecopy-g34-kernels}); PK5-4 does not check the sharp identity.
\item[DP5-1] The character count of the five-copy moment space, the trace-adjoint flip, the coset count and the target calibration of the direct probes (Lemma~\ref{lem:fivecopy-direct-probes}, Proposition~\ref{prop:fivecopy-finite-probes}).
\item[FP5-1] Verified-arithmetic certificate of Theorem~\ref{thm:fivecopy-sharp-certificate}: evaluation of the $3424$ rational direct probes with single-column norm companions, the rank-$3283$ right inverse in $3754$ independent symmetric coordinates with $\|I-LAR\|_2<0.070$, and the positive repair of the identity Grams, with correction below $0.324$. The triangular dyadic baseline, frames and probes occupy $441612$ bytes.
\item[FP5-2] An independent exact check, on scalar probes, of the baseline trace and of all $8965$ evaluator coefficients in the width-$16$ frames, whose first $r_b\le7$ columns give the $3754$ coordinates of FP5-1.
\item[FP5-3] An independent reimplementation of the probe evaluator (physical local transport, Gram transport and
symmetric powers rebuilt from the stored data, all coset pairs summed) reproduces all twelve face traces and the $8965$
width-$16$ evaluator coefficients inside the evaluator's error radii on twelve non-scalar probes, and refutes a wrong-convention
control.
\item[FP5-4] Exact orthogonality and fingerprints of the seed-generated repair frames, action and norm bound of the independent symmetric coordinates, rigorous rotation radii and a wrong-frame control.
\item[TT-1] Exact contractions of the three unflipped all-$V$ faces over the
$14{,}432$ allowed support five-multisets check disjoint even/odd column supports,
as do the four stored IK5 arrays. The stored physical kernels annihilate both
field components, as predicted by the column-parity proof in
Proposition~\ref{prop:ico-fivecopy-g34-kernels}.
\item[FO-1] Integer one-body, pair and spin-flip frame contractions for the entire real icosahedral
family; exact symbolic graph, skewness and frame reconstruction identities on $1/32<y<1/12$ (proved analytically from the computational-basis marginals and spin-flip parity in Theorem~\ref{thm:moment-orbit-rigidity}); the affine invariant $H(\psi_y)=155+96y$ and its tangent-square defect $96^2t^2$ in Theorem~\ref{thm:fivecopy-family-equality}.
\item[LH-1--LH-4] Exact stationarity of $\Psiico$ (LH-1); the integer field decomposition $H=(A+\sqrt{555}B)/18340$
of the real $256\times256$ Lagrange Hessian (LH-2); the $22$ local-unitary and phase tangents in its kernel (LH-3);
and the exact symmetry-adapted integer block decomposition into $1\times1$, $2\times2$, and $4\times4$ blocks with
closed-form characteristic polynomials and exact minimum transverse curvature $(350-6\sqrt{2005})/4585>80/4585>78/4585$
(LH-4), for Theorem~\ref{thm:icolocal}.
\item[QB-1--QB-3] The rational remainder margin $285123/14672000000$ (QB-1), the overlap form of the orbit distance (QB-2),
and the diagonal phases of the $32$ nonzero amplitudes (QB-3), for Theorem~\ref{thm:icoquantbasin}.
\item[EO-1] Exhaustive integer switching/complement classification of the $720$ ordinary-site
permutations: six classes of size $120$, giving $42$ labelled LU orbits after the distinguished-site
choice (Corollary~\ref{cor:icosahedral-labelled-orbits}).
\item[MO-1, MO-2] Independent purity-parameter checks of the pair tensors, tetrads, skewness arithmetic, frame overlaps and reconstruction coefficients in Theorem~\ref{thm:moment-orbit-rigidity}.
\item[SF-1, SF-5] The focused sextic covariants, complete scalar-gradient basis controls,
and a coefficient check of the closed zero-focus Gram with determinant $72x(1-2x)$
in Section~\ref{sec:sextic-general}. Completeness and rank throughout the family are analytic.
\item[HI-1--HI-4] Exact copy-invariant counts from \eqref{eq:fqe-invariant-dimension}, used for the free
coordinates in Theorem~\ref{thm:sextic-sos-obstruction}.
\item[LP-0--LP-2, GS-1--GS-3] Exact Rains-LP dual identities for $4\le n\le12$, the exact primal point attaining $L_2=1/8$ at
$n=7$ (the AME weight distribution), the graph-state cut ranks
of Table~\ref{tab:graphs}, the weight distribution of $C_{10}(1,2,4)$ and the Petersen purity
(Theorem~\ref{thm:lptable}, Remark~\ref{rem:ten}).
\item[LP-14, LP-15, GS-14, GS-15, LP-13, QP-13] The same at $n=14,15$ (Theorem~\ref{thm:fourteenfifteen}); at $n=13$, the
exact Rains dual for $25/1408$ (LP-13) and the exact stabilizer count of the $X$-measured fourteen-qubit witness,
with all $14\times3$ one-qubit Pauli measurements compared (QP-13), for Proposition~\ref{prop:thirteeninterval}.
\item[LP-19, GS-19, NG-1--NG-3] The exact Rains dual at $n=19$ (LP-19), the starting graph witness (GS-19), and the
improved graph with its stabilizer enumeration and independent cut-rank count (NG-1--NG-3), for
Theorem~\ref{thm:strictLP1319}.
\item[SW-0--SW-4] The three-copy Schur--Weyl certificates of Theorem~\ref{thm:threecopy-sw}: the per-site algebra
(SW-0); regressions against the Haar moment and random states at $n=4,5$, not used in the proof (SW-1); exact
positive definiteness of the rational multipliers (SW-2) and of every $\omega$-block of the exact residual (SW-3);
and the resulting purity bounds and code exclusions (SW-4). Runtime about $7$ minutes, dominated by $n=25$.
\item[TA-1] Numerical only: the three-copy level compared with the Rains linear programme for $4\le n\le20$, in the
paragraph after Theorem~\ref{thm:threecopy-sw}.
\item[E13L-1--E13L-2] The thirteen-qubit entropy witness of Corollary~\ref{cor:thirteen-entropy-witness}: exact
Gaussian-integer amplitudes and norm by a Walsh transform, the exact Pauli-code block compression of
Lemma~\ref{lem:pauli-code-compression} at all $1716$ balanced cuts with every block eigenvalue enclosed in Arb above
$1/2100^2$ (E13L-1), and the Arb enclosure of $E_6$ (E13L-2). The same script's Hessian checks E13L-0, E13L-3--E13L-5 are not used.
Runtime about $15$ minutes.
\item[GL-1, GL-2] The same at $n=16,17$ (Theorem~\ref{thm:sixteenseventeen}).
\item[E18-1--E18-4] The exact Rains dual and the attaining graph state at $n=18$ (Theorem~\ref{thm:eighteen}).
\item[TY-1--TY-2] The circulant graph states $C_{16}(2,3,4,8)$ and $C_{20}(3,4,5,10)$ of
Corollary~\ref{cor:typeII-sixteen-twenty}: odd degree at every vertex ($\tau_n=1$), the exact $\mathbb F_2$ ranks of all balanced
cuts, and the purities $1/143$ and $131/77792$.
\item[ST-1--ST-6] The rational multipliers and residual signs of \eqref{eq:tradeoff-dual} for each
$n\in\{4,8,10,14,16,20\}$, and the exact purity $L_n+c_n$ of a $\tau_n=1$ graph state at each $n$
(Theorem~\ref{thm:purity-spinflip-tradeoff}).
\item[ST-7, ST-8] The closed-form multipliers and weights of \eqref{eq:odd-tradeoff}, checked against an exact elimination for odd $n=5,\dots,41$, and the
exact evaluation of both sides on Gaussian-integer states at $n=5,7,9$ (Theorem~\ref{thm:odd-spinflip-tradeoff}).
\item[ST-9] The witness $\mathrm{GHZ}_{n-2}\otimes\Phi^+$ for the optimality of $c_n$ over all pure states: its Pauli weights,
$O=1/(2n)$, and $\pi_m$ from the subsystem-purity generating polynomials, with $\pi_m+c_nO=\tfrac12$ exactly for odd
$n=5,\dots,43$ and a negative control.
\item[SDS-1--SDS-3] Exact Pauli selection in spin-flip-doubled states (Theorem~\ref{thm:spinflip-double},
Corollary~\ref{cor:ordinary-triples}).
\item[CC-1] The quartic identity~\eqref{eq:complexsos}, all $226$ coefficients, in formal variables
(proved analytically by Hamming-weight block decomposition and the Hafnian--Gram identity in Theorem~\ref{thm:complexsos}).
\item[CP-4] The algebraic matching-pattern identity in Proposition~\ref{prop:confphase} (proved analytically from $B\mathbf{1}=0$ and $B\overline B=5I_5-J_5$).
\end{description}